\documentclass[12pt]{article}
\usepackage{epsf}
\usepackage{graphicx}
\usepackage{graphics}
\usepackage{cite}
\usepackage{pgf}
\usepackage{epsfig}
\usepackage{booktabs}
\usepackage{pdfpages}
\usepackage{caption}
\usepackage{lipsum}
\usepackage{float}

\usepackage{multirow,booktabs}
\usepackage[all]{pstricks}
\usepackage{pst-plot,pst-math,pst-grad}

\def\ber{\begin{eqnarray}}
\def\eer{ \end{eqnarray}}  
\def\be{\begin{equation}}
\def\ee{\end{equation}}
\def\la{\label} 
\def\lan{\langle} 
\def\ran{\rangle}

 \def\n{\nonumber\\}

\def\bigone{\hbox{1\kern -.23em {\rm l}}}     
\def\ZZ{\hbox{\zfont Z\kern-.4emZ}}

\begin{document}
	
	\noindent
	
	\begin{titlepage}
		
		\begin{center}
			
			{\Large {\bf
				         Quantum entanglement measures from  Hyperscaling violating  geometries with finite radial cut off  at general 
d,  $\theta$ from the emergent global symmetry 
 				 }\\
			}
			
			\vspace*{5mm}
			\vspace*{1mm}
			 Chandrima Paul
	
			\vspace*{5mm}
			
			{ \it 
				Department of Theoretical Physics, The institute of Mathematical Science\\
                                4th Cross Street, C.I.T Campus, Tharamani, Tamilnadu-600013, India
			}
\footnote{Email \, :  \, chandrimap@imsc.res.in, \,plchandrima@gmail.com}

		\end{center}
		
\begin{abstract}
Recently the study of $T{\overline{T}}$ deformed field theory on the boundary of some spacetime regime and its gravity dual with bulk geometry at finite radial cut-off $\rho_c$,  given by deformation coefficient $\mu$,  has drawn great attention.   Since $T{\overline{T}}$ deformation is a solvable irrelevant deformation,    one expects the boundary theory have a smooth global description over the complete parameter-regime of $\mu$ and along all the spacetime directions.    Consequently any measure of the quantum entanglement, for single or disjoint interval of length l on boundary, should must have a global description over the complete parameter-regime of $(l, \mu)$ or on 2D $(l,\rho_c)$ plane for its gravity dual.  Here  we  tried to  find these quantum-measures  through RT prescription from Hyperscaling violating bulk geometry with finite radial cut off  and found  mathematically it is impossible to find such global form of these measures of quantum entanglement because the turning point $\rho_0 (l,\rho_c)$  is never solvable globally either in its exact or in any perturbative form and we have shown the perturbative solutions can be found at most locally which can give the quantum-measures defined locally only over certain regime of $(l, \mu)$ parameter space or 2D  $(l,\rho_c)$ plane in gravity!  So, in our attempt to find the globally defined quantum-measures holographically,  we have further shown that,  the concerned bulk geometry,  on application of RT formalism,  showing an emergent global symmetry structure on these parameter-space  which when combined with the global boundary conditions and other consistency conditions can completely fix the complete global structure  of the turning point for the most general dimension (d) and the Hyperscaling violating factor ($\theta$), which is shown to consists of three types of the symmetry-invariant terms where we found that one of them can be completely determined, shown to be much much more dominating over two others in the regime  $l >> \rho_c $ and  $\rho_c >> l$,  so that the turning point $\rho_0 (l,\rho_c)$  constructed from this type of term can be considered as exact over the regime  $l >> \rho_c$  and  $\rho_c >> l$ and an interpolating expression between the two,   away from the concerned regime where it is also found this interpolating expression is  quite close to the exact one!  Consequently, here we have constructed the gravity dual of three measures of the quantum entanglement, namely HEE,  HMI, EWCS, based on this $\rho_0 (l,\rho_c)$, where we also both from field theory as well as from gravity side argued intutively about the expected properties of these three quantum-measures in the deformed theory and found they are behaving completely accordingly for all $(d, \theta)$  proving the validity of their global expressions.  We have further discussed the impact of this emergent global symmetry on the quantum-measures and also explored its possible spacetime origin although the issue is subjected to a proper  and detailed study.     
\end{abstract}
\end{titlepage}

\newpage
\tableofcontents

\newpage

\section{Introduction}

The purpose of AdS/CFT correspondence is to realize the quantum information of a physical system.  Since for most of the physical systems, in order to understand the quantum theory,  we need to deal with a strongly coupled field theory when in many occasions it may not be feasible to get the nonperturbative information about such system, the theory of gravity, through the principle of holography, is being used as a tool to understand those systems.  One interesting fact behind this holographic correspondence is that,   while in a quantum system, the number of qubits stored over a region of space 
is related to the respective surface area(in Planck units),  in the theory of quantum gravity,  the number of the degrees of freedom scales like area, e.g the entropy of a black hole, given by the area of the horizon. This is indeed a key-oject to extend the holographic correspondence both sides!
\vskip0.5mm
A useful measure of the quantum information of any physical system is given by entanglement of entropy.  This concerns the question that if we decompose any physical system into two parts, say A and its complement $ A^c $, then how the degrees of freedom of these two parts are effectively correlated to each other or better to say what is the exact quantum-entanglement  between these two subsystems?   To understand the concept, we consider QFTs including CFTs. We consider the theory is defined at zero temperature, so that it is at its pure ground state $|\Psi\ran$

The density matrix of the complete system is given by\footnote{We assume no degeneracy of the ground state}

\be
\rho_{\rm tot}  = |\Psi\ran \lan \Psi \ran
\la{densitymatrix}
\ee

Next we decompose the complete system in two subsystem A , B, so that  the complete Hilbert space ${\mathcal{H}}$ is given by 
$ {\mathcal {H}}_{\rm tot} = {\mathcal {H}}_A \otimes {\mathcal {H}}_B $.  So the reduced density matrix of the subsystem A is
\be
\rho_A = {\rm Tr}_B {\rho_{\rm tot}}
\la{reduced density matrix}
\ee
So the entanglement of entropy of the subsystem A is defined as 
\be
S_A =  - {\rm tr}_A \, \rho_A \, {\rm log} \,\rho_A 
\la{ee}
\ee

\begin{figure}[H]
\includegraphics[width=.65\textwidth]{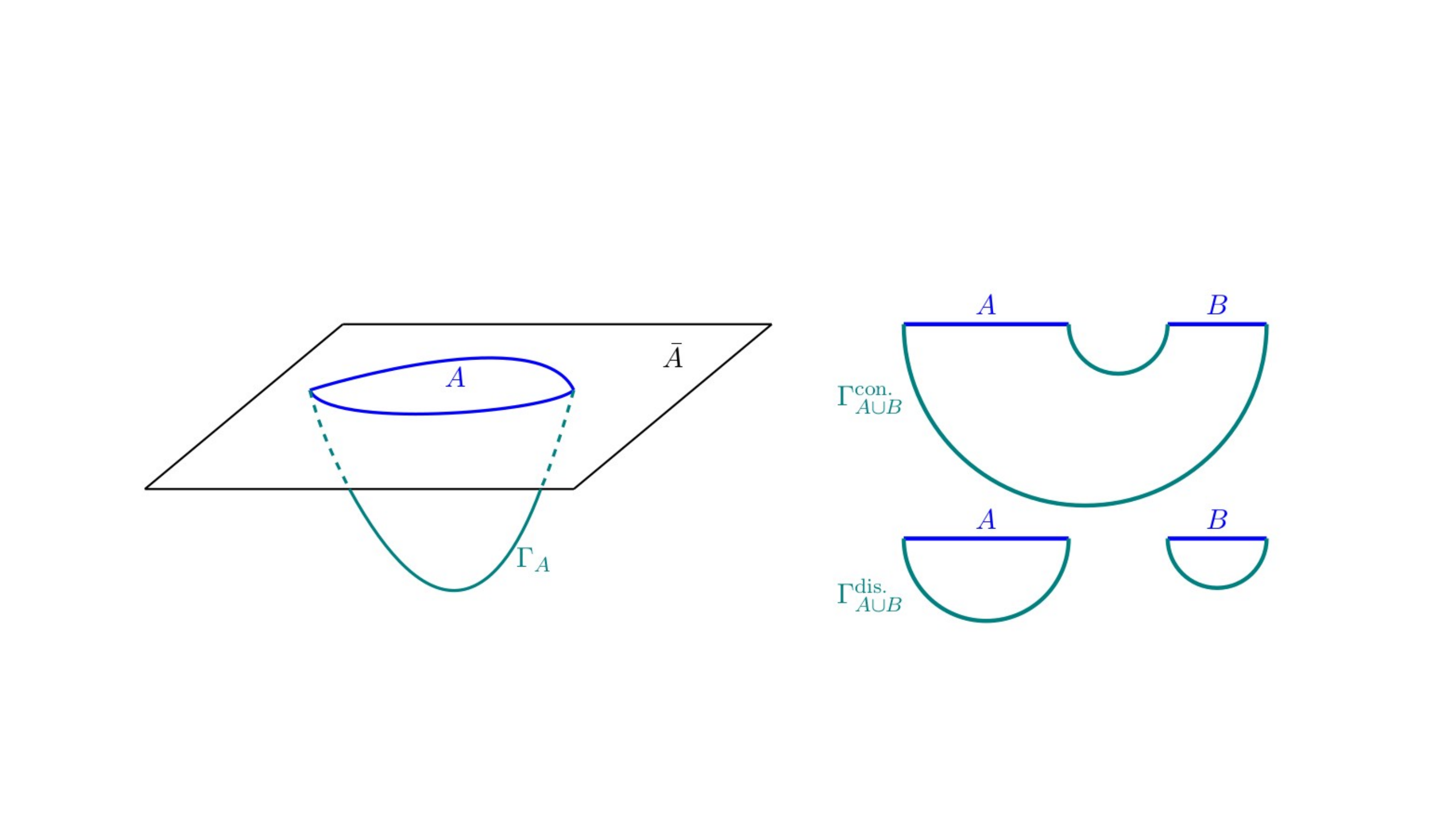}
\caption{  The schematic image of the gravity realization of EE and HMI \,,\, (left) \,:\, The gravity dual of EE,  realized through the entangling surface $\Gamma_A$ \,\,,\,\, (right) \,:\, The gravity dual of MI,  realized through the connected  and disconnected entangling surface  $\Gamma^{\rm con}_{A\cup B}$ and $\Gamma^{\rm dis}_{A\cup B}$ 
}
\label{heehmi}
\end{figure}

This (\ref{ee}) is an effective measure that how closely entangled the wave function $|\Psi\ran$ is !  $S_A$ is always an UV dicergent quantity and strikingly the coefficient of the UV diverging term turns out to be proportional to area of the boundary $\partial A$ and given by
\be
S_A = \gamma \cdot {\frac{{\rm Area} \partial A}{a^{d -1}}} + \, {\rm Subleading} \,\, {\rm terms} \,,
\la{arealaw}
\ee
with $\gamma$ is a system dependent constant and   "a"   is UV cut off,   as first pointed out in \cite{area1, area2}.   In reality, in field theory side, one can evaluate EE for very simple cases like, for the single interval for vacuum of the two-dimensional CFT \cite{eesimplecase}.   For the holographic understanding,    
Ryu-Takayanagi proposed   the holographic dual of EE \cite{rt, rtoriginal},   which is as following\, :\,
\vskip0.5mm
Let for the subregion A, a codimension-2 spacelike minimal hypersurface, whose boundary matches with the boundary of A, i.e $\partial{\Gamma}_A = \partial A$.
Then on the gravity side the EE is defined as

\be
S= {\frac{{\rm Area}^{\rm min} ({\Gamma}_A) }   {4 \pi G_N}} \,,
\la{koutuhal}
\ee

with $G_N$ is Newton constant.   Actually RT formula in (\ref{koutuhal})  is a leading order term  in  $G_N$ expansion in HEE,  which in dual field theory(if it is large N gauuge theory)    corresponds to the order $(N^2)$ term.  A quantum correction to this expression, i.e the order $G_N^0$ $(N^{0})$  was proposed in \cite{quantumcorrection}.    We made a schematic representation of RT prescription in Fig.(\ref{heehmi})
RT proposal has gone through various consistency checks.   It was extended to higher derivative gravity \cite{higherderivative1, higherderivative2, higherderivative3, higherderivative4  }.  Since RT formalism is developed for time indxependent case,   a covariant generalzation was proposed in 
\cite{covariantentropy},  where the authors, based on light-sheet formalism developed by Bousso in  \cite{lightsheet1,lightsheet2,lightsheet3} to describe the entropy bound which is a bound  which describes a limit on  maximum entropy/information can pass through any space time region and further explored in \cite{interestingapplication} being used up to describe the covariant generalization of the entangling surface as proposed in \cite{rt}, \cite{rtoriginal}.

\vskip0.5mm

Once we understood EE and HEE both from the field theory and the gravity side, we are again stuck with two questions!  Firstly it is evident from (\ref{arealaw}) that EE is a UV  divergent quantity, One possible regularaization as can be thought of is lattice regularaization, i.e in more general sense,  replacing the single object by disjoint objects, where the total EE of the complete set of such objects is finite and the divergence appears when they are again brought together to form a single one! Secondly,  the immediate question invokes that while the Ryu Takayanagi-surface, as given in(\ref{koutuhal}), what should be the gravity dual of CFT with disjoint interval?
\vskip0.5mm
The first issue was considered in \cite{Swingle}, where two disjoint  subsystem, A and B are being considered and the concept of mutual information is being introduced.   The mutual information of two system A, B is given by the correlation 
\be
I(A,B) = S(A) + S(B) - S(A \cup B)\,
\la{mutualinformation}
\ee 
with S denotes EE.
 while  EE is UV divergemt, $I(A,B)$ is UV finite and also gives an upper bound to the correlator between two operator  $O_A$ and $ O_B$ defined over the subregion 
A,B respectively,  given by
\be
I(A,B) \ge {\frac{{ \left(\lan O_A O_B \ran - \lan O_A \ran \lan O_B\ran     \right)}^2   }{2{||O_A||}^2   {||O_B||}^2   }}
\la{correlation}
\ee

Clearly,  if we keep increasing the separation between A and B,   at one point they will supposed to be disentantangled,   where thdse correlators (\ref{correlation}) are supposed to vanish and the system is supposed to undergo a phase transition,  which was indeed explored in \cite{headrick}. Next it was shown in \cite{quantumcorrection} that this phase transition is  just a leading order effect! and it does not exist if we reach to $O(N^{0})$ or $O(G_N^0)$ order!

I(A,B) at finite temperature can also shown to satisfy an area law,  like (\ref{arealaw}) and that way shares the property like EE.  Since originally it was introduced as a regularization tool  to EE, clearly once the separation between A and B go away, it will diverge.

 Regarding the gravity dual of mutual information (\ref{mutualinformation}),  one applies RT proposal (\ref{koutuhal}).    When we bring A, B together,   the minimal entangling area  of the combined object  will contribute to I(A,B).    From  Fig.(\ref{heehmi}) it is evident that when the two objects are quite distant,  the minimal area entangling surface is given by the combination of their individual entangling surface, i.e $S(A \cup B)  = S(A) + S(B)$.  However once we bring them more and more closer,  the system-topology will change at one point, when the connected entangling surface( as shown in Fig.(\ref{heehmi}) will have less minimal area compared to the disconnected configuration,  given by  $S(A \cup B)  \ne S(A) + S(B)$,  when the system undergoes a phase  transition, which has to be a  first order phase transition \cite{headrick} because this is coming from the competition from two completely different types of surface,  when at the transition-point $HMI = 0$ and alongwith the role of  their dominance change!  Since for the disconnected configuration $I(A,B) = 0$ as evident from (\ref{mutualinformation}),  so clearly   $I(A,B)$ acts as an effective measure of the quantum entanglement between A and B.  The holographic dual of the mutual information was studied in 
\cite{headrick, Swingle}.   HMI in the context of black hole geometry was studied in  \cite{tonni}  and the complete finite temperature generalization was studied in \cite{Fischler}.

However,  while EE can give quantum entanglement between two subsystems which are in pure state it is not the right measure of the quantum entanglement of the mixed state!    This follows  from our discussion  at  very beginning  that say we start  with a system with a pure state  which when decomposed into two subsystems, A and B with density matrix  $\rho_A \otimes \rho_B $, where the states involved there coming from both pure state and mixed state where the second refers to the states coming from the entanglement of the d.o.f  between A and B.  However from the expression EE  (\ref{ee})  its evident that even those d.o.f of A and B which do no couple with each other and so unentangled,  make nonzero  contribution to EE!
 So  there are many new measures to describe the quantum entanglement of the mixed state started to  appear where the significant work done in this regard 
can be found in \cite{newmeasure1, newmeasure2}.   There were many holographic measures of the quantum informations were on the way like the relative entropy
\cite{relativeentropy},  quantum information metric \cite{quantuminformation}  and computational complexity  \cite{computationalcomplexity1,  computationalcomplexity2}.
\vskip0.5mm
Now given a physical system is in a mixed state, one cam always introduce a  fictitious auxiliary system  of states so that combined system is pure and this process is called the purification of the original mixed state.  A good measure of the quantum entanglement of the mixed state is the  entanglement of purification (EoP) 
which is defined as follows:
\vskip0.5mm
Let $\rho_{AB}$ is the density matrix on a biparite system given by Hilbert space ${\mathcal{H}}_A\otimes {\mathcal{H}}_B $   with A, B are in mixed state and let $A^\prime$,   $B^\prime$ are purifier to A,B.    If  $|\psi\ran \in {\mathcal{H}}_{A A^\prime} \otimes {\mathcal{H}}_{B B^\prime}$ be a purification of $\rho_{AB}$
Since there are many possible ways of purifying $\rho_{AB}$,  so EoP of $\rho$ is given by

\be
E_p (\rho) = \min_{\psi, A^\prime} S_{A A^\prime} \,
\la{eopdefinition}
\ee

where we minimize over $\psi$ i.e choosing a least possible path to purify  $\rho_{AB}$ and also minimizing over ${A^\prime}$ , i.e  minimizing over all possible ways of  partitioning the purification ${A^\prime} {B^\prime}$  and $ S_{A A^\prime}$ is the Von Neumann entropy of the reduced density matrix  ${\rm Tr}_{B B^\prime} |\psi\ran\lan \psi|$.  

\vskip0.5mm

\begin{figure}[H]
\includegraphics[width=.65\textwidth]{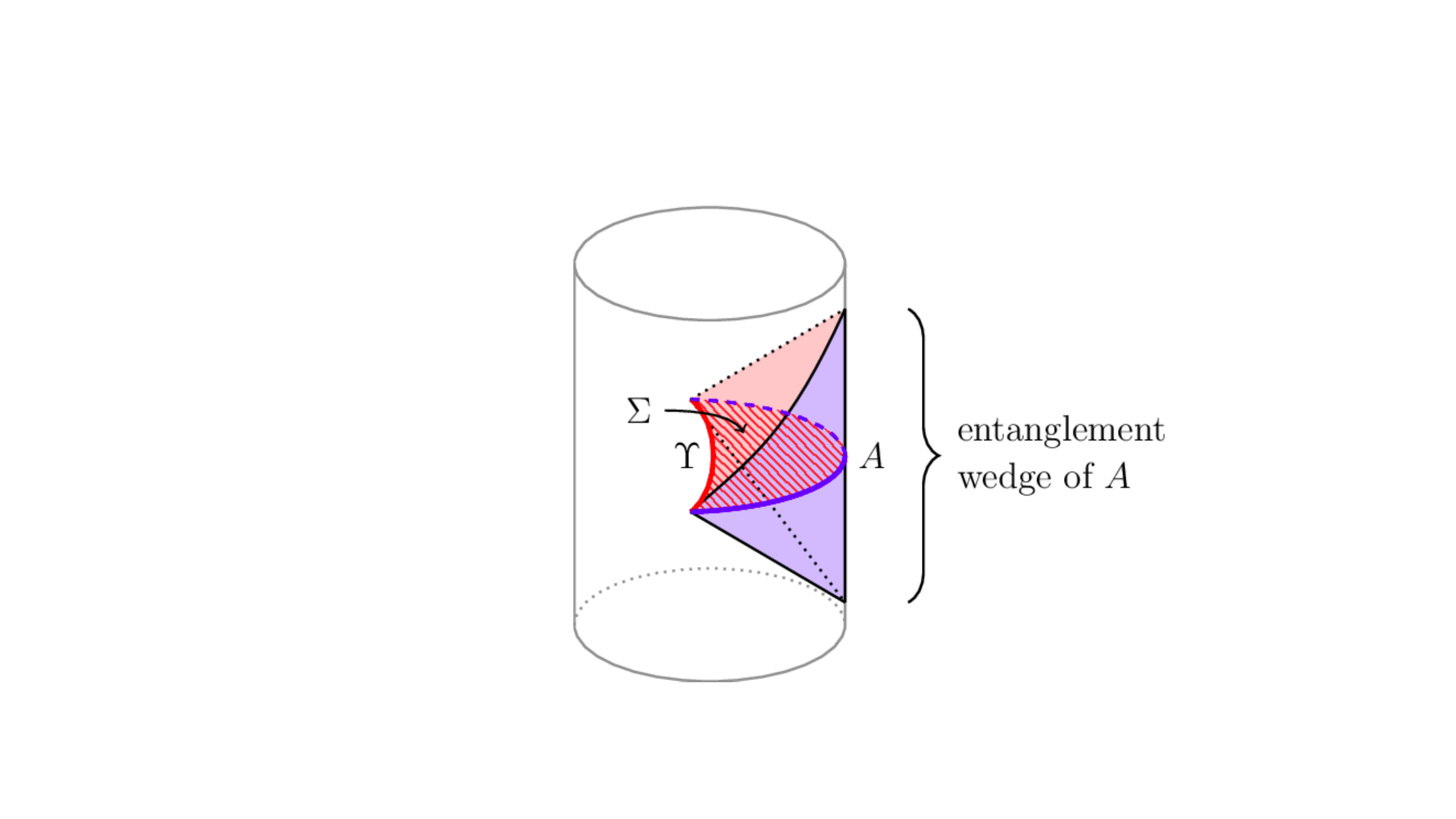}
\caption{ A schematic representation of the entanglement wedge where A\,:\, Boundary subregion, \, $\gamma$ \, : \, Entangling surface,\, $\Sigma \equiv R_A$ as we discussed.   The entanglement wedge of A is given by the bulk domain of dependence of $\Sigma$
}
\label{entanglementwedge}
\end{figure}

To understand the holographic dual of EoP,  first we  need to understand what is the holographic dual of the density matrix.   A clear answer to that came from
\cite{gravdualdensity1, gravdualdensity2, gravdualdensity3} when it was found that the holographic dual of the density matrix is given by entanglement wedge.   
This is defined as if we consider a boundary subregion A, in time-independent or in covariantly generalized case and the entangling surface is given by $\Gamma_A $,  then there always exist a spacelike region $R_A$,  shown to be the part of bulk Cauchy slice \cite{gravdualdensity3},  so that 
$\partial R_A = A \cup \Gamma_A $.   The entanglement wedge of the subregion A,  denoted by  $( E_W (A))$ is the bulk domain of dependence of $R_A$,  i.e $ E_W (A)) = D(R_A)$.  A schematic representation of entanglement wedge is given in Fig.(\ref{entanglementwedge})

\begin{figure}[H]
\includegraphics[width=.65\textwidth]{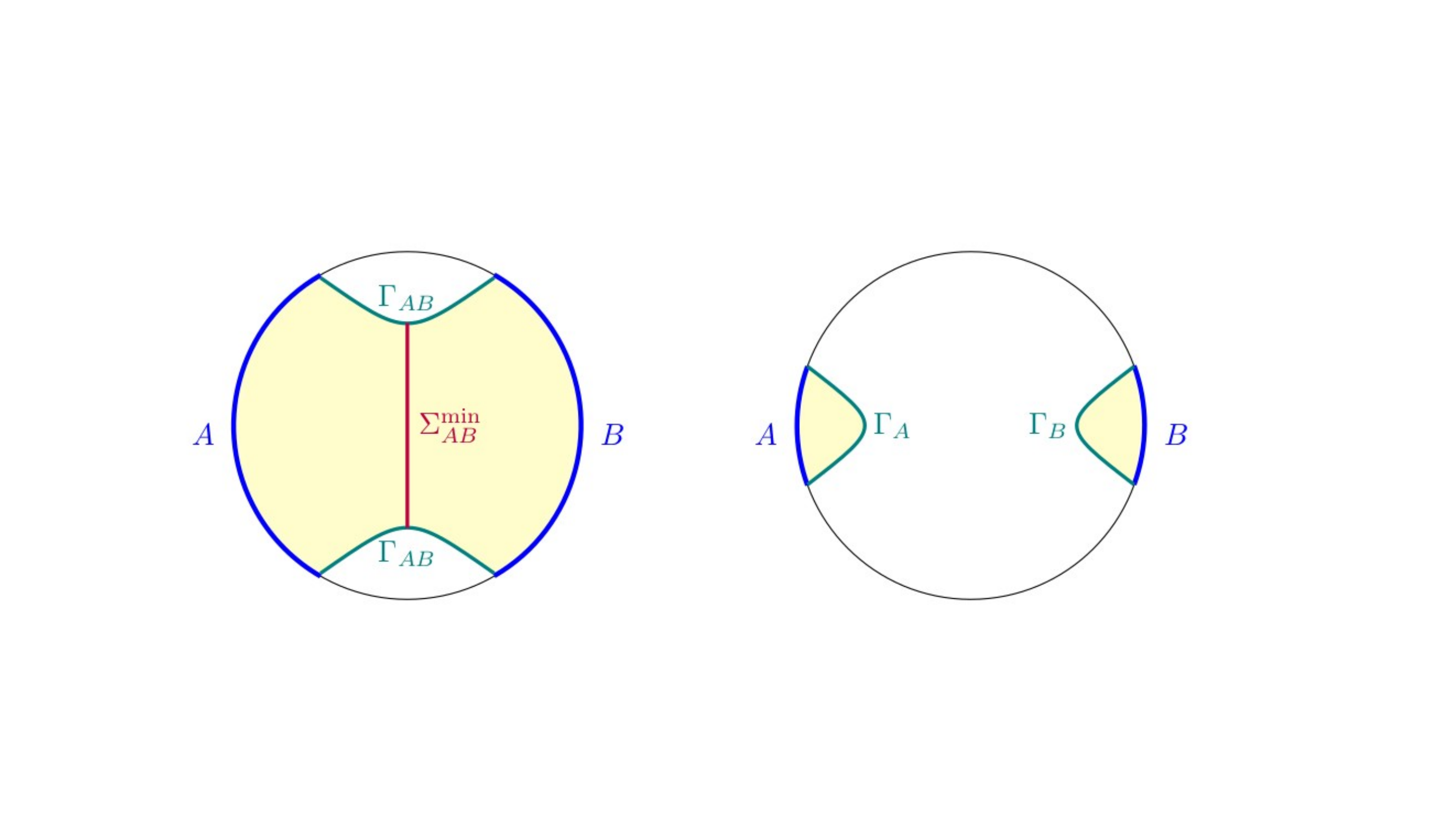}
\caption{ A schematic representation of the entanglement wedge cross section for connected(left) and disconnected(right) phase where for the disconnected phase it vanishes
}
\label{ewcs}
\end{figure}

The holographic dual of EoP was proposed in \cite{taka, holoewcs}.   This can be understood as follows:
\vskip0.5mm
Let us find a combined system of two boundary subregion A, B.  We would like to evaluate their entanglement of purification which we denote by   $E_{\rm ph} (A,B)$.  
Let $\Gamma_{AB}$ denotes the entangling surface which for connected phase the two entangling surfaces starts from the boundary of one subregion and end on the boundary of the other and   for the disconnected phase  $\Gamma_{AB}$ gives two entrangling surface on A,B separately.  as denoted in Fig.(\ref{ewcs}).   From 
Fig.(\ref{ewcs})  is easy to visualize that that the region bounded by A,B, $\Gamma_{AB}$ gives the entanglement wedge.  The holographic dual of EoP is given by the area of the minimum-area surface $\Sigma_{AB}^{\rm min}$, starts from the one entangling surface and ends on the other, basically consists of all the geodesics connecting the two,  which divides the entanglement wedge in two parts.  So we have, 
\be
E_{\rm ph} (A,B) =  {\frac{\rm Area \left(\Sigma_{AB}^{\rm min}  \right)}{4G_N}}
\la{wedge}
\ee

Clearly, at the disconnected phase $E_{\rm ph} (A,B)$ vanishes!

  So exactly at the point   where mutual information is taking a first order phase transition 
$E_{\rm ph} (A,B)$  also undergoes a discontinuous phase transition there,  i.e will drop from a finite value to zero!

So far we have discussed about the measures of quantum entanglement of  pure and mixed states of the  field theory on the boundary of a manifold and its gravity dual.  A lots of recent interest have grown in the direction on $T{\overline{T}}$ deformed field theory and its holographic dual.  The $T{\overline{T}}$ deformation  to the 2D CFT  was originally introduced by Zamolochikov \cite{zamolodchikov}, with

\be
T{\overline{T}}(x,y)  =  \left( T_{ij}(x) T^{ij}(y) - {T}^i_i(x)  \right) \, ,
\la{ttbar11112222}
\ee
 where T is the energy momentum tensor so that we have conservation law  $\nabla_a \lan T^{ab} \ran = 0$.

$T{\overline{T}}(x,y)$ is a bilocal operator and posses a OPE 

\be
T{\overline{T}} (x,y) = \lim_{x \rm \to y} \left( T_{ij}(x) T^{ij}(y) - {T}^i_i(x)  {T}^i_i(y)  \right) + \displaystyle\sum_\alpha A_\alpha (x - y) \nabla_y O_\alpha (x) \,
\la{ttbarope}
\ee
with $O_\alpha$ is a local operator $A_\alpha (x - y)$, diverges as $x \rightarrow y$, 

The v.e.v of the energy momentum tensor undergoes a flow equation \cite{entanglettbar1}

\be
\lan T^a_a \ran = - {\frac{c}{24 \pi}} R - {\frac{\mu}{4}}\left( T_{ij}(x) T^{ij}(y) - {T}^i_i(x)  \right) \, ,
\la{ttbarflow}
\ee

with c is the central charge.    It is an irrevalent deformation!  	Now $T{\overline{T}}$ deformed theory has come to its focal point when the author of \cite{movingcft}  found the holographic dual of the $T{\overline{T}}$ deformed CFT on the   2D boundary, is given by a 3D bulk geometry with a  finite radial cut-off $r_c$.  So writing the   $T{\overline{T}}$ deformed CFT  
as $S = S_{\rm CFT} + \mu\int d^2 x T{\overline{T}} $,  the exact relation between $\mu$ and $r_c$ is given by 
\be
\mu = {\frac{16\pi G}{r_c^2}}
\la{murcrelation}
\ee
A higher dimensional generalization of the holographic dual of $T{\overline{T}}$ deformed theory was explored in \cite{marika}. The next question came into the picture how will these holographic construction of the measure of quantum entanglement, as we discuscussed so far, will be modified since now the boundary field theory lies on the cut off surface, defined within the bulk geometry at $r = r_c$?  It was shown in \cite{entanglettbar1} that as long as we consider the time-independent case, the RT proposal will be followed as usual.    Following that,   the construction and the aspect of EE in holographic dual theory  of of $T{\overline{T}}$ deformed  field theory
were studied in couple of articles including \cite{alisuggested1, alisuggested2,  alisuggested3 }.  It was further shown in \cite{dualttbarcovariant} that above view of anchoring the entangling surfac on the boundary cut off surface  cannot be generalized for the covariant case \cite{covariantentropy} because the HRT surface is violating strong subadditivity and consequently an alternative prescription was proposed therein!  Recently a lots of effort has been made to understand the measures of quantum entanglement from the   $T{\overline{T}}$  deformed field theory which includes \cite{ alisuggested1, alisuggested2,alisuggested3, holographyttbar1,  holographyttbar2, holographyttbar3,  holographyttbar4, holographyttbar5 } 
 
\vskip0.5mm
Next we come to the point,  since we need to find the quantum entanglement properties of some real interesting physical system, we must need to find an appropriate dual geometry.  
Recently there is lots of effort to understand condensed matter systems from gravity through holographic principle where the particular system of interest  are those which does not obey Fermi liquid behaviour and shows strange anomalous scaling property!  These systems undergoing  the  quantum phase transition where at the critical point of such  transition the system shows spacetime scaling symmetry which through holographic principle realized in gravity through some dual bulk theory with the metric obeys such symmetry!  The involved symmetries are mostly characterized by anisotrpic scaling symmetry in space and time direction!  Indeed there is a more general class of symmetries obeyed by more generalized geometries which has wide application in CMT system,  called Hyperscaling violalating  geometries.  The general form of the metric and its consequent symmetry structure is described by:

\ber
ds^2 &=& L^2 {\left({\frac{r}{R}} \right)}^{-{\frac{2\theta}{d}}}\left( - {\frac{dt^2}{r^{2z}}} + {\frac{d{{\vec{x}}_d^2} + dr^2}{r^2}} \right)
\la{hyperscalingvkolating}
\eer

\be
\left\lbrace t, {\vec{x}}. r \right\rbrace \rightarrow \left\lbrace  \lambda^z t, \lambda{\vec{x}}, \lambda r\right\rbrace \quad;\quad ds^2 \rightarrow \lambda^{\frac{2\theta}{d}}ds^2
\la{hyperscalingsymmetry}
\ee

When $z \ne 1$ it gives anisotropic scaling symmetry in time direction from the space direction.  $\theta$ is known to be Hyperscaling violating factor.  Among many interesting CMT application of such geometry, one is, in \cite{karch}   it was shown this geometry gives right dual description of the strange metallic phase of a high temperature superconductor!  The  entanglement of entropy, without any boundary $T{\overline{T}}$ deformation, was hologhraphically evaluated  for such geometris for the zero and finite temperature \cite{lifshitz, chargedbrane} 

\vskip0.5mm

In this work we will consider the Hyperscaling violating geometries with finite radial cut off ($\rho_c$)   and try to obtain some holographic measures of the quantum entanglement
namely  holographic entanglement of entropy, holographic mutual information and the entanglement wedge cross section for the boundary physical system for boundary subregion in the form of the strip of length l where the boundary system  has smooth behaviour over complete parameter regime $(\mu,l)$ ($\mu$ is the $T{\overline{T}}$ deformation coefficient  related to the cut off through (\ref{murcrelation}  ))  and  consequently try to obtain the expression of these quantum measures from this Hyperscaling violating  geometry,  which are dual to that system,  where we must need to find the global expression of these holographic quantum measures, i.e they should must be be defined globally over 2D $(l,\rho_c)$ plane since in the boundary system they are globally defined over $(\mu, l)$ parameter regime!  The biggest problem to obtain such global description of these quantum measures through RT prescription is that the turning point $\rho_0(l,\rho_c)$ can never be solvable globally either in exact or in perturbative form and we are going to show that these  perturbative solutions   can be defined at most locally.  So in this article we will find them in two alternative ways,  first from the basic speculative approach and next from the consideration of the global symmetry, which we found to be  emerging from the application of RT formalism to such geometries!
\vskip0.5mm
Indeed in one very recent article \cite{aspect1} the authors constructed these quantum measures holographically which most unfortunately have some limitations including that the quantum-measures constructed therein are defined locally over certain regime of $(l. \rho_c)$ plane,  through perturbative expression of the turning point,  when indeed in section 2.2 we are going to discuss the limitations of these perturbative expressions.   Consequently the quantum measures constructed therein while these can well describe  the quantum entanglement of some possible physical system,  which when taken through the variation over the parameter regime $\mu, l$, they do not have the smooth flow from one parameter regime to the other but cannot describe a physical system which can be smoothly defined over complete parameter regime of $(\mu, l)$ or in other words give the global description of the quantum-measures over complete $(\mu, l)$ parameter-regime or in gravity, throughout 2D $(l,\rho_c)$ plane, for which one needs a different prescription, which is the purpose of this present work.

\vskip0.5mm
It is also intersting to see that while the methods used in \cite{aspect1} to evaluate the measures of the quantum entanglement of those possible physical systems on the boundary ,  which have some discontinuity within complete parameter regime of $(\mu, l)$  and that way cannot take a smooth transition from one regime to the other,  our metho, as we are going to develop here, works exclusively for thoe boundary systems which have a smooth decription over complete $(\mu, l)$ parameter regigime,  can take a smooth transition from one regime to any other,  where our method cannot describe those systems which have some possible discontinuity,  somewhere in  the parameter regime of $(\mu, l)$  and from that perspective,  the two descriptions are in some sense, compensating to each other!

The organization of this article is as follows:
\vskip0.5mm
 
In section 2 we will  evaluate HEE for Hyperscaling violating geometry with finite radial cutoff and present a speculative approach to find the global expression of the turning point $\rho_0(l,\rho_c)$.    The section 2.1 concerns the explicit evaluation for zero temperature while in section 2.2 we will discuss the problem with the perturbative solutions i.e why they cannot be defined globally over the complete $(l, \rho_c)$ plane.   We will also present there our speculative approach to find the global expression of the turning point $\rho_0 (l, \rho_c)$ where all the explicit calculations and the consequent plots we have presented in Appendix A. 
In section 3 we will explore   the emergent global symmetry of the theory  and show how this symmetry-aspect can give the global  expression for the turning point $\rho_0 (l, \rho_c)$ which will shown to be exact for $l >> \rho_c$ and $\rho_c >> l$ and for rest of the regime it is an interpolating expression between these two  where our consistency checking plots will show that these interpolating expressions are quite close to the exact one.  In section 3.1 we will find this emergent global symmetry and show that it will fix the complete structure of the global solution of the turning point .  In section 3.2 we will explore the most generalized term as predicted by this symmetry structure and find the most dominant solution in $l>> \rho_c$ and $\rho_c >> l$ regime.  In section 3.3 we will present all the consistency checking plots for this global solution for $\rho_0(l,\rho_c)$.   In section 3.4 we will present the plots showing behaviour of $\rho_0(l,\rho_c)$.  In section 4, based on the field theory scenario and the gravity scenario we will present our intutive argument for the expected behaviour  of the holographic entanglement of entropy  in the presence of $T{\overline{T}}$ deformation or the finite radial cut off,  and in the consequent subsections therein,  we will poresent different 3D and 2D plots to show that these expected properties do hold over the complete parameter regime of 
 $d - \theta$.  In section 5, 6 we will construct the most simplified expression of HEE for $l>> \rho_c$ and $\rho_c >> l$ regime and through    different 3D plots we will show that the indeed merge with the exact one in these respective regime for all possible $d - \theta$ with $d - \theta \ne 1$ since that is already in its simplest form.   In section 7 and 8  again based on field theory and gravity scenario    we will give some intutive arguments for the expected behaviour of     holographic mutual information and entanglement wedge cross section  
in the presence of $T{\overline{T}}$ deformation or the finite radial cut off, respectively   and in the consequent subsections therein,  we will poresent different 3D and 2D plots to show that these expected properties do hold over the complete parameter regime of $ d - \theta$.
Finally in section 9 we will come to conclusion.   In the appendix A we will present all the explicit calculations and findings of the global solution of the turning point based on our speculative approach as presented in section 2.2 and show the consistency-checking-plots.

\section{Evaluation of holographic entanglement of entropy in hyperscaling violating Lifshiftz background for zero temperature case for finite cut off}

Here in this section, we will apply Ryu-Takayanagi formalism \cite{rtoriginal, rt }, to evaluate the holographic entanglement of entropy, with entangling region as a strip, for a  bulk hyperscaling violating geometry with finite radial cut off for most general dimension d and the most general Hyperscaling violating factor $\theta$ when one is imposed with the condition $\theta \le d$.  

\subsection{The holographic entanglement of entropy}  

We start with the Hyperscaling violating Lifshitz geometry is given by \cite{lifshitz} described in d+2 dimension

\be
ds_{d+2}^2 =  r^{- \frac{2\theta}{d}}\left( - r^{2z} dt^2   + {\frac{dr^2}{r^2}} + r^2 \displaystyle\sum_{i=1}^{d} dx_i^2 \right)
\la{finalmetric}
\ee

with $\theta$, the Hyperscaling violating factor.

As we mentioned in the introduction,    in \cite{movingcft},  $T {\overline{T}}$ deformation \cite{zamolodchikov} was originally  introduced in the context of AdS/CFT     for two dimensional boundary CFT, and then for its generalization to the higher dimensional boundary field theory  was explored in \cite{marika}.   It was discussed  in \cite{movingcft,marika}  that the holographic dual of the boundary theory with $T {\overline{T}}$ deformation with general boundary dimension,    is given by a bulk geometry with finite radial cut off towards asymptotics. at $r = r_c$, where $r_c$ is related to the  $T {\overline{T}}$ deformation parameter $\mu$ by (\ref{murcrelation}).   So we consider the theory where the bulk is described by (\ref{finalmetric}) which can be dual to some appropriate undeformed   field theory on boundary  and  then under application of $T {\overline{T}}$ deformation to the same,  left us with a bulk dual given by 
(\ref{finalmetric}) with a finite radial cut off given by $r =  r_c $.

So the metric on the boundary is given by
\be
ds_{d+2}^2 = {r^{- \frac{2\theta}{d}}_c}\left(  - {r^{2z}_c} dt^2    + r^2_c  \displaystyle\sum_{i = 1}^d dx_i^2\right)
\la{boundarymetric}
\ee

In order  to evaluate holographic entanglement of entropy,

we specify a boundary subregion, which must be a subregion on a spatial surface on the boundary at $r=r_c$, must be given by $t=t_o$, which we consider as following 
\be  
-l/2 \ge x_d \ge l/2,   \quad\quad   0 \ge x_i \ge L,  \quad\quad   ({\rm for} \quad i = 1,  ....d-1) \quad\quad   r = r_c,  \quad\quad     t= t_o 
\la{consideredsubregion}
\ee

So that geometric distance between the two endpoints in $x_d$ directiion
\be
d_{\rm proper} ={(r_c)}^{1 - {\frac{\theta}{d}} } \int_{- {\frac{l}{2}}}^{ {\frac{l}{2}}} d x_d = {(r_c)}^{1 - {\frac{\theta}{d}} } l
\la{proper}
\ee

Now according to the Ryu-Takayanagi proposal the holographic entanglement of entropy is proportional to the surface area of the minimal surface, whose boundary coincides with the boundary of the above subregion.  The surface will be given by $ x_d = x_d (r)$,  where one can write the expression of the induced metric on this surface

\be
ds_{\rm ind}^2 = r^{2 - \frac{2\theta}{d}} \left\lbrack  \left( x_d^{\prime 2}  +{\frac{1}{r^4} }\right)dr^2  + \displaystyle\sum_{i =1}^{d-1} dx_i^2  \right\rbrack
\la{inducedmetric}
\ee
So the area of the hypersurface over the subregion (\ref{consideredsubregion}) is given by

\be
A = L^{d-1} \int dr r^{d  -\theta}  {\sqrt{x_d^{\prime2} + {\frac{1}{r^4}}  }}\, ,
\la{areaekn}
\ee

with the requirement that the boundary of this surface,  is given by the  boundary of the subregion (\ref{consideredsubregion}) and since the surface is described by the profile $x_d  =  x_d (r)$.   In order to minimize the area (\ref{areaekn}),  we treat the area-integral as the action,  so that its minimum value is given by the equation of motion, which implies the canonical conjugate momenta of $x_d$ is a constant of motion, since the integrand in (\ref{areaekn})  does not explicitly depends on $x_d$.  Consequently we have
\be
r^{d  -\theta } {\frac{x_d^\prime}{\sqrt{x_d^{\prime 2} + {\frac{1}{r^4}} }}} = r^{d  -\theta }_0 \, ,
\la{turningpoint}
\ee
where $r_0$ is a constant,  actually considered to be the turning point which is evident from the fact
\be
{x_d^\prime} = {\frac{1}{r^2}}{\frac{r^{d - \theta }_0}{\sqrt{ r^{2(d  - \theta )} -  r^{2(d  - \theta )}_0 }}}\, ,
\la{newmomentumnow}
\ee
so that
$ x_d^{\prime}(r_0) $ diverge.
Clearly $x_d^\prime$ will not be imaginary and that way we will have solution for  minimal surface for $r \ge r_0$.  Since  $T \overline{T}$ deformation is equivalent to putting a boundary cut off at $r = r_c$.  So we have 

\be
{\frac{l}{2}} =  \int_{x_d (r_0)}^{x_d(r_c)} dx_d = \int_{r= r_0}^{r = r_c} dr {x_d^\prime}\, ,
\la{integralcharm}
\ee

First we note that (\ref{integralcharm}), this is giving one global boundary condition

\be
{\rm for} \,\,\,l = 0 \,,\, r_0 = r_c
\la{bc2}
\ee

Next we reexpress the integral in (\ref{integralcharm}) as
\ber
{\frac{l}{2}} &=&  \int_{r_0}^{r = r_c} dr {x_d^\prime}\n
                     &=& - \int_{\rho_c}^{\rho_0} d\rho  {\frac{dx_d}{d\rho}}  \n
                     &=& \int_{\rho_c}^{\rho_0} d\rho  {\frac{\rho^{\theta   -d }_0}{\sqrt{ \rho^{2( \theta  - d )} -  \rho^{2( \theta  - d )}_0 }}}\n
                     &=& \rho_0 \int_{\frac{\rho_c}{\rho_0}}^{1} d\xi {\frac{\xi^{ d - \theta }}{\sqrt{ 1 -  \xi^{2( d - \theta   )} }}}\n
                      &=& -  {\frac{ \rho_0  \,\, {{}_2 F_1}   \left\lbrack {\frac{1}{2}}, {\frac{1}{2}}(1 +{ \frac{1}{d - \theta }}), {\frac{1}{2}}(3 +{ \frac{1}{d - \theta }}) , 1 \right \rbrack }{  \theta - d  - 1   }}\n
                      &+& \left({\frac{\rho_c}{\rho_0}}\right)^{d - \theta +1} {\frac{ \rho_0  \,\, {{}_2 F_1}   \left\lbrack {\frac{1}{2}}, {\frac{1}{2}}(1 +{ \frac{1}{d - \theta}}), {\frac{1}{2}}(3 +{ \frac{1}{d - \theta}}) , \left({\frac{\rho_c}{\rho_0}}\right)^{2(d - \theta)}  \right \rbrack }{  \theta - d  - 1   }} \, ,
                 \la{tanmoy12}
                 \eer
 where in the second step of the above(\ref{tanmoy12})   we considered a coordinate transformation $\rho = {\frac{1}{r}}$   and also in the third step we use (\ref{newmomentumnow}) where we implement this coordinate transformation to the both side of this equation.   Note for this coordinate transformation,  which also gives $\rho_c  =  {\frac{1}{r_c}}$,  the relation  (\ref{murcrelation}) can be rewritten as

\be
\mu = {16\pi G}{\rho_c^2}
\la{murcrelationew}
\ee

Also the hypergeometric function, used in (\ref{tanmoy12})    for $|z| \le 1$,   $   {{}_2 F_1}(a,b,c,z)  $ is defined as

\be
 {{}_2 F_1} (a,b,c,z)  = \displaystyle\sum_{k = 0}^\infty {\frac{(a)_k (b)_k z^k}{c_k  k!}}\, ,
 \la{hypergeometric}
 \ee
 
 where
 \be
 (a)_n = 1, \quad  \quad {\rm for} \quad   n=0 \quad  \quad; \quad  \quad (a)_n = a(a+1)(a+2).......(a +n -1) ,\quad  \quad {\rm for} \quad n > 0\,  ,
 \la{hypergeometriccondition}
 \ee

Finally we are left with the governing equation of the turning point $\rho_0 (l,\rho_c)$as

\ber
 {\frac{l}{2 \rho_0 \left( l, \rho_c \right)  }} &=&  {\frac{   \,\, {{}_2 F_1}   \left\lbrack {\frac{1}{2}}, {\frac{1}{2}}(1 +{ \frac{1}{d - \theta }}), {\frac{1}{2}}(3 +{ \frac{1}{d - \theta }}) , 1 \right \rbrack }{ d+1-  \theta   }}\n
                      &-& \left({\frac{\rho_c}{\rho_0 \left( l, \rho_c \right) }}\right)^{d - \theta +1} {\frac{   \,\, {{}_2 F_1}   \left\lbrack {\frac{1}{2}}, {\frac{1}{2}}(1 +{ \frac{1}{d - \theta}}), {\frac{1}{2}}(3 +{ \frac{1}{d - \theta}}) , \left({\frac{\rho_c}{\rho_0}}\right)^{2(d - \theta)}  \right \rbrack }{ d + 1 - \theta   }} \n
                 \la{tanmoy}
                 \eer

 Finally, in order to evaluate the holographic entanglement of entropy, we need to evalute the minimal area of hypersurface, as given in (\ref{areaekn}), with the expression (\ref{newmomentumnow}) to obtain

   \ber
   A   &=& L^{d-1} \int_{r_0}^{r_c} dr r^{d  -\theta}  {\sqrt{x_d^{\prime2} + {\frac{1}{r^4}}  }}\n
        &=&  L^{d-1} \int_{\rho_c}^{\rho_0} d\rho    {\frac{1}{\rho^2}}     \rho^{\theta - d }  {\sqrt{ \rho^4    \left( {\frac{dx}{d\rho}}\right)^2 + \rho^4  }} \n
        &=&    L^{d-1} \int_{\rho_c}^{\rho_0} d\rho        \rho^{\theta - d }  {\sqrt{ \left(  {\frac{dx}{d\rho}}\right)^2 + 1  } }\n
         &=&    L^{d-1} \int_{\rho_c}^{\rho_0} d\rho        \rho^{\theta - d }  {\sqrt{ \left(  {\frac{dx}{d\rho}}\right)^2 + 1  } }\n          
            &=&    L^{d-1} \int_{\rho_c}^{\rho_0} d\rho        \rho^{\theta - d }  {\sqrt{ \left(  {\frac{\rho^{\theta   -d }_0}{\sqrt{ \rho^{2( \theta  - d )} -  \rho^{2( \theta  - d )}_0 }}}    \right)^2 + 1  } }\n
              &=&    L^{d-1} \int_{\rho_c}^{\rho_0} d\rho  \rho^{\theta - d }  {\frac{\rho^{\theta   -d }}{\sqrt{ \rho^{2( \theta  - d )} -  \rho^{2( \theta  - d )}_0 }}} \n 
              &=&    L^{d-1} (\rho_0)^{\theta - d +1} \int_{{\frac{\rho_c}{\rho_0}}}^{1} d\xi  {\frac{  \xi^{\theta - d }  }{\sqrt{ 1-  \xi^{2( d  - \theta )} }}}\n
              &=&  {\frac{      L^{d-1} (\rho_0)^{\theta - d +1} \,\, {{}_2 F_1}   \left\lbrack {\frac{1}{2}}, {\frac{1}{2}}( - 1 +{ \frac{1}{d - \theta }}), {\frac{1}{2}}(1 +{ \frac{1}{d - \theta }}) , 1 \right \rbrack }{  \theta + 1- d     }}\n
                      &- & \left({\frac{\rho_c}{\rho_0}}\right)^{ \theta +1- d} {\frac{     L^{d-1} (\rho_0)^{\theta - d +1}  \,\, {{}_2 F_1}   \left\lbrack {\frac{1}{2}}, {\frac{1}{2}}( - 1 +{ \frac{1}{d - \theta}}), {\frac{1}{2}}(1 +{ \frac{1}{d - \theta}}) ,
                       \left({\frac{\rho_c}{\rho_0}}\right)^{2(d - \theta)}  \right \rbrack }{  \theta - d  + 1   }} \n 
\la{areaexpression}
  \eer                                   
                      
Finally one can write the expression of holographic entanglement of entropy as
\ber
      S &=&   {\frac{      L^{d-1} (\rho_0)^{\theta - d +1} \,\, {{}_2 F_1}   \left\lbrack {\frac{1}{2}}, {\frac{1}{2}}( - 1 +{ \frac{1}{d - \theta }}), {\frac{1}{2}}(1 +{ \frac{1}{d - \theta }}) , 1 \right \rbrack }{  4 G_N (\theta + 1- d )    }}\n
                      &- & \left({\frac{\rho_c}{\rho_0}}\right)^{ \theta +1- d} {\frac{     L^{d-1} (\rho_0)^{\theta - d +1}  \,\, {{}_2 F_1}   \left\lbrack {\frac{1}{2}}, {\frac{1}{2}}( - 1 +{ \frac{1}{d - \theta }}), {\frac{1}{2}}(1 +{ \frac{1}{d - \theta}}) , \left({\frac{\rho_c}{\rho_0}}\right)^{2(d - \theta)}  \right \rbrack }{  4 G_N( \theta - d  + 1)   }}
                      \la{entropy}
                      \eer

One can evaluate the entanglement of entropy (\ref{entropy}), provided one has an expression of the turning point  $\rho_0(l,\rho_c)$ ,  which is a solution of  (\ref{tanmoy}), as we proceed with in the next section.

\subsection{ A speculative study of the turning point.   the regime of its exact solvability in 2D \,\,$(l,\rho_c )$  plane  and the problem with perturbative solution \, : \,}

Here we start with the governing equation (\ref{tanmoy}) for $\rho_0(l,\rho_c)$.   Since this is never solvable  analytically,   globally,  so in this section   we will explore various possibilities.    First we note,  (\ref{tanmoy})  gives the boundary condition at its two boundaries $(l, \rho_c = 0)$ and $(l = 0 , \rho_c)$, given by

\be
\rho_0( l, \rho_c)|_{l = 0}  =\rho_c\quad; \quad \rho_0(l,\rho_c)|_{\rho_c = 0} ={\frac{A_{10} l}{2}}\quad; \quad \Rightarrow\quad  {\frac{\rho_c}{\rho_0}} \le 1\quad;\quad {\frac{A_{10} l}{2\rho_0}} \le 1 \, ,
\la{basicrelation}
\ee

where $A_{10}$ is the proportionality constant appears in the case without $T \overline{T}$  deformation, $\rho_0(l) = {\frac{A_{10}{l}}{2}}$, as obttained in \cite{chargedbrane},  with $A_{10}$ is given by

\ber
A_{10}     &=& {\frac{\Gamma\left\lbrack{\frac{1}{2(d - \theta)}}  \right\rbrack}{{\sqrt{\pi}}\Gamma\left\lbrack {\frac{1}{2}}\left(1 + {\frac{1}{d- \theta}}\right) \right\rbrack}}
\la{proportionalityconstantgen}
\eer
Also one of this boundary condition,  we already explored in (\ref{bc2}) which arised from the integral expression of (\ref{tanmoy}).

Note that (\ref{basicrelation}) is basically the solutions of (\ref{tanmoy})  at the two boundaries $ and (l,\rho_c = 0)$  and $(l = 0 , \rho_c )$ respectively.

Next we consider the fact that in general $\rho_0(l, \rho_c)$ can be expressed as $\rho_0 (l,\rho_c) = \displaystyle\sum_{m,n} A_{m n} l^m \rho_c^n$ with (m,n) may not be integer!  Substituting that in (\ref{tanmoy}), if we consider the two limits $l \rightarrow \infty \, , \, \rho_c \, {\rm finite}$ and $\rho_c \rightarrow \infty \, , \, l \, {\rm finite}$,  then it is easy to see that (\ref{tanmoy}) will be consistent, i.e l.h.s = r.h.s,  at this two limits,  implies the fact

\ber 
&& {\rm for}\, l \rightarrow \infty\,,\, \rho_c \, {\rm finite}  \quad  \Rightarrow \quad {\frac{\rho_c}{\rho_0}} << 1,\quad;\quad {\frac{A_{10} l}{2\rho_0}} \sim 1 \n
 && {\rm  for} \, \rho_c \rightarrow \infty\,,\, l  \, {\rm finite} \quad  \Rightarrow \quad  {\frac{A_{10} l}{2\rho_0}} << 1\quad;\quad {\frac{\rho_c}{\rho_0}} \sim 1
 \la{eithror}
 \eer
  
The implication of (\ref{eithror}) is 

\ber
& & {\rm for}\, , l >> \rho_c \,,\, \rho_0 (l,\rho_c) \sim {\frac{A_{10} l}{2}}\n
& & {\rm for}\, , \rho_c >> l \,,\, \rho_0 (l,\rho_c) \sim \rho_c
\la{compare}
\eer

The equation (\ref{compare}) enable us to make a guess that although (\ref{tanmoy}) is not solvable in general however perhaps one can expect  in the regime $l >> \rho_c$, and $\rho_c >> l$, one can determine the solutution exactly!  The next question is, since the expression of $\rho_0 (l,\rho_c)$. as given in (\ref{compare})  
are actually the exact solution of (\ref{tanmoy}) at $\rho_c = 0$ and $l = 0$ respectively, so can we have a perturbation around them, to find a global  expression of the turning point $\rho_0 (l,\rho_c)$ defined over the complete 2D $(l.\rho_c)$ plane(actually the first quadrant of it), even at the perturbative level?
However if we start from either regime, say $\rho_c >> l$,  where the exact solution at $ l = 0 $ is given by $\rho_0 (l,\rho_c) = \rho_c$,  a possible perturbative expansion around the same is given by 
\be
\rho_0(l,\rho_c) = \rho_c + \displaystyle\sum_n a_n O\left(l^n \right) + \displaystyle\sum_n  b_n O\left( {\left({\frac{l}{\rho_c}}\right)}^n \right) + .......
\la{perturbativerho0}
\ee

However this perturbative expansion (\ref{perturbativerho0}), while work well in $\rho_c >> l$ regime but no more work in the regime $l >> \rho_c$ where the terms will largely grow!  Moreover, we   are going to  explicitly show in the next section,  that, since  the theory is given with two boundary conditions (\ref{boundarycondition}),    consequently, if for either regime of the two,  i.e  $l>> \rho_c$ and $\rho_c >> l $,   we write a perturbative expansion for $\rho_0(l,\rho_c)$ just like (\ref{perturbativerho0}) and try to think that whether the two perturbative solutions,  written in the two regimes can be combined together  as the part of a bigger mathematical-series-expression,  which can be thought as the complete global solution for $\rho_0(l, \rho_c)$,  that really  does not work since the either perturbative solution in the concerned regime is adapted with the local boundary condition in that regime (as one can verify from (\ref{perturbativerho0})), they are not adapted with the global boundary conditions, i.e both the boundary conditions in  (\ref{boundarycondition}),    whereas a global solution for $\rho_0(l,\rho_c)$,  as defined over the complete $(l,\rho_c)$  plane, has to be adapted with both the boundary conditions!   So to summarize, we neither by solving (\ref{tanmoy}) globally,  not with any perturbative approximation, can obtain the global solutiion for the same! 
\vskip0.5mm
Here we need to mention, this perturbative expression of the turning point as given in (\ref{perturbativerho0}), when constructed as local solution of (\ref{tanmoy}), i.e the  solution over either regime,  they can indeed give the genuine expression of turning point from RT formalism locally over that regime and that way can give the well behaved expression of HEE and the other measures of quantum entanglement over that regime however their main drawback is they can no way give a complete interpolating expression of the turning point and HEE
between the two regime $l>>\rho_c$ and $\rho_c >> l$ or in other words a complete global expression  of $\rho_0(l,\rho_c)$ over 2D $(l,\rho_c)$ plane! Hence on the dual boundary they cannot describe a physical system which is smoothly defined  over complete paramer regime of $(\mu , l)$
(as follows from (\ref{murcrelationew})),  or more specifically when these dual boundary systems, which are specified by such locally defined quantum-measures,  when undergoes the variation of the parameters $(\mu, l)$, they cannot take a smooth transition from the one parameter regime to the other!   Also,  another most unfortunate drawback  of these perturbative solutions are that since they can  be obtained from Taylor-expasion of the concerned  function  where the successive terms in the series are taken through derivatives,   in our case this  will yield a term of the form ${\left( {\frac{l}{\rho_0}}\right)}^{d - \theta - 1}$ or  ${\left( {\frac{ \rho_c}{\rho_0}}\right)}^{d - \theta - 1}$ at first order of Taylor expansion,   which on either boundary, given by $(l, \rho_c = 0)$  or $(l = 0, \rho_c )$,  gives divergence for $d - \theta < 1$,  make Taylor expansion invalid!  So these perturbative solutions, one can obtain for $d - \theta \ge 1$ only.   
    As we mentioned in the introductions, quantum-measures based on this perturbative solutions  is studied in \cite{aspect1},  very recently!
So, to get, a proper insight about the global expression of the turning point $\rho_0(l, \rho_c)$,  which also can be described for the complete parameter regime of $(d,\theta)$,  with $\theta \le d$,  we will   consider the only exactly solvable case with $ d - \theta = 1$,   where we have

,\ber
{\frac{l}{2}} &=& \rho_0 \sqrt{1 - {\frac{\rho_c^2}{\rho_0^2}}}\n
\Rightarrow {\frac{ l}{2 \rho_0}} &=&  \sqrt{1 - {\frac{\rho_c^2}{\rho_0^2}}} \quad \quad {\rm since}\quad {\rm for}\quad  d - \theta = 1, A_{10} = 1 \n
\Rightarrow {\frac{\left(A_{10}l\right)^2}{4 \rho_0^2}} + {\frac{\rho_c^2}{\rho_0^2}} &=& 1\n
\Rightarrow \rho_0 &=& \left\lbrace \left(  {\frac{A_{10} l}{2}} \right)^2 + \left(\rho_c\right)^{2})\right\rbrace^{\frac{1}{2}} 
\la{dthetra1}
\eer 

Note the expression of $\rho_0$ above satisfies all the criteria at the limit   $l >> \rho_c$ or $\rho_c >> l$  regime,    which we discussed above.
The equation (\ref{dthetra1}),   which is defined for $ d - \theta =1$ can have natural generalization for general $d, \theta $, at least in  the limit   ${\frac{A_{10} l}{2}} >>\rho_c$ or $\rho_c >>{\frac{A_{10} l}{2}}$ as of the form
 
 \be
  \left({\frac{A_{10} l}{2\rho_0}}\right)^{2(d-\theta)} + \left( {\frac{\rho_c}{\rho_0}}\right)^{2(d-\theta) } = 1
 \la{firstexpectation}
 \ee
 
 or  
 \be
  \left({\frac{A_{10} l}{2\rho_0}}\right)^{(d-\theta +1)} + \left( {\frac{\rho_c}{\rho_0}}\right)^{(d-\theta + 1)}  = 1
 \la{secondexpectation}
 \ee

We will proceed with this expression of $\rho_0(l,\rho_c)$, as given in  (\ref{firstexpectation}, \ref{secondexpectation}) to check whether it gives a global solution for the turning point $\rho_0 (l,\rho_c)$ of
 (\ref{tanmoy}), most effective over the regime $l>> \rho_c$ and $\rho_c >> l$! We have presented all our derivations and the consequent plots im Appendix-A.  In the next section, we will derive the same global solution of $\rho_0 (l,\rho_c)$ in more fundamental way, i.e from its emergent global symmetry point of view, where we see the above mentioned  global solution for  $\rho_0 (l,\rho_c)$ is, as given in (\ref{firstexpectation}, \ref{secondexpectation}),   describes the    the most dominant part of $\rho_0 (l,\rho_c)$,  over the regime  $l>> \rho_c$ and $\rho_c >> l$, will give a more stronger proof for this said global solution.

\section{ The  global solution for turning point $\rho_0(l,\rho_c)$ from the  emergent global symmetry }

Here in this section we will proceed to understand the global solution for the turning point $\rho_0 (l, \rho_c) $.  The governing equation  for $\rho_0(l,\rho_c)$, as given by (\ref{tanmoy}), because of its very much complicated and highly nonlinear   mathematical structure, not really solvable globally.  However, in this section we are going to show, this equation (\ref{tanmoy}) and consequently the underlying theory has certain global symmetry structure, where the symmetry  emerges on the application of RT formalism to the given geometry with finite radial cut off and also  which when combined with the global boundary condition  ( which we are going to state more specifically in (\ref{boundarycondition}) )  and other consistency conditions of the theory, it can completely reveal the complete structure of the full global solution of $\rho_0(l,\rho_c)$
and can completely determine the part of the solution for $\rho_0(l, \rho_c )$, which is the most dominating part of the  solution of $\rho_0(l,\rho_c)$ in the regime. $l>>\rho_c$ and $\rho_c >> l$.  Here we will describe it and proceed with the subsequent analysis.

\subsection{The  global symmetry structure and its consequences }

The governing equation, which expresses  the turning point $\rho_0$,  as a function of $l, \rho_c$, is given by (\ref{tanmoy})

Now we have
\ber
{{}_2 F_1}   \left\lbrack {\frac{1}{2}}, {\frac{1}{2}}(1 +{ \frac{1}{d - \theta }}), {\frac{1}{2}}(3 +{ \frac{1}{d - \theta }}) , 1 \right \rbrack  
&=& \left( d -\theta + 1 \right)\times\n
& &{\frac{{\sqrt{\pi}}{\Gamma\left\lbrack {\frac{1}{2}}\left(1 + {\frac{1}{d - \theta}}\right)\right\rbrack}}{\Gamma\left\lbrack {\frac{1}{2\left(d - \theta\right)}}\right\rbrack}}
\la{f}
\eer

Also note, for the case without $T\overline{T}$ deformation, we have ${\frac{A_{10} l}{ 2 \rho_0}} = 1$, with $A_{10} = {\frac{\Gamma\left\lbrack{\frac{1}{2(d - \theta)}}  \right\rbrack}{{\sqrt{\pi}}\Gamma\left\lbrack {\frac{1}{2}}\left(1 + {\frac{1}{d- \theta}}\right) \right\rbrack}}$

So we have
\be
{{}_2 F_1}   \left\lbrack {\frac{1}{2}}, {\frac{1}{2}}(1 +{ \frac{1}{d - \theta }}), {\frac{1}{2}}(3 +{ \frac{1}{d - \theta }}) , 1 \right \rbrack 
= {\frac{\left( d -\theta + 1 \right)}{A_{10}}}
\la{1relation}
\ee

Substituting (\ref{1relation}) in  (\ref{tanmoy}) we get 
\ber
{\frac{l}{2 \rho_0}} &=& {\frac{1}{A_{10}}}
                      - \left({\frac{\rho_c}{\rho_0}}\right)^{d - \theta +1} {\frac{   \,\, {{}_2 F_1}   \left\lbrack {\frac{1}{2}}, {\frac{1}{2}}(1 +{ \frac{1}{d - \theta}}), {\frac{1}{2}}(3 +{ \frac{1}{d - \theta}}) , \left({\frac{\rho_c}{\rho_0}}\right)^{2(d - \theta)}  \right \rbrack }{ d + 1 - \theta   }} \n
                 \la{tanmoynewform}
                 \eer

This equation (\ref{tanmoynewform}), giving two boundary conditon

\vskip4mm
\textbf{I\,.\, Boundary Condition}
\ber
\rho_0 &=& \rho_0 \left( l ,\rho_c\right)\n
{\rm For}& & \rho_c = 0 \quad \Rightarrow \quad {\frac{A_{10} l}{ 2 \rho_0}} = 1\n
{\rm For}& & l = 0 \quad \Rightarrow \quad  \rho_0 = \rho_c
\la{boundarycondition}
\eer
\vskip6mm
Apart from the above boundary conditions, we also have 

\vskip4mm
\textbf{II\,.\, Consistency  Condition for \,\, $d - \theta = 1$}

\be
\rho_0 =  \left\lbrace \left(  {\frac{A_{10} l}{2}} \right)^2 + \left(\rho_c\right)^{2})\right\rbrace^{\frac{1}{2}} \quad;\quad A_{10} = 1 
\la{dtheta1}
\ee 
\vskip6mm
Also the existence of a  global solution of $\rho_0$ from  (\ref{tanmoynewform}), defined at each and every point in the first quadrant of the  $l,\rho_c$ plane,  desires  the following basic facts to be valid which we will check now to visualize the global nature of S and consequently the global nature of the turning point $\rho_0$, i.e whether at all a global and unique solution of $\rho_0$ exist or not: 
\vskip4mm
\textbf{III\,.\, Basic Facts}
\vskip2.5mm
1. $\rho_0 = \rho_o \left(l , \rho_c \right)$ is an unique function
\vskip1mm
2.The unique function $\rho_0 = \rho_o \left(l , \rho_c \right)$ is the solution for (\ref{tanmoynewform}) for every  possibles values of $(\rho_c , l)$
\vskip1mm
3. The unique function $\rho_0 = \rho_o \left(l , \rho_c \right)$ is the solution for (\ref{tanmoynewform}) for every  possibles values of $d - \theta$, i.e,  $d - \theta> 1, d - \theta = 1, d -\theta < 1$
\vskip6mm
Along with $\textbf{ I, II}$ we will follow $\textbf{III}$, will proceed to understand the global nature of the turning point $\rho_0 (l, \rho_c)$
\vskip1mm

Here we wrote \textbf{III} as our proposal or the basic assumptions to obtain a global solution for $\rho_0 (l,\rho_c )$, where as we proceed, we will see its relevance and will realize it is an essential condition!

Let us consider the fact that  (\ref{tanmoynewform}) can be expressed as

\be
f\left( {\frac{A_{10}l}{2 \rho_0 \left( l, \rho_c \right) }}\, ,\, {\frac{\rho_c}{\rho_0 \left( l, \rho_c \right)}}\, ,\, d - \theta\right) =  1
\la{form}
\ee

To get a realization of the possible symmetry structure of (\ref{tanmoynewform}), we consider (\ref{form}) and also consider it on the boundary, $(l, \rho_c = 0)$\,: 

\be
f\left( {\frac{A_{10}l}{2 \rho_0 \left( l, 0  \right) }}\, ,\, 0 , \, d - \theta \right) =  1
\la{formrhoc0}
\ee 

On this (\ref{formrhoc0}),    we need to apply the features of  the concerned boundary condition. i.e the features of first equation of  (\ref{boundarycondition})

Please note  (\ref{formrhoc0}) concerns the case without $T \overline{T}$ deformation and true for every l and also for every $d - \theta$.   Now also note, while in 2D $(l, \rho_c)$ plane,  we can generate the whole plane, from a single point by  translation or scaling transformation by using two parameters, while coming to boundary, we can generate the whole boundary line from a single point, by using single parameter, e.g for l-axis, i.e the line in $(l , \rho_c )$ plane, set with $\rho_c = 0 $ , starting from a point $l = l_0 (l_0 \ne 0)$, we can generate the whole l-axis by a scaling transformation 
\be
l \rightarrow ml
\la{scalingaxis}
\ee

Under the same (\ref{formrhoc0}) can be rewritten as

\be
f\left( {\frac{A_{10}(kl)}{2 \rho_0 \left( kl, 0  \right) }}\, ,\, 0 , \, d - \theta \right) =  1
\la{formrhoc0re}
\ee

Since we have (\ref{formrhoc0})  is true for every l, and also every $d - \theta$,   so we must have
\be
\rho_0 \left( kl, 0, d-\theta  \right) = k \rho_0 \left( l, 0, d-\theta  \right) 
\la{scalinglaxis}
\ee
Now, starting from the boundary point $(l, \rho_c = 0)$, we can move very little inside, think of a line parallel line to l-axis, given by $\rho_c = \epsilon$, which we can obtain from l-axis by the translation 
\be
(l, \rho_c = 0 ) \rightarrow (l, \rho_c + \epsilon ) =  (l,\epsilon )\,,
\la{lineparallel}
\ee

and raise the question, whether the symmetry in the boundary (\ref{formrhoc0re}), has any consequence in the closest-bulk-region (\ref{lineparallel}) or in otherwords, is there any chance that the symmetry on the boundary (\ref{scalinglaxis}), is a direct consequence of any existing global symmetry in the bulk(i.e 2D, $(l , \rho_c)$ plane, with either variable not equal to zero)?
\vskip2mm
 It is not in general.
However, if we consider the governing equation (\ref{tanmoynewform}), strikingly, it is invariant under a   global symmetry  

\ber
& & (l, \rho_c)\rightarrow (k l \, ; k \rho_c ) \Rightarrow \quad   \rho_0 (kl, k\rho_c) \rightarrow k \rho_0 (l, \rho_c) \, \,\n
& & {\rm for}\,\, {\rm all}\,\,  d - \theta \, ,
\la{2dsymmetry}
\eer

Since $\rho_0 (l,\rho_c)$ is scalar function and under any symmetry transformation,$$ (l\,,\rho_c) \rightarrow (l^\prime\,,\,\rho_c^\prime)\,\,;\,\,\rho_0 (l.\rho_c) \rightarrow  \rho_0^\prime(l^\prime \,,\,\rho_c^\prime)\,;\, {\rm with}\, \rho_0^\prime(l^\prime \,,\,\rho_c^\prime) = \rho_0 (l,\rho_c)$$

\vskip0.5mm

So with the view of the above transformation

\be
\rho_0^\prime = {\frac{1}{k}} \rho_0
\la{transformed.function}
\ee

In other words, we see that,  under the transformation  $(l, \rho_c) \, \rightarrow \, (kl , k\rho_c)$, \,\,  $\rho_0(l.\rho_c)$and ${\frac{\rho_0 (kl, k\rho_c)}{k}}$ are the solution of identical algebraic equation and hence must be identical to each other
 
\be'
\rho_0(l.\rho_c) =  \rho_0^\prime(l^\prime \,,\,\rho_c^\prime)  = {\frac{\rho_0 (kl, k\rho_c)}{k}}
\la{identical}
\ee

Next we look at the other boundary $( l = 0 \,,\, \rho_c)$, where $\rho_0(l,\rho_c)$ obeys the boundary condition given by the second equation of   (\ref{boundarycondition}).

Since   $\rho_0 ( l = 0,   \,\, \rho_c) = \rho_c$,  again we have the same scaling symmetry

\be
(l ,\rho_c )\rightarrow (kl,  k\rho_c ) \,\, ; \,\, \rho_0 ( l = 0, \rho_c \,;\, d  - \theta) \rightarrow 
\rho_0 ( kl = 0,  \,\, k\rho_c\,;\, d - \theta) = k\rho_0 ( l = 0, \,\, \rho_c \,;\, d- \theta)
\la{scalingotherboundary}
\ee
However the above scaling symmetry (\ref{scalingotherboundary}), again leads to the same bulk-scaling-symmetry-realization as in (\ref{2dsymmetry})) and hence all the above arguments, which we made w.r.t the boundary at $(l. \rho_c = 0)$ will also follow  w.r.t the boundary $(l = 0. \rho_c )$ as well!

\vskip1mm
 However in spite of the above symmetry (\ref{2dsymmetry})  and its consistency with the existing symmetry in both the boundary,$(l, \rho_c = 0)\,,\,(l = 0 \,,\, \rho_c)$,  we are still not   in a position to say very firmly that the global solution of    (\ref{tanmoynewform}), given by $\rho_0(l,\rho_c)$, will posses this symmetry at each and every point of $(l,\rho_c)$ plane, for the following reason:

\vskip0.5mm

 The global symmetry in our case (\ref{2dsymmetry}), is actually an one-parameter symmetry in 2D plane, the specific transformation $(l,\rho_c) \rightarrow (kl,k\rho_c)$ actually generates the transformation from one point to another point along the individual radial  lines which emerges  from the origin, which certainly not like 2D transformation symmetry which can generate the symmetry-transformation   along any arbid path on the whole 2D plane!
\vskip0.5mm

 So from the very much complicated and nonlinear nature of the governing equation ({\ref{tanmoynewform} ). there may always be a possibility, that we have some regime in $(l, \rho_c)$ plane where any solution for $\rho_0 (l, \rho_c)$ does not exist! 
\vskip0.5mm
 In our specific case,  which generates the transformation-symmetry on the individual radial lines and if in case on this line there is a finite region of discontinuity, where any solution for $\rho_0(l,\rho_c)$ does not exist,  then,  like the 2D-transformation-symmetry-case, we really do not find any alternative path to generate the symmetry-transformation between the same two spacetime points  avoiding the finite region of discontinuity! In other words, to claim, the general solution of ({\ref{tanmoynewform} ), given by $\rho_0 (l,\rho_c)$,  obey the global symmetry  (\ref{2dsymmetry}),  the proposal $\textbf{III}$, has to be valid!
So, to make an  a priori  check of the proposal $\textbf{III} $, first we consider, the complete expression of $\rho_0 (l,\rho_c)$ is a sum of the symmetry-invariant and the symmetry-non-invariant-part,
and towards the end of this section we will show that the symmetry-non-invariant part does not exist!

\vskip1mm
To proceed with, we see there are few very impotant point to note:
\vskip1mm
$\textbf{P1}$\,:\, The one-parameter-bulk-symmetry, as given in (\ref{2dsymmetry}), is the only possible 2D generalization of 1D-boundary-symmetry (\ref{scalinglaxis}), and (\ref{scalingotherboundary}) where one can also see it in other way around, i.e if one extends one-parameter-bulk-symmetry (\ref{2dsymmetry}), to the boundary, set by $\rho_c = 0$, this indeed gives the boundary-symmetry (\ref{scalinglaxis}) and by setting $l = 0$, this gives the symmetry (\ref{scalingotherboundary})!

\vskip2mm
 
$\textbf{P2}$\,:\,  This scaling symmetry, as can be seen from  (\ref{2dsymmetry}), is completely independent of $d-\theta$ and equally true for all $d-\theta$, for any k.

\vskip2mm

$\textbf{P3}$\,:\,  If we consider the case for $d-\theta = 1$, from (\ref{dtheta1}), we see that $\rho_0(l, \rho_c)$ for $d-\theta = 1$ indeed obeys this scaling symmetry.  However since this is the only exactly solvable case, so this solution is valid over complete $(l, \rho_c)$ plane! 
,
\vskip2mm

$\textbf{P4}$\,:\, Since $\rho_0 (l, \rho_c)$, which is the solution of  (\ref{tanmoynewform}),  for any general $d - \theta$ must be consistent with (\ref{dtheta1}), i.e when we substitute $d-\theta = 1$ in that, it reproduces the solution of $\rho_0 (l, \rho_c)$ for $d-\theta = 1$!  So the solution for $\rho_0 (l, \rho_c)$ for $d-\theta = 1$,  must be  present in its most generalized form for general $d - \theta$, as a part of the complete 
solution of $\rho_0 (l, \rho_c)$ for a general $d - \theta$, where it is evident from the structure of $\rho_0$ solution for (\ref{dtheta1}), its most generalized form for other $d - \theta$,  which is nonvanishing on substitution of $d-\theta = 1$ is given by (\ref{firstexpectation}) and (\ref{secondexpectation}).
, where on substitution of $d-\theta = 1$ it reproduces (\ref{dtheta1}), must  obey the symmetry (\ref{2dsymmetry}) (as we are going to show explicitly in the next section)!

\vskip0.5mm

 Along with the above facts, now we will proceed to understand,  the general structure of the global solution for $\rho_0 (l.\rho_c)$.
\vskip0.5mm

Since the complete solution for $\rho_0(l. \rho_c )$ must be consistent with both the boundary conditions in (\ref{boundarycondition}), so we must have, the complete solution of $\rho_0(l,\rho_c)$ consists of four parts\,:
\vskip2mm
1.\, $\rho_0^1$ \,:\, The scaling symmetry invariant part nonvanishing on either or both the boundaries, with the boundaries given by $(l\,, \rho_c = 0)$ and $(l = 0\, , \rho_c)$!
\vskip1mm

2.\,$\rho_0^2$ \,:\, The scaling symmetry noninvariant part nonvanishing on either or both the boundaries!

\vskip1mm

3.\, $\rho_0^3$ \,:\, The scaling symmetry invariant part vanishing  on  both the boundaries or  vanishing on either boundary and nonvanishing on the  other one!
\vskip1mm

4.\, $\rho_0^4$ \,:\, The scaling symmetry noninvariant part vanishing  on  both the boundaries or  vanishing on either boundary and nonvanishing on the  other one!
\vskip4mm

According to the fact that $\rho_0 (l, \rho_c )$ is a global solution,  .  we must have,  all these four parts of solutions are consistent with both the boundary conditions (\ref{boundarycondition}) and with (\ref{dtheta1}).  
\vskip2mm

First we concern about $\rho_0^2$, i.e   The scaling symmetry noninvariant part nonvanishing on either or both the boundaries!
\vskip0.5mm
This part must be given by  $\rho_0^2 (l , \rho_c)$, so that we have

\be 
 \rho_0^2 (k l , k \rho_c) \ne k \rho_0^2 (l ,  \rho_c) 
\la{nonvanishing}
\ee

However this implies\, :
\vskip0.5mm

Since in this case all the product terms  within the solution $\rho_0 (l, \rho_c)$,  of the form $l^m \rho_c^n$ with $ m \ne 0, n \ne 0$, must vanishes  on both the boundaries, so this condition, must be implemented with the condition

Either

\be 
 \rho_0^2 (k l , k \rho_c = 0) \ne k \rho_0^2 (l , \rho_c = 0 ) \quad {\rm while} \quad\rho_0^2 (k l = 0 , k \rho_c ) = k \rho_0^2 (l = 0 , \rho_c  )
\la{nonvanishing1}
\ee

Or

\be 
 \rho_0^2 (k l = 0, k \rho_c ) \ne k \rho_0^2 (l = 0 , \rho_c ) \quad {\rm while} \quad \rho_0^2 (k l , k \rho_c  = 0) = k \rho_0^2 (l  , \rho_c = 0 )
\la{nonvanishing2}
\ee

Or

\be 
 \rho_0^2 (k l = 0, k \rho_c ) \ne k \rho_0^2 (l = 0 , \rho_c ) \quad {\rm and} \quad \rho_0^2 (k l , k \rho_c  = 0) \ne k \rho_0^2 (l  , \rho_c = 0 )
\la{nonvanishing3}
\ee

However $\rho_0 (l,\rho_c)$ as a global  solution of (\ref{tanmoynewform}), must satisfy both (\ref{scalinglaxis}, \ref{scalingotherboundary}) simultaneously, which implies that the either of the above (\ref{nonvanishing1}, \ref{nonvanishing2}, \ref{nonvanishing3}) cannot be hold as a part of complete expression of $\rho_0 (l,\rho_c)$   which is a complete global solution of (\ref{tanmoynewform})! So we come to a conclusion:
\vskip1mm
  $\rho_0^2 (l ,  \rho_c) $ can never be a part of the complete global solution $\rho_0 (l , \rho_c)$!

Regarding $\rho_0^1$,  we will decide on next section.

To get insight on the term  $\rho_0^3$, $\rho_0^4$,  first we rewrite (\ref{tanmoynewform}) in the following way!

\ber
{\frac{A_{10} l}{2 \rho_0}} &=& 1
                      - A_{10} \left({\frac{\rho_c}{\rho_0}}\right)^{d - \theta +1} {\frac{   \,\, {{}_2 F_1}   \left\lbrack {\frac{1}{2}}, {\frac{1}{2}}(1 +{ \frac{1}{d - \theta}}), {\frac{1}{2}}(3 +{ \frac{1}{d - \theta}}) , \left({\frac{\rho_c}{\rho_0}}\right)^{2(d - \theta)}  \right \rbrack }{ d + 1 - \theta   }} \n
                 \la{tanmoynewform12}
                 \eer

Next we consider the proposal $\textbf{III}$ that $\rho_0 (l, \rho_c)$ must be an unique solution of the governing equation at each and every point of $(l,\rho_c)$ plane! 

Now, following the above discussion,  we have 
\be
\rho_0 (l,\rho_c)  = \rho_0^1(l,\rho_c) + \rho_0^3(l,\rho_c) + \rho_0^4(l,\rho_c)\, ,
\la{solution}
\ee
since $ \rho_0^2 $ does not exist!
We can write
\be
\rho_0^3 + \rho_0^4 = \displaystyle\sum_{m.n} a_{m,n} l^m \rho_c^n \,
\la{rho034}
\ee
where m.n are any positive integer or noninteger with $(m = 1\,,\, n = 0)$ and  $(m = 0\,,\, n = 1)$ excluded! Note, that m,n, has to be positive, as otherwise $\rho_0 (l, \rho_c )$  are divergent on  either bounary and then it is not a consistent solution! Also the said values of m,n are excluded from the series,  those are being considered as part of  $\rho_0^1$, i.e scaling  symmetry invariant (since $(m+n = 1)$) and nonvanishing on either boundary and will be considered in the next section.
\vskip0.5mm

Next following, (\ref{solution}) and (\ref{rho034}), we substitute the complete form of the solution of $\rho_0 ( l, \rho_c)$ in (\ref{tanmoynewform12}).   as we obtained:
\vskip0.5mm

\ber
& & {\frac{A_{10} l}{2(\rho_0^1 +  \displaystyle\sum_{m.n} a_{m,n} l^m \rho_c^n )}} = 1\n
&-& A_{10} \left({\frac{\rho_c}{(\rho_0^1 +  \displaystyle\sum_{m.n} a_{m,n} l^m \rho_c^n )}}\right)^{d - \theta +1} {\frac{   \,\, {{}_2 F_1}   \left\lbrack {\frac{1}{2}}, {\frac{1}{2}}(1 +{ \frac{1}{d - \theta}}), {\frac{1}{2}}(3 +{ \frac{1}{d - \theta}}) , \left({\frac{\rho_c}{(\rho_0^1 +  \displaystyle\sum_{m.n} a_{m,n} l^m \rho_c^n )}}\right)^{2(d - \theta)}  \right \rbrack }{ d + 1 - \theta   }} \n
                 \la{tanmoynewform34}
                 \eer

 Next, in order to get an idea about $\rho_{03} + \rho_{04}$ we consider (\ref{tanmoynewform34}) in different regime of $(l, \rho_c)$ plane :  
\vskip4mm
\textbf{1.\,\, $ l \rightarrow \infty$, \,\,\,\,$\rho_c$  \,\, finite \,\,,\,\,  $l >> \rho_c$ \,\,,\,\,The regime continued to the boundary 
$(l\,,\,\rho_c = 0)$ }
\vskip4mm
Let us consider the most dominatimg term  of $l^m$ in (\ref{rho034}),  i.e the term with  highest m, among all the terms there ,  which is considered to be the most dominating term in the expression of $\rho_{03} + \rho_{04}$, in this regime.    Considering $\rho_0$ (\ref{solution}),      
since this is the condition corresponds to $l>> \rho_c$, i.e a term very close to and continued from the boundary $(l, \rho_c = 0)$, so according to the previous discussion(\ref{scalinglaxis},\ref{boundarycondition}), the scaling symmetry invariant term $\rho_0^1 \sim l$ at this concerned regime.   
So for the leading term, in $l^m$, in (\ref{rho034}), if  $m > 1$, then this is the most dominating term in the  denominator of the l.h.s of (\ref{tanmoynewform34}) , 
i.e within the expression of $\rho_0$,  in our concerned regime of $(l,\rho_c)$ plane, we have  ${\frac{\rho_c}{\rho_0}} \sim 0 $ ,  ${\frac{l}{\rho_0}} \sim 0 $,  the l.h.s is zero, while, in r.h.s, since ${\frac{\rho_c}{\rho_0}} \sim 0$ in this regime,  we have r.h.s = 1,  which is inconsistent! So we must need the boundary behaviour of $\rho_0(l, \rho_c) \sim l$, in the concerned boundary given by $(l,\rho_c = 0)$,  to maintain the consistency of (\ref{tanmoy}) and to have a consistent solution for $\rho_0 (l,\rho_c)$ and that way we must have
\be
m \le 1
\la{m1}
\ee

Next we consider the limit, 
\vskip4mm
$\textbf{2}$ .\,\, $ \rho_c \rightarrow \infty$, \,\,\,\,$l \,\,{\rm finite}$,\,\, $\rho_c >> l$, $\textbf{the regime is continued to the boundary}$ $(l = 0,\, \rho_c)$
\vskip4mm
From the exactly similar consideration  for the previous case,  in the series (\ref{rho034}), the most dominating term in the series is given by   maximum n.  So if $n > 1$, then in the complete expression of $\rho_0 (l,\rho_c)$, this term is even dominant over the scaling invariant term  $\rho_0^1$, which again according to  (\ref{scalingotherboundary}, \ref{boundarycondition}), will scale as $\rho_0^1 \sim \rho_c$, at the concerned  regime since this regime is continuously connected to the boundary boundary  $(l = 0 , \rho_c)$,  
so that in this case 
${\frac{\rho_c}{\rho_0}} \sim 0 $ ,  ${\frac{l}{\rho_0}} \sim 0 $,  consequently we have the r.h.s of (\ref{tanmoynewform34}) $\sim$ 1, the l.h.s $\sim$ 0, which is inconsistent!  So we must have

\be 
n \le 1 
\la{n1}
\ee

Next we consider the limit, 
\vskip4mm
\textbf{3.\,\, $ \rho_c \rightarrow \infty$, \,\, $l \rightarrow \infty$   }
\vskip4mm

Since at this limit, we have $l \sim \rho_c  \sim \infty $ and since also the the scaling invariant term $\rho_0^1$,  varies linearly with $l \sim \rho_c$, so as long as, we have $m+n \le 1 $ we have consistency because for $m + n < 1$, this term from (\ref{rho034}) is less dominant compared to $\rho_0^1$ and, so the both side of (\ref{tanmoynewform34}) are well behaved and it is a consistent equation. However for $m+n > 1$, since in both side of  (\ref{rho034}), we have either 
${\frac{l}{\rho_0}}$ or ${\frac{\rho_c}{\rho_0}}$ term, since the numerator is linear in l or $\rho_c$, so in this limit, we have ${\frac{l}{\rho_0}} \sim {\frac{\rho_c}{\rho_0}}\sim 0$, which, as we already discuss that under the substitution of the same, the l.h.s is equal to 0 and r.h.s is equal to 1, which is inconsistent! So to maintain consistency, we must have
\be
m + n \le 1\, ,
\la{mn}
\ee

Next, we consider the limit

\vskip4mm
\textbf{4.\,\, $ \rho_c \rightarrow 0 $, \,\,, $l \rightarrow 0$   }
\vskip4mm

Let us consider the denominator of the l.h.s of (\ref{tanmoynewform34}).  In the expression of $\rho_0 $, since the scaling symmetry invariant term must be linear in $l\sim \rho_c$ in this limit,  so if $m+n >1 $, this term is irrelevant  compared to the scaling invariant term in this  limit !  However if $ m + n < 0 $, the term from (\ref{rho034}), becomes relevant compared to, linearly varyimg  
scaling invariant term $\rho_0^1$ and we come up with a situation ${\frac{l}{\rho_0}} \sim  {\frac{\rho_c}{\rho_0}} \sim 0$, which, we just discussed, is an inconsistent solution of  (\ref{tanmoynewform34}).  So, to maintain consistency we must have

\be
m+ n \ge 1
\la{secondcondition}
\ee

Combining (\ref{mn}, \ref{secondcondition}, \ref{m1} , \ref{n1}), we come to conclusion

\be
m+n = 1\,\,;\,\, m\le 1\,;\, n \le 1\,,
\la{mplusn}
\ee

However this (\ref{mplusn}) implies scaling symmetry invariance (\ref{2dsymmetry})!  So we conclude $\rho_0^4$ does not exist!
\vskip0.5mm
One more point to note that in the definition of $\rho_0^3$, and $\rho_0^4$, we mentioned that the scaling-invariant and scaling noninvariant term, vanishing on either or on the both of the boundary, whereas the series for the sum of both the terms we have shown in (\ref{rho034}) represents the term which are vanishing on both the boundaries! Actually, for $\rho_0^3$ these are the scaling symmetry invariant term with $(m = 1 \,, \, n = 0)$ or $(m = 0 \,,\,n=1)$ in (\ref{rho034}), where we have explicitly mentioned that these terms are excluded from the series because they are the part of $\rho_0^1$.  Next in $\rho_0^4$, which are symmetry-noninvariant terms,  vanishing on either boundary instead of on the both,  given by the term of the form $l^m$ or $\rho_c^n$, with $m,n <1$. However, since the terms are inconsistent with the boundary conditionsse 
(\ref{boundarycondition}), we conclude that they do not exist!  
\vskip0.5mm
\vskip0.5mm
So as a summary of all the above conditions, we see that, for all those scaling invariant terms  which vanishes on both the boundaries, given by (\ref{rho034}),   
we have
\be
m \le 1\,,\,n\le 1\,, \,m + n =  1\, ,
\la{conditionnonscaling}
\ee

with $(m = 0, n = 1 \, , \, m = 1, n = 0)$ excluded, because, as mentioned this can be considered as part of $\rho_0^1$, will be discussed next section!
However, we note,   that  this condition (\ref{conditionnonscaling}), from (\ref{rho034}) gives

\ber
\rho_0^3 &=& \displaystyle\sum_{m.n} a_{m,n}\left(d-\theta \right)\, {\left({\frac{A_{10}l}{2}}\right)}^m \rho_c^n\n
{\rm with} & &  m + n = 1\,\,;\,\,m \le 1\,,\,n \le 1\,\,,\,\, {\rm with} \,\,(m = 0, n = 1 )\,\, , \,\,( m = 1, n = 0)\,\, {\rm excluded}
\la{rho03}
\eer

(with  $a_{m,n}\left(d-\theta \right)$ expressing its functional dependence of $a_{mn}$  on $d - \theta$).
\vskip0.5mm
and $\rho_0^4$ does not exist!

Here we mention that m, n are dependent on $ d - \theta$ about which we have explicitly discussed in the next section.

\vskip0.5mm
\vskip1mm
In the next  sections, we will try to understand these two!  There we also show that the first part can be written as a sum of two parts and so the complete global expression of $\rho_0 (l,\rho_c)$ cosists of three parts!    

Finally we see, as we discussed, previously, that we found, that the complete global solution of (\ref{tanmoynewform}) as given by $\rho_0 (l , \rho_c)$ bears scaling symmetry invariance at each and every point of $(l,\rho_c)$ plane!

\vskip1mm
This implies, the proposal $\textbf{III}$ must be true!

\subsection{The derivation  of  the most dominant part of  $\rho_0(l,\rho_c)$ over the regime $l >> \rho_c$ and $\rho_c >> l$}

Here we gather all the information in the previous subsection, consider  the contributing terms in the global solution of $\rho_0(l,\rho_c)$, as given by $\rho_0^1$ and $\rho_0^3$, express it in more generalized way.   to see the most generalized structure of the  global solution for $\rho_0^{\rm soln}  (l, \rho_c)$,  which  will be sum  of 4 parts

\ber
\rho_0^{\rm soln}  (l, \rho_c) &=& \rho_{01}^{\rm soln}  (l, \rho_c) + \rho_{02}^{\rm soln}  (l, \rho_c)  + \rho_{03}^{\rm soln}  (l, \rho_c) + \rho_{04}^{\rm soln}  (l, \rho_c)  \n
\la{rho0soln}
\eer

First we consider $\rho_{01}$, which is the sum of all possible type of the scaling-symmetry-invariant terms, which are nonvanishing and nondivergent on both the boundaries, consistent with (\ref{boundarycondition}),(\ref{dtheta1}),    given by a sum :

\ber
\rho_{01}  (l, \rho_c) &=& A_1\left(d-\theta \right){\left( {\left({\frac{A_{10} l}{ 2 }}\right)}^{2  \left( d - \theta\right)} + {\left( \rho_c \right)}^{2  \left( d - \theta\right)} \right)}^{\frac{1}{2(d - \theta)}}\n
&+& C_1\left(d-\theta \right){\left( {\left({\frac{A_{10} l}{ 2 }}\right)}^{  \left( d - \theta + 1 \right)} + {\left( \rho_c \right)}^{  \left( d - \theta + 1 \right)} \right)}^{\frac{1}{(d - \theta + 1)}}\n
&+& \displaystyle\sum_{n_1 > 0 \, , \ne 1}^\infty  \displaystyle\sum_{n_2 > 0}^\infty  A_{n_1 n_2}\left(d-\theta \right)\left\lbrack \displaystyle\sum_{m > 0}^\infty 
\,\,  B_{n_1 n_2 m}\left(d-\theta \right) {\left\lbrace \left( d - \theta\right) - 1 \right\rbrace}^ m \right\rbrack \times\n
& &{\left( {\left({\frac{A_{10} l}{ 2 }}\right)}^{\left( 2 n_1 \left( d - \theta\right) + n_2 \right)}  + 
{\left( \rho_c \right)}^{\left( 2 n_1 \left( d - \theta\right) + n_2 \right)} \right)}^{\frac{1}{2 n_1 (d - \theta)  + n_2 }}\n
&+& \displaystyle\sum_{n_1 > 0 \, \ne 1}^\infty  \displaystyle\sum_{n_2 > 0}^\infty  C_{n_1 n_2}\left(d-\theta \right)\left\lbrack \displaystyle\sum_{m > 0}^\infty  D_{n_1 n_2 m} {\left\lbrace \left( d - \theta\right) - 1 \right\rbrace}^ m \right\rbrack \times \n
& & {\left( {\left({\frac{A_{10} l}{ 2 }}\right)}^{\left(  n_1 \left( d - \theta  + 1 \right) + n_2 \right)}  
+ {\left( \rho_c \right)}^{\left(  n_1 \left( d - \theta + 1 \right) + n_2 \right)}  \right)}^{\frac{1}{ n_1 (d - \theta + 1)  + n_2 }}\n
&+& \displaystyle\sum_{n_1 > 0 \, , \ne 1}^\infty  \displaystyle\sum_{n_2 > 0}^\infty  A_{n_1 n_2}\left\lbrack \displaystyle\sum_{m > 0}^\infty  B_{n_1 n_2 m} {\left\lbrace \left( d - \theta\right) - 1 \right\rbrace}^ m\right\rbrack \times\n
& & \lbrack {\left({\frac{A_{10} l}{ 2 }}\right)}^{\left( 2 n_1 \left( d - \theta\right) + n_2 \right)}  + \displaystyle\sum_{k_1} a_{k_1} 
{\left({\frac{A_{10} l}{ 2 }}\right)}^{\left( 2 n_1 \left( d - \theta\right) + n_2 - {k_1}\right)} {\left( \rho_c \right)}^{k_1} + ....................\n
& &........ .......... ..+\displaystyle\sum_{k_2}  a_{k_2} 
{\left(\rho_c\right)}^{\left( 2 n_1 \left( d - \theta  \right) + n_2 - {k_2}\right)} {\left({\frac{A_{10} l}{ 2 }}  \right)}^{k_1} +
{\left( \rho_c \right)}^{\left( 2 n_1 \left( d - \theta\right) + n_2 \right)}\rbrack^{\frac{1}{2 n_1 (d - \theta)  + n_2 }} +...  \n
\la{firstsoln}
\eer

We see in (\ref{firstsoln}),  there are basically two type of terms in the equation (\ref{firstsoln}).  The first type, which we denote by $\rho_{01a}$,   are the  terms which are direct generalization of (\ref{dtheta1}) to general $d - \theta$ (reproduce the same on substitution of $d - \theta = 1$  ), as given in (\ref{firstexpectation}, \ref{secondexpectation}). 
.  Since one can make this generalization in two possible ways so we see in (\ref{firstsoln}) that there are two such terms as given in the first two lines of (\ref{firstsoln}).
   
  The rest of the terms in  (\ref{firstsoln}) are, although consistent with  (\ref{boundarycondition}) however are absent from  (\ref{dtheta1}), thats why we expressed them with a coefficient as a function of $(d - \theta - 1)$, e.g $B_{n_1 n_2 m} {\left\lbrace \left( d - \theta\right) - 1 \right\rbrace}^ m$,  so that they will vanish on substitution of  $ d - \theta = 1$  and that way show a consistency with (\ref{dtheta1}).   We denote the sum of this type of terms within the expression of $\rho_{01}$   by $\rho_{01b}$. 

Indeed in (\ref{firstsoln}) there could be many more possible terms of the second type, i.e $\rho_{02b}$,  so that we expressed the equation   as an indefinite sum.

The consistency of (\ref{firstsoln}) with (\ref{boundarycondition}) (i.e when we consider the same equation on the boundary)   implies that the sum of the coefficients must be equal to one
\ber
& &\displaystyle\sum_{f} A_f \left(d - \theta \right) = 1\n
&\Rightarrow& A_1\left( d - \theta\right) + C_1\left( d - \theta\right) + \displaystyle\sum_{n_1 > 0 \, , \ne 1}^\infty  \displaystyle\sum_{n_2 > 0}^\infty  A_{n_1 n_2}\left\lbrack \displaystyle\sum_{m > 0}^\infty 
 B_{n_1 n_2 m} {\left\lbrace \left( d - \theta\right) - 1 \right\rbrace}^ m \right\rbrack\n
&+&\displaystyle\sum_{n_1 > 0 \, \ne 1}^\infty  \displaystyle\sum_{n_2 > 0}^\infty  C_{n_1 n_2}\left\lbrack \displaystyle\sum_{m > 0}^\infty  D_{n_1 n_2 m} {\left\lbrace \left( d - \theta\right) - 1 \right\rbrace}^ m\right\rbrack + .....................  = 1 \, ,
\la{coefficientboundary}
\eer
where the sum $\displaystyle\sum_{f}$  refers to the sum of the coefficients of all possible functions "f"  in (\ref{firstsoln}). 

Here in order to analyze  it further,   we argue that,  within the consistency condition (\ref{coefficientboundary}),  the contribution from the sum of the coefficients
from the first type of terms ($\rho_{01a}$)   and the second type of terms ($\rho_{01b}$),  in (\ref{firstsoln}), as we mentioned above,  should must be treated completely independently! Note,  if this is not the case then following our argument as we mentioned above,    since  the sum of their coefficient  has to be one,  just because once we express $\rho_0(l, \rho_c)$ over either boundary, $(l, \rho_c = 0)$ or $(l = 0, \rho_c)$  it must has to be consistent with (\ref{boundarycondition})    and also at the same time this consistency has to maintained for all $d - \theta$, i.e their sum has to be one for all $d - \theta$
so these coefficients as the explicit function of $d - \theta$ must has to be related by some trivial identity (e.g just the same way we relate the squares of different independent tangent vectors on sphere or hyperboloid so that the respective sum always give one irrespective of the value of the angular parameters)
So to see whether our argument is valid or not, let us, just for example, consider a situation, where (\ref{coefficientboundary}) consists of four terms,$ X_1 \left(d-\theta\right) \,,\, X_2 \left(d-\theta\right) \,;\, Y_1 \left(d-\theta\right) \,,\, Y_2 \left(d-\theta\right)$  where $X_1 \,,\, X_2$ are the coefficient of the first type of terms which are nonvanishing on substitution of  $d - \theta = 1$  and reproduce  (\ref{dtheta1})  whereas   $Y_1 , Y_2 $. are the coefficient of the second type of terms which are vanishing under the substitution of $d - \theta = 1$!
\vskip0.5mm
So, (\ref{coefficientboundary}) must be realized  in our exaample as 

\be
X_1 \left(d-\theta\right) + X_2 \left(d-\theta\right) + Y_1 \left(d-\theta\right) + Y_2 \left(d-\theta\right) = 1
\la{toymodel}
\ee

Since the above must has to be true for each and every $( d - \theta )$, it must has to be a trivial identity, like
\ber
X_1 \left(d-\theta\right) = \cos^2 (d - \theta - 1) \cos^2 (d - \theta)\,\,&;&\,\, X_2 \left(d-\theta\right) =  \cos^2 (d - \theta - 1) \sin^2 (d - \theta) \n
Y_1 \left(d-\theta\right) = \sin^2 (d - \theta - 1) \cos^2 (d - \theta) &;&  Y_2\left(d-\theta\right) = \sin^2 (d - \theta - 1) \sin^2 (d - \theta)\,
\la{example1}
\eer

where one can see $Y_1 \left(d-\theta\right) \,, \,Y_2\left(d-\theta\right)$ indeed vanishes on substitution of $d - \theta = 1$ and  sum of the four gives the identity (\ref{toymodel}).  Indeed many more such identities are feasible where we picked up just one!

Next we see  from (\ref{example1}),    that sum of the coefficient of the first type of terms( i.e the terms nonvanishing on substitution of  $d - \theta = 1$  and reproduces (\ref{dtheta1}) on substitution  of the same ), i.e $ X_1  + X_2$ vanishes on substitution of  $(d - \theta - 1) = {\frac{\pi}{2}}$!              
\vskip0.5mm

Here we claim that the sum of the coefficient of the first type($\rho_{01a}$)  of terms must be remain nonvanishing for any value of $ d - \theta$.

Let us explain, that why we cannot have the sum of the coefficient of those terms, as given by $\rho_{01a}$ vanishing at certain value of $d - \theta$!  .   Firstly we note that the sum of the coefficients, in the above example, of both the first type  $\rho_{01a}$ and second type of term $\rho_{01b}$ is actually appears as an overall coefficient of the particular type of the terms, within  the complete expression of $\rho_0(l, \rho_c)$ and vanishing of the same for any value of $d-\theta$ implies that the absence of that type  of terms from the expression of $\rho_0(l,\rho_c)$, for that particular value of $d - \theta$!  Now in the last section we have seen that the structure of the complete solution is decided by the symmetry structure (\ref{2dsymmetry}) and the symmetry structure is completely irrespective of $d-\theta$, as we mentioned!  Now from the discussion of the last section and the present section it is evident that the different category of terms within the complete expression of $\rho_0 (l,\rho_c)$
e.g we denote it by $\rho_{01a} , \rho_{01b}$ as given in (\ref{firstsoln}), or $\rho_{03}$ from (\ref{rho03}), are of different specific structure and accordingly different category of terms are chareacterized by different structure and so they will be domnant over the different specific regime of $(l,\rho_c)$ plane, as specified by relative ratio of $(l,\rho_c)$(i.e as example $l>>\rho_c $, $l \sim \rho_c $ etc. !). Now if the above scenario is true, i.e $\rho_{01a}$ disappears from the complete solution for $\rho_{0} (l. \rho_c)$ for certain $d -\theta$. then in case, even over some specific regime of $(l,\rho_c)$ plane it is a dominant solution
then for the specific value of $d - \theta$, for which it is vanishing, it will not be the dominant solution over the same specified regime for the same $d - \theta$ and all the other values of $d - \theta$   close to it!!
However this is not only going against the prediction of the symmetry-structure (\ref{2dsymmetry}), which is saying the regime of dominance for certain category of terms within the complete solution of $\rho_{0} (l, \rho_c)$, must be decided by its structure only and should must be irrespective of $d - \theta$, but it is also the governing equation (\ref{tanmoynewform}) not making any such indication, i.e, it is not taking any special shape for any particular $d - \theta$ (except its extreme value) where this kind of twist can happen!

In other words, the complete structure of $\rho_0 (l, \rho_c)$, should must be irrespective of $d - \theta$ and should must be same for all $d - \theta$!

Finally let us tell the fact that here we have considered a very  simple identity (\ref{example1}) when given the fact that the second type of term, $\rho_{01b}$
can have infinitely many terms, the real situation is much more generalized! However we know,  we can always handle such case as well by using the trivial identity using more and more angular variables for the coefficients of the second type of terms $\rho_{01b}$ whereas the first type of terms $\rho_{01a}$,  have always two terms, so the picture will remain same! 

One can try to avoid this problem by choosing the solution for (\ref{toymodel}) through another identity

\ber
X_1 \left(d-\theta\right) = \cosh^2 (d - \theta - 1) \cos^2 (d - \theta)\,\,&;&\,\, X_2 \left(d-\theta\right) =  \cosh^2 (d - \theta - 1) \sin^2 (d - \theta) \n
Y_1 \left(d-\theta\right) = -\sinh^2 (d - \theta - 1) \cos^2 (d - \theta)\,\,&;&\,\, Y_2\left(d-\theta\right) = -\sinh^2 (d - \theta - 1) \sin^2 (d - \theta)\,
\la{example2}
\eer

`However (\ref{example2}) is showing that the part of $\rho_0 (l,\rho_c)$, which is nonvanishing on substitution of $d - \theta = 1$   and reproduces 
(\ref{dtheta1}) on the substitution of the same,  getting an extra overall functional dependence $\cosh^2 (d - \theta - 1)$, where one can observe from the following plot Fig.\ref{coshsinh}, that the same function diverge around $d - \theta - 1  = 80$, which is finite, although the complete $\rho_{01} = \rho_{01a} + \rho_{01b} $
is finite!

\begin{figure}[H]
\includegraphics[width=.65\textwidth]{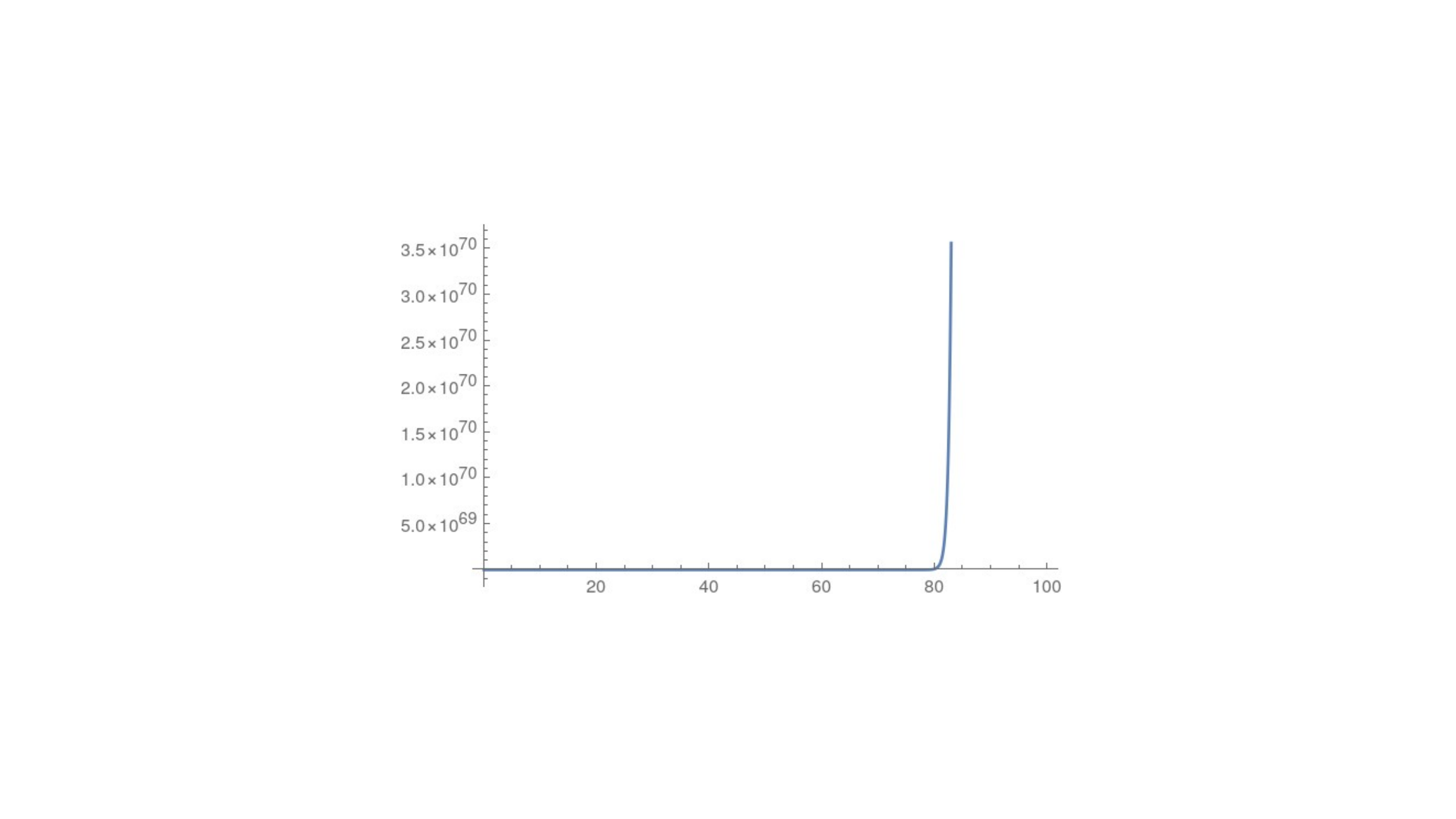}
\includegraphics[width=.65\textwidth]{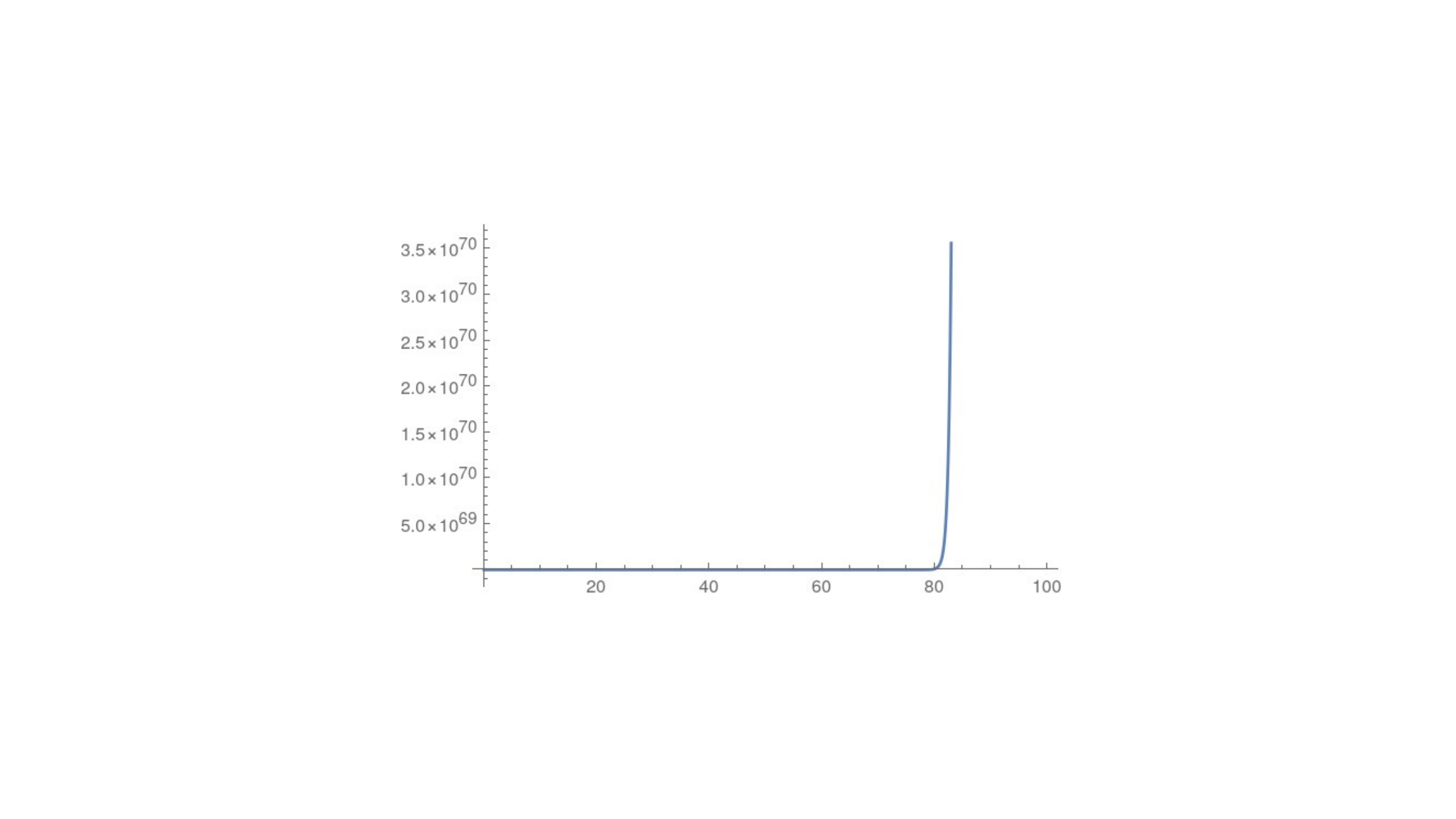}	
\caption{The plot of $\cosh(d - \theta - 1)$ \, (in the left) \,\,\, and \,\,\, $\sinh(d - \theta - 1)$\,\, (in the right)\,\,  vs $(d - \theta)$ }
\label{coshsinh}
\end{figure}

Clearly, while string theory does make some restriction on $d - \theta$, say in bosonic string we have $ d = 26 $ etc., however as far as the governing equation 
(\ref{tanmoynewform}) is concerned, it should not make any restriction on $d - \theta$,   whereas according to Fig.(\ref{coshsinh}),   we see that both the part of $\rho_{01}$, i.e the part nonvanishing on substitution of $d - \theta = 1$ and reproduces(\ref{dtheta1})  and the part vanishing on substitutiopn of $d - \theta = 1$  separately become divergent around  $d - \theta \sim 80$ (which is finite) though combinedly they may be finite!  Actually these kind of functions, which are monotonically increasing or decreasing with $d - \theta$ and nonvanishing for any value of $d - \theta$, can show this blowing up behaviour at some point!

However since in different regime of $(l, \rho_c)$ plane different part of the complete solution of $\rho_0 (l,\rho_c)$ (like $\rho_{01a}, \rho_{01b} $) can dominate where the above type function, as an overall coefficient of the part of complete expression of $\rho_0(l,\rho_c)$, which represents  particular category, can show abnormal behaviour to $\rho_0(l,\rho_c)$, in the regime where such part of the solution is dominating!
\vskip0.5mm

This gives us a view that the overall coefficient, which is the sum of the coefficient in the present case, of a particular part of $\rho_0(l,\rho_c)$, which represents a particular category, must be independent of $d-\theta$, as is expected from the symmetry-structure (\ref{2dsymmetry}) which is irrespective of $d-\theta$
 as we discussed in details! 
\vskip0.5mm
So one can try a possibility, that the sum of the coefficient of $\rho_{01a}$ and $\rho_{01b}$ can be given by "c" and "1-c" ! However the only exactly solvable case 
for $d- \theta = 1$, showing $c = 1$!

\vskip0.5mm

 This leads us to conclude that in the basic relation $ \displaystyle\sum_{f} A_f \left(d - \theta \right) = 1$, in 
(\ref{coefficientboundary}), if we express this as a sum over two parts, i.e 
$\displaystyle\sum_{f} A_f = \displaystyle\sum_{f_1}A_{f_1\,; {\rho_{01a}}} + 
\displaystyle\sum_{f_2} A_{f_2\,; {\rho_{01b}}}$, where the first one refers to the collection of coefficient of those terms, nonvanishing at $d-\theta = 1$, the other one vanishes at $d - \theta = 1$, we must have the condition that (\ref{coefficientboundary}) to be expressed as
\ber
\displaystyle\sum_{f} A_f &=& \displaystyle\sum_{f_1}A_{f_1\,; {\rho_{01a}}} + 
\displaystyle\sum_{f_2} A_{f_2\,; {\rho_{01b}}} \n
\displaystyle\sum_{f_1} A_{f_1\,; {\rho_{01a}}} &=& 1\n
\displaystyle\sum_{f_2} A_{f_2\,; {\rho_{01b}}}&=& 0
\la{furtherconditions}
\eer

Here we proceed by discussing the two type of terms, $\rho_{01a} , \rho_{01b}$ separately!
First we consider the fact that from (\ref{firstsoln}) we have

\ber
\rho_{o 1a} &=& A_1\left(d-\theta \right){\left( {\left({\frac{A_{10} l}{ 2 }}\right)}^{2  \left( d - \theta\right)} + {\left( \rho_c \right)}^{2  \left( d - \theta\right)} \right)}^{\frac{1}{2(d - \theta)}}\n
&+& C_1\left(d-\theta \right){\left( {\left({\frac{A_{10} l}{ 2 }}\right)}^{  \left( d - \theta + 1 \right)} + {\left( \rho_c \right)}^{  \left( d - \theta + 1 \right)} \right)}^{\frac{1}{(d - \theta + 1)}}
\la{completenonvanishing}
\eer
with the consistency condition,  (\ref{coefficientboundary}), according to (\ref{furtherconditions}), reduced to

\be
A_1\left(d-\theta \right) + C_1\left(d-\theta \right) = 1
\la{reduced-consistency}
\ee

The question is, which term from the above two terms in (\ref{completenonvanishing}) is more relevant near either bounadary-region or both are relevant? In order to find an answer let us note a few facts:
\vskip1mm

   Firstly the two terms in (\ref{completenonvanishing}), can be re-expressed as:
\ber
& & A_1 \left(d-\theta \right) {\left({\frac{A_{10} l}{ 2 }}\right)}{\left( 1 + {\left({\frac{ \rho_c }{{\frac{A_{10} l}{ 2 }}}}\right)}^{2  \left( d - \theta\right)}  \right)}^{\frac{1}{2(d - \theta)}} ;\n
& & C_1 \left(d-\theta \right){{\left({\frac{A_{10} l}{ 2 }}\right)}\left(1 + {\left({\frac{ \rho_c }{{\frac{A_{10} l}{ 2 }}}}\right)}^{  \left( d - \theta + 1\right)} \right)}^{\frac{1}{(d - \theta + 1)}}
\la{comparison}
\eer

Next we note that 

\be
{\rm For}\,\,\, d - \theta > \,\, 1 \quad, \quad  2(d - \theta) > d - \theta + 1  \quad; \quad  {\rm For}\,\,\, d - \theta < 1 \quad, \quad  d - \theta + 1 >  2(d - \theta) 
\la{dthetanew}
\ee

Now if we consider a specific regime near the boundary given by $(l, \rho_c = 0)$,  where we specify  the regime $l >> \rho_c $, we see that due to (\ref{dthetanew}),
the first order correction to the linear expression of $\rho_0$, where this linear expression is given by as in (\ref{boundarycondition}), $\rho_0 = {\frac{A_{10} l}{ 2 }}$, as evident from (\ref{comparison}),  is  much less in the first expression in (\ref{comparison}) compared to the second expression for $ d - \theta > 1$. 
However for $d-\theta < 1$ the situation is reversed, i.e the first order correction to the linear behaviour of $\rho_0$, in l, is much less in the second expression in (\ref{comparison}), compared to the the first one! Two are exactly identical for $d - \theta = 1$.

\vskip0.5mm
If we look at the other boundary given by $(l = 0 \,,\, \rho_c  )$, and specify the regime $\rho_c >> l$,  then pulling out the $\rho_c$-part rather than l - part, from the complete expression in (\ref{completenonvanishing}), and construct the exactly similar expression  as in (\ref{comparison}), however now $\rho_0(l,\rho_c)$ is  linear in $\rho_c$, as expected from (\ref{boundarycondition}), we again reach to the exactly same scenario, as we considered, in the case of the other boundary! 

\vskip0.5mm
Clearly then it appears that   $A_1\left(d-\theta \right){\left( {\left({\frac{A_{10} l}{ 2 }}\right)}^{2  \left( d - \theta\right)} + {\left( \rho_c \right)}^{2  \left( d - \theta\right)} \right)}^{\frac{1}{2(d - \theta)} }  $   appears to be a better solution for $\rho_{01a}$  for  $d - \theta > 1$ in the regime $l >> \rho_c$ and $\rho_c >> l$,  whereas $C_1\left(d-\theta \right){\left( {\left({\frac{A_{10} l}{ 2 }}\right)}^{  \left( d - \theta + 1 \right)} + {\left( \rho_c \right)}^{  \left( d - \theta + 1 \right)} \right)}^{\frac{1}{(d - \theta + 1)}}    $ appears to be a better solution for $\rho_{01a}$ for $d - \theta < 1$ for the regime 
$l >> \rho_c$ and $\rho_c >> l$!  
\vskip0.5mm

Now,  the question is,   as appears in (\ref{firstsoln}),  can we have the combination of the two, as a common solution for all values of $(d - \theta$?  
\vskip0.5mm

Firstly,  $  A_1\left(d-\theta \right){\left( {\left({\frac{A_{10} l}{ 2 }}\right)}^{2  \left( d - \theta\right)} + {\left( \rho_c \right)}^{2  \left( d - \theta\right)} \right)}^{\frac{1}{2(d - \theta)} }     $  is undefined for $d - \theta = 0$, so it can be used as a solution for $d - \theta >  1$!  Also from 
(\ref{comparison}) its evident that if the above solution for $\rho_{01a}$ is being used for $d - \theta < 1$,  then more and more $d - \theta$ will be smaller, the leading order correction to the linear term in $l>> \rho_c$ and $\rho_c >> l$ will increase and exactly at $ d - \theta = 0 $ it becomes equal to the linear term itself! So it can never appear in the solution for $\rho_{01a}$ for $d - \theta < 1$!  On the otherhand $C_1 {\left( {\left({\frac{A_{10} l}{ 2 }}\right)}^{  \left( d - \theta + 1 \right)} + {\left( \rho_c \right)}^{  \left( d - \theta + 1 \right)} \right)}^{\frac{1}{(d - \theta + 1)}}$ appears to be a better solution for $\rho_{01a}$ for $d - \theta < 1$ for the regime $l >> \rho_c$ and $\rho_c >> l$  because one can see that for any value of $d - \theta < 1$, the leading order correction is always much much less than the leading order term!
\vskip0.5mm
So we conclude for the solution for $\rho_{01a}$, it is given by

\ber
{\rm for} \, \, \, d - \theta > 1\,\,\,\, \rho_{01a} &=& {\left( {\left({\frac{A_{10} l}{ 2 }}\right)}^{2  \left( d - \theta\right)} + {\left( \rho_c \right)}^{2  \left( d - \theta\right)} \right)}^{\frac{1}{2(d - \theta)}}\n
{\rm for} \, \, \, d - \theta < 1 \, ; \, \rho_{0 1 b} &=& {\left( {\left({\frac{A_{10} l}{ 2 }}\right)}^{  \left( d - \theta + 1 \right)} + {\left( \rho_c \right)}^{  \left( d - \theta + 1 \right)} \right)}^{\frac{1}{(d - \theta + 1)}}\n
\la{ultimaterho0nonvanishing}
\eer

Along with the boundary consistency condition, with (\ref{reduced-consistency}) reduced to

\ber
{\rm for} \, \, \, d - \theta > 1\,\,\,A_1\left(d-\theta \right)  &=& 1\,\, , \,\,C_1\left(d-\theta \right) = 0\n
{\rm for} \, \, \, d - \theta < 1\,\,\,A_1\left(d-\theta \right)  &=& 0\,\, , \,\,C_1\left(d-\theta \right) = 1\n
\la{boundary-consistency-redefined}
\eer

Note that both the solutions in (\ref{ultimaterho0nonvanishing}) becomes exactly identical at $d-\theta = 1$ and also become identical with (\ref{dtheta1})!
Note that, here, all the arguments we have used to select the solution for $d-\theta > 1$ amd $d - \theta <1$, has been cross checked  in the next subsection, to show its numerical consistency.  We numerically found, if we proceed in the reverse way, i.e the solution we have selected for $ d-\theta > 1$, try to see it as a solution for $d - \theta < 1$ or vice versa, we will see, that is not consistent!
\vskip0.5mm

Next, following the above discussion, we write the most general solution for $\rho_{01b}$, i.e the part of $\rho_0 (l,\rho_c)$, nonvanishing on both the boundary and vanishes on substitution of $ d -\theta = 1$!  Following the above discussion these terms must be of two types, i.e one type for $d - \theta > 1$ and the other type for $ d - \theta < 1$ and the both should be identical at $ d - \theta = 1$.   Now these terms,  since according to the symmetry structure (\ref{2dsymmetry}) should must be of the same structure for every $d - \theta$ so these terms will must be a continuous funxtion of $d - \theta$! So although these functions vanishes at $d - \theta = 1$ however the functional continuity  at $d - \theta = 1$ demands that at the limit  
\be
\rho_{01b} {|}_ {d - \theta = 1 + \epsilon} = \rho_{01b} {|}_ {d - \theta = 1 - \epsilon} \quad; \quad {\rm with} \quad \epsilon \rightarrow 0
\la{continuity}
\ee

Along with that we have the general form of $\rho_{01b}$ given by

\vskip2mm
\textbf{For}  $d - \theta > 1$

\vskip2mm

\ber
\rho_{01b}  (l, \rho_c) &=& 
 \displaystyle\sum_{n > 1 }^\infty   A_{n}\left(d-\theta \right)\left\lbrack \displaystyle\sum_{m > 0}^\infty 
\,\,  B_{n m}\left(d-\theta \right) {\left\lbrace \left( d - \theta\right) - 1 \right\rbrace}^ m \right\rbrack \times\n
& & {\left( {\left({\frac{A_{10} l}{ 2 }}\right)}^{ 2 n \left( d - \theta\right)}    + 
{\left( \rho_c \right)}^{2 n \left( d - \theta\right)}   \right)}^{\frac{1}{2 n (d - \theta)   }}                  \n
&+& \displaystyle\sum_{p > 1}^\infty    A_{p}\left\lbrack \displaystyle\sum_{m > 0}^\infty  B_{p m} {\left\lbrace \left( d - \theta\right) - 1 \right\rbrace}^ m\right\rbrack \times\n
& & {\left\lbrack {\left({\frac{A_{10} l}{ 2 }}\right)}^{  2 p \left( d - \theta  \right)}   + \displaystyle\sum_{k} a_{k} 
{\left({\frac{A_{10} l}{ 2 }}\right)}^{ \left\lbrace  2 p\left( d - \theta \right)  -  2 k \left(d - \theta \right)\right\rbrace } {\left( \rho_c \right)}^{k}  + 
........ .. +
{\left( \rho_c \right)}^{  2p \left( d - \theta \right)  } \right \rbrack}^{\frac{1}{ 2p (d - \theta )  }} +...\,,
\la{rho01b1}
\eer

where n any positive integer/noninteger greater than one!

\vskip2mm
\textbf{For}  $d - \theta < 1$

\vskip2mm

\ber
\rho_{01b}  (l, \rho_c) &=& 
 \displaystyle\sum_{n > 1 }^\infty   A_{n}\left(d-\theta \right)\left\lbrack \displaystyle\sum_{m > 0}^\infty 
\,\,  B_{n m}\left(d-\theta \right) {\left\lbrace \left( d - \theta\right) - 1 \right\rbrace}^ m \right\rbrack \times\n
& & {\left( {\left({\frac{A_{10} l}{ 2 }}\right)}^{ n \left( d - \theta  + 1\right)}    + 
{\left( \rho_c \right)}^{ n \left( d - \theta + 1\right)}   \right)}^{\frac{1}{ n (d - \theta + 1)   }}                 \n
&+& \displaystyle\sum_{p > 1}^\infty    A_{p}\left\lbrack \displaystyle\sum_{m > 0}^\infty  B_{p m} {\left\lbrace \left( d - \theta\right) - 1 \right\rbrace}^ m\right\rbrack \times\n
& & {\left\lbrack {\left({\frac{A_{10} l}{ 2 }}\right)}^{  p \left( d - \theta + 1 \right)}   + \displaystyle\sum_{k} a_{k} 
{\left({\frac{A_{10} l}{ 2 }}\right)}^{ \left\lbrace  p\left( d - \theta + 1 \right)  -  k \left(d - \theta + 1 \right)\right\rbrace } {\left( \rho_c \right)}^{k}  + 
........ .. +
{\left( \rho_c \right)}^{  p \left( d - \theta + 1\right)  } \right \rbrack}^{\frac{1}{ p (d - \theta + 1)  }} +...\,,
\la{rho01b2}
\eer

Along with, following (\ref{furtherconditions} , \ref{coefficientboundary})   the sum of the coefficients will give zero.....

\ber
& &\displaystyle\sum_{n > 1 }^\infty   A_{n}\left(d-\theta \right)\left\lbrack \displaystyle\sum_{m > 0}^\infty 
\,\,  B_{n m}\left(d-\theta \right) {\left\lbrace \left( d - \theta\right) - 1 \right\rbrace}^ m \right\rbrack +\n
& &\displaystyle\sum_{p > 1}^\infty    A_{p}\left\lbrack \displaystyle\sum_{m > 0}^\infty  B_{p m} {\left\lbrace \left( d - \theta\right) - 1 \right\rbrace}^ m\right\rbrack = 0
\la{coefficient}
\eer

Next from (\ref{rho0soln}), we consider $\rho_{02}$, which is within the symmetry invariant part of $\rho_0$, gives vanishing contribution on one boundary and remain nonvanishing on the other.  Note that these are also the kind of terms, vanishes on the substitution of $d - \theta = 1$.

\vskip0.5mm
Clearly these type of terms given by the sum of the two terms
\be
\rho_{02} = A_{\rm vanishing}\left( d - \theta \right) l  \,\, ; \,\,  B_{\rm vanishing} \left( d - \theta \right)\rho_c \,
\la{rho02}
\ee

where $A_{\rm vanishing}\left( d - \theta \right)$ and $B_{\rm vanishing} \left( d - \theta \right)$ are the coefficients which vanishes on substitution of $d-\theta = 1$
Clearly, these type of term, if exists, their coefficient will contribute to the total sum of coefficient in (\ref{vanishing sum}) in one boundary and not in the other, since (\ref{boundarycondition}) must have to hold!  However this cannot happen in practice, because the terms within the sum of the coefficients, as given by 
$\displaystyle\sum_{f_2} A_{f_2}\left(d - \theta \right) = 0$   is completely fixed as a trivial identity and the mentioned sum must be same w.r.t both the boundaries, since $\rho_0 (l,\rho_c)$ is a global function!    

\vskip0.5mm
So we conclude, the terms, given by the category (\ref{rho02}) cannot exist and so in (\ref{rho0soln}), the second term $\rho_{02}^{\rm soln}$  does not exist !

\vskip1mm
Next, from (\ref{rho0soln}), we choose the third category of the terms, which vanishes on both the boundaries, discussed in the last section in (\ref{rho03} ), which we mention here again :

\ber
\rho_0^3 &=& \displaystyle\sum_{m.n} a_{m,n} {\left({\frac{A_{10}l}{2}}\right)}^m \rho_c^n \n
& & {\rm with} \,\,m + n = 1\,\,;\,\,m \le 1\,,\,n\le 1\,\,,\,\, {\rm with} \,\,(m = 0, n = 1 )\,\, , \,\,( m = 1, n = 0)\,\, {\rm excluded}\n
\la{rho03again}
\eer

Here m, n must be dependent on $d -\theta$, although we do not know their exact form!
A possible form, admitting the restriction on m,n from (\ref{rho03again}),  a possible form of this this kind of terms can be

\ber
\rho_0^3 &=& \displaystyle\sum_{m.n} a_{m,n} {\left({\frac{A_{10}l}{2}}\right)}^{\frac{m}{2(d-\theta)}} \rho_c^{\frac{m}{2(d-\theta)}}, \,\, {\rm for}\,\, d - \theta > 1\n
&=& \displaystyle\sum_{m.n} a_{m,n} {\left({\frac{A_{10}l}{2}}\right)}^{\frac{m}{(d-\theta + 1)}} \rho_c^{\frac{n}{(d-\theta + 1)}}, \, \, {\rm for}\,\,  d - \theta < 1 \n
\la{rho03againpossibletype}
\eer

Now let us consider the expression of $\rho_0 (l , \rho_c)$ from (\ref{solution}).  Clearly this has three type of terms,,\,\,  $\rho_{01}$ \,:\, Scaling-symmetry-invariant, nonvanishing on both the boundaries given by $(l,\rho_c = 0)$ and $(l = 0, \rho_c)$ ,\,\, which is the sum of two type of terms  $\rho_{01a}$ and $\rho_{01b} $,  and also a part  $\rho_{03}$\,:\, Scaling-symmetry-invariant, vanishing on both  the boundaries given by $(l,\rho_c = 0)$ and $(l = 0, \rho_c)$\,\,, 

\vskip0.5mm
Let us focus on a specific regime of $(l,\rho_c)$ plane, given by $l>>\rho_c$ and  $\rho_c >> l$ .   Based on all the previous analysis, in order to understand the leading term of $\rho_0 (l, \rho_c)$ over this regime,  here we are going to write the leading contributions of all this 3 parts, $\rho_{01a},\rho_{01b}, \rho_{03}$ ,    First we note  the  part, which is nonvanishing on the above-mentioned boundaries  $\rho_{01}$  is the sum of two parts

\be
\rho_{01} = \rho_{01a} + \rho_{01b}
\la{rho01ab}
\ee

\vskip0.5mm
1. $\rho_{01a}$ \, : \, These are the terms  in $\rho_0$ which are nonvanishing on the boundary $(l,\rho_c = 0)$ and $(l = 0, \rho_c)$ and denoted by $\rho_{01}$,  which are direct generalization of the terms in the case of $d-\theta = 1$, (\ref{dtheta1}) exactly as mentioned in (\ref{ultimaterho0nonvanishing}). Clearly from  (\ref{ultimaterho0nonvanishing}), it is evident that, 
\ber
{\rm for}\, l>>\rho_c \,\, ; \,\, \rho_{01a} &\sim& {\frac{A_{10} l}{2}}\n
{\rm for} \,\rho_c >>l \,\, ; \,\,\rho_{01a}&\sim&  \rho_c
\la{rho01a1}
\eer

\vskip0.5mm
2. $\rho_{01b}$ \, : \, These are the terms in $\rho_0$ which are nonvanishing on the boundary $(l,\rho_c = 0)$ and $(l = 0, \rho_c)$ and denoted by $\rho_{01}$, which are not the direct generalization the terms in the case of $d-\theta = 1$, (\ref{dtheta1}) and hence given by general form as given in (\ref{rho01b2}),

In order to understand their contribution in the regime $ l >> \rho_c$ and $\rho_c >> l$  first we recall (\ref{coefficient}) where it was shown that 
the sum of their coefficient is vanishing which indeed gives the most dominant contribution from these terms to be zero!
To see the first order correction to these terms for $d - \theta > 1$ and $d - \theta < 1$

\ber
& &\rho_{01b} = \displaystyle\sum_{f_2} A_{f_2} \left(d - \theta \right){\left( {\left({\frac{A_{10} l}{ 2 }}\right)}^{f \left( d - \theta\right)} + {\left( \rho_c \right)}^{f  \left( d - \theta\right)} \right)}^{\frac{1}{f(d - \theta)}}\n
\Rightarrow &=& \displaystyle\sum_{f_2} A_{f_2} \left(d-\theta \right) {\left({\frac{A_{10} l}{ 2 }}\right)}{\left( 1 + {\left({\frac{ \rho_c }{{\frac{A_{10} l}{ 2 }}}}\right)}^{f_2  \left( d - \theta\right)}  \right)}^{\frac{1}{2f_2 (d - \theta)}} \,\, ; \, {\rm near}\,\, {\rm boundary} \,\,(l, \rho_c = 0)\n
&=& \displaystyle\sum_{f_2} A_{f_2} \left(d-\theta \right) {\rho_c}{\left( 1 +  {\left({\frac{{\frac{A_{10} l}{ 2 }}  }{\rho_c}}\right)}^{f_2  \left( d - \theta\right)}  \right)}^{\frac{1}{2f_2 (d - \theta)}} \,\, ; \, {\rm near}\,\, {\rm boundary} \,\,(l = 0, \rho_c )\n
\la{vanishing sum}
\eer

with $\displaystyle\sum_{f_2} A_{f_2} \left( d - \theta)\right) = 0$

Here by $f_2 \left(d -\theta\right)$ we mean $2 m \left( d -\theta \right)$ for $d - \theta > 1$ and  $m (d - \theta + 1)$     for $d- \theta < 1$ and the two are identical at $d - \theta = 1$ for m any positive integer/noninteger greater than one!

So if we focus on the regime $l>> \rho_c$ and $\rho_c >> l$, since it follows from (\ref{vanishing sum})\, :

1. The leading-order term vanishes in this specific regime   because the sum of its total coefficient vanishes!
\vskip0.5mm
2.  The first order correction in this regime,  as evident from (\ref{vanishing sum}),  is negligible compared to the leading term in $\rho_{01a}$ as given in 
 (\ref{rho01a1}) ! 
\vskip0.5mm

So we conclude, over this specific regime, in the complete   solution for $\rho_0 (l , \rho_c)$ ,  the specific part, $\rho_{01b}$ is negligible compared to $\rho_{01a}$

Finally for $\rho_{03}$, we see the expression are of ${\left({\frac{A_{10} l}{2}}\right)}^m \rho_c^n$ with $ 0 < m, n < 1$  as given in (\ref{rho03again})!

Now, if we recall (\ref{rho01a1}) since the dominant contribution at the regime $l>> \rho_c $ can be written in the form
$${\left({\frac{A_{10} l}{2}}\right)}^m {\left({\frac{A_{10} l}{2}}\right)}^n$$  since $m + n = 1$.  So clearly  the terms within $\rho_{03}$ is much much less compared to the one in $\rho_{01a}$ because , in this regime

$$ \rho_c^n << {\left({\frac{A_{10} l}{2}}\right)}^n $$ !    Same argument can be made for the regime $$\rho_c >> l$$ as well!

\vskip1mm

Before going to conclusion, we recall, that in (\ref{rho0soln}) we mentioned about a fourth type of tern  $\rho_{04}$, which can be of the form

\be
\rho_{04} = a \, f_1 \left({\frac{\rho_c}{l}}\right) + b\,f_2 \left({\frac{l}{\rho_c}}\right)
\la{4}
\ee
where a,b are constant and ${f_1}, {f_2}$ are any function! However since these type of terms are divergent on either boundary so they cannot be present in global solution for $\rho_0(l , \rho_c)$.    So we conclude, in the global solution for $\rho_0 (l,\rho_c)$, in the regime given by $l>> \rho_c$ and $\rho_c >> l$, it is $\rho_{01a}$, which is given by (\ref{ultimaterho0nonvanishing})  is considered to be unique solution for $\rho_0 (l , \rho_c)$ since the other contribution are negligible compared to the same over this regime!

\vskip0.5mm

Finally we will come to the question that the global symmetry,  as we have defined on the parameter space of the dual of $T{\overline{T}}$ deformed boundary theory, does it have any possible spacetime origin?   To find an answer we first consider the impact of this symmetry in the context of Holographic mutual information in 
(\ref{symmetryhmi}),  where when we introduce two strip on the boundary, each of length l and introduce the separation between them as h,  the scaling symmetry requires an extended form $(l,\rho_c, h) \rightarrow (kl, k\rho_c, kh) $!  Indeed we can introduce n number of boundary subregion with (n - 1) number of separations between them on any specific x-axis,   say $x_d$ in (\ref{finalmetric}),  this scaling symmetry must be extended to each of the n  strip length and each of the (n - 1) subregion h \, !  This clearly gives an impression that this scaling symmetry is the symmetry od $x_d$ axis itself, i.e $x_d \rightarrow k x_d$  which can be a part of larger global symmetry  like $x_i \rightarrow k x_i$  with $i = 1,2, .....,d$!   Next, from (\ref{murcrelationew}),  its evident that $\rho_c \rightarrow k\rho_c  \, \Rightarrow r_c \rightarrow {\frac{r_c}{k}} \,\, {\rm or} \,\, r \rightarrow {\frac{r}{k}}$.  However   the symmetry given by    

\be
x_i \rightarrow k x_i \, , \, r_c \rightarrow {\frac{r_c}{k}} 
\la{possiblesymmetry}
\ee
is not really a symmetry of bulk-geometry or the cut-off boundary of the Hyperscaling violating background as given by (\ref{finalmetric}) and (\ref{boundarymetric})    even on a fixed time-slice  because of the presence of the overall factor $r^{- \frac{2\theta}{d}}$ or $r^{- {\frac{2\theta}{d}}}_c$  in the bulk   and boundary  metric respectively!   Thatswhy we are calling (\ref{possiblesymmetry})  as the emergent symmetry,   which is not there in the original theory but just emerging on  the application of RT formalism on such  geometry with finite radial cut off!  Perhaps a possible way to visualize the symmetry (\ref{possiblesymmetry}) in our context is that if we demand that the holography should persist
even in the presence of finite radial cut off and that way perhaps one can think of the boundary surface at $r = r_c$,  given by the metric (\ref{boundarymetric}),  have a CFT description so that perhaps one can remove this overall factor $r^{- \frac{2\theta}{d}}$, through a possible conformal transformation and next if we consider a fixed time slice on the respective boundary(since RT proposal is for time-independent case),   it would be given by   $ r^2  \displaystyle\sum_{i = 1}^d  d x_i^2 $,  which  indeed obeys such symmetry (\ref{possiblesymmetry}) !   However one really need to explore this direction in depth, subjected to further study!

\subsection{Mathematica plot to  cross-check the  most dominant part of  $\rho_0(l,\rho_c)$ over the regime $l >> \rho_c$ and $\rho_c >> l$ both for $d - \theta > 1$ and $d - \theta < 1$}
 
 Here we rewrite the basic relation (\ref{tanmoy}) as
  
\ber 
 {\frac{l}{2 \rho_0}}
  &+& \left({\frac{\rho_c}{\rho_0}}\right)^{d - \theta +1} {\frac{   \,\, {{}_2 F_1}   \left\lbrack {\frac{1}{2}}, {\frac{1}{2}}(1 +{ \frac{1}{d - \theta}}), {\frac{1}{2}}(3 +{ \frac{1}{d - \theta}}) , \left({\frac{\rho_c}{\rho_0}}\right)^{2(d - \theta)}  \right \rbrack }{ d + 1 - \theta   }}\n
 &-&  {\frac{   \,\, {{}_2 F_1}   \left\lbrack {\frac{1}{2}}, {\frac{1}{2}}(1 +{ \frac{1}{d - \theta }}), {\frac{1}{2}}(3 +{ \frac{1}{d - \theta }}) , 1 \right \rbrack }{ d+1-  \theta   }} = 0
  \la{rewritebasic}
  \eer
 
 Next we plot the l.h.s of ( \ref{rewritebasic})   as the function of $(\rho_c , l)$, for $d - \theta > 1$  as well as for $d - \theta < 1$,   with the substitution of the expression pf $\rho_0$ from (\ref{ultimaterho0nonvanishing})  and check  the plots whether the function, which is the l.h.s of  (\ref{rewritebasic}) , is actually going to zero, for given functional form of $\rho_0 (\rho_c , l)$

\vskip5mm
\textbf{The numerical consistency for the global solution $\rho_0$ for $d-\theta > 1$}

\vskip5mm

In this parameter regime we have the expression for the global solution for $\rho_0$ is given by

\ber
{\rm for} \, \, \, d - \theta > 1\,\,\,\, \rho_{0\, , \, {\rm scaling \, invariant}} &=& {\left( {\left({\frac{A_{10} l}{ 2 }}\right)}^{2  \left( d - \theta\right)} + {\left( \rho_c \right)}^{2  \left( d - \theta\right)} \right)}^{\frac{1}{2(d - \theta)}}\n
\la{firstultimaterho0solution}
\eer

We will substitute (\ref{firstultimaterho0solution}) in the l.h.s of (\ref{rewritebasic}), draw a 3D plot with l.h.s of (\ref{rewritebasic}), l ,$\rho_c$, to check whether plot is indeed going to zero for $l >> \rho_c$ and $\rho_c >> l$ regime.

\begin{figure}[H]
\begin{center}
\textbf{ For $ d - \theta = {\frac{5}{3}}$  }
\end{center}
\vskip2mm
\includegraphics[width=.65\textwidth]{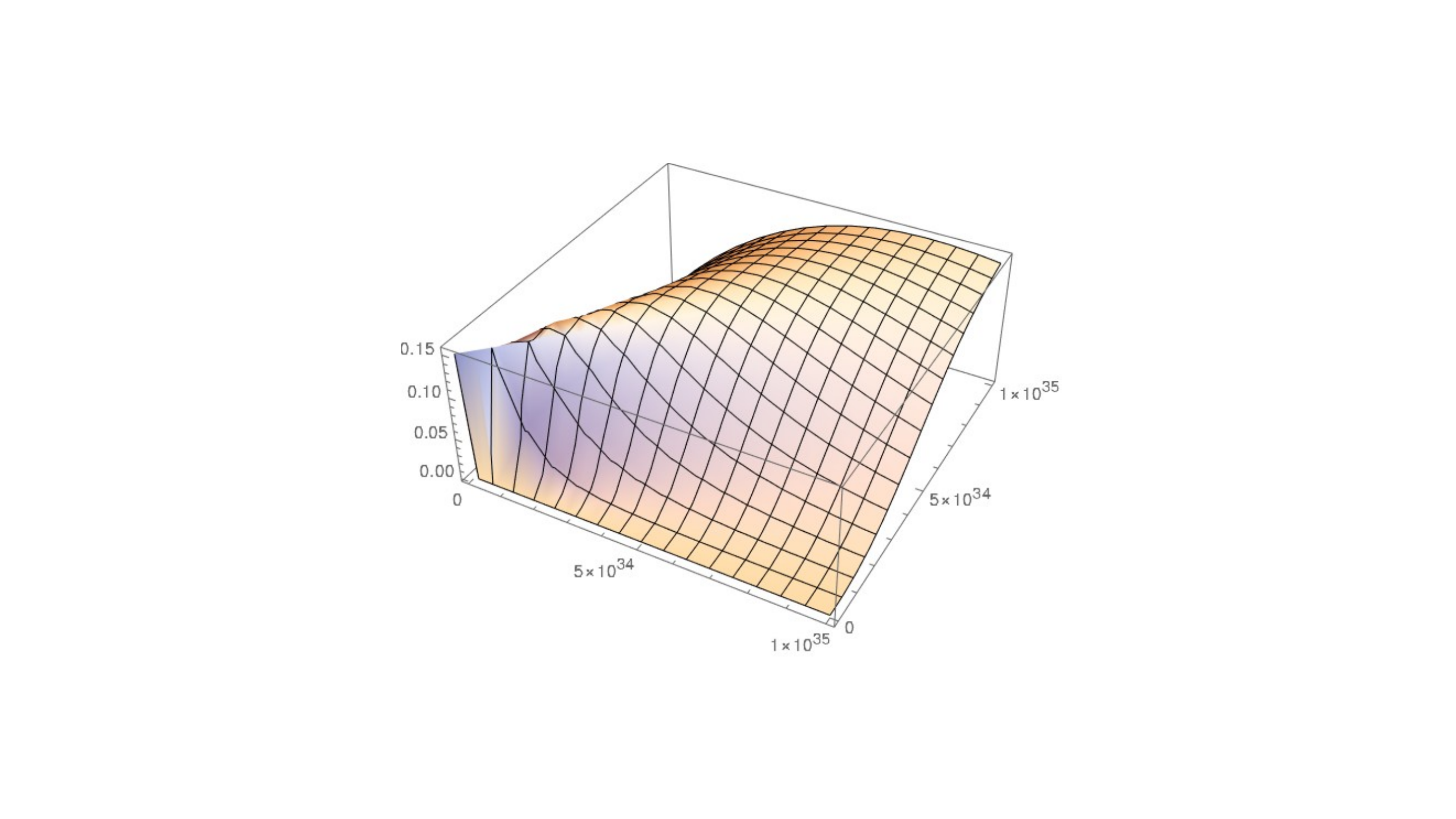}
\includegraphics[width=.65\textwidth]{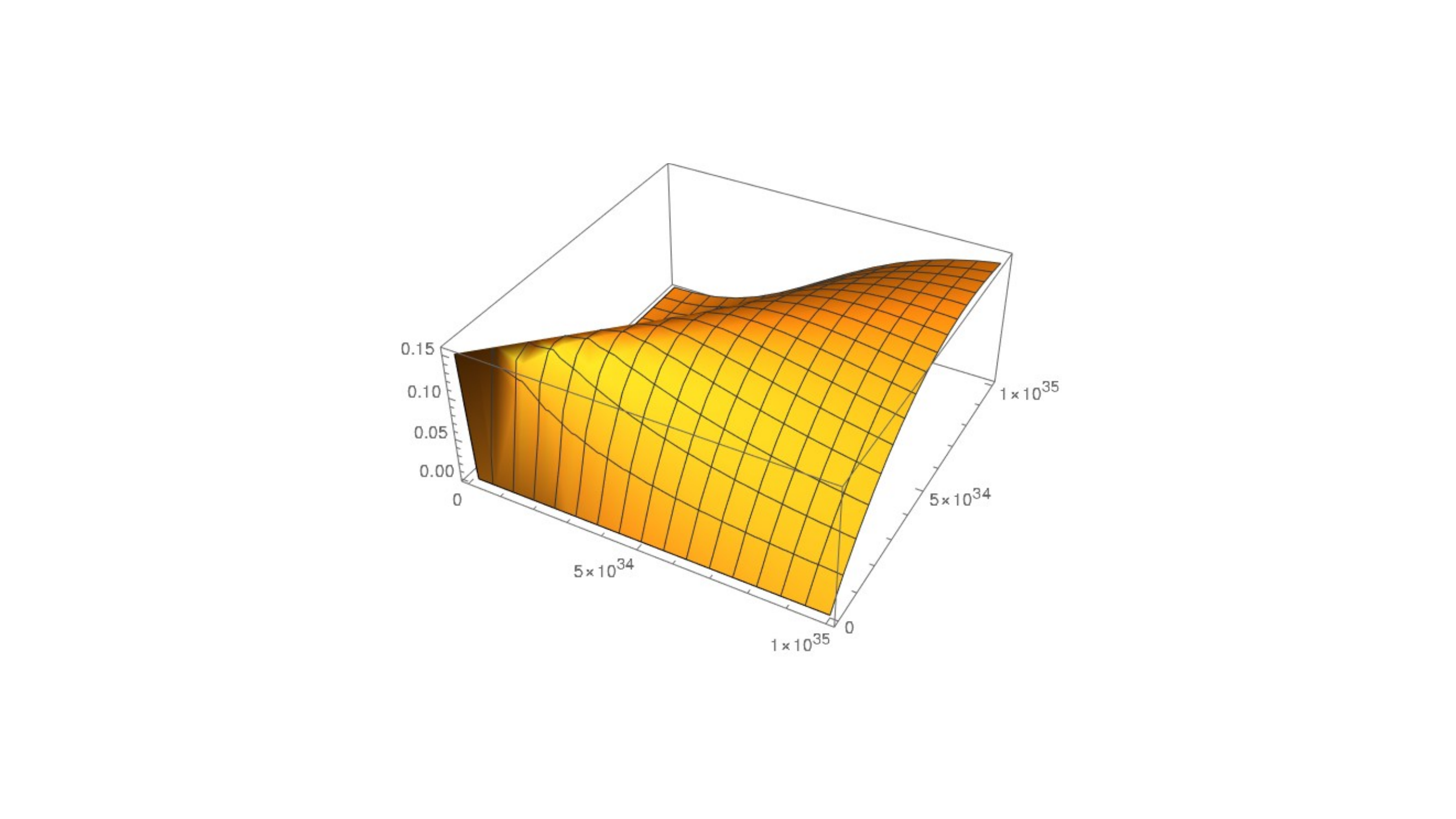}	
\caption{ The 3D plot for the said crosschecking of the part of the solution for $\rho_{0} (l,\rho_c)$ which is the main solution in the regime  $l>>\rho_c$ and $\rho_c >> l$, as one obtain from  (\ref{ultimaterho0nonvanishing}) and much much more dominant over any other contribution to $\rho_{0} (l,\rho_c)$  in this regime,  \quad;\quad  (In the left)\,:\, The 3D plots with x-axis \,:\,l, y-axis $\rho_c$, showing consistency for $l >> \rho_c$ regime \quad;\quad         (In the right)\,:\, The same 3D plots with x-axis\,:\, $\rho_c$\,,\, y-axis\,:\, l.\,\,showing the consistency for $\rho_c >> l$ regime }
\label{consistency 5by3}
\end{figure}

\begin{figure}[H]
\begin{center}
\textbf{ For $ d - \theta = {\frac{11}{4}}$  }
\end{center}
\vskip2mm
\includegraphics[width=.65\textwidth]{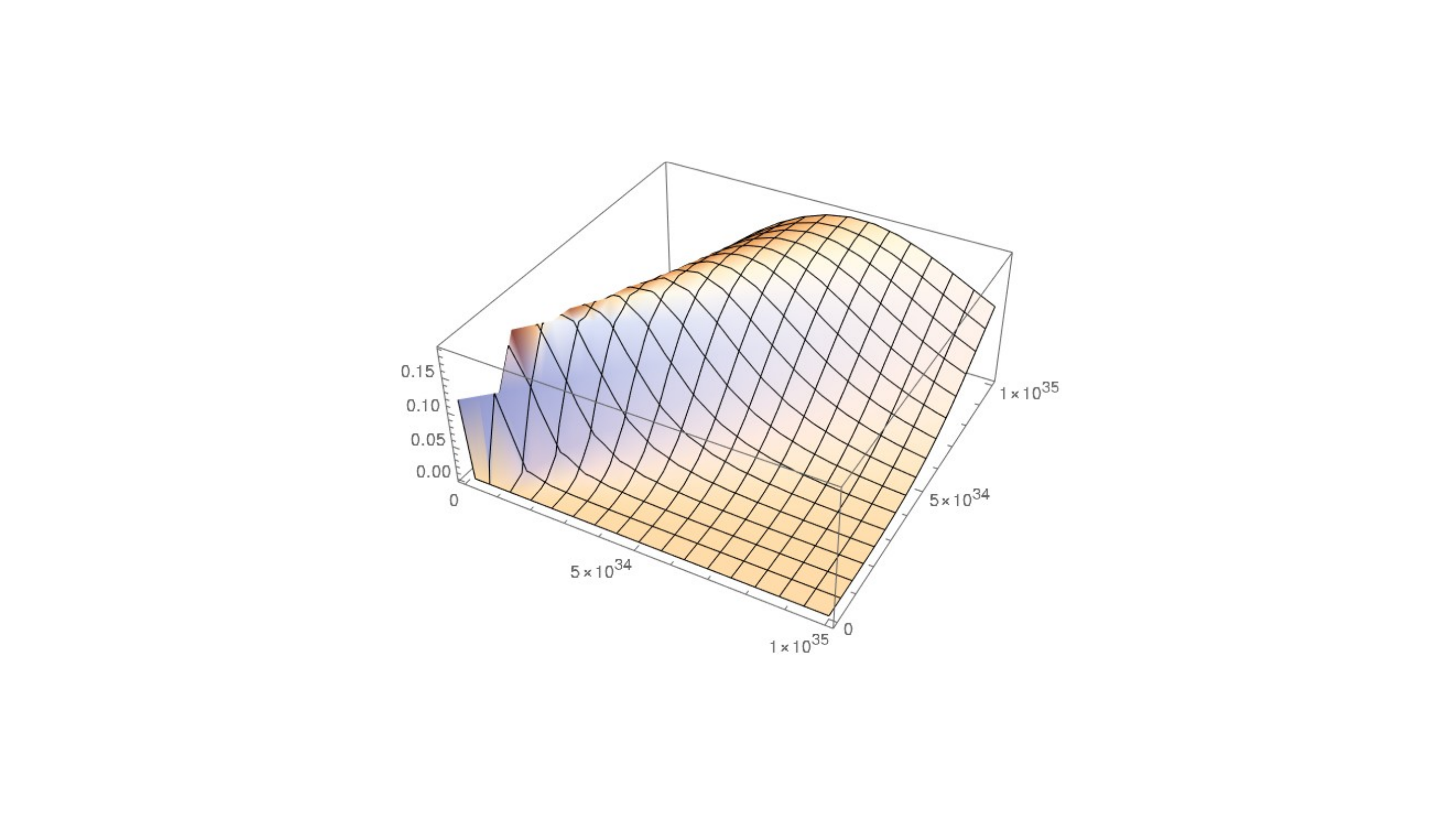}
\includegraphics[width=.65\textwidth]{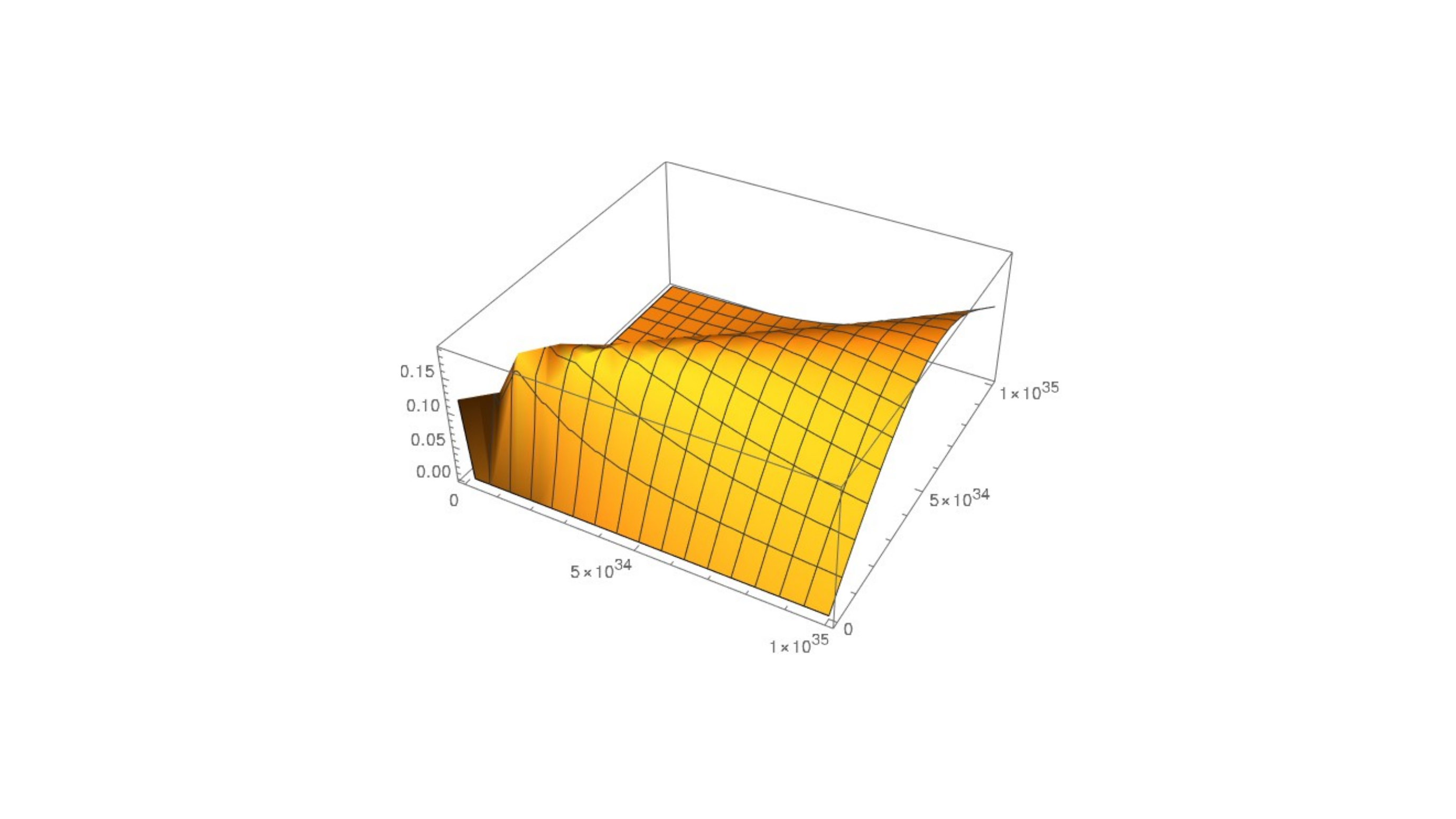}	
\caption{ The 3D plot for the said crosschecking of the part of the solution for $\rho_{0} (l,\rho_c)$ which is the main solution in the regime  $l>>\rho_c$ and $\rho_c >> l$, as one obtain from  (\ref{ultimaterho0nonvanishing}) and much much more dominant over any other contribution to $\rho_{0} (l,\rho_c)$  in this regime,  \quad;\quad  (In the left)\,:\, The 3D plots with x-axis \,:\,l, y-axis $\rho_c$, showing consistency for $l >> \rho_c$ regime \quad;\quad         (In the right)\,:\, The same 3D plots with x-axis\,:\, $\rho_c$\,,\, y-acis\,:\, l.\,\,showing the consistency for $\rho_c >> l$ regime }
\label{consistency 11by4}
\end{figure}

\begin{figure}[H]
\begin{center}
\textbf{ For $ d - \theta = {\frac{5}{9}}$  }
\end{center}
\vskip2mm
\includegraphics[width=.65\textwidth]{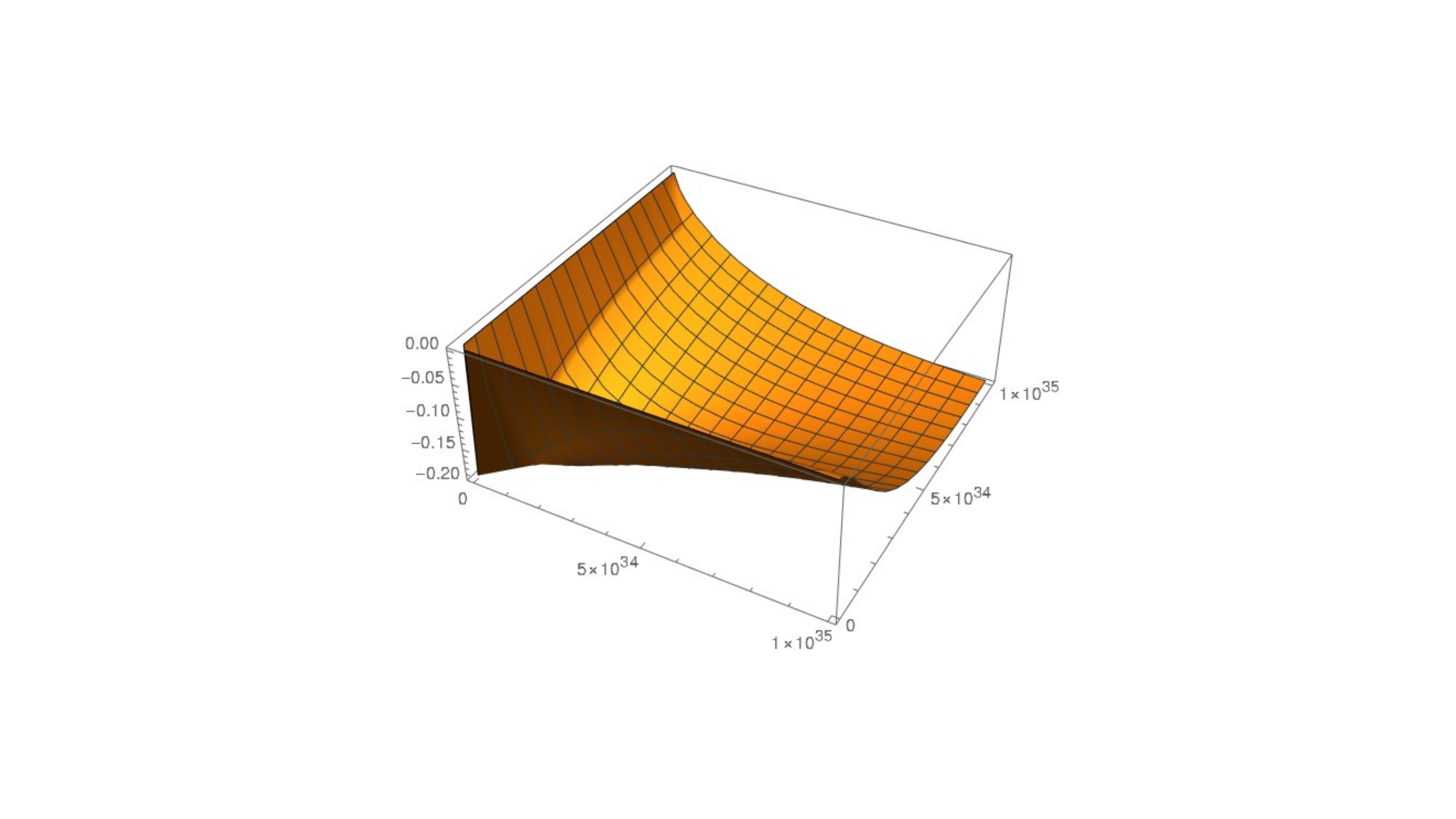}
	
\caption{ The 3D plot for the said crosschecking of the part of the solution for $\rho_{0} (l,\rho_c)$ which is the main solution in the regime  $l>>\rho_c$ and $\rho_c >> l$, as one obtain from  (\ref{ultimaterho0nonvanishing})  for $d - \theta < 1$   and much much more dominant over any other contribution to $\rho_{0} (l,\rho_c)$  in this regime)
\quad;\quad   The image showing the 3D plots with x-axis \,:\,l, y-axis $\rho_c$, showing consistency for $l >> \rho_c$  and $\rho_c >> l $ regime 
}
\label{consistency 5by9}
\end{figure}

\begin{figure}[H]
\begin{center}
\textbf{ For $ d - \theta = {\frac{1}{5}}$  }
\end{center}
\vskip2mm
\includegraphics[width=.65\textwidth]{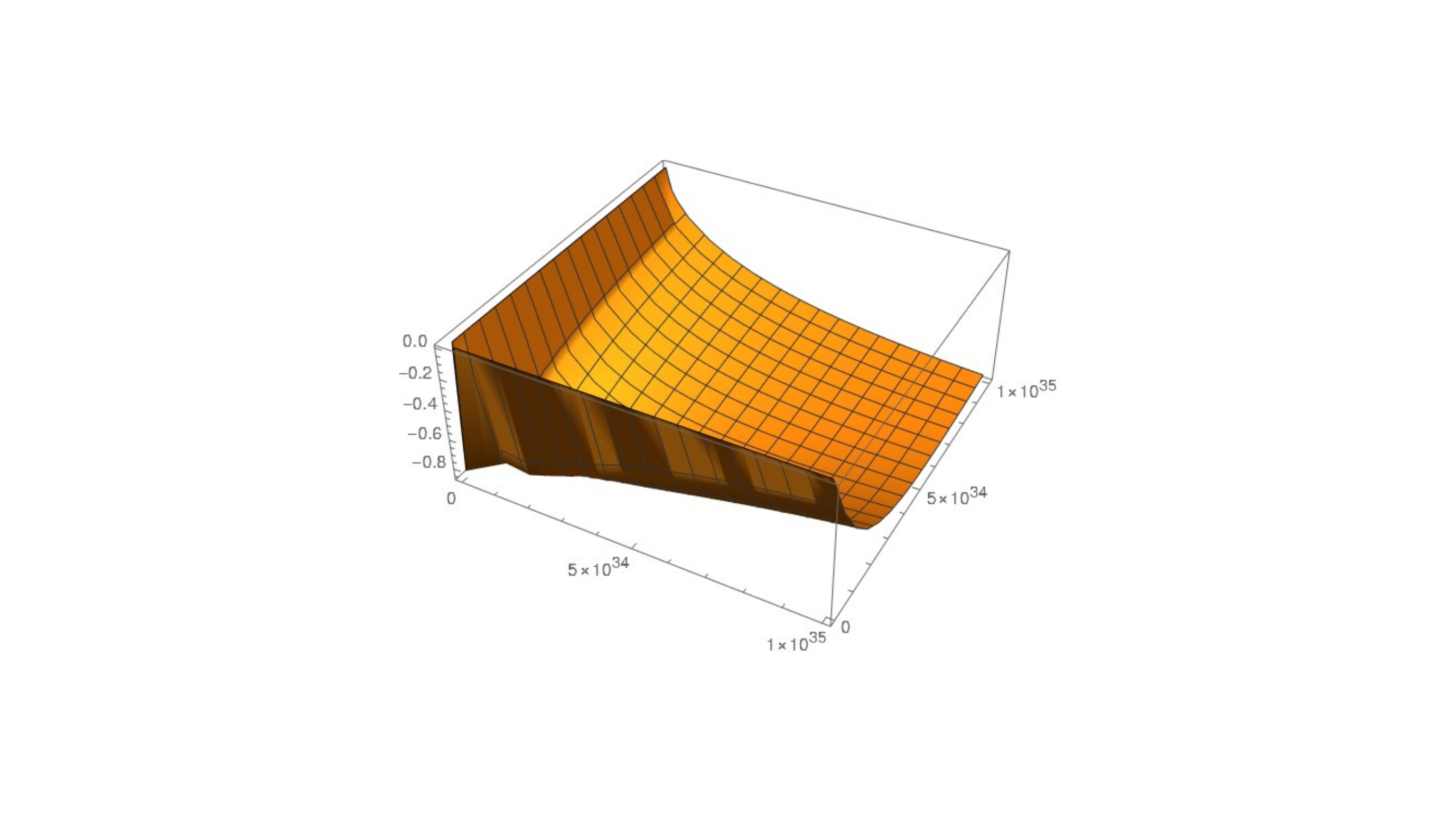}
	
\caption{ The 3D plot for the said crosschecking of the part of the solution for $\rho_{0} (l,\rho_c)$ which is the main solution in the regime  $l >> \rho_c$ and 
$\rho_c >> l$, as one obtain from  (\ref{ultimaterho0nonvanishing})  for $d - \theta < 1$   and much much more dominant over any other contribution to 
$\rho_{0} (l,\rho_c)$  in this regime, \quad;\quad   The image showing the 3D plots with x-axis \,:\,l, y-axis $\rho_c$, showing consistency for $l >> \rho_c$  and $\rho_c >> l $ regime}

\label{consistency 1by5}
\end{figure}

\begin{figure}[H]
\begin{center}
\textbf{ For $ d - \theta = {\frac{1}{9}}$  }
\end{center}
\vskip2mm
\includegraphics[width=.65\textwidth]{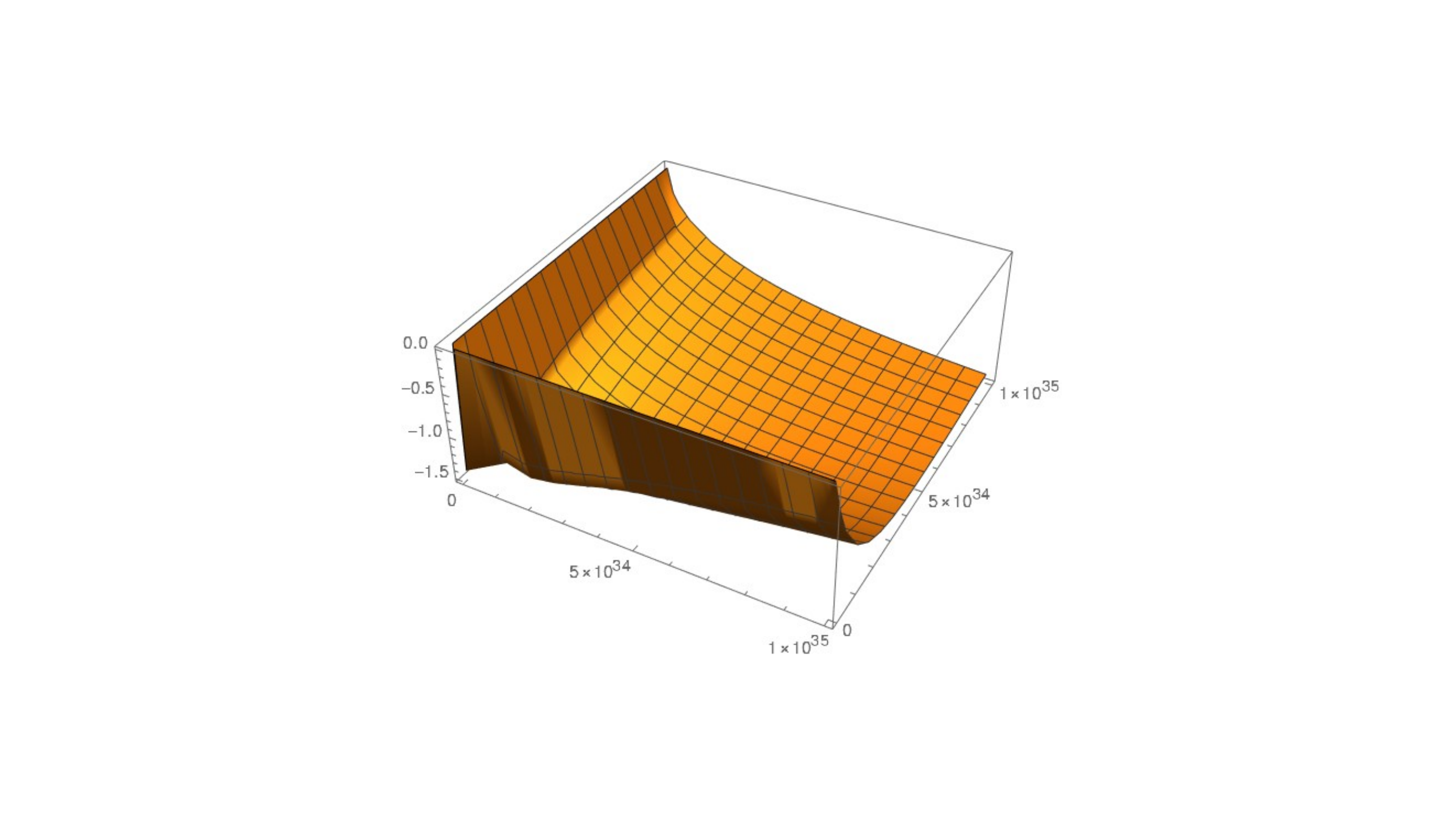}
\caption{ The 3D plot for the said crosschecking of the part of the solution for $\rho_{0} (l,\rho_c)$ which is the main solution in the regime  $l >> \rho_c$ and 
$\rho_c >> l$, as one obtain from  (\ref{ultimaterho0nonvanishing})  for $d - \theta < 1$   and much much more dominant over any other contribution to 
$\rho_{0} (l,\rho_c)$  in this regime, \quad;\quad   The image showing the 3D plots with x-axis \,:\,l, y-axis $\rho_c$, showing a comparatively large deviation from the value of l.h.s of (\ref{rewritebasic}),  which is  expected in the regime  $l >> \rho_c$  and $\rho_c >> l $ regime}
\label{consistency 1by9}
\end{figure}

Here we must comment,    all the consistency checking plots gave us a view that   we  took a very long range of $(l,\rho_c)$ namely $ 0 \, {\rm to} \, {10}^{35} $ just for the purpose that these regimes can show up as finite space within 2D $(l,\rho_c)$ plane and consequently in our plot  and indeed found that our global solution for the turning point is an exact solution for $l>>\rho_c$ and $\rho_c >> l$ regime,  although in the second one it holds for a very short space compared to  the first regime.   However we observed, even we move away from these regimes,  the maximum deviation of the l.h.s of (\ref{rewritebasic})  from zero, which corresponds to the exact solution,  is going upto $ \sim 0.8 $ for $ d - \theta > 1$ as well as for $d - \theta < 1$  upto $(d - \theta) \sim {\frac{1}{5}}$.  While we go below from $ d - \theta = {\frac{1}{5}} $,  as we considered the case $d - \theta = {\frac{1}{9}}$  in Fig.(\ref{consistency 1by9}  ) which is going upto 1.5!   These values,  upto for the case $ d - \theta \le {\frac{1}{5}}$ are actually quite less compared to the available range of  values of the Hypergeometric function when its argument varies between zero to one,   giving us a view that the expression of  $\rho_0(l,\rho_c)$ as we found in (\ref{ultimaterho0nonvanishing}),  is quite close to the exact particularly for $ \sim d - \theta \ge {\frac{1}{5}}$,  at any point of $(l, \rho_c)$ plane even if we are away from $l >> \rho_c$ and $\rho_c >> l$ regime!  Finally while we go to $\sim  d - \theta <  {\frac{1}{5}}$ we found a very large deviation because the theory is not conistent there!   This can be found from Fig.(\ref{hypergeometricconsistencyconsistency} ),  which is giving a view, in order to get consistent expression of the quantum measures, we should must restrict ourself $ \sim (d  - \theta) \ge {\frac{1}{5}}$, as below that, the theory is not appeared to be consistent,  because the involed Hypergeometric function in (\ref{tanmoy}) diverges in the $l >> \rho_c$ and $\rho_c >> l$ regime, as we have shown in Fig.(\ref{hypergeometricconsistencyconsistency})!

\begin{figure}[H]
\begin{center}
\textbf{ The Hypergeometric function  $ {{}_2 F_1}   \left\lbrack {\frac{1}{2}}, {\frac{1}{2}}(1 +{ \frac{1}{d - \theta}}), {\frac{1}{2}}(3 +{ \frac{1}{d - \theta}}) , \left({\frac{\rho_c}{\rho_0}}\right)^{2(d - \theta)}  \right \rbrack $  vs $ d - \theta$ for specific values of ${\frac{\rho_c}{\rho_0}}$ in the regime $ l >> \rho_c$ and also in the regime $\rho_c >> l$  }
\end{center}
\vskip2mm
\includegraphics[width=.55\textwidth]{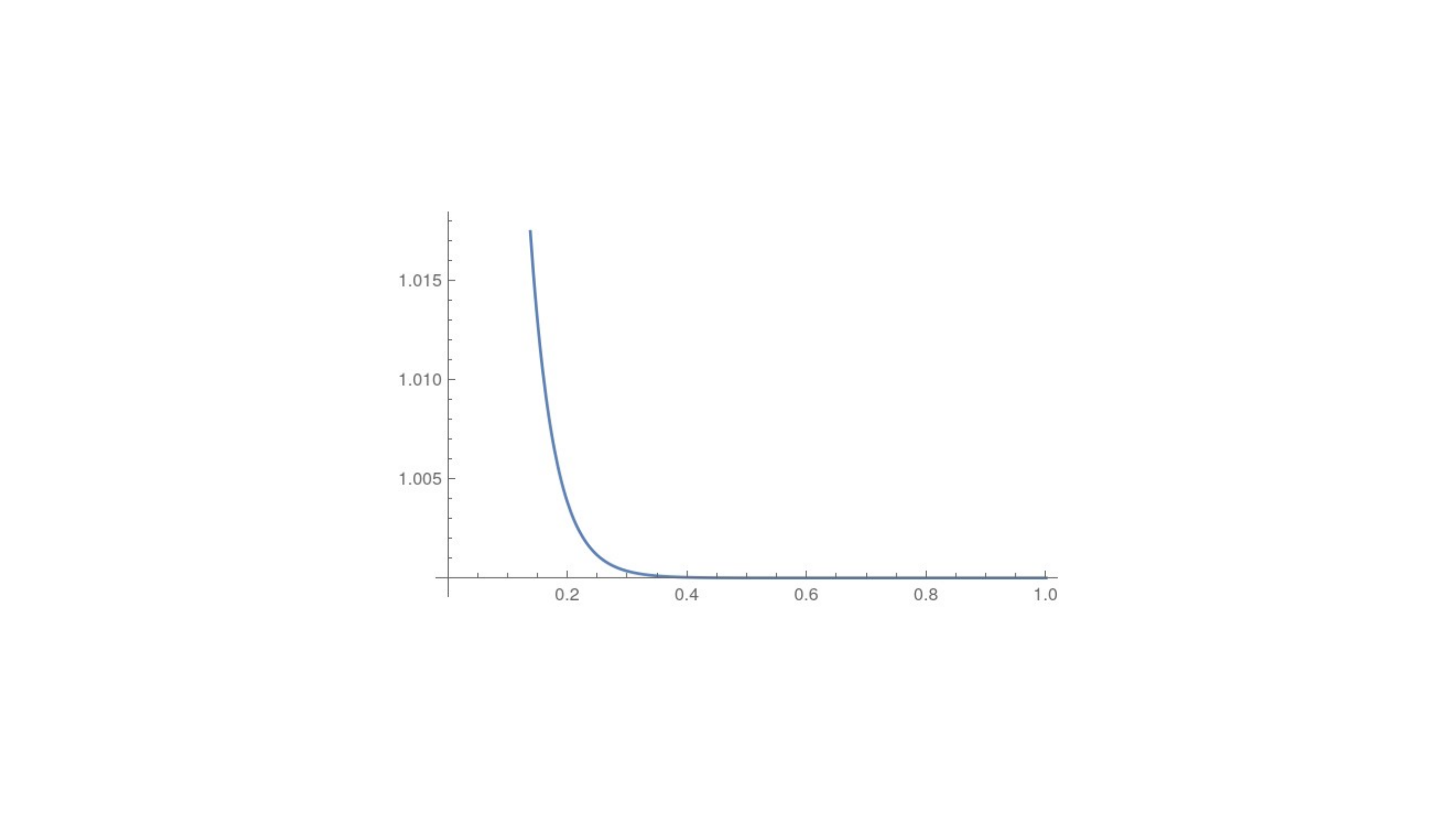}
\includegraphics[width=.55\textwidth]{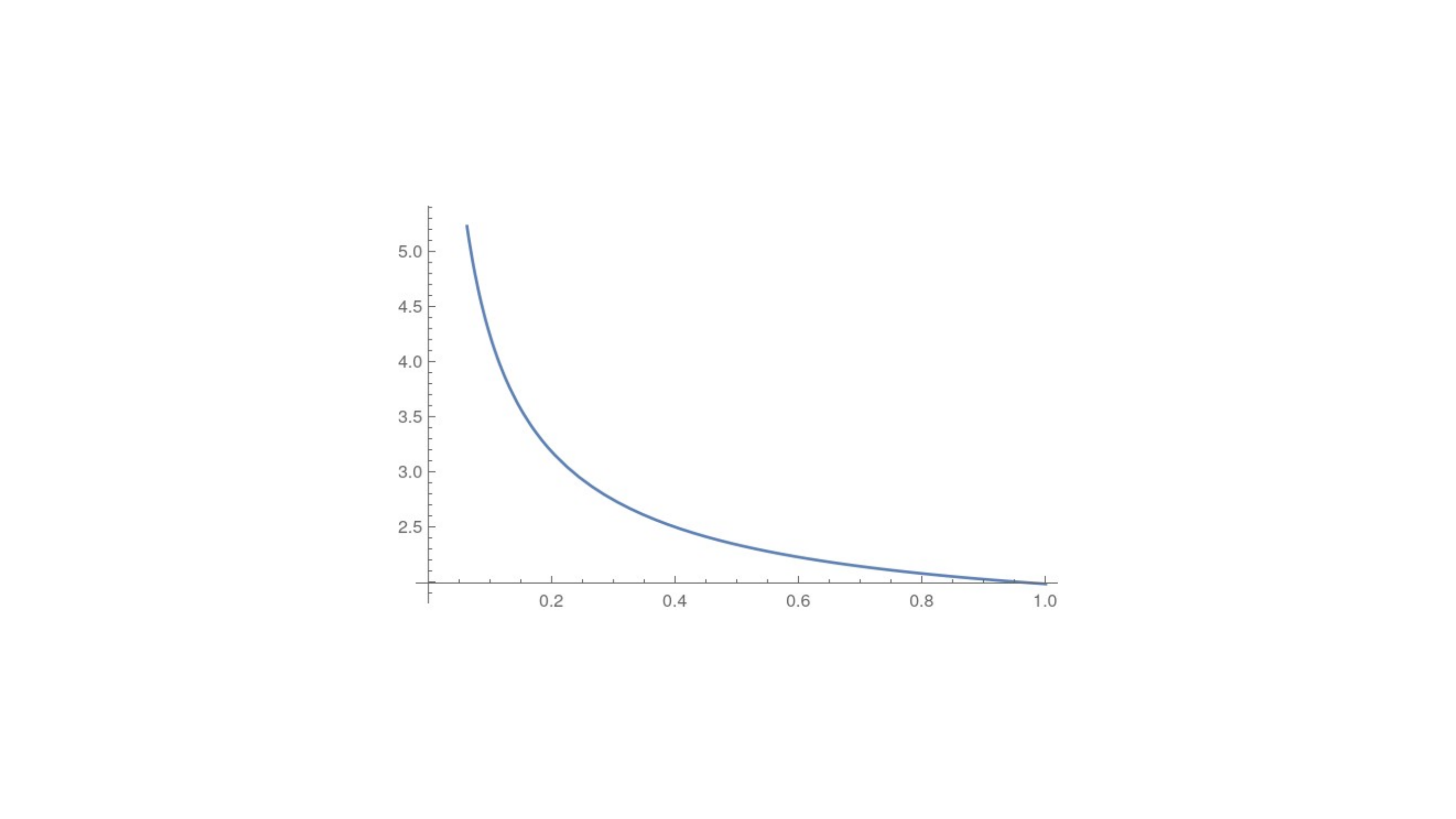}
	
\caption{ The Hypergeometric function involved in the governing equation for turning point (\ref{tanmoy}) is plotted against  $d - \theta$ for. \quad \,:\, (left)\,:\, $l>>\rho_c$  regime with ${\frac{\rho_c}{\rho_0}} = 0.00001$; \quad (right)\,:\, $\rho_c >> l$ regime with ${\frac{\rho_c}{\rho_0}} = 0.99999$ ; \,\,  Both the plots are showing divergence below $d - \theta \sim {\frac{1}{5}}$  showing the governing equation is not consistent there  }

\label{hypergeometricconsistencyconsistency}
\end{figure}

So, to summarize, here we found, the global solution for turning point from  (\ref{ultimaterho0nonvanishing}), indeed can be considered as the genuine solution for the 
(\ref{tanmoy}) in the regime $l>>\rho_c$ and $\rho_c >> l$ and even away from that regime,   the uppermost-variation-limit of these consistency-check  plots are showing that the global solution for the turning point is just too close to    (\ref{ultimaterho0nonvanishing}),  except the case $ \sim d-\theta < {\frac{1}{5}}$,  
since in this regime the Hypergeometric function involved in (\ref{tanmoy}) showing a divergence in the regime ${\frac{\rho_c}{\rho_0}} \sim 1$ and    also in ${\frac{\rho_c}{\rho_0}} \sim 0$,   the theory is not well-defined there!

\subsection{ The turning point as a function of $l,\rho_c$ and its evolution with $ d - \theta$ for $d-\theta < 1$ as well as  $d-\theta < 1$ }

Here we present some 3D plots, showing the features   of  the turning point  for $d - \theta > 1$, in (\ref{tttturn3by2})  and for $d - \theta < 1$ in (\ref{evturn1by8})

\begin{figure}[H]
\begin{center}
\textbf{ For $d-\theta > 1$, the plots for turning point with $(l,\rho_c)$ for different $ d - \theta$ showing their evolution with $d - \theta$ for $d-\theta > 1$  and also their variation with $(l, \rho_c)$}
 
\end{center}
\includegraphics[width=.32\textwidth]{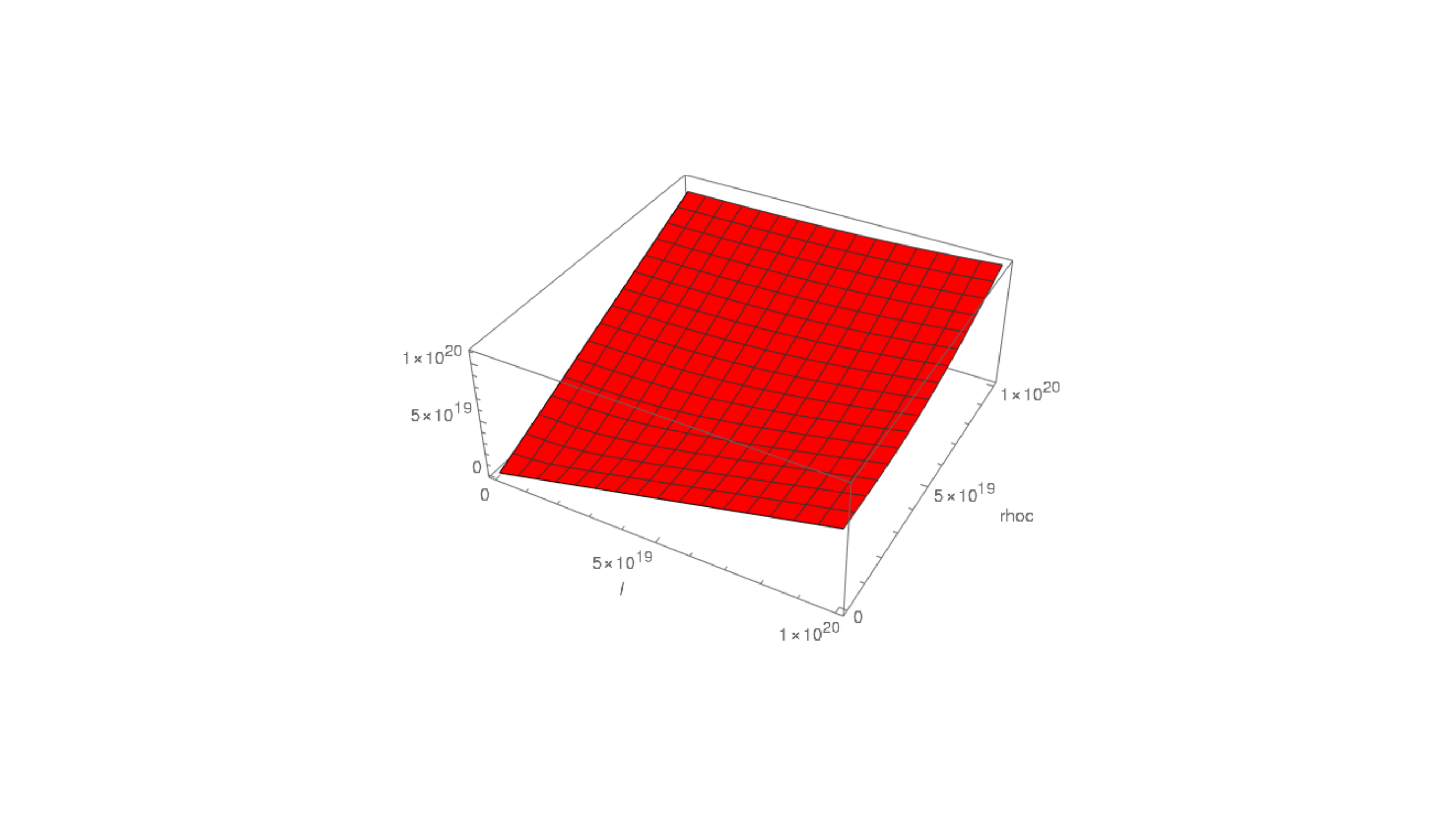}
\includegraphics[width=.32\textwidth]{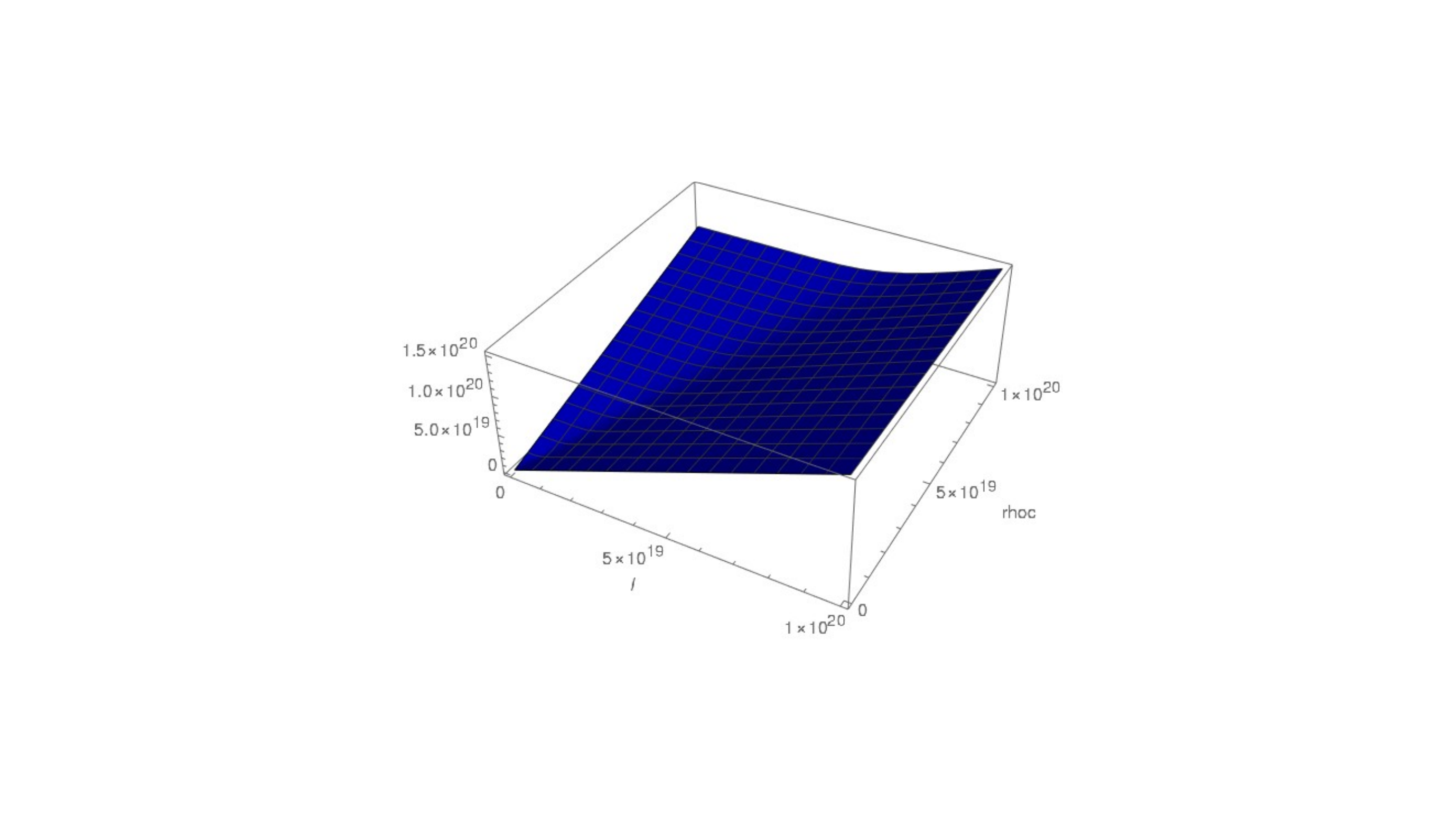}
\includegraphics[width=.32\textwidth]{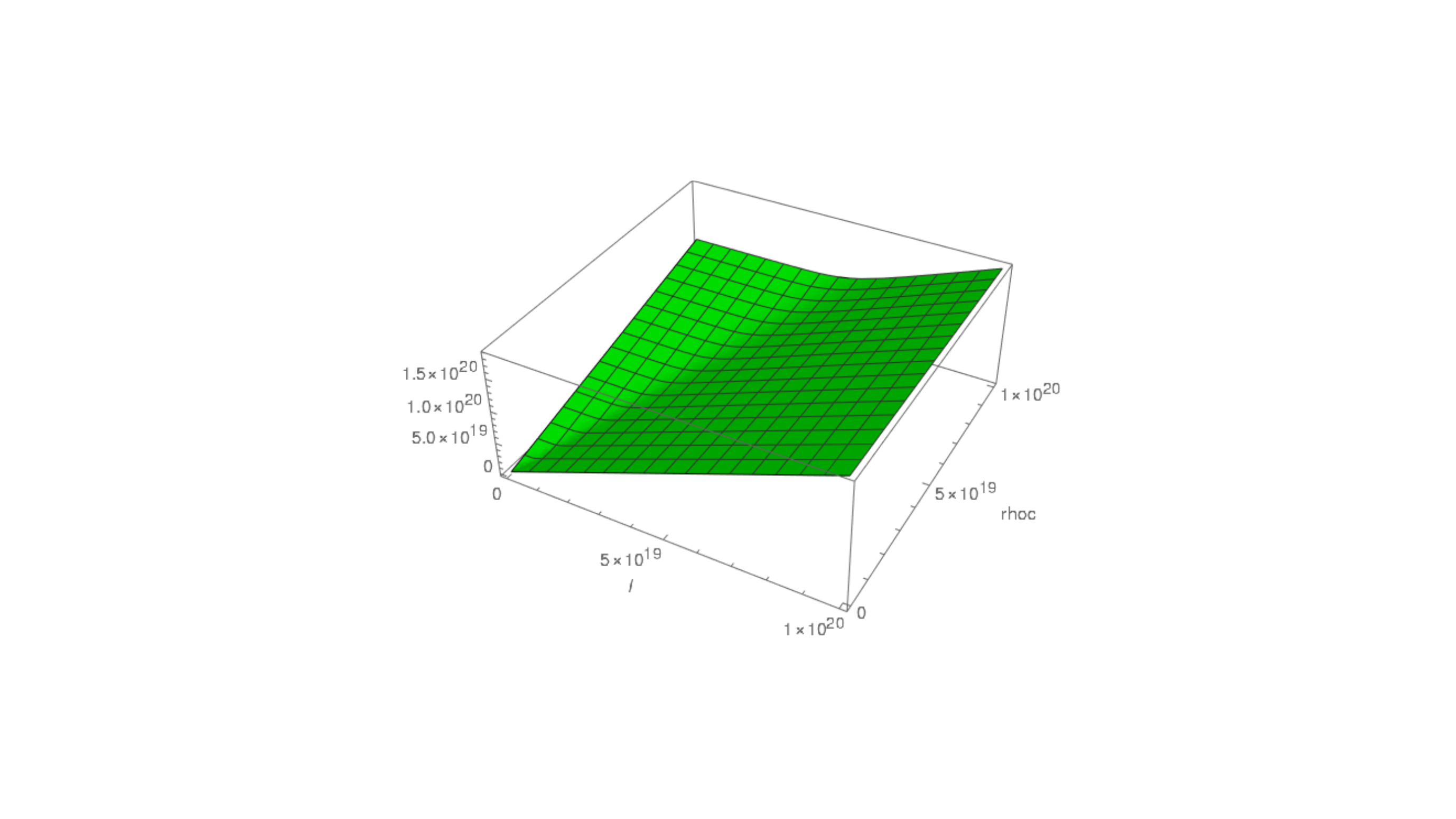}
\includegraphics[width=.32\textwidth]{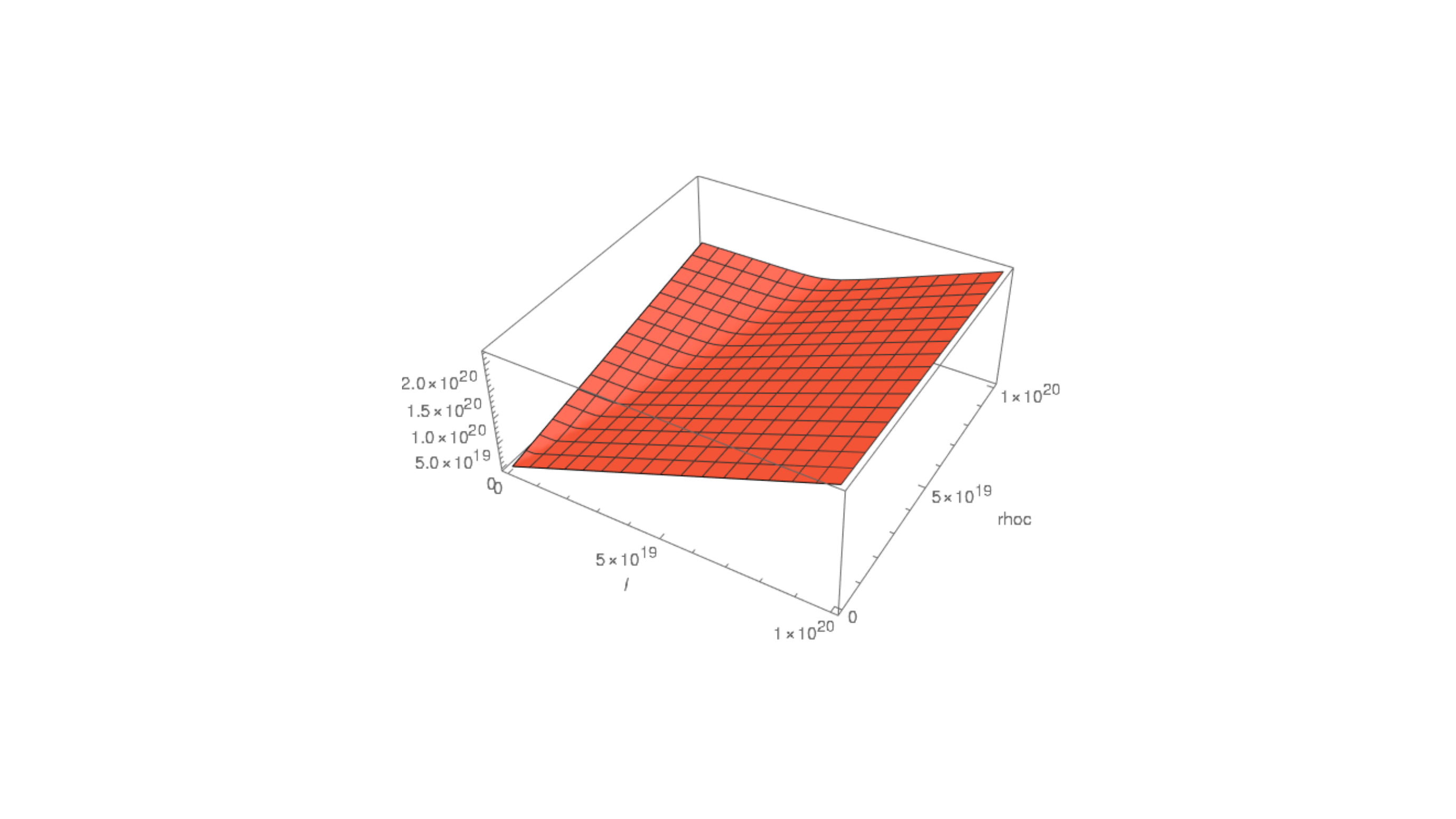}
\includegraphics[width=.32\textwidth]{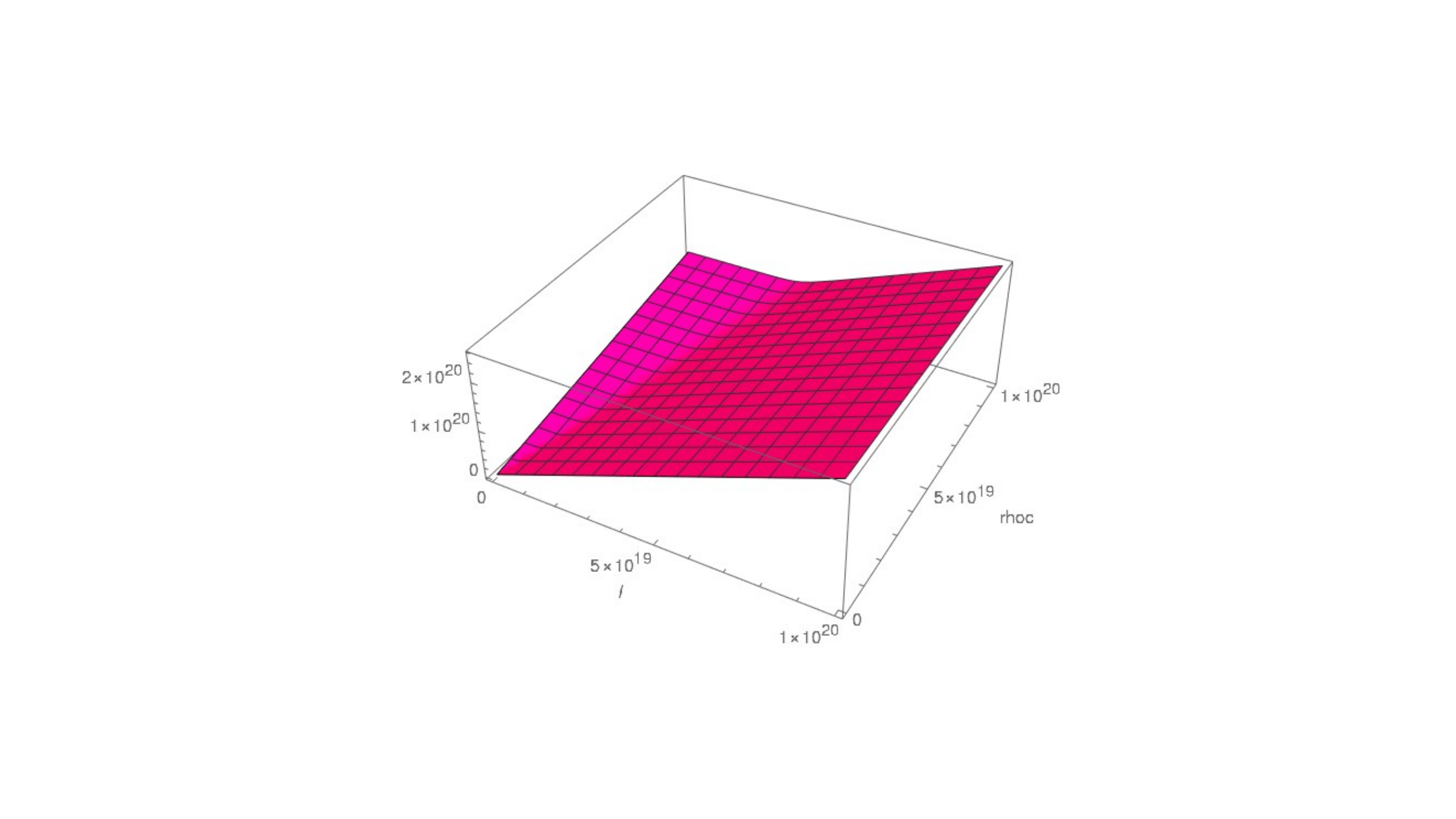}
\includegraphics[width=.32\textwidth]{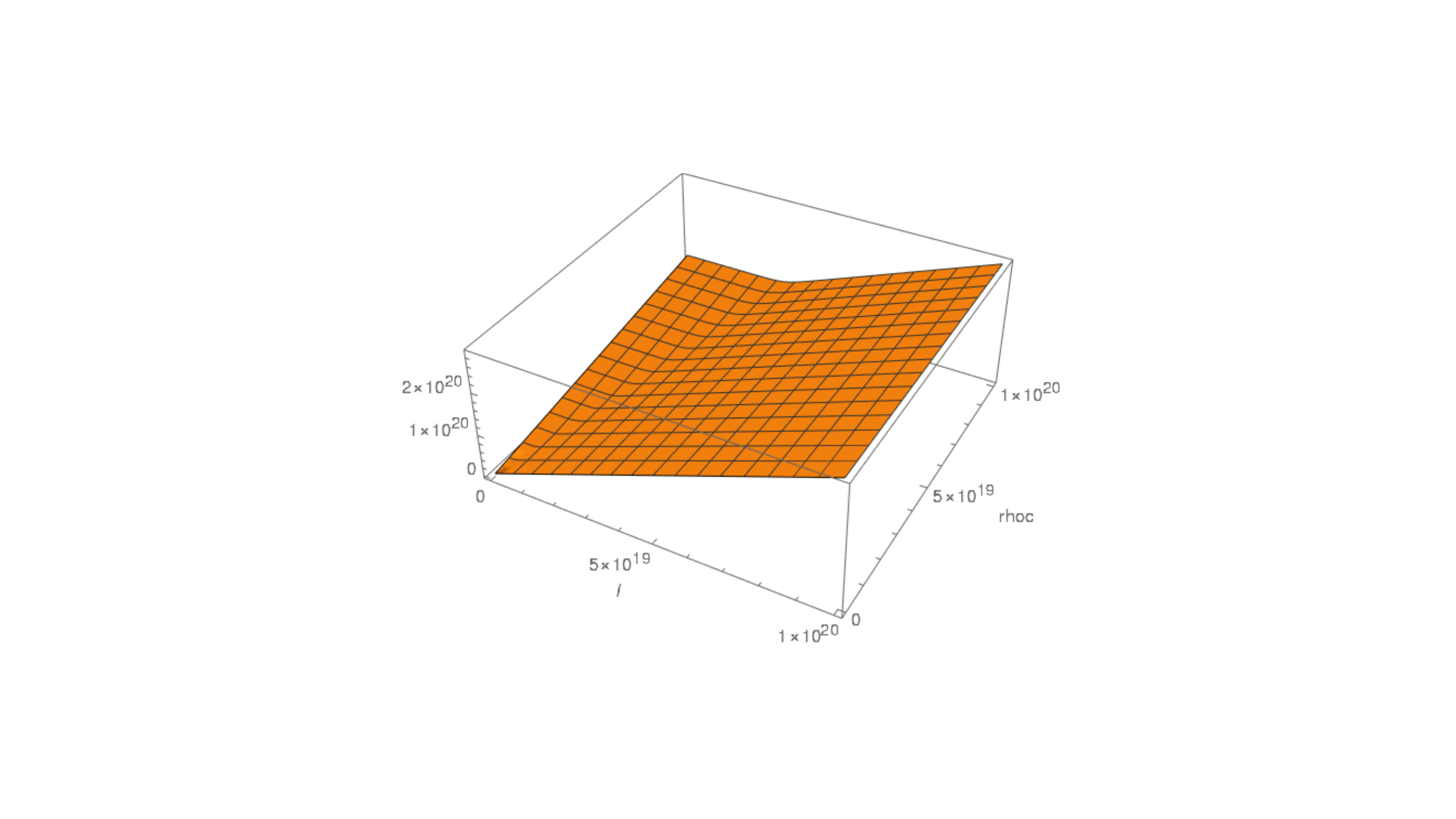}

\begin{center}	
\includegraphics[width=.65\textwidth]{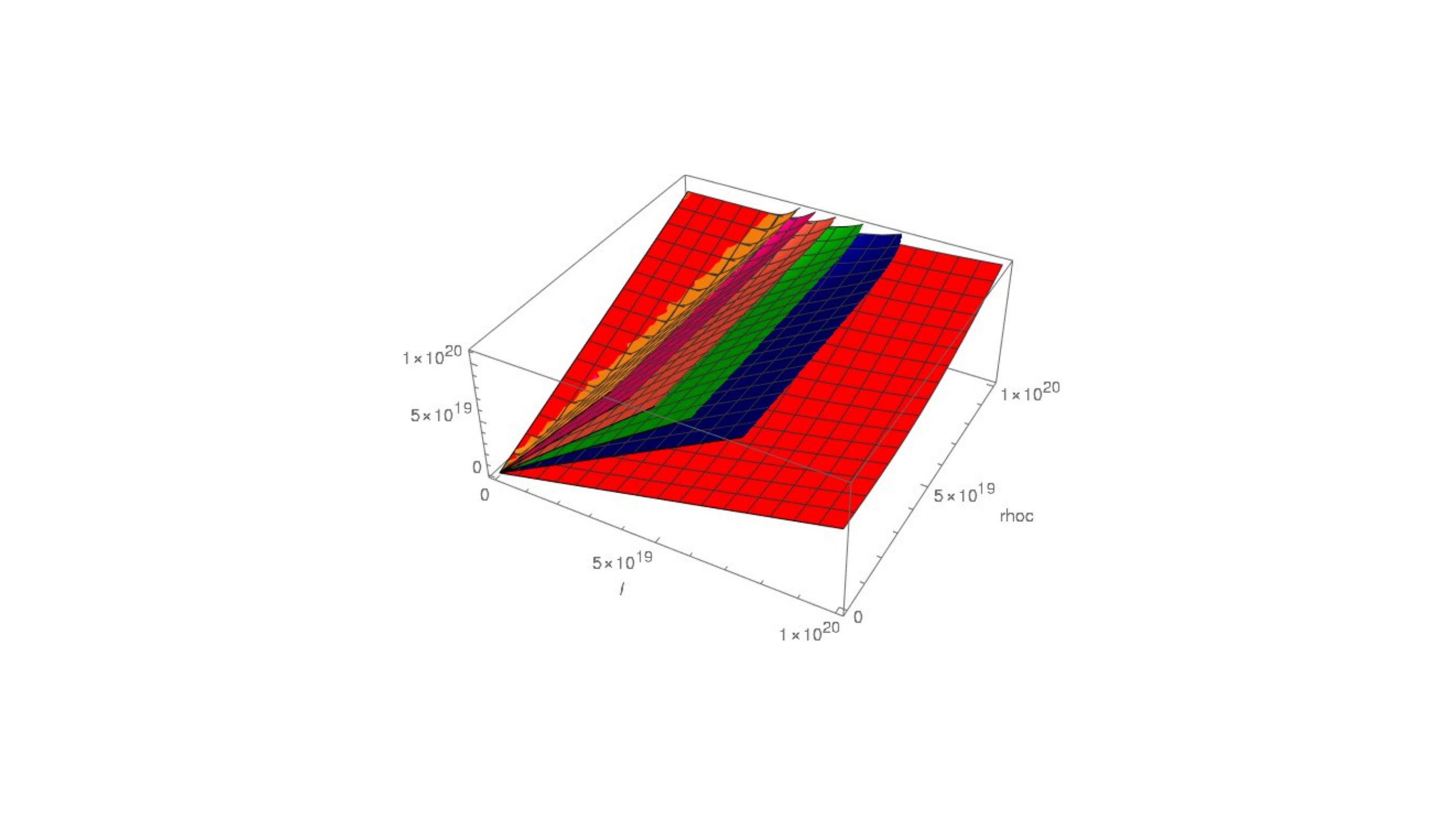}
\end{center}

\caption{First row \,\,:\,\,( From  left to right) \,,\, Turning point plotted as a function of $(l,\rho_c)$ for $d - \theta = 1.5 \,,\, d - \theta =  4.1\,,\,d - \theta = 5.2 $\quad;\quad (Middle row ) \,\,:\,\,( From  left to right) \,,\, Turning point plotted as a function of $(l,\rho_c)$ for $d - \theta = 6.3\,,\, 
d - \theta = 7.4 \,,\,\,,\,d - \theta = 8.5 $\quad;\quad (Last row)   \, ; \, The combination of all, showing  showing evolution of turning point with $d - \theta$.
The plots are also showing that the turning point increases with both $(l, \rho_c)$   }
\la{tttturn3by2}
\end{figure}

\begin{figure}[H]
\begin{center}
\textbf{  For $d-\theta < 1$, the plots for turning point with $(l,\rho_c)$ for different $ d - \theta$ showing their evolution with $d - \theta$ for $d-\theta < 1$ 
 and also their variation with $(l, \rho_c)$}
\end{center}
\includegraphics[width=.32\textwidth]{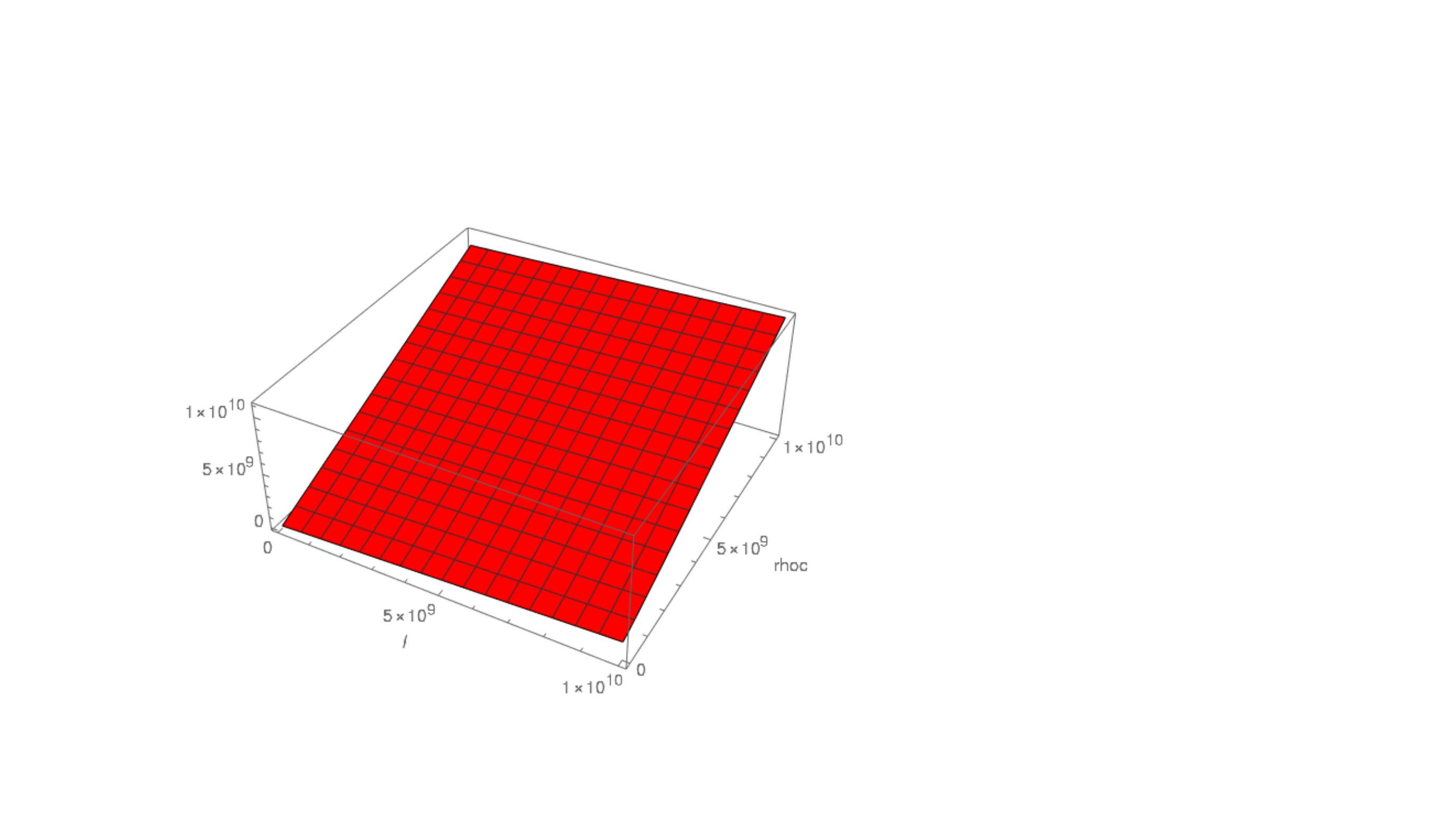}
\includegraphics[width=.32\textwidth]{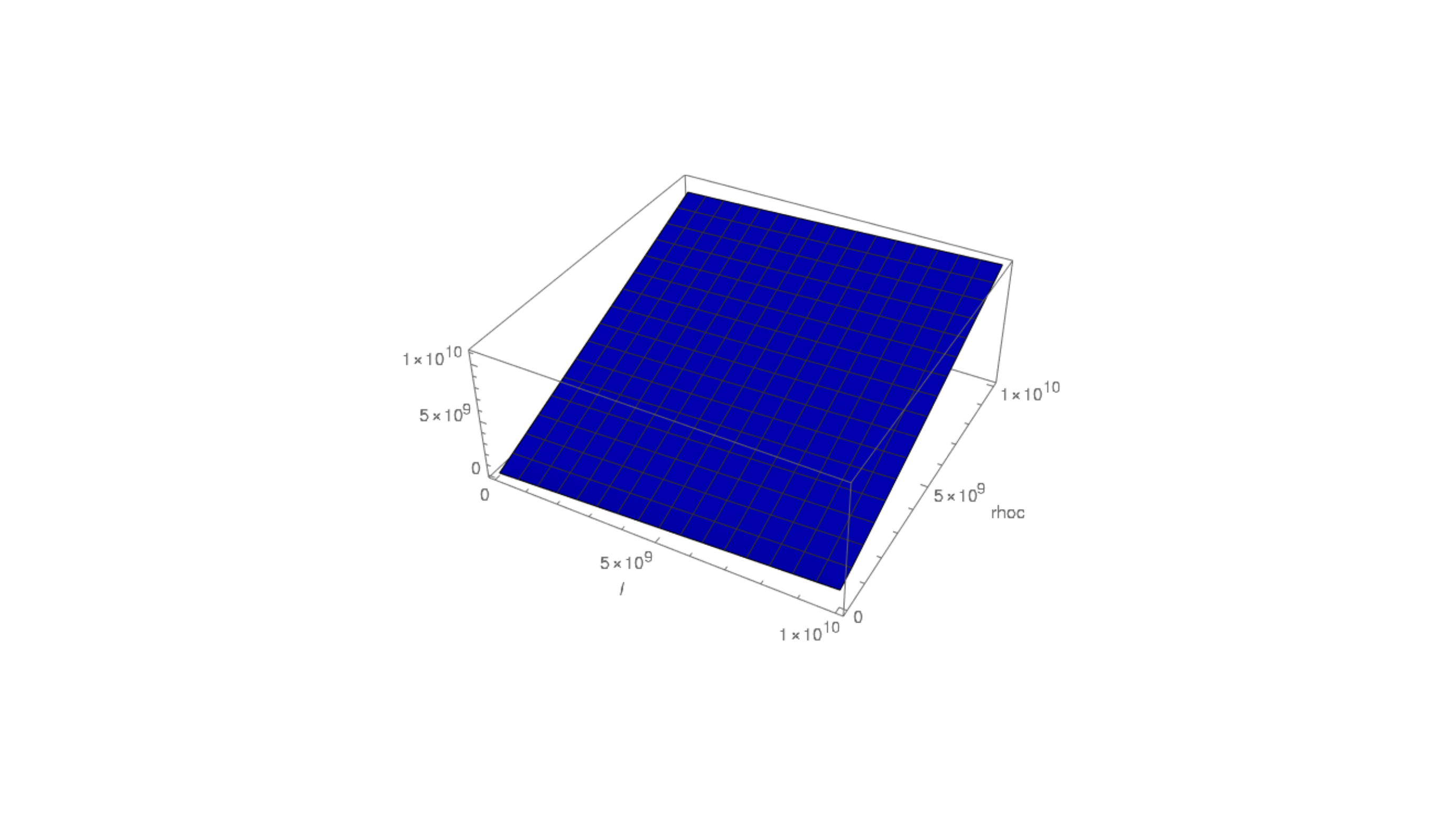}
\includegraphics[width=.32\textwidth]{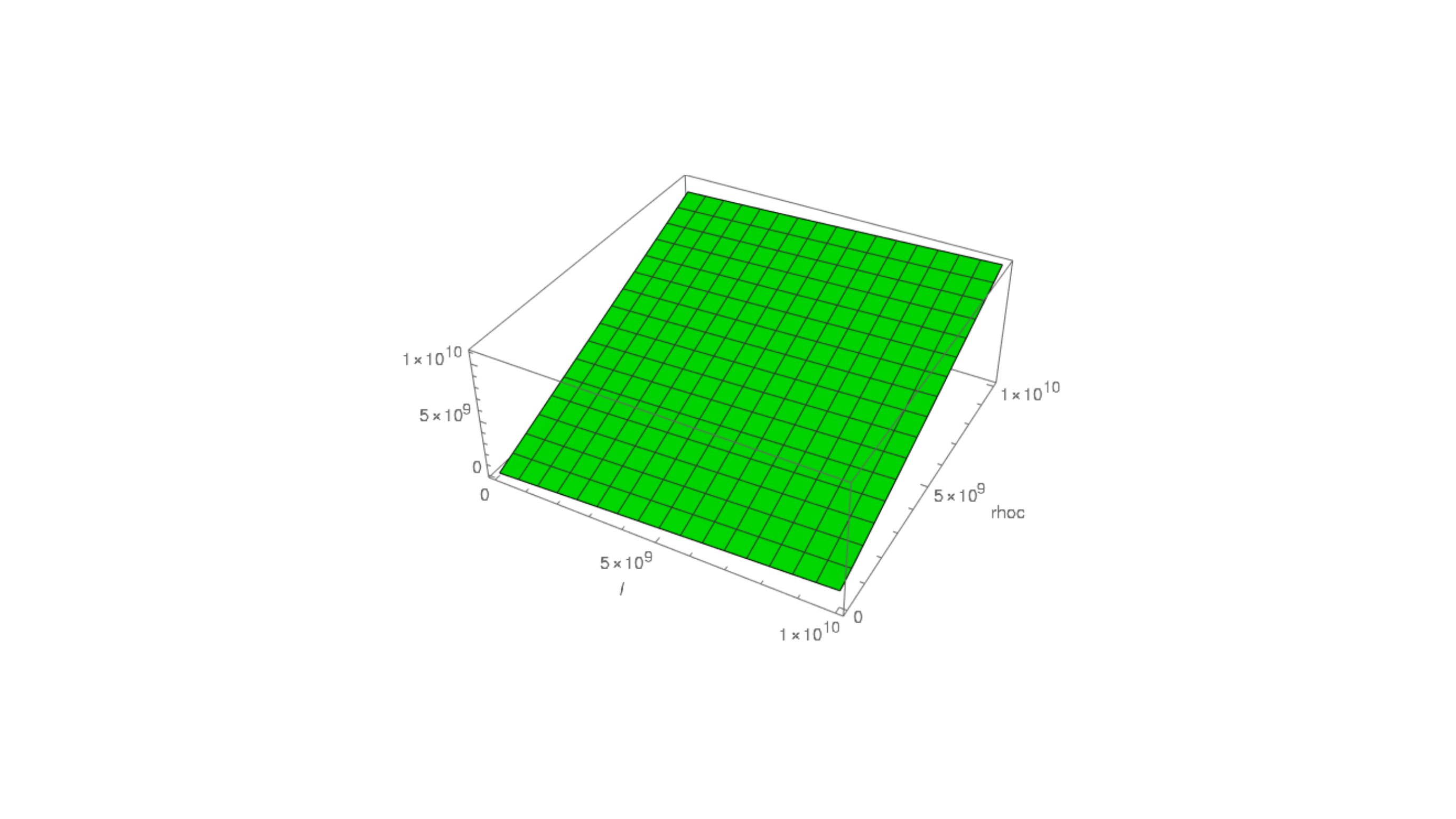}
\includegraphics[width=.32\textwidth]{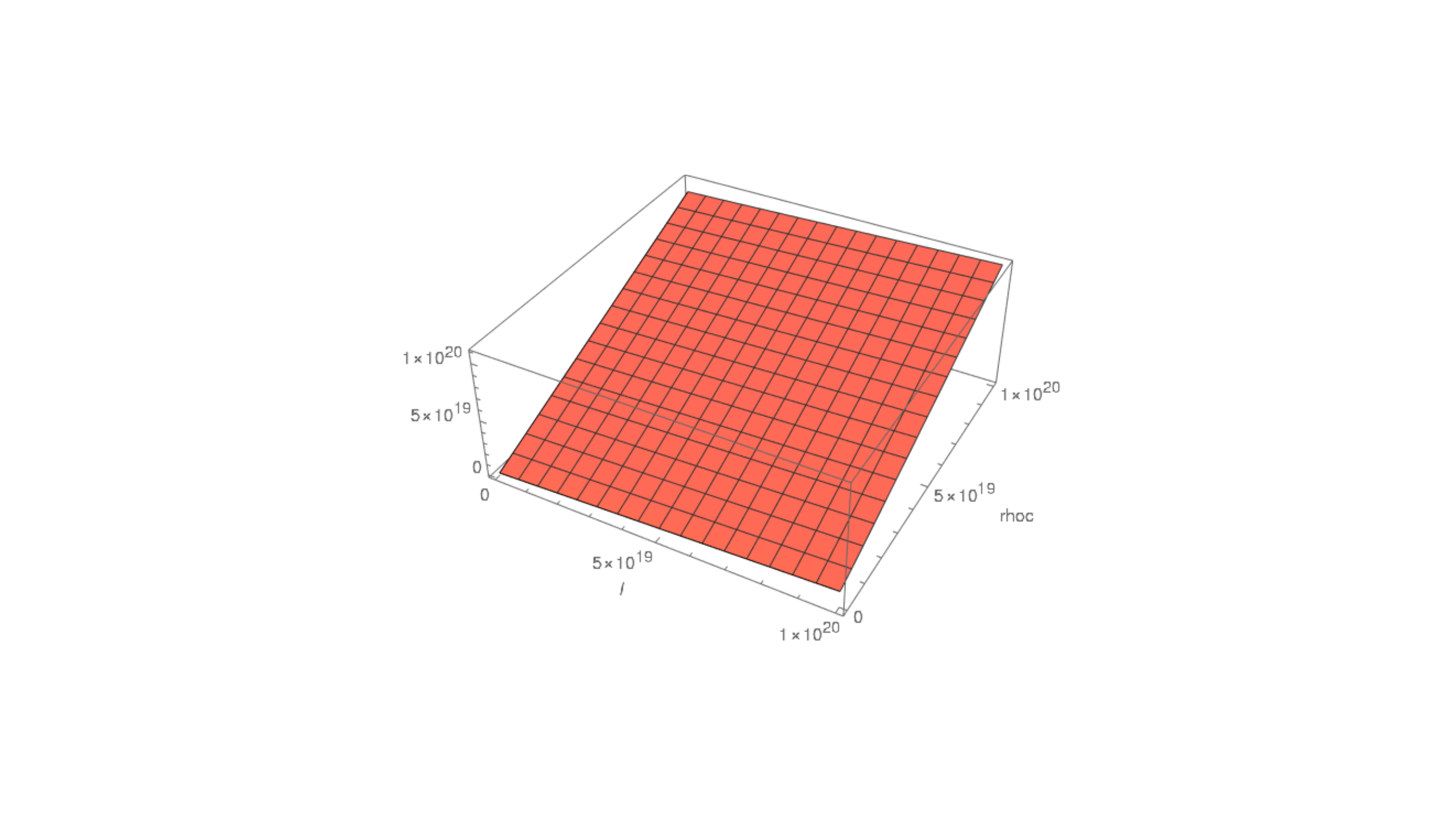}
\includegraphics[width=.32\textwidth]{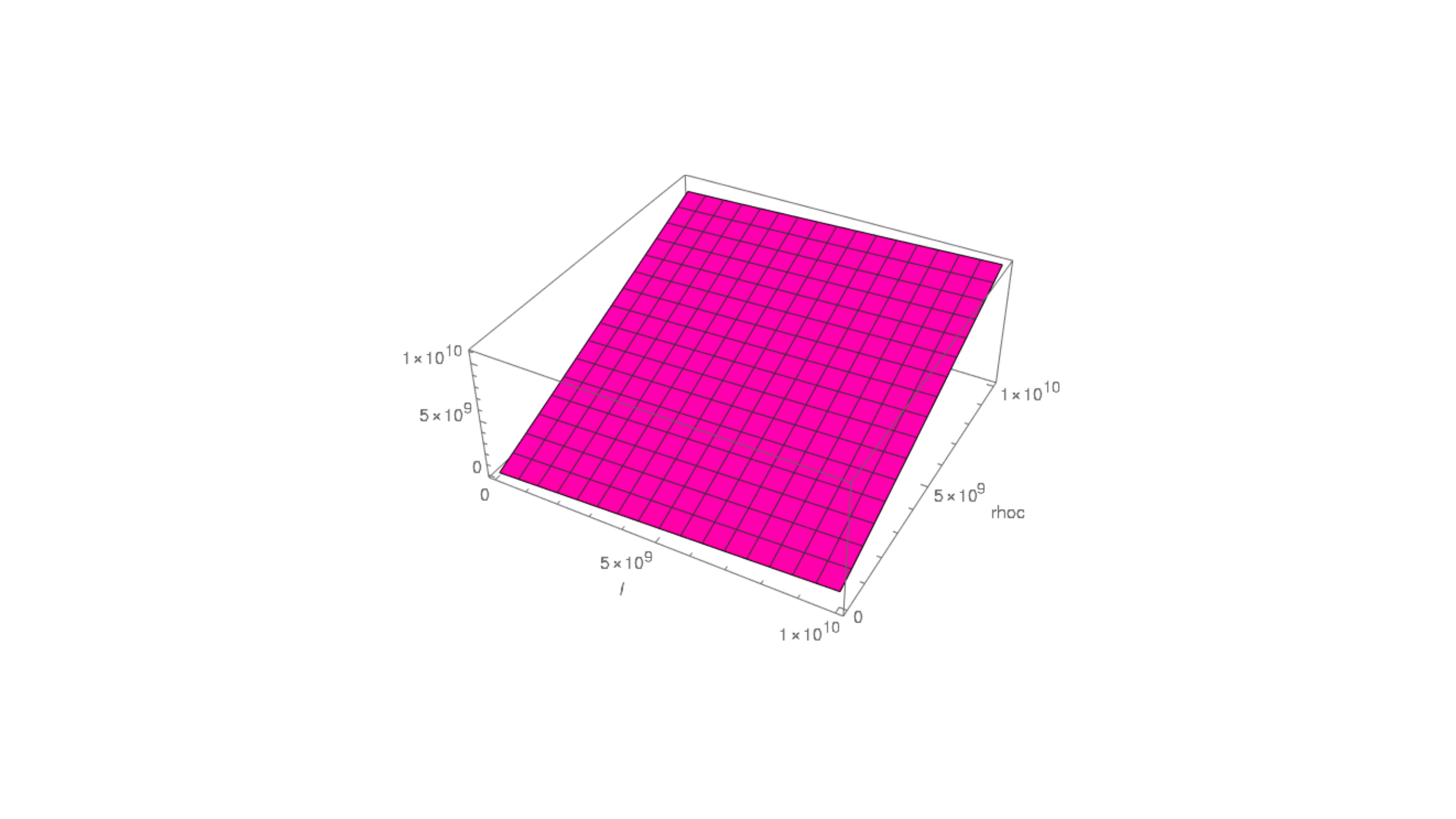}
\includegraphics[width=.32\textwidth]{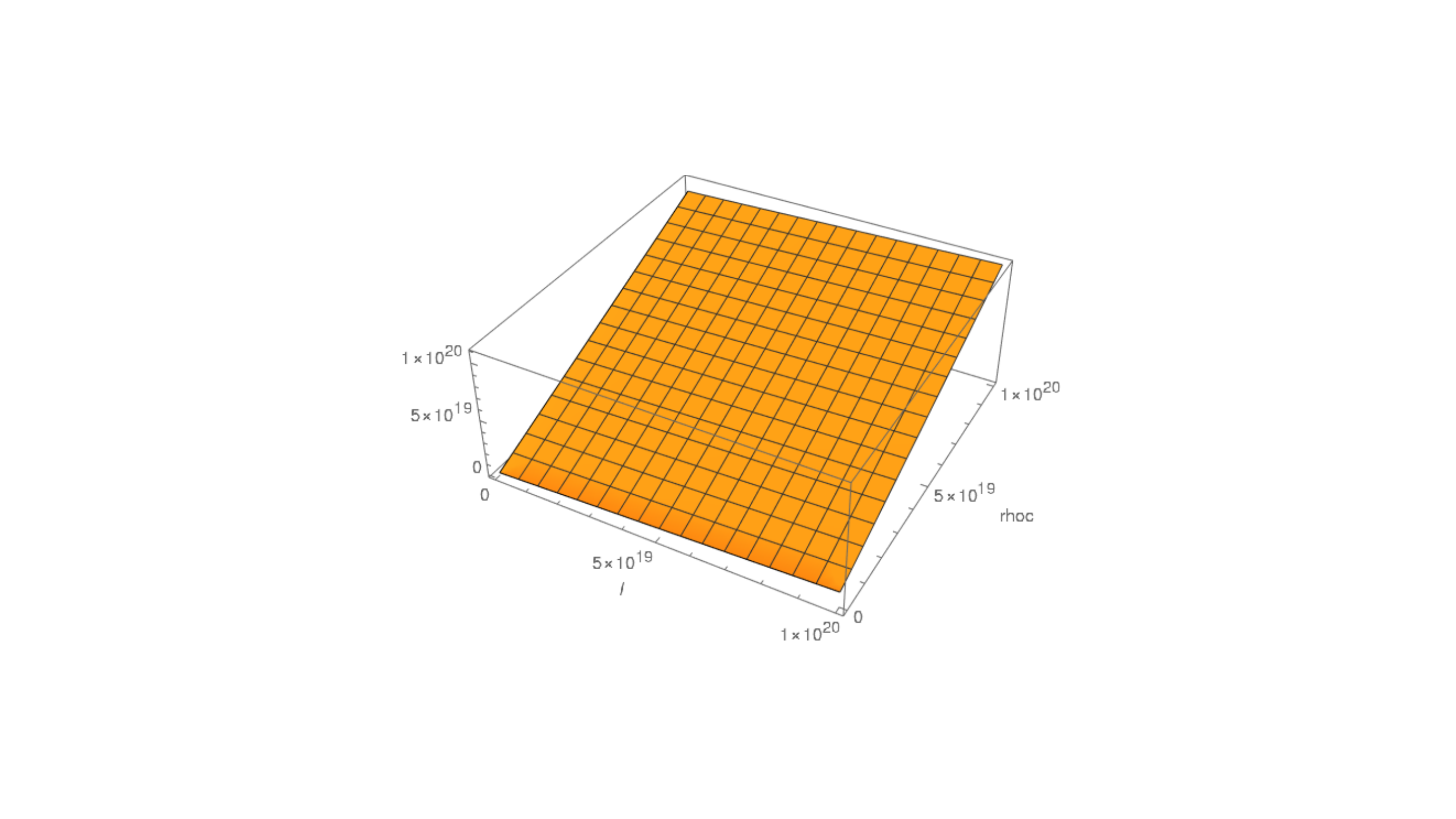}
\begin{center}	
\includegraphics[width=.65\textwidth]{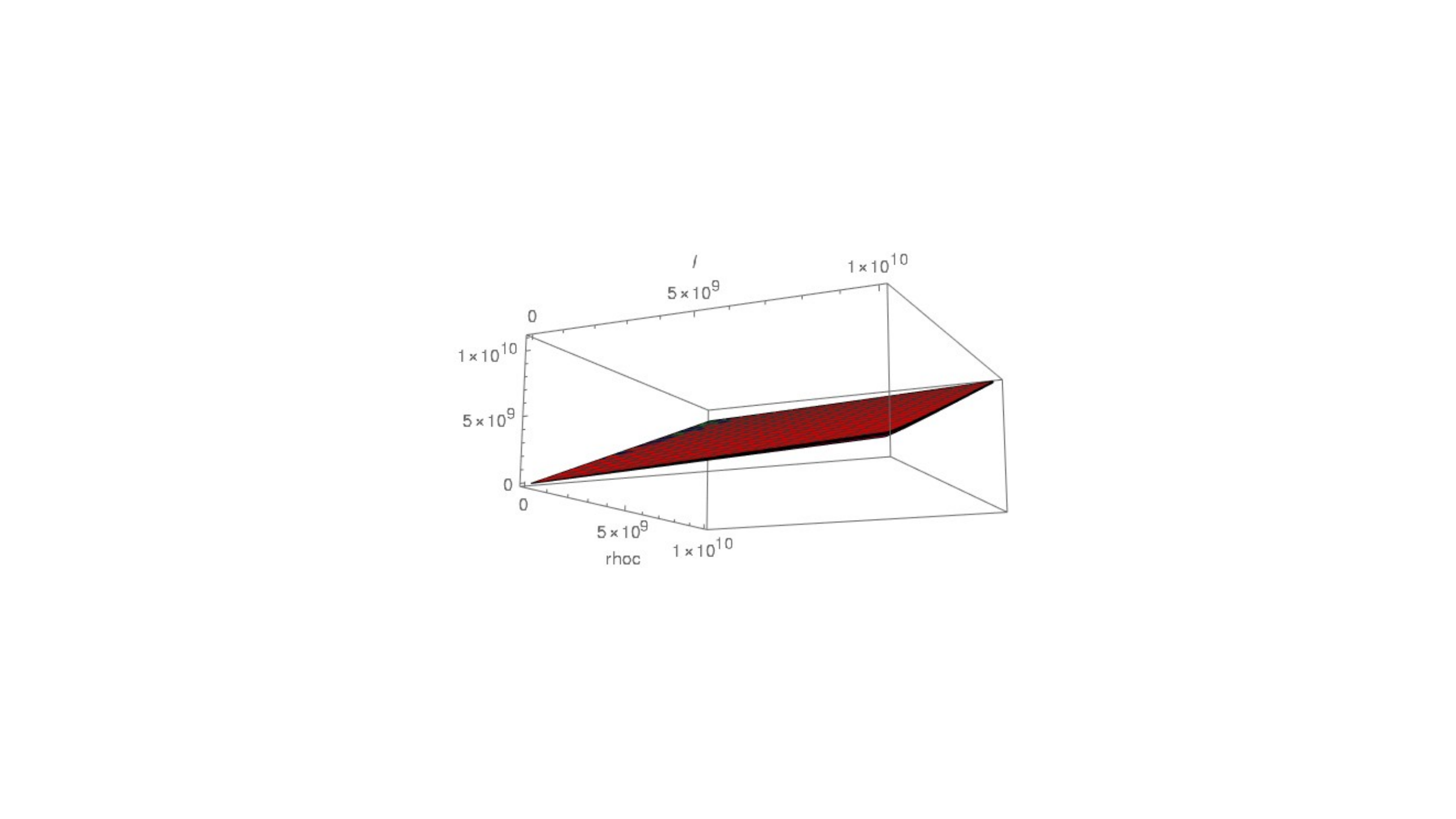}
\end{center}

\caption{First row \,\,:\,\,( From  left to right) \,,\, Turning point  plotted as a function of $(l,\rho_c)$ for $d - \theta = {\frac{1}{8}}\,,\, d - \theta = {\frac{1}{9}}\,,\,d - \theta = {\frac{1}{10}}$\quad;\quad (Middle row)  \,\,:\,\,( From  left to right) \,,\, Turning point plotted as a function of $(l,\rho_c)$ for $d - \theta = {\frac{1}{11}}\,,\, d - \theta = {\frac{1}{12}}\,,\,d - \theta = {\frac{1}{13}}$\quad;\quad Last row   \, ; \, The combination of all, showing the turning point, showing its evolution with $ d - \theta$ for $d - \theta < 1$.   The plots are also showing that the turning point increases with both $(l, \rho_c)$
 }
\la{evturn1by8}
\end{figure}

\section{Holographic entanglement of entropy in the presence of finite radial cut off}

We recall the expression of the holographic entanglement of entropy (\ref{entropy}). Here in this section we will first from field theory as well as from gravity side,  intutively argue about its expected  properties in the presence of finite radial cut off  and show that our constructed H.E.E based on the global solution of turning point (\ref{ultimaterho0nonvanishing}), maintaining the desired behaviours over complete $(l,\rho_c)$ plane for complete pararameter-regime of $d - \theta$ except the case $d - \theta << 1$, where the theory  is itself not very consistent,  as we discussed in the previous section!
However first we need to know the possible singularity structure. For the case without $T{\overline{T}}$ deformation in the bulk Hyperscaling violating geometry, as studied in \cite{lifshitz, chargedbrane}, H.E.E has a singularity at $\rho_c \rightarrow 0$ for $d-\theta \ne 1$, where the necessary regularization and renormalization was done therein. H.E.E  also showed a logarithmic divergence\cite{chargedbrane} at $\rho_c \rightarrow 0$ for $d - \theta = 1$.  However  for nonzero cutoff $\rho_c$,  as we found here,   H.E.E is free from both the singularities and consequently no renormalization is required. Now coming to the basic properties of H.E.E in the $T{\overline{T}}$ deformed theory
we first recall the case without $T{\overline{T}}$ deformation or the bulk radial cut-off.   According to (\ref{arealaw}),  clearly once   we increase the 
strip-length l, the EE will increase! Now to get field theory insight on the behaviour of EE in the deformed theory we consider \cite{entanglettbar1}.
In CFT one evaluate EE by replica trick \cite{rtoriginal}  where one evaluates S on an n-sheeted cover geometry of the boundary manifold and finally take $n \rightarrow 1$

\ber
S &=& \lim_{n \to 0} {\frac{{\rm tr}_A \rho_A^n  - 1}{1 -n}}\n
&=& - {\frac{\partial}{\partial n}}\,{\rm tr}_A \rho_A^n|_{n = 1}\n
&=& - {\frac{\partial}{\partial n}}\, {\rm log}\,{\rm tr}_A \rho_A^n|_{n = 1}\, ,
\la{replicatrick}
\eer
where the last line came from the fact that we use normalization  ${\rm tr}_A \rho_A = 1$.
The partition function on an n-sheeted cover geometry,  where we replace the  original  time by Euclidean time  as usual in replica-evaluation, with  the replacement of periodicity $\tau \equiv \tau + 2\pi$  by  
$\tau \equiv \tau + 2 n \pi$,  it  can be written as
\ber
Z_n &=&  
{\rm Tr} \left\lbrack  P e^{- \int_0^{2 \pi n} d\tau H_{n}(\tau) }    \right\rbrack = {\rm Tr} \left\lbrack\rho_n^n \right\rbrack\n
\rho_n &=& P e^{- \int_0^{2 \pi } d\tau H_{n}(\tau) } \, ,
\la{endless}
\eer
where in the second equality we have used the fact $H(\tau) = H(2\pi + \tau)$  since we are replacing the replica Hilbert space with the direct product of n copies of Hilbert space.  

Finally substituting (\ref{endless}) in (\ref{replicatrick})  and also following \cite{entanglettbar1, {holographyttbar2}  }  we can write the Entanglement of entropy as
\be
S = \left(1` - n{\frac{d}{dn}}\right) \log Z|_{n = 1} \,, 
\la{eereplica}
\ee
where Z is evaluated on n-sheeted cover of the boundary manifold
and S is EE.

Now in \cite{entanglettbar1} the authors considered the boundary theory with large central charge so that following (\ref{ttbarflow}), T can be effectively treated as a classical field.  Also since we can    write the partion function in terms of the path integral with lagrangian description,   in this classical limit,  the classical term will dominate.  Finally in  \cite{entanglettbar1} the author have evaluated it for a couple of boundary-manifold geometries.  Here we will
 consider the example one of these, a sphere, where the n sheeted  
cover geometry is given by

\be
ds^2 = r^2 \left\lbrack r^2 d\theta^2  + n^2 \sin^2 \theta d\phi^2 \right\rbrack
\la{nsheetedcover}
\ee

Under a change of n the partition function changes as
\be
{\frac{d \log Z}{d n}} |_{n = 1} = - \int \sqrt{g} T^{\phi}_{\phi}
\la{variationpart}
\ee

One can substitute (\ref{variationpart}) in (\ref{eereplica}) to obtain

\ber
S &=& \log Z|_{n = 1}   +   {n\left\lbrace \int \sqrt{g} T^{\phi}_{\phi}  \right\rbrace}_{n = 1}  \n
{\rm with} & & \delta \log Z = - {\frac{1}{2}} \int d^2 x {\sqrt{g}} \lan T^{ab} \ran \delta g_{ab}
\la{eereplicarewrite}
\eer

The equation (\ref{eereplicarewrite}) is constructed for  a specific case and the specific geometry,  which can be generalized to other geometries and also can be generalized to higher dimension.   Moreover,   once we deviate from the large central charge limit, the quantum correction to the energy momentum tensor can be added
as well!
Firstly, in order to understand the behaviour of EE under variation of  $T{\overline{T}}$ deformation coefficient $\mu$ or the cut-off  $\rho_c$,   one indeed needs to evaluate EE in the case of our Hyperscaling violating bulk  geometry,   on  its  n-sheeted boundary-cover-geometry    with finally $ n \rightarrow 1$, which is beyond the scope of the present article!  However here we will give an intutive reasoning both from field theory and gravity side,  to predict the behaviour of EE with the cut-off!
\vskip0.5mm
Clearly the partition function Z in (\ref{eereplicarewrite})  can be written in a path integral formulation \cite{rtoriginal},  which alongwith  fact that the general form of the $T{\overline{T}}$  deformed  action can be written in the form $S = S_0  + \mu \int d^2 x T{\overline{T}} $,     we can write
\be
Z = \int \prod d[{\rm fields}] \,\,e^{- S_0 - \mu \int d^2 x T{\overline{T}} }\,;
\la{hostile}
\ee
where $S_0$ is  the action for the undeformed CFT  and we always choose $\mu$ to be positive and finally we are considering Eucledean time $\tau$ in the action so that the path integral (\ref{hostile}) is of the above form!  This can be generalized to the higher dimension as well!
\vskip 0.5mm

To understand the behaviour of EE with $\mu$,  which according to   (\ref{murcrelationew})   will give  an understanding,  the behaviour of HEE with the cut off $\rho_c$, we need to  understand the variation of the partition function Z with $\mu$. Firstly for large central charge limit,  we just have the classical term where the positivity of $(T,\,  \overline{T})$ on the conformal plane implies we are left with $e^{-\mu \int T { \overline{T}}}$ which clearly decreases with the increase of $\mu$ and at $\mu \rightarrow \infty$ it vanishes!   Coming to quantum correction,   once we move away from the large central charge limit,  following \cite{holographyttbar2},   the bare partition function will diverge and one can think of possible counterterm. However here we will focuss on specifically the $\mu$   dependence! Next, since we are focussing on 2D,  
it  was shown in \cite{marika}, that the higher dimensional generalization implies a term in CFT
$ {\mathcal{T}} = \lim_{x \to y} \left(T_{ij}(x)T^{ij}(y) - {\frac{1}{d - 1}} {(T_i^i)}(x){(T_j^j)}(y)\right)$ which is in accordance with (\ref{ttbarope}),  where indeed the quantum corrections coming in  terms of one point function 
$ \lim_{x \to y}  \lan\left(T_{ij}(x)T^{ij}(y) - {\frac{1}{d - 1}} {(T_i^i)}(x){(T_j^j)}(y)\right)\ran $.  However it was further shown in \cite{marika} that in a d-dimensional CFT, $ \lan{(T_i^i)}(x){(T_j^j)}(0) \ran = 0 $ (one can always take $y = 0$), indicating the positive definiteness of $\lan \mu T { \overline{T}}\ran$   so that even the one point function,  giving a contribution to Z of the form 
$\sim e^{-\mu \lan T{ \overline{T}}\ran}$ which again falls with the increase of $\mu$ and eventually becomes zero at $\mu  \rightarrow \infty$.!  Next for the contribution for more and more higher point function to Z,   we consider the  factorization property of  $(T \overline{T})$  deformation  \cite{zamolodchikov}, they can be expressed in terms of one point function and so the same logic can be applied to the contribution from these term to Z as well! 
 Moreover,  more and more we consider quantum correction to Z, it follows from (\ref{eereplicarewrite}) that the energy momentum tensor will also be modified accordingly and at $\mu \rightarrow \infty$ it vanishes! So the substitution of all the quantum corrected expression from Z and $T_{ab}$ in  the expression of EE 
(\ref{eereplicarewrite}) implies that EE will decrease with the increase of $\mu$ and eventually become zero at $\mu \rightarrow \infty$.  However according to
 (\ref{murcrelationew}),  the increase in $\mu$ implies the increase in the cut off $\rho_c$  and so from field theory side we see intutiuvely that with the increase of cut off $\rho_c$,  EE will decrease and will go to zero at $\rho_c >> l$.

\vskip0.5mm
In the gravity side,  the behaviour of HEE can be read directly from the original   RT proposal (\ref{koutuhal}).   First we must recall (\ref{murcrelationew}) which implies  that the increase in $\mu$  mean,  increase of the cut off $\rho_c$  in the dual gravity  and so   we can   expect   that HEE will decrease with the increase of cut off $\rho_c$.  Clearly,  since with the increase of strip-length l,  the area of the boundary of the  boundary-subregion will increase,  where the entangling surface  $\Gamma_A$ is anchored in,  so clearly $\Gamma_A$  will increase with increase of l.   To understand the effect of cut-off $\rho_c$ on $\Gamma_A$, we recall   the fact that  in order to define $\Gamma_A$,  since its boundary is anchored to the boundary of the strip,   which is extended  along (say) x direction,  so we define the entangling surface through the embedding  $x = x(r)$,        
 where the  area of entangling surface  will also depend on the effective length in the radial direction available.   More the cut-off $\rho_c$ will increace
this  effective length will more and more decrease.   We have explained through  Fig.(\ref{entanglingsurface}), where in the left we have the normal Ryu-Takayanagi surface where the minimal area giving HEE.  In the right of Fig.(\ref{entanglingsurface}) we have introduced the cut off surface into the bulk, which clearly trying to decrease the entangling surface area.  However at the same time, the tip of the surface,  being more and more pushed towards inside because it depends on the turning point $\rho_0(l,\rho_c)$ and which also increases with the cut off $\rho_c$, so that there is a competition between the two!  However, since with the increase of cut off $\rho_c$,  it is $\rho_c$  always increases faster  than  $\rho_0(l,\rho_c)$,  as evident from   (\ref{ultimaterho0nonvanishing})  so the entangling area gradually shrinks down and finally at $\rho_c >> l$,
 the respective area and consequently HEE becomes zero since at this regime we are having $\rho_0 \simeq \rho_c$ (in general it follows from (\ref{bc2}), since with $\rho_c>> l$,  effectively in the extreme limit one can think $ l = 0$).  

\vskip0.5mm

Also from (\ref{ultimaterho0nonvanishing}) it is evident that for $l >> \rho_c$ regime we will have the most dominent term in  the expression of EE in (\ref{entropy}) is given by $ \sim {\left\lbrace\rho_0 (l, \rho_c) \right\rbrace}^{1 - d + \theta} \sim l^{1 - d + \theta}$
  so that HEE, in this regime is expected to decrease with the increase of $(d - \theta)$ for fixed $(l,\rho_c)$ in spite of the decrease of the denominator, given by 
$1 - ( d- \theta)$ which will be less effective in large l regime!  On the otherhand,  once we move to the regime $\rho_c >> l$   we will have the dominant contribution of HEE in this regime will be given by zero (as follows on substitution of (\ref{ultimaterho0nonvanishing}) in (\ref{entropy})) and remain invariant on variation of $d - \theta$!    Combining the two aspects,  one can see that when we define the global expression  over $(l, \rho_c)$ plane,  HEE  will decrease with the increase of $(d - \theta)$,  when the difference in their expression for different  $(d - \theta)$ will be maximum in $l >> \rho_c $ regime and minimum in $\rho_c >> l$ regime, where they all merge together to zero!  Indeed this is the behaviour of HEE we are going to establish in the next few sections! 

\vskip0.5mm

Finally,  coming to the question of the impact of the global symmetry (\ref{2dsymmetry}) on HEE,    following  (\ref{entropy})
\be
(l,  \rho_c) \rightarrow (kl, k\rho_c) \Rightarrow {S}(kl, k\rho_c) = {k}^{1 - (d - \theta)}{S}(kl,  k\rho_c)
\la{symmetryee}
\ee

\vskip0.5mm
Here along with our constructed expression of HEE  we will see whether these properties holds for full parameter-regime of $d - \theta$, i.e $d- \theta = 1$, $d - \theta > 1 $ and $d - \theta < 1$.  Here finally we must comment on the fact that when the $\rho_c = 0$, the symmetry,  as expressed in (\ref{symmetryee}),  do hold 
as shown in \cite{lifshitz, chargedbrane }.   However there it was a normal mathematical  functional dependence and not really corresponds to any spacetime symmetry with the geometrical origin.   This fact actually supports the view of our discussion in section 3 that  the emergent bulk symmetry can really be viewed as generalization of boundary scaling symmetry or vice versa.

\begin{figure}[H]
\begin{center}
\textbf{ Entangling surface  with and without cut off }
\end{center}
\vskip2mm
\includegraphics[width=.65\textwidth]{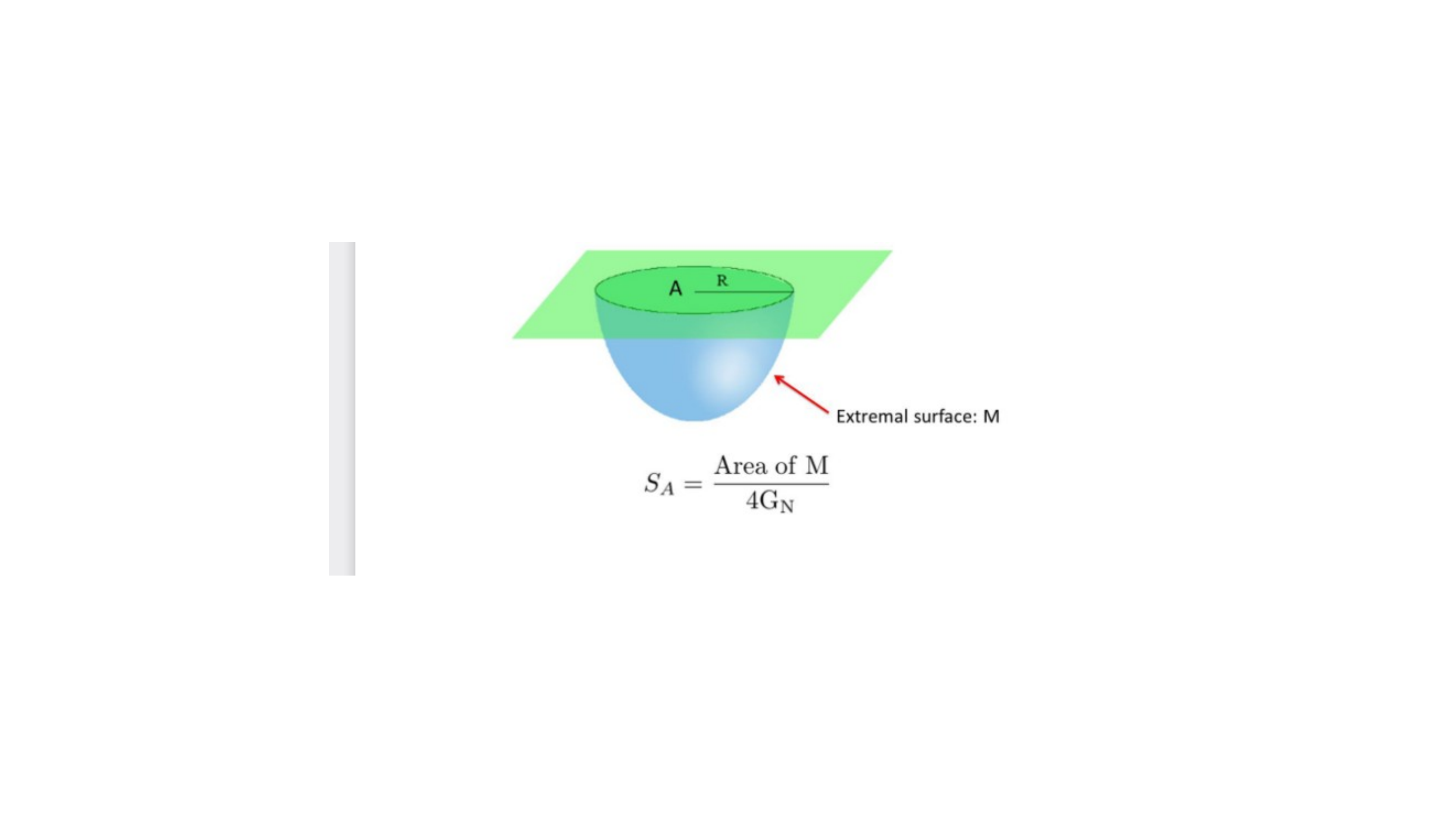}
\includegraphics[width=.65\textwidth]{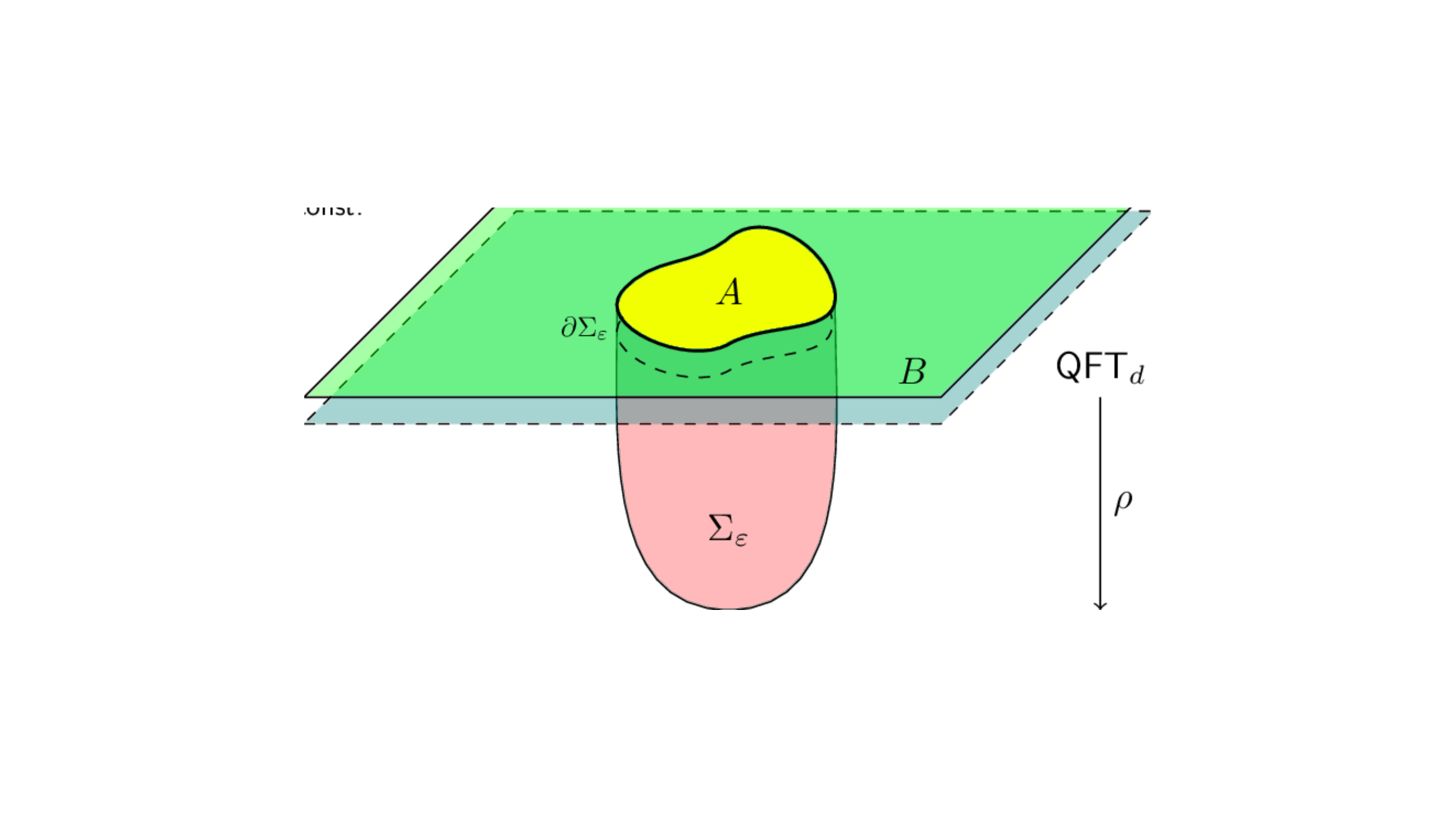}
\caption{(left)\,:\, Entangling surface from RT proposal, no cut off, \,\,\, (right)\,:\,Entangling surface from RT proposal in the presence of a cut off }
\la{entanglingsurface}
\end{figure}

\subsection{Holographic entanglement of entropy for $ d - \theta = 1$}

We start with the expression of holographic entanglement of entropy for $ d - \theta = 1$ which is given by
\be
S_{d - \theta = 1} = \log \left\lbrack {\frac{\rho_0 + \sqrt{\rho_0^2 - \rho_c^2}}{\rho_c}}\right\rbrack
\la{s0L}
\ee

 We recall at $d - \theta = 1$ , we have the turning point $\rho_0$ given by (\ref{dthetra1})

\be
\rho_0^2 = {\left( {\frac{l}{2}}  \right)}^2 + \rho_c^2
\la{rho0expression}
\ee

Substituting (\ref{rho0expression})  in (\ref{s0L}) we obtain, for $d - \theta = 1$

\be
S_{d - \theta = 1} = {\frac{R^d L^{d - 1}}{4G_N}} \log \left\lbrack {\frac{\rho_0 +  {\frac{l}{2}}}{\rho_c}}\right\rbrack
\la{s0complete}
\ee

So the plot for H.E.E vs $(l,\rho_c)$   with ${\frac{R^d L^{d - 1}}{4G_N}}$ as unit,     is shown in in Fig.(\ref{heeb1})

\begin{figure}[H]
\begin{center}
\textbf{ For $ d - \theta = 1$, H.E.E\, vs \,$(l,\rho_c)$, plotted for long range of $(l,\rho_c)$ to probe $l >> \rho_c$ and $\rho_c >> l$ regime  where the expression of H.E.E is exact }
\end{center}
\vskip2mm
\includegraphics[width=.85\textwidth]{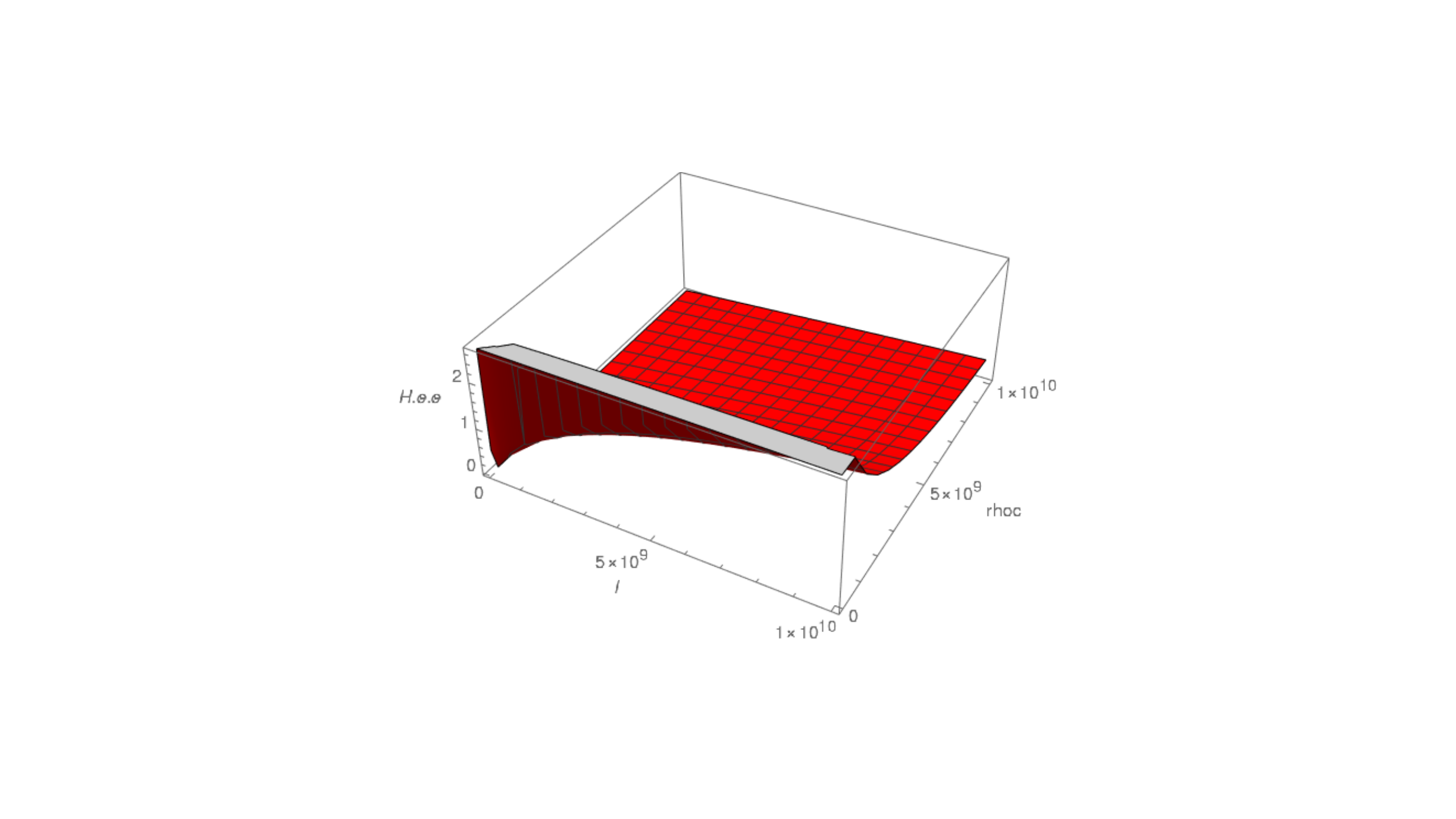}
\includegraphics[width=.55\textwidth]{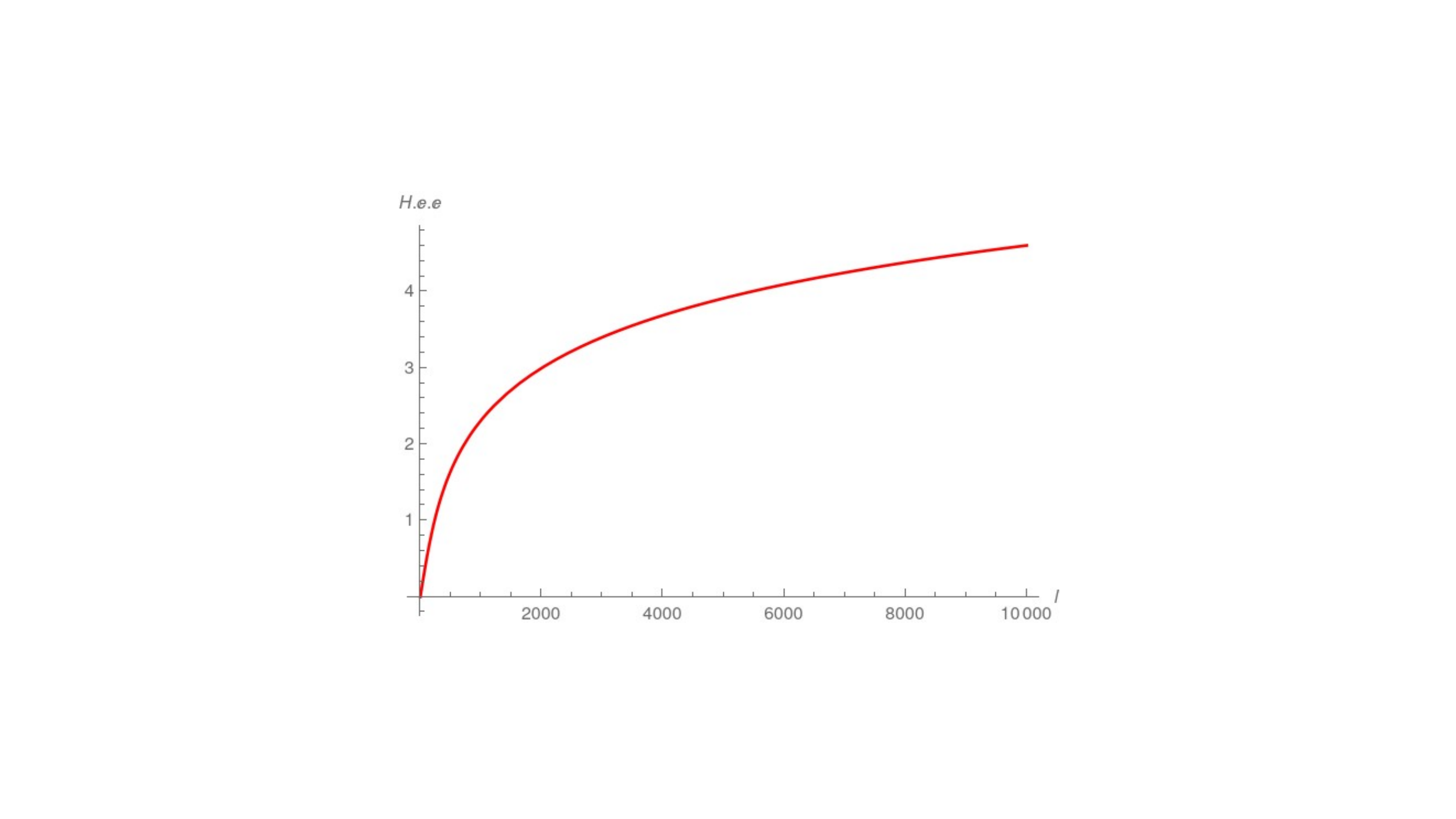}
\includegraphics[width=.55\textwidth]{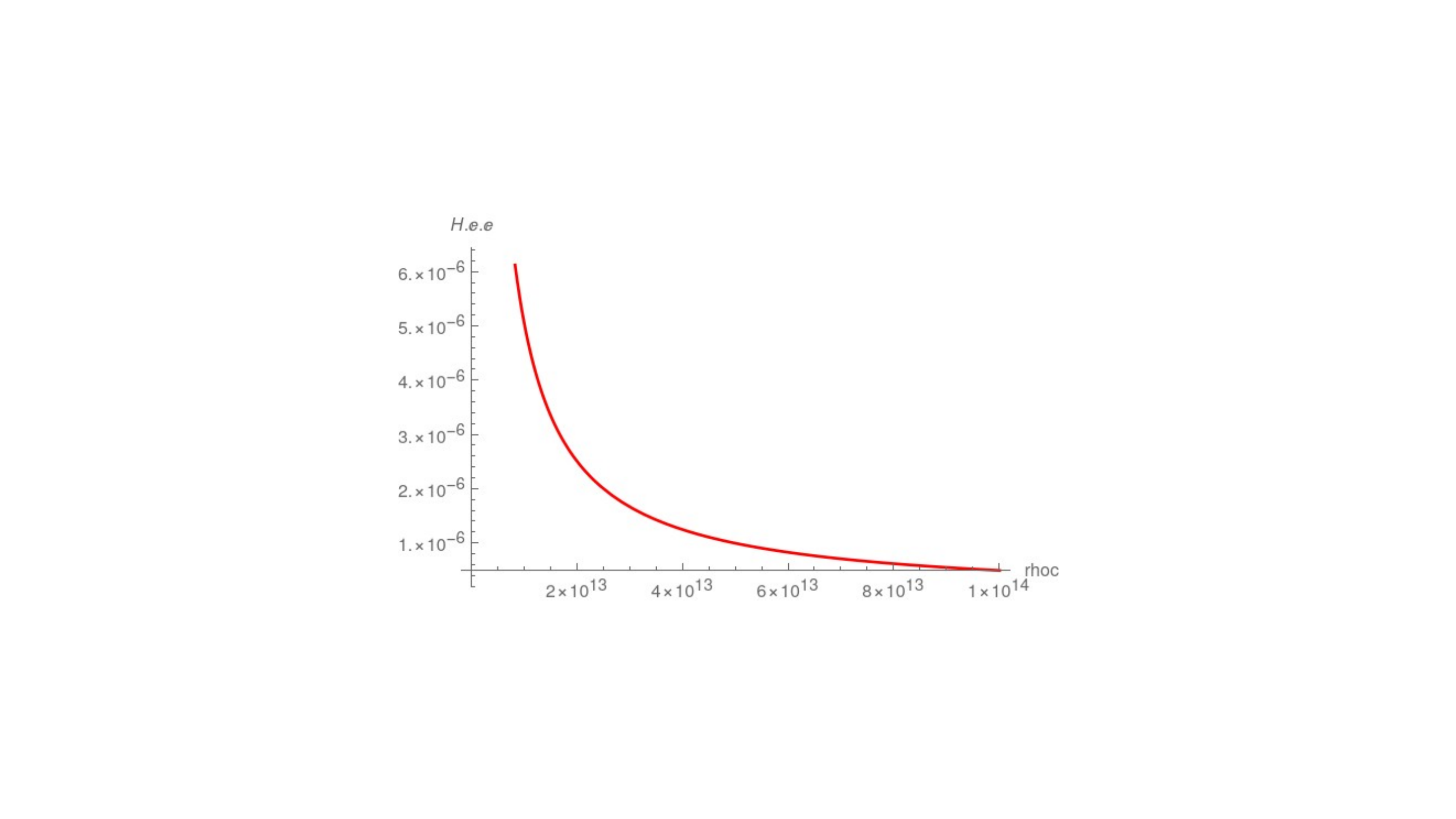}
\caption{(First row)\, : \, H.E.E for $d - \theta = 1$, vs $(l,\rho_c)$ \quad;\quad (Last row)\,: \,  (left) \, : \. H.E.E  vs l for  $\rho_c =  100$ \,\,,\,\, (right) \,,\,: H.E.E vs $\rho_c$ plot for $ l = (10)^8 $, \,\, All the plots are showing that H.E.E, for $ d - \theta = 1$,  increases with the increase in l and decreases with the increase in $\rho_c$ and goes to zero in $\rho_c >> l$ regime, as expected }
\la{heeb1}
\end{figure}

\subsection{ H.E.E for $d - \theta > 1 $   }

Here we first recall  the global  solution for the turning point $\rho_0(l, \rho_c)$,  as given in (\ref{ultimaterho0nonvanishing}),  which, according to the analysis of the last section,  can be considered as the exact solution over $l  >> \rho_c$ and $\rho_c >> l$ 
regime,  while for the rest of the regime of $(l , \rho_c)$ plane, it gives  an interpolating expression between these two regimes, when all our consistency-check plots, for this global solution  in the last section, showing, even this interpolating expressions are quite close to the exact one  and that way we will see that they are good enough to give all the expected properties of H.E.E, H.M.I, E.W.C.S,  when they  are used up to construct these respective expressions,  throughout the complete regime of $(l,\rho_c)$!  
Consequently, when we substitute $\rho_0 (l,\rho_c)$ from  (\ref{ultimaterho0nonvanishing})  in    (\ref{entropy}), the consequent expression of S can be considered as the exact expression in the $l>> \rho_c$ and $\rho_c >> l$ regime, while the same expression in the compensating regime of $(l \,, \,\rho_c)$ plane can be considered as the interpolating expression of S between these two regime with the interpolating expression is quite close to the exact one!

For $d - \theta>1$, we have the expression of the holographic entanglement of entropy for $d - \theta > 1$ given by

\ber
      S &=&   {\frac{      L^{d-1} {\left({\left( {\left({\frac{A_{10} l}{ 2 }}\right)}^{2  \left( d - \theta\right)} + {\left( \rho_c \right)}^{2  \left( d - \theta\right)} \right)}^{\frac{1}{2(d - \theta)}}\right)}^{\theta - d +1} \,\, {{}_2 F_1}   \left\lbrack {\frac{1}{2}}, {\frac{1}{2}}( - 1 +{ \frac{1}{d - \theta }}), {\frac{1}{2}}(1 +{ \frac{1}{d - \theta }}) , 1 \right \rbrack }{  4 G_N (\theta + 1- d )    }}\n
                      &- & \left(\rho_c \right)^{ \theta +1- d} {\frac{     L^{d-1}  \,\, {{}_2 F_1}   \left\lbrack {\frac{1}{2}}, {\frac{1}{2}}( - 1 +{ \frac{1}{d - c}}), {\frac{1}{2}}(1 +{ \frac{1}{d - \theta}}) , \left({\frac{\rho_c}{\left\lbrack    
{\left( {\left({\frac{A_{10} l}{ 2 }}\right)}^{2  \left( d - \theta\right)} + {\left( \rho_c \right)}^{2  \left( d - \theta\right)} \right)}^{\frac{1}{2(d - \theta)}} \right\rbrack }}\right)^{2(d - \theta)}  \right \rbrack }{  4 G_N( \theta - d  + 1)   }}
                      \la{entropylrhoc1}
                      \eer

 Since the expression of S is unique and  identical in both $l>> \rho_c$ and $\rho_c >> l$ and we have the complete interpolating expression between them, so essentially here we will present 3D and 2D  plots $(S, l,\rho_c)$ for $d - \theta > 1$ to establish its expected properties. Here we mainly present the plots 
over very long range of $(l , \rho_c)$ to visualize $l >> \rho_c $ and $\rho_c >> l $ regime where the expression of H.E.E is exact.    However we also present some plots over the short range to establish the fact that even in the regime of $(l,\rho_c)$ plane away from the $l >> \rho_c $ and $\rho_c >> l $,  the interpolating expression is close to the exact one and consequently showing the expected behaviour!  Here we have shown the plots for two values of $d - \theta$.  The plots for $d - \theta = {\frac{5}{3}}$ in Fig.(\ref{g5by33Ds1}, \ref{g5by33Dl1}, \ref{5by32Ds1})   and the plots for $d - \theta = {\frac{9}{2}}$ in Fig.( \ref{g9by23Ds1}, \ref{g9by23Dl1}, \ref{g9by22Ds1},  \ref{g9by22Dl1}  )

\begin{figure}[H]
\begin{center}
\textbf{ For $ d - \theta = {\frac{5}{3}}$,  Presentation in short range for l,$\rho_c$  }
\end{center}
\vskip2mm
\includegraphics[width=.65\textwidth]{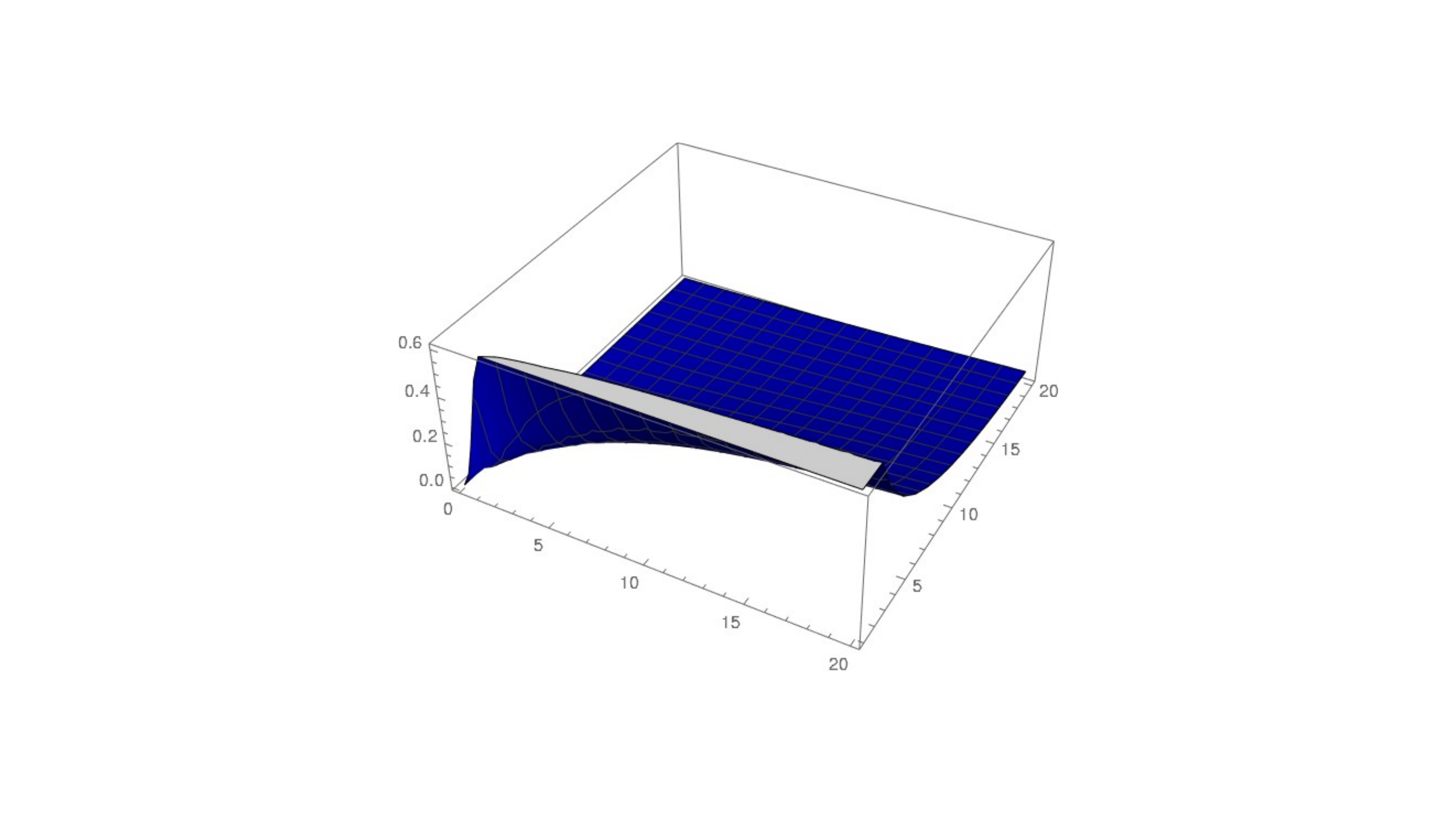}
\includegraphics[width=.65\textwidth]{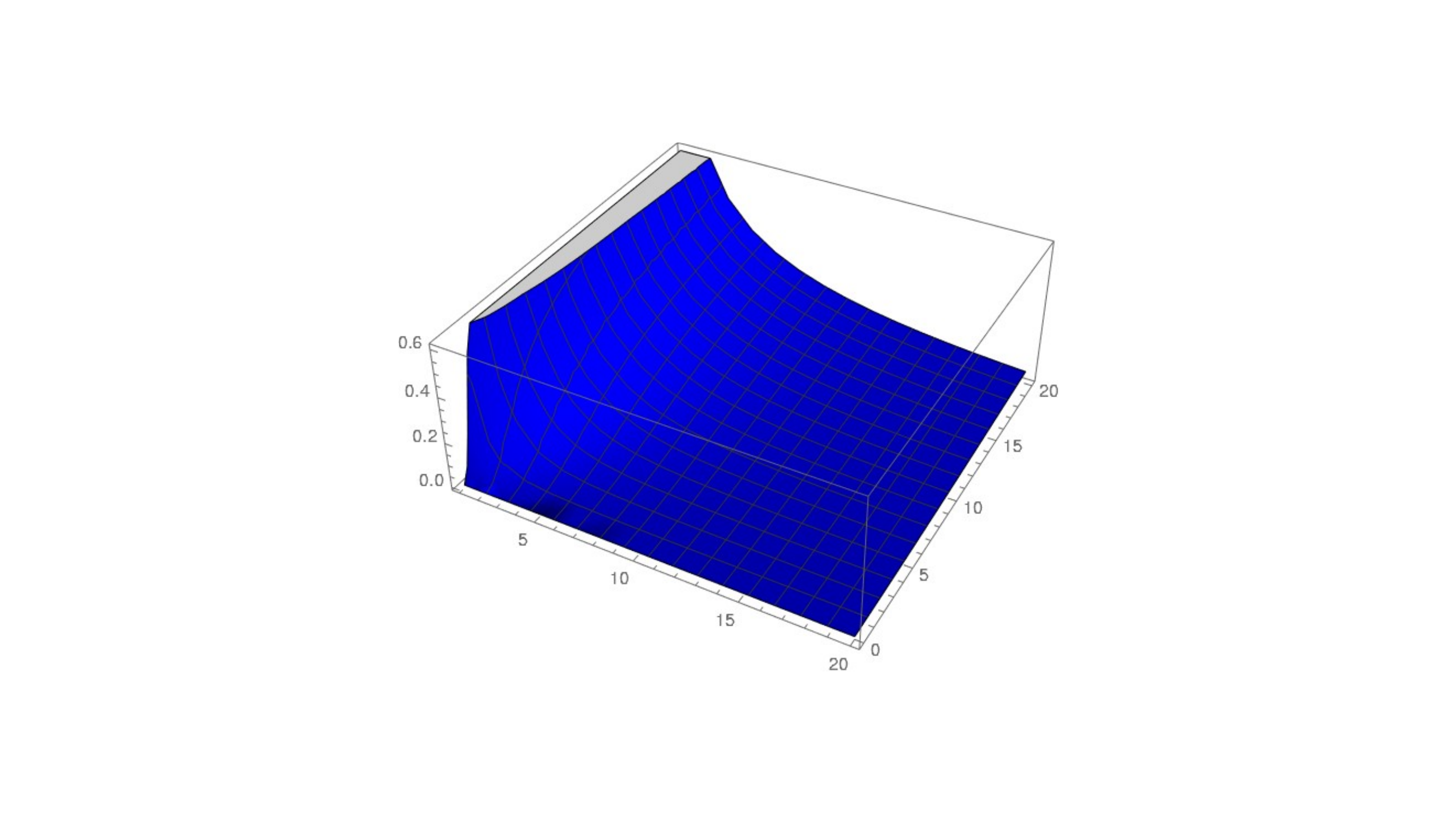}	
\caption{ S as a function of l, $\rho_c$, which is exact in $l  >> \rho_c$ and $\rho_c>> l$ and in rest of the regime its an interpolating function\quad;\quad (left) x-axis\, : l\, y-axis $\rho_c$\,\, (right) x-axis and y-axis are interchanged\,\, : \,\,  The plots are showing that even we are far away from  $l >> \rho_c $ and $\rho_c >> l $ regime still the expression of H.E.E is showing its expected behaviour, i.e increasing with l and falling with the increase of cut-off $\rho_c$ establishing the fact that in the interpolating regime the expression of S is close to the exact one }
	
\label{g5by33Ds1}
\end{figure}

\begin{figure}[H]
\begin{center}
\textbf{ For $ d - \theta = {\frac{5}{3}}$,  Presentation in long range for l,$\rho_c$  }
\end{center}
\vskip2mm
\includegraphics[width=.65\textwidth]{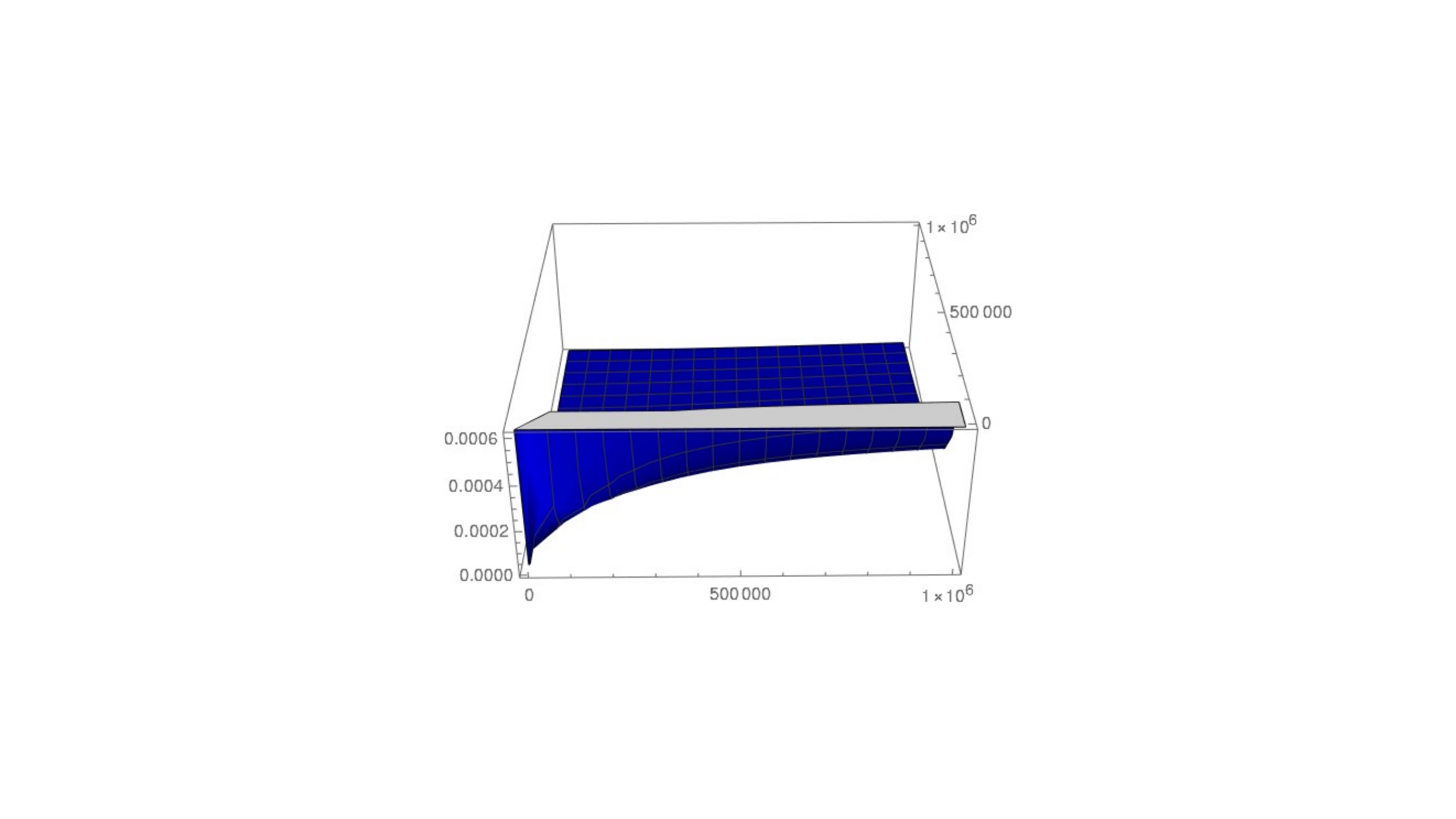}
\includegraphics[width=.65\textwidth]{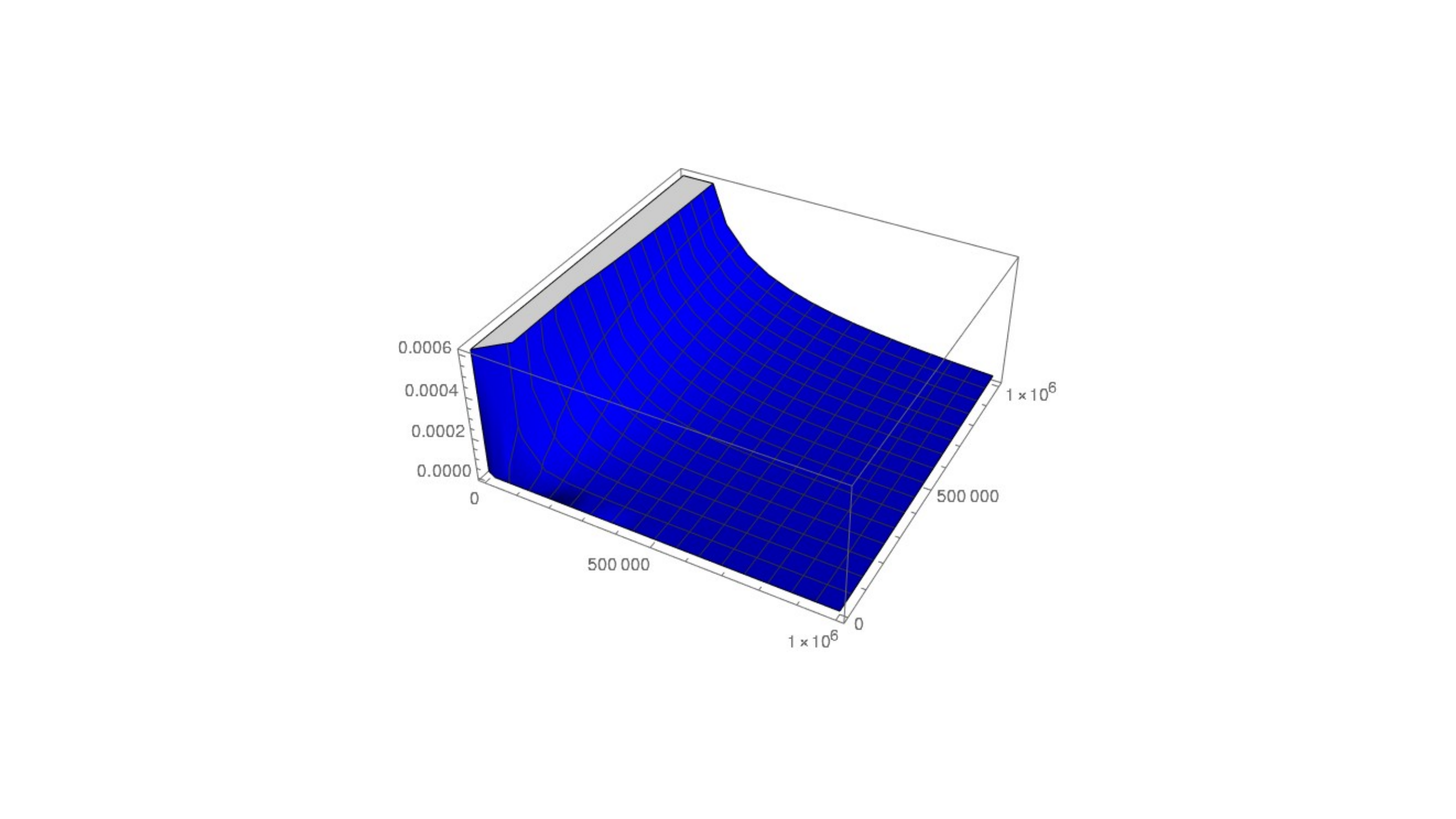}	
\caption{ S as a function of l, $\rho_c$, considered over long range  to probe  $l>>\rho_c$ and $\rho_c>> l$  regime where the expression of H.E.E is exact and in rest of the regime its an interpolating function\quad;\quad (left) x-axis\, : l\, y-axis $\rho_c$\,\, (right)\, x-axis and y-axis are interchanged \,\,: \,\, Both the plots are showing that H.E.E maintains its expected behaviour throughout $(l,\rho_c)$ plane, i.e increasing with l and falling with the increase of $\rho_c$}
	
\label{g5by33Dl1}
\end{figure}

\begin{figure}[H]
\begin{center}
\textbf{ For $ d - \theta = {\frac{5}{3}}$ \, , \, Presentation at long range  in 2D}
\end{center}
\vskip2mm
\includegraphics[width=.65\textwidth]{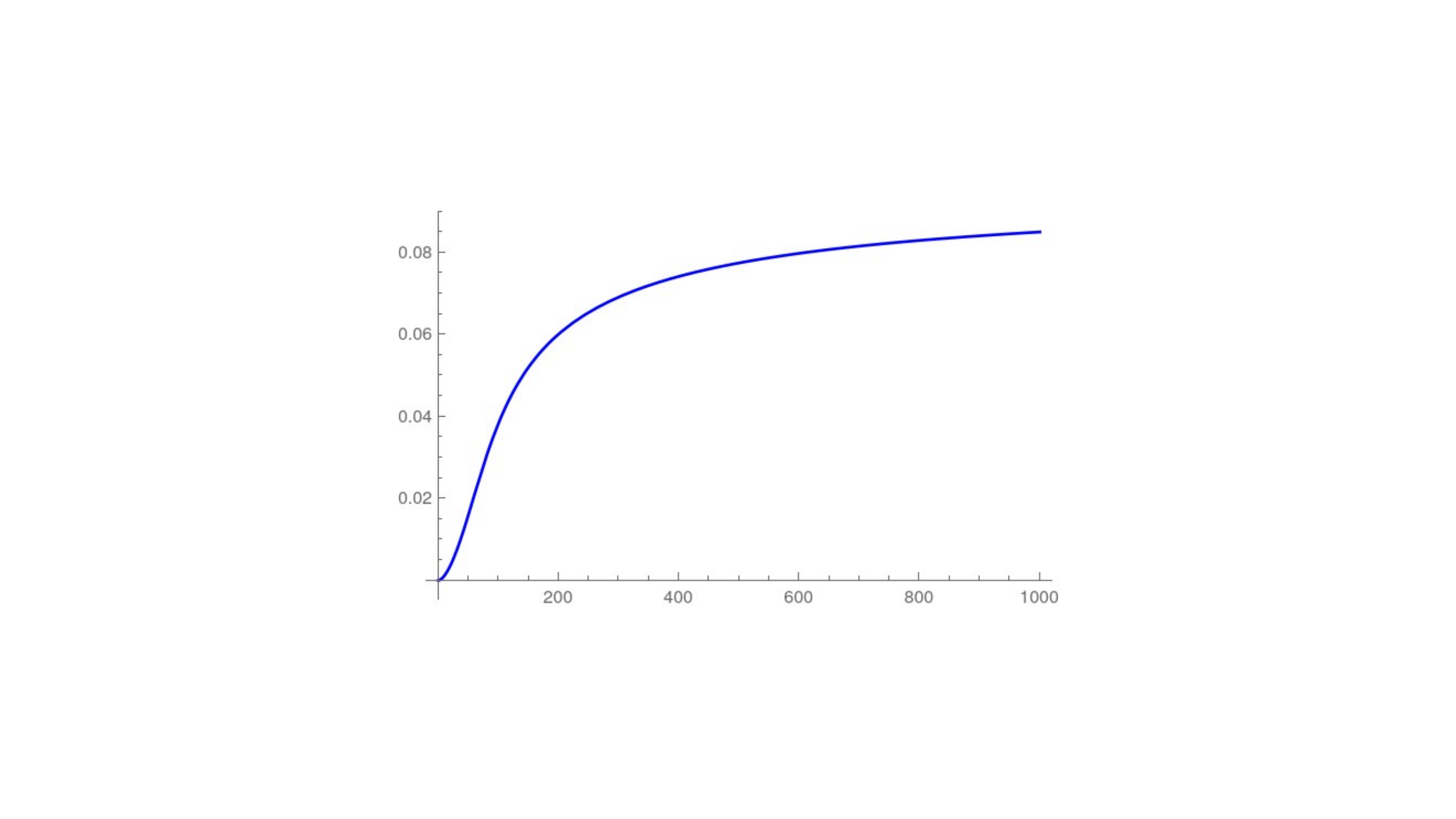}
\includegraphics[width=.65\textwidth]{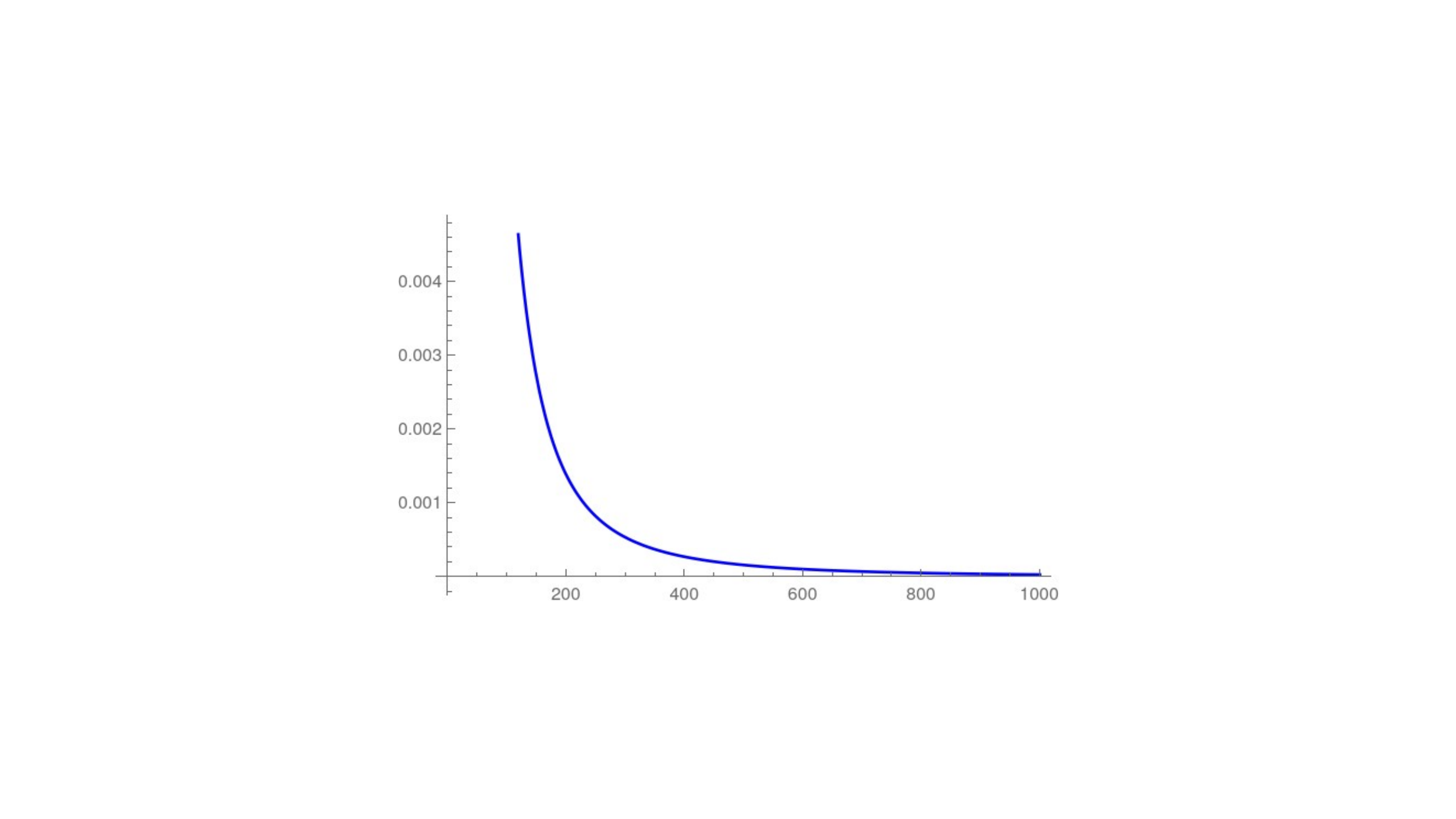}	
\caption{2D view \quad;\quad (left) S as a function of l. for $\rho_c = 60$   \,\, (right) \, S as a function of $\rho_c$. for $ l = 60$ \,\,: \,\, Both the plots are showing that H.E.E maintains its expected behaviour  i.e increasing with l and falling with the increase of $\rho_c$ }
\label{5by32Ds1}
\end{figure}

\begin{figure}[H]
\begin{center}
\textbf{ For $ d - \theta = {\frac{9}{2}}$,  Presentation in short range for l,$\rho_c$  }
\end{center}
\vskip2mm
\includegraphics[width=.65\textwidth]{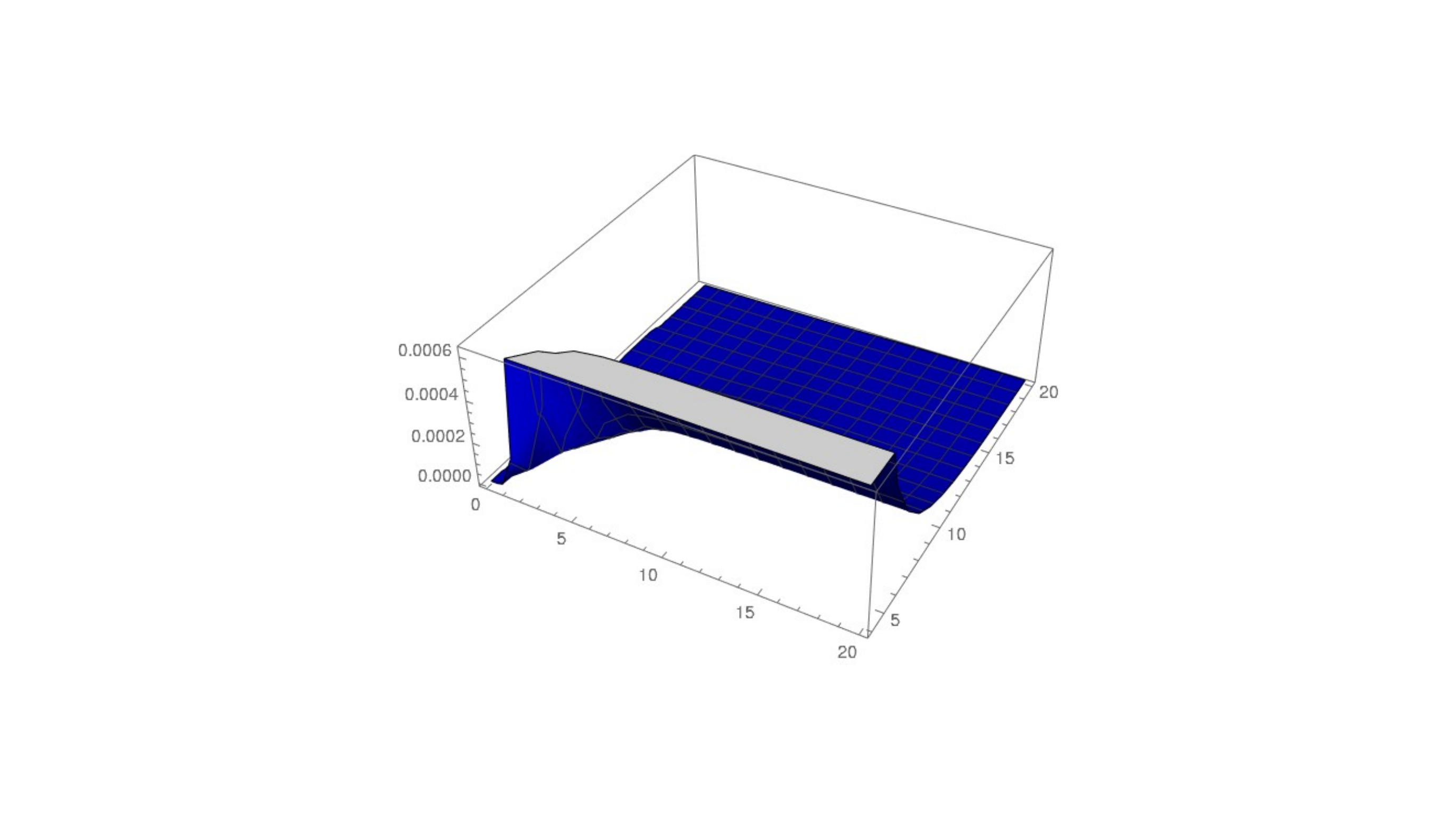}
\includegraphics[width=.65\textwidth]{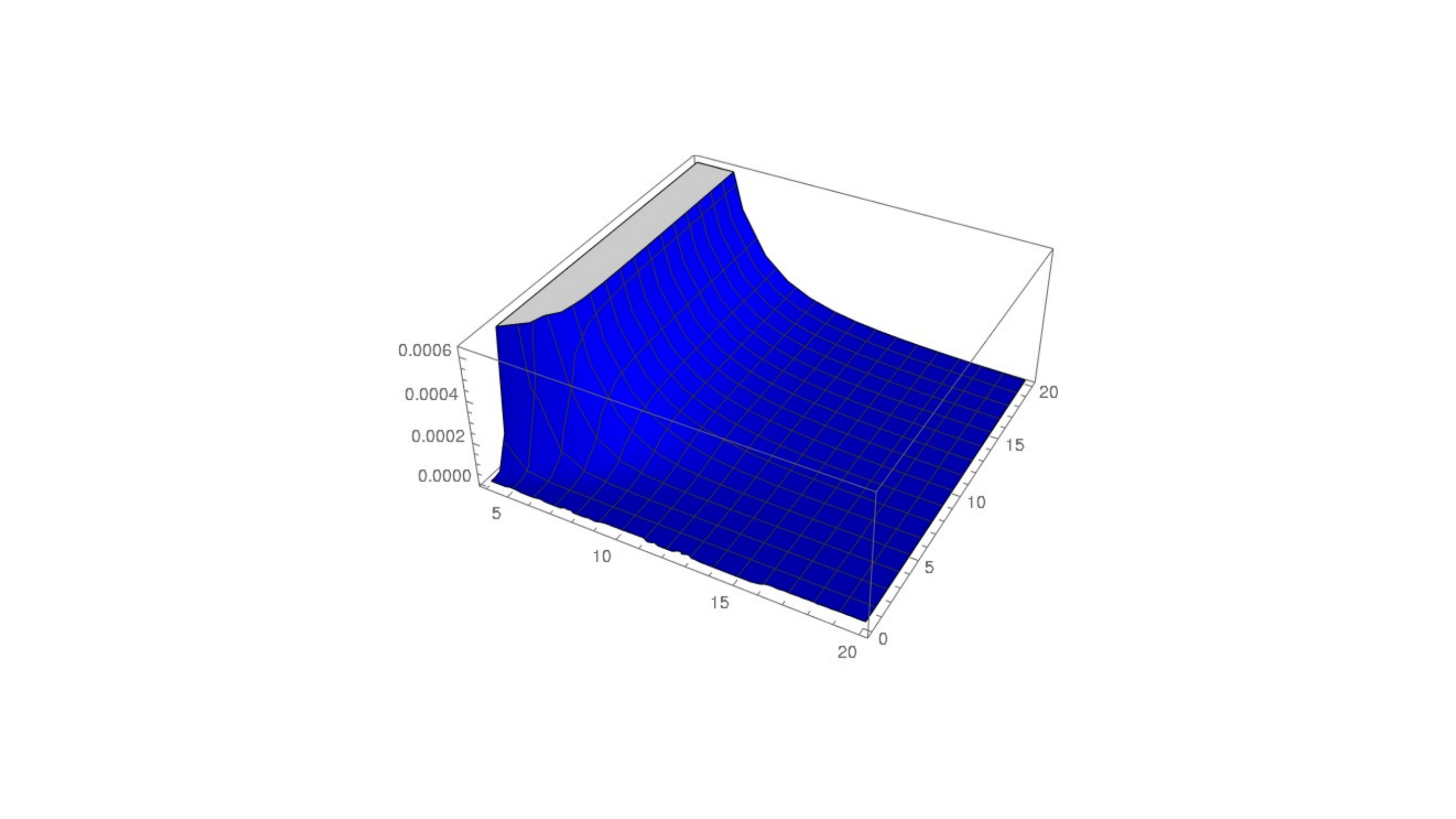}	
\caption{S as a function of l, $\rho_c$,  which is exact in $l >>  \rho_c$ and $ \rho_c >> l$ and in rest of the regime its an interpolating function\quad;\quad (left) x-axis\, : l\, y-axis $\rho_c$\,\, (right) x-axis and y-axis are interchanged\,\, : \,\, The plots are showing that even we are far away from  $l >> \rho_c $ and $\rho_c >> l $ regime still the expression of H.E.E is showing its expected behaviour, i.e increasing with l and falling with the increase of cut-off $\rho_c$  establishing the fact that in the interpolating regime the expression of S is close to the exact one   }
\label{g9by23Ds1}
\end{figure}

\begin{figure}[H]
\begin{center}
\textbf{ For $ d - \theta = {\frac{9}{2}}$,  Presentation in long range for l,$\rho_c$  }
\end{center}
\vskip2mm
\includegraphics[width=.65\textwidth]{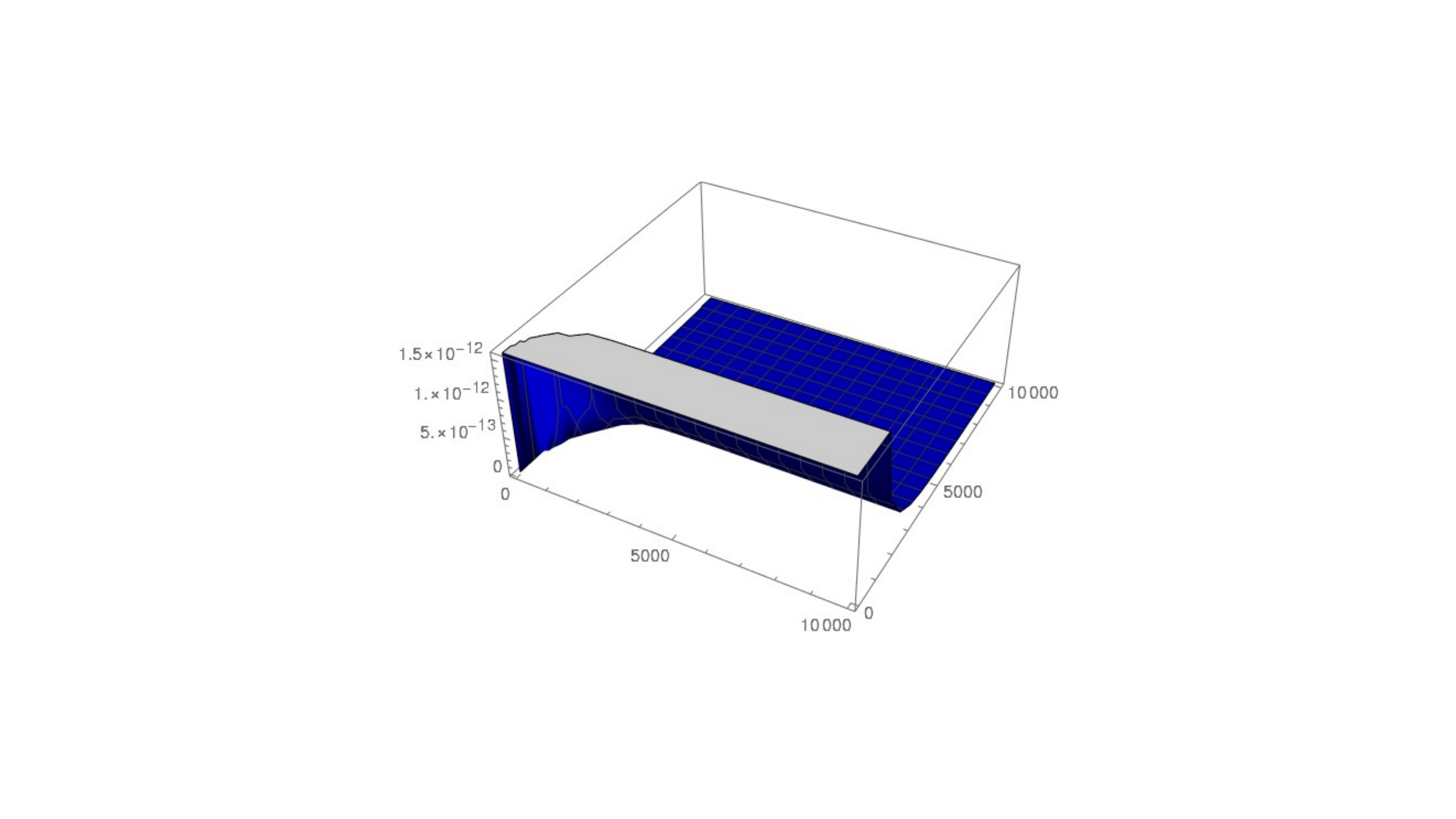}
\includegraphics[width=.65\textwidth]{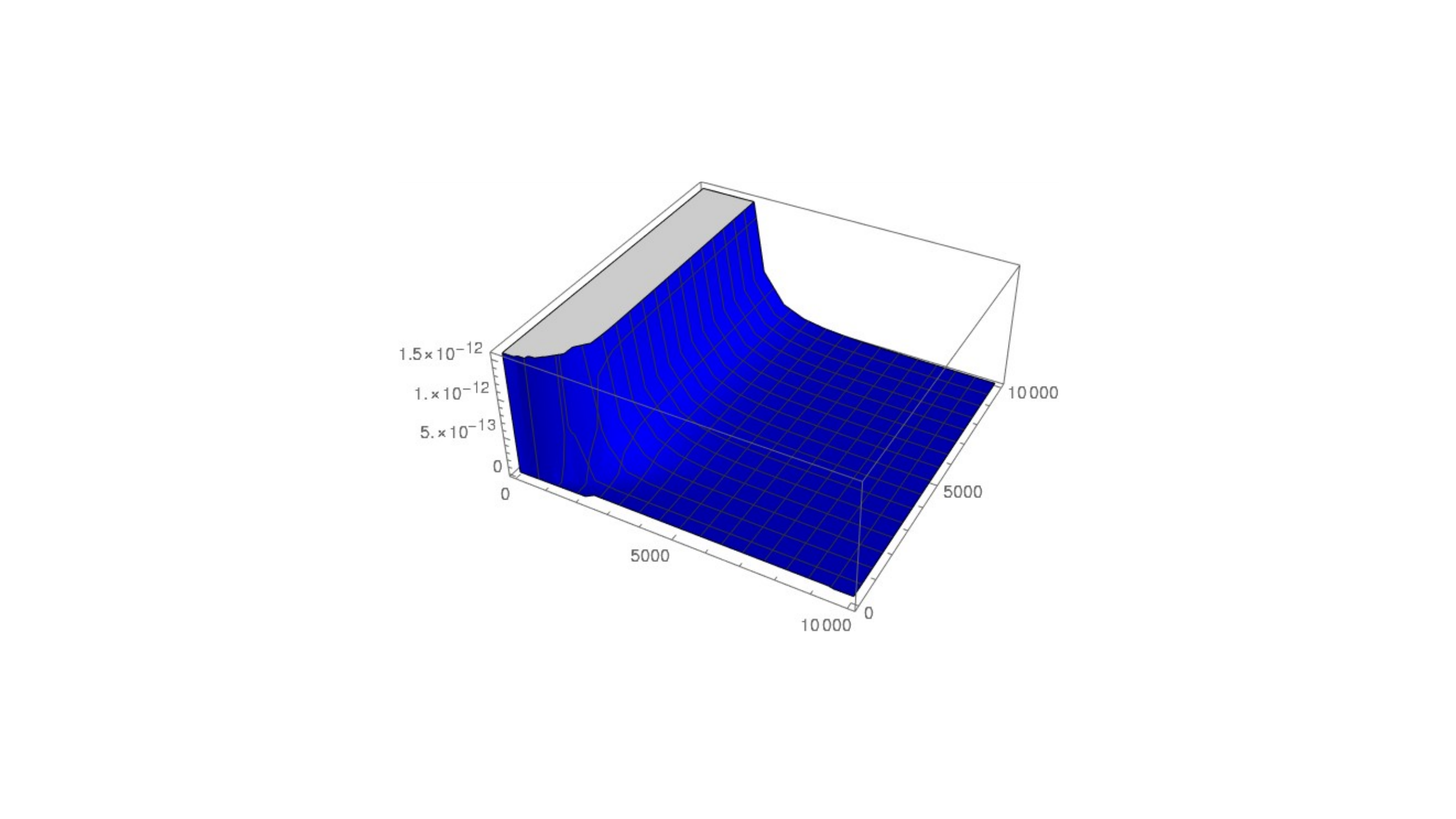}	
\caption{S as a function of l, $\rho_c$, considered over long range  to probe  $l >> \rho_c$ and $\rho_c >> l$  regime where the expression of H.E.E is exact and in rest of the regime its an interpolating function \quad;\quad (left) x-axis \, : l\, y-axis $\rho_c$ \,\, (right)\, x-axis and y-axis are interchanged \,\,: \,\, Both the plots are showing that H.E.E maintains its expected behaviour throughout $(l,\rho_c)$ plane,  i.e increasing with l and falling with the increase of $\rho_c$  }
\label{g9by23Dl1}
\end{figure}

\begin{figure}[H]
\begin{center}
\textbf{ For $ d - \theta = {\frac{9}{2}}$ \, , \, Presentation at short range  in 2D}
\end{center}
\vskip2mm
\includegraphics[width=.65\textwidth]{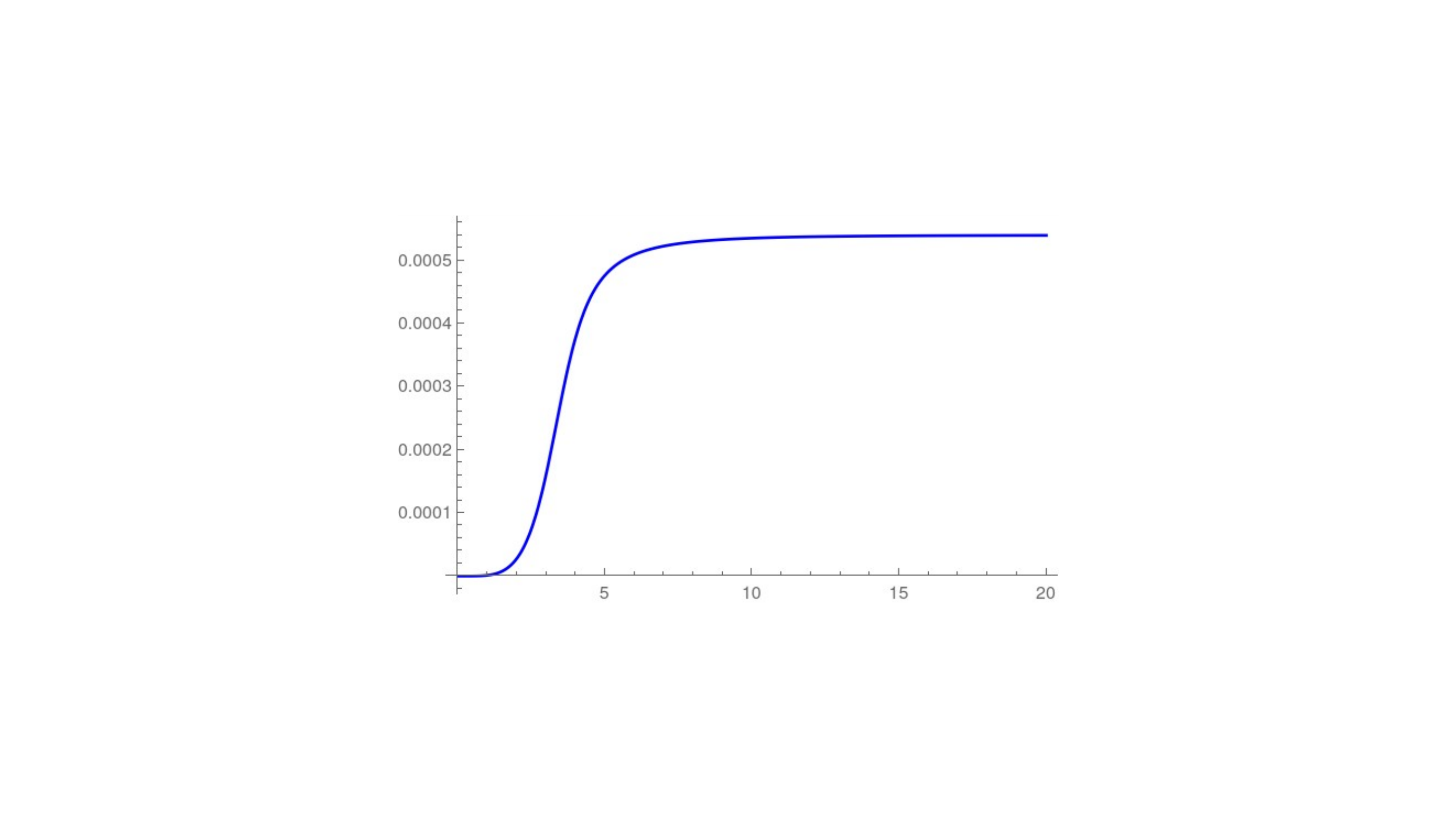}
\includegraphics[width=.65\textwidth]{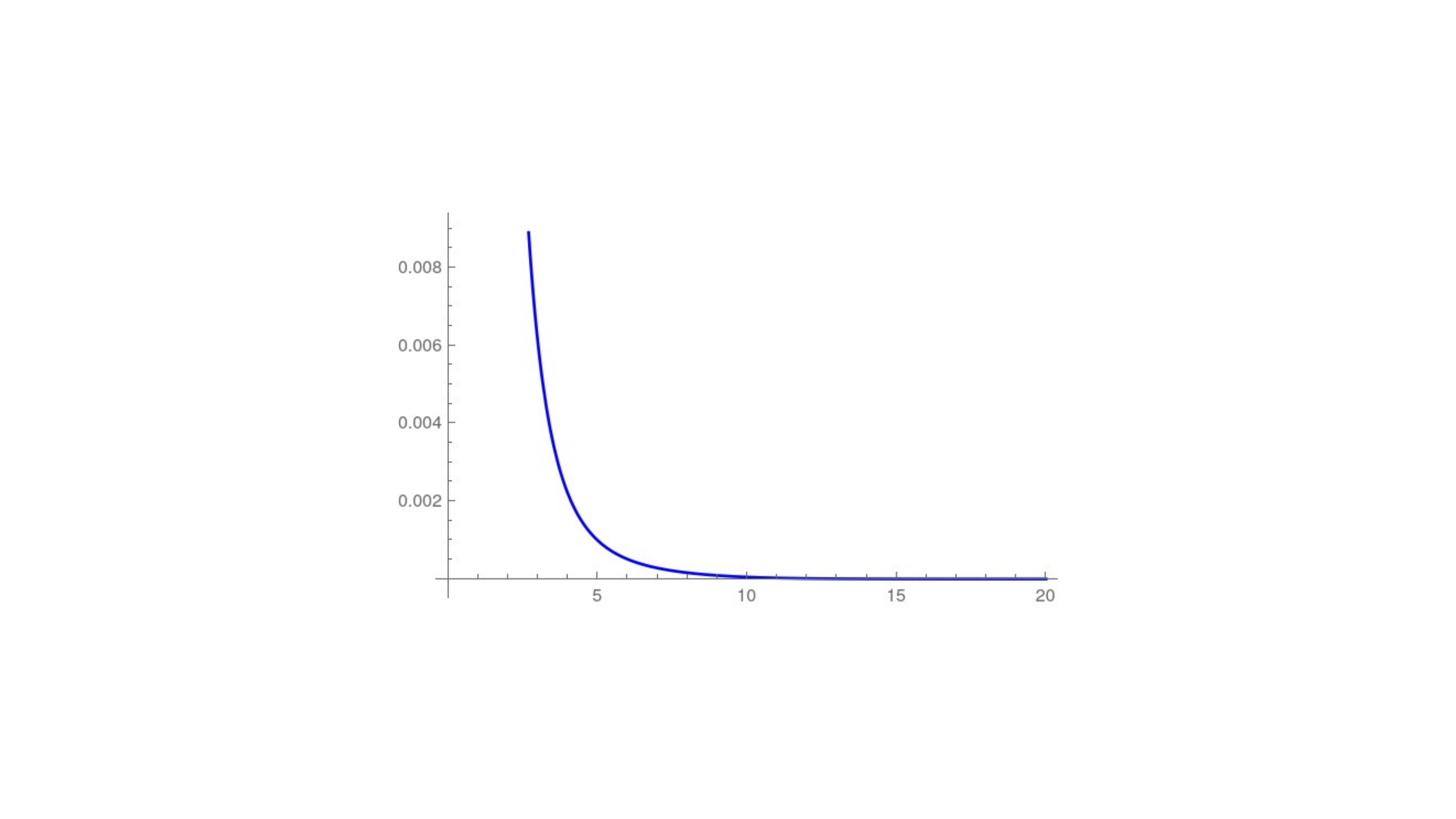}	
\caption{2D view \quad;\quad (left) S as a function of l,  for $\rho_c = 6$   \,\, (right) \, S as a function of $\rho_c$. for $ l = 6$ \,\, : \,\, Both the plots are showing that H.E.E maintains its expected behaviour, i.e increasing with l and falling with the increase of $\rho_c$, even in the regime far away from $l >> \rho_c$ and  $\rho_c >> l $ regime of  $(l,\rho_c)$ plane
   }
\label{g9by22Ds1}
\end{figure}

\begin{figure}[H]
\begin{center}
\textbf{ For $ d - \theta = {\frac{9}{2}}$ \, , \, Presentation at long range  in 2D}
\end{center}
\vskip2mm
\includegraphics[width=.65\textwidth]{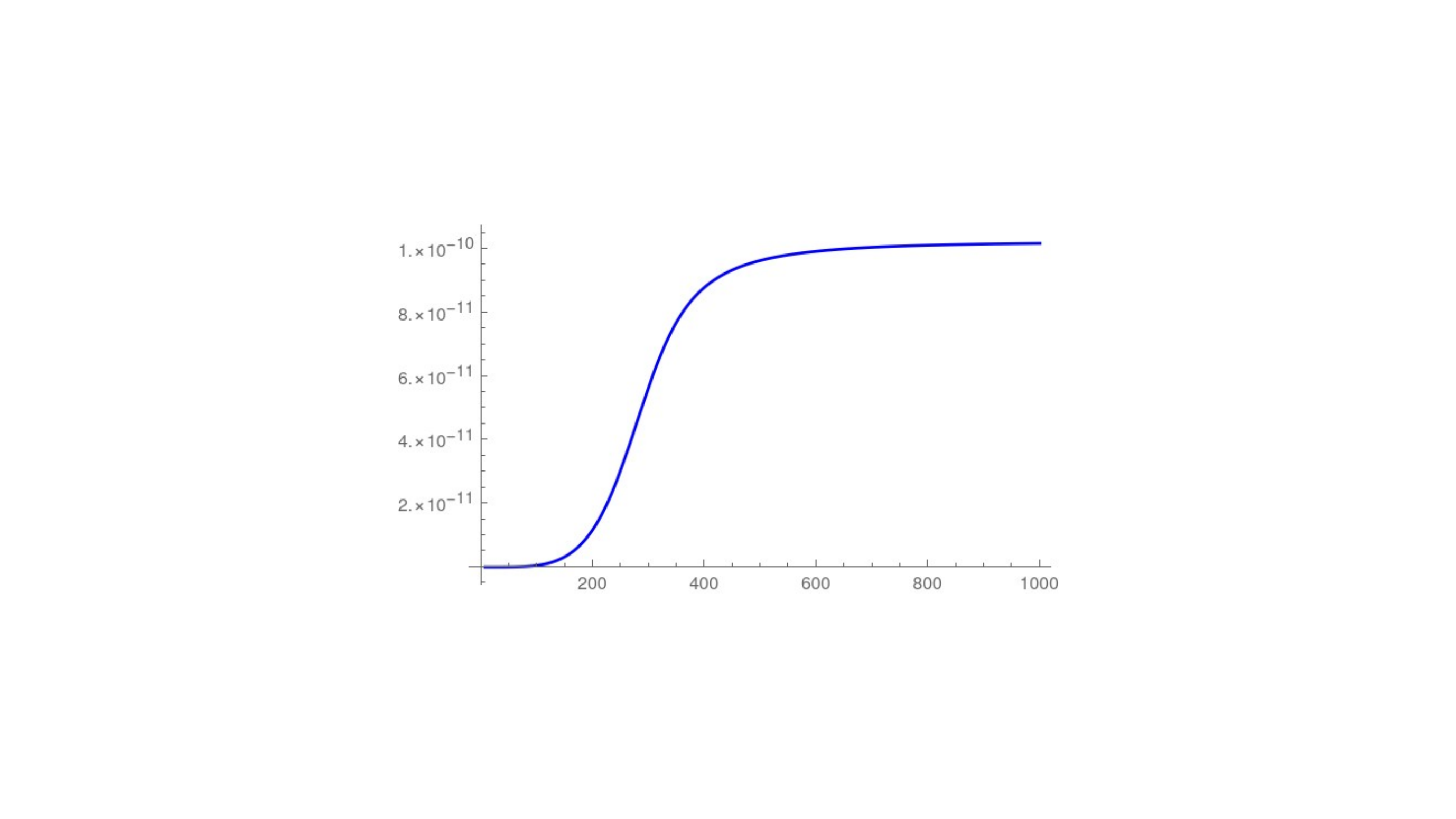}
\includegraphics[width=.65\textwidth]{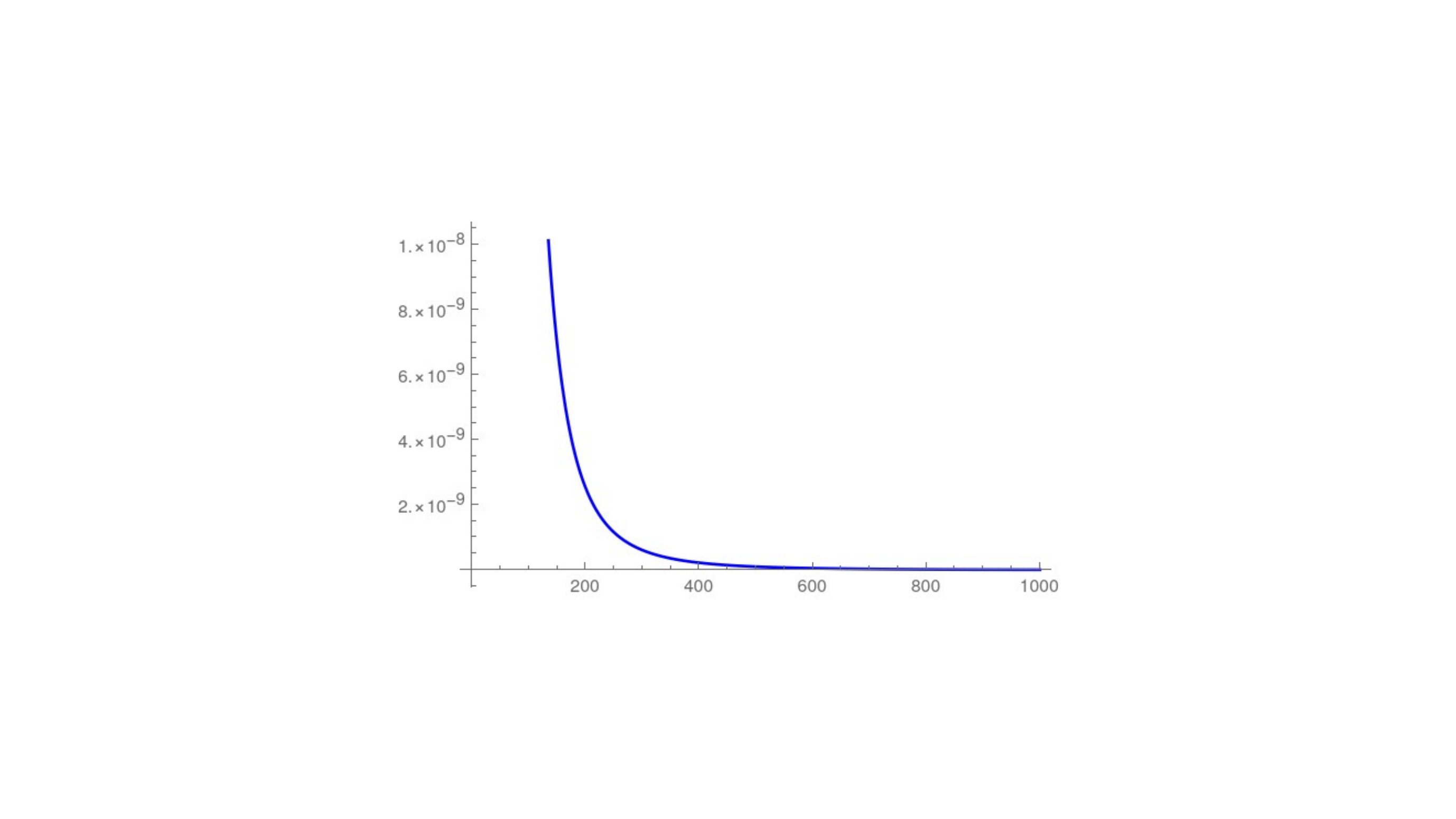}	
\caption{2D view \quad;\quad (left)\, S as a function of l,  for $\rho_c = 500$   \,\, (right) \, S as a function of $\rho_c$. for $ l = 500$ \,\, : \,\, Both the plots are showing that H.E.E maintains its expected behaviour, i.e increasing with l and falling with the increase of $\rho_c$   }
\label{g9by22Dl1}
\end{figure}

\subsection{H.E.E for $ d - \theta < 1$ }

The expression of S for $d - \theta < 1$, following (\ref{entropy}) and (\ref{ultimaterho0nonvanishing})

\ber
      S &=&   {\frac{      L^{d-1} {\left(  {\left( {\left({\frac{A_{10} l}{ 2 }}\right)}^{  \left( d - \theta + 1 \right)} + {\left( \rho_c \right)}^{  \left( d - \theta + 1 \right)} \right)}^{\frac{1}{(d - \theta + 1)}}   \right)}^{\theta - d +1} \,\, {{}_2 F_1}   \left\lbrack {\frac{1}{2}}, {\frac{1}{2}}( - 1 +{ \frac{1}{d - \theta }}), {\frac{1}{2}}(1 +{ \frac{1}{d - \theta }}) , 1 \right \rbrack }{  4 G_N (\theta + 1- d )    }}\n
                      &- & \left({\rho_c}\right)^{ \theta +1- d} {\frac{     L^{d-1}   \,\, {{}_2 F_1}   \left\lbrack {\frac{1}{2}}, {\frac{1}{2}}( - 1 +{ \frac{1}{d - c}}), {\frac{1}{2}}(1 +{ \frac{1}{d - \theta}}) , \left({\frac{\rho_c}{\left\lbrack  {\left( {\left({\frac{A_{10} l}{ 2 }}\right\rbrack}^{  \left( d - \theta + 1 \right\rbrack} + {\left( \rho_c \right)}^{  \left( d - \theta + 1 \right)} \right)}^{\frac{1}{(d - \theta + 1)}}  \right)}}\right)^{2(d - \theta)}  \right \rbrack }{  4 G_N( \theta - d  + 1)   }}
                      \la{entropylrhoc2}
                      \eer

Clearly, just like the case for $d - \theta > 1$ here also for  $ d - \theta < 1$   we have the above expression of S (\ref{entropylrhoc2}), gives the exact expression of S for $ l >> \rho_c$ and $ \rho_c  >> l $ and for the rest of the regime of $(l, \rho_c)$ plane, it can be thought as an interpolating expression between these two regime!

Here again, we consider ${\frac{L^{d-1} }{  4 G_N  }}$ as unit and evaluate HEE in this  unit.  Here we present the plots for $d - \theta =  {\frac{4}{9}}\,,\, {\frac{2}{9}}$.

\begin{figure}[H]
\begin{center}
\textbf{ For $ d - \theta = {\frac{4}{9}}$, for short range of $l, \rho_c$,  \, \,; \,\, 3D view   }
\end{center}
\vskip2mm
\includegraphics[width=.65\textwidth]{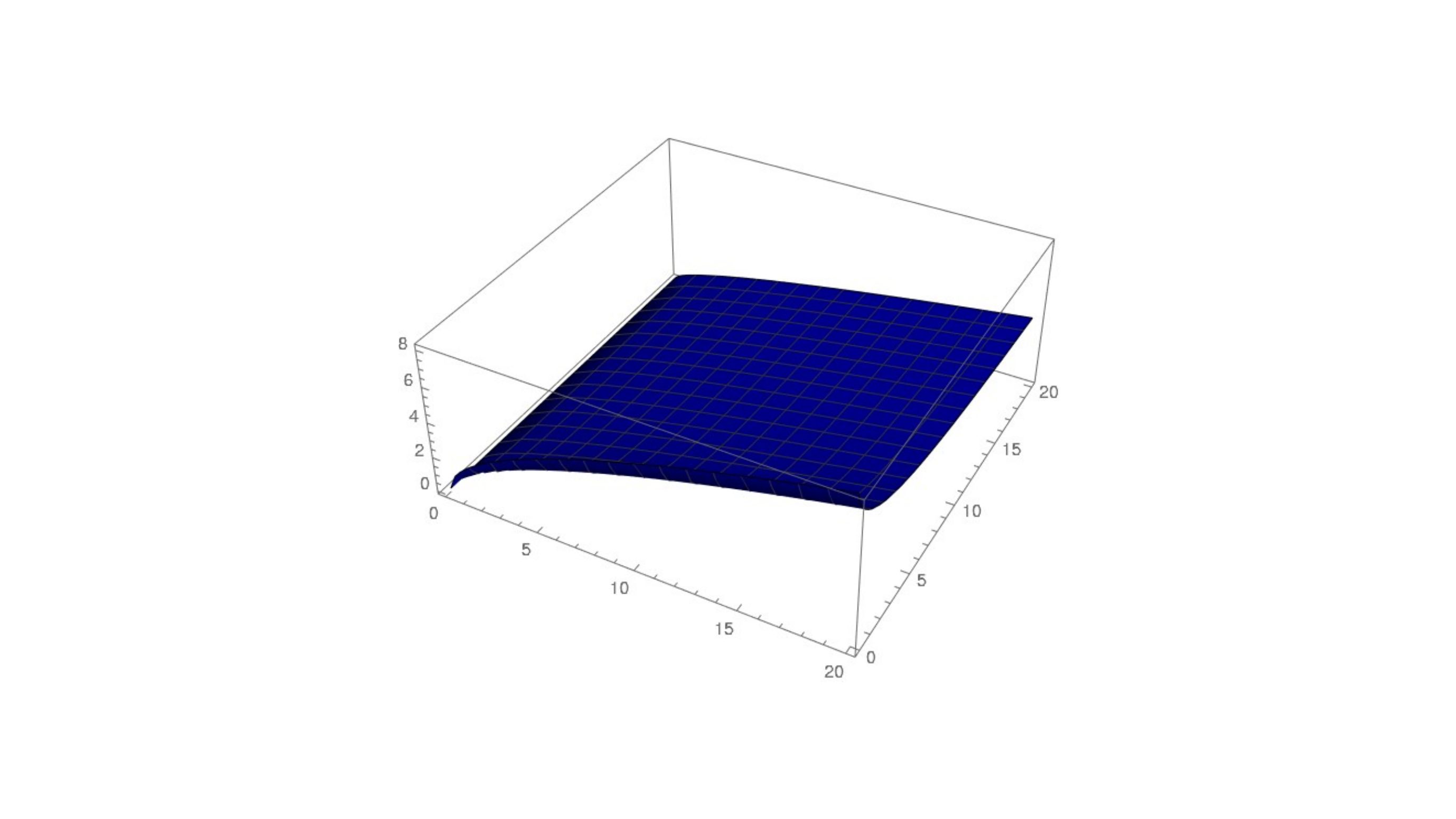}
\includegraphics[width=.65\textwidth]{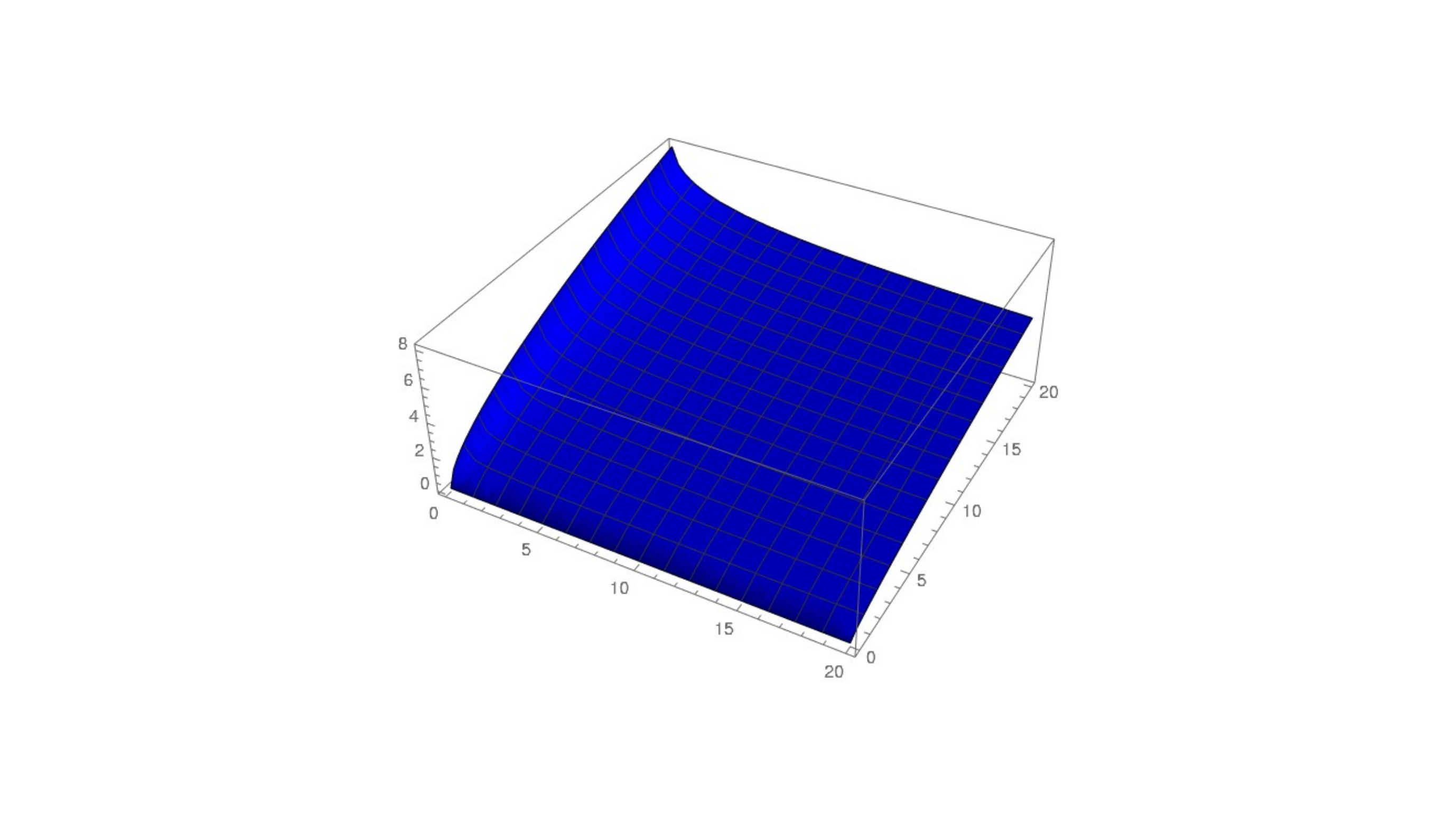}	
\caption{ S as a function of l, $\rho_c$, which is exact in $l  >> \rho_c$ and $\rho_c>> l$ and in rest of the regime its an interpolating function\quad;\quad (left) x-axis\, : l\, y-axis $\rho_c$\,\, (right) x-axis and y-axis are interchanged\,\, : \,\,  The plots are showing that even we are far away from  $l >> \rho_c $ and $\rho_c >> l $ regime still the expression of H.E.E is showing its expected behaviour, i.e increasing with l and falling with the increase of cut-off $\rho_c$ establishing the fact that in the interpolating regime the expression of S is close to the exact one}
\la{l4by93Ds1 }
\end{figure}

\begin{figure}[H]
\begin{center}
\textbf{ For $ d - \theta = {\frac{4}{9}}$, for short range of $l, \rho_c$,  \, \,; \,\,  2D view  }
\end{center}
\vskip2mm
\includegraphics[width=.65\textwidth]{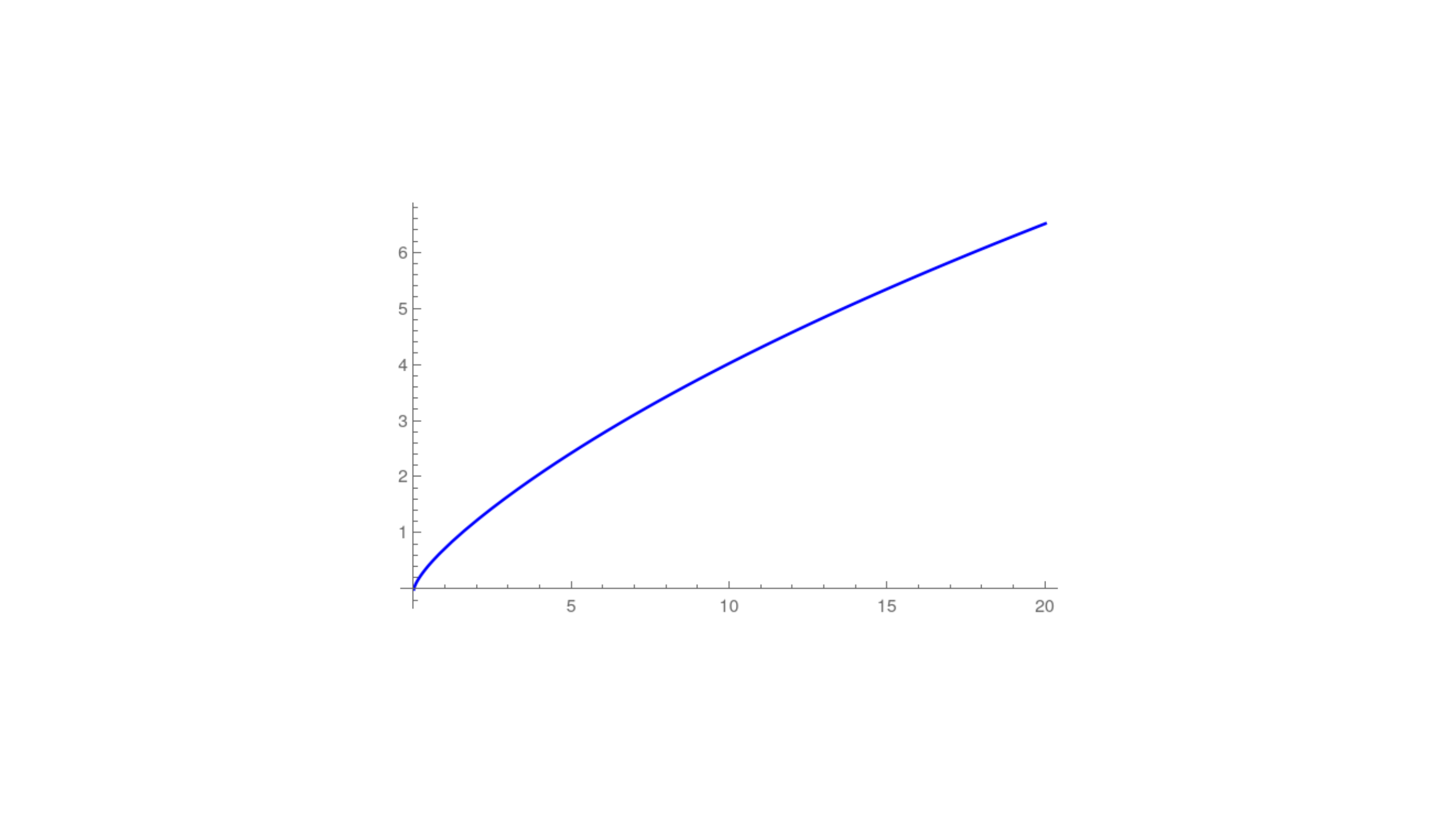}
\includegraphics[width=.65\textwidth]{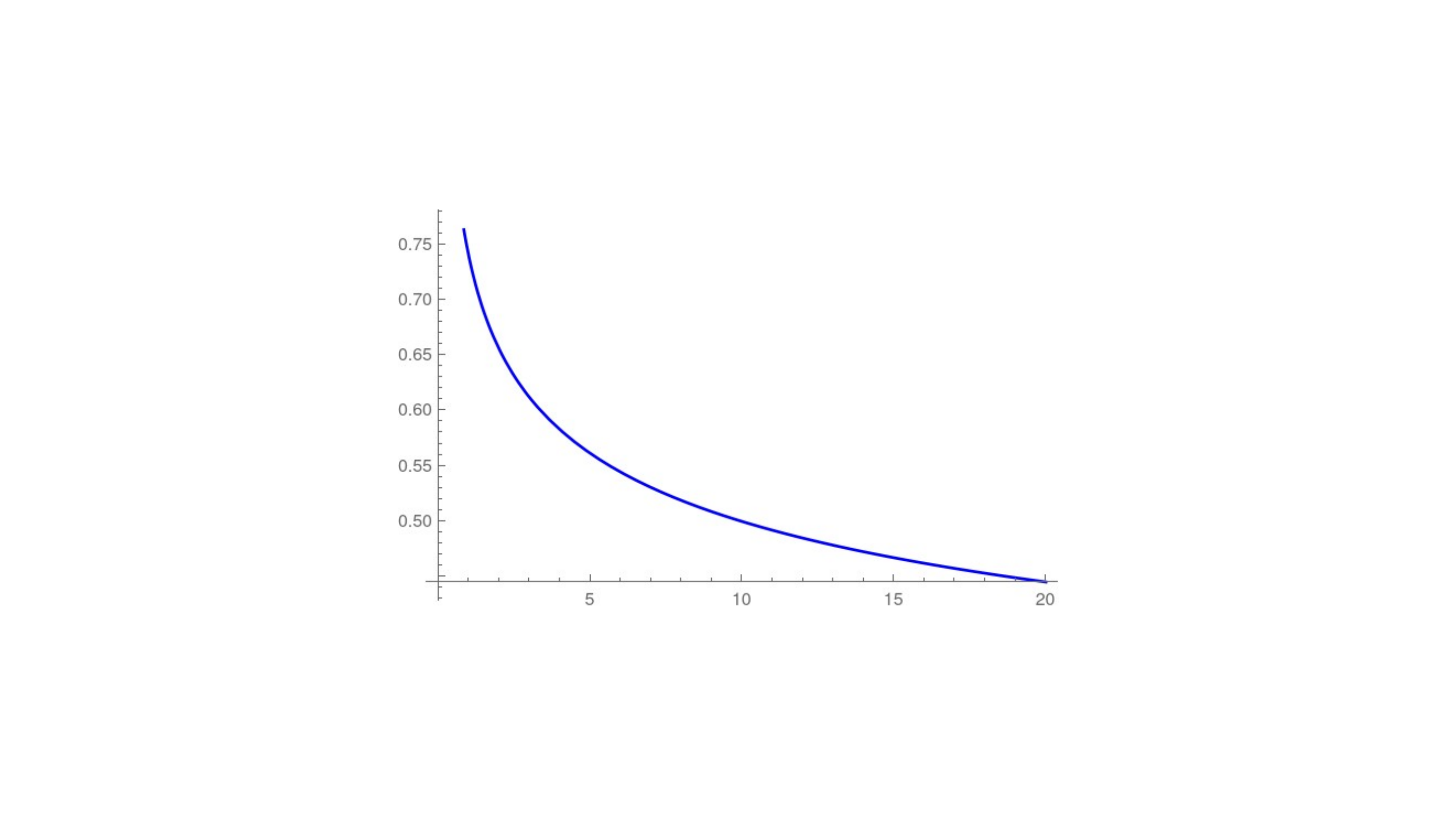}	
\caption{2D view \quad;\quad (left) S as a function of l. for $\rho_c = 1$   \,\, (right)S as a function of $\rho_c$. for $ l = 1$ \,\, : \,\, Both the plots are showing that H.E.E maintains its expected behaviour, i.e increasing with l and falling with the increase of $\rho_c$, even in the regime far away from $l >> \rho_c$ and  $\rho_c >> l $ regime of  $(l,\rho_c)$ plane  }
\la{l4by92Ds1}
\end{figure}

\begin{figure}[H]
\begin{center}
\textbf{ For $ d - \theta = {\frac{4}{9}}$, for long range of $l, \rho_c$,  \, \,; \,\, 3D view   }
\end{center}
\vskip2mm
\includegraphics[width=.65\textwidth]{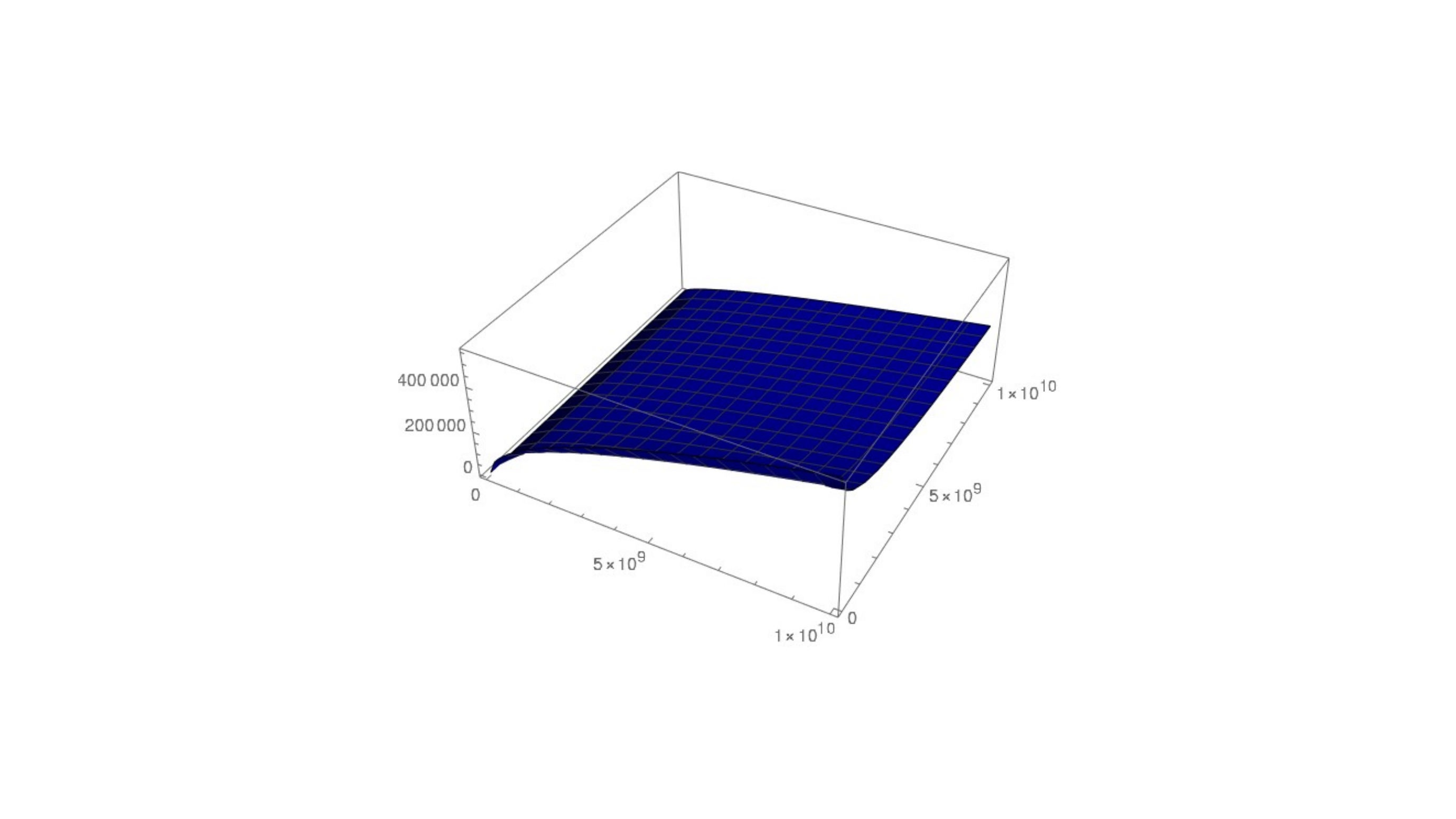}
\includegraphics[width=.65\textwidth]{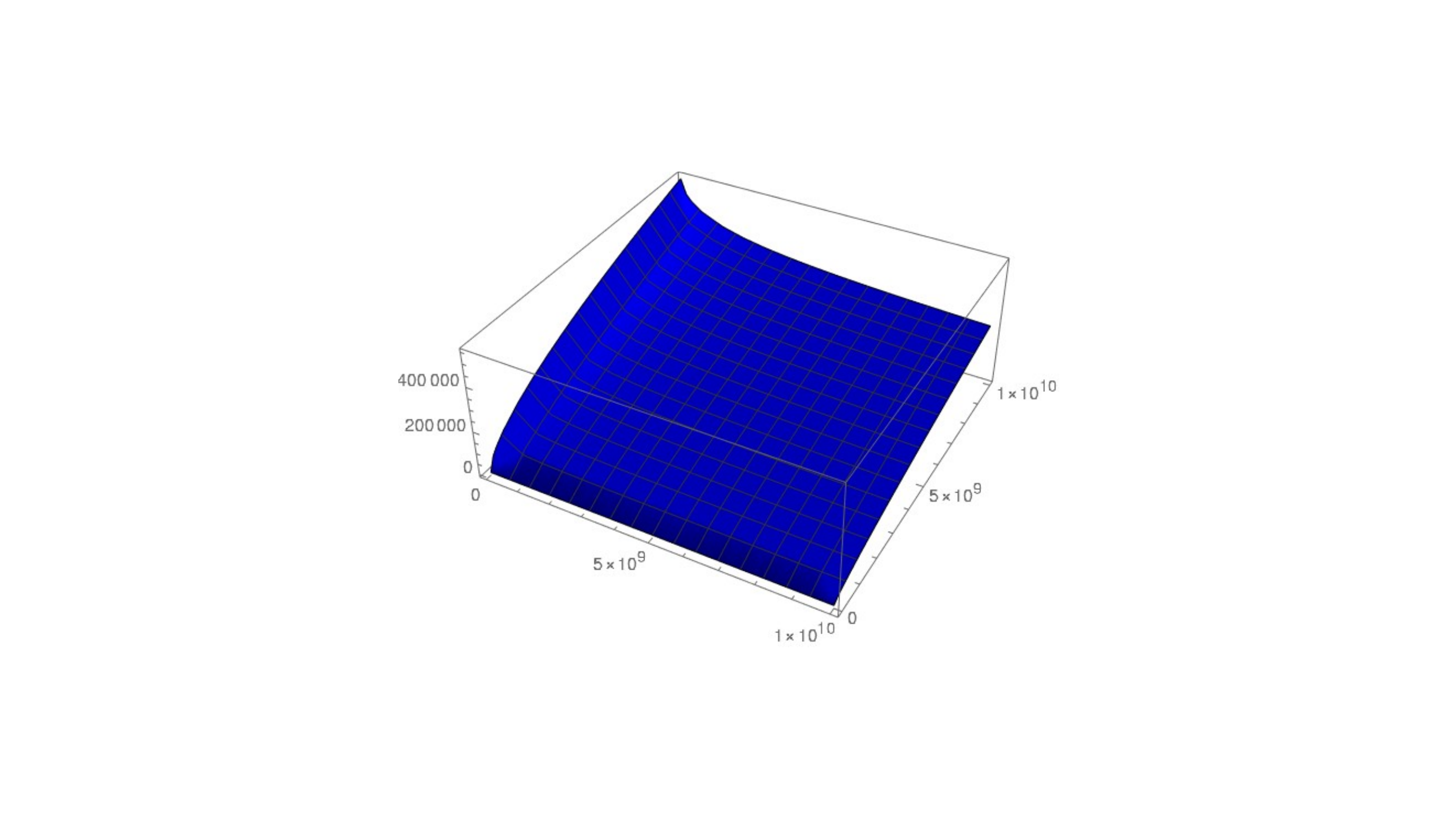}	
\caption{ S as a function of l, $\rho_c$, considered over long range  to probe  $l >> \rho_c$ and $\rho_c >> l$  regime where the expression of H.E.E is exact and in rest of the regime its an interpolating function \quad;\quad (left) x-axis \, : l\, y-axis $\rho_c$ \,\, (right)\, x-axis and y-axis are interchanged \,\,: \,\, Both the plots are showing that H.E.E maintains its expected behaviour throughout $(l,\rho_c)$ plane,  i.e increasing with l and falling with the increase of $\rho_c$   }
\la{l4by93Dl1}
\end{figure}

\begin{figure}[H]
\begin{center}
\textbf{ For $ d - \theta = {\frac{4}{9}}$, for long range of $l, \rho_c$,  \, \,; \,\,  2D view  }
\end{center}
\vskip2mm
\includegraphics[width=.65\textwidth]{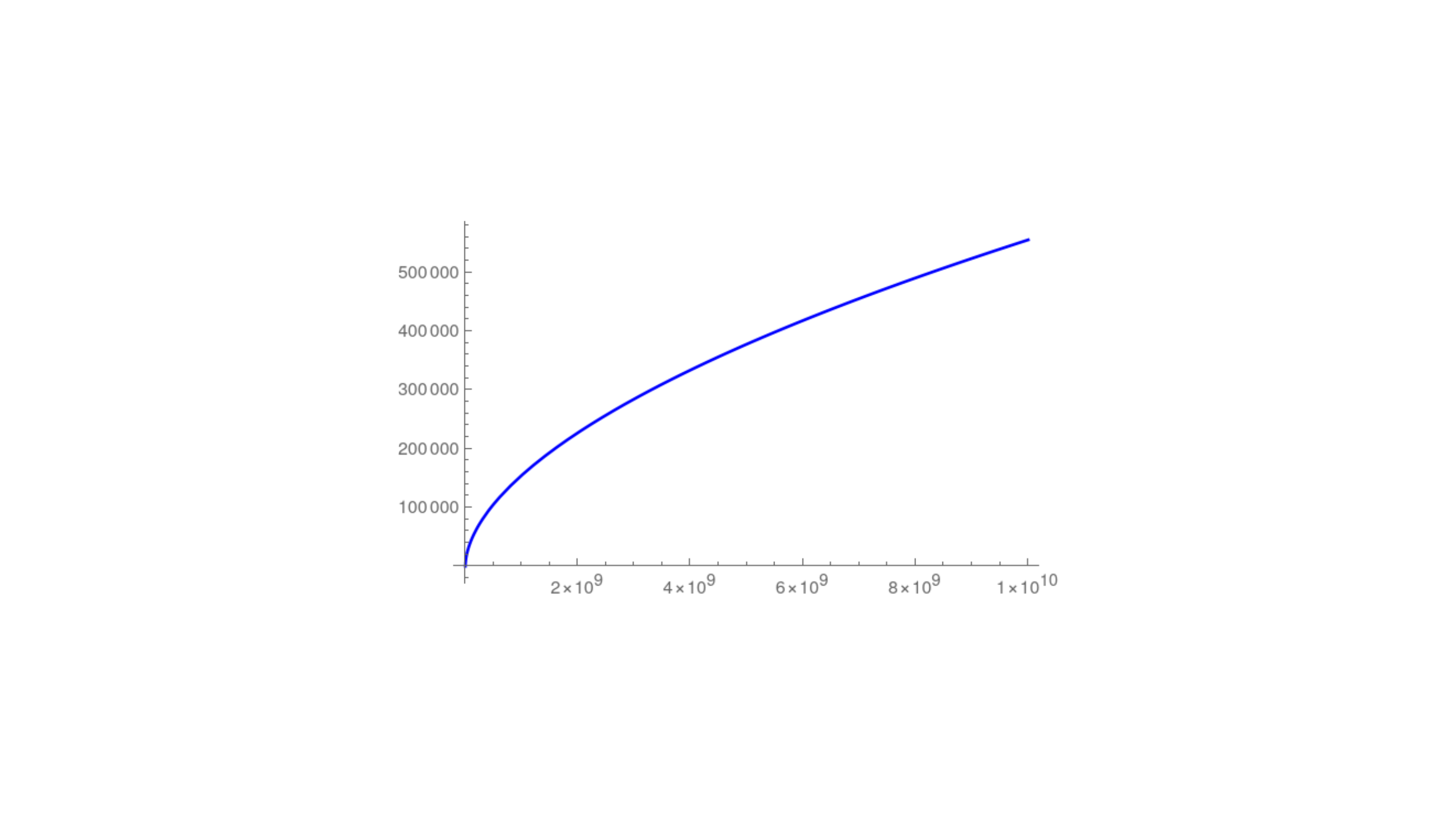}
\includegraphics[width=.65\textwidth]{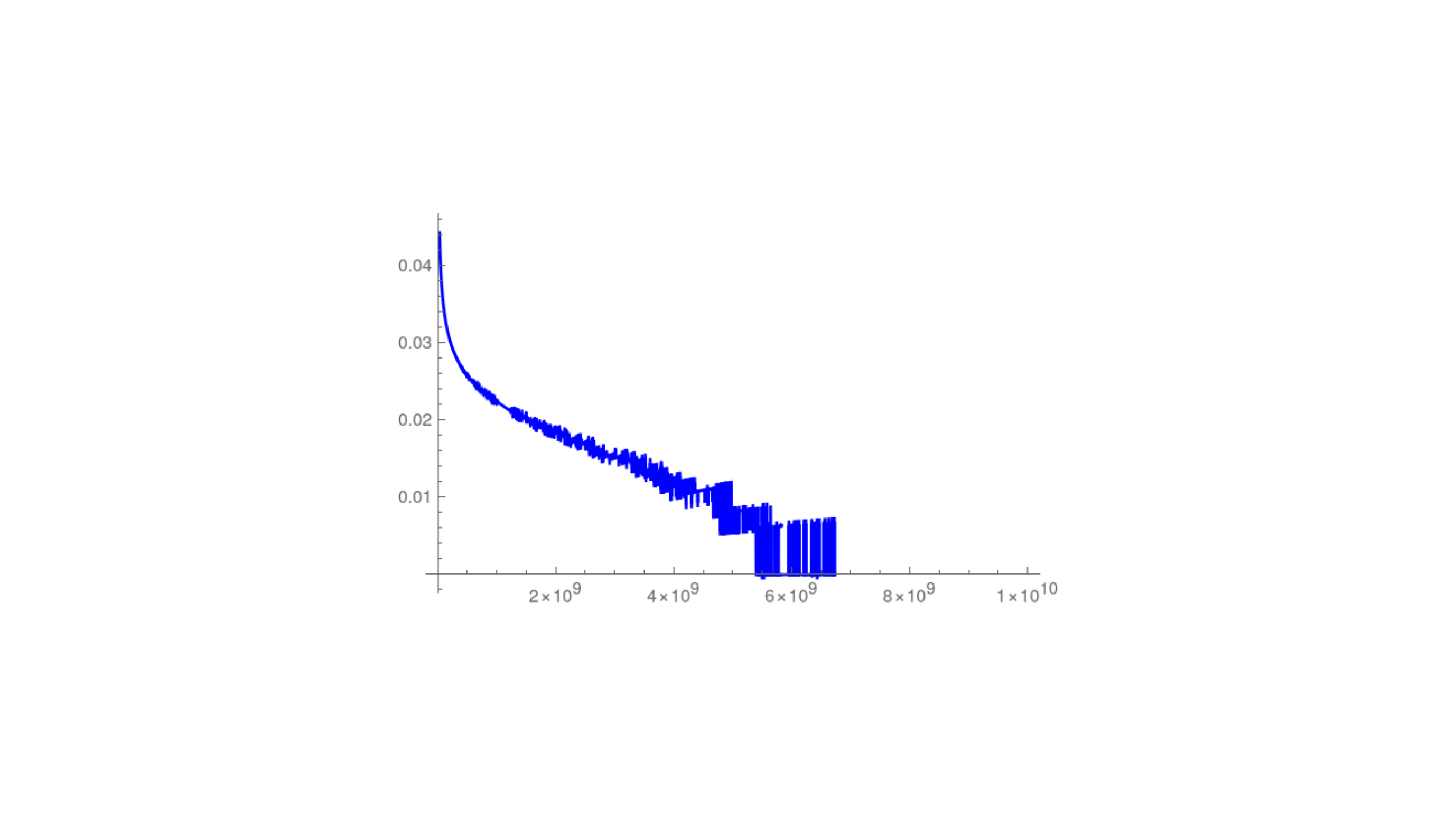}	
\caption{2D view \quad;\quad (left) S as a function of l. for $\rho_c = 1$   \,\, (right)S as a function of $\rho_c$. for $ l = 1$ \,\, : \,\, Both the plots are showing that H.E.E maintains its expected behaviour, i.e increasing with l and falling with the increase of $\rho_c$   }
\la{l4by92Dl1}
\end{figure}

\begin{figure}[H]
\begin{center}
\textbf{ For $ d - \theta = {\frac{2}{9}}$, for short range of $l, \rho_c$,  \, \,; \,\, 3D view   }
\end{center}
\vskip2mm
\includegraphics[width=.65\textwidth]{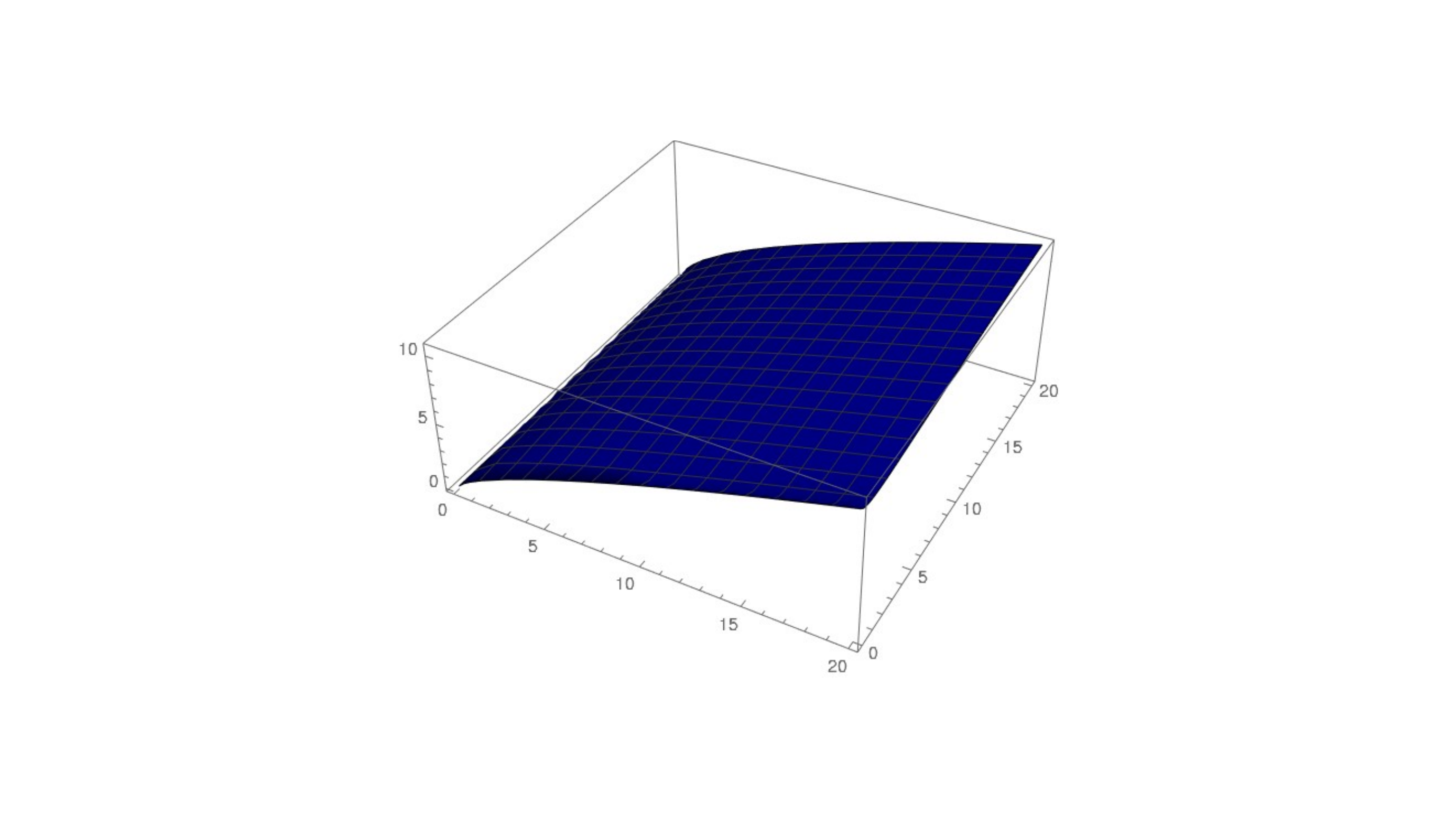}
\includegraphics[width=.65\textwidth]{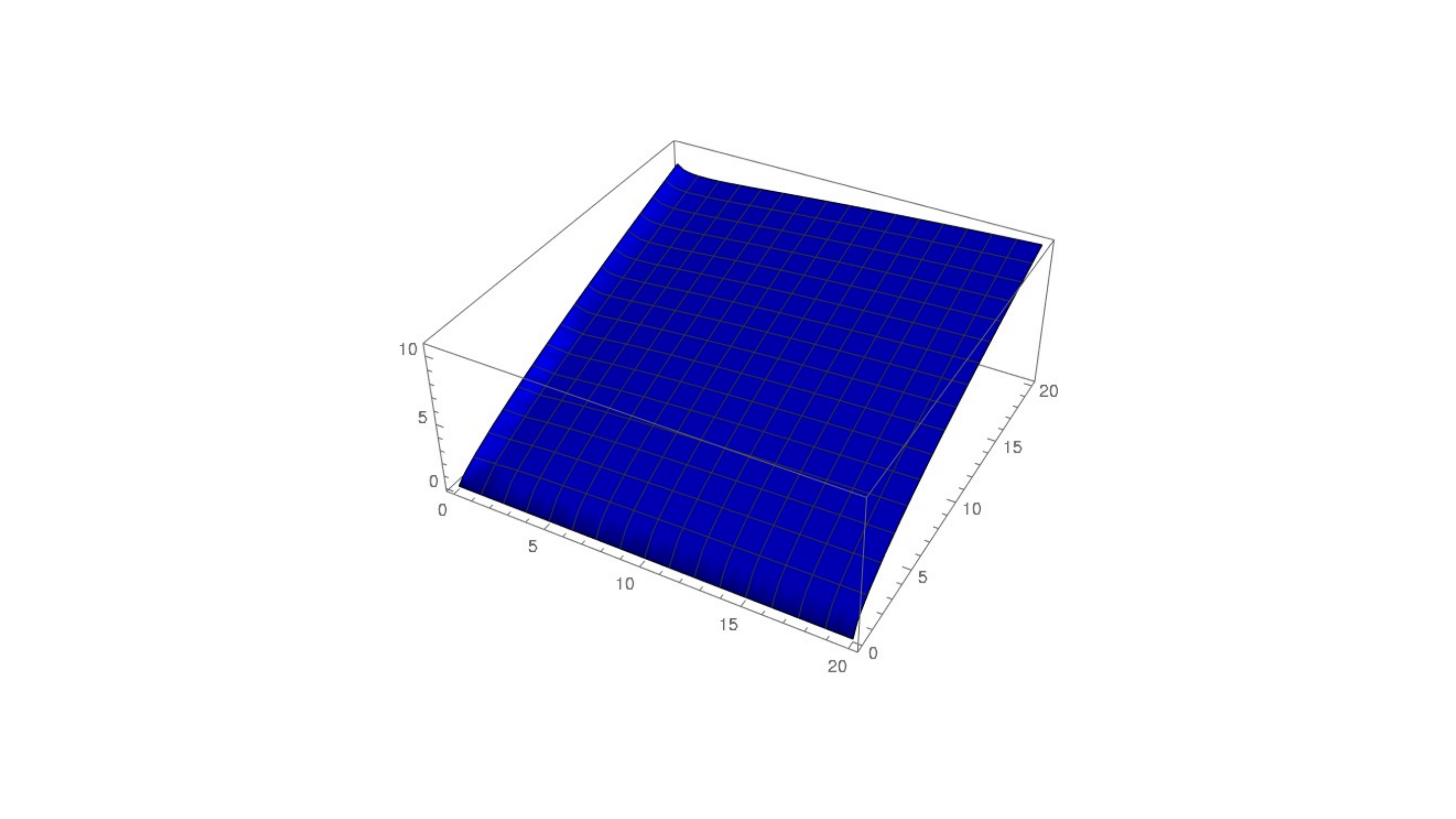}	
\caption{S as a function of l, $\rho_c$, which is exact in $l  >> \rho_c$ and $\rho_c>> l$ and in rest of the regime its an interpolating function\quad;\quad (left) x-axis\, : l\, y-axis $\rho_c$\,\, (right) x-axis and y-axis are interchanged\,\, : \,\,  The plots are showing that even we are far away from  $l >> \rho_c $ and $\rho_c >> l $ regime still the expression of H.E.E is showing its expected behaviour, i.e increasing with l and falling with the increase of cut-off $\rho_c$ establishing the fact that in the interpolating regime the expression of S is close to the exact one    }
\la{l2by93Ds1 }
\end{figure}

\begin{figure}[H]
\begin{center}
\textbf{ For $ d - \theta = {\frac{2}{9}}$, for short range of $l, \rho_c$,  \, \,; \,\,  2D view  }
\end{center}
\vskip2mm
\includegraphics[width=.65\textwidth]{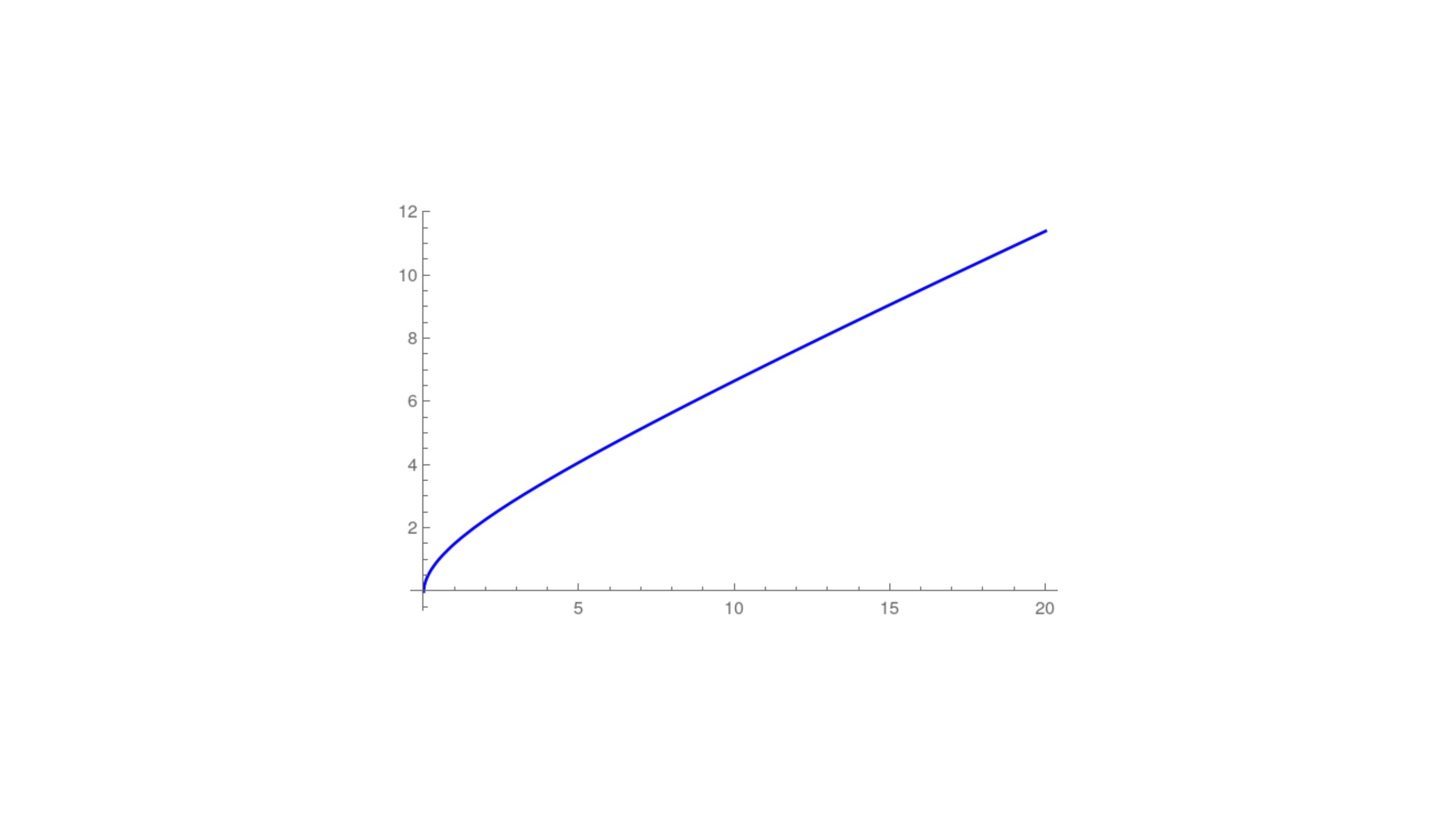}
\includegraphics[width=.65\textwidth]{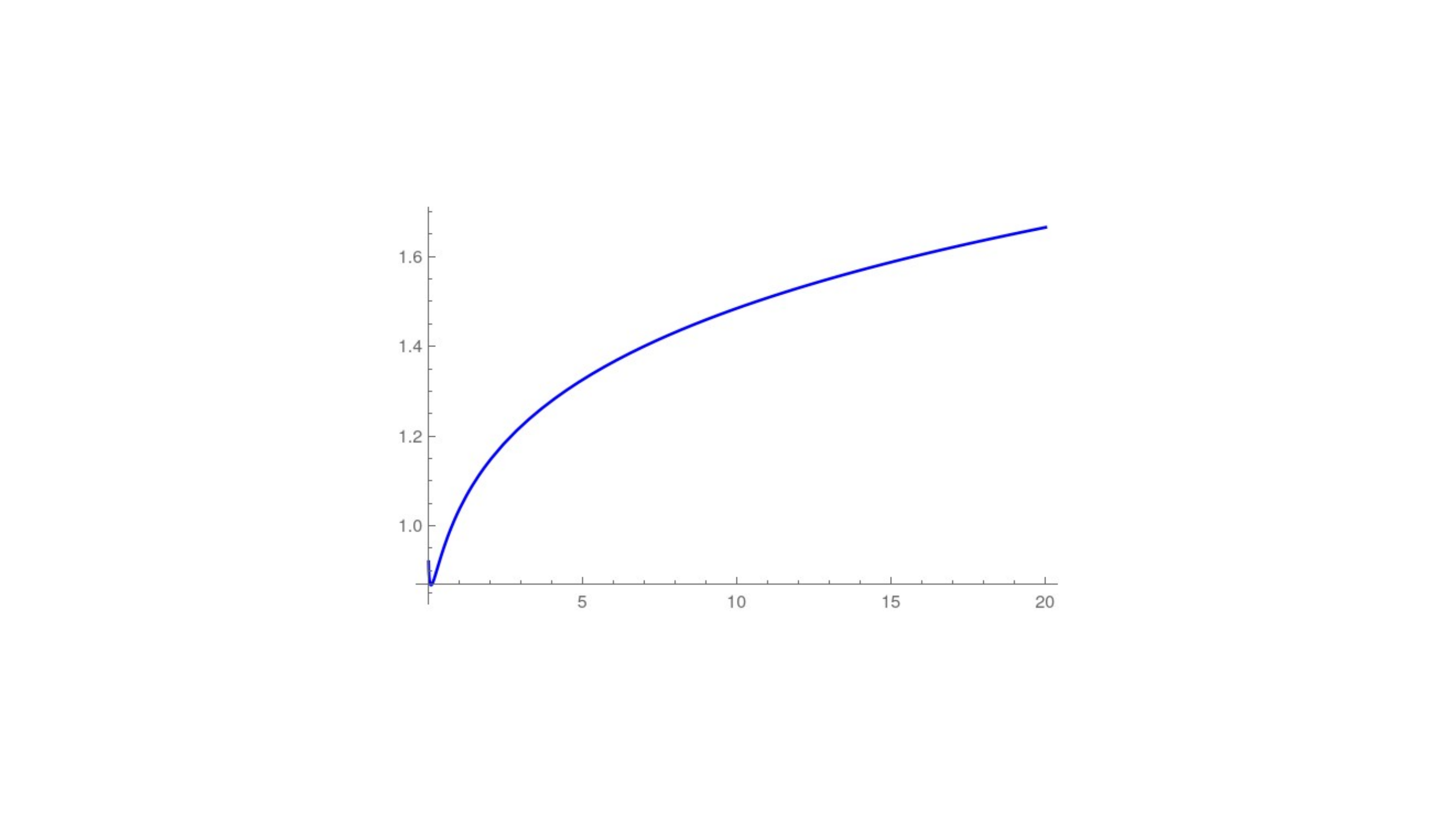}	
\caption{2D view \quad;\quad (left) S as a function of l. for $\rho_c = 1$   \,\, (right) S as a function of $\rho_c$. for $ l = 1$\,\,:\,\,   \,\,  \,\, : \,\, Both the plots are showing that H.E.E maintains its expected behaviour, i.e increasing with l and falling with the increase of $\rho_c$, even in the regime far away from $l >> \rho_c$ and  $\rho_c >> l $ regime of  $(l,\rho_c)$ plane  }
\la{l2by92Ds1}
\end{figure}

\begin{figure}[H]
\begin{center}
\textbf{ For $ d - \theta = {\frac{2}{9}}$, for long range of $l, \rho_c$,  \, \,; \,\, 3D view   }
\end{center}
\vskip2mm
\includegraphics[width=.65\textwidth]{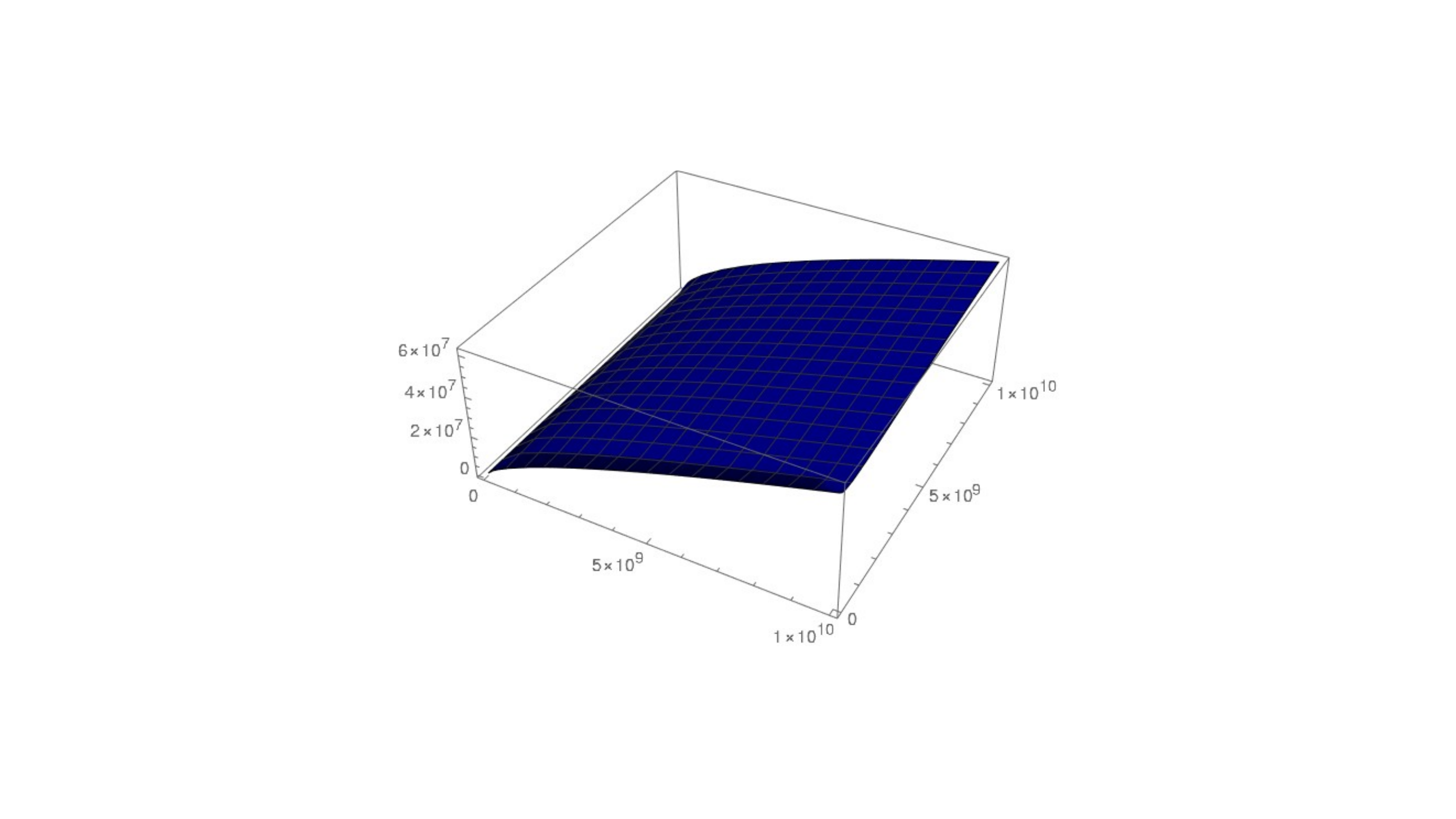}
\includegraphics[width=.65\textwidth]{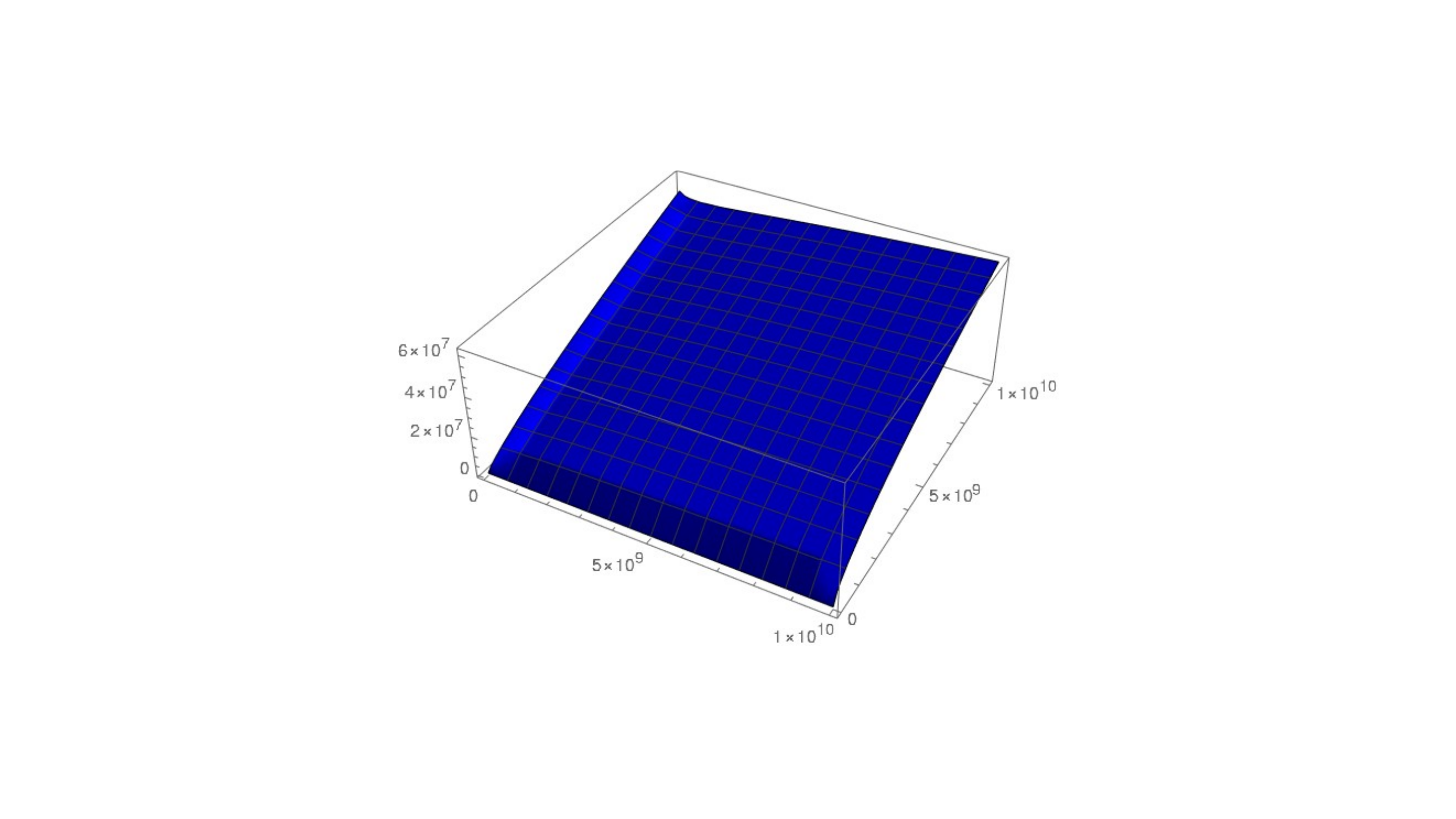}	
\caption{ S as a function of l, $\rho_c$, considered over long range  to probe  $l >> \rho_c$ and $\rho_c >> l$  regime where the expression of H.E.E is exact and in rest of the regime its an interpolating function \quad;\quad (left) x-axis \, : l\, y-axis $\rho_c$ \,\, (right)\, x-axis and y-axis are interchanged \,\,: \,\, Both the plots are showing that H.E.E maintains its expected behaviour throughout $(l,\rho_c)$ plane,  i.e increasing with l and falling with the increase of $\rho_c$  }
\la{l2by93Dl1 }
\end{figure}

\begin{figure}[H]
\begin{center}
\textbf{ For $ d - \theta = {\frac{2}{9}}$, for long range of $l, \rho_c$,  \, \,; \,\,  2D view  }
\end{center}
\vskip2mm
\includegraphics[width=.65\textwidth]{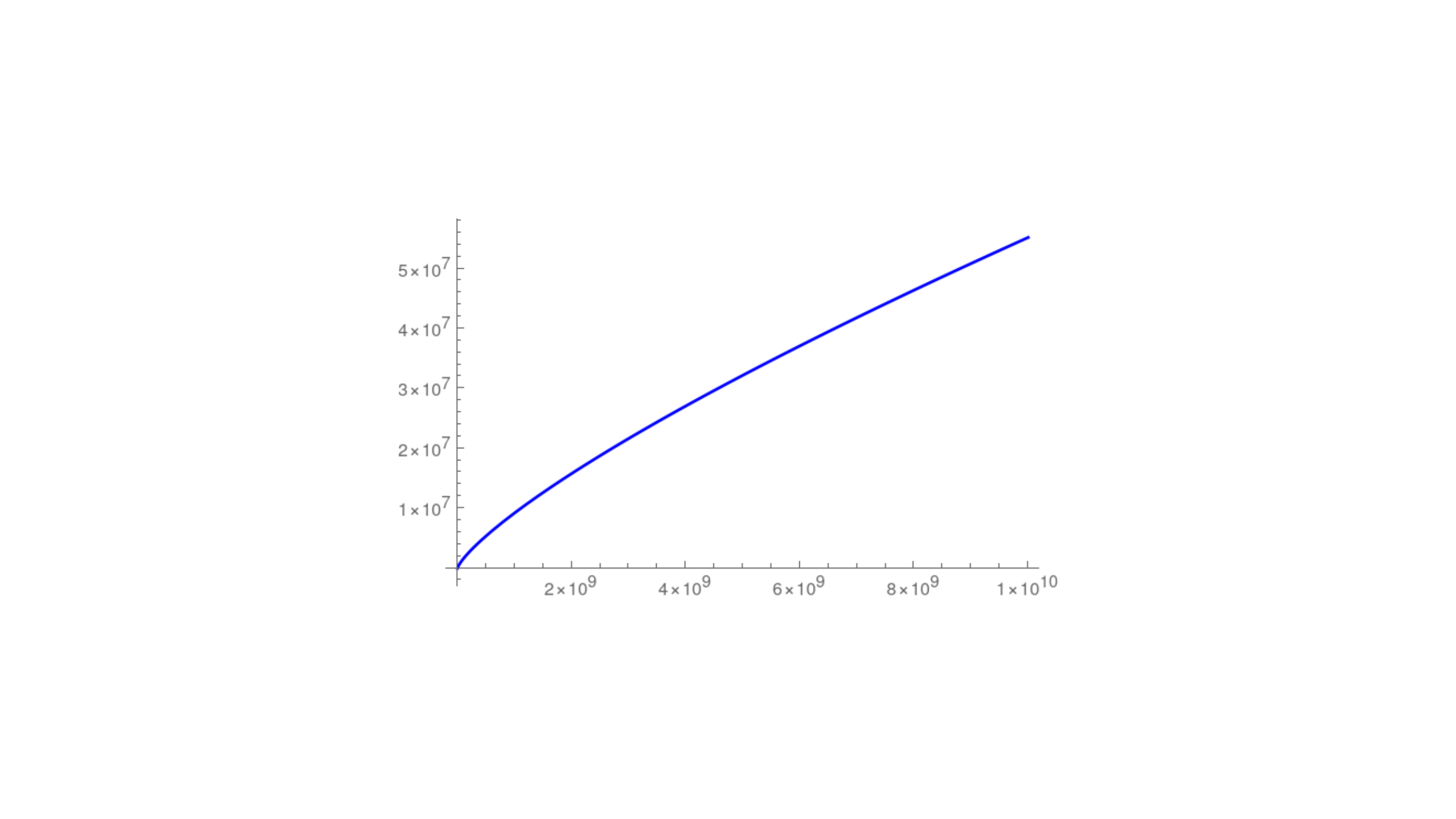}
\includegraphics[width=.65\textwidth]{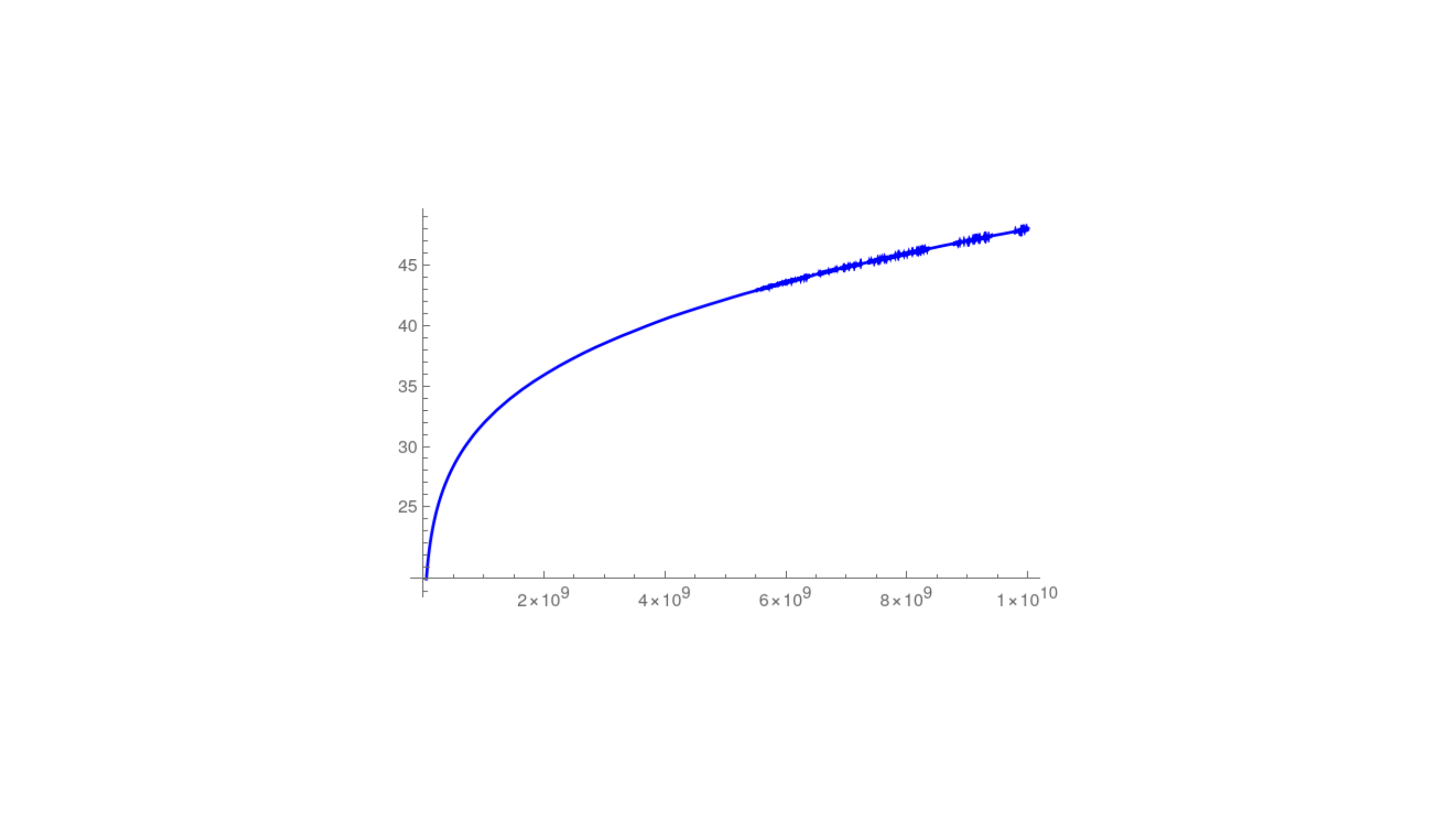}	
\caption{2D view \quad;\quad (left) S as a function of l. for $\rho_c = 1$   \,\, (right)S as a function of $\rho_c$. for $ l = 1$ \,\, : \,\, Both the plots are showing that H.E.E maintains its expected behaviour, i.e increasing with l and falling with the increase of $\rho_c$  }
\la{l2by92Dl1}
\end{figure}

\subsection{Evolution of holographic entanglement of entropy for with $d - \theta$ for both $d - \theta > 1$ and $d - \theta < 1$}

Here in this section we will see how the holographic entanglement of entropy evolve with $d - \theta$, when considered over the same range of $(l,\rho_c)$, where here we must restrict ourselves to very long range in order to find $l >> \rho_c$  and $\rho_c >> l$   regime as a finite regime which can be seen only when we consider it on very long range only!
Here first we consider $d - \theta < 1$, shown the evolution over very long range, $l, \rho_c, \,\,{\rm From } \,0  \, {\rm to}\, {10}^{20}$, we have shown this overlap in Fig.(\ref{exact1by8}). Next we show the evolution of S with $d - \theta$ for the case of $ d - \theta > 1$ given in \, ,\,  Fig.(\ref{exact25 }). 
Finally we show the evolution of H.E.E from the pararameter regime $d-\theta < 1$ to $ d- \theta > 1$ through the case $d- \theta = 1$ in Fig.(\ref{evlessgreatequal})
Here we must comment that Fig. (\ref{evlessgreatequal}) emphasizes on the fact that The evolution of H.E.E with $(d -\theta)$, including the case $d - \theta = 1$ is just as consistent as the case excluding it,  as long as the cut off  $\rho_c$ is nonzero.  This is actually in contrast with the case without $T {\overline{T}}$ deformation, which considers the situation $\rho_c = 0$ and in that case, H.E.E at $d - \theta = 1$ shows a logarithmic divergence \cite{chargedbrane} !

\begin{figure}[H]
\begin{center}
\textbf{ For $d-\theta < 1$,  combination of the 3D plots of the Holographic entanglement of entropy vs $(l,\rho_c)$ and their combination showing the evolution of HEE with $d - \theta$ for $d - \theta < 1$, considered over very long range $l, \rho_c, \,\,{\rm From } \,0  \, {\rm to}\, {10}^{20}$}
 
\end{center}
\includegraphics[width=.32\textwidth]{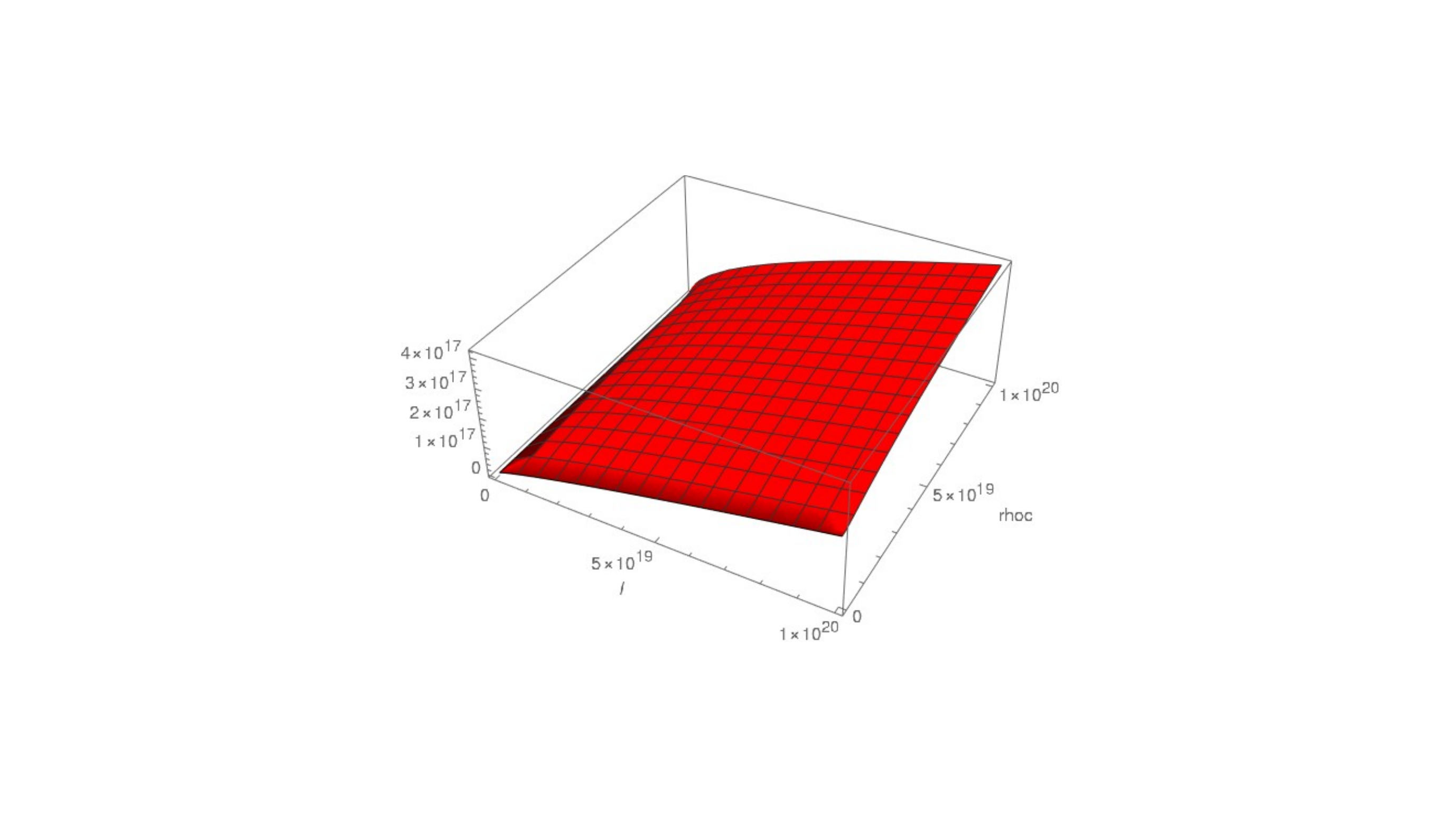}
\includegraphics[width=.32\textwidth]{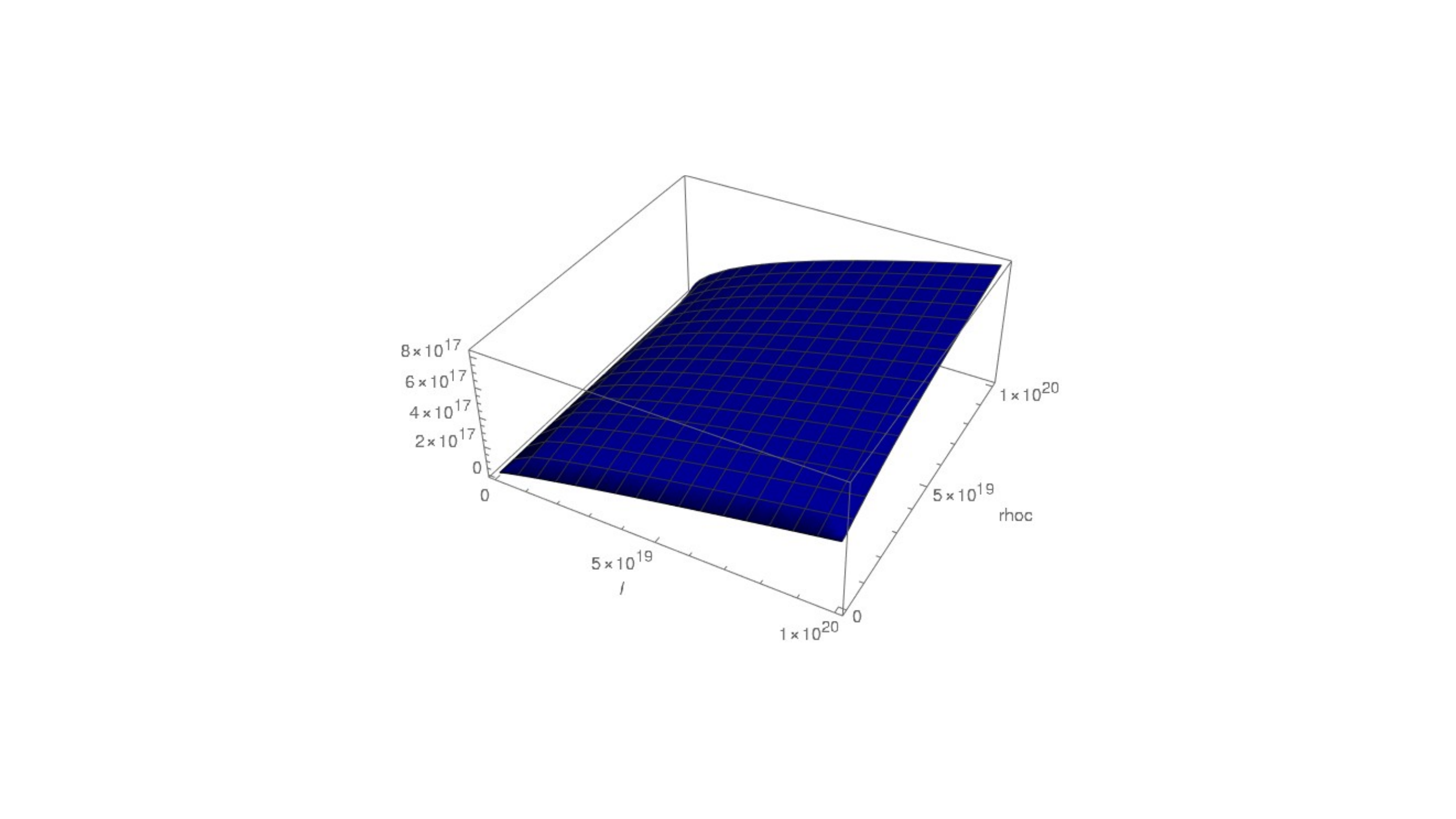}
\includegraphics[width=.32\textwidth]{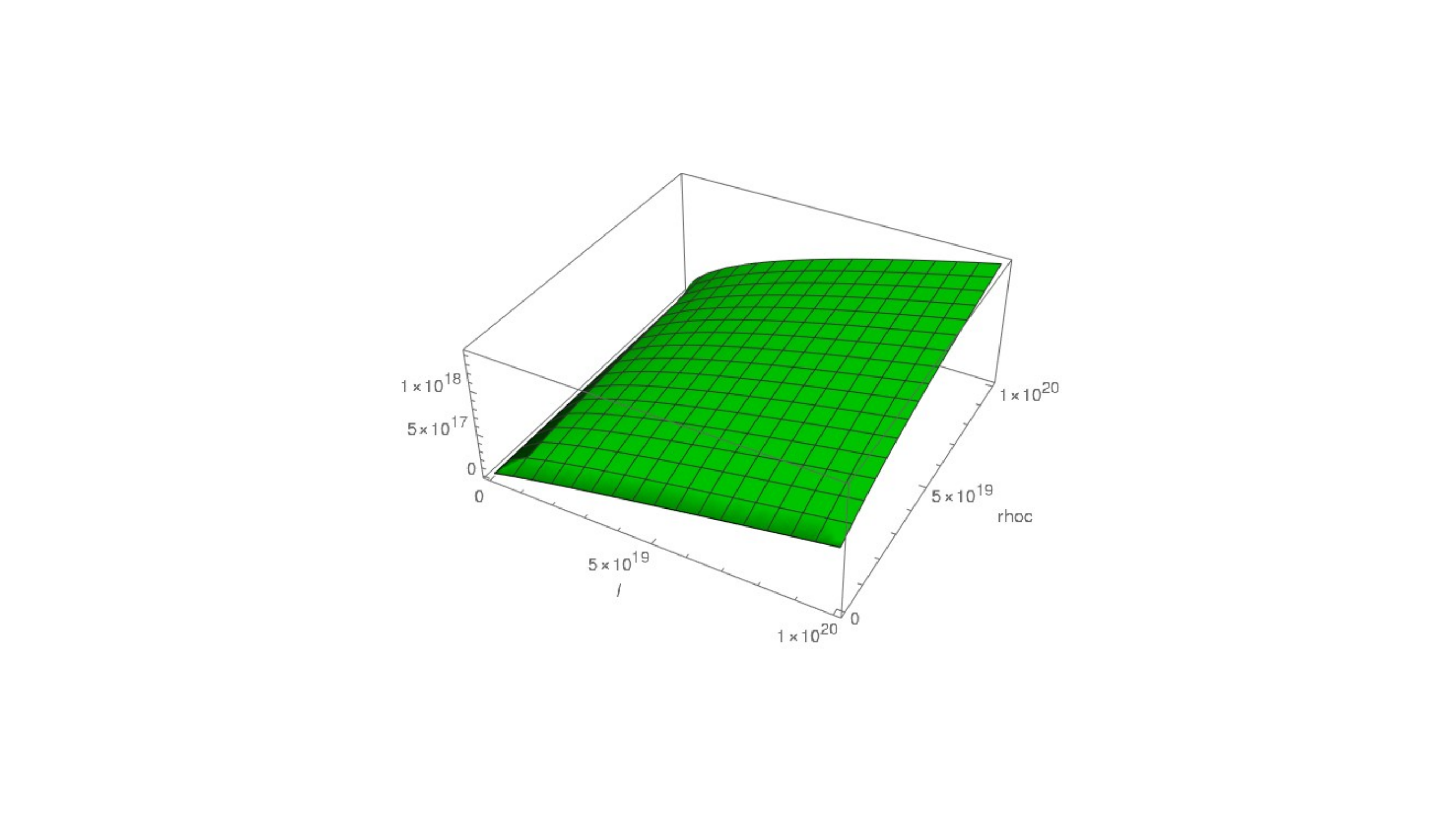}
\includegraphics[width=.32\textwidth]{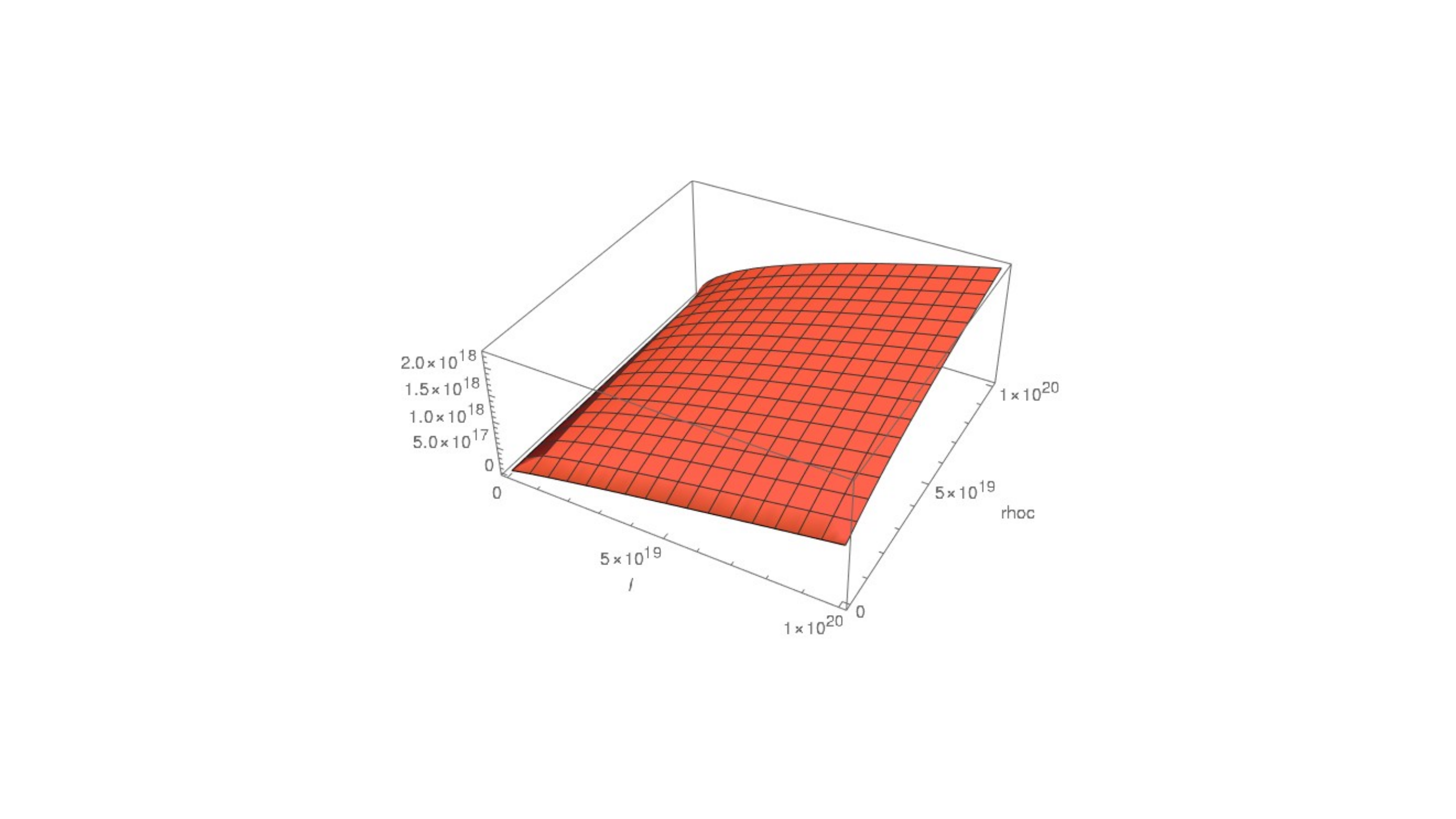}
\includegraphics[width=.32\textwidth]{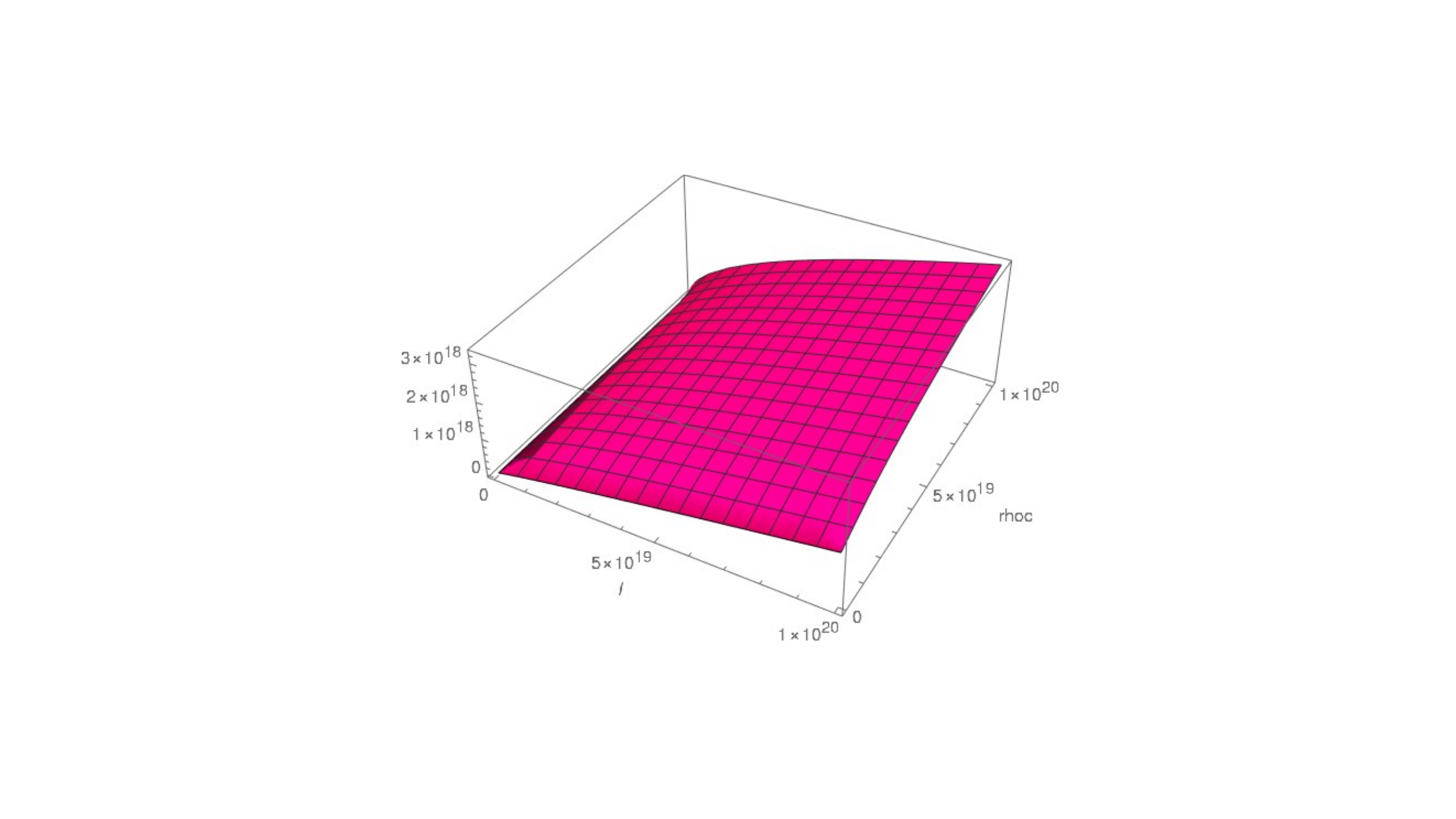}
\includegraphics[width=.32\textwidth]{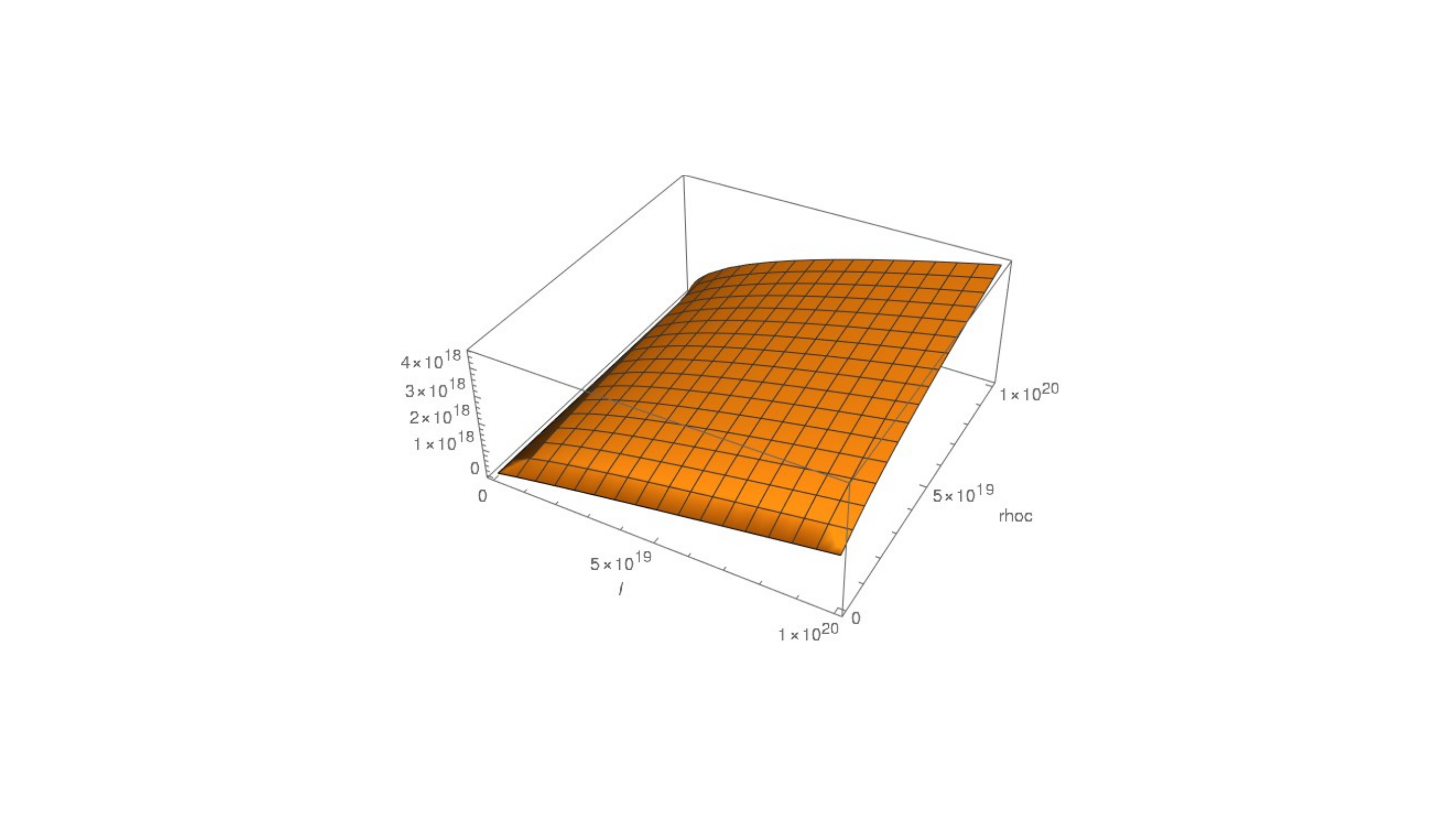}
\begin{center}	
\includegraphics[width=.65\textwidth]{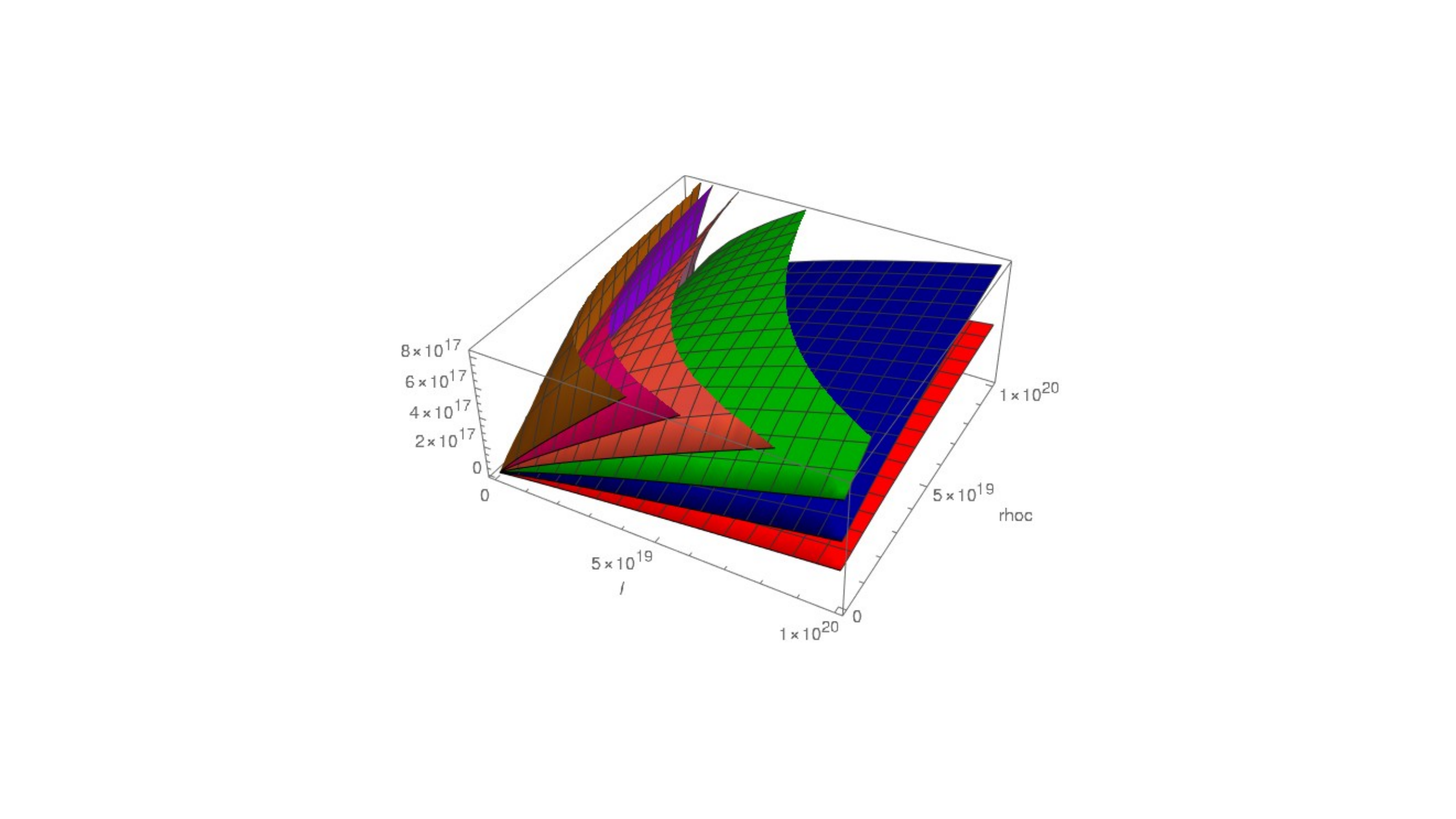}
\end{center}

\caption{First row \,\,:\,\,( From  left to right) \,,\, HEE plotted as a function of $(l,\rho_c)$ for $d - \theta = {\frac{1}{8}}\,,\, d - \theta = {\frac{1}{9}}\,,\,d - \theta = {\frac{1}{10}}$\quad;\quad Middle row  \,\,:\,\,( From  left to right) \,,\, HEE plotted as a function of $(l,\rho_c)$ for $d - \theta = {\frac{1}{11}}\,,\, d - \theta = {\frac{1}{12}}\,,\,d - \theta = {\frac{1}{13}}$\quad;\quad Last row   \, ; \, The combination of all, showing the HEE  increases for the decrease of $d - \theta$
 }
\la{exact1by8}
\end{figure}

\begin{figure}[H]
\begin{center}
\textbf{ For $d-\theta > 1$,  combination of the 3D plots of the Holographic entanglement of entropy vs $(l,\rho_c)$ and their combination showing the evolution of HEE with $d - \theta$ for $d - \theta > 1$, considered over very long range $l, \rho_c, \,\,{\rm From } \,0  \, {\rm to}\, {10}^{10} $}
 \end{center}
\includegraphics[width=.32\textwidth]{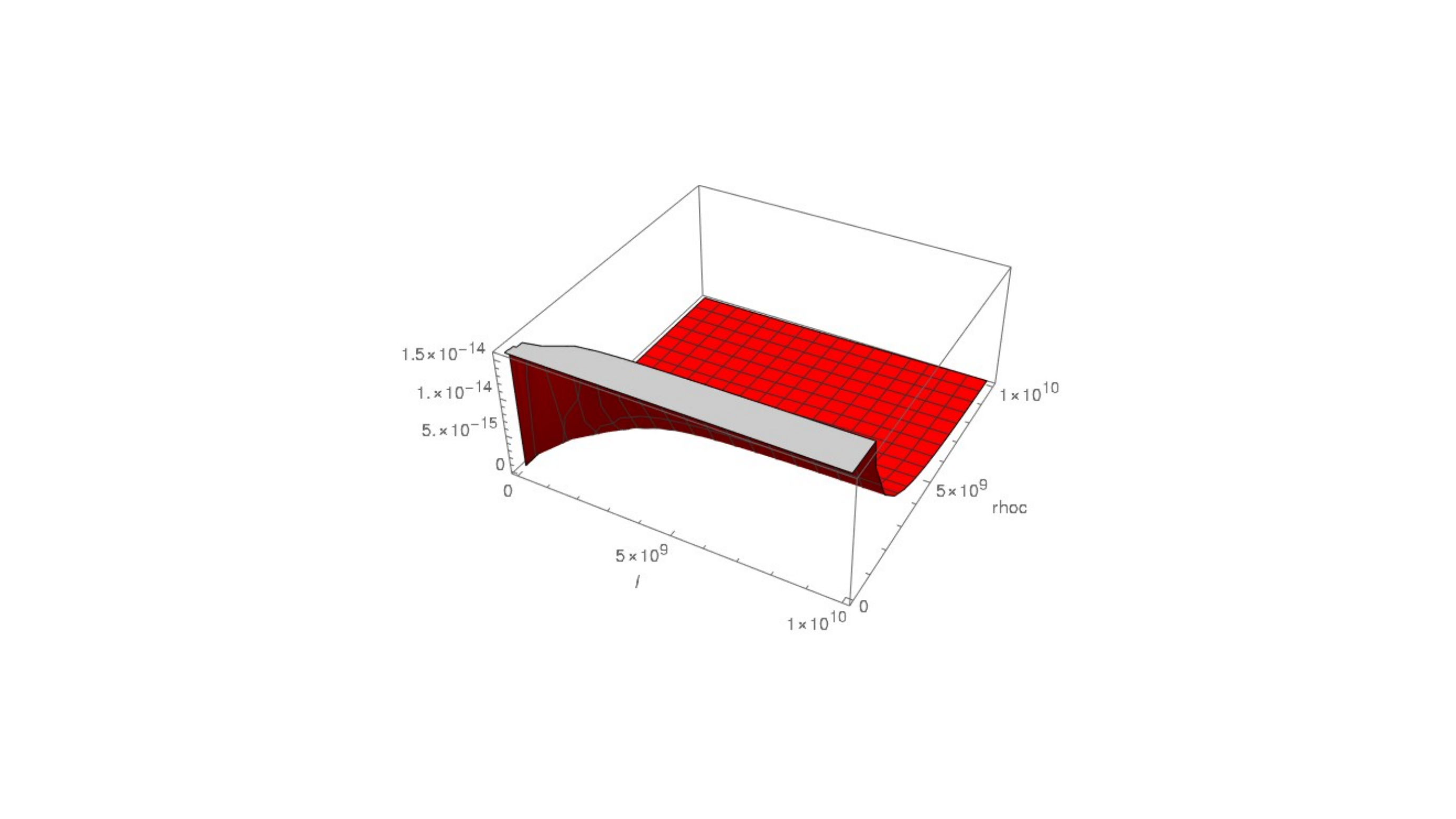}
\includegraphics[width=.32\textwidth]{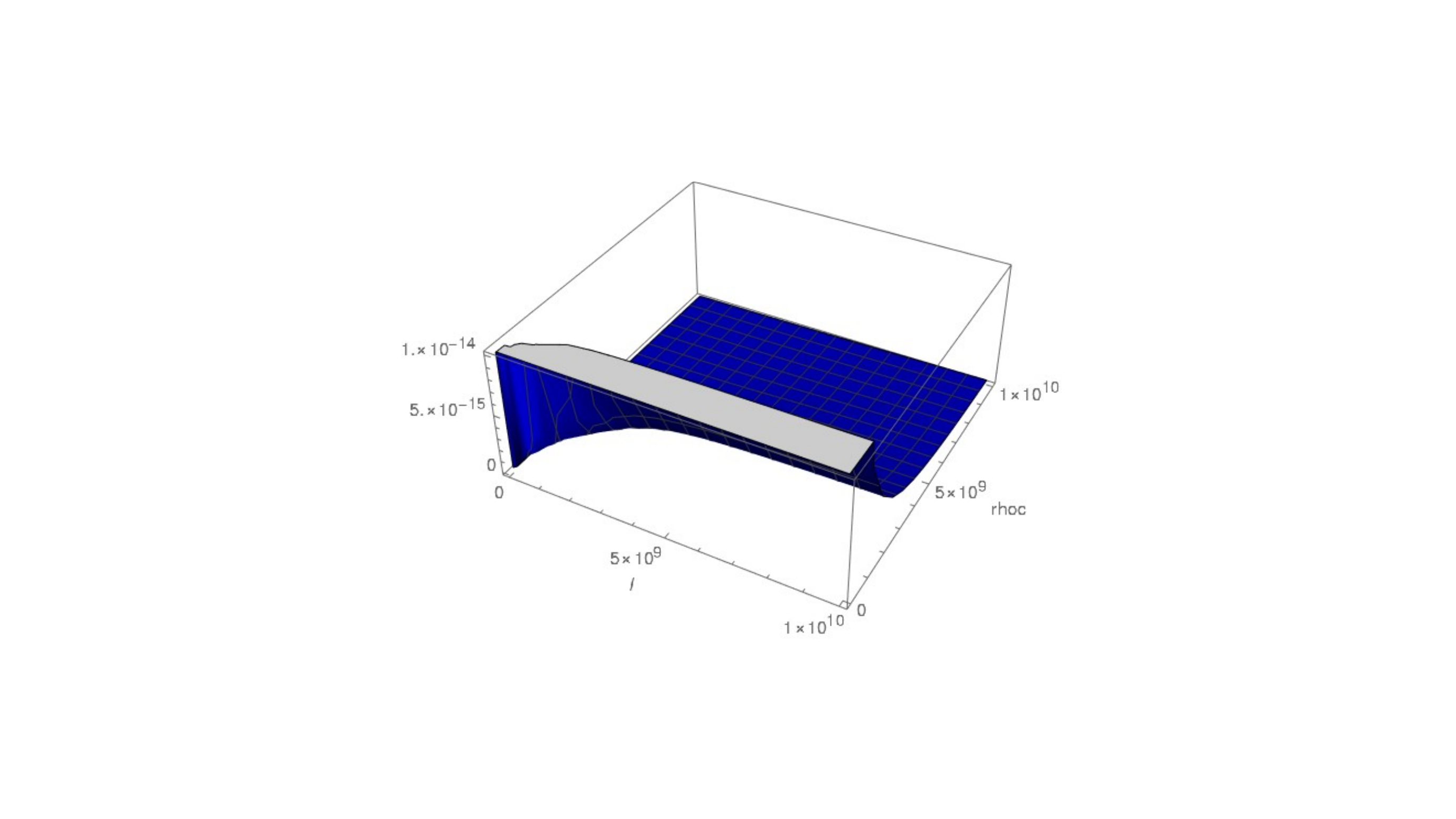}
\includegraphics[width=.32\textwidth]{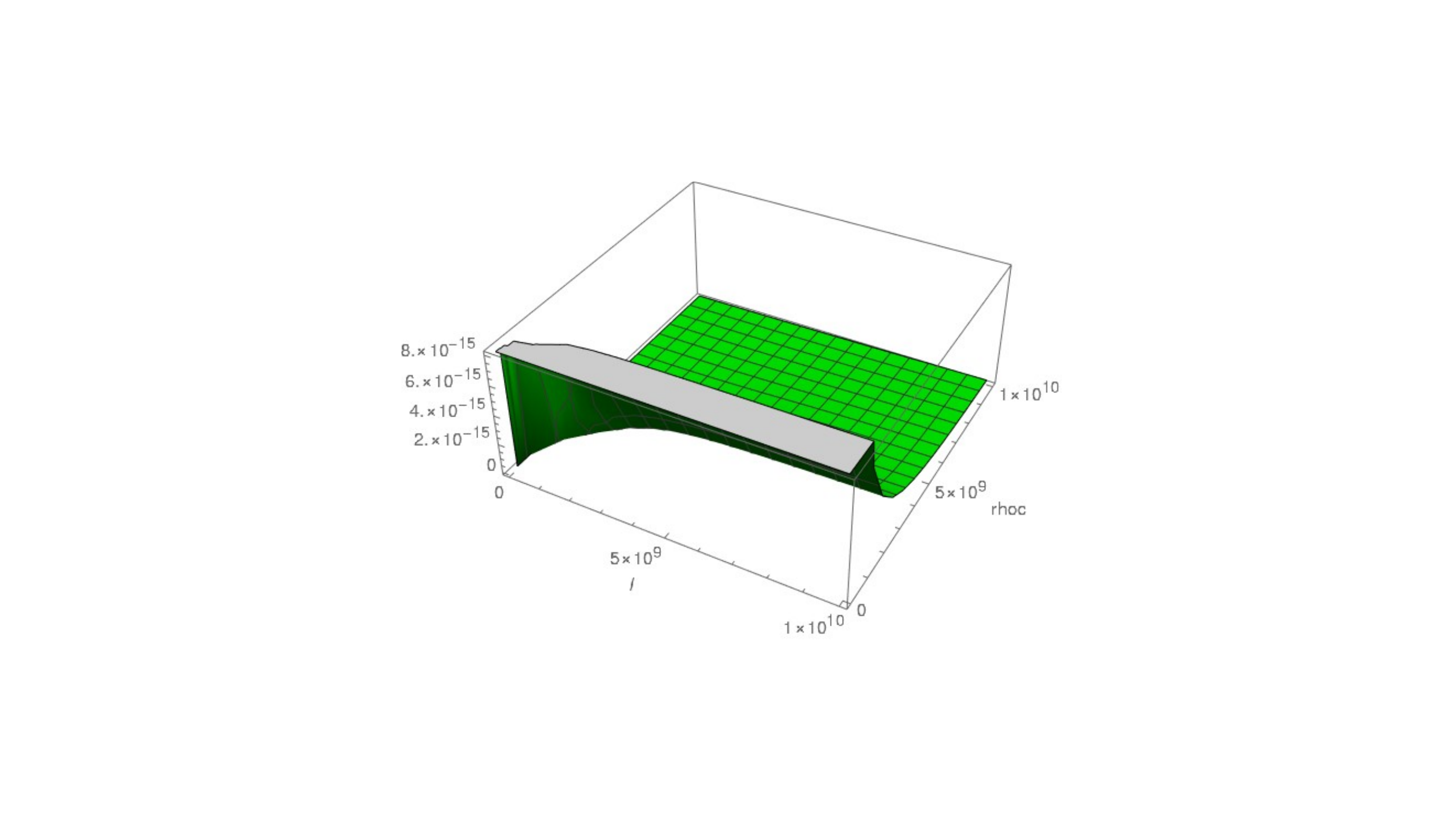}
\includegraphics[width=.32\textwidth]{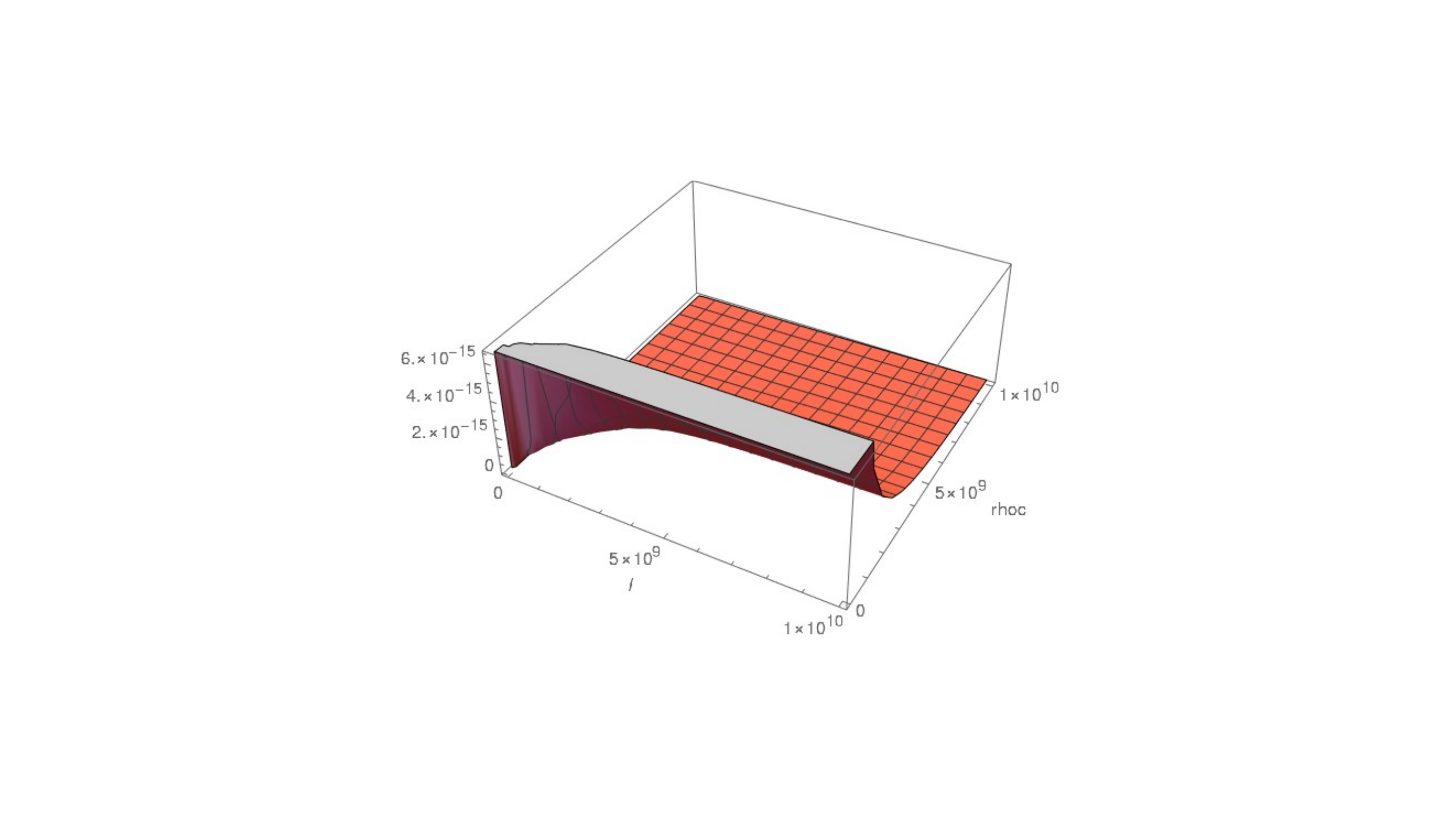}
\includegraphics[width=.32\textwidth]{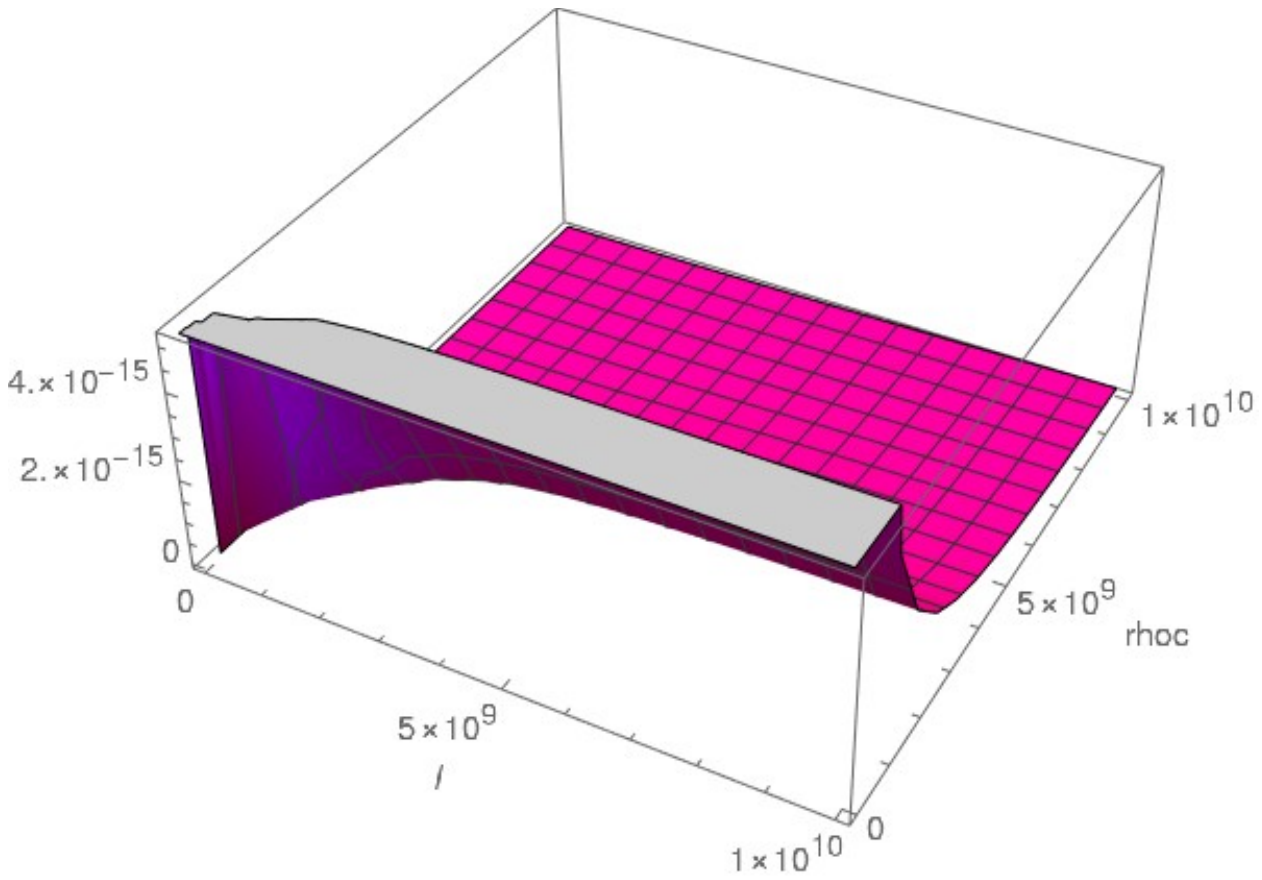}
\includegraphics[width=.32\textwidth]{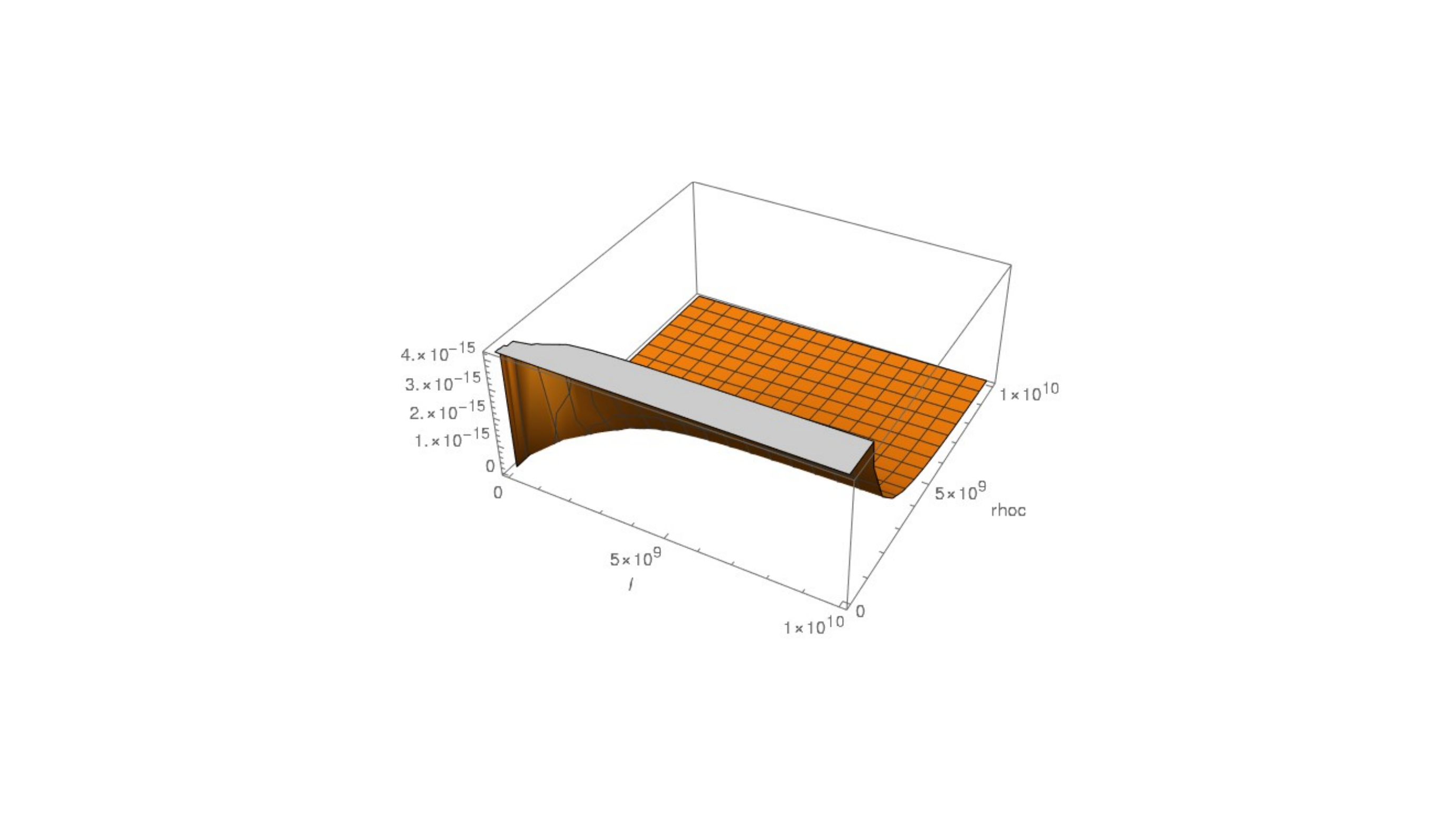}
\begin{center}	
\includegraphics[width=.65\textwidth]{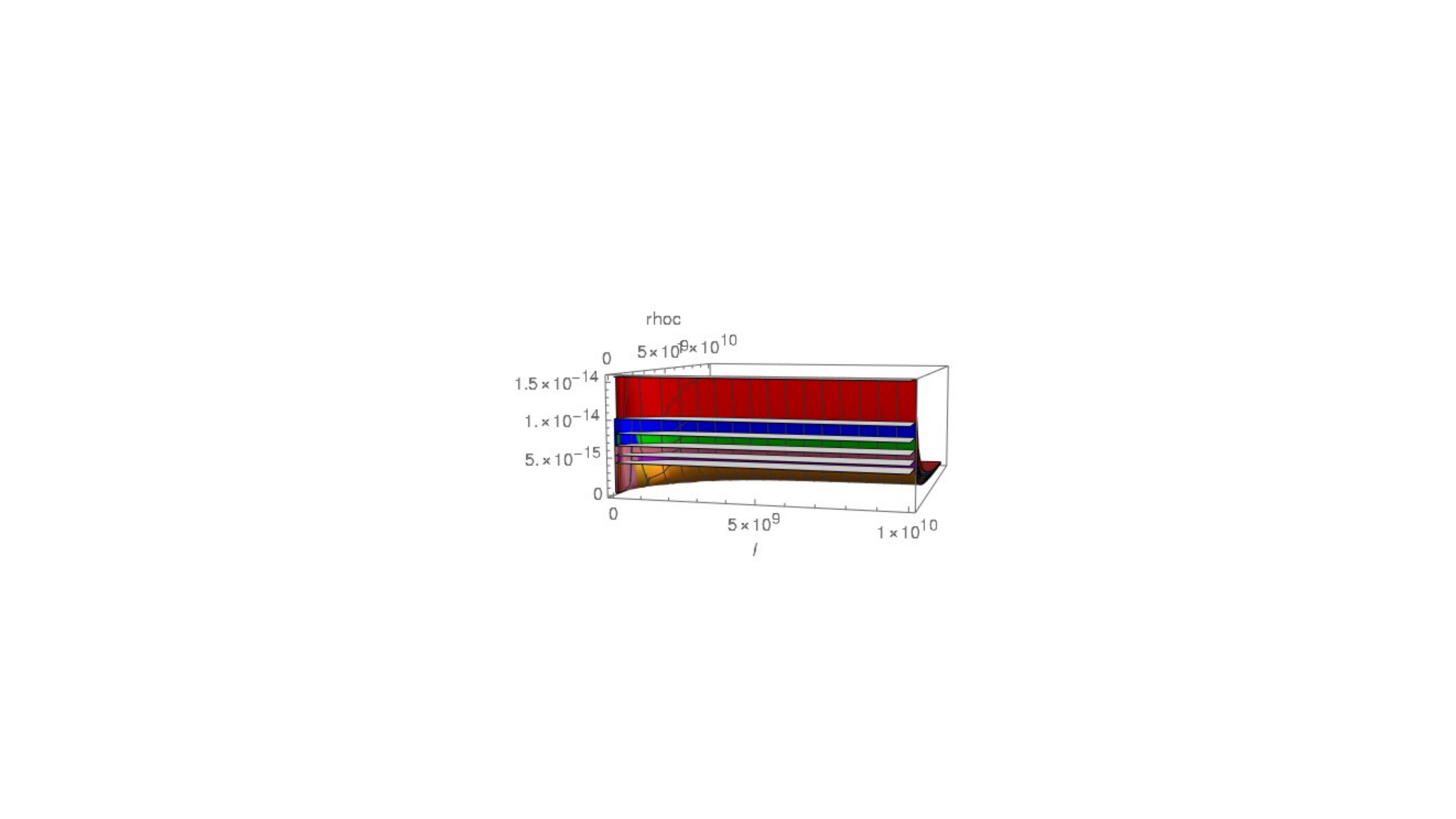}
\end{center}
\caption{First row \,\,:\,\,( From  left to right) \,,\, HEE plotted as a function of $(l,\rho_c)$ for $d - \theta = 2.50\,,\, d - \theta = 2.52\,,\,d - \theta = 2.53 $\quad;\quad (Second row)  \,\,:\,\,( From  left to right) \,,\, HEE plotted as a function of $(l,\rho_c)$ for $d - \theta = 2.54  \, ,\,
 d - \theta = 2.55 \,,\,d - \theta = 2.56 $  \quad;\quad (last row)  \, , \, The combination of all, showing the HEE  increases with  the decrease of $d - \theta$}
\la{exact25 }
\end{figure}

\begin{figure}[H]
\begin{center}
\textbf{ Evolution of H.E.E from $ d - \theta < 1 $ to $d- \theta > 1$ through $ d- \theta = 1$ }
\end{center}
\includegraphics[width=.36\textwidth]{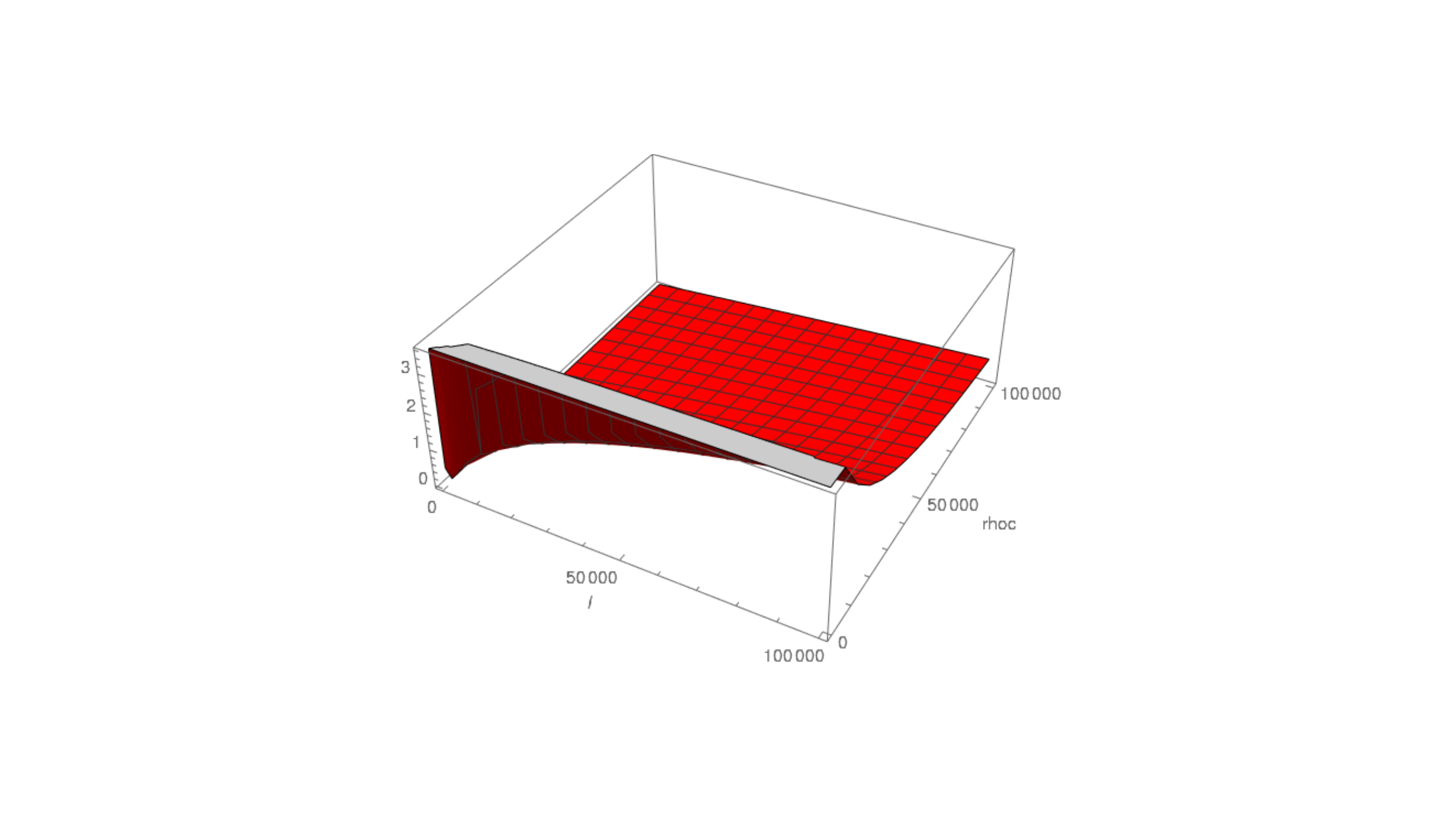}
\includegraphics[width=.36\textwidth]{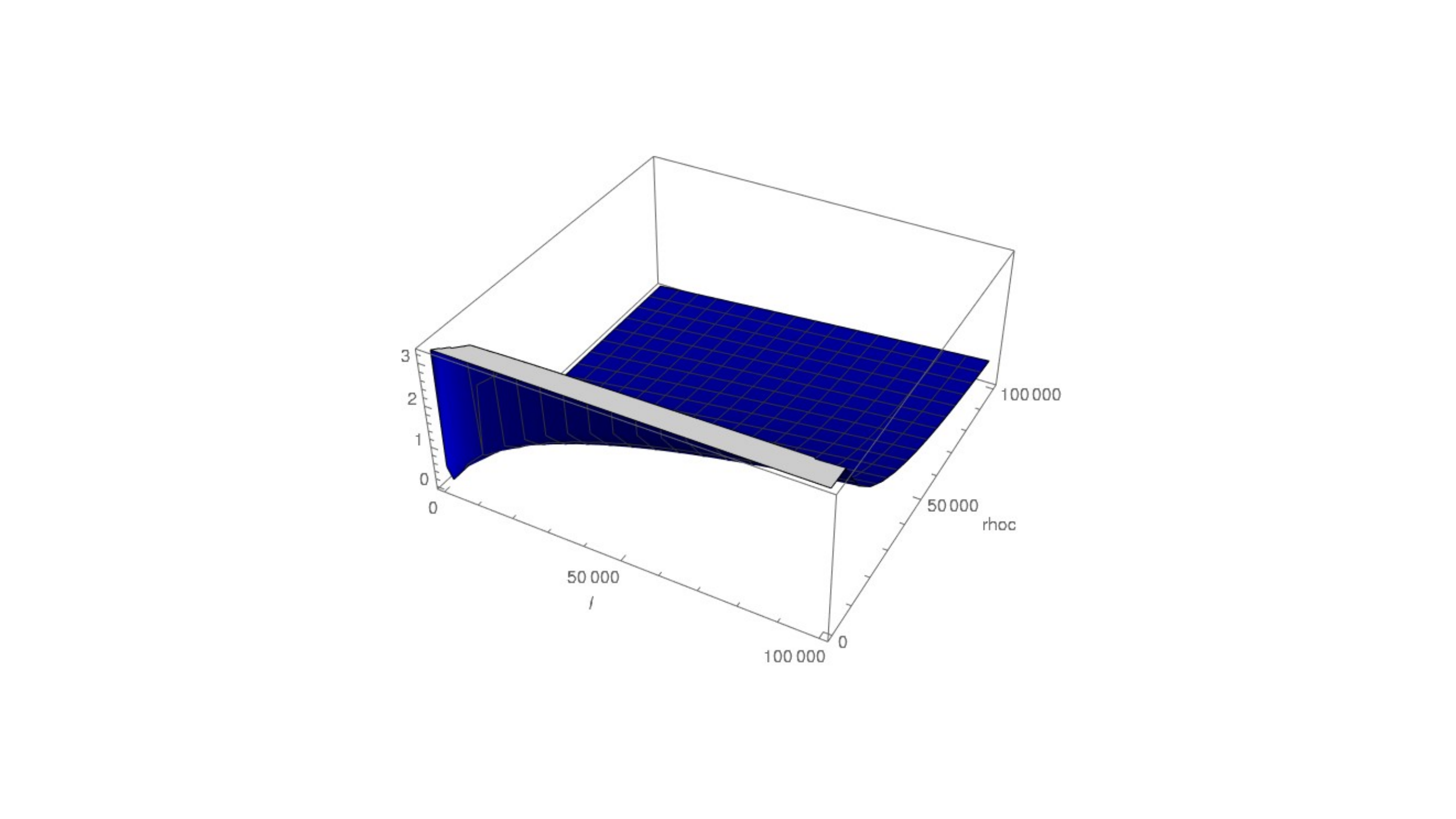}
\includegraphics[width=.36\textwidth]{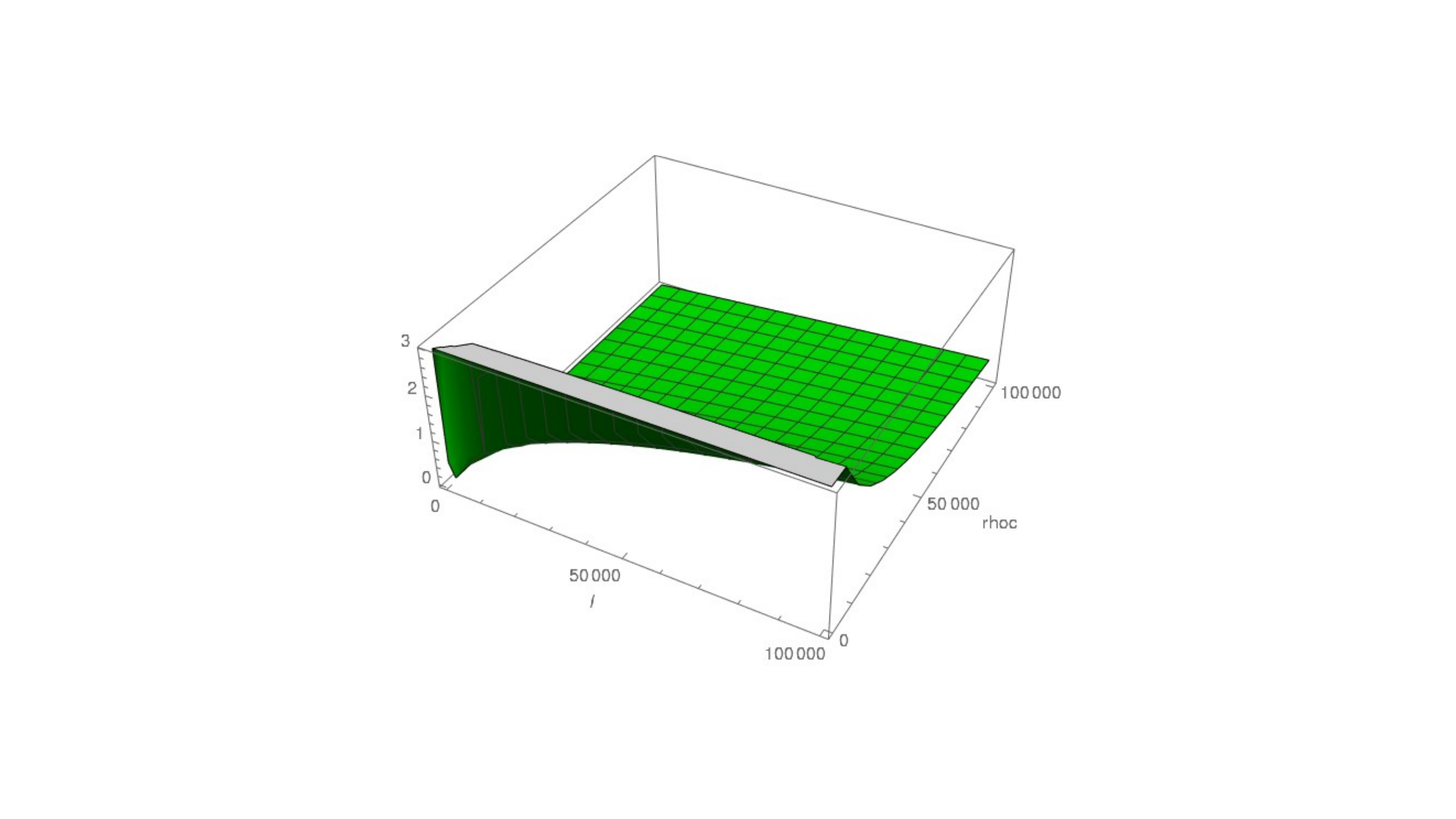}
\includegraphics[width=.36\textwidth]{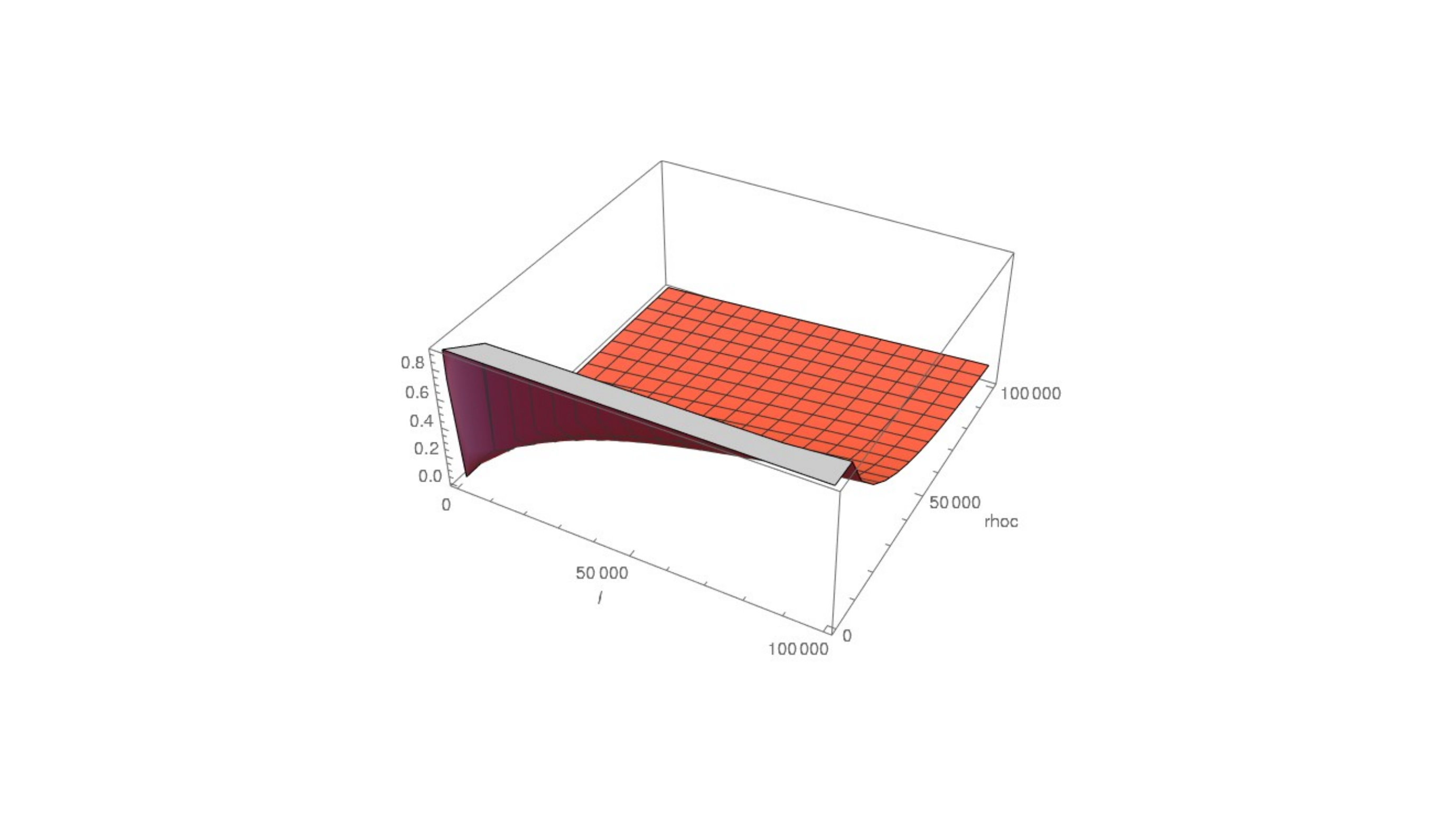}
\includegraphics[width=.36\textwidth]{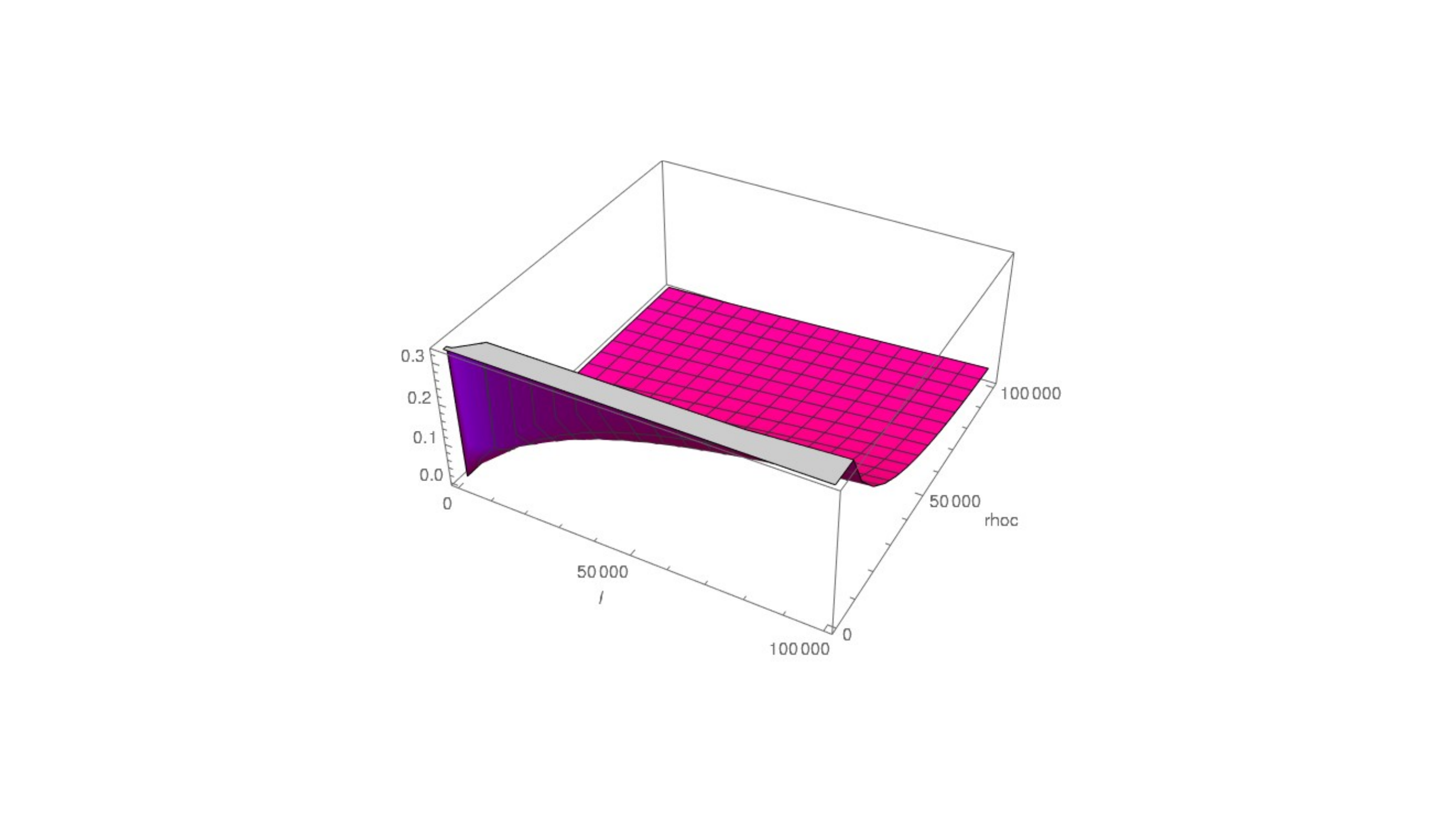}
\includegraphics[width=.36\textwidth]{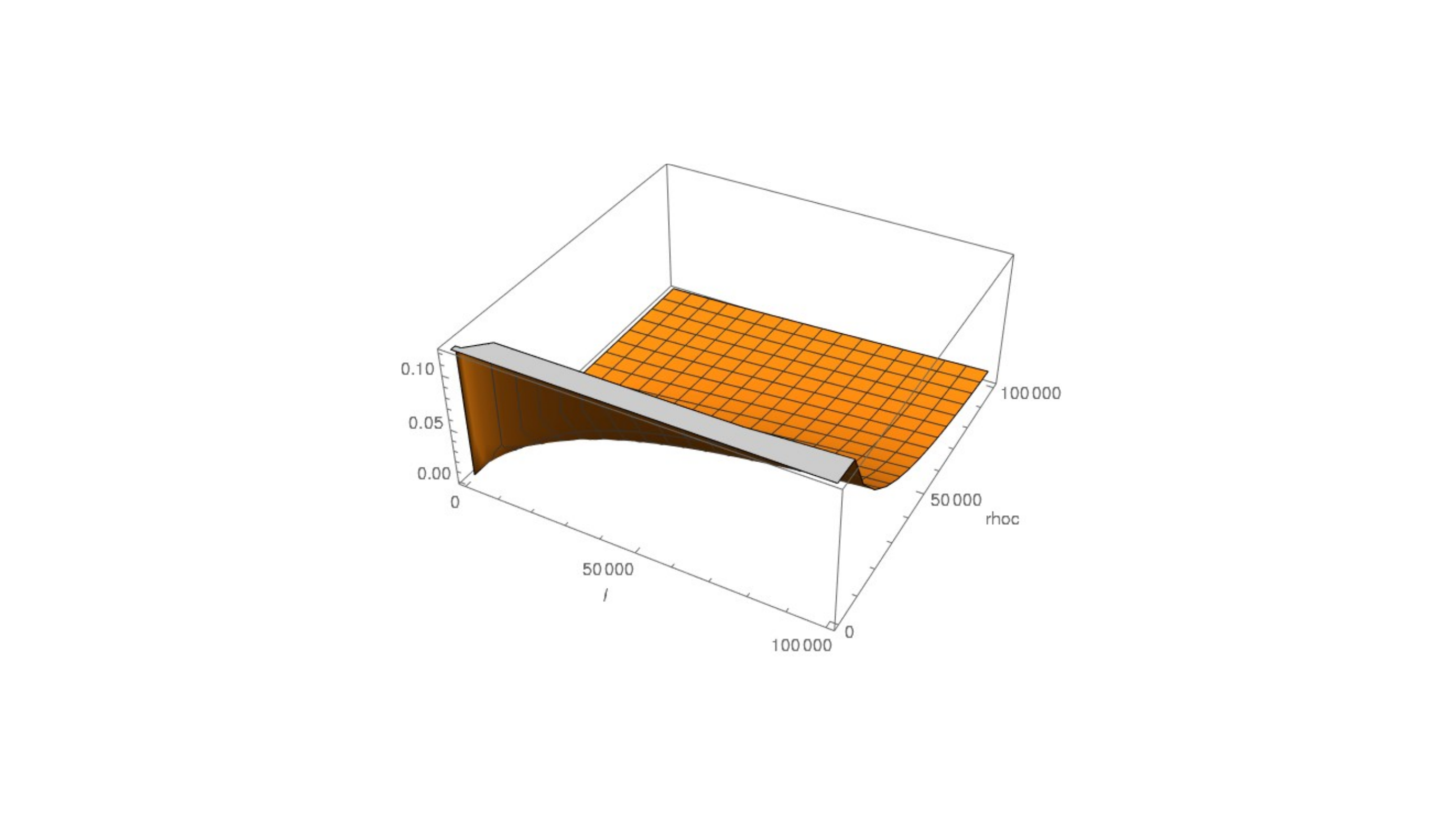}
\includegraphics[width=.36\textwidth]{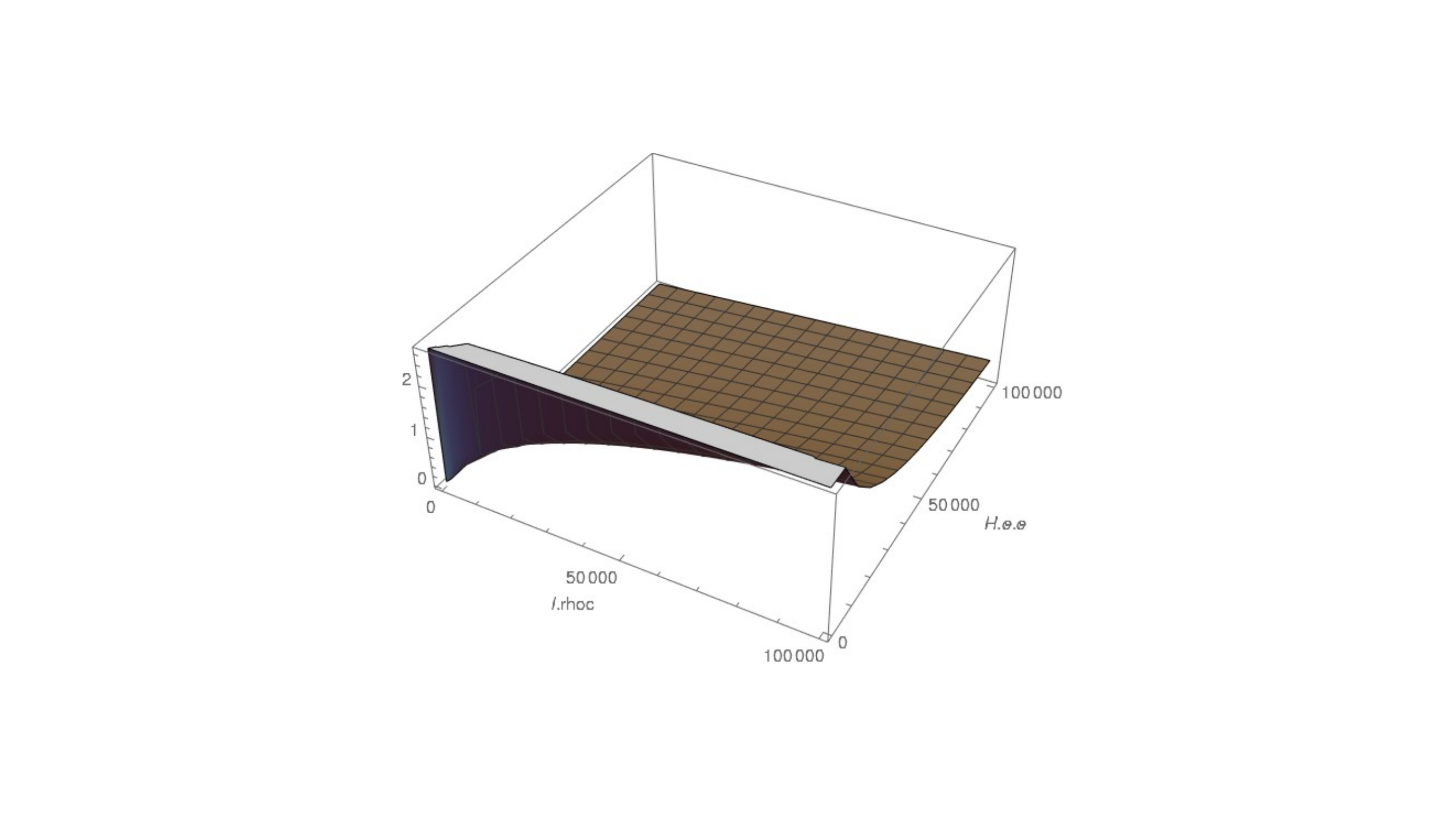}
\includegraphics[width=.45\textwidth]{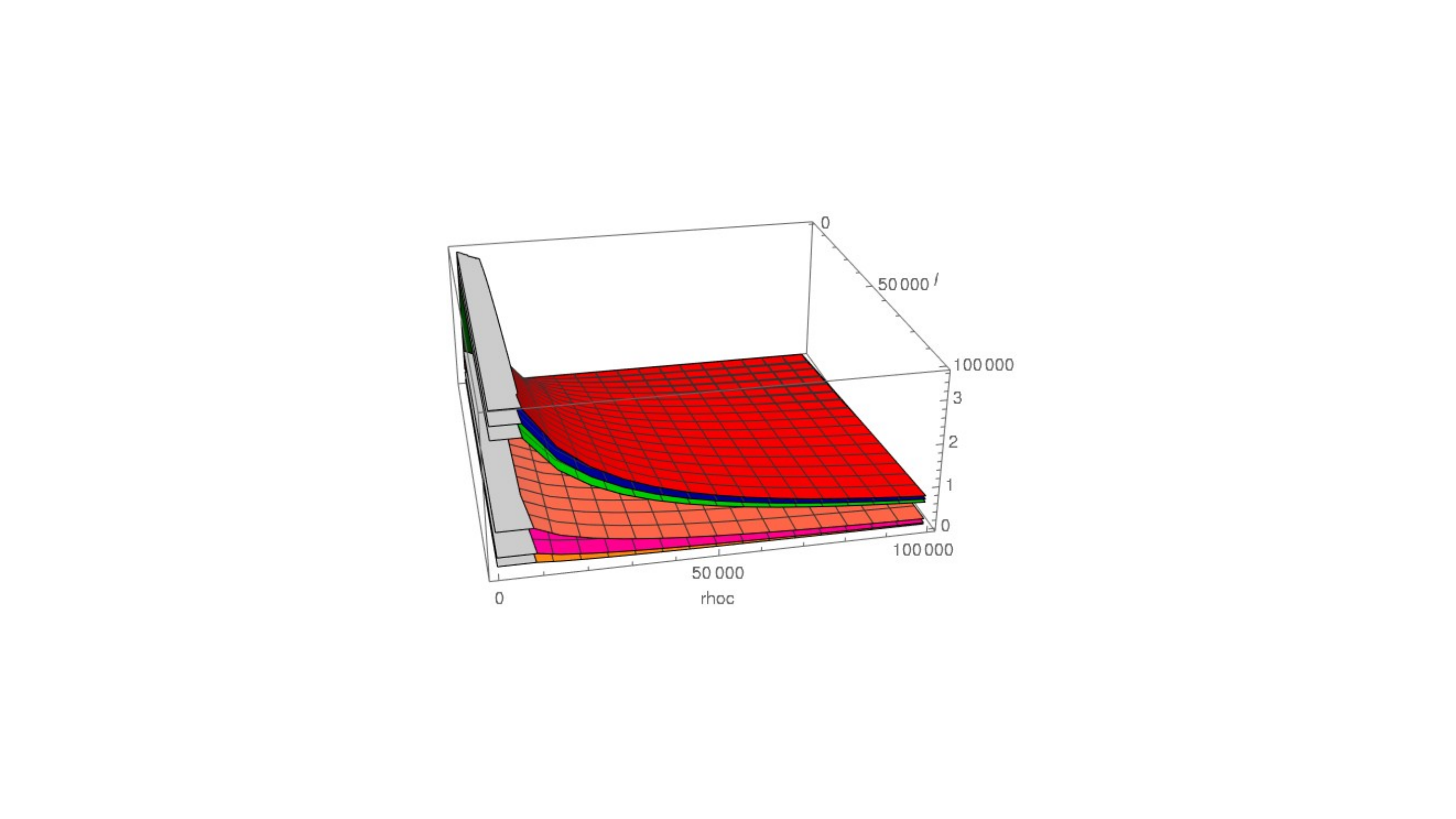}
\begin{center}	
\includegraphics[width=.60\textwidth]{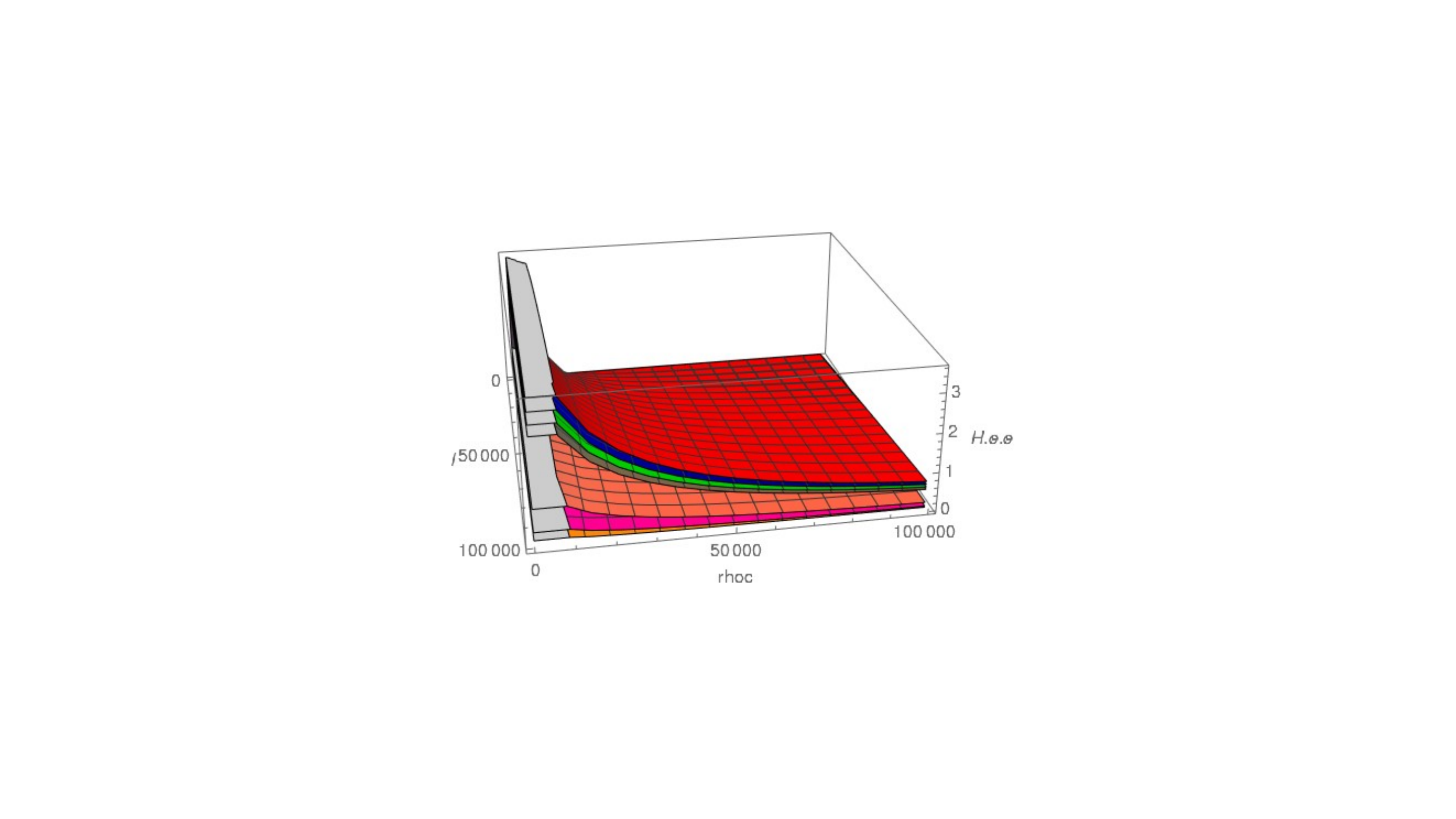}
\end{center}
\caption{  HEE plotted with  $(l,\rho_c)$   \, is exact only in $l>> \rho_c$ and $\rho_c >> l$,\, (first row)\, (left)\,,\, for $d - \theta = 0.97 $\,,\,(right)\,,\,$ d - \theta = 0,98 $,
 \,;\, (Second row)\,,\, (left)\,:\, $d - \theta =0.99$ \, , \, (right)\,:\, $d - \theta = 1.1$ 
 \, ;\,    (Third row )\,,\, (left)\,:\, $d - \theta = 1.2$, \,,\, (right) \,: \, $ d - \theta = 1,3 $ \quad;\quad 
(Fourth row)\,,\, (left) \,:\, HEE plotted as a function of $(l,\rho_c)$ for $d- \theta =1$ \,,\, (right) \,:\. The overlap of all, excluding the case $d - \theta = 1$
showing H.E.E at zero temperature decreses with increase of $d - \theta$ \,,\,
(Last row) \,,\, The combination of all including H.E.E at $d- \theta =1$ showing HEE decreases as a smooth function of $d - \theta$ from the regime $d - \theta < 1$
to $d - \theta > 1$ through $ d - \theta = 1$}
\la{evlessgreatequal}
\end{figure}

\section{  Simplified expression of holographic entanglement of entropy for $l >> \rho_c$ for $ d - \theta > 1$ and $d - \theta < 1$ }

Our lesson from the last section that the expression of H.E.E, for $d - \theta \ne 1$, as given by (\ref{entropylrhoc1}, \ref{entropylrhoc2}) are too complicated to give any physics-insight.  Given the fact, one can have the exact expression of H.E.E in the regime $l >> \rho_c$ and $\rho_c >> l$, one can always try to have a simplified expression of the same.  Accordingly, in this section we will proceed to find the simplified expression of H.E.E for $l >> \rho_c$ regime and present the    3D plots for this approximated expression alongwith the exact expression over long range for $(l, \rho_c )$ to show that these two matches exactly in $l >> \rho_c$ regime.

Here first we recall  the expression of H.E.E from (\ref{entropy})

The gaussian  Hypergeometricfunction by definition can be written as

\ber
{}_2 F_1 \left\lbrack  a\, , \, b, \, c \, , z \right\rbrack  &=&  1 + {\frac{a b}{1! c}}z + {\frac{a (a + 1) b (b + 1)}{2! c (c + 1)}}z^2 +...\n
\la{simplified}
\eer
Since for $l >> \rho_c$  we have ${\frac{\rho_c}{\rho_0}} \sim 0 $ , so around that point, we can expand the Hypergeometric function 
as

\ber
& & {}_2 {F_1}   \left\lbrack {\frac{1}{2}}, {\frac{1}{2}}( - 1 +{ \frac{1}{d - \theta }}), {\frac{1}{2}}(1 +{ \frac{1}{d - \theta }}) ,  \right \rbrack  \n
&=&1+ {\frac{   \left(   {\frac{1}{2}}\right) \cdot\left(  {\frac{1}{2}}( - 1 +{ \frac{1}{d - \theta }} )  \right)  }{1! {\frac{1}{2}}(1 +{ \frac{1}{d - \theta }} ) }}
{\left({\frac{\rho_c}{\rho_0}}\right)}^{2(d - \theta)} \n
&+& {\frac{  \left(  {\frac{1}{2}} \cdot  \left(    {\frac{1}{2}}  + 1   \right) \right)\cdot\left({\frac{1}{2}}( - 1 +{ \frac{1}{d - \theta }})  \cdot( {\frac{1}{2}}( - 1 +
{\frac{1}{d - \theta }})  + 1)  \right)}
{2!\left(  {\frac{1}{2}}(1 +{ \frac{1}{d - \theta }} )  \cdot (({\frac{1}{2}}(1 +{ \frac{1}{d - \theta }}) + 1)   \right)}}{\left({\frac{\rho_c}{\rho_0}}\right)}^{4(d - \theta)}  +..............\n
&=& 1+ {\frac{   \left(     {\frac{1}{2}}( - 1 +{ \frac{1}{d - \theta }} )  \right)  }{ {(1 +{ \frac{1}{d - \theta }} ) }}}
{\left({\frac{\rho_c}{\rho_0}}\right)}^{2(d - \theta)}\n
 &+& {\frac{ {\frac{3}{8}} \cdot\left(( - 1 +{ \frac{1}{d - \theta }} )   \cdot ( 1 +
{\frac{1}{d - \theta }} )    \right)}
{\left(  (1 +{ \frac{1}{d - \theta }} )  \cdot (3 +{\frac{1}{d - \theta }})    \right)}}{\left({\frac{\rho_c}{\rho_0}}\right)}^{4(d - \theta)}  +..............\n
&=& 1+ {\frac{ \left( {\frac{1}{2}}( - 1 +{ \frac{1}{d - \theta }} )  \right)  } {(1 +{ \frac{1}{d - \theta }} ) }}
{\left({\frac{\rho_c}{\rho_0}}\right)}^{2(d - \theta)}\n
&+& {\frac{    {\frac{3}{8}} \cdot\left(( - 1 +{ \frac{1}{d - \theta }})  \cdot( ( 1 +
{\frac{1}{d - \theta }})  )  \right)}
{\left(  (1 +{ \frac{1}{d - \theta }} )  \cdot ((3 +{ \frac{1}{d - \theta }}) )   \right)}}{\left({\frac{\rho_c}{\rho_0}}\right)}^{4(d - \theta)}  +..............\n
&=&1+ {\frac{   \left(     {\frac{1}{2}}( - 1 +{ \frac{1}{d - \theta }} )  \right)  }{(1 +{ \frac{1}{d - \theta }} ) }}
{\left({\frac{\rho_c}{\rho_0}}\right)}^{2(d - \theta)}\n
&+& {\frac{    {\frac{3}{8}} ( - 1 +{ \frac{1}{d - \theta }})  }
{ (3 +{ \frac{1}{d - \theta }} )   }}{\left({\frac{\rho_c}{\rho_0}}\right)}^{4(d - \theta)}  +...
\la{hyperexpansion0}
\eer

Finally we substitute The expression of the Hypergeometric function from (\ref{hyperexpansion0}) in the expression of HEE as given in (\ref{entropy}), to obtain
the simplified expression of the HEE

\ber
  S &=&   {\frac{      L^{d-1} (\rho_0)^{\theta - d +1} \,\, {{}_2 F_1}   \left\lbrack {\frac{1}{2}}, {\frac{1}{2}}( - 1 +{ \frac{1}{d - \theta }}), {\frac{1}{2}}(1 +{ \frac{1}{d - \theta }}) , 1 \right \rbrack }{  4 G_N (\theta + 1- d )    }}\n
&-& \left(\rho_c \right)^{ \theta +1- d} {\frac{     L^{d-1 } }{  4 G_N( \theta - d  + 1)   }}  \times \left\lbrack 1+ {\frac{  \left(     {\frac{1}{2}}( - 1 +{ \frac{1}{d - \theta }} )  \right)  }{ {(1 +{ \frac{1}{d - \theta }} ) }}}\right\rbrack\n
&-& {\left(\rho_c\right)}^{ \theta +1- d} {\frac{     L^{d-1}} {  4 G_N ( \theta - d  + 1)  }}  
\left\lbrack  {\frac{    {\frac{3}{8}} ( - 1 +{ \frac{1}{d - \theta }})  }
{ (3 +{ \frac{1}{d - \theta }} )   }}{\left({\frac{\rho_c}{\rho_0}}\right)}^{4(d - \theta)} 
  \right\rbrack + ...... 
\la{entropyrewrite}
\eer

Next, to simplify it further we recall the expression of the turning point from (\ref{ultimaterho0nonvanishing})

For the regime $l>> \rho_c$, we have the simplified expression of the turning poing, 
given by for $d - \theta > 1$

\ber
\rho_0 &=& {\left( {\left({\frac{A_{10} l}{ 2 }}\right)}^{2  \left( d - \theta\right)} + {\left( \rho_c \right)}^{2  \left( d - \theta\right)} \right)}^{\frac{1}{2(d - \theta)}} \n
&=&    {\left({\frac{A_{10} l}{ 2 }}\right)} {\left( 1 + {\frac{1}{2(d - \theta)}} {\frac { {{\left( \rho_c \right)}^{2  \left( d - \theta\right)} }}{{\left({\frac{A_{10} l}{ 2 }}\right)}^{2  \left( d - \theta\right)} } }     \right)} \n 
\la{nnn}
\eer
For $d - \theta < 1$, we have 

\ber
\rho_0 &=& {\left( {\left({\frac{A_{10} l}{ 2 }}\right)}^{2  \left( d - \theta\right)} + {\left( \rho_c \right)}^{2  \left( d - \theta\right)} \right)}^{\frac{1}{2(d - \theta)}} \n
&=&    {\left({\frac{A_{10} l}{ 2 }}\right)} {\left( 1 + {\frac{1}{(d - \theta + 1)}} {\frac { {{\left( \rho_c \right)}^{  \left( d - \theta  + 1 \right)} }}{{\left({\frac{A_{10} l}{ 2 }}\right)}^{  \left( d - \theta  + 1 \right)} } }     \right)} \n 
\la{nnn11}
\eer

One can substitute (\ref{nnn}, \ref{nnn11}) in (\ref{ultimaterho0nonvanishing}) to obtain the simplified expression of HEE as can be considered the expression of HEE , in $l>>\rho_c$ and $\rho_c >> l$, when can be considered over very long range!
\vskip2mm

The plots for $ d - \theta > 1 $, are preaented in ( \ref{slgr5by3greatl1},    \ref{slgr8by3greatl1}) ,

  The Plots for $d - \theta < 1$ ,   are presented in  (\ref{lgreatrhoc4by9l1} ,\ref{lgreatrhoc1by12l1 } )

\begin{figure}[H]
\begin{center}
\textbf{ For $d-\theta > 1$,  with  $ d - \theta = {\frac{5}{3}}$, \,\,:\,\,  Comparison   between the simplified expression of HEE for $ l >>  \rho_c $ with the exact expression of HEE in this regime, can be realized only in long range of  long range of $l, \rho_c$, }
\end{center}
\includegraphics[width=.32\textwidth]{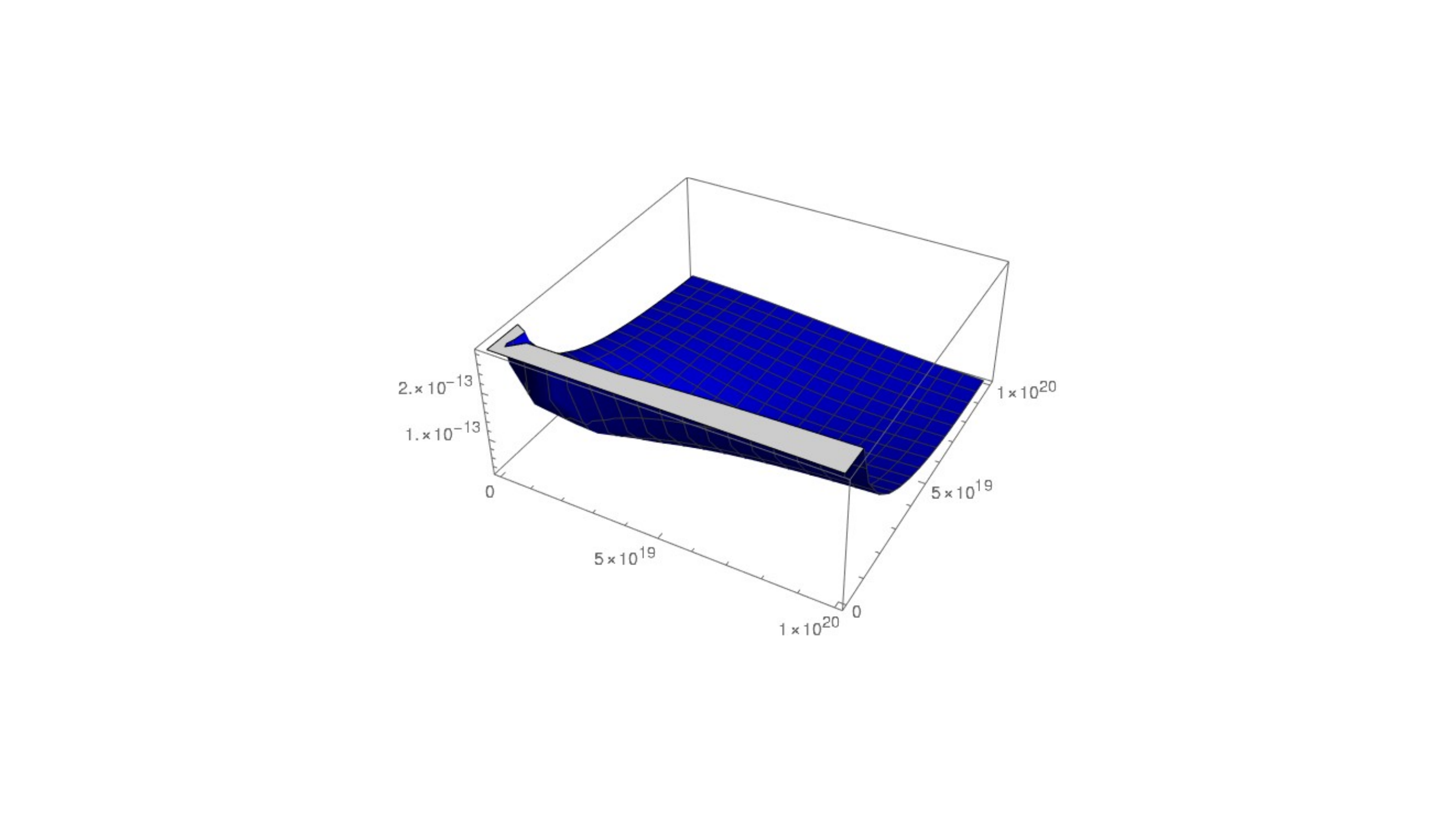}
\includegraphics[width=.32\textwidth]{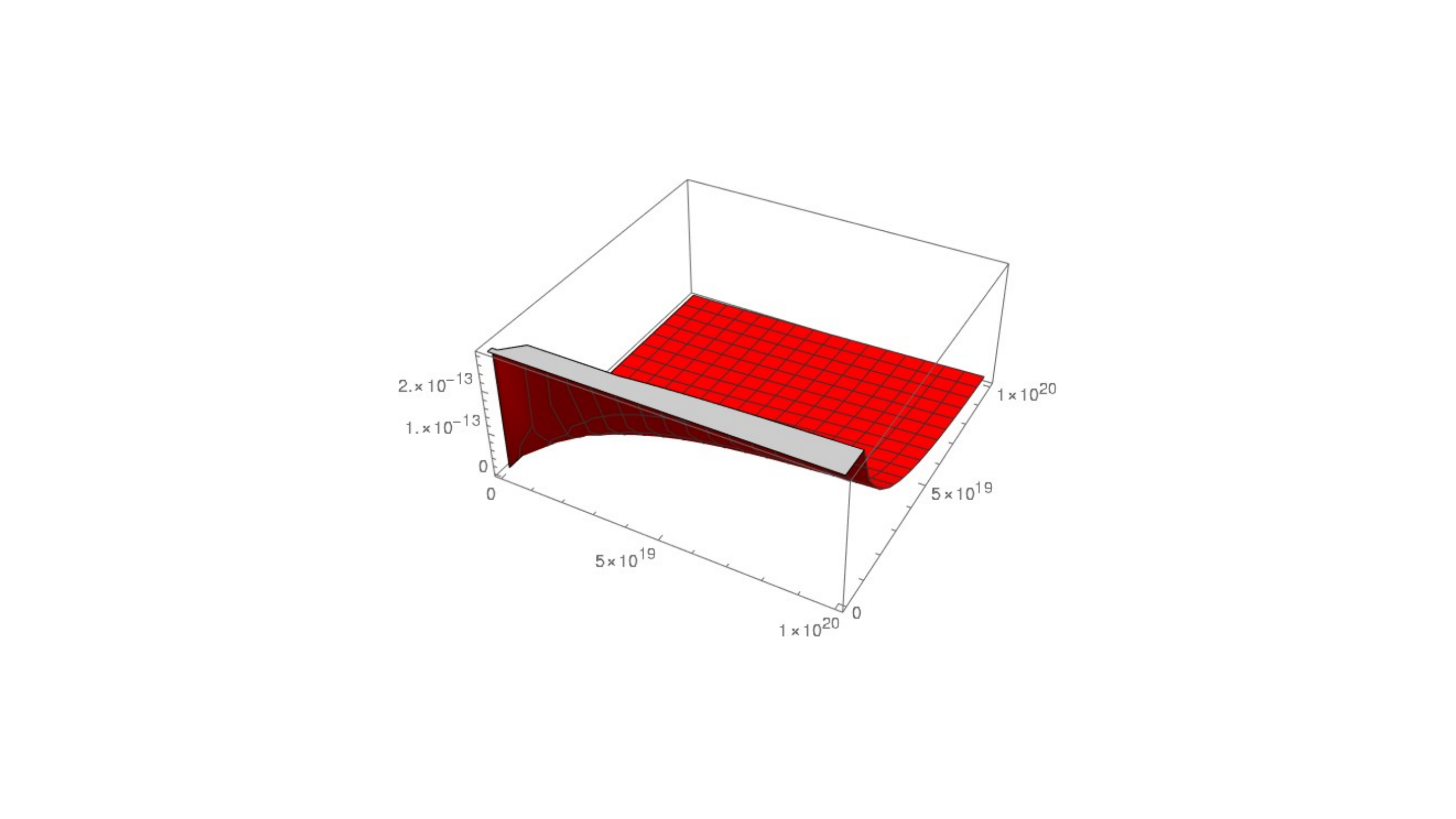}
\includegraphics[width=.32\textwidth]{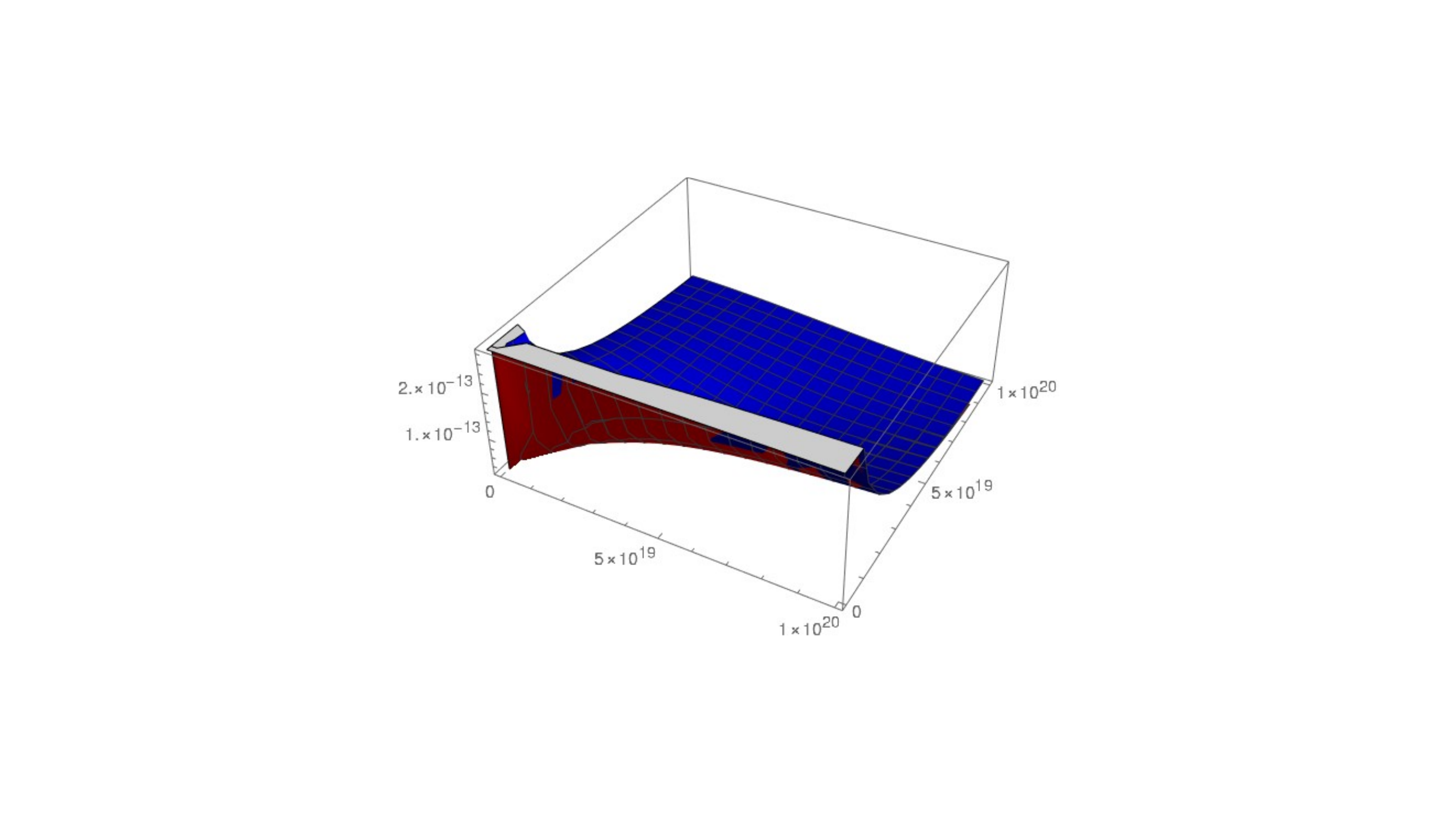}
\includegraphics[width=.32\textwidth]{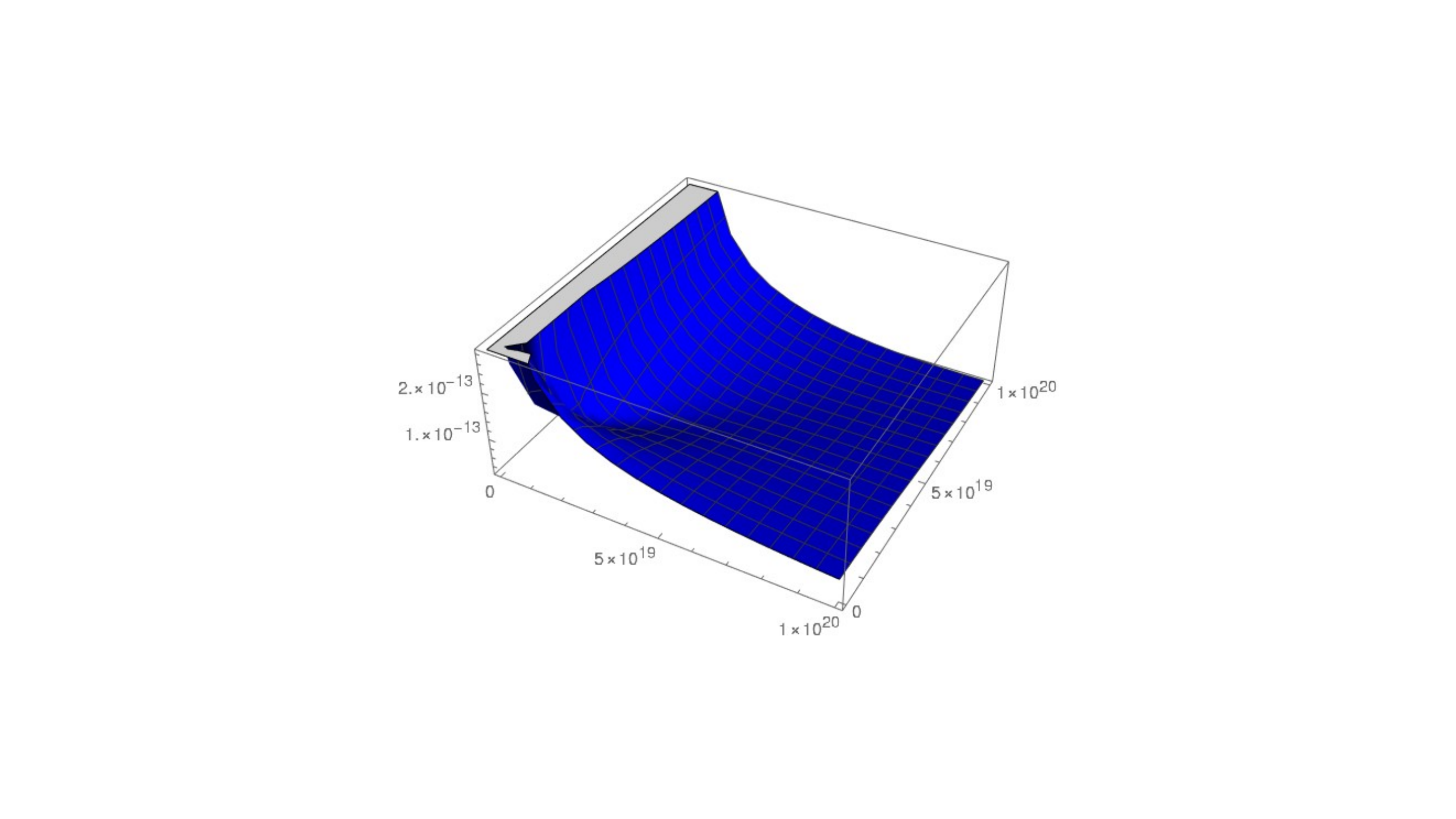}
\includegraphics[width=.32\textwidth]{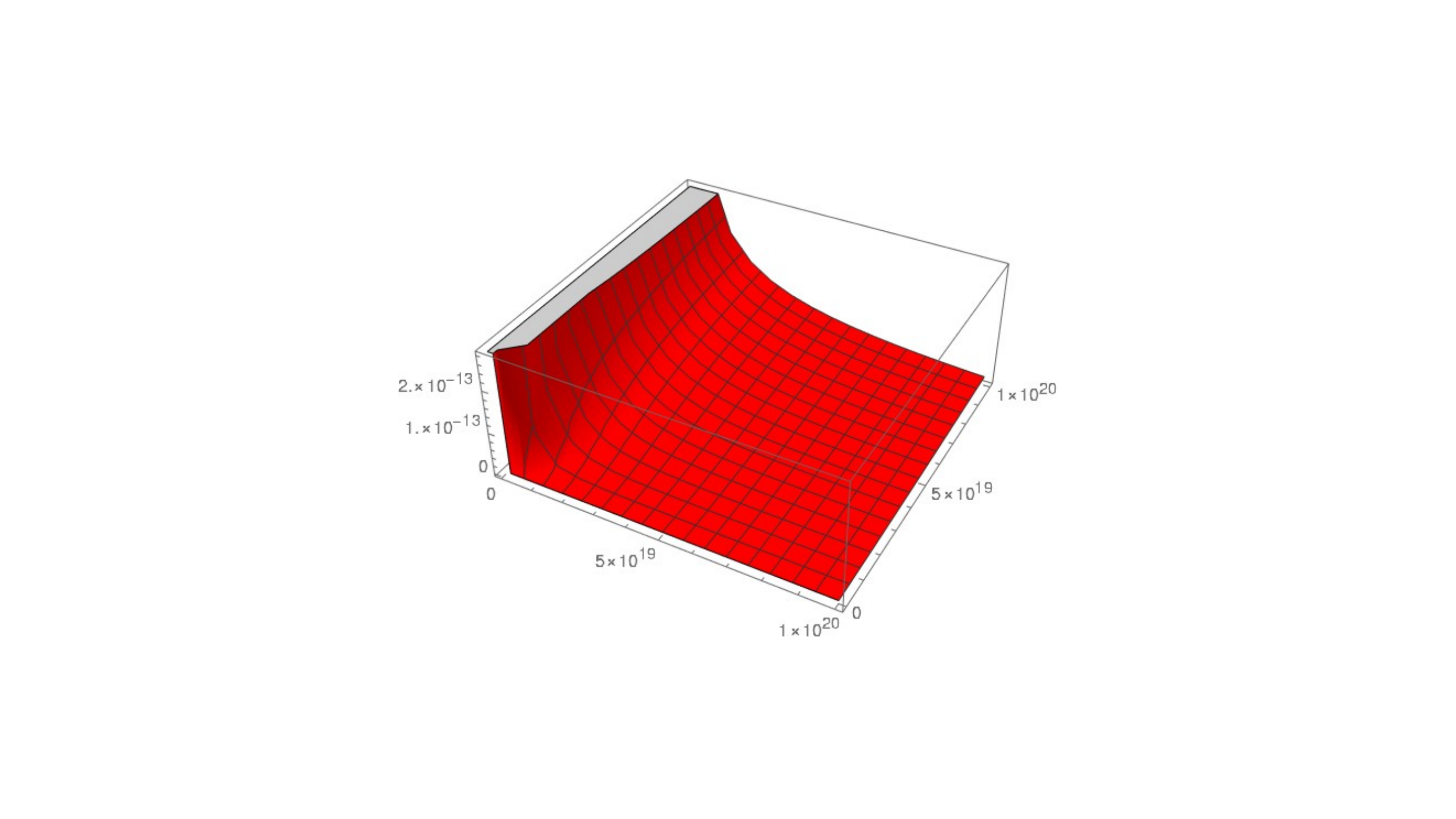}
\includegraphics[width=.32\textwidth]{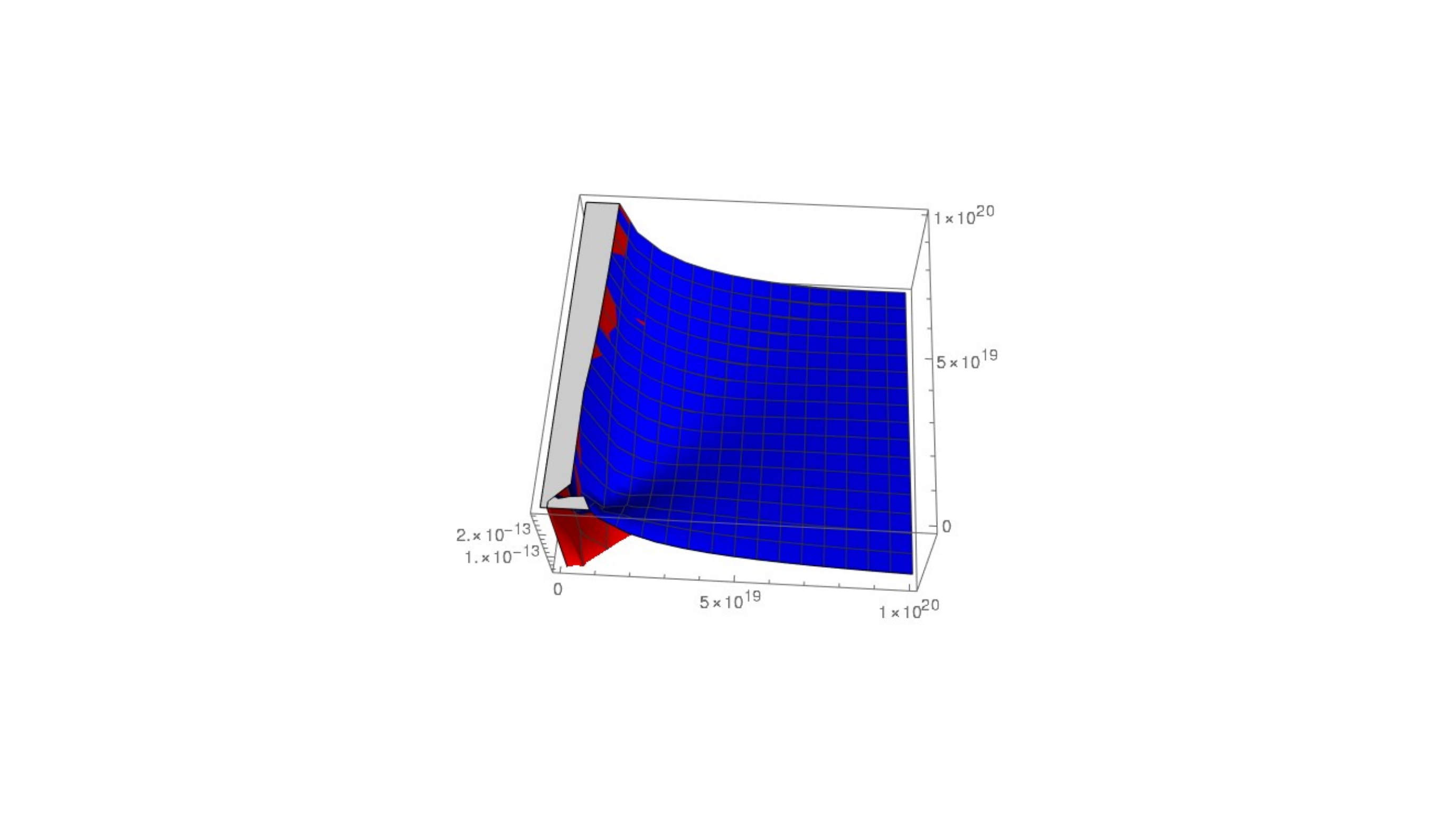}
\begin{center}
\includegraphics[width=.55\textwidth]{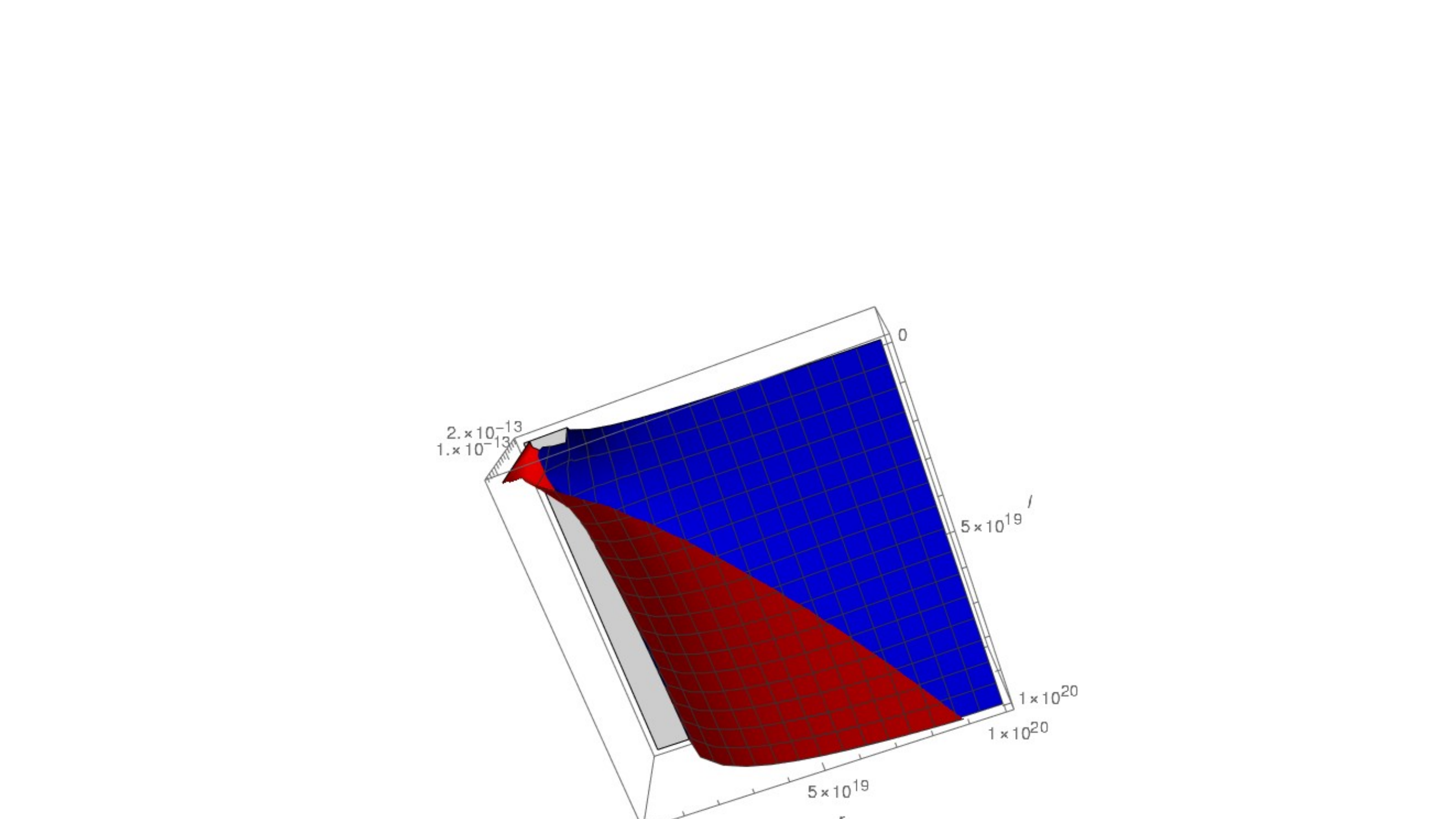}
\end{center}	
\caption{First row \, ; \,\, x-axis\,:\,l; y-axis\,:\,$\rho_c$\,\,:\,\,  (left)  \,:\,l; Simplified expression of S as a function of $l\,,\,\rho_c$ \quad,\quad (Middle)\,\, : \,\,Exact expression of S as a function of $l\,,\,\rho_c$  ; (right) The overlap between the two \,,\, The figures are showing the two matches eonly in the regime $l  >> \rho_c$
\quad;\quad
Middle  row \,:\, All the above with x-axis and y-axis interchanged
\quad;\quad (Last row)\,;\, Backside view of the overlap the simplified(given in blue) and exact expression(given in red) of HEE, showing that two merges in $l>>\rho_c$ regime only }
\la{slgr5by3greatl1}
\end{figure}

\begin{figure}[H]
\begin{center}
\textbf{ For $ d-\theta > 1$,  with  $ d - \theta = {\frac{8}{3}}$, \,\,:\,\,  Comparison   between the simplified expression of HEE for $ l >>  \rho_c $ with the exact expression of HEE in this regime, can be realized only in long range of  long range of $l, \rho_c$ }
\end{center}
\includegraphics[width=.32\textwidth]{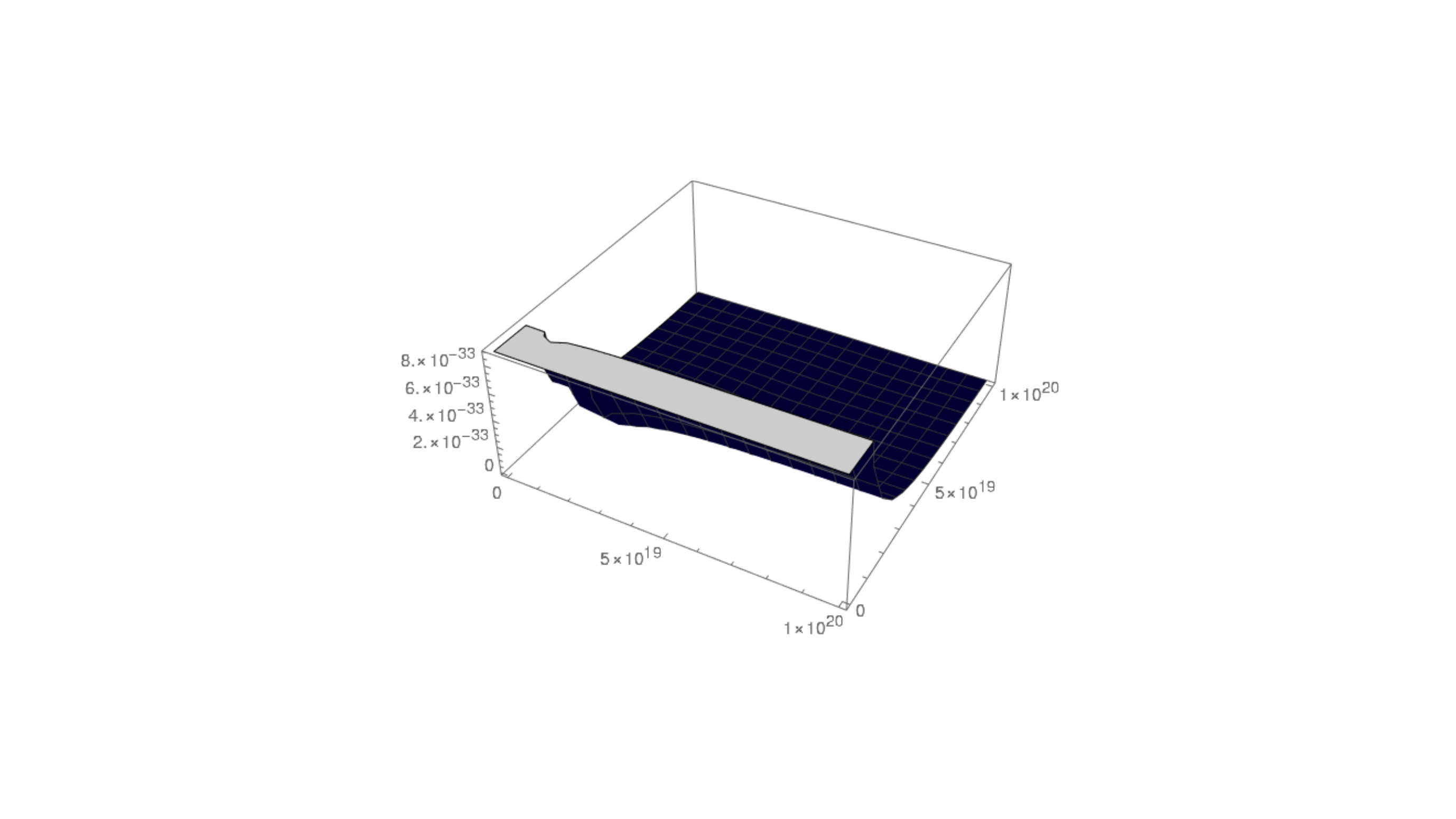}
\includegraphics[width=.32\textwidth]{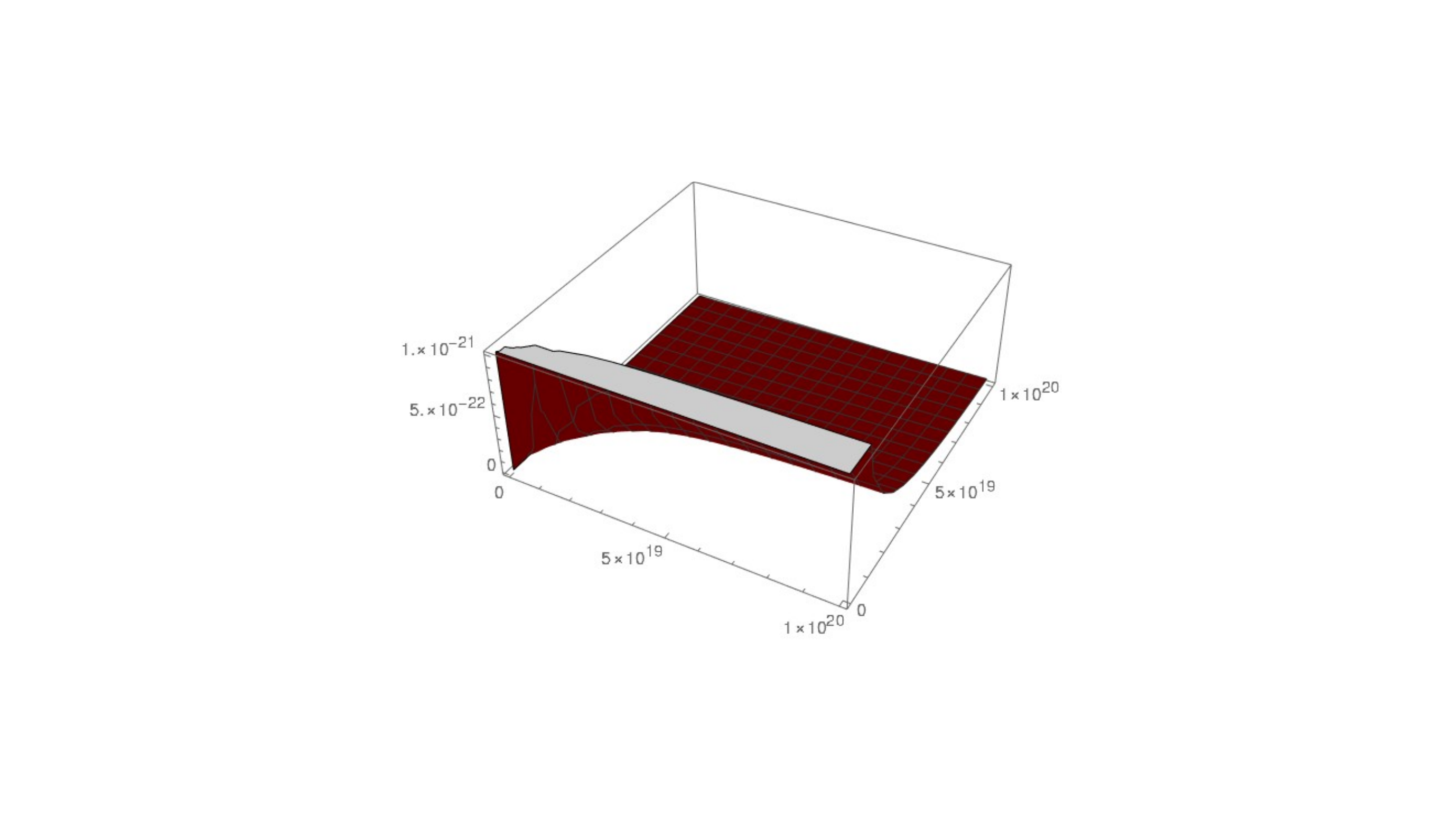}
\includegraphics[width=.32\textwidth]{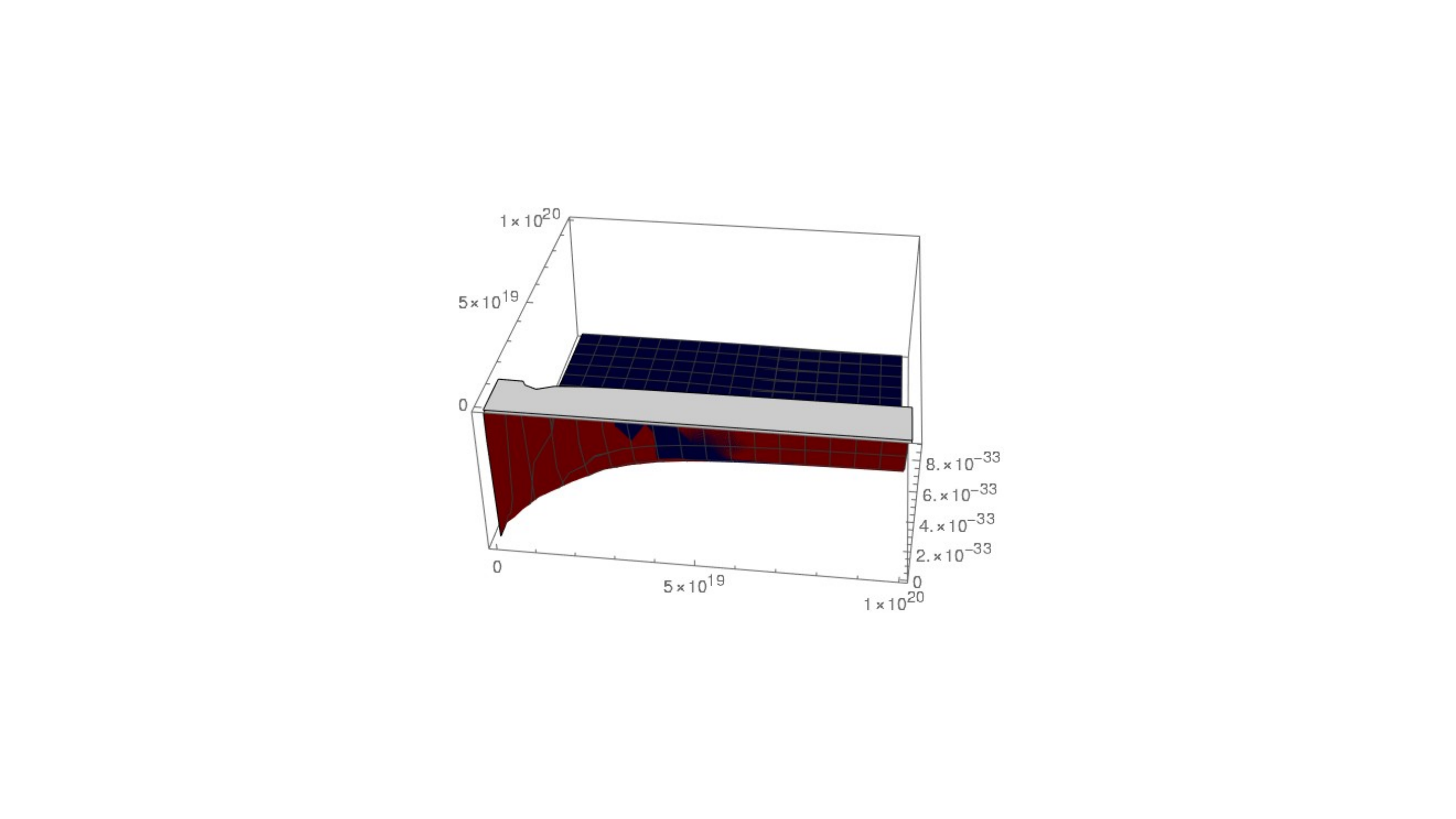}
\includegraphics[width=.32\textwidth]{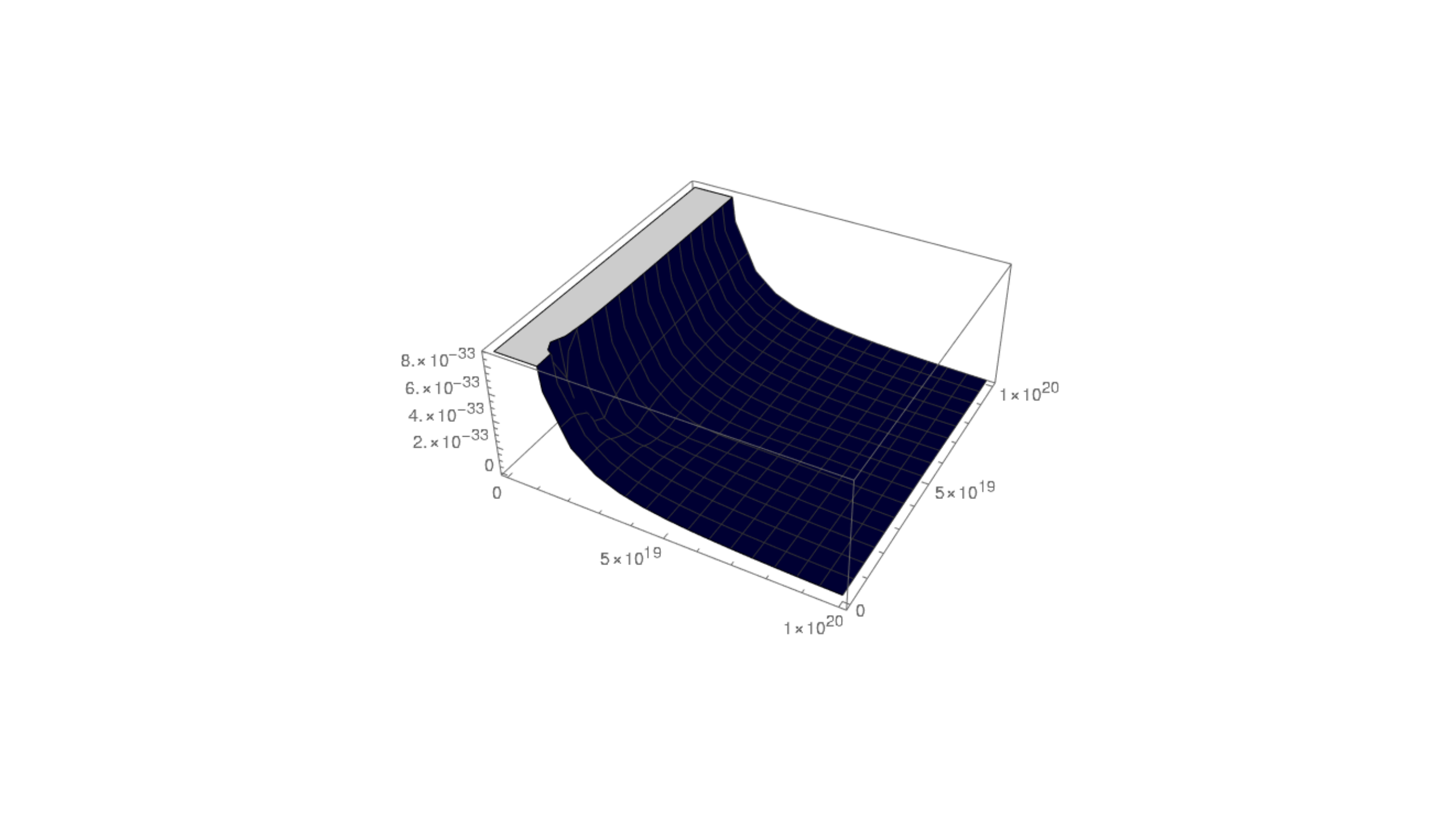}
\includegraphics[width=.32\textwidth]{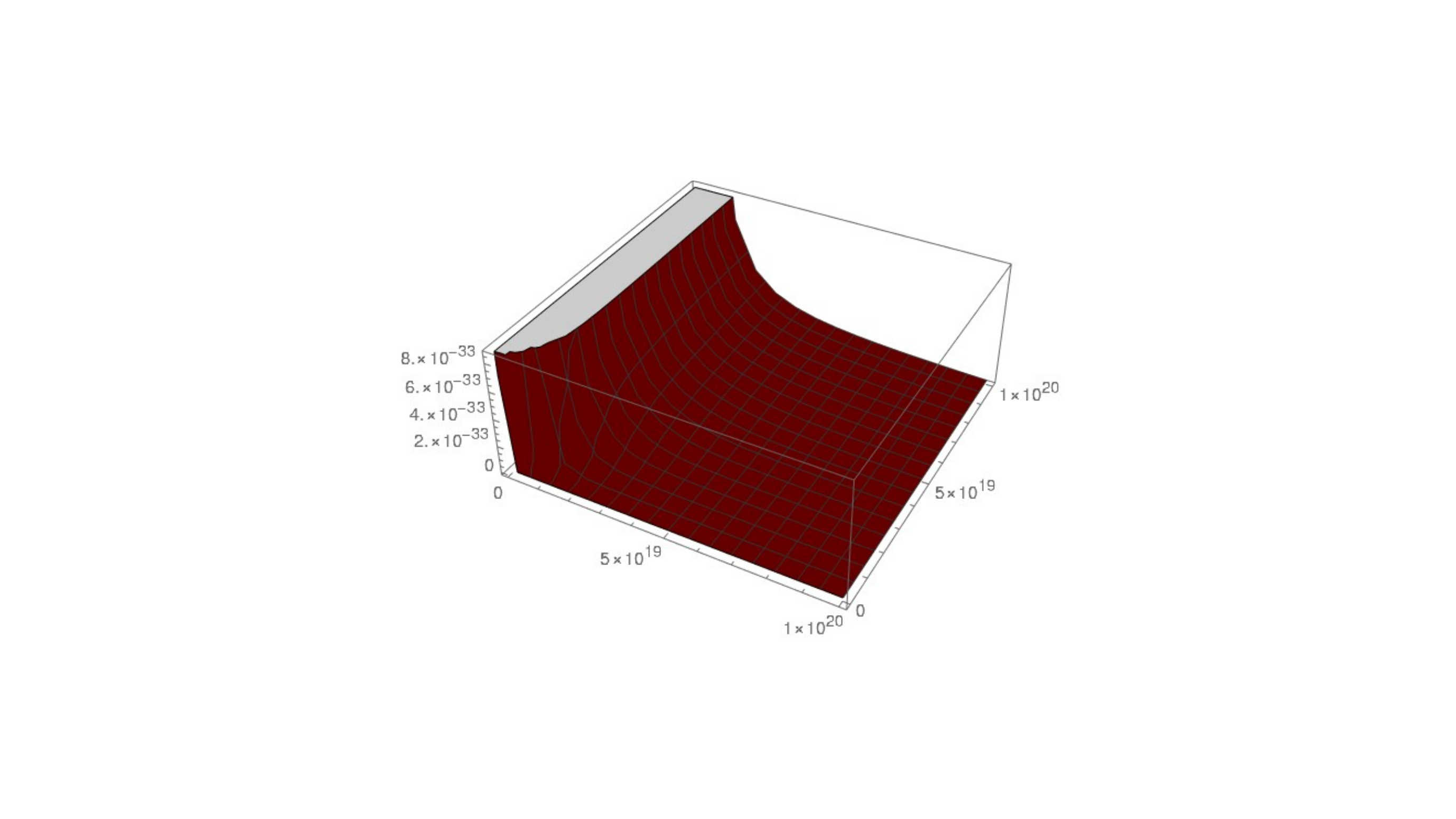}
\includegraphics[width=.32\textwidth]{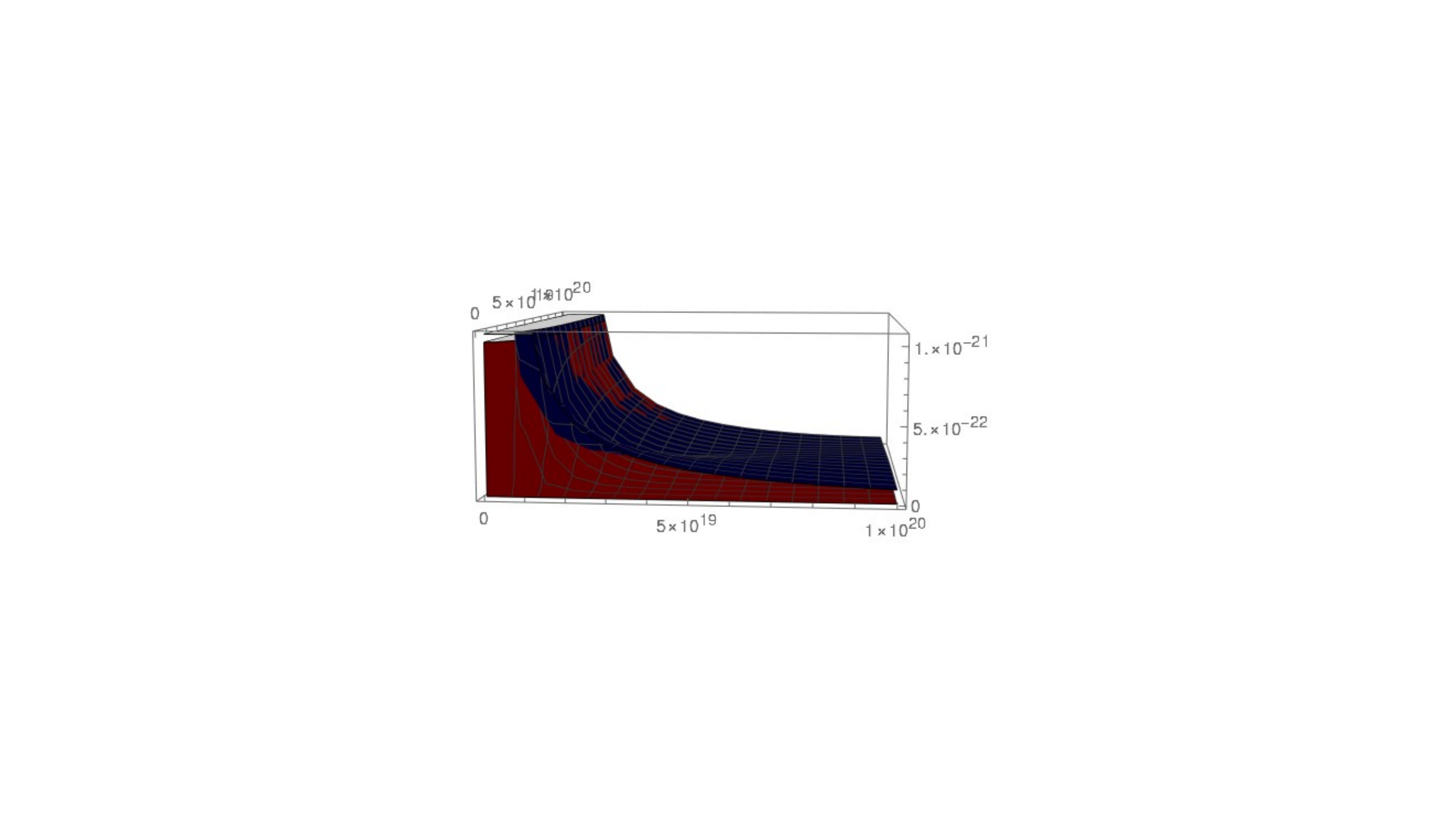}
\caption{First row \,;\,\, x-axis\,:\,l; y-axis\,:\,$\rho_c$\,\,:\,\,  (left)  \,:\,l; Simplified expression of S as a function of $l\,,\,\rho_c$\,;\, (Middle)\,:\, \, Exact expression of S as a function of $l,\,\rho_c$\,;\, (right) The overlap between the two\,:\, The figures are showing the two matches only in the regime $l  >> \rho_c$
\quad;\quad
Last row\,:\, All the above with x-axis and y-axis interchanged
 }
\la{slgr8by3greatl1}
\end{figure}

\begin{figure}[H]
\begin{center}
\textbf{ For $d-\theta < 1$,  with  $ d - \theta = {\frac{4}{9}}$, \,\,:\,\,  Comparison   between the simplified expression of HEE for $ l >>  \rho_c $ with the exact expression of HEE in this regime, can be realized only in long range of  long range of $l, \rho_c$ }
\end{center}
\includegraphics[width=.32\textwidth]{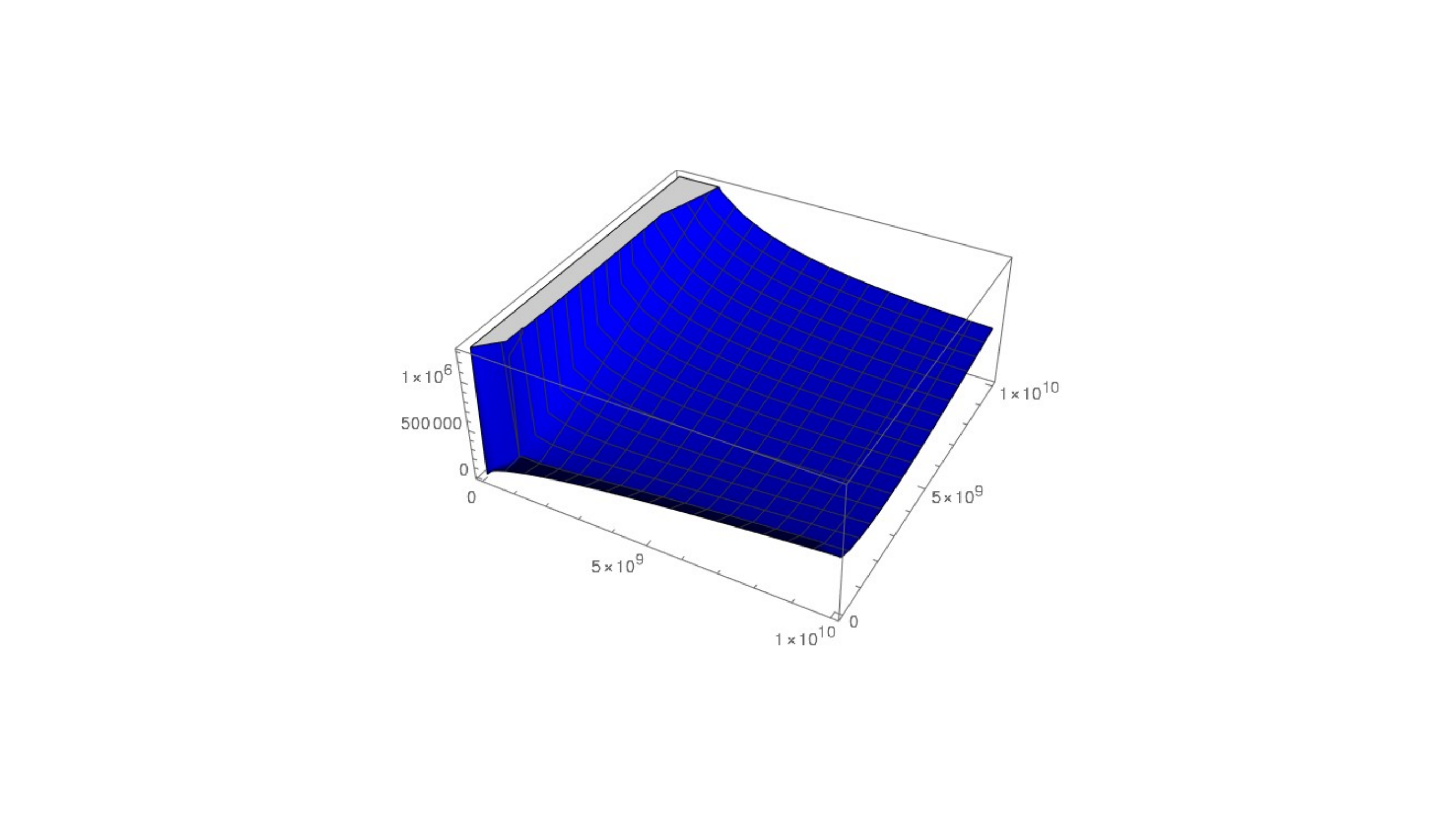}
\includegraphics[width=.32\textwidth]{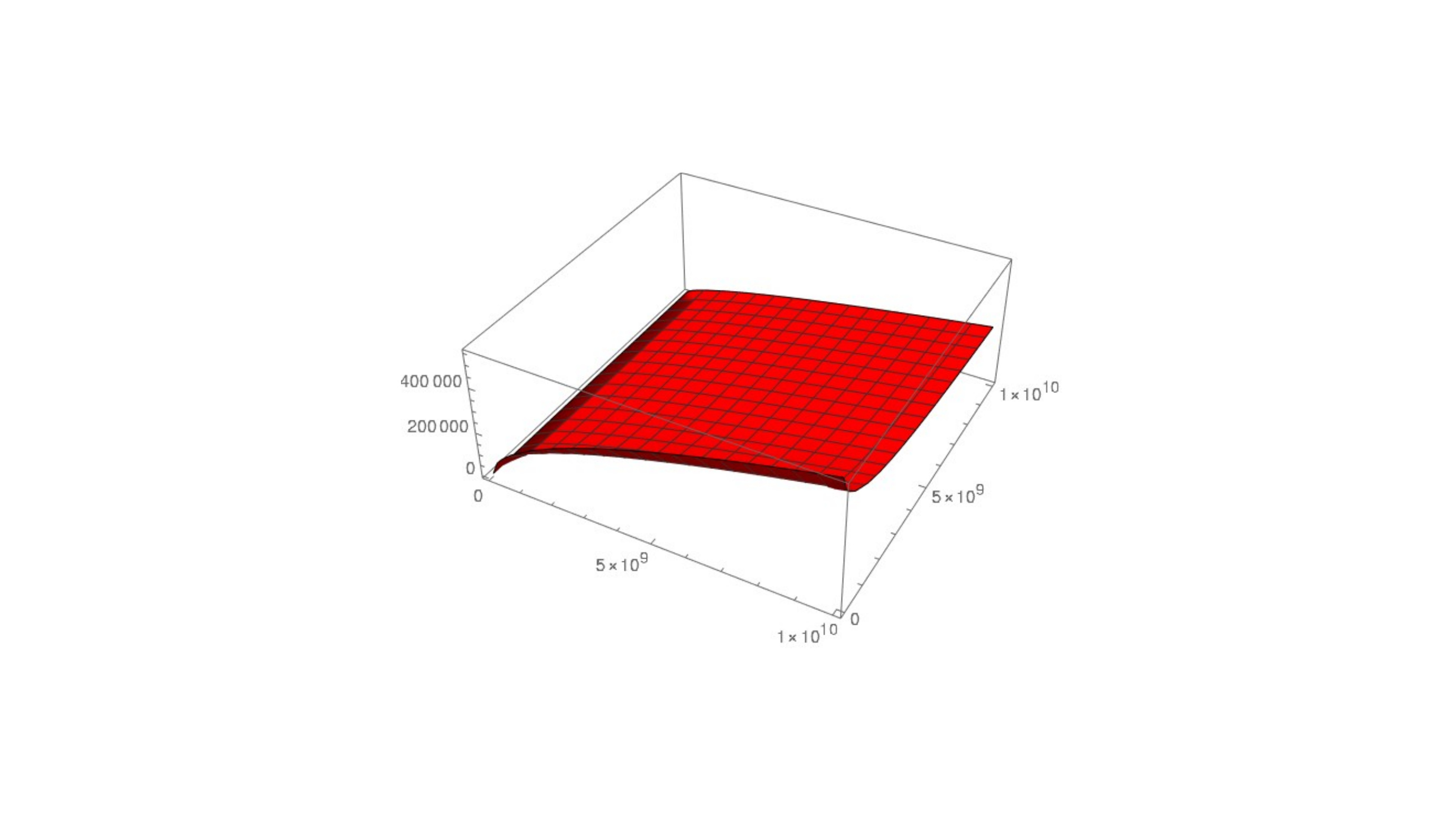}
\includegraphics[width=.32\textwidth]{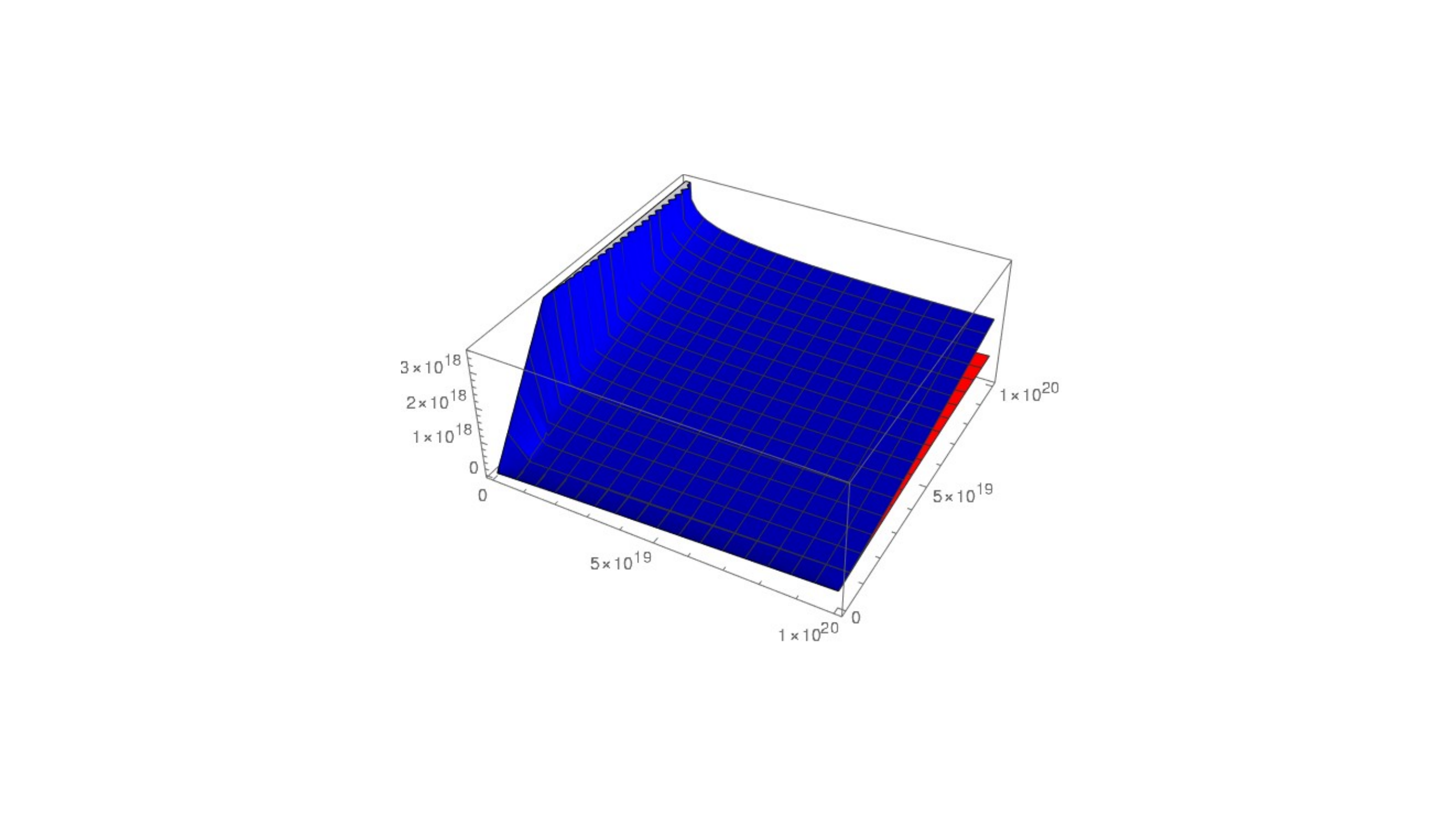}
\includegraphics[width=.32\textwidth]{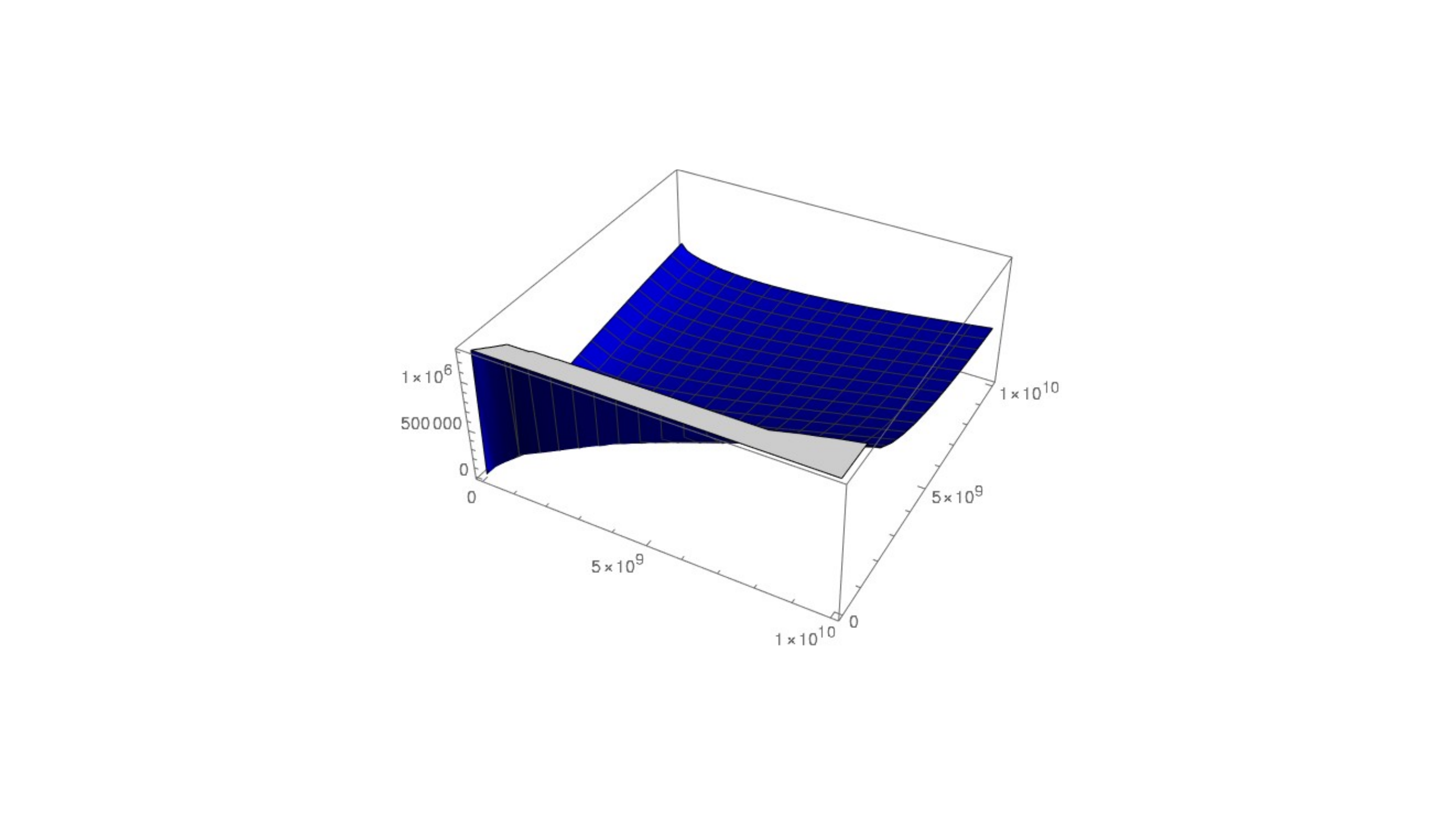}
\includegraphics[width=.32\textwidth]{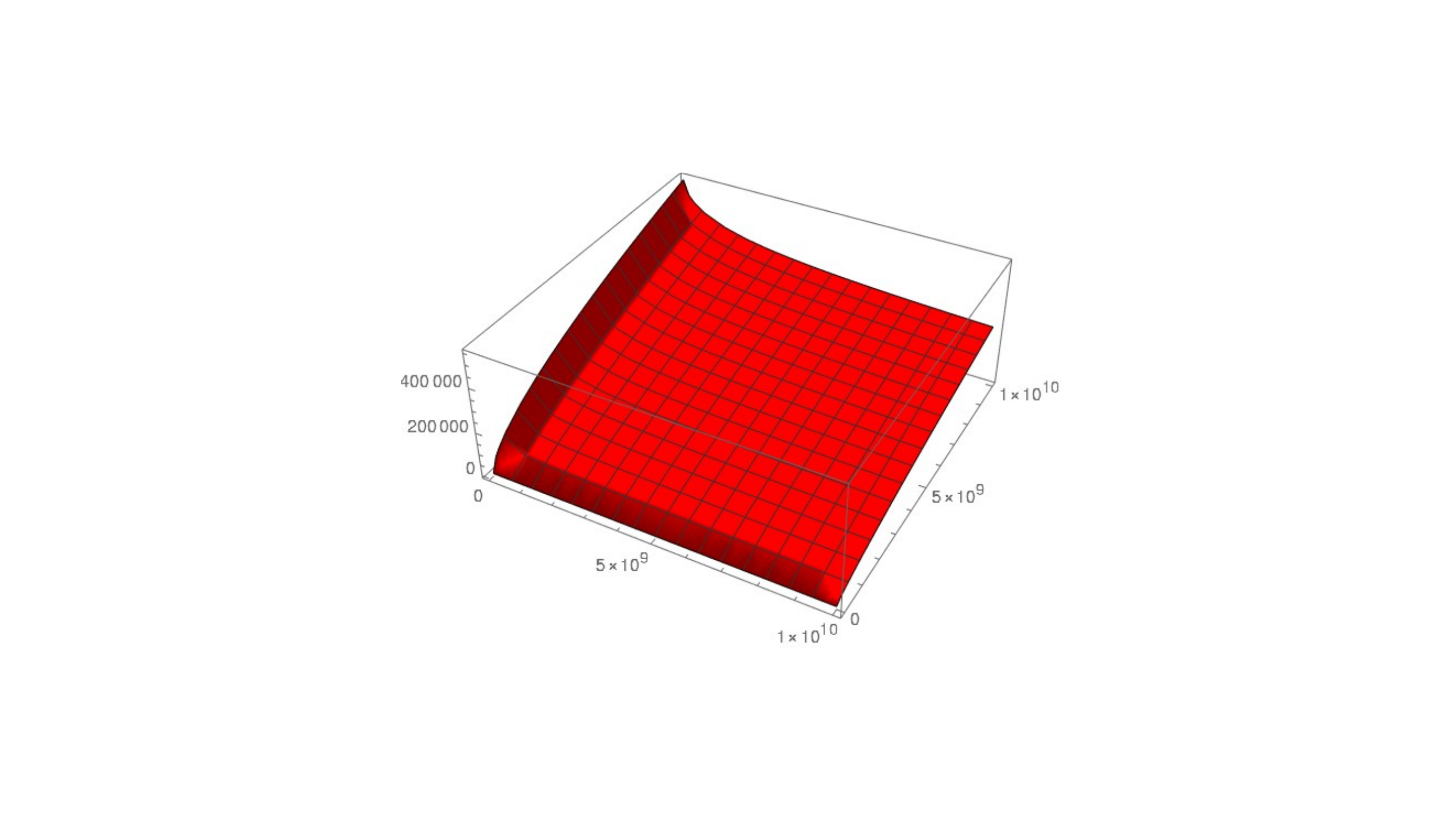}
\includegraphics[width=.32\textwidth]{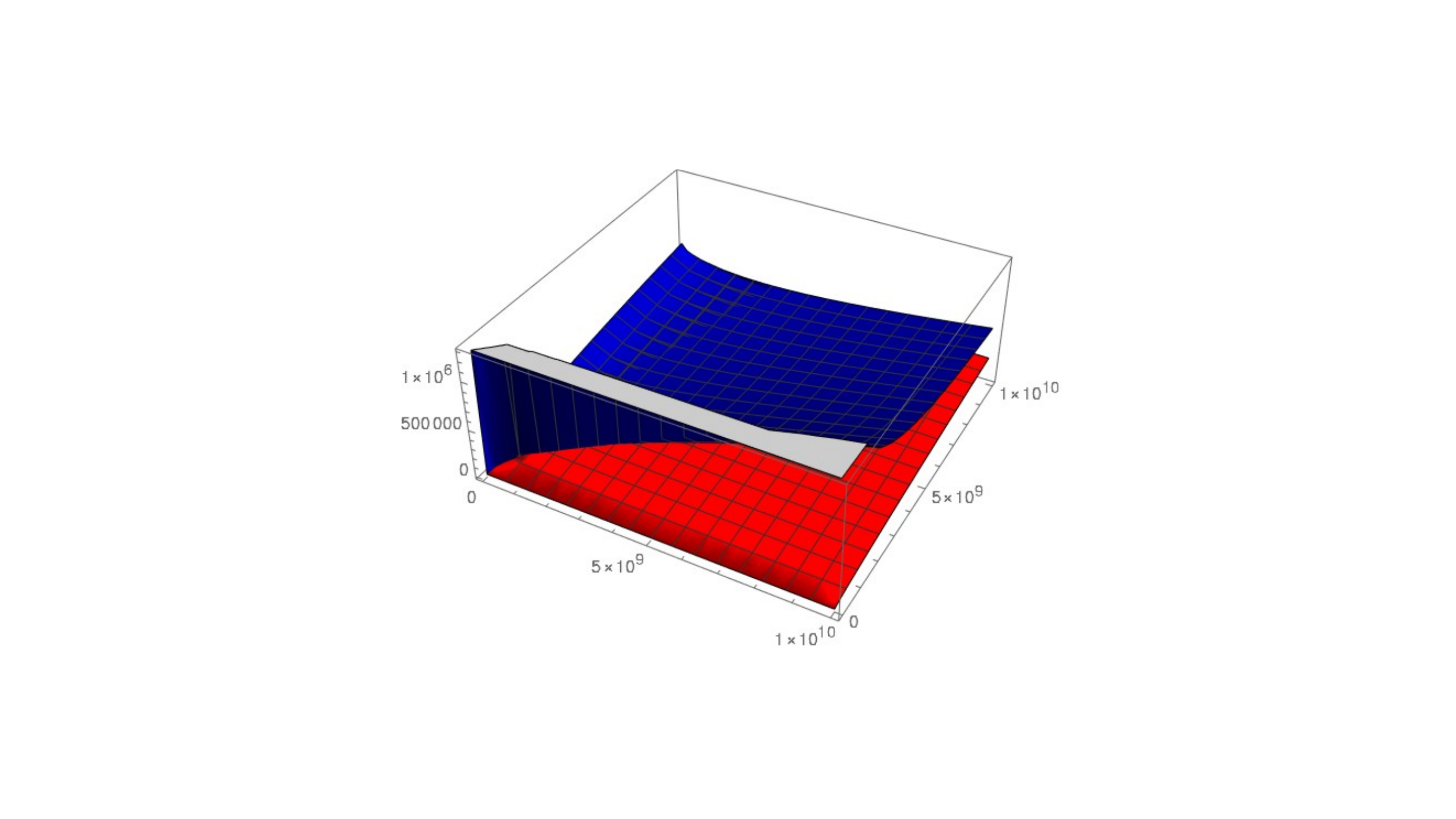}
\caption{First row\, :\,\, x-axis\,:\,l; y-axis\,:\,$\rho_c$  :  (left)  \,:\,l; Simplified expression of S as a function of $l, \rho_c$ ; (Middle) : \, Exact expression of S as a function of $l\,,\,\rho_c$\,; (right) The overlap between the two\,:\, The figures are showing the two matches eonly in the regime $l  >> \rho_c$
\quad;\quad Last row \,:\, All the above with x-axis and y-axis interchanged}
\la{lgreatrhoc4by9l1}
\end{figure}

\begin{figure}[H]
\begin{center}
\textbf{ For $d-\theta < 1$,  with  $d - \theta = {\frac{1}{12}}$, \,\,:\,\,  Comparison   between the simplified expression of HEE for $ l >>  \rho_c $ with the exact expression of HEE in this regime, can be realized only in long range of  long range of $l, \rho_c$ }
\end{center}
\includegraphics[width=.32\textwidth]{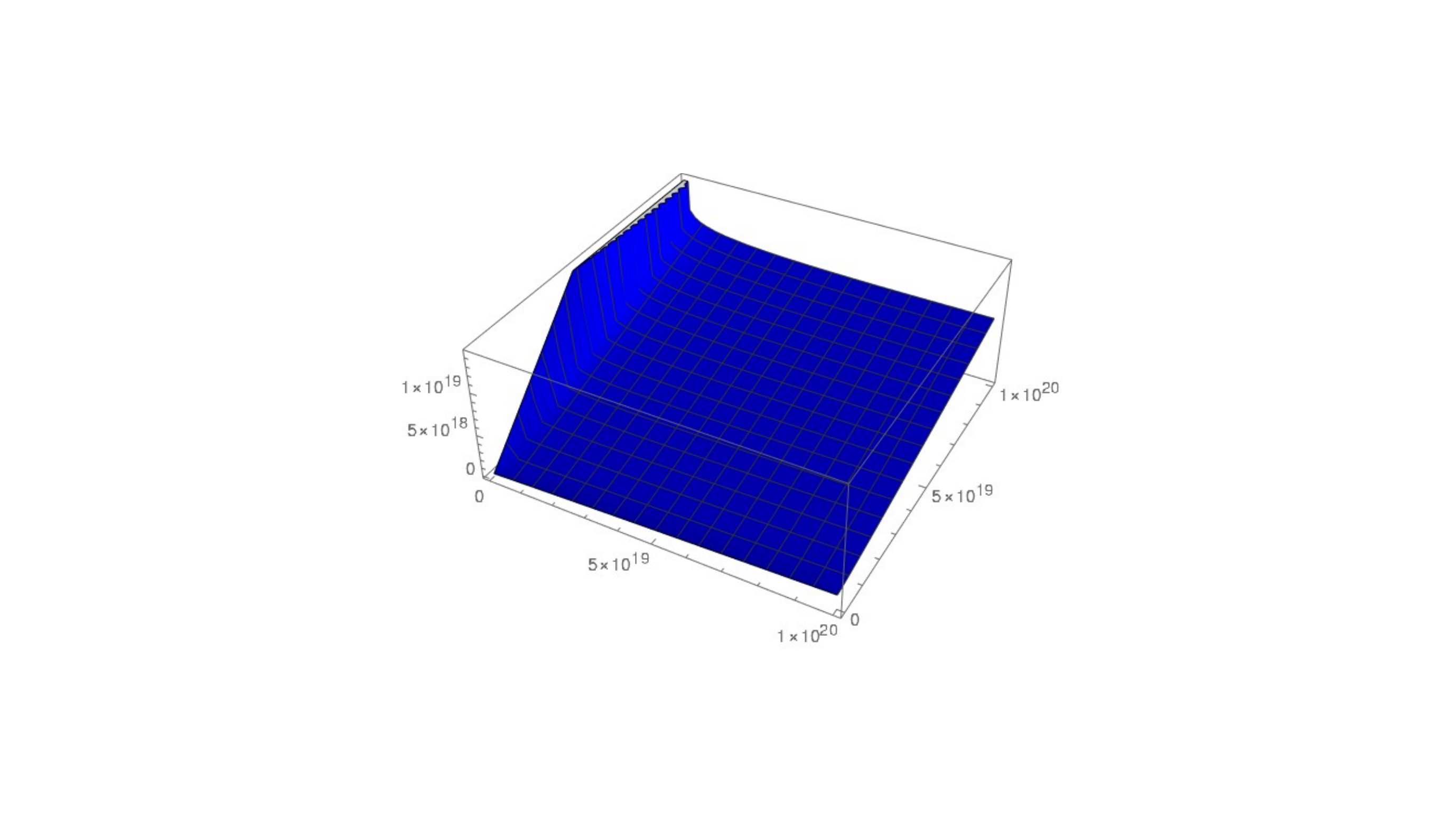}
\includegraphics[width=.32\textwidth]{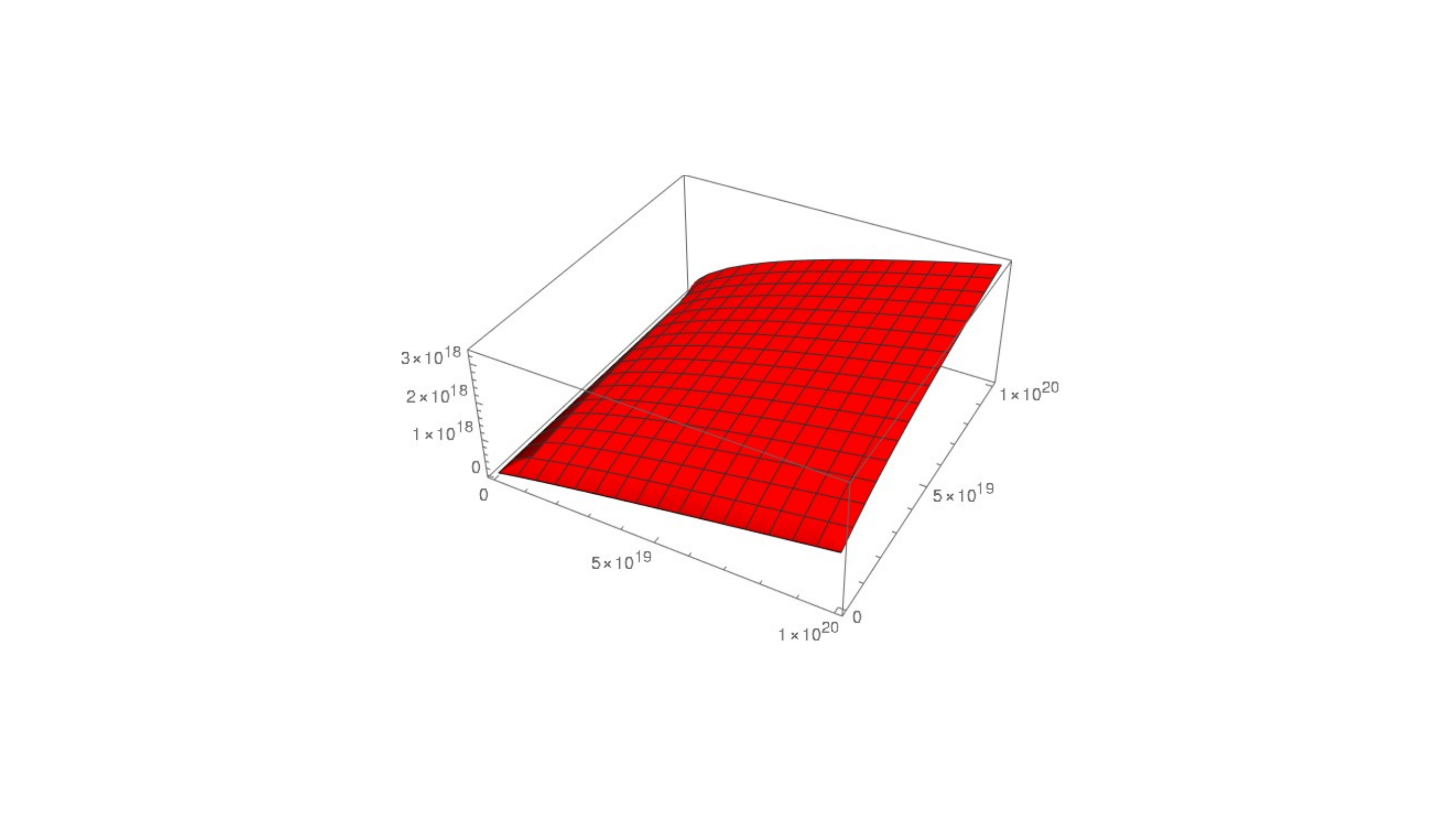}
\includegraphics[width=.32\textwidth]{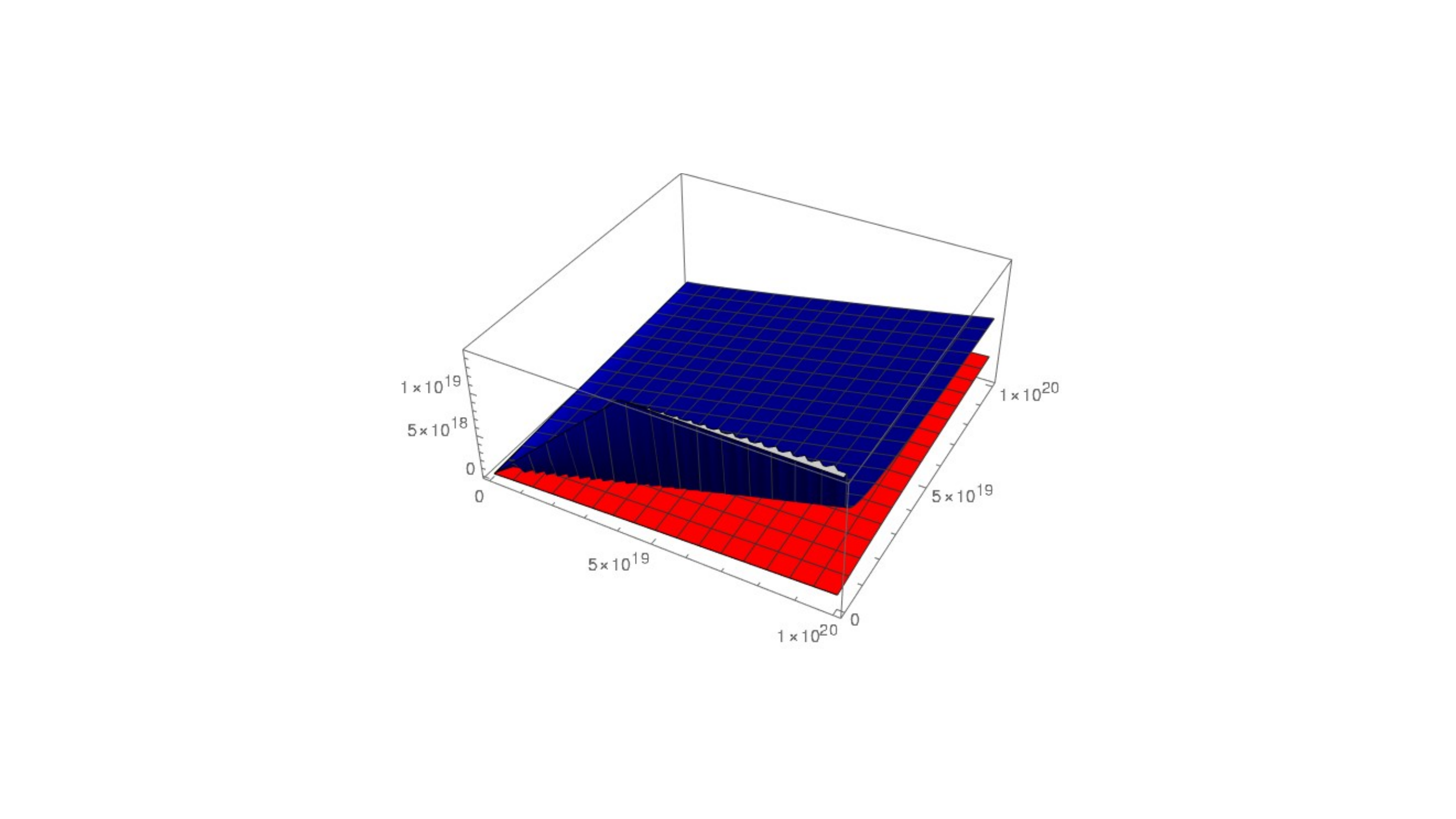}
\includegraphics[width=.32\textwidth]{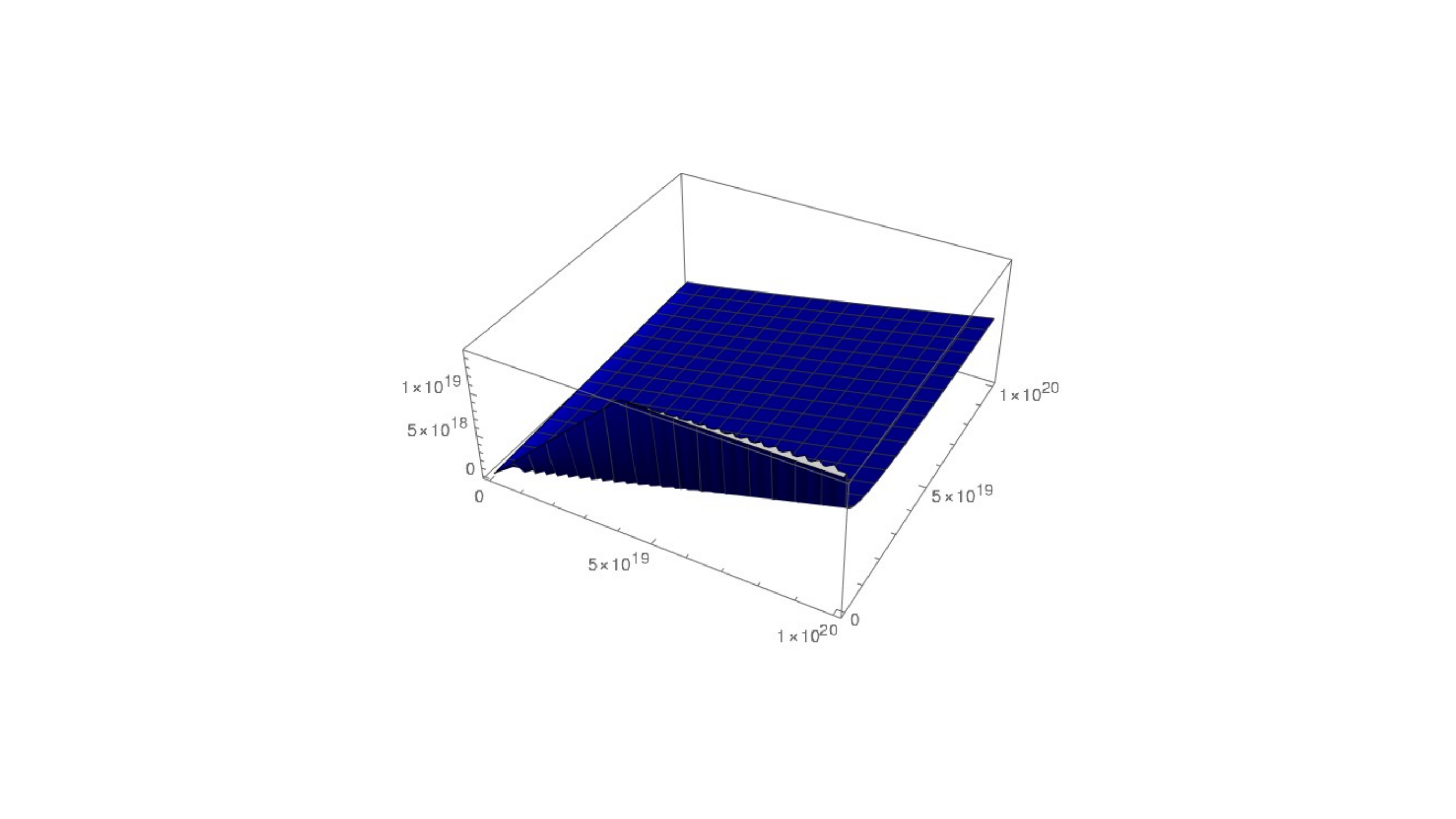}
\includegraphics[width=.32\textwidth]{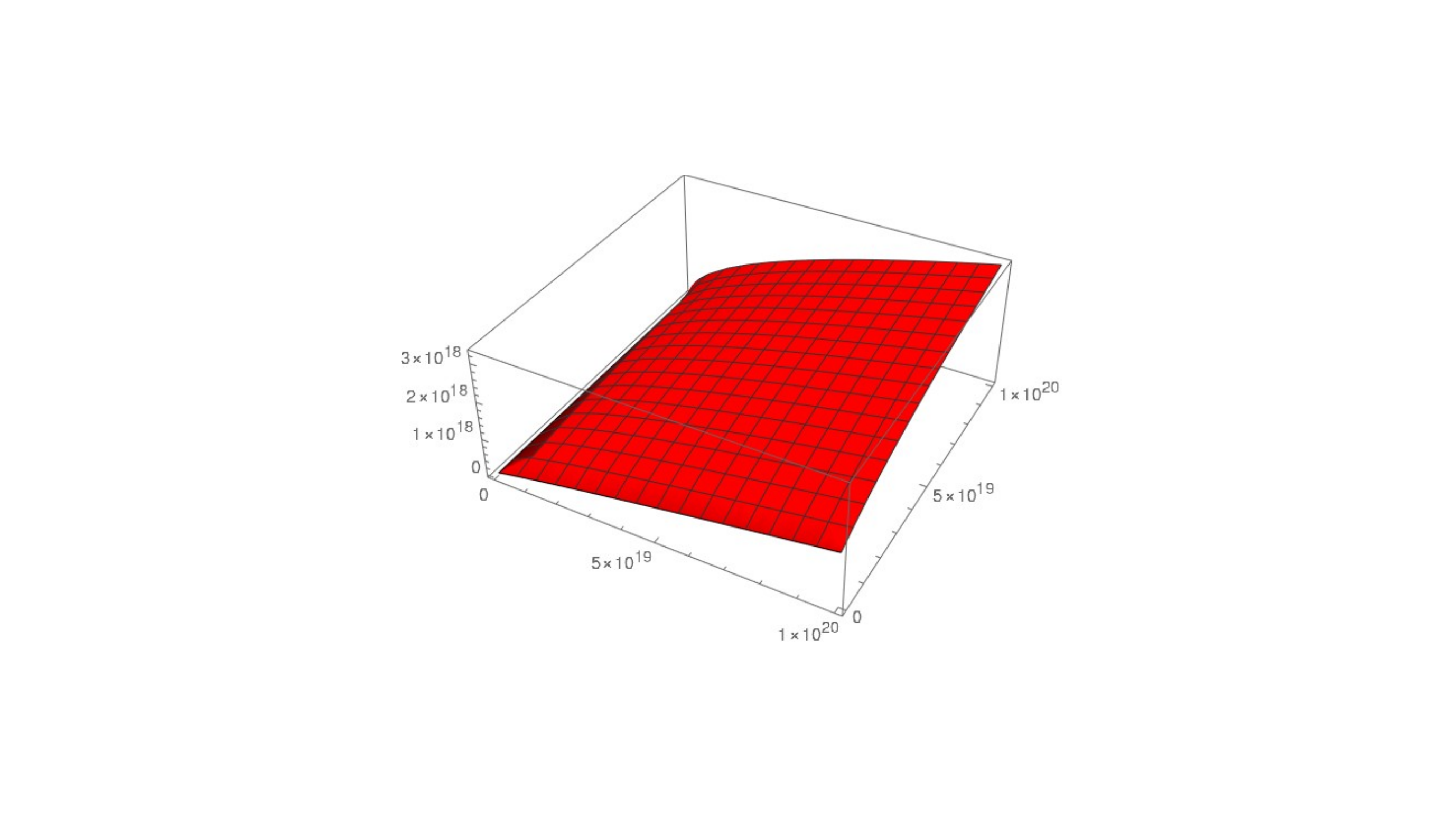}
\includegraphics[width=.32\textwidth]{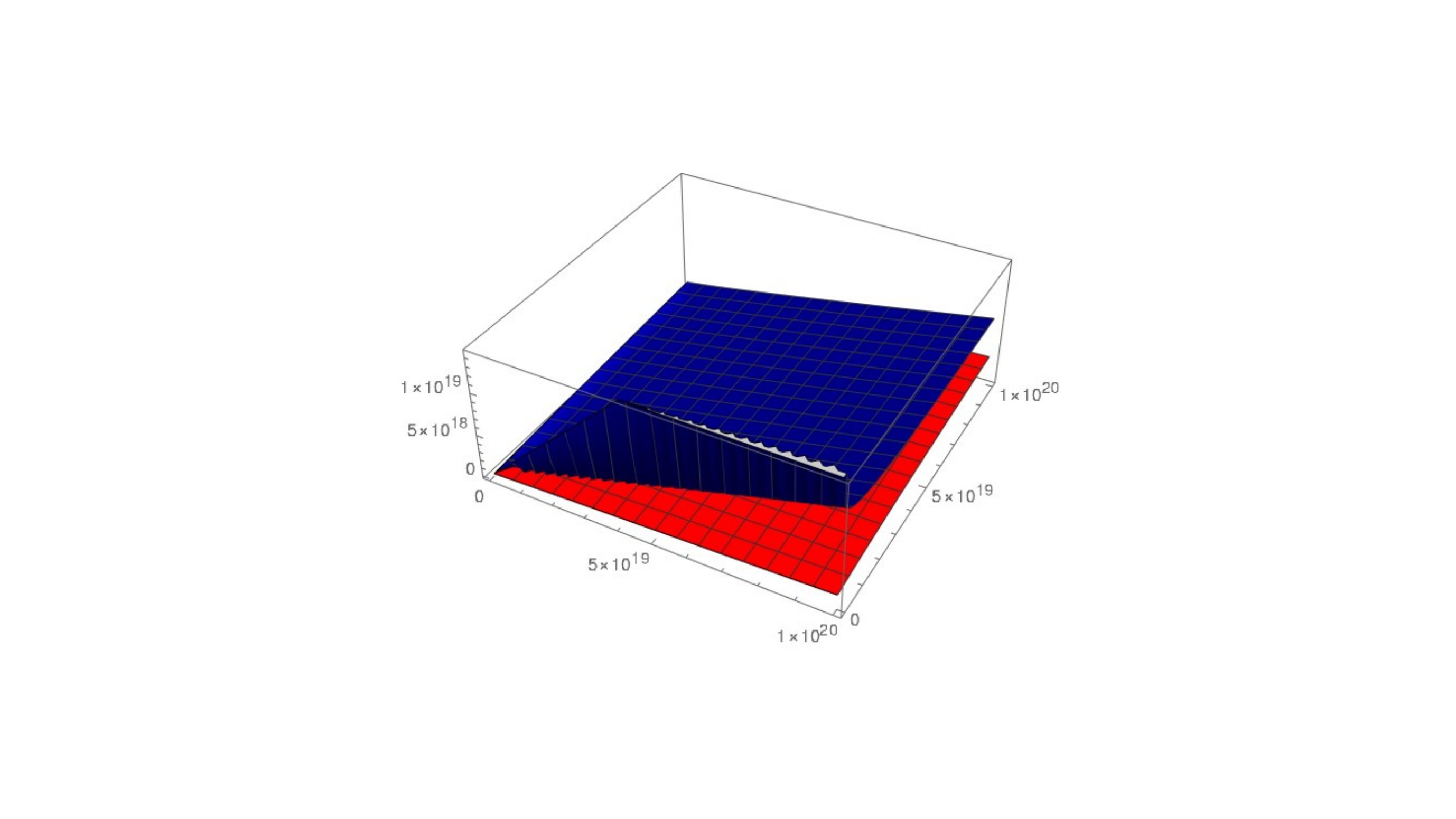}
\caption{First row \, ;\,\, x-axis\,:\,l; y-axis\,:\,$\rho_c$,  (left)  \,;\,l; Simplified expression of S as a function of $l\, \rho_c$\,; \, (Middle)\,\, : \,\,Exact expression of S as a function of $l,\,\rho_c$ ; (right) The overlap between the two : The figures are showing the two matches eonly in the regime $l  >> \rho_c$
\quad;\quad
Last row\,:\, All the above with x-axis and y-axis interchanged}
\la{lgreatrhoc1by12l1 }
\end{figure}

\section{  Simplified expression of holographic entanglement of entropy for $\rho_c>> l $   $d- \theta \ne 1$ and $d - \theta < 1$ }

Here we recall the  expression of the holographic entanglement of entropy as given in (\ref{entropy}),  where we would like to have the simplified expression of  H.E.E in the regime ${\frac{\rho_c}{\rho_0}} \sim 1$.  However since the Hypergeometric function is not Taylor-expandable around this point, so we use rhe identity\footnote{I would like to thank F. Omidi  for explaining this point to me}

\ber
{}_2 F_1 \bigg[a,b,c,z \bigg] &=& A  {}_2 F_1 \Big[a,b,1+a+b-c , 1-z \Big]
\cr && \cr 
&& +B (1-z)^{c-a-b} \, {}_2 F_1 \Bigg[ c-a , c-b, 1+c-a-b , 1-z \Bigg]\,
\la{identitity}
\eer

with   A, B,  given by

\ber
A &=& {\frac{\Gamma(c) \Gamma(c-a-b)}{\Gamma(c-a) \Gamma(c-b)}}\n 
B &=& {\frac{\Gamma(c) \Gamma(a+b-c)}{\Gamma(a)\Gamma(b)}}
\la{AB}
\eer 

So we have the Hypergeometric function, as involed in the expression of HEE is 

\ber
& &{{}_2 F_1} 
  \left\lbrack {\frac{1}{2}}, {\frac{1}{2}}( - 1 +{ \frac{1}{d - \theta}}), {\frac{1}{2}}(1 +{ \frac{1}{d - \theta}}) , \left({\frac{\rho_c}{\rho_0}}\right)^{2(d - \theta)}  \right \rbrack\n
&=&{\frac{\Gamma\left\lbrack {\frac{1}{2}}(1 + { \frac{1}{d - \theta}})   \right\rbrack \Gamma\left\lbrack {\frac{1}{2}} \right\rbrack}
{\Gamma\left\lbrack {\frac{1}{2(d - \theta)}} \right\rbrack \Gamma \left\lbrack 1 \right\rbrack}} {}_2 F_1 \left\lbrack {\frac{1}{2}}, {\frac{1}{2}}( - 1 +{ \frac{1}{d - \theta}}), {\frac{1}{2}},   1 - {\left({\frac{\rho_c}{\rho_0}}\right)}^{2(d - \theta)}\right\rbrack\n
&+& {\left( 1 - {\left({\frac{\rho_c}{\rho_0}}\right)}^{2(d - \theta)} \right)}^{\frac{1}{2}} {\frac{\Gamma\left\lbrack {\frac{1}{2}}(1 + { \frac{1}{d - \theta}})   \right\rbrack \Gamma\left\lbrack - {\frac{1}{2}} \right\rbrack}
{\Gamma\left\lbrack {\frac{1}{2}} \right\rbrack \Gamma \left\lbrack  {\frac{1}{2}}( - 1 +{ \frac{1}{d - \theta}}),   \right\rbrack}}    {}_2 F_1 \left\lbrack {\frac{1}{2( d - \theta)}}, 1, {\frac{3}{2}},   1 - {\left({\frac{\rho_c}{\rho_0}}\right)}^{2(d - \theta)}\right\rbrack\n
&=& {\frac{\Gamma\left\lbrack {\frac{1}{2}}(1 + { \frac{1}{d - \theta}})   \right\rbrack \Gamma\left\lbrack {\frac{1}{2}} \right\rbrack}
{\Gamma\left\lbrack  {\frac{1}{2(d - \theta)}}  \right\rbrack \Gamma \left\lbrack 1 \right\rbrack}} \times\n
& &  \left\lbrack  1 + {\frac{1}{2}}\left(- 1 +  { \frac{1}{  (d - \theta)}}\right)
{\left( 1 - {\left({\frac{\rho_c}{\rho_0}}\right)}^{2(d - \theta)}\right)}
+ {\frac{1}{8}}{\left( 1 +   { \frac{1}{  (d - \theta)} } \right)}{\left( -1 +   { \frac{1}{  (d - \theta)} } \right)} {\left( 1 - {\left({\frac{\rho_c}{\rho_0}}\right)}^{2(d - \theta)} \right)}^2\right\rbrack \n
&+&{\left( 1 - {\left({\frac{\rho_c}{\rho_0}}\right)}^{2(d - \theta)}\right)}^{\frac{1}{2}}     {\frac{\Gamma\left\lbrack {\frac{1}{2}}(1 + { \frac{1}{d - \theta}})   \right\rbrack \Gamma\left\lbrack - {\frac{1}{2}} \right\rbrack}
{\Gamma\left\lbrack {\frac{1}{2}} \right\rbrack \Gamma \left\lbrack  {\frac{1}{2}}( - 1 +{ \frac{1}{d - \theta}}),   \right\rbrack}}\times\n
& &\left\lbrack 1 + {\frac{1}{3(d - \theta)}}{\left( 1 - {\left({\frac{\rho_c}{\rho_0}}\right)}^{2(d - \theta)}\right)}  + {\frac{1}{6(d - \theta)}}\cdot{\frac{4 + {\frac{2}{(d - \theta)}}   }{5}}{\left( 1 - {\left({\frac{\rho_c}{\rho_0}}\right)}^{2(d - \theta)}\right)}^2 \right\rbrack\n
\la{promise}
\eer

Substituting the above in the expression of Holographic entanglement of entropy we get

\ber
      S &=&   {\frac{      L^{d-1} (\rho_0)^{\theta - d +1} \,\, {{}_2 F_1}   \left\lbrack {\frac{1}{2}}, {\frac{1}{2}}( - 1 +{ \frac{1}{d - \theta }}), {\frac{1}{2}}(1 +{ \frac{1}{d - \theta }}) , 1 \right \rbrack }{  4 G_N (\theta + 1- d )    }}\n
&-& {\left({\frac{\rho_c}{\rho_0}}\right)}^{ \theta +1- d} {\frac{     L^{d-1} (\rho_0)^{\theta - d +1}  }{  4 G_N( \theta - d  + 1)   }}\times\n
& & \left\lbrack { 1 + {\frac{1}{2}}\left(- 1 +  {\frac{1}{  (d - \theta)}}\right)}
{\left( 1 - {\left({\frac{\rho_c}{\rho_0}}\right)}^{2(d - \theta)}\right)}
+ {\frac{1}{8}}{\left( 1 +   { \frac{1}{  (d - \theta)} } \right)}{\left( -1 +   { \frac{1}{  (d - \theta)} } \right)} {\left( 1 - {\left({\frac{\rho_c}{\rho_0}}\right)}^{2(d - \theta)} \right)}^2   \right\rbrack  \n
&-& {\left({\frac{\rho_c}{\rho_0}}\right)}^{ \theta +1- d} {\frac{     L^{d-1} (\rho_0)^{\theta - d +1}  }{  4 G_N( \theta - d  + 1)   }}\times\n
& &{\left( 1 - {\left({\frac{\rho_c}{\rho_0}}\right)}^{2(d - \theta)}\right)}^{\frac{1}{2}}     {\frac{\Gamma\left\lbrack {\frac{1}{2}}(1 + { \frac{1}{d - \theta}})   \right\rbrack \Gamma\left\lbrack - {\frac{1}{2}} \right\rbrack}
{\Gamma\left\lbrack {\frac{1}{2}} \right\rbrack \Gamma \left\lbrack  {\frac{1}{2}}( - 1 +{ \frac{1}{d - \theta}}),   \right\rbrack}}\times\n
& &\left\lbrack 1 + {\frac{1}{3(d - \theta)}}{\left( 1 - {\left({\frac{\rho_c}{\rho_0}}\right)}^{2(d - \theta)}\right)}  + {\frac{1}{6(d - \theta)}}\cdot{\frac{4 + {\frac{2}{(d - \theta)}}   }{5}}{\left( 1 - {\left({\frac{\rho_c}{\rho_0}}\right)}^{2(d - \theta)}\right)}^2 \right\rbrack\n
\la{rewr}
\eer

Now to simplify further,  we recall  (\ref{ultimaterho0nonvanishing}) and write the simplified expression of $\rho_0(l, \rho_c)$ in $\rho_c >> l$ regime

\ber
{\rm for} \, \, \, d - \theta > 1\,\,\,\, \rho_{0} &=&  {\left( {\left({\frac{A_{10} l}{ 2 }}\right)}^{2  \left( d - \theta\right)} + {\left( \rho_c \right)}^{2  \left( d - \theta\right)} \right)}^{\frac{1}{2(d - \theta)}}\n
                                                   &=& {\left( \rho_c \right)} {\left(  1  +  {\frac{1}{ 2(d - \theta )}}  {\frac{   {\left({\frac{A_{10} l}{ 2 }}\right)}^{ 2 \left( d - \theta\right)} }{ {\left( \rho_c \right)}^{ { 2\left( d - \theta\right)} } }}  \right)}\n    
\la{purpose1}
\eer

\ber
{\rm for} \, \, \, d - \theta < 1\,\, , \,\, \rho_{0} &=&  {\left( {\left({\frac{A_{10} l}{ 2 }}\right)}^{  \left( d - \theta + 1 \right)} + {\left( \rho_c \right)}^{  \left( d - \theta + 1 \right)} \right)}^{\frac{1}{(d - \theta + 1)}}\n
 &=& {\left( \rho_c \right)} {\left(  1  +  {\frac{1}{(d - \theta  + 1)}}  {\frac{   {\left({\frac{A_{10} l}{ 2 }}\right)}^{  \left( d - \theta  + 1\right)} }{ {\left( \rho_c \right)}^{ { \left( d - \theta  + 1\right)} } } }  \right)}\n                                               
\la{purpose2}
\eer

Substituting (\ref{purpose1})  and (\ref{purpose2}) in (\ref{entropy}), one gets the simplified expression of HEE in the regime $\rho_c >> L$  for $d - \theta > 1$ and $d - \theta < 1$

Here we present the plots with the comparison between the exact expression and simplified expression for H.E.E for $\rho_c >> l$ .   The plots for $d - \theta > 1$ are given in Fig.( \ref{rhocgreatl5by2l1} , \ref{rhocgreatl8by3l1 } ).  The plots for  $d - \theta < 1$ are given in (\ref{rhocgreatl1by9l1 }, \ref{rhocgreatl1by12l1} )

\begin{figure}[H]
\begin{center}
\textbf{ For $d-\theta > 1$,  with  $ d - \theta = {\frac{5}{2}}$, \,\,:\,\,  Comparison   between the simplified expression of HEE for $ \rho_c >>  l $ with the exact expression of HEE in this regime, can be realized only in long range of   $l, \rho_c$, }
\end{center}
\includegraphics[width=.32\textwidth]{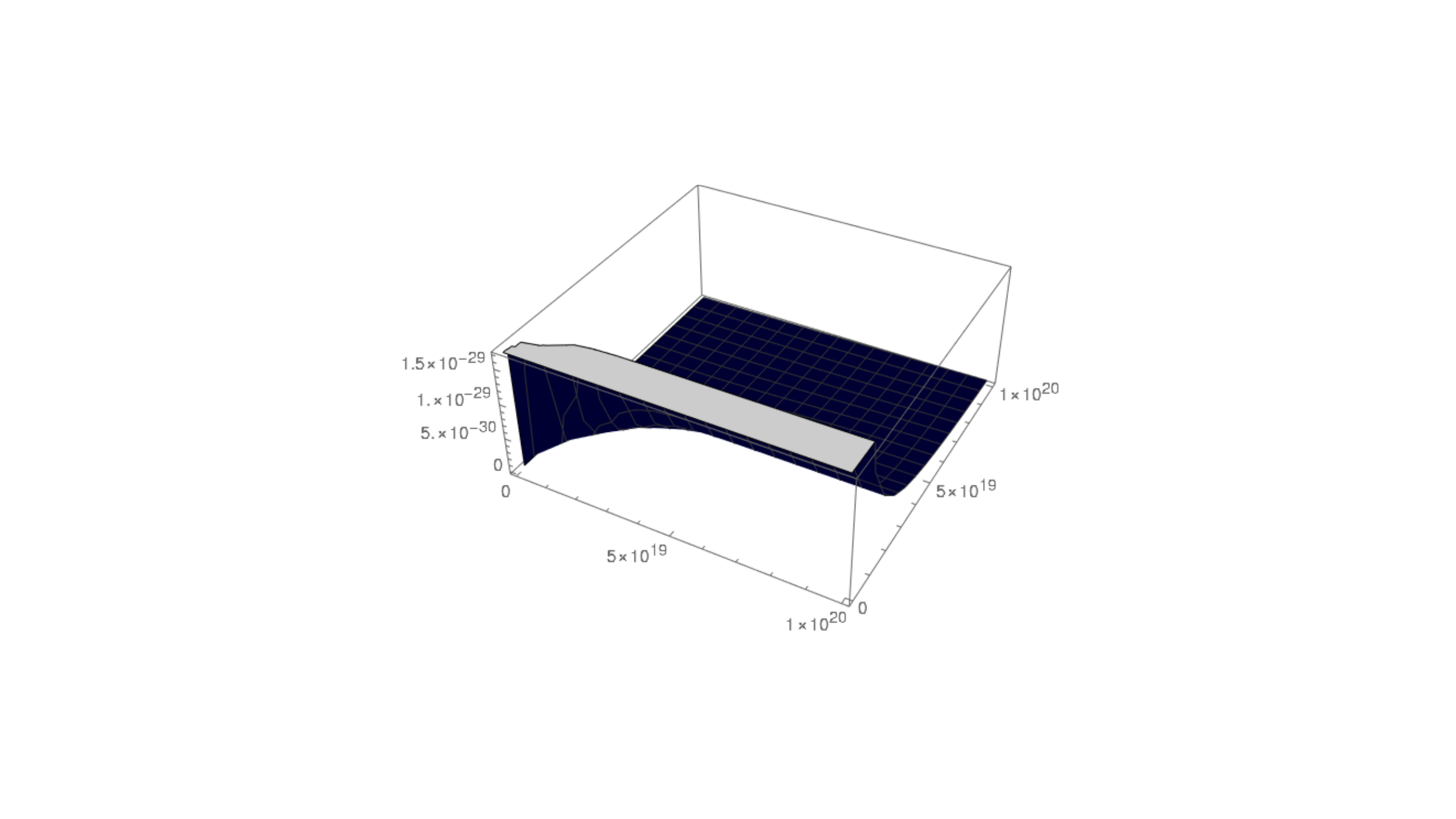}
\includegraphics[width=.32\textwidth]{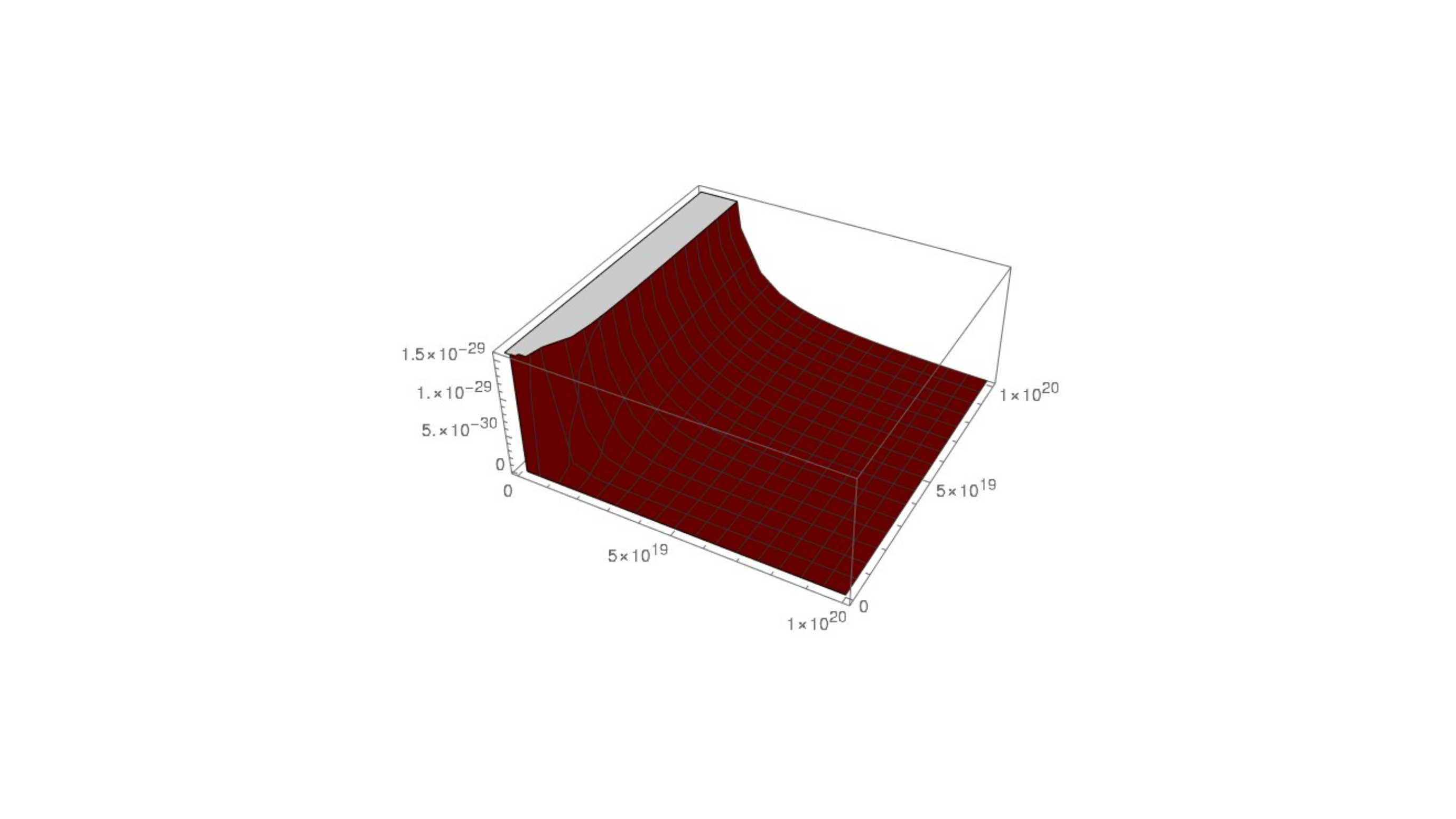}
\includegraphics[width=.32\textwidth]{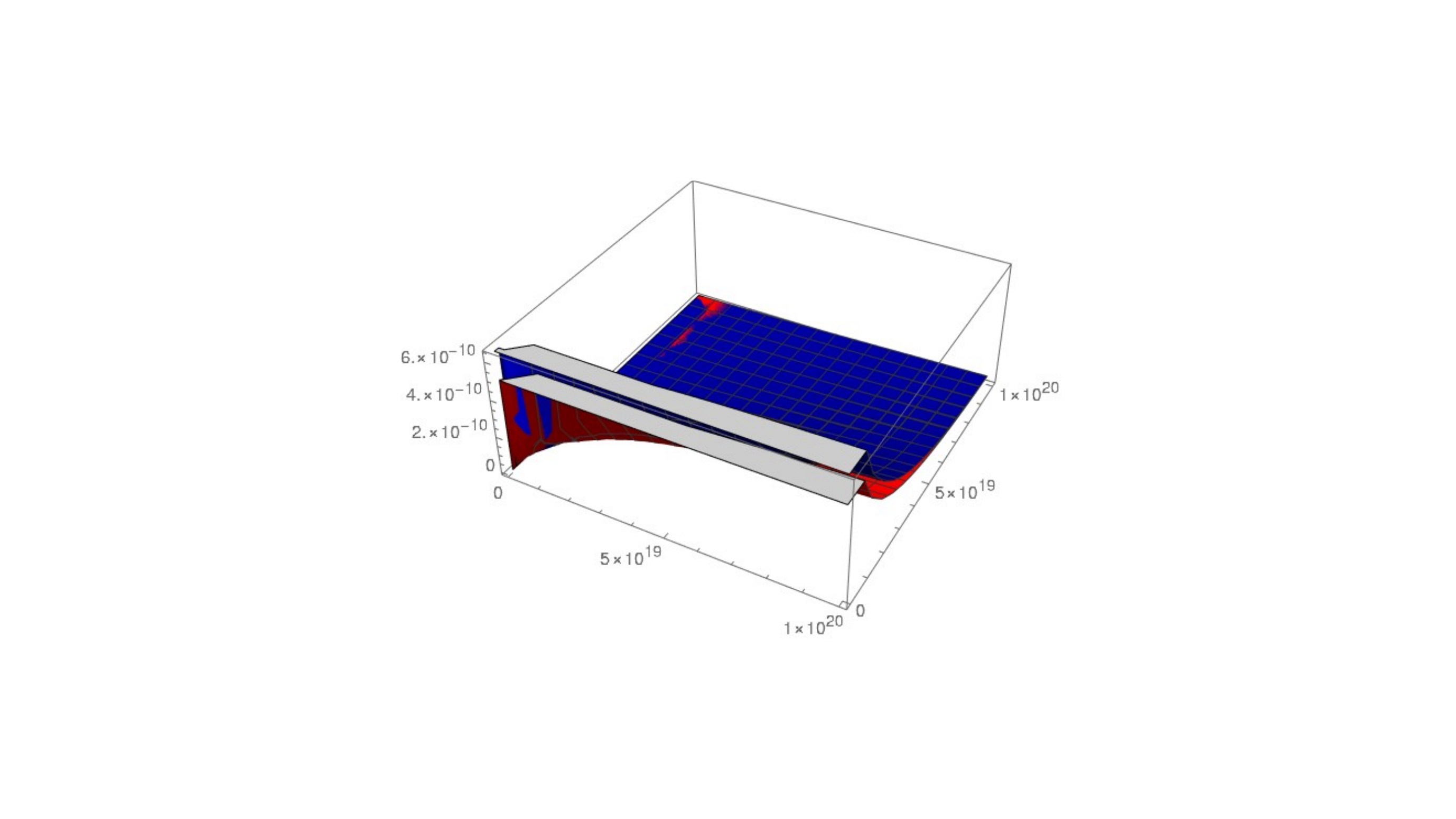}
\includegraphics[width=.32\textwidth]{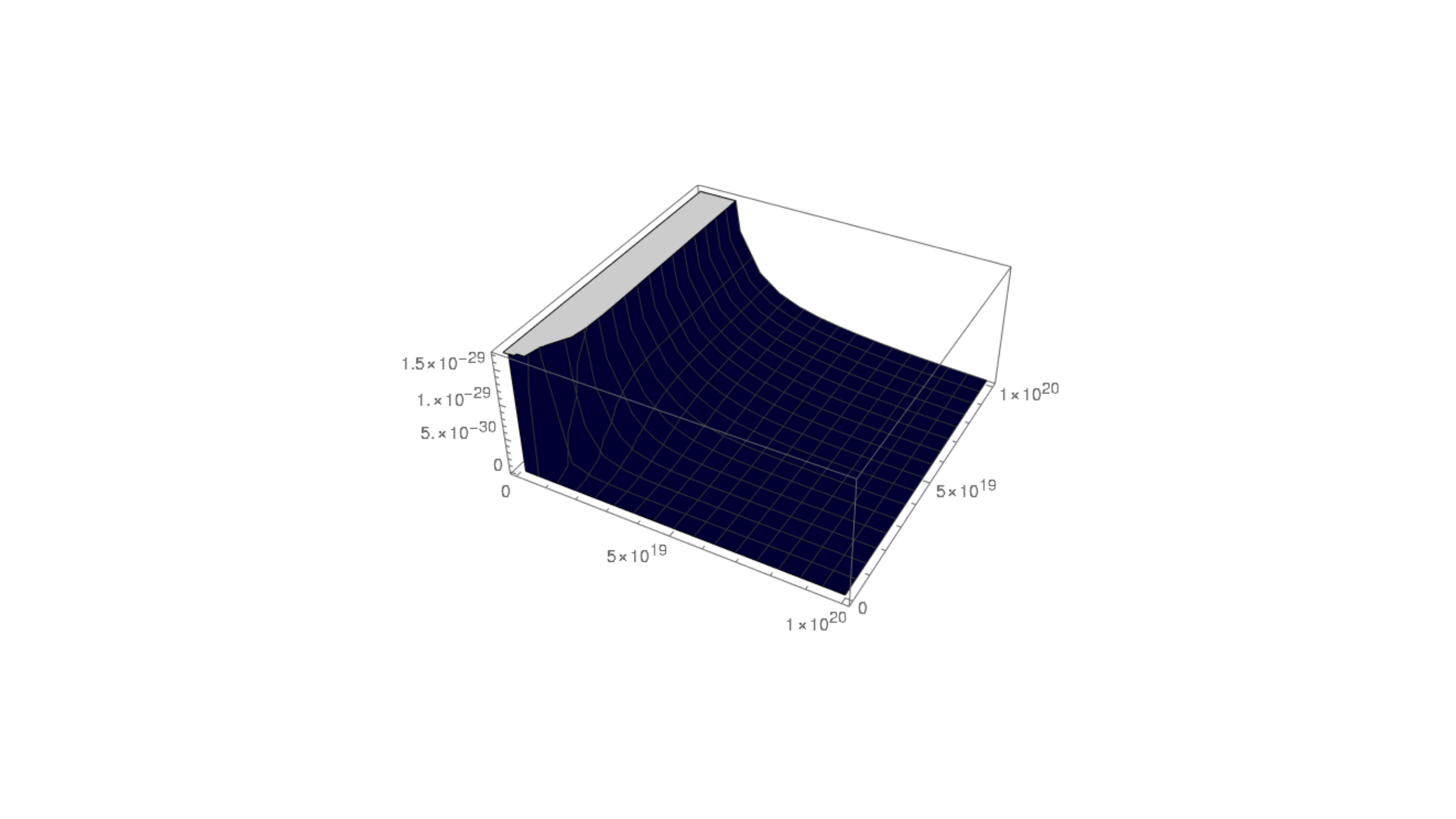}
\includegraphics[width=.32\textwidth]{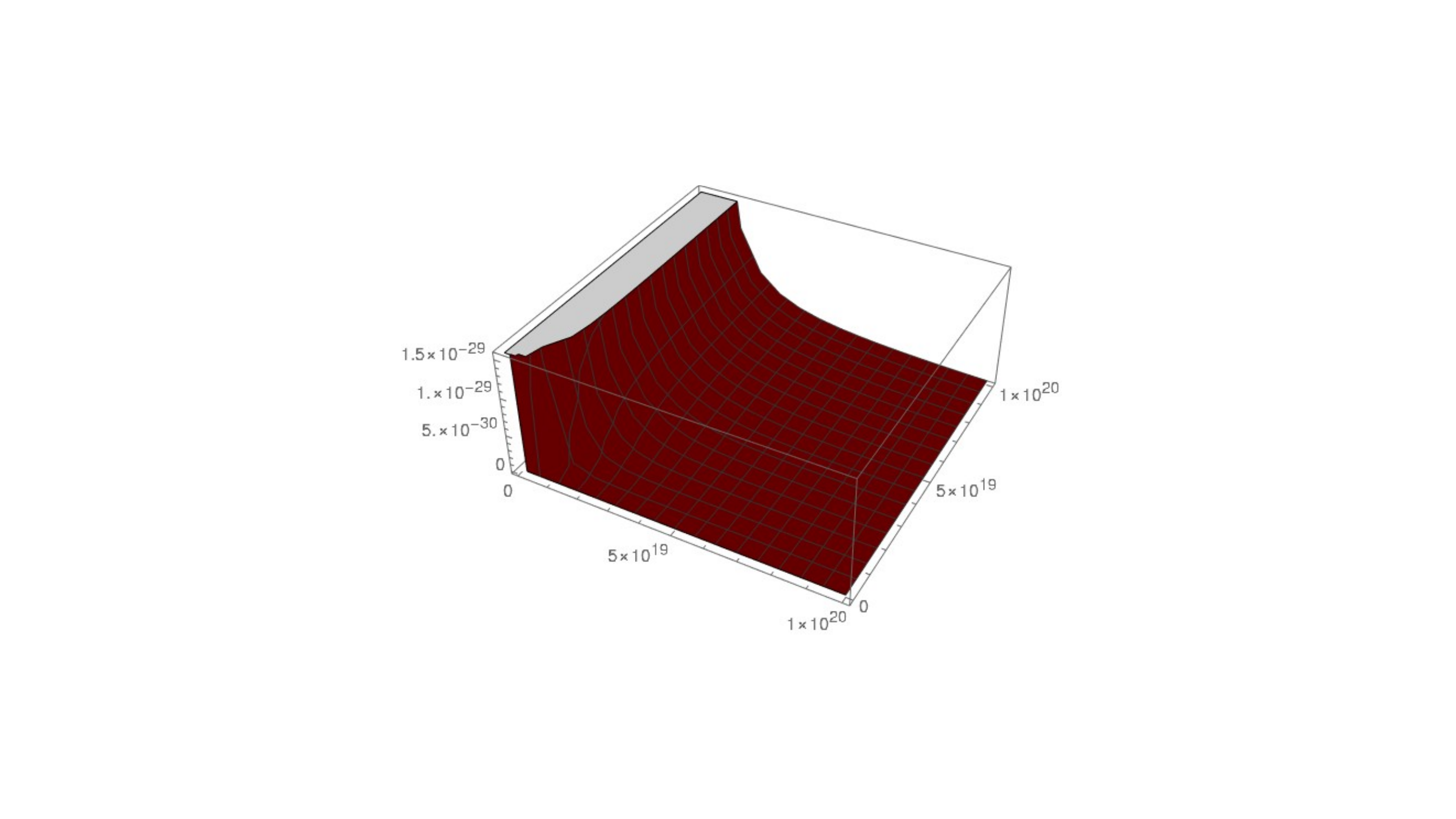}
\includegraphics[width=.32\textwidth]{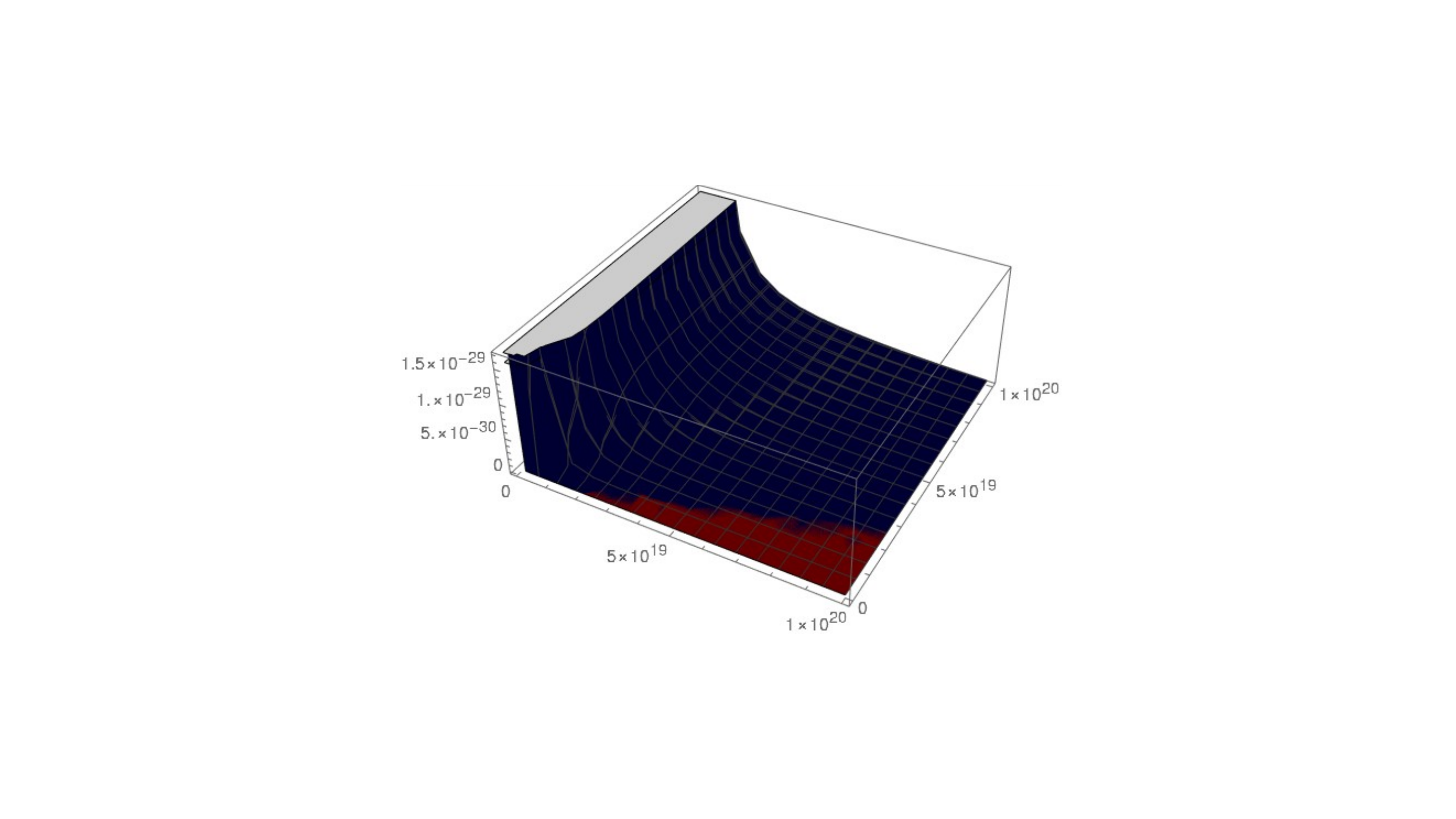}
\begin{center}
\includegraphics[width=.55\textwidth]{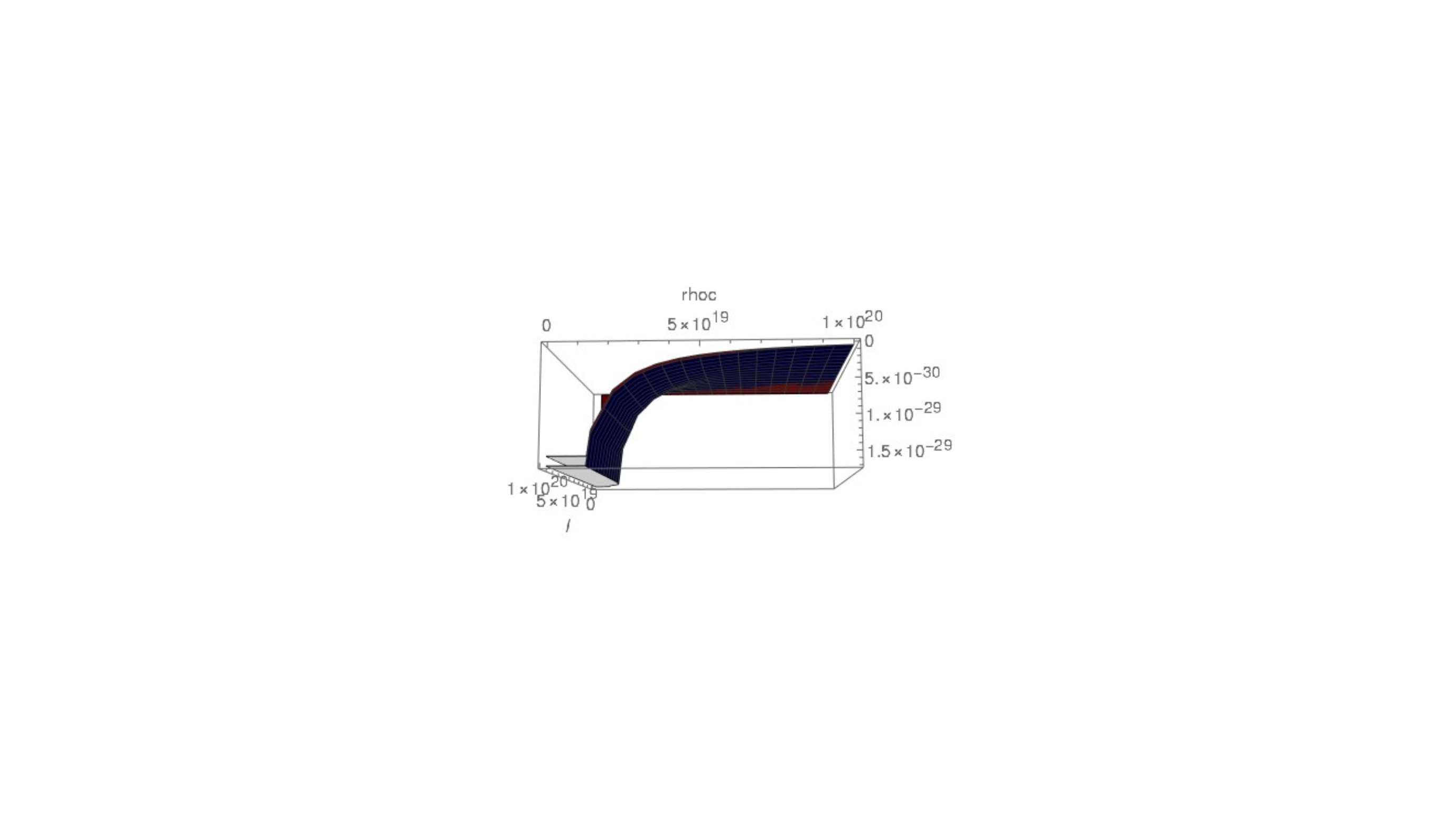}
\end{center}	
\caption{First row\,\,;\,\, x-axis\,:\,l; y-axis\,:\,$\rho_c$\,\,:\,\,  (left)  \,:\, Simplified expression of S as a function of $l\, ,\rho_c$  \,;\,\,\, (Middle)\,\, : \,\,Exact expression of S as a function of $l\, , \,\rho_c$\,\,\,;\,\,\, (right) The overlap between the two ; The figures are showing the two matches only in the regime $\rho_c  >> l$
\quad;\quad
Middle  row \,:\, All the above with x-axis and y-axis interchanged
\quad;\quad  Last row \,;\, Backside view of the overlap the simplified(given in blue) and exact expression(given in red) of HEE, showing clearly that  two merges in 
$\rho_c >> l$ regime only }
\la{rhocgreatl5by2l1}
\end{figure}

\begin{figure}[H]
\begin{center}
\textbf{ For $d-\theta > 1$,  with  $ d - \theta = {\frac{8}{3}}$, \,\,:\,\,  Comparison   between the simplified expression of HEE for $ \rho_c >>  l $ with the exact expression of HEE in this regime, can be realized only in long range of  $l, \rho_c$, }
\end{center}
\includegraphics[width=.32\textwidth]{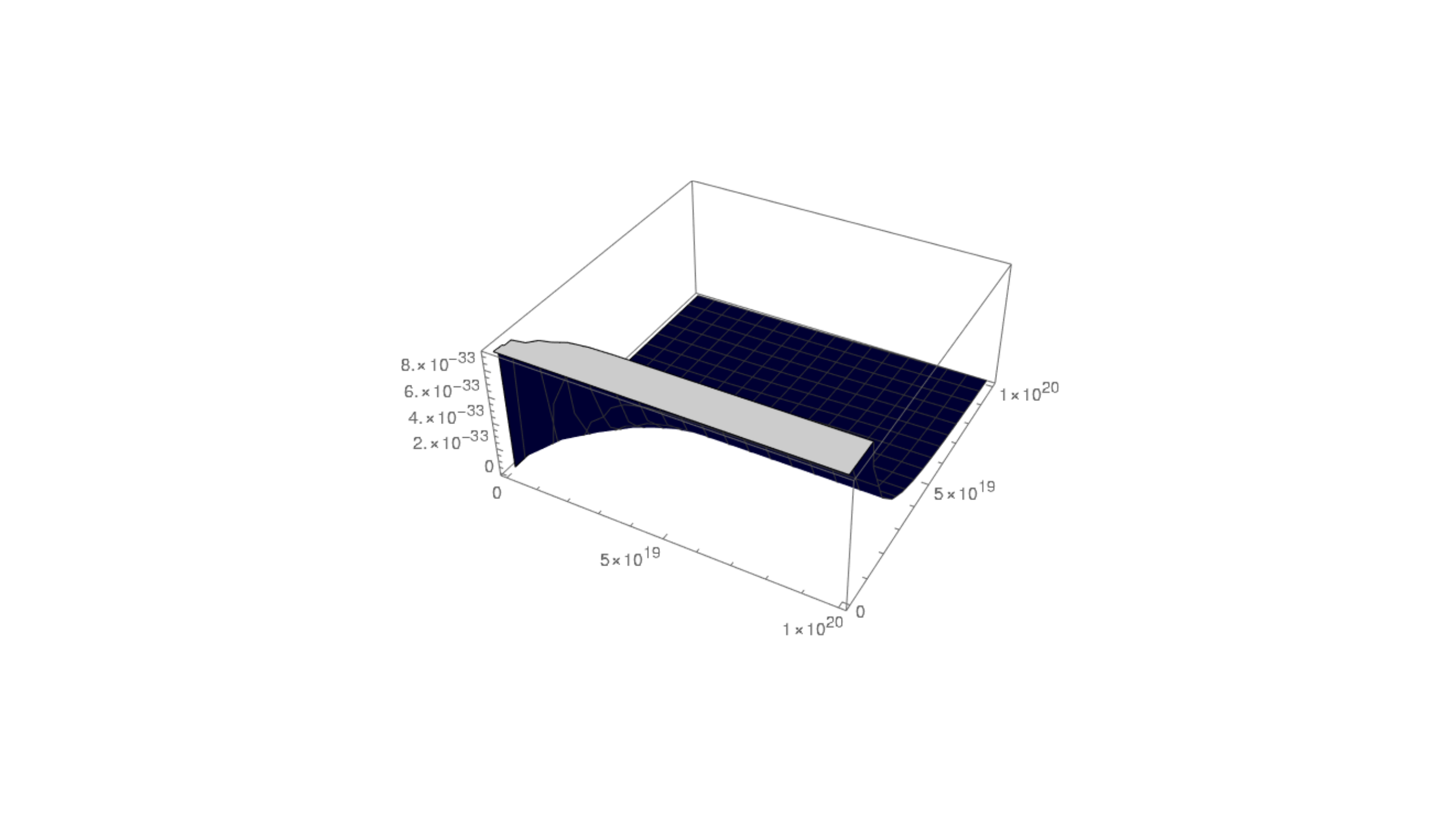}
\includegraphics[width=.32\textwidth]{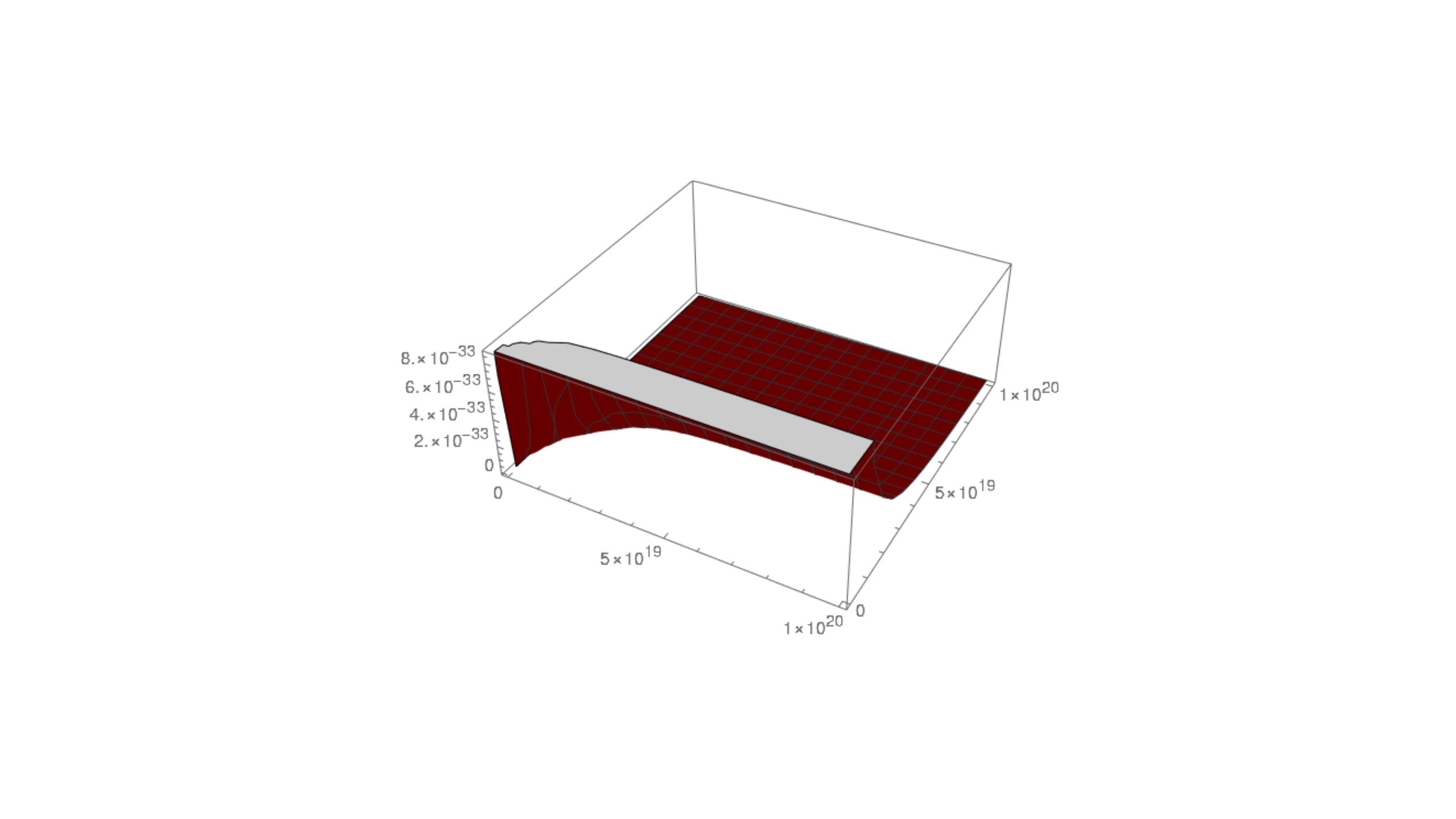}
\includegraphics[width=.32\textwidth]{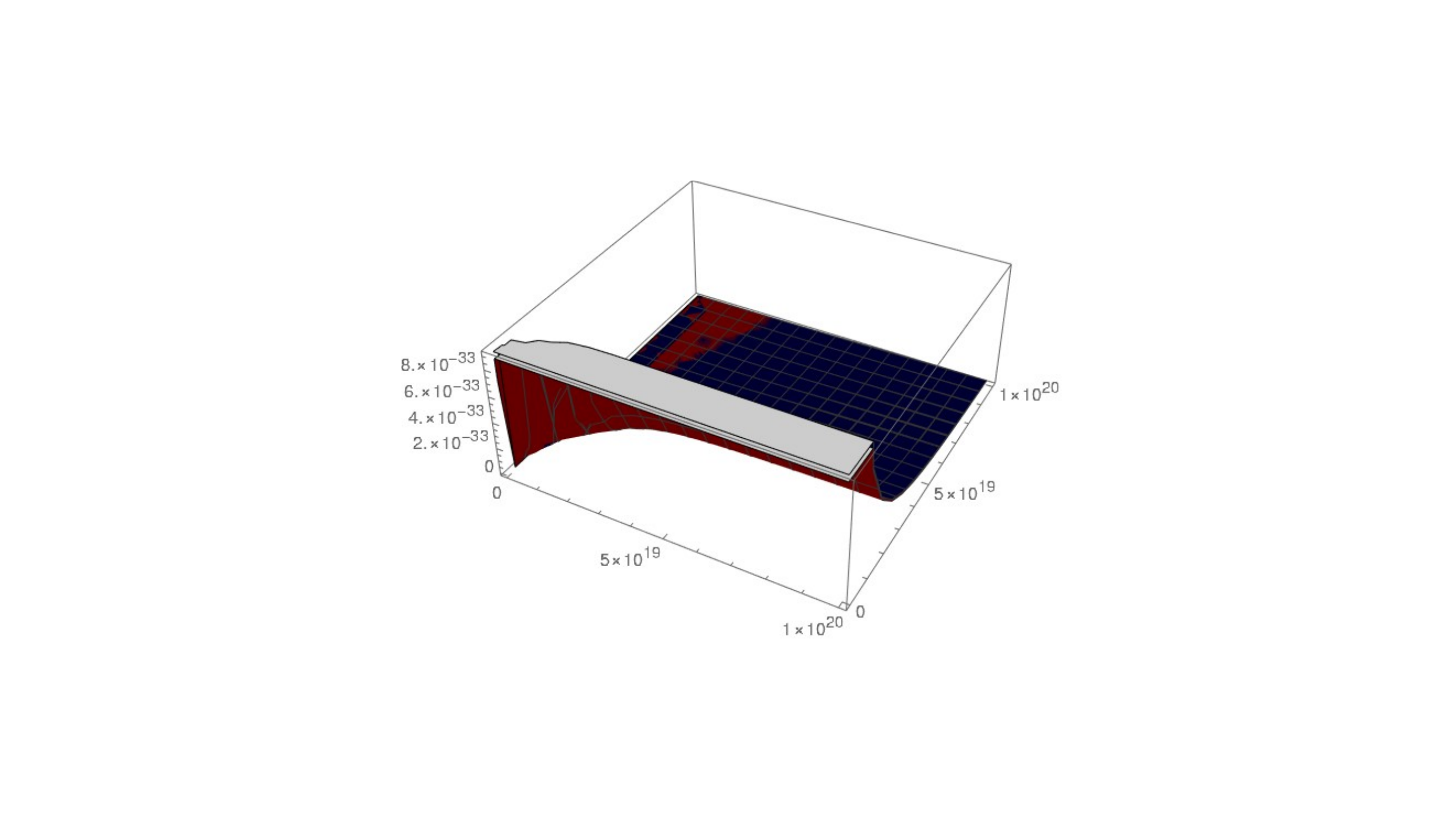}
\includegraphics[width=.32\textwidth]{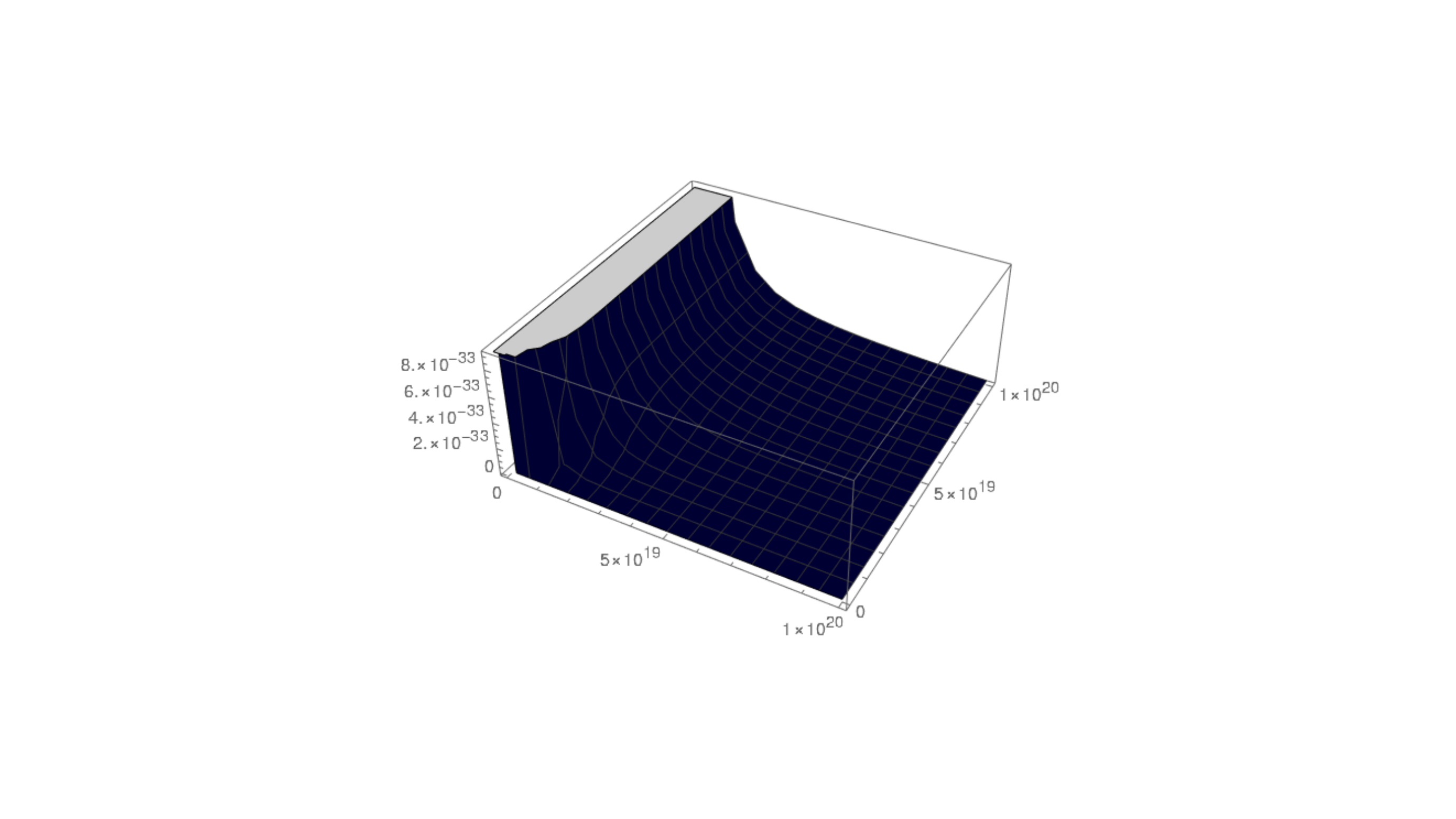}
\includegraphics[width=.32\textwidth]{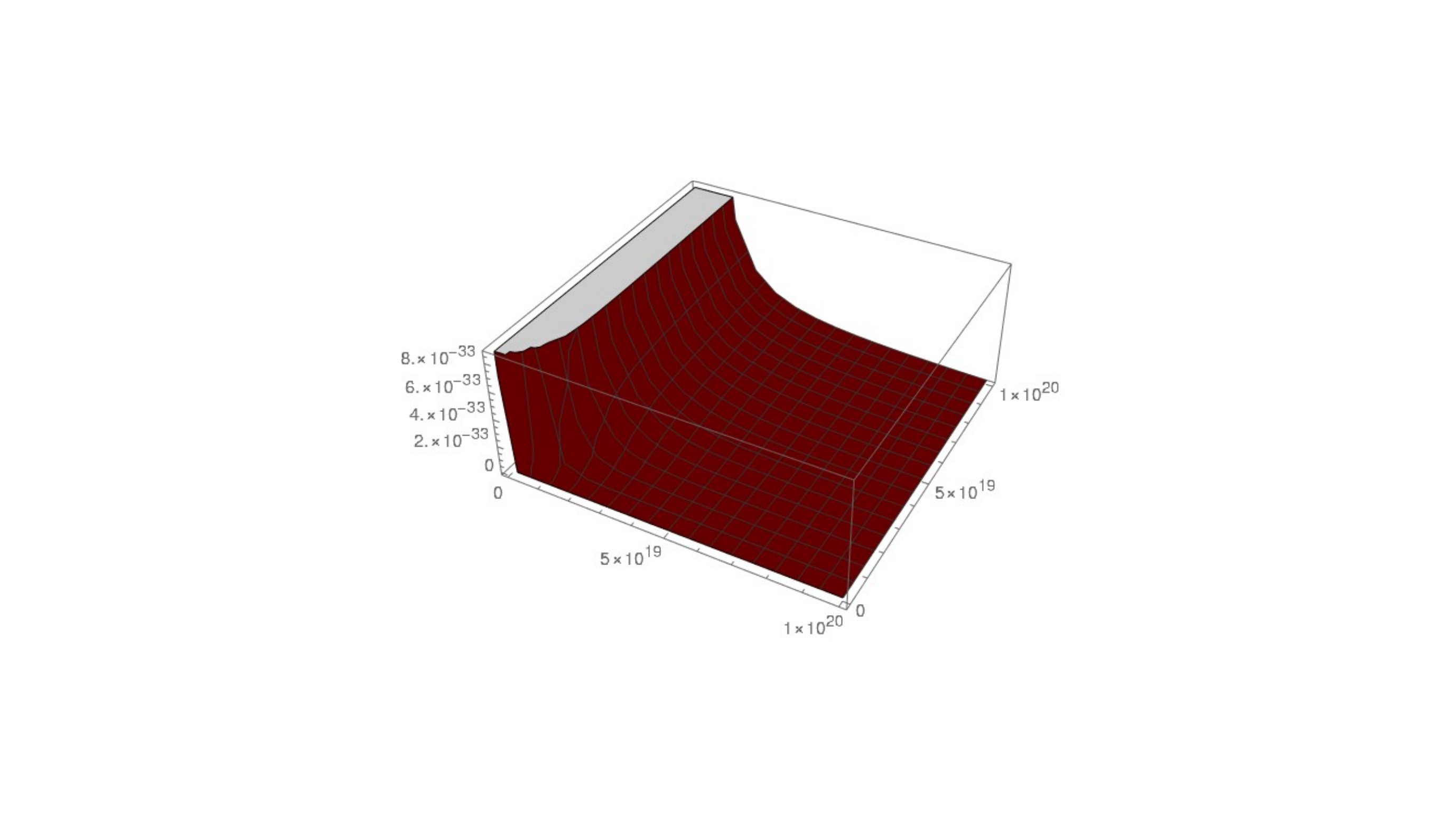}
\includegraphics[width=.32\textwidth]{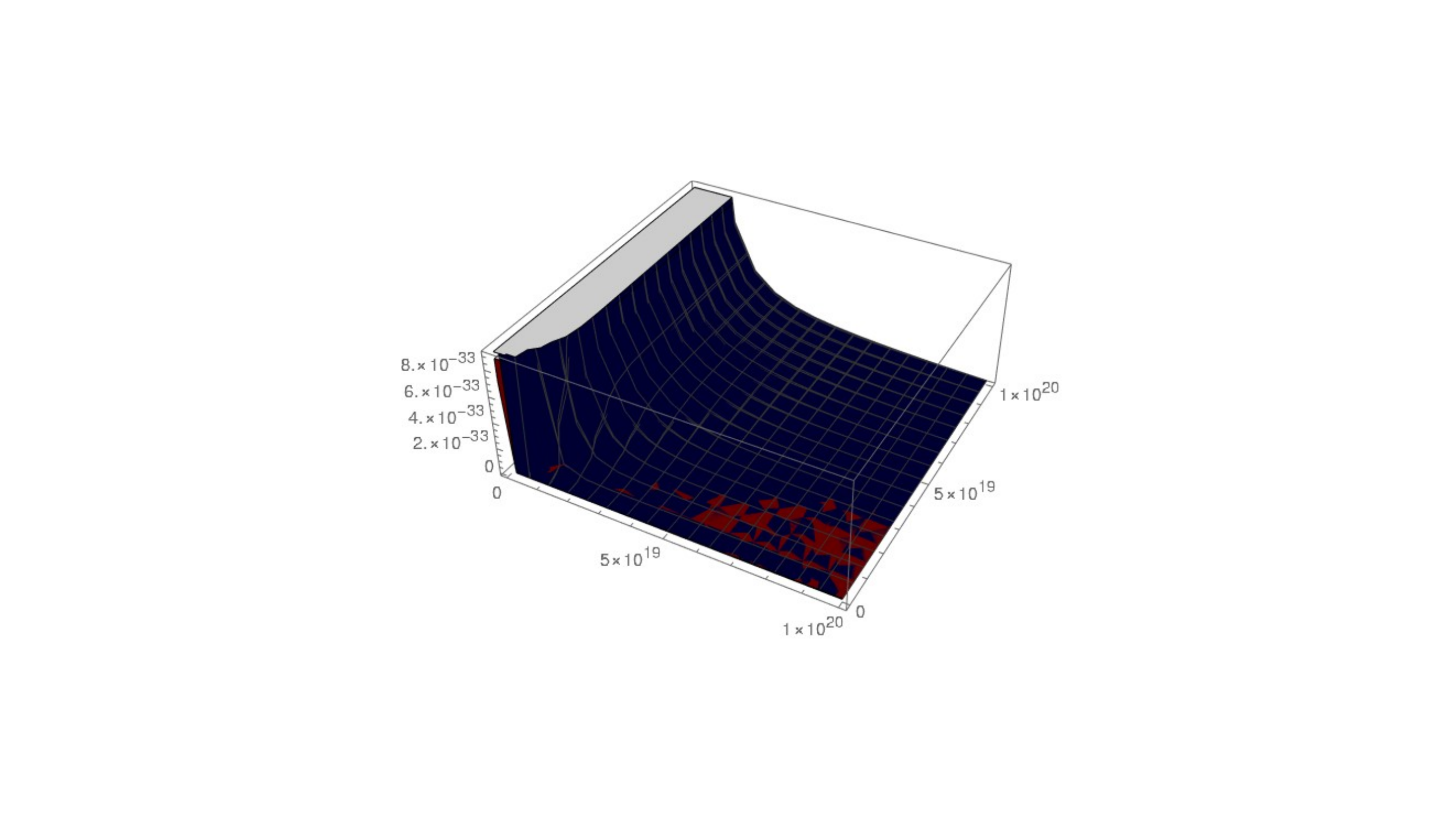}
\begin{center}
\includegraphics[width=.55\textwidth]{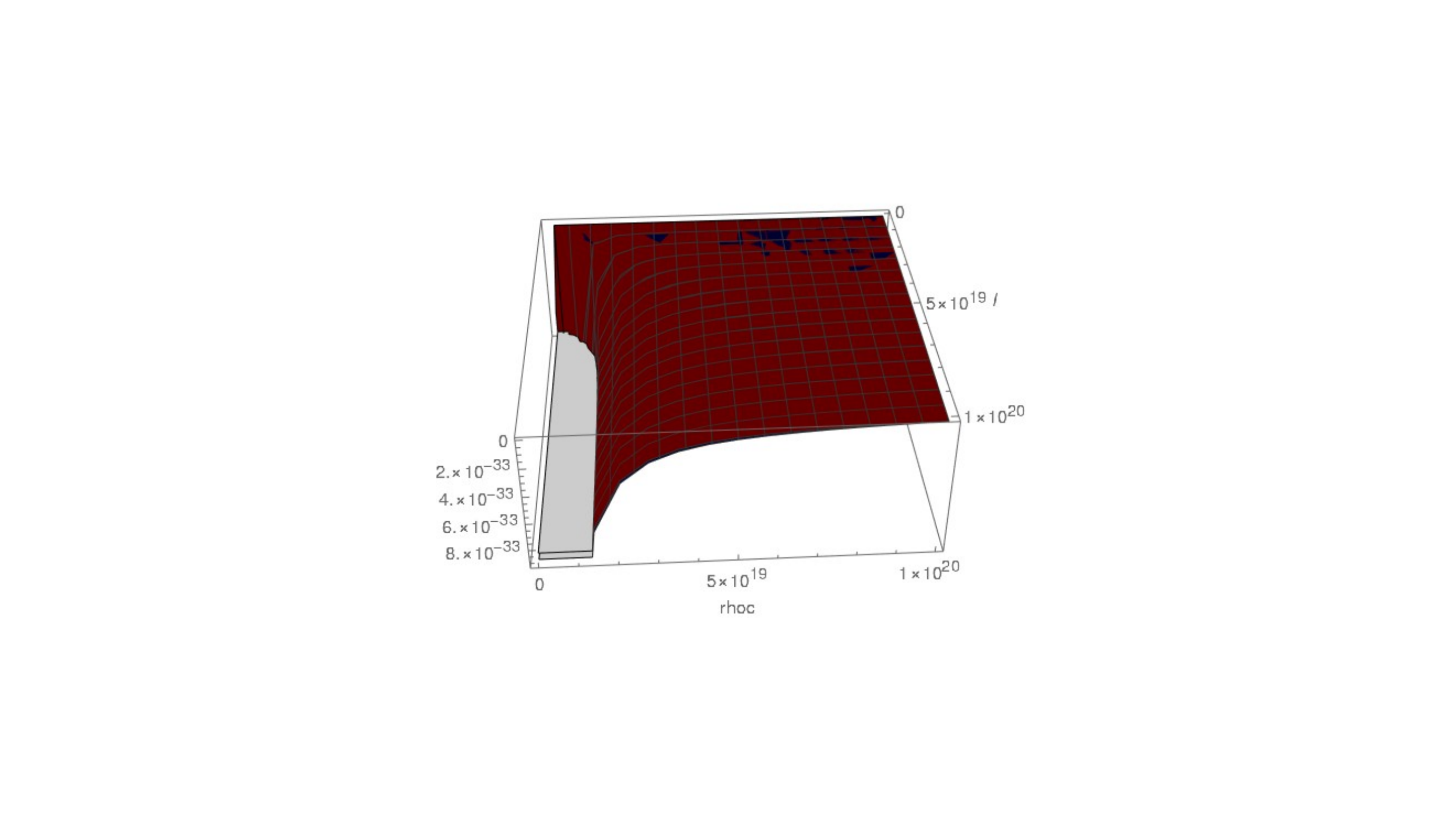}
\end{center}	
\caption{First row\,\,;\,\, x-axis\,:\,l; y-axis\,:\,$\rho_c$\,\,:\,\,  (left)  , Simplified expression of S as a function of $l,\rho_c$ ; (Middle)   \, , \, Exact expression of S as a function of $l\, .\,\rho_c$\,;\, (right) The overlap between the two  \, ;\, The figures are showing the two matches only in the regime $\rho_c  >> l$
\quad;\quad
Middle  row \,:\, All the above with x-axis and y-axis interchanged
\quad;\quad  Last row\,;\, Backside view of the overlap the simplified(given in blue) and exact expression(given in red) of HEE, showing clearly that  two merges in 
$\rho_c >> l$ regime only }
\la{rhocgreatl8by3l1 }
\end{figure}

\begin{figure}[H]
\begin{center}
\textbf{ For $d-\theta < 1$,  with  $ d - \theta = {\frac{1}{9}}$, \,\,:\,\,  Comparison   between the simplified expression of HEE for $ \rho_c >>  l $ with the exact expression of HEE in this regime, can be realized only in   long range of $l, \rho_c$, }
\end{center}
\includegraphics[width=.32\textwidth]{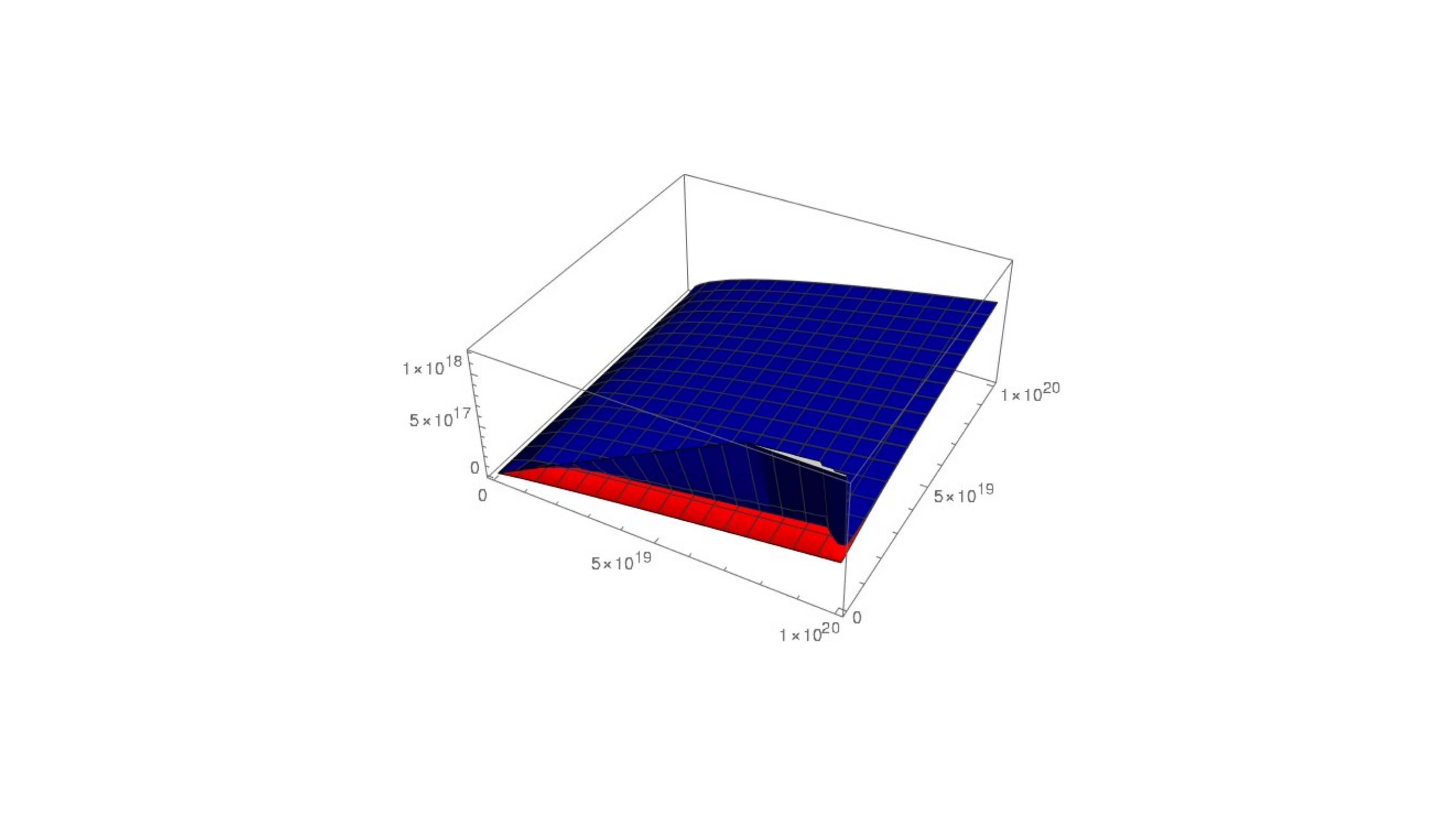}
\includegraphics[width=.32\textwidth]{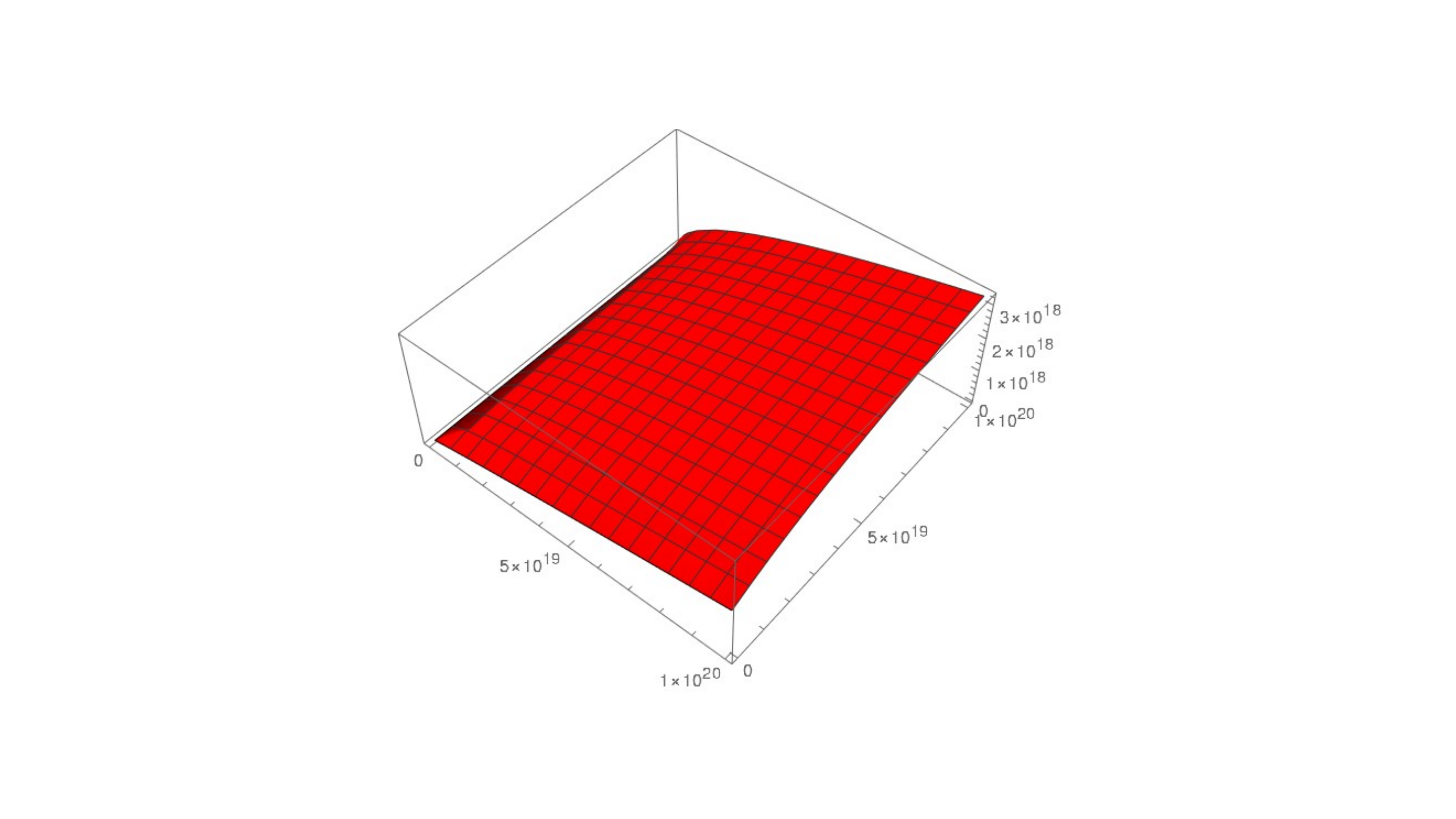}
\includegraphics[width=.32\textwidth]{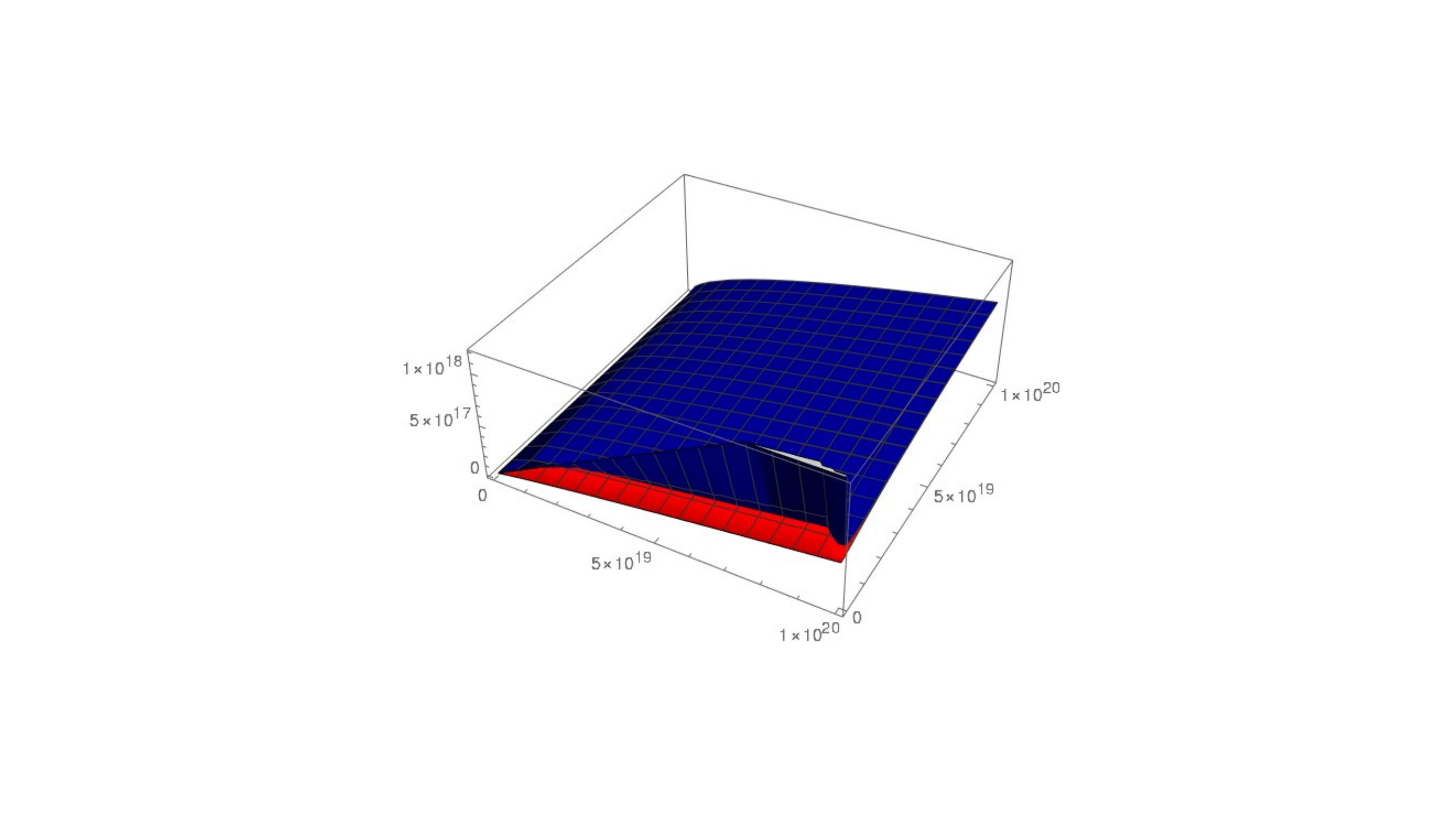}
\includegraphics[width=.32\textwidth]{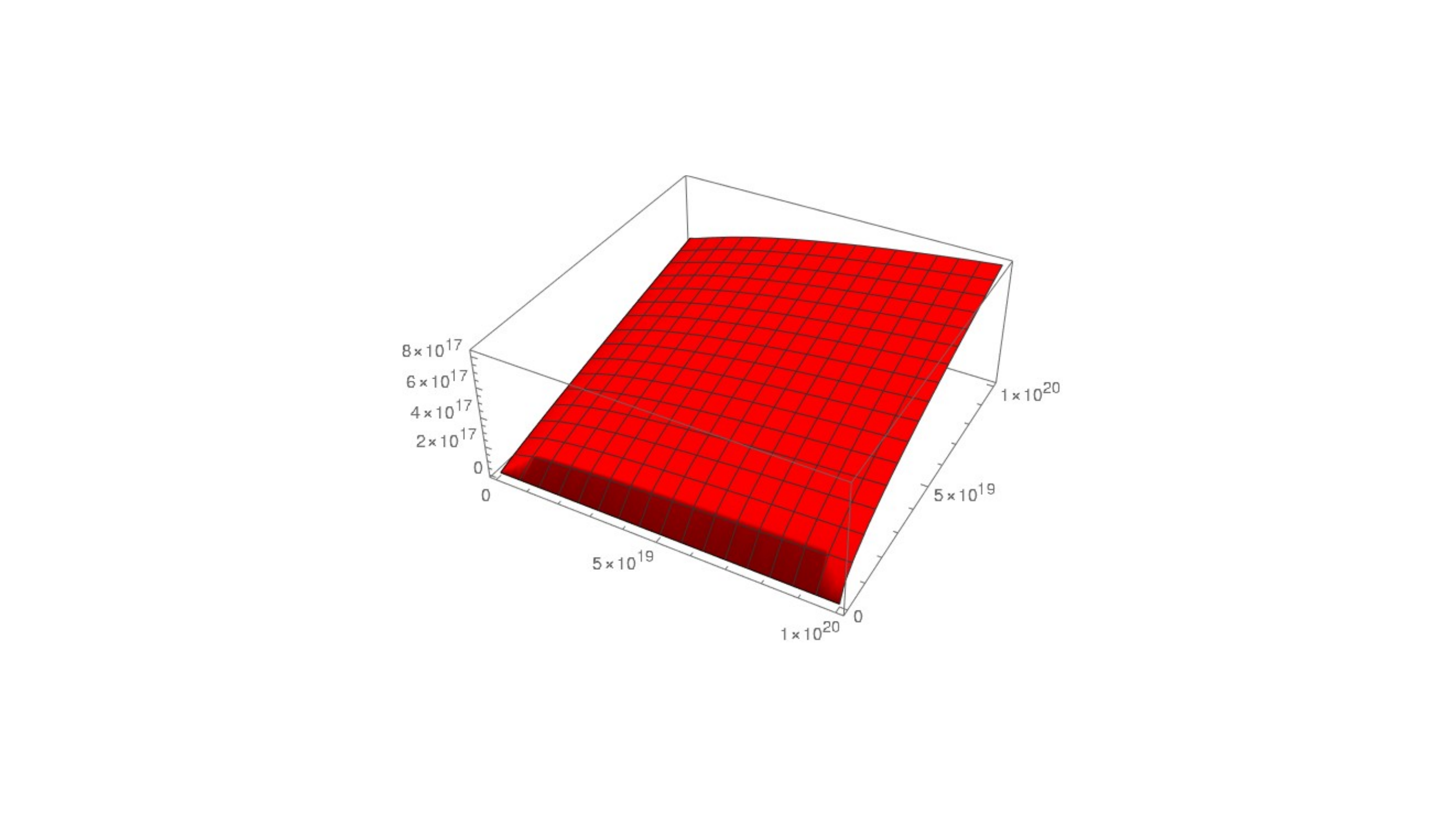}
\includegraphics[width=.32\textwidth]{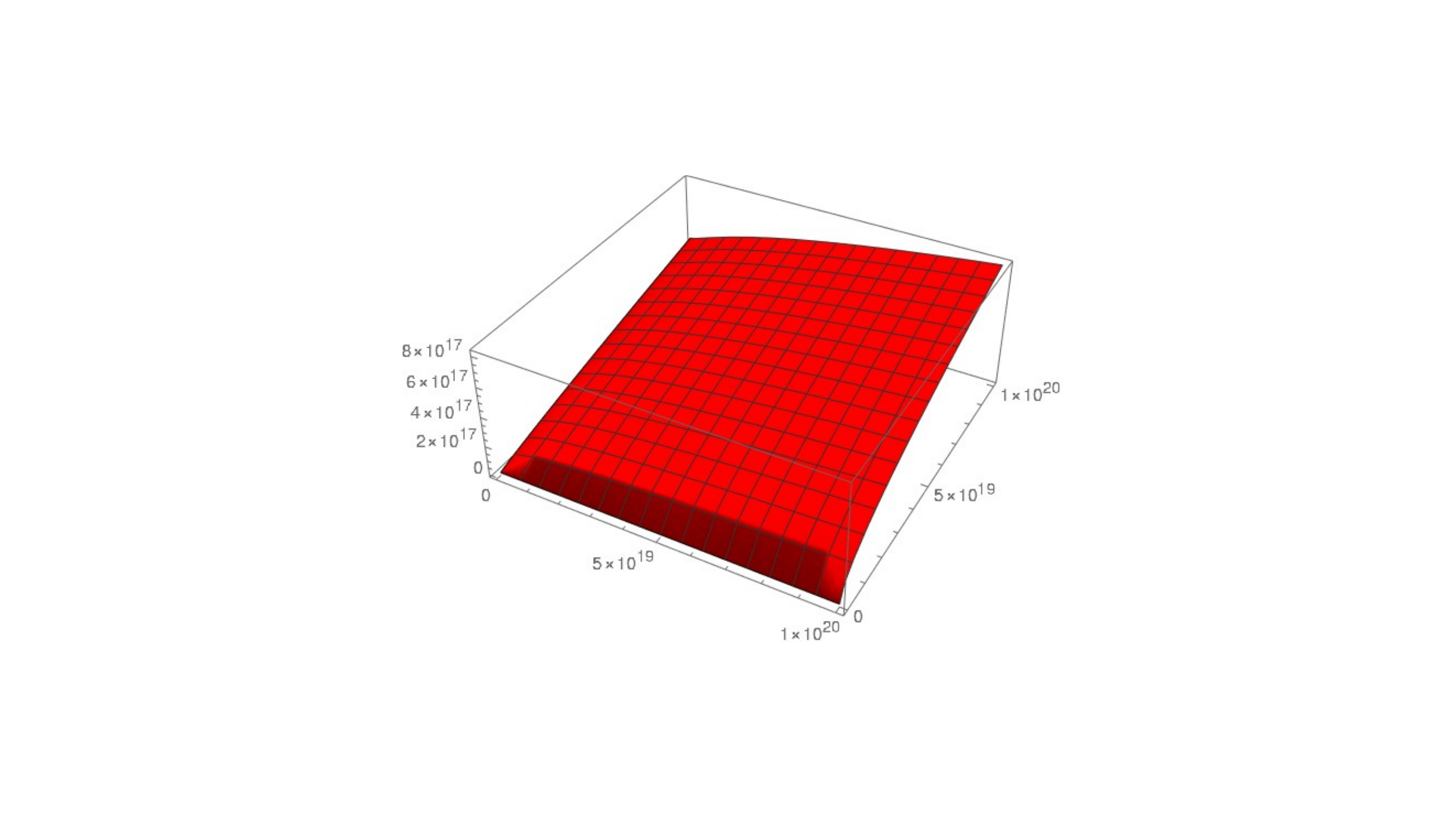}
\includegraphics[width=.32\textwidth]{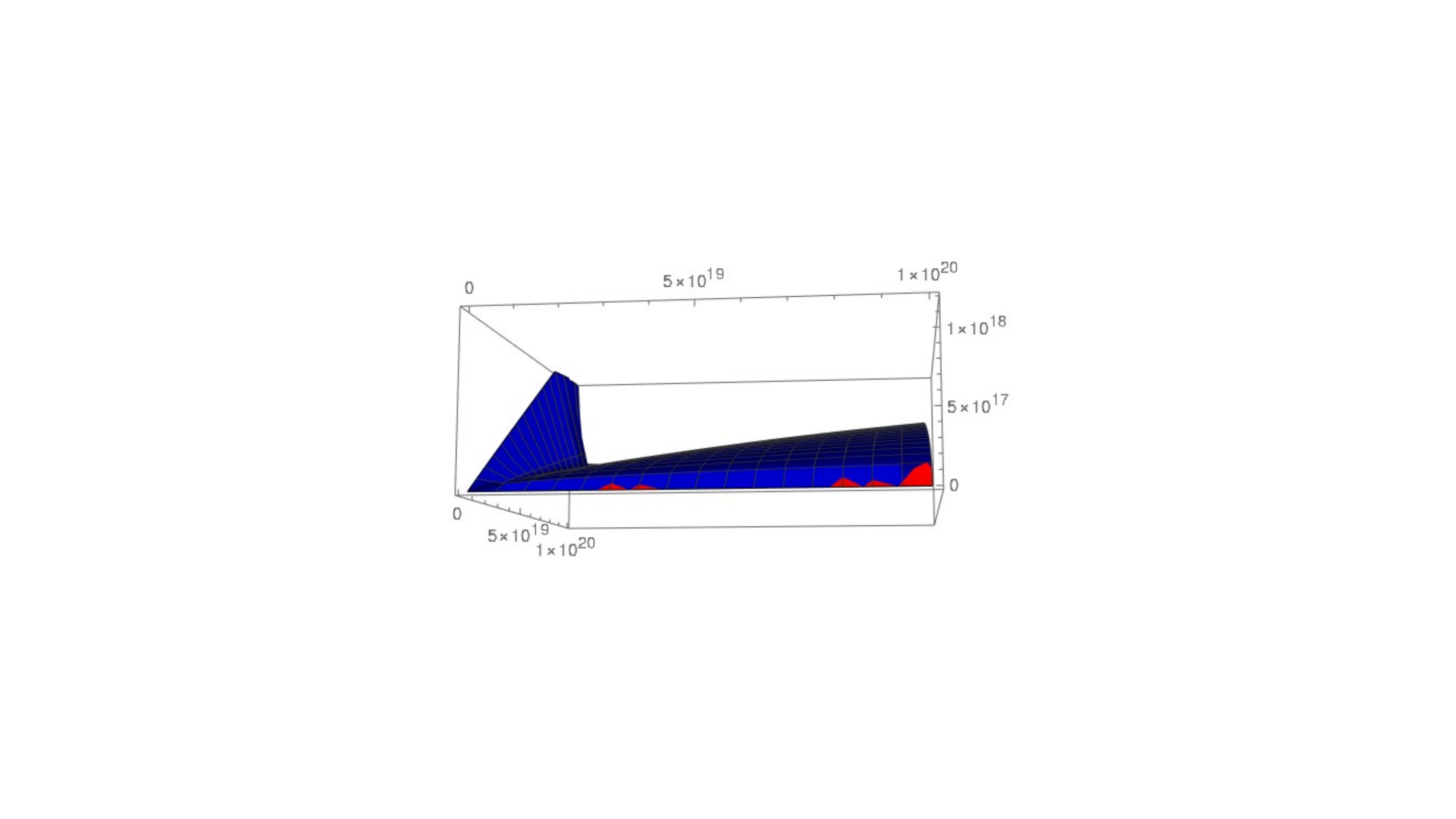}
	
\caption{First row\,\,;\,\, x-axis\,:\,l; y-axis\,:\,$\rho_c$\,\,:\,\,  (left)  ; Simplified expression of S as a function of $l ;\rho_c$  ; (Middle)\,\, : \, Exact expression of S as a function of $l\, ; \,\rho_c$\,\,\,;\,\,\, (right) The overlap between the two\, ; \, The figures are showing the two matches only in the regime $\rho_c  >> l$
\quad;\quad
Last row \,:\, All the above with x-axis and y-axis interchanged
 }
\la{rhocgreatl1by9l1 }
\end{figure}

\begin{figure}[H]
\begin{center}
\textbf{ For $d-\theta < 1$,  with  $ d - \theta = {\frac{1}{12}}$, \,\,:\,\,  Comparison   between the simplified expression of HEE for $ \rho_c >>  l $ with the exact expression of HEE in this regime, can be realized only in long range of  long range of $l, \rho_c$, }
\end{center}
\includegraphics[width=.32\textwidth]{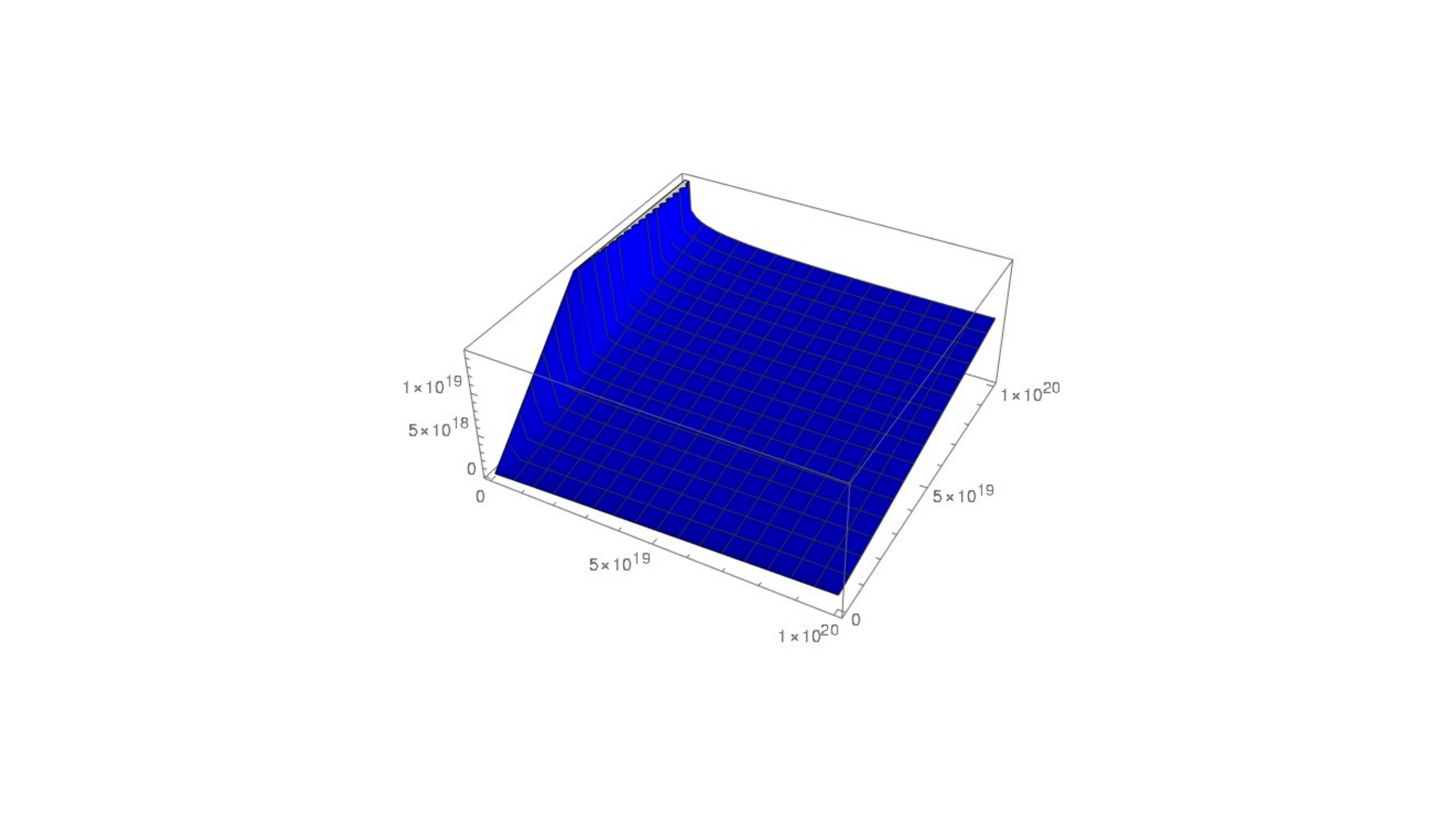}
\includegraphics[width=.32\textwidth]{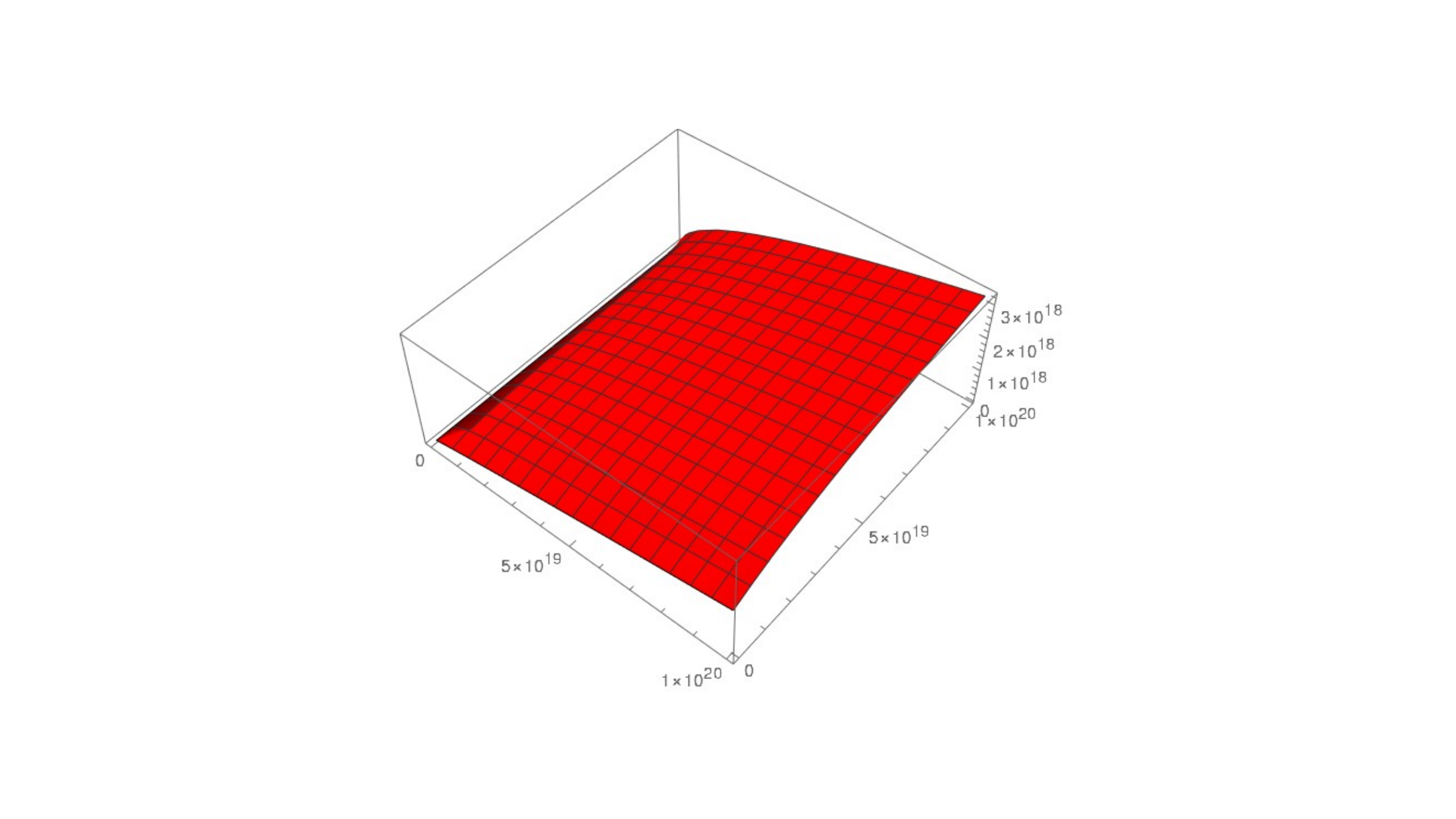}
\includegraphics[width=.32\textwidth]{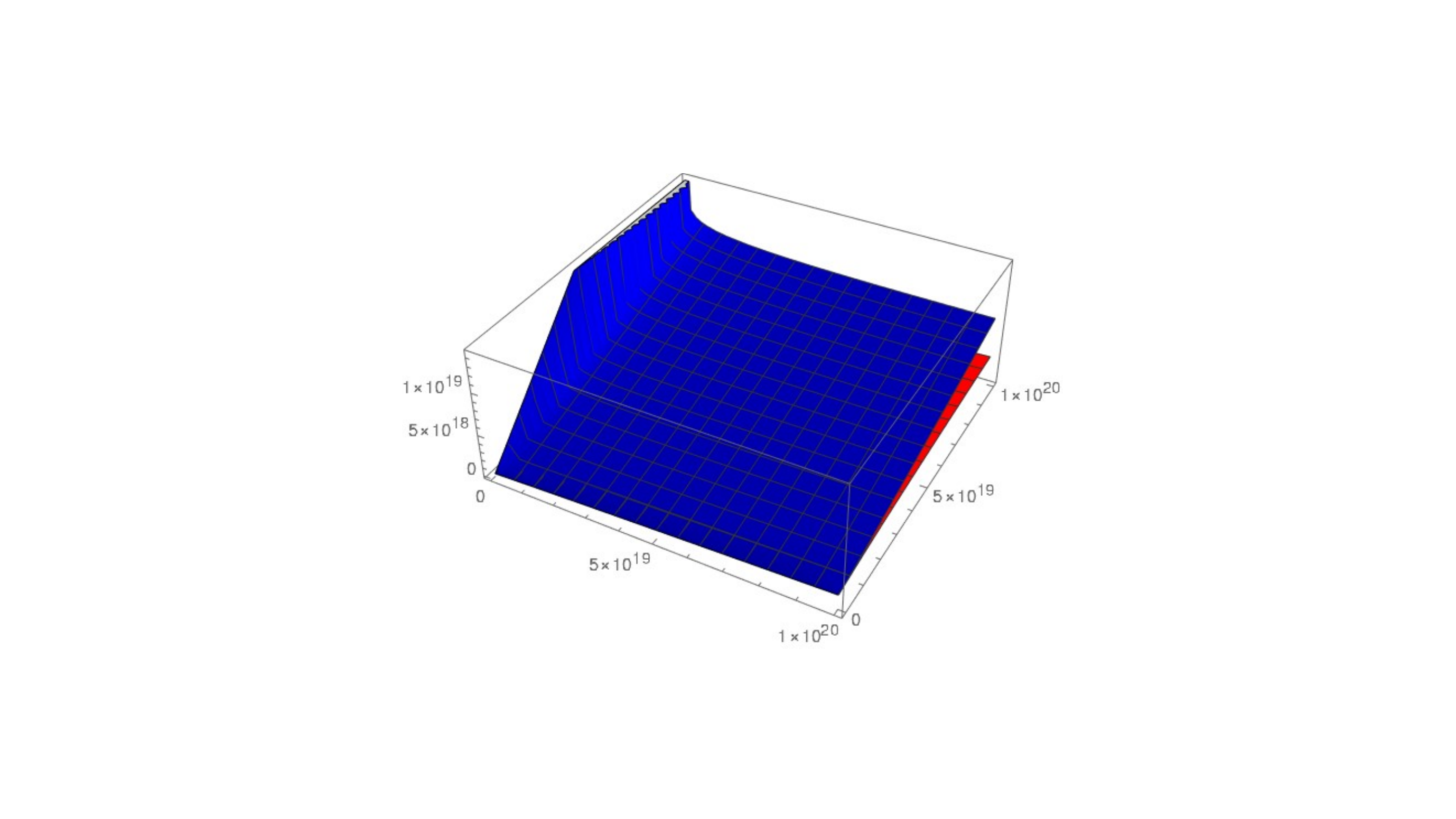}
\includegraphics[width=.32\textwidth]{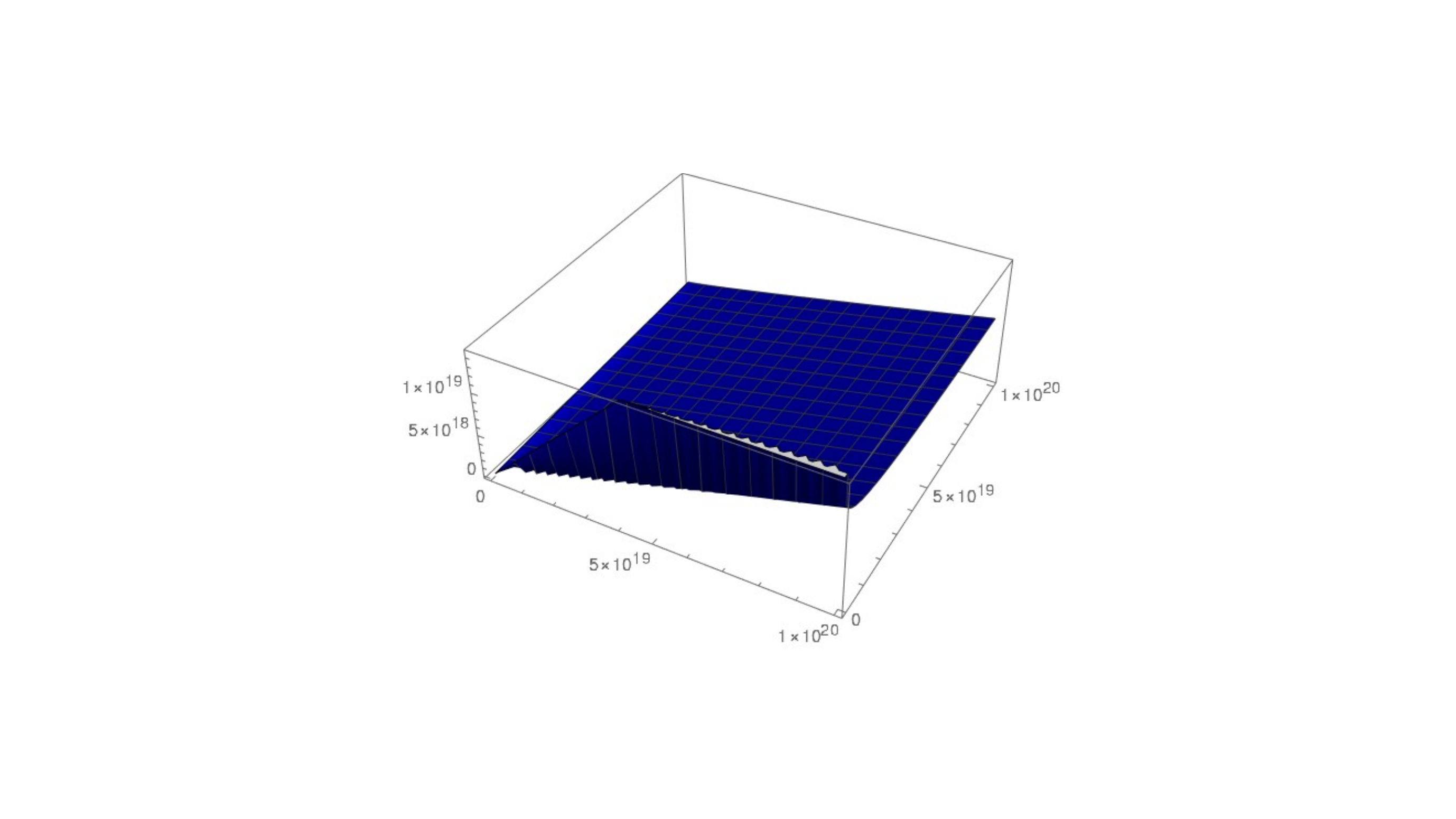}
\includegraphics[width=.32\textwidth]{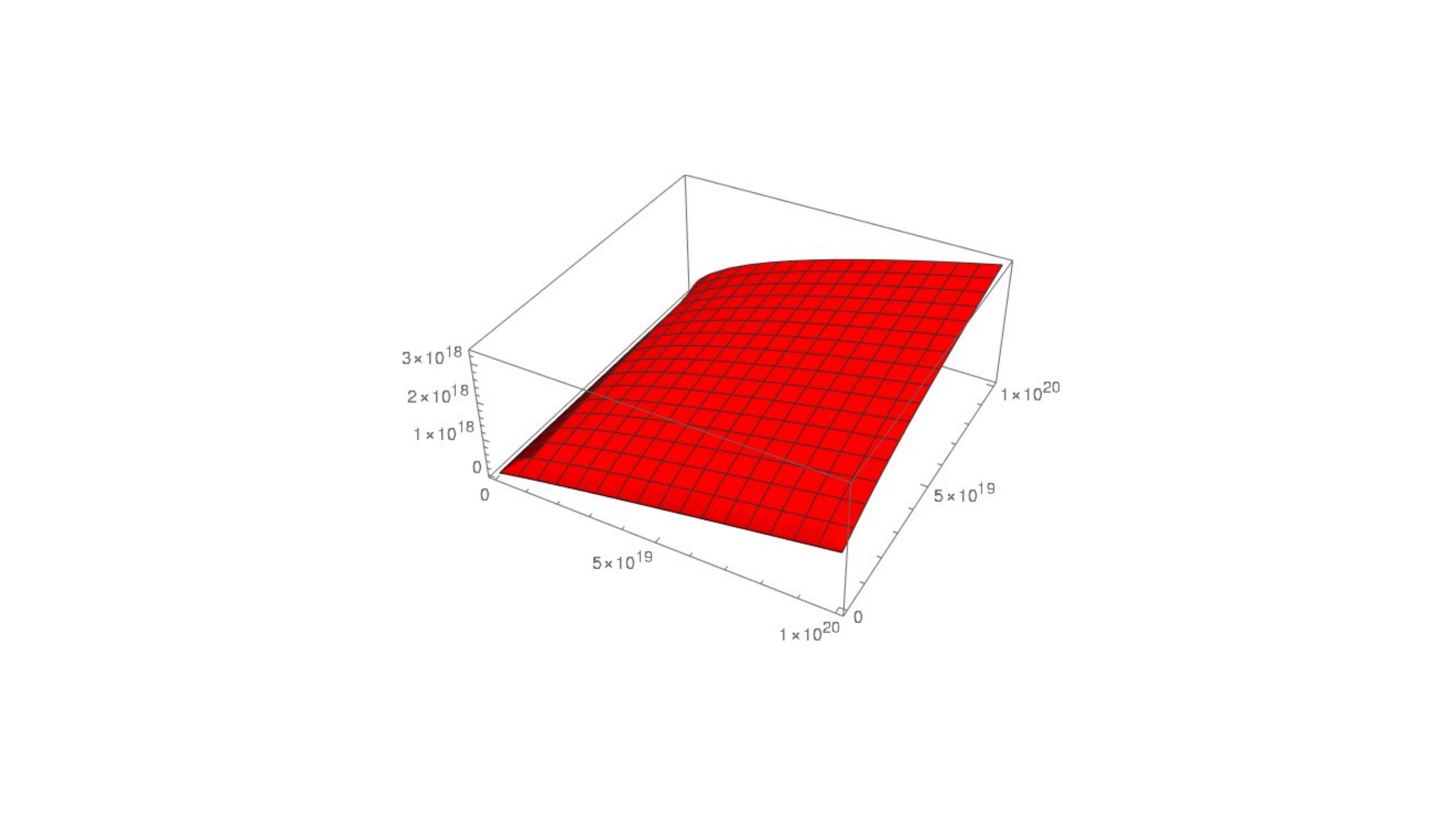}
\includegraphics[width=.32\textwidth]{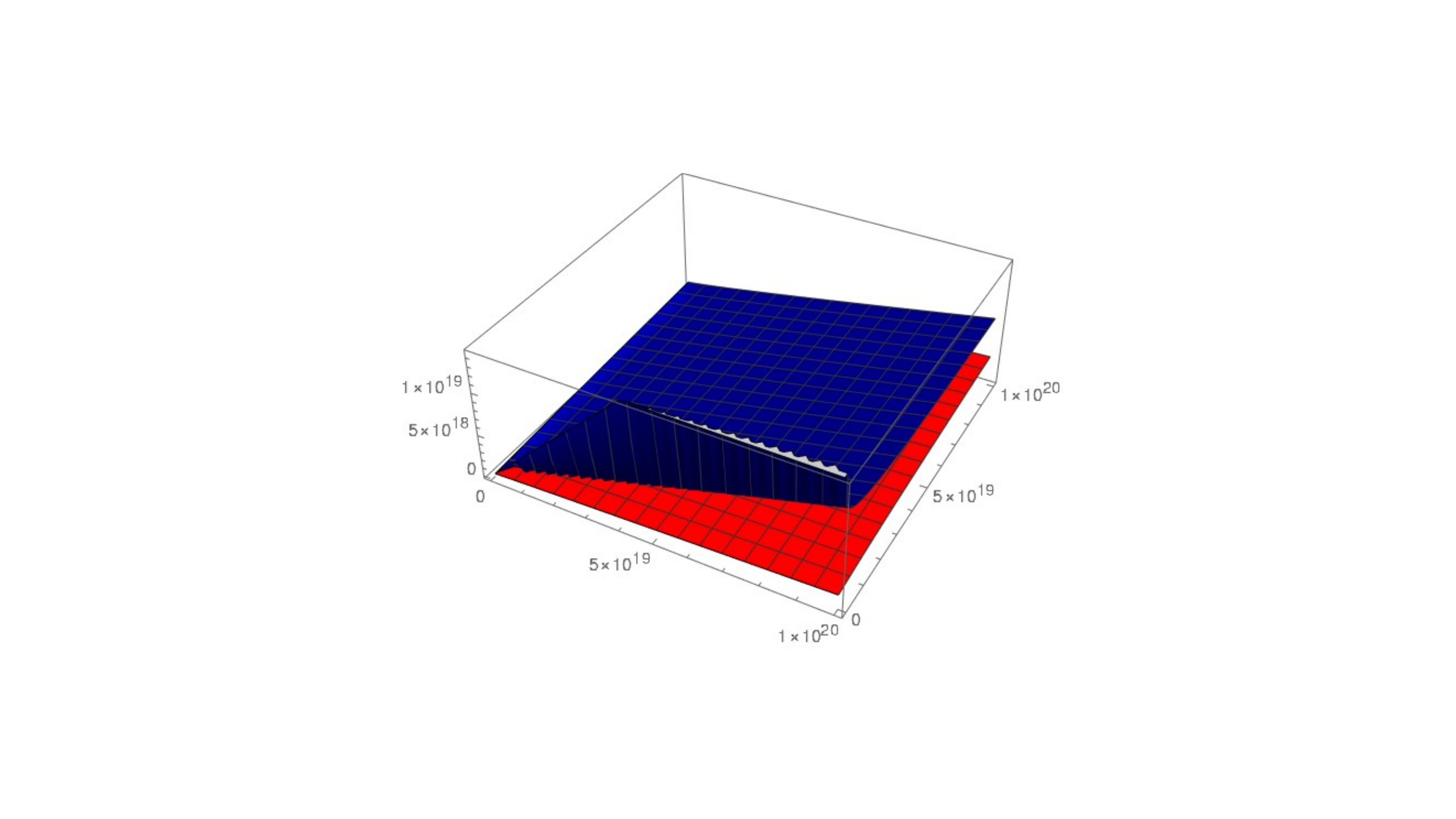}
	
\caption{First row \,: \,   x-axis ;  y-axis\,:\,$\rho_c$\,\,:\,\,  (left)  \,:\, Simplified expression of S as a function of $l \,   ; \rho_c$ \, ;\, (Middle)\,\, : Exact expression of S as a function of $l , \rho_c $  \, , \, (right) The overlap between the two \,;\, The figures are showing the two matches only in the regime $\rho_c  >> l$
\quad;\quad
Last row \,:\, All the above with x-axis and y-axis interchanged
 }
\la{rhocgreatl1by12l1}
\end{figure}

\section{Holographic mutual information in the presence of finite radial cut off}
\vskip1mm

As  we already stated,  the concept of  mutual information and its holographic dual was originally introduced in \cite{Swingle, headrick}  as an attempt to get gravity dual of  of disjoint interval of  CFT and also to remedy the problem involving UV divergence of EE.       Here we describe the effective measure w.r.t two strips,  each of length l,  placed at a distance h, as shown in Fig.(\ref{hmi3}).   

\begin{figure}[H]
\includegraphics[width=.65\textwidth]{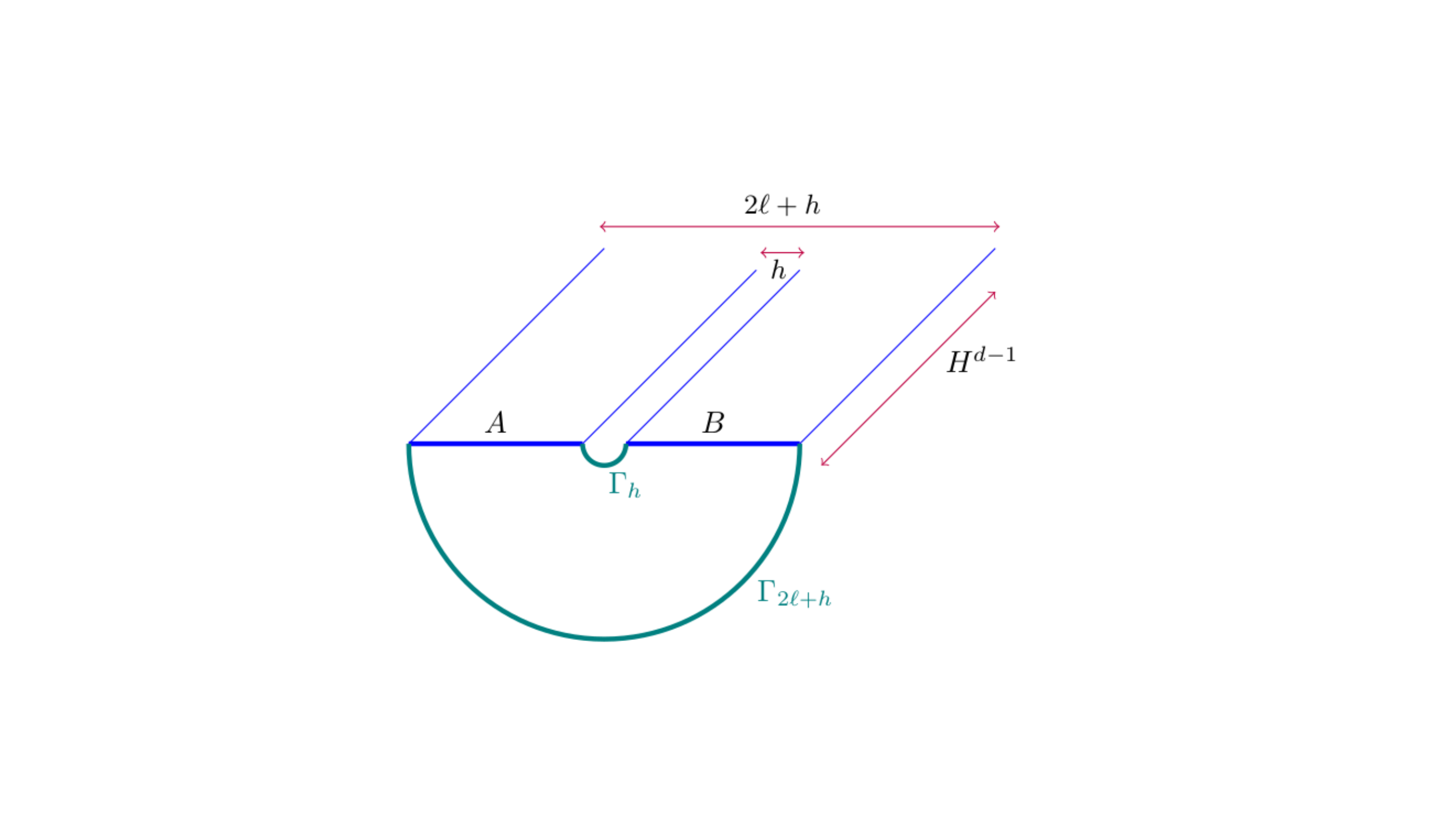}
\caption{A schematic representation   of two stripes in their connected phase
 }
\la{hmi3}
\end{figure}

 Firstly when the systems are too close,  they can be thought as sharing common entangling surface, as shown in Fig.(\ref{hmi3}),   which starts from the  boundary of one these strip of length l, end on the boundary of the others.   The entanglement of entropy is given by $  S_{\rm connected} = S(A\cup B)$.  In other scenario 
the objects  are too distant, can be thought as disconnected.  In either case we have the mutual information is given by
 
$ I(A,B) = {\rm Min} ( S_{\rm connected} , S_{\rm disconnected})$  and explicitly given by (\ref{correlation})

Here  for our specific case we have two possibilities:

1.  \,\,   $ l<< h$ \, : \, Two systems are isolated, $\Rightarrow S(A\cup B) = S(A) + S(B)$.
This implies

\be
I(A,B) = 0 \quad;\quad  {\rm for}  \quad l << h  
\la{dis}
\ee

The other regime, 

2. \,\,  $l >>  h $\, : \,  Two systems are entangled, $\Rightarrow S(A\cup B) = S(2l + h) + S(h)$

This implies

\be
I(A,B) =   2S(l)  -  S(2l + h)  - S(h)  \quad;\quad {\rm for}  \quad l << h
\la{H}
\ee

Now to predict the properties of H.M.I in the presence of the cut off,  we recall the analysis in the previous section  for H.E.E where from (\ref{eereplicarewrite})  it was argued that HEE will decrease with the increase of the cut off $\rho_c$.   Next,  if  we substitute the first equation of  (\ref{eereplicarewrite}) in its most generalized form,   i.e generalized to higher dimension, generalized to other geometries, etc.,  in (\ref{H}),   then from the argument that H.M.I will always be positive definite,  it is easy to see that with the increase  of cut off $\rho_c$(i.e with the increase of $T{\overline{T}}$ deformation coefficient $\mu$),  H.M.I will decrease,  as follows from all the arguments we made in the  section 4  regarding the reason  EE should decrease with the increase of $\mu$.    To understand the phase transition from the field theory side,   we consider the fact when A and B are  separated  by long distance the mutual information between them can be expressed as the operator product expansion, where we replace A and B by sum of the local operator through state operator mapping \cite{headrick,hmiopexpansion1,hmiopexpansion2,hmiopexpansion3} .  The leading term will come from the  exchange of the pair of operator   each of weight  $\triangle$ giving by \cite{quantumcorrection}  
\be
I(A,B) \sim \displaystyle\sum_{\triangle} {\frac{C_\triangle}{r^{4 \triangle}}} \, ,
\la{iab}
\ee

with $C_{\triangle}$ is the square of o.p.e coefficient,  appearing when we replace the region A and B in the replica space by sum of the local operators defined on the n copies of CFT through the expression  $\displaystyle \sum  C_O^A O $,  with the O is the local operator, defined on n copies of CFT.    The local operators,  defined on the n copies CFT or on the n-th sheet replica,  can be expressed as the product of operators of original CFT living on differen replicas.    Once we have their OPE coefficient,  we can find
 
\be
C_{\triangle}  = \partial_n \left\lbrack \displaystyle\sum  C_O^A C_O^B\right\rbrack_{n = 1}
\la{operator}
\ee
where the sum is done over all the operator contributing at leading order (\ref{iab}).  Now,  as we mentioned.   the local operator, defined on n copies of CFT,  are actually given by the product of the operators on different replicas.     Consequently,  when we set $n = 1$,  the nth sheet replica space will turn into the unreplicated space where the normal CFT is defined,  so that $ C_O^A,   C_O^B $  will just be the normal one point function of the CFT lying over this regimes and so clearly,  they will vanish,  making the Mutual Information I(A,B)  ,   at least to its leading order term,   to vanish,  as evident from substitution of  (\ref{operator}) to (\ref{iab})!   However this is certainly not the scenario  
when these objects A and B  lies too close to each other since in that case we cannot express them as the sum of the local operatoirs through state-operator mapping because they are supposed to be entangled in that case so clearly this free-CFT-operator-expansion cannot go!   So clearly when we keep increase the separation between A and B,   there must be a phase transition in between!
\vskip0.5mm
Now to see this scenario in the case of $T{\overline{T}}$ deformed case,   we need to evaluate $C_O^A$ explicitly in this case,  which is beyond the scope of the present article!   However one can argue intutively as well!
Following \cite{movingcft} and \cite{ttbarenergyspectrum1, ttbarenergyspectrum2,ttbarenergyspectrum3 } ,  $T{\overline{T}}$ deformation, given with coefficient $\mu$,  is a solvable deformation  and it actually changes the energy spectrum of a CFT state, denoted by conformal dimension  $(\triangle_n, \triangle_{\tilde{n}})$.  For the specific case of a cylinder,  it is given by
\ber
E_n (\mu, L) L &=& {\frac{2\pi}{\tilde{\mu}}}\left( 1 - \sqrt{1 - \sqrt{1 - 2{\tilde{\mu}} M_n + {\tilde{\mu}}^2 J_n^2}} \right) \n
& & {\rm with} \,\,\,{\tilde{\mu}} = {\frac{\pi\mu}{L^2}},\,\, M_n = \triangle_n + \triangle_{\tilde{n}} - {\frac{c}{12}},\,\, J_n = {\triangle_n} -  \triangle_{\tilde{n}}
\la{shiftedspectrum}
\eer

Clearly   the relation (\ref{shiftedspectrum})   implies,  the states in the  $T{\overline{T}}$ deformed CFT	 can come from one to one mapping from the undeformed CFT!   In that case,  we can expect one can expand two separate intervals A and B as the sum of local operators through the state operator mapping in that respective CFT,  just like undeformed case  \cite{headrick,hmiopexpansion1,hmiopexpansion2,hmiopexpansion3}.    All the discussions we just made for undeformed case will remain valid for $T{\overline{T}}$ deformed case as well and clearly since the one point function on unreplicated space   must vanish in that respective CFT as well,   from the same formal CFT principle,  so one can see  the phase transition when the separation between A and B will reach to a certain critical limit,  even in the deformed case!  Clearly in that case.  for a given strip-length (l) and given cut-off $\rho_c$,  the H.M.I  will fall with the increase of the separation of two disjoint intervals, given by h  and ultimately goes to zero at the critical point!  However . this phase transition is a leading order scenario in $G_N$,  or in $N^2$ for large N dual field theory where was it was shown for the undeformed CFT in \cite{quantumcorrection} that this phase transition does not exist if we consider the first order quantum correction!   However,  we are yet to see this feature for our  $T{\overline{T}}$ deformed case!
\vskip0.5mm

Moreover in HMI is an UV finite quantity as long as the separation h is nonzero because the singularity-factor appears on regularization,   cancels out on subtraction in its expression! However, when they brought together,  i.e they shares a common boundary it will become divergent!  In the presence of a cut-off since it acts as an effective regulator, making EE non-UV-divergent so HMI  will remain finite even at zero separation but divergence will show up  if we make $\rho_c = 0$!

Now regarding the expected behaviour of H.M.I  in the presence of a finite radial cut off,  one can view the scenario from gravity side more easily!  
In the  section 4 we explained with Fig.(\ref{entanglingsurface}),   since the cut off $\rho_c$ itself increases faster tham turning point $\rho_0 (l, \rho_c)$   where the tip of the entangling surface is given by the turning point,   so EE decreases with the increase of the cut-off.  Next we consider Fig.(\ref{hmi5}).

Next we write $I(A,B) = \left( S(l) - S(2l + h) \right) +  \left( S(l) - S( h) \right) $ and consider the fact that for connected phase $l >> h$.  
Now,  once we introduce the cut off surface $\rho_c$ and increase $\rho_c$,  as follows from  Fig.(\ref{entanglingsurface}),    as the combined effect of the both \,:\,I.\,,  pushing of cut-off surface more and more inside the bulk and \,\, II.\,  the entangling surface will tend to grow due to increase of $\rho_c$ because its tip , given by the turning point $\rho_0$ will grow (follows from (\ref{ultimaterho0nonvanishing}) ),  since it is evident that due to the first effect  $ S(2l + h)$ will fall faster than $S(l)$ which even fall faster than $S(h)$  and due to the second effect $S(2l + h)$ will grow slower than $S(l)$ which either grow slower than 
$S(h)$ (follows from (\ref{ultimaterho0nonvanishing}) ),  so in the expression of I(A,B),    $\left( S(l) - S( h) \right) $ will fall while $|\left( S(l) - S(2l + h) \right)  |$ will grow, however since its a negative quantity, it will fall again!  As a consequence, H.M.I will decrease with the increase of cut off $\rho_c$!
\vskip0.5mm
For,  fixed $(l,\rho_c)$,  if we increase h,  it is evident from Fig.( \ref{hmi3} ),   both S(h) and S(2l + h) will grow because the tip of the entangling surface given by the turning point will grow (follows from (\ref{ultimaterho0nonvanishing}) )  while S(l) remain fixed,  so from (\ref{H}), I(A,B) will decrease and eventually become zero at  $h = h_{\rm crit}$, when the disconnected phase will start to dominate!
\vskip0.5mm

Also since HMI is  positive definite,   so it is just in the same way we argue for HEE,   HMI will also decrease with the increase of $ d - \theta$,  with the maximum difference between two different  $d - \theta$ expression in   $l >> \rho_c$ regime and all merge together to zero in $\rho_c >> l$  regime, as indeed we are going to see!

\vskip0.5mm

 Finally,  as evident from (\ref{entropy},\ref{dis}, \ref{H}),  if we try to extend the symmetry (\ref{2dsymmetry}) as
$(l, \rho_c, h) \rightarrow  (kl, k\rho_c, kh) $,  we have

\ber
& & (l, \rho_c, h) \rightarrow  (kl, k\rho_c, kh)\n
& & \Rightarrow I(A,B)  \rightarrow  {(k)}^{1 -d +  \theta } I(A,B)\,,
\la{symmetryhmi}
\eer

Here finally we must comment on the fact that when the $\rho_c = 0$, the symmetry,  as expressed in (\ref{symmetryhmi}),  do hold 
as shown in \cite{lifshitz, chargedbrane }.   However there it was a normal mathematical  functional dependence and not really corresponds to any spacetime symmetry with the geometrical origin.   This fact actually supports the view of our discussion in section 3 that  the emergent bulk symmetry can really be viewed as generalization of boundary scaling symmetry or vice versa. 

\begin{figure}[H]
\includegraphics[width=.75\textwidth]{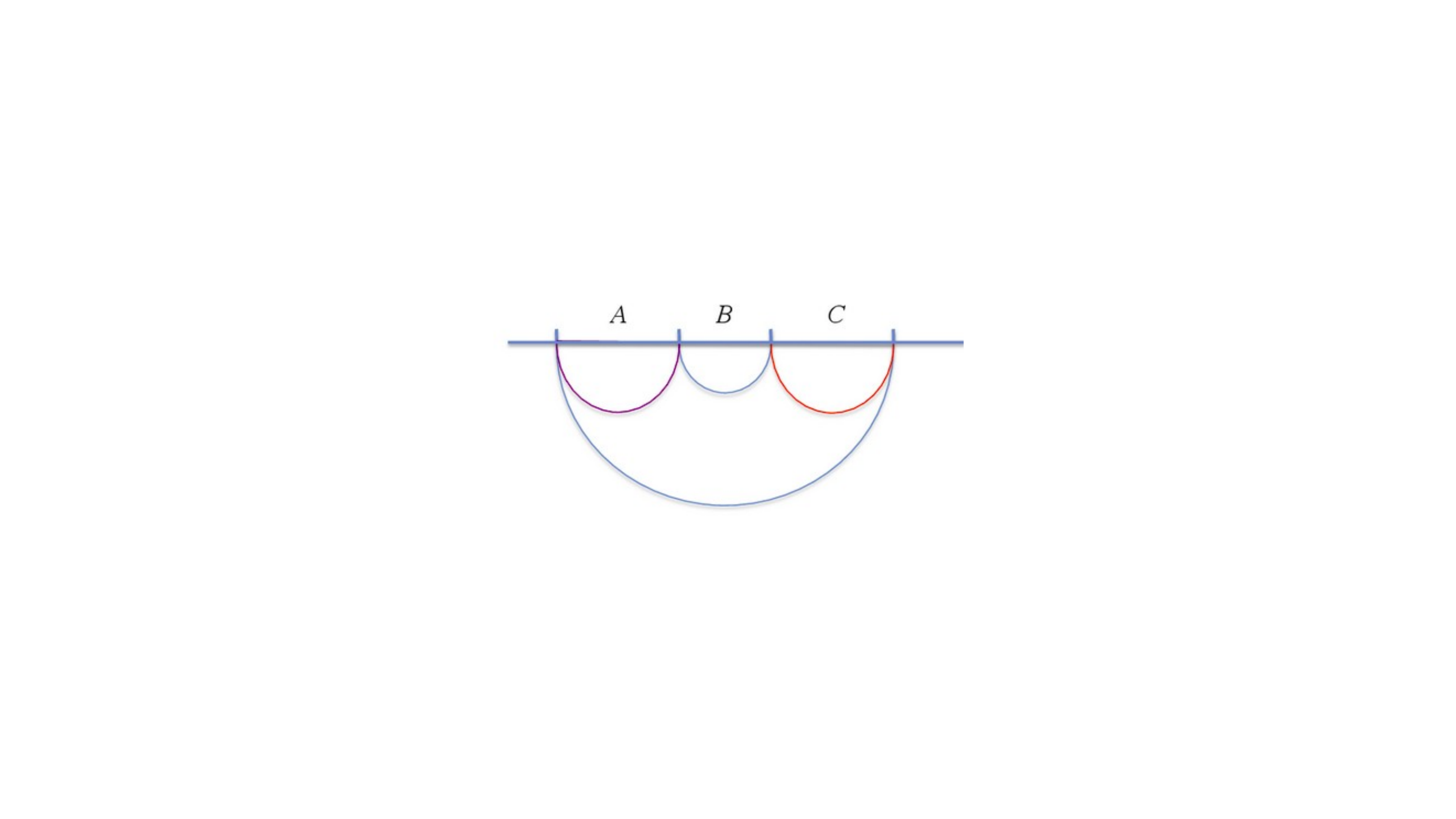}
\caption{H.M.I for two strips with A and C denoting two strips of length l along with their entangling surface  $\Gamma_l$.  The intermediate region B can be treated as a separation between them given by h,  with the entangling surface $\Gamma_h$.  The uppermost hemisphere will denote the entangling surface of the complete connected region denoted by $\Gamma_{2l + h}$   }
\la{hmi5}
\end{figure}

Here in the next few sections, we will show,  the expression of the I(A,B) as we have constructed based on the expression of EE (\ref{entropy}),  depends on  the turning point  (\ref{ultimaterho0nonvanishing}) ), as we have constructed from the global symmetry structure, indeed showing all these prperties, for complete range of $(l,\rho_c)$.  for most general $d,\theta$ with $\theta \le d $!  Moreover, since due to the complication of our expression of H.M.I  we cannot exactly determiune $h_{\rm crit}$, i.e the value of h for a given $(l.\rho_c)$ where the phase transition takes place,  which is given by $H.M.I = 0$ ,  so in the relevant plots in the next few subsection, we will show the phase transition is taking place, without numerical details of  $h_{\rm crit}$.

\subsection{HMI for $d-\theta =  1$}

Here we recall H.E.E for $d - \theta = 1$ (\ref{s0complete}), so clearly H.M.I is given by

\ber
S_{d - \theta = 1} &=& 2{\frac{R^d L^{d - 1}}{4G_N}} \log \left\lbrack {\frac{\rho_0\left( l,\rho_c \right)  +  {\frac{l}{2}}}{\rho_c}}\right\rbrack\n
                   &-& {\frac{R^d L^{d - 1}}{4G_N}} \log \left\lbrack {\frac{\rho_0 \left(   (2l + h) , \rho_c \right)  +  {\frac{(2l + h)}{2}}}{\rho_c}}\right\rbrack\n
                   &-& {\frac{R^d L^{d - 1}}{4G_N}} \log \left\lbrack {\frac{\rho_0 \left(    h , \rho_c \right)  +  {\frac{( h)}{2}}}{\rho_c}}\right\rbrack\n
\la{s0hmi}
\eer

Here we present the basic plots to probe the nature of 	H.M.I at $b = 1$, given in (\ref{hmib11}, \ref{hmib1hl}, \ref{hmirhoch1b1}).   Also in these plots we consider very long range of $(l,\rho_c)$ to probe $l >> \rho_c$ and $\rho_c >> l$ regime.

\begin{figure}[H]
\begin{center}
\textbf{For $ d - \theta = 1$ : \,H.M.I vs $(l,\rho_c) $, H.M.I vs l and H.M.I vs $\rho_c$ plots for fixed h, showing the basic
properties of H.M.I.  }
\end{center}
\vskip2mm
\includegraphics[width=.45\textwidth]{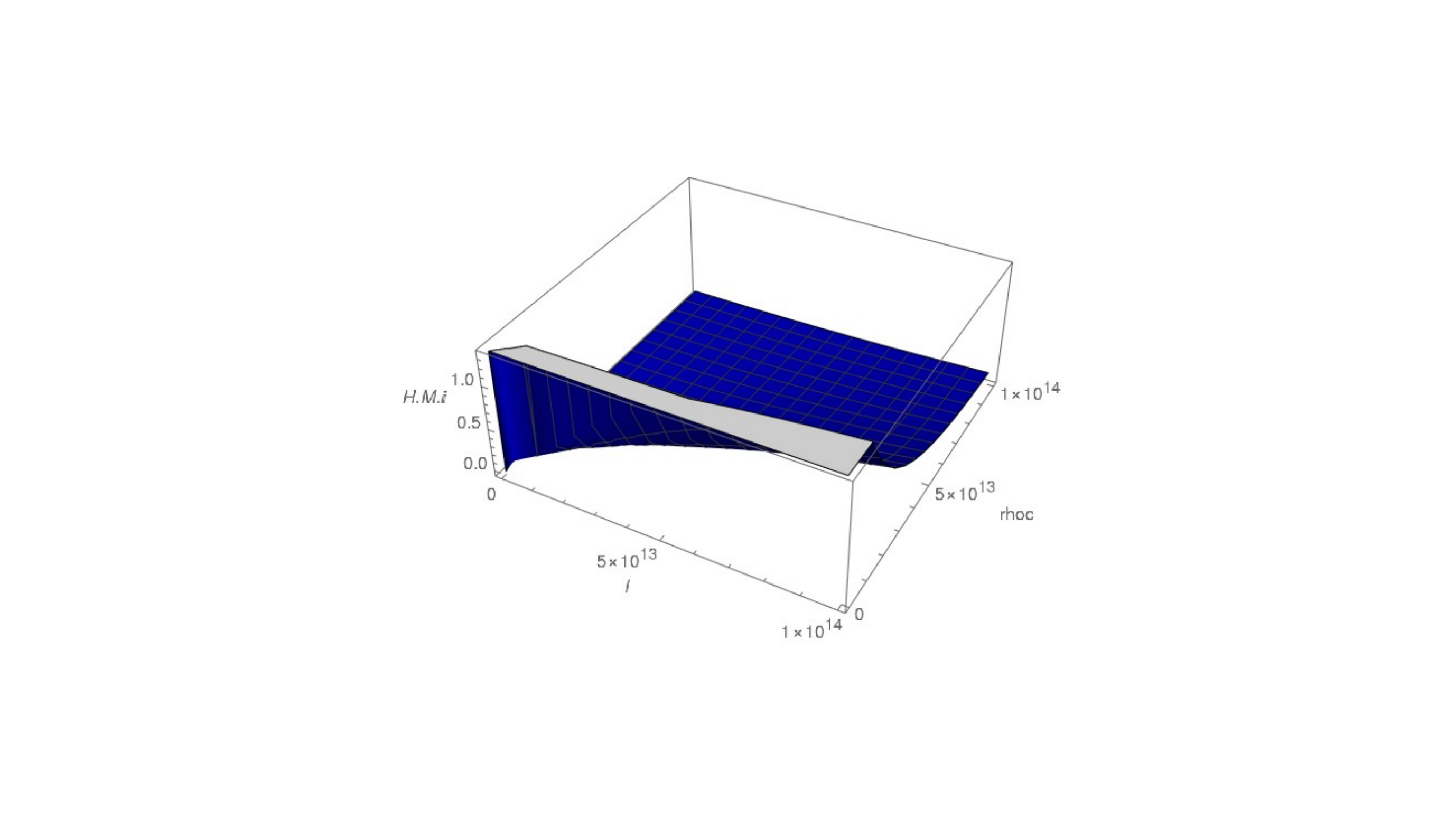}
\includegraphics[width=.45\textwidth]{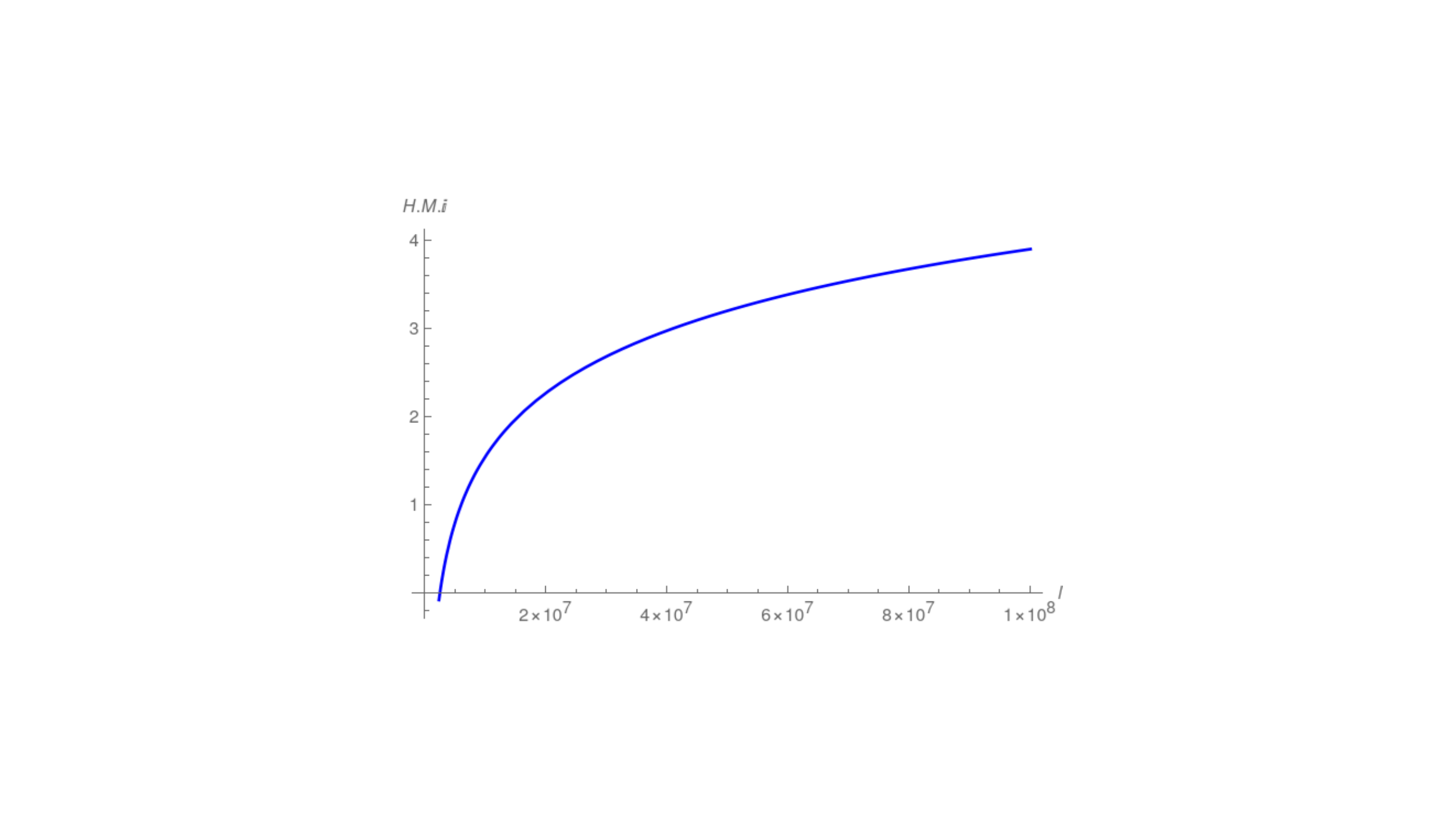}
\includegraphics[width=.45\textwidth]{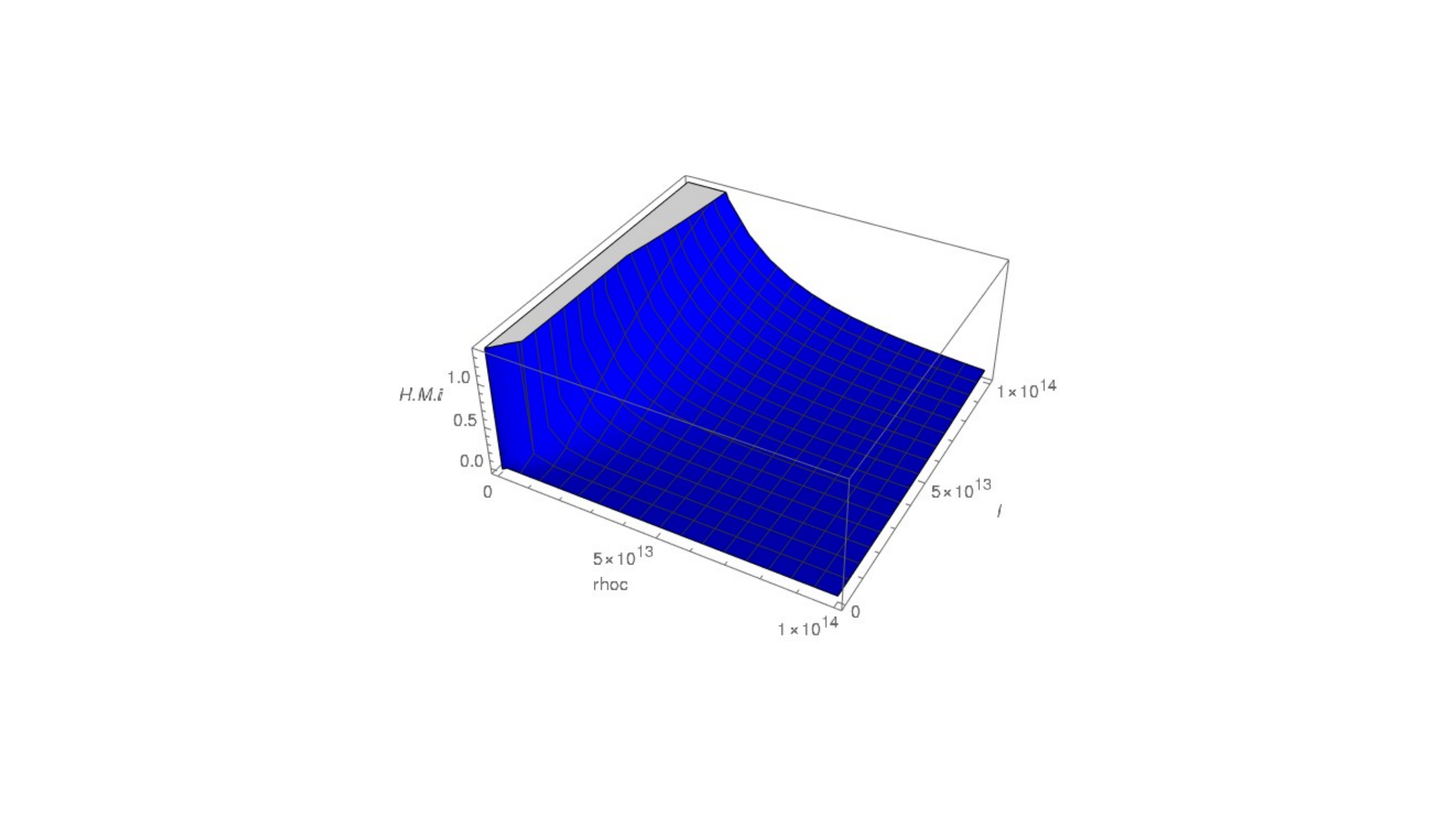}
\includegraphics[width=.45\textwidth]{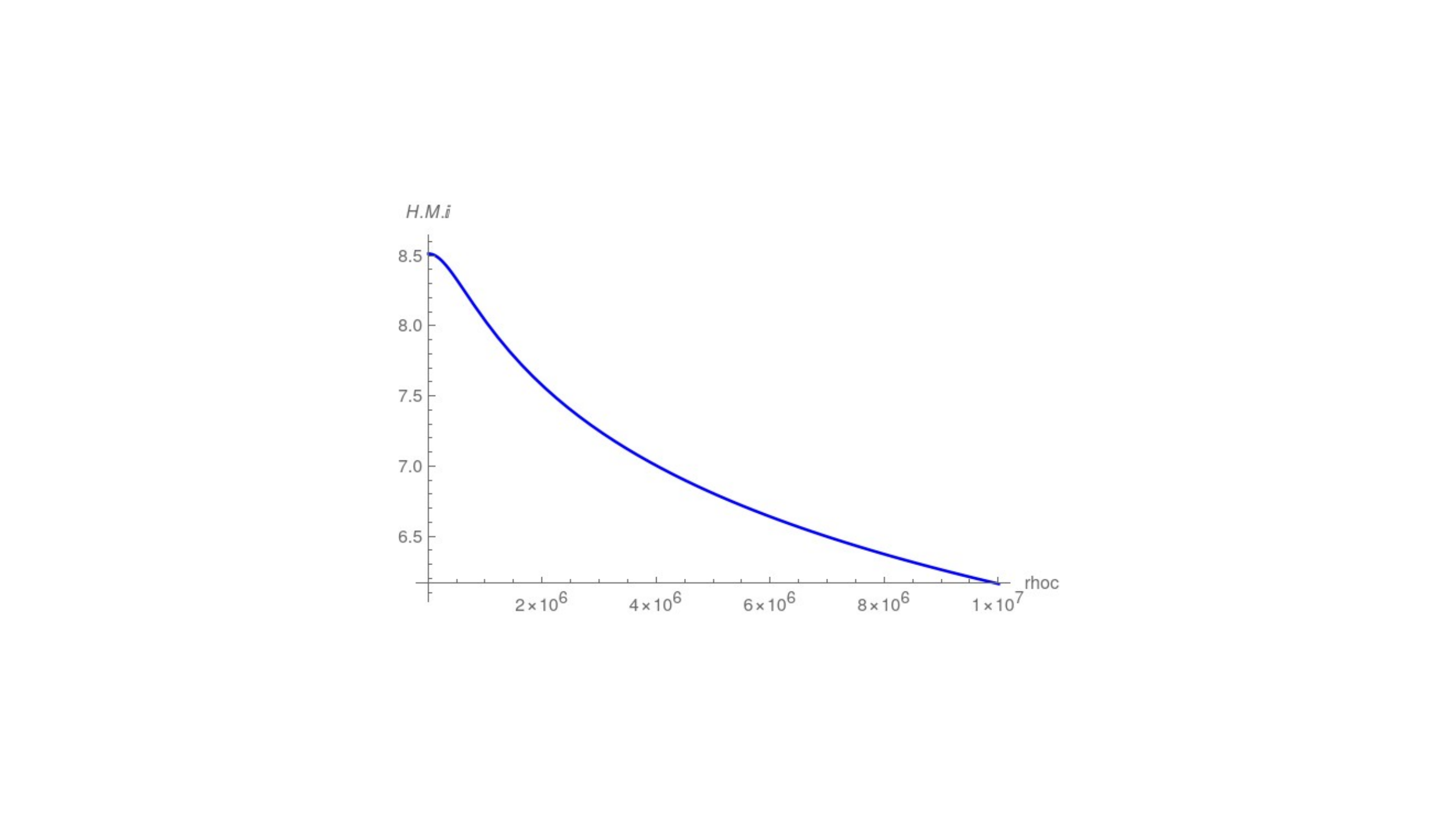}
\includegraphics[width=.45\textwidth]{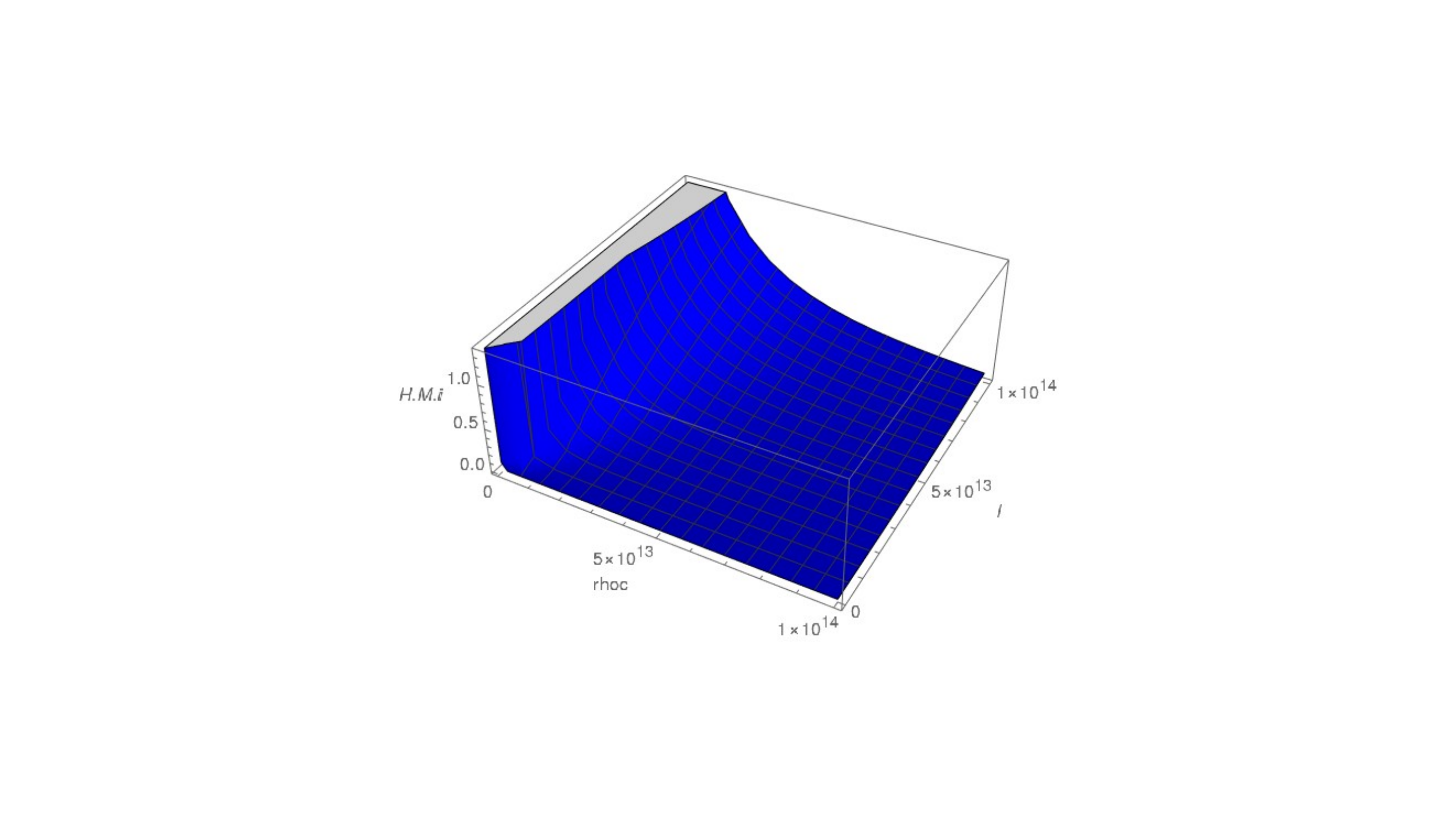}
\includegraphics[width=.45\textwidth]{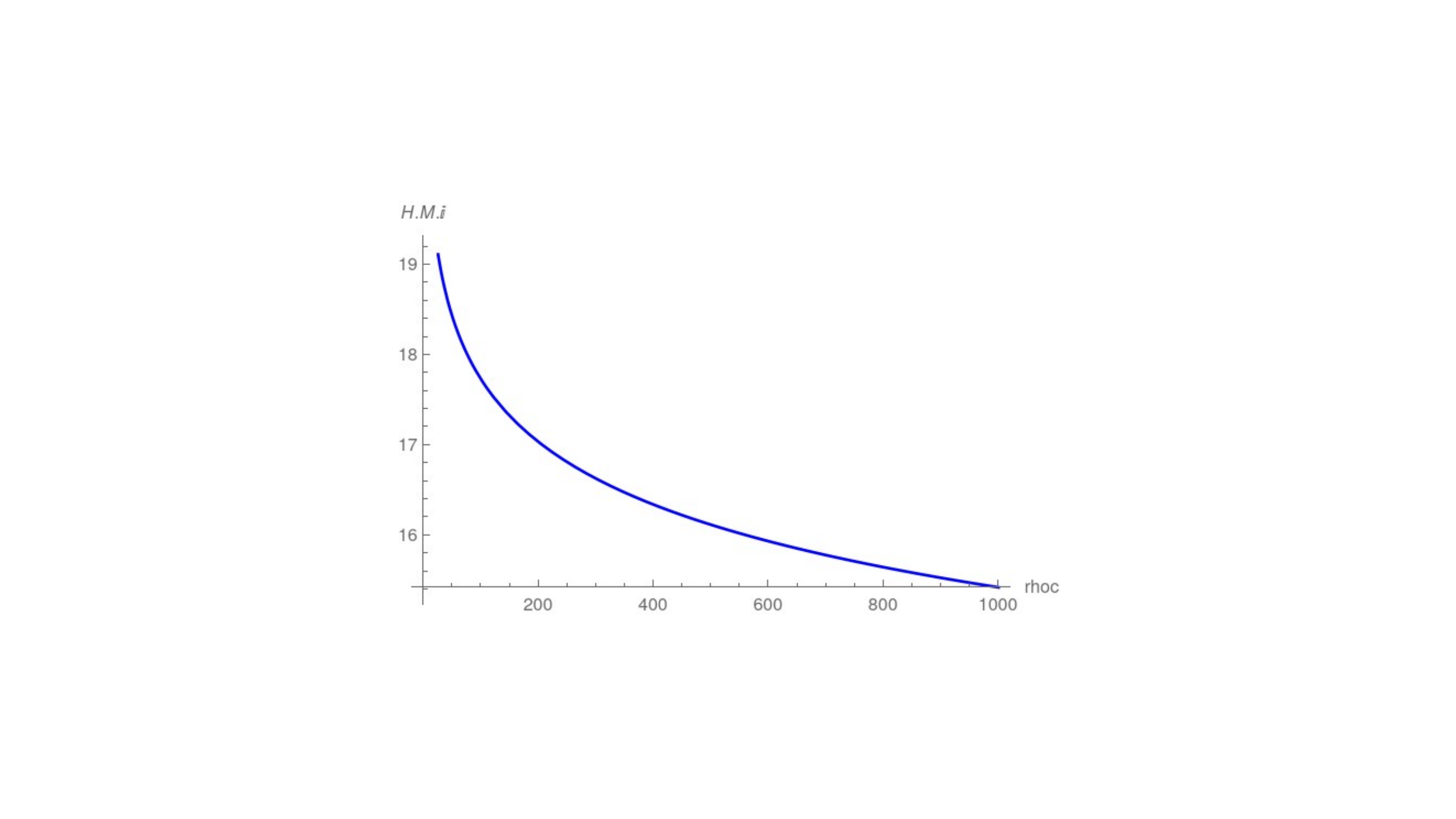}
\caption{(First row) \,\,: (left)\,:\, H.M.I as a function of $(l,\rho_c)$ for  $h = (10)^6$ \,,\, (right)\, : \, H.M.I  as a function of l with $\rho_c = 100$,  $h = (10)^6$  \,:\,  Both the plots are showing when we increase l gradually from zero, for fixed h,   at certain point given in terms of $h_{\rm crit}$, H.M.I is undergoing a first order phase transition and also for a given $\rho_c$,   H.M.I increases with the increase of l
  \quad;\quad (Second row) \,:\,  ( left)\, : \, H.M.I  as a function of $(\rho_c , l)$ for  $h = (10)^6$\,, (right) \,:\, H.M.I  as a function of $\rho_c$  for $l = {(10)}^{15}$,$h = {(10)}^6$,  this value of l is chosen to probe $ l >> \rho_c $ regime where for $\rho_c >> l$ H.M.I is zero \, :\,  Both the 3D and 2D plots are showing, for a given l, H.M.I falls with the increase of cut off $\rho_c$ and goes to zero for $\rho_c >> l $ regime.  Also for nonzero h,    H.M.I is finite at 
$\rho_c = 0$
 \quad;\quad (Last row) \,:\,  ( left)\, : \, H.M.I  as a function of $(\rho_c , l)$ for  $h = 0 $\,, (right) \,:\, H.M.I  as a function of $\rho_c$  for $l = {10}^{15}$,$h = 0$,   \, :\,  Both the 3D and 2D plots are showing for $h = 0$  H.M.I diverge at $\rho_c = 0$ as expected }
\la{hmib11}
\end{figure}

\begin{figure}[H]
\begin{center}
\textbf{ For $d - \theta =1$, HMI vs $(h, l)$ plot  for fixed $\rho_c$}
\end{center}
\vskip2mm
\includegraphics[width=.65\textwidth]{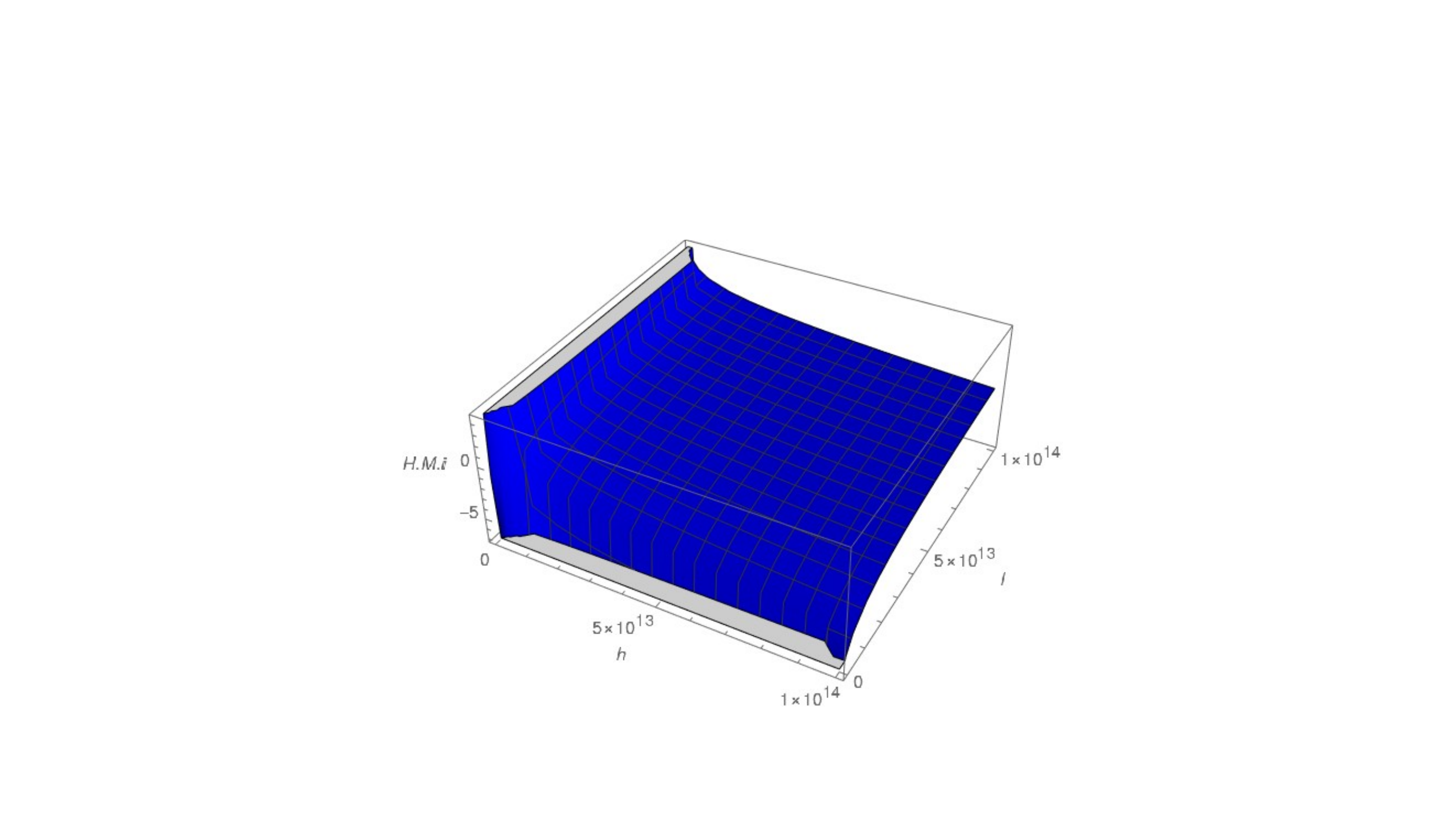}
\caption{H.M.I vs $(h, l)$ plot  for $\rho_c = 10$ \,:\,showing that it increases with l, decreases with h  and not divergent at $h = 0$ for nonzero cutoff $\rho_c$  }
\la{hmib1hl}
\end{figure}

\begin{figure}[H]
\begin{center}
\textbf{H.M.I as a function of ($\rho_c$ , h ) for $d - \theta = 1$  }
\end{center}
\vskip2mm
\includegraphics[width=.65\textwidth]{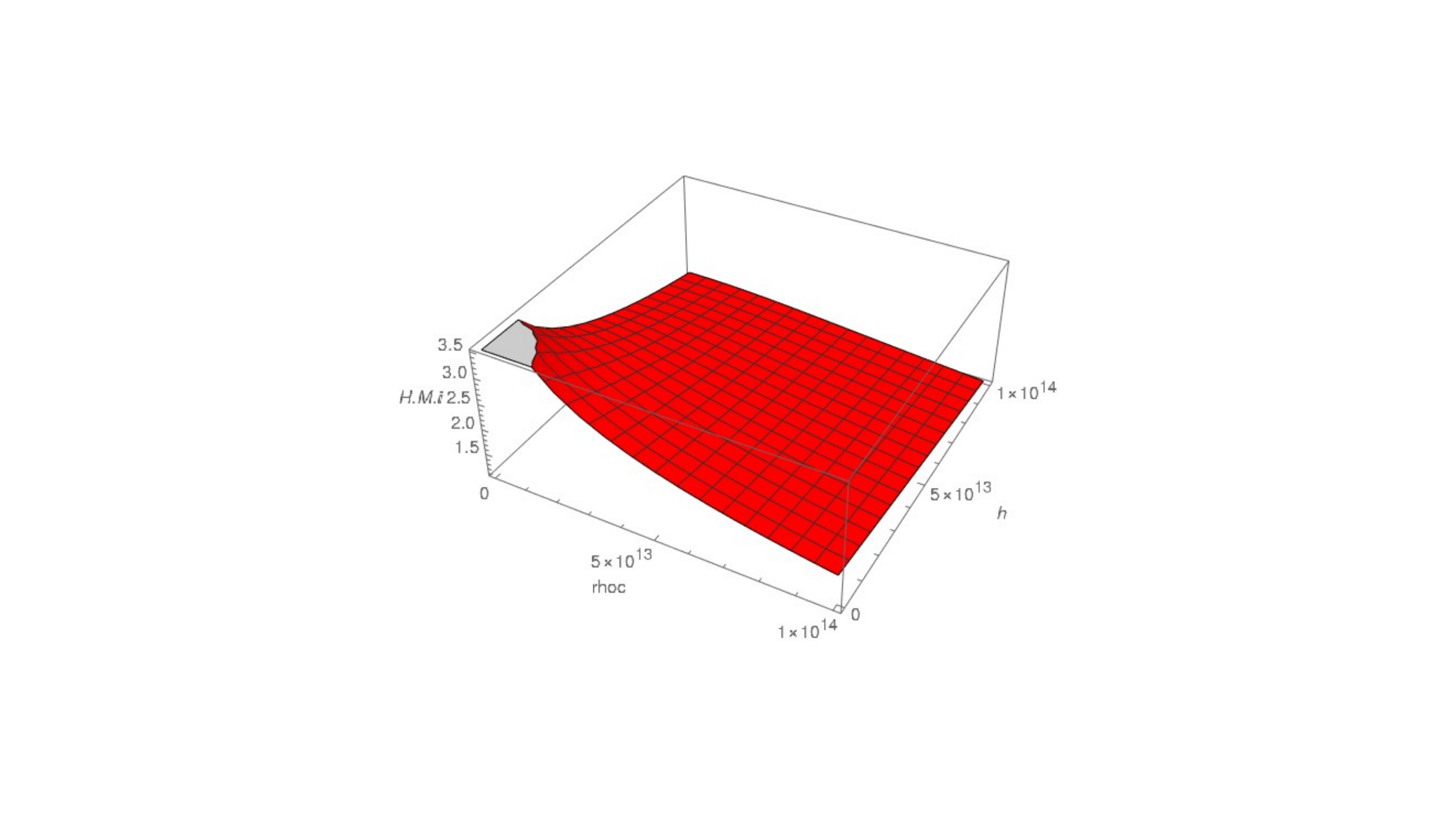}
\includegraphics[width=.65\textwidth]{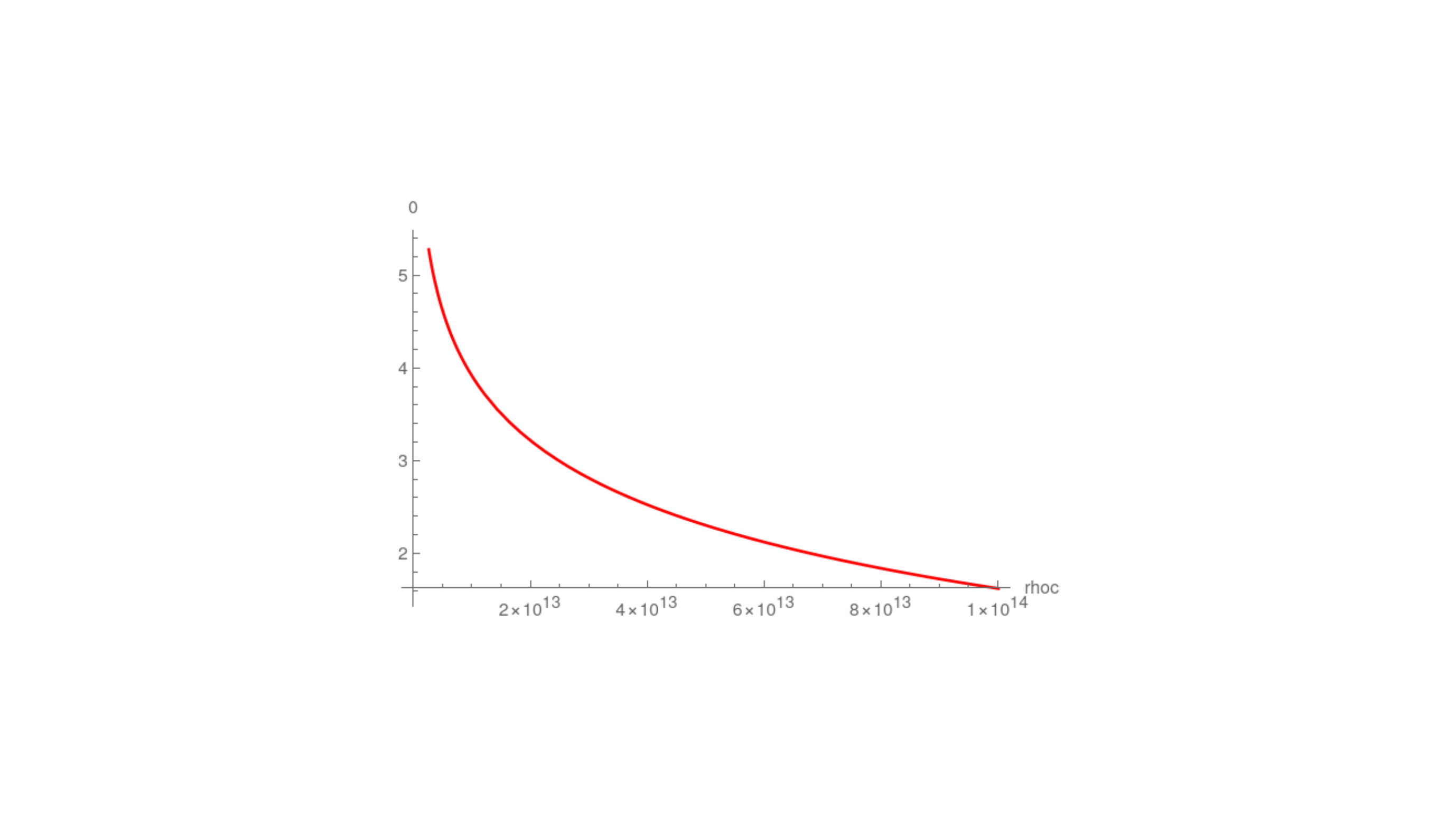}	

\caption{(left) \,:\, H.M.I as a function of ( $\rho_c$,h),  $l = {10}^{15}$\,\,,\,\, (right)\,:\, H.M.I as a function of $\rho_c$ for $l= {10}^{15}$, $h=0$\,:\,  Both the plots are showing at $h = 0$,  H.M.I diverge at zero cut off,  however become finite for nonzero $\rho_c$	
}
\label{hmirhoch1b1}
\end{figure}

\subsection{HMI for $d-\theta > 1$}

To write the expression of H.M.I first we recall the expression of HEE for $d - \theta > 1$ (\ref{entropylrhoc1}).   Substituting the same in the expression of HMI  for two strips of length l,  in their connected phase (\ref{H}),  we obtain the expression of HMI for $d - \theta > 1$

\ber
& &H.M.I\n
&=&   2{\frac{      R^{d-1}    L^{d-1} {\left({\left( {\left({\frac{A_{10} l}{ 2 }}\right)}^{2  \left( d - \theta\right)} + {\left( \rho_c \right)}^{2  \left( d - \theta\right)} \right)}^{\frac{1}{2(d - \theta)}}\right)}^{\theta - d +1} \,\, {{}_2 F_1}   \left\lbrack {\frac{1}{2}}, {\frac{1}{2}}( - 1 +{ \frac{1}{d - \theta }}), {\frac{1}{2}}(1 +{ \frac{1}{d - \theta }}) , 1 \right \rbrack }{  4 G_N (\theta + 1- d )    }}\n
                      &- & 2\left(\rho_c \right)^{ \theta +1- d} {\frac{  R^{d-1}      L^{d-1}  \,\, {{}_2 F_1}   \left\lbrack {\frac{1}{2}}, {\frac{1}{2}}( - 1 +{ \frac{1}{d - c}}), {\frac{1}{2}}(1 +{ \frac{1}{d - \theta}}) , \left({\frac{\rho_c}{\left\lbrack    
{\left( {\left({\frac{A_{10} l}{ 2 }}\right)}^{2  \left( d - \theta\right)} + {\left( \rho_c \right)}^{2  \left( d - \theta\right)} \right)}^{\frac{1}{2(d - \theta)}} \right\rbrack }}\right)^{2(d - \theta)}  \right \rbrack }{  4 G_N( \theta - d  + 1)   }}\n
&-&   {\frac{     R^{d-1}    L^{d-1} {\left({\left( {\left({\frac{A_{10} (2l + h)}{ 2 }}\right)}^{2  \left( d - \theta\right)} + {\left( \rho_c \right)}^{2  \left( d - \theta\right)} \right)}^{\frac{1}{2(d - \theta)}}\right)}^{\theta - d +1} \,\, {{}_2 F_1}   \left\lbrack {\frac{1}{2}}, {\frac{1}{2}}( - 1 +{ \frac{1}{d - \theta }}), {\frac{1}{2}}(1 +{ \frac{1}{d - \theta }}) , 1 \right \rbrack }{  4 G_N (\theta + 1- d )    }}\n
 &+ & \left(\rho_c \right)^{ \theta +1- d} {\frac{    R^{d-1}    L^{d-1}  \,\, {{}_2 F_1}   \left\lbrack {\frac{1}{2}}, {\frac{1}{2}}( - 1 +{ \frac{1}{d - \theta }}), {\frac{1}{2}}(1 +{ \frac{1}{d - \theta}}) , \left({\frac{\rho_c}{\left\lbrack    
{\left( {\left({\frac{A_{10} (2l + h)}{ 2 }}\right)}^{2  \left( d - \theta\right)} + {\left( \rho_c \right)}^{2  \left( d - \theta\right)} \right)}^{\frac{1}{2(d - \theta)}} \right\rbrack }}\right)^{2(d - \theta)}  \right \rbrack }{  4 G_N( \theta - d  + 1)   }}\n
&-&    {\frac{   R^{d-1}      L^{d-1} {\left({\left( {\left({\frac{A_{10} h}{ 2 }}\right)}^{2  \left( d - \theta\right)} + {\left( \rho_c \right)}^{2  \left( d - \theta\right)} \right)}^{\frac{1}{2(d - \theta)}}\right)}^{\theta - d +1} \,\, {{}_2 F_1}   \left\lbrack {\frac{1}{2}}, {\frac{1}{2}}( - 1 +{ \frac{1}{d - \theta }}), {\frac{1}{2}}(1 +{ \frac{1}{d - \theta }}) , 1 \right \rbrack }{  4 G_N (\theta + 1- d )    }}\n
&+& \left(\rho_c \right)^{ \theta +1- d} {\frac{    R^{d-1}    L^{d-1}  \,\, {{}_2 F_1}   \left\lbrack {\frac{1}{2}}, {\frac{1}{2}}( - 1 +{ \frac{1}{d - \theta}}), {\frac{1}{2}}(1 +{ \frac{1}{d - \theta}}) , \left({\frac{\rho_c}{\left\lbrack    
{\left( {\left({\frac{A_{10} h}{ 2 }}\right)}^{2  \left( d - \theta\right)} + {\left( \rho_c \right)}^{2  \left( d - \theta\right)} \right)}^{\frac{1}{2(d - \theta)}} \right\rbrack }}\right)^{2(d - \theta)}  \right \rbrack }{  4 G_N( \theta - d  + 1)   }}
\la{hmibgreaterthan1}
\eer

Given the fact that the global expression for the turning point, is exact in the regime $l>> \rho_c$ and $\rho_c>> l$ only,  so here we will use (\ref{hmibgreaterthan1}),  obtain the 3D and 2D plots of H.M.I vs $(l,\rho_c)$ for some fixed h    and H.M.I vs $(h,  l)$ for some fixed cut off $\rho_c$,   and different $ d- \theta$ , study its feature in the regime  $l>> \rho_c$ and $\rho_c>> l$, while we understand in the rest of the regime in $(l,\rho_c)$ plane it is an interpolating functiom between these two regime.  Here we present all the relevant plots in Fig.(\ref{hmibasic3by2}, \ref{hmibasic7by3}, \ref{hmibasich1122}, \ref{hmirhoch3by2},  \ref{hmievbgreater1} ),  will show HMI is giving the expected behaviour and new features.  

\begin{figure}[H]
\begin{center}
\textbf{ For $ d - \theta > 1$, with  $ d - \theta  = {\frac{3}{2}}$\,: \,H.M.I vs $(l,\rho_c) $, H.M.I vs l and H.M.I vs $\rho_c$ plots for fixed h}
\end{center}
\vskip2mm
\includegraphics[width=.55\textwidth]{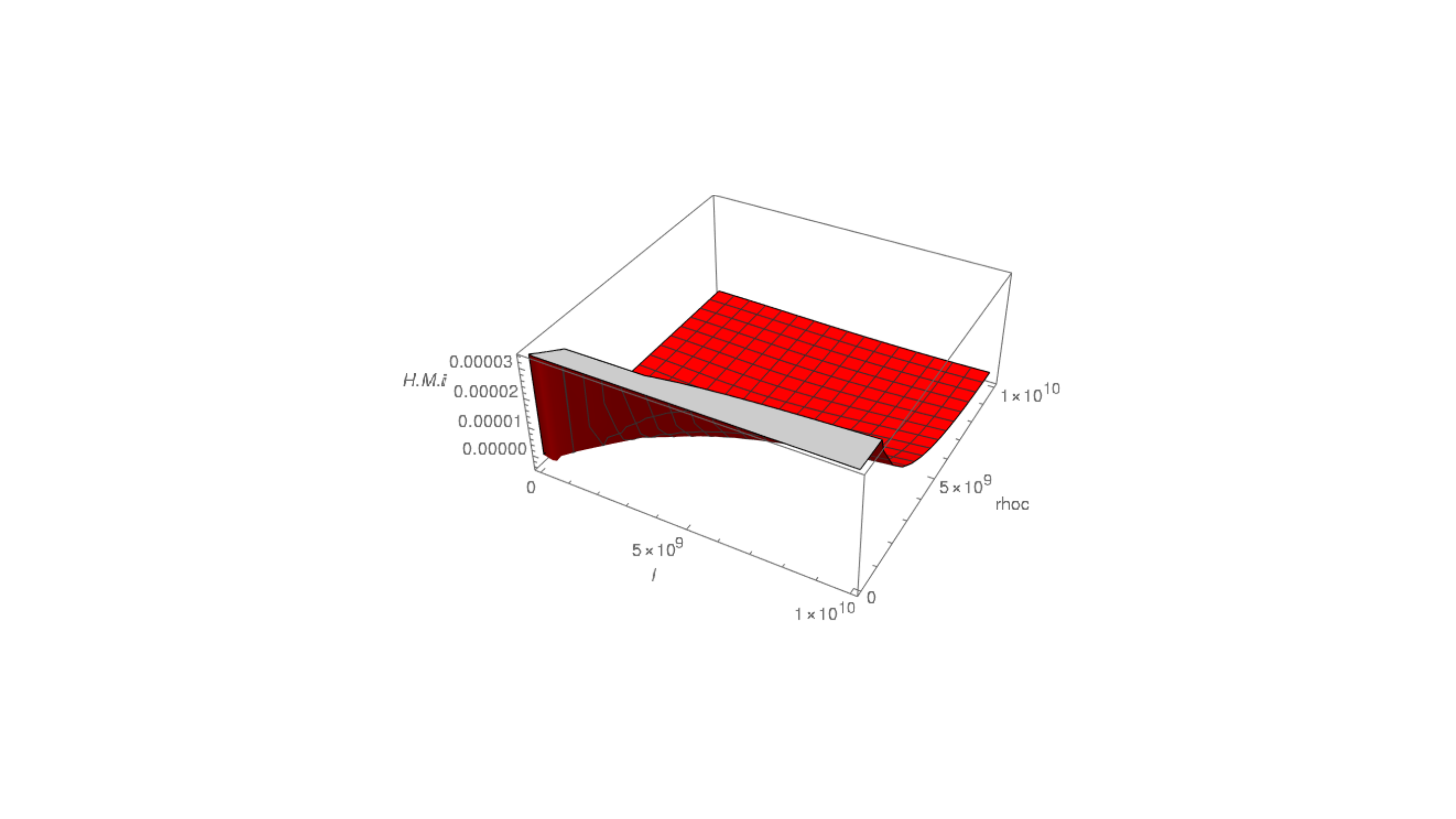}
\includegraphics[width=.55\textwidth]{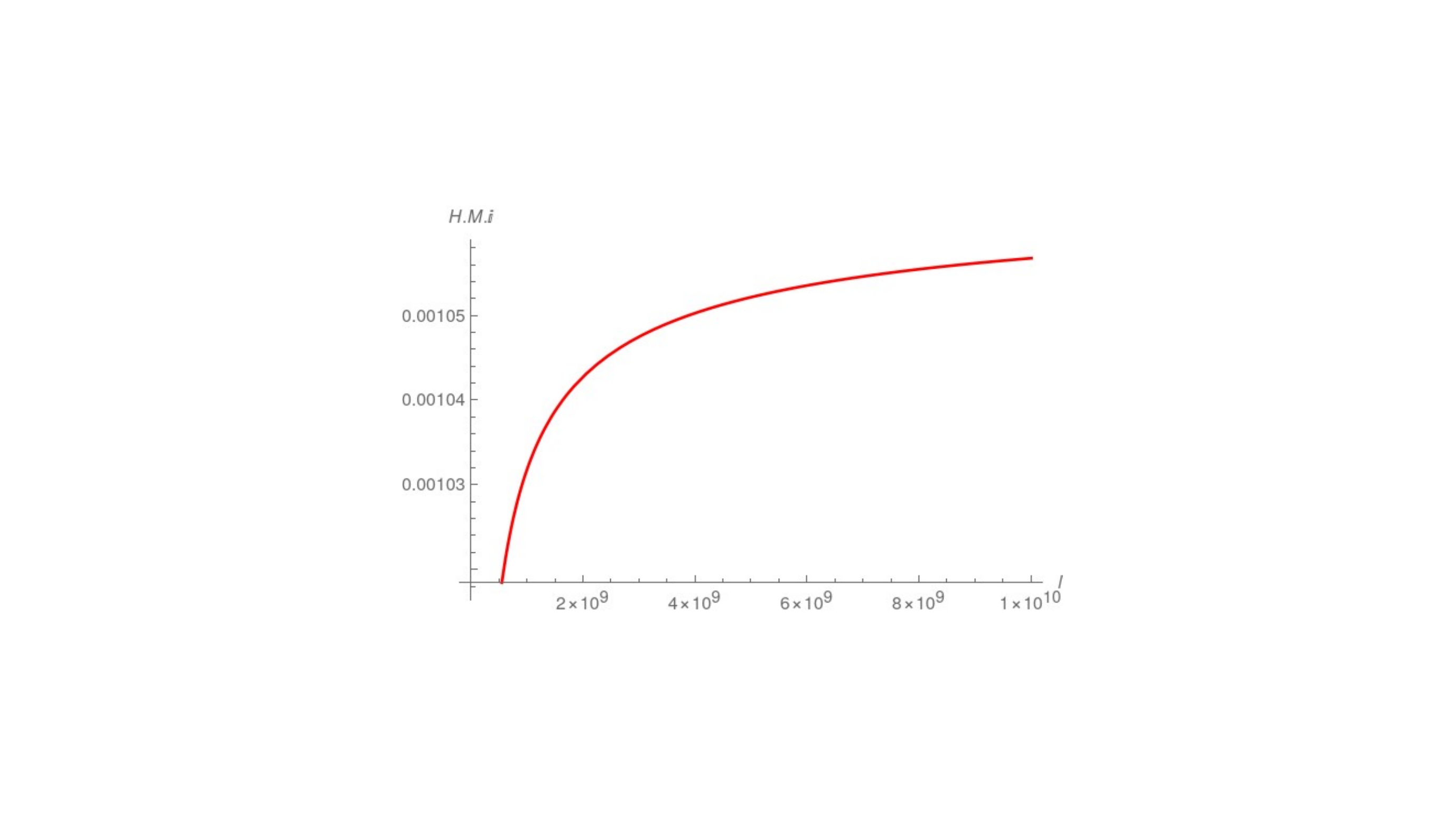}	
\includegraphics[width=.55\textwidth]{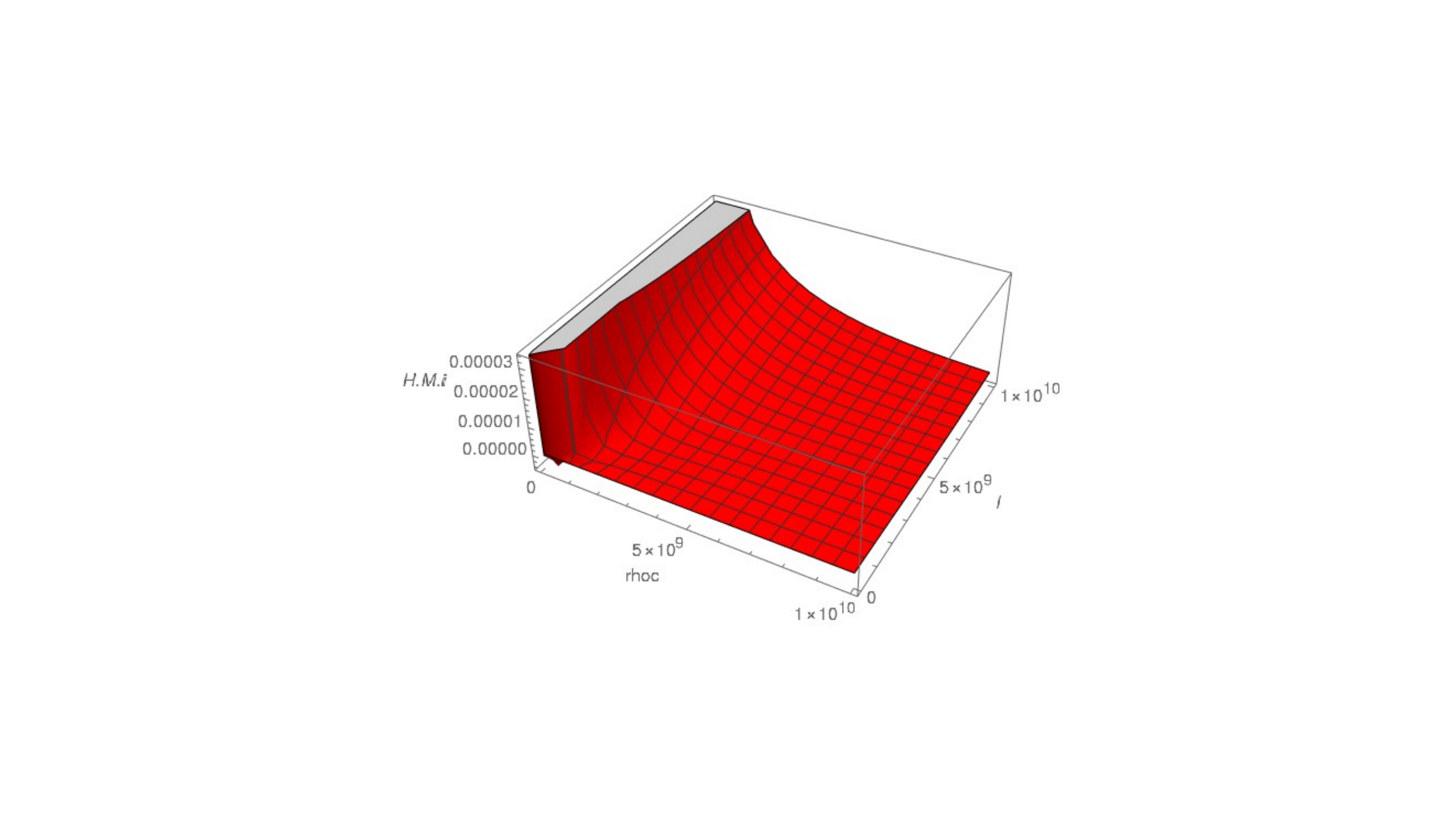}
\includegraphics[width=.55\textwidth]{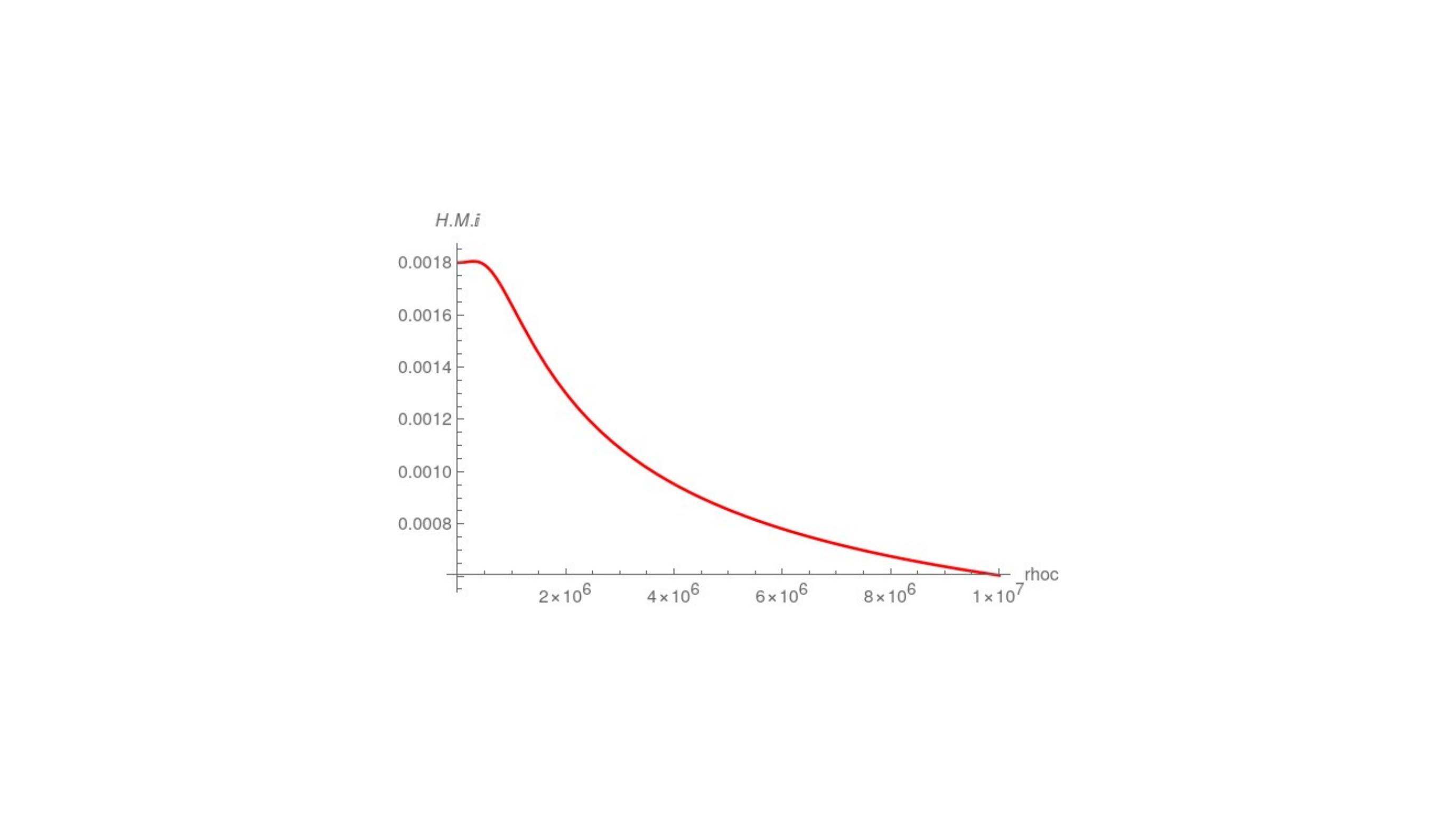}
\includegraphics[width=.55\textwidth]{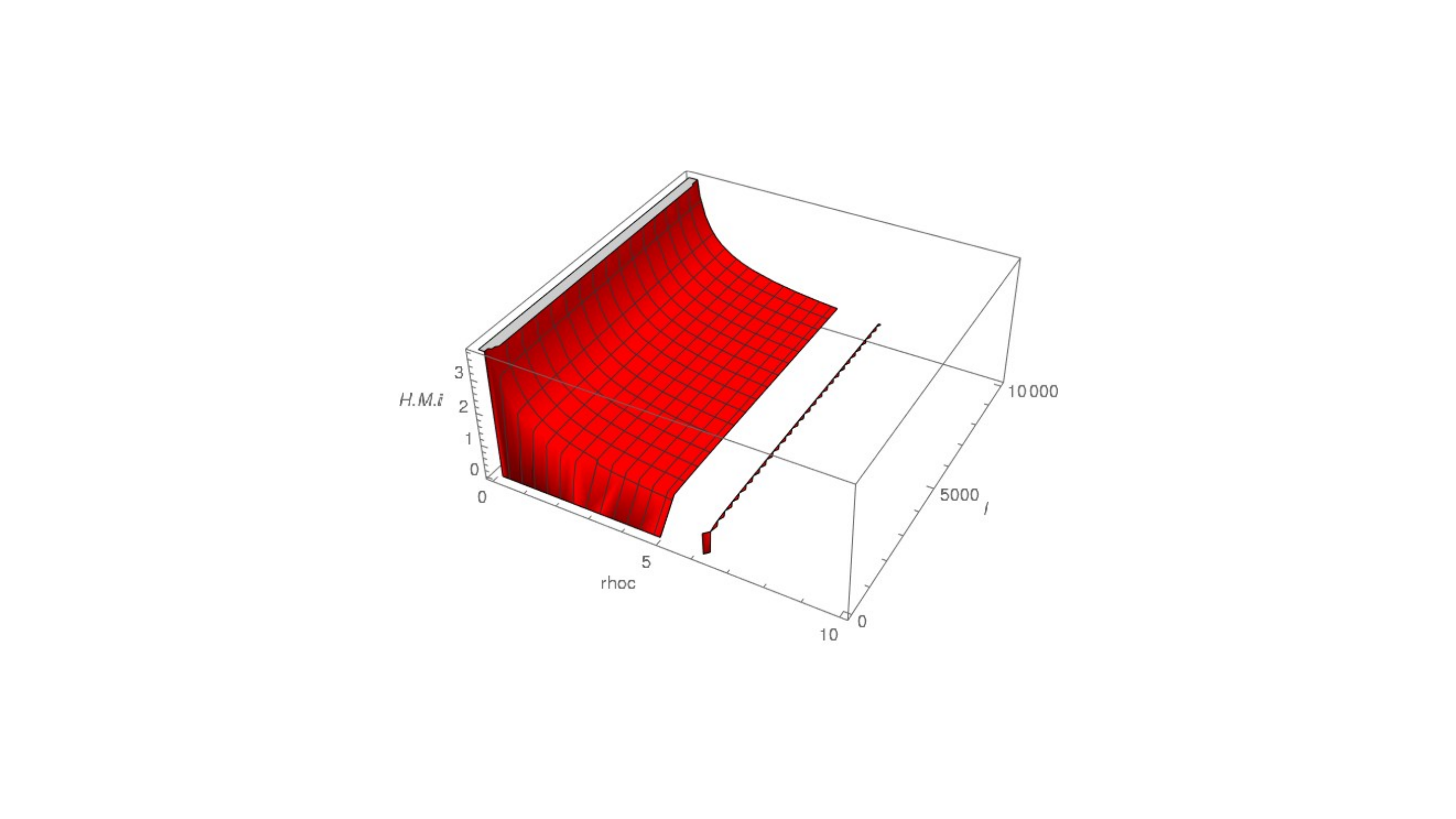}
\includegraphics[width=.55\textwidth]{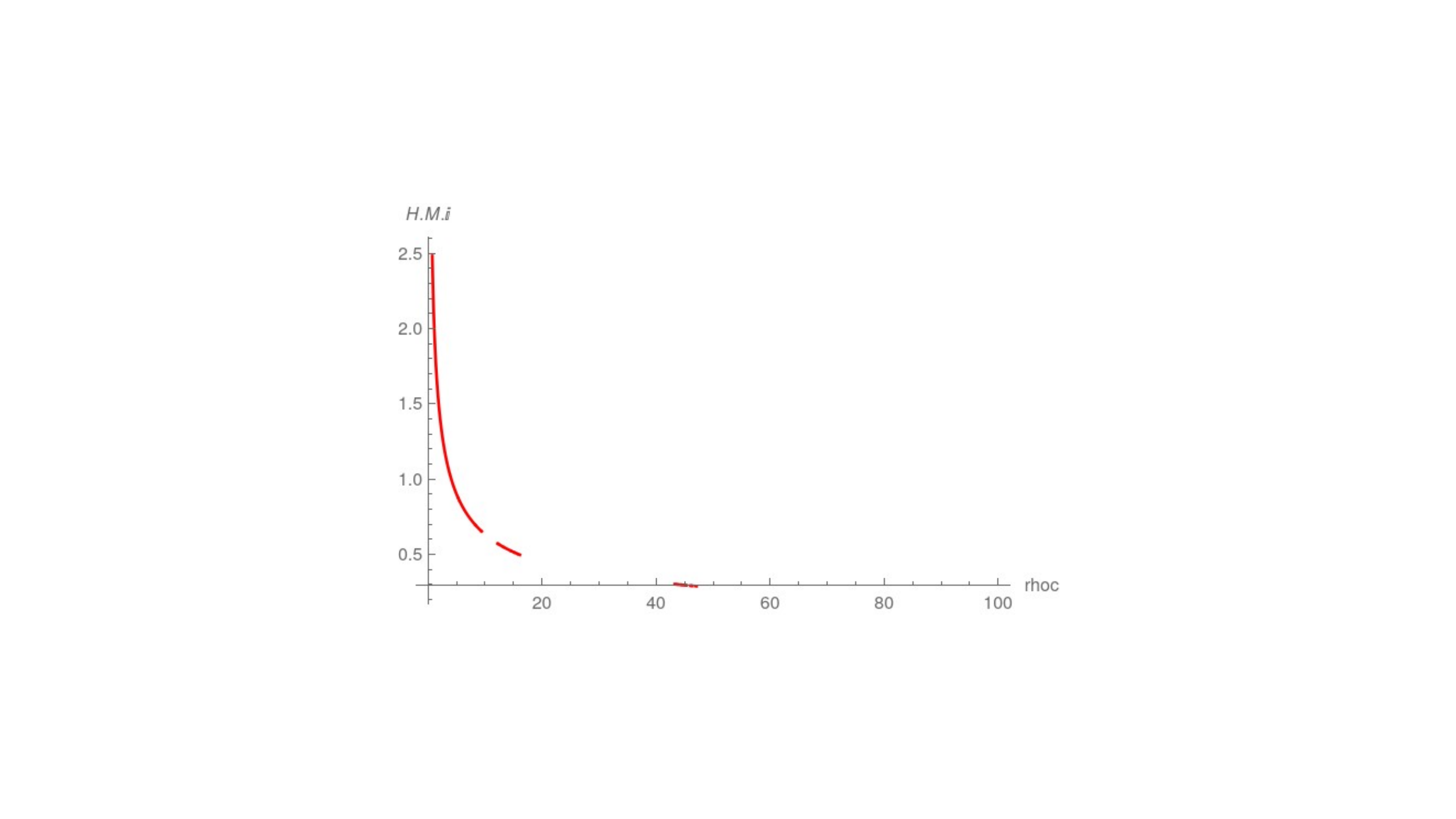}
\caption{(First row) \,\,: (left)\,:\, H.M.I as a function of $(l,\rho_c)$ for  $h = (10)^6$ \,,\, (right)\, : \, H.M.I  as a function of l with $\rho_c = 1$,  $h = (10)^6$  \,:\,  Both the plots are showing when we increase l gradually from zero, for fixed h,   at certain point given in terms of $h_{\rm crit}$, H.M.I is undergoing a first order phase transition and also for a given $\rho_c$,   H.M.I increases with the increase of l
  \quad;\quad (Second row) \,:\,  ( left)\, : \, H.M.I  as a function of $(\rho_c , l)$ for  $h = (10)^6$\,, (right) \,:\, H.M.I  as a function of $\rho_c$  for $l = {(10)}^{10}$,$h = {(10)}^6$,  this value of l is chosen to probe $l>> \rho_c$ regime where for $\rho_c >> l$ H.M.I is zero \, :\,  Both the 3D and 2D plots are showing, for a given l, H.M.I falls with the increase of cut off $\rho_c$ and goes to zero for $\rho_c >> l $ regime.  Also for nonzero h,    H.M.I is finite at $\rho_c = 0$
 \quad;\quad (Last row) \,:\,  ( left)\, : \, H.M.I  as a function of $(\rho_c , l)$ for  $h = 0 $\,, (right) \,:\, H.M.I  as a function of $\rho_c$  for $l = {10}^{10}$,$h = 0$,   \, :\,  Both the 3D and 2D plots are showing for $h = 0$  H.M.I diverge at $\rho_c = 0$ as expected}
\label{hmibasic3by2}
\end{figure}

\begin{figure}[H]
\begin{center}
\textbf{For $ d - \theta > 1$, with  $ d - \theta  = {\frac{7}{3}}$\,: \,H.M.I vs $(l,\rho_c) $, H.M.I vs l and H.M.I vs $\rho_c$ plots for fixed h}
\end{center}
\vskip2mm
\includegraphics[width=.55\textwidth]{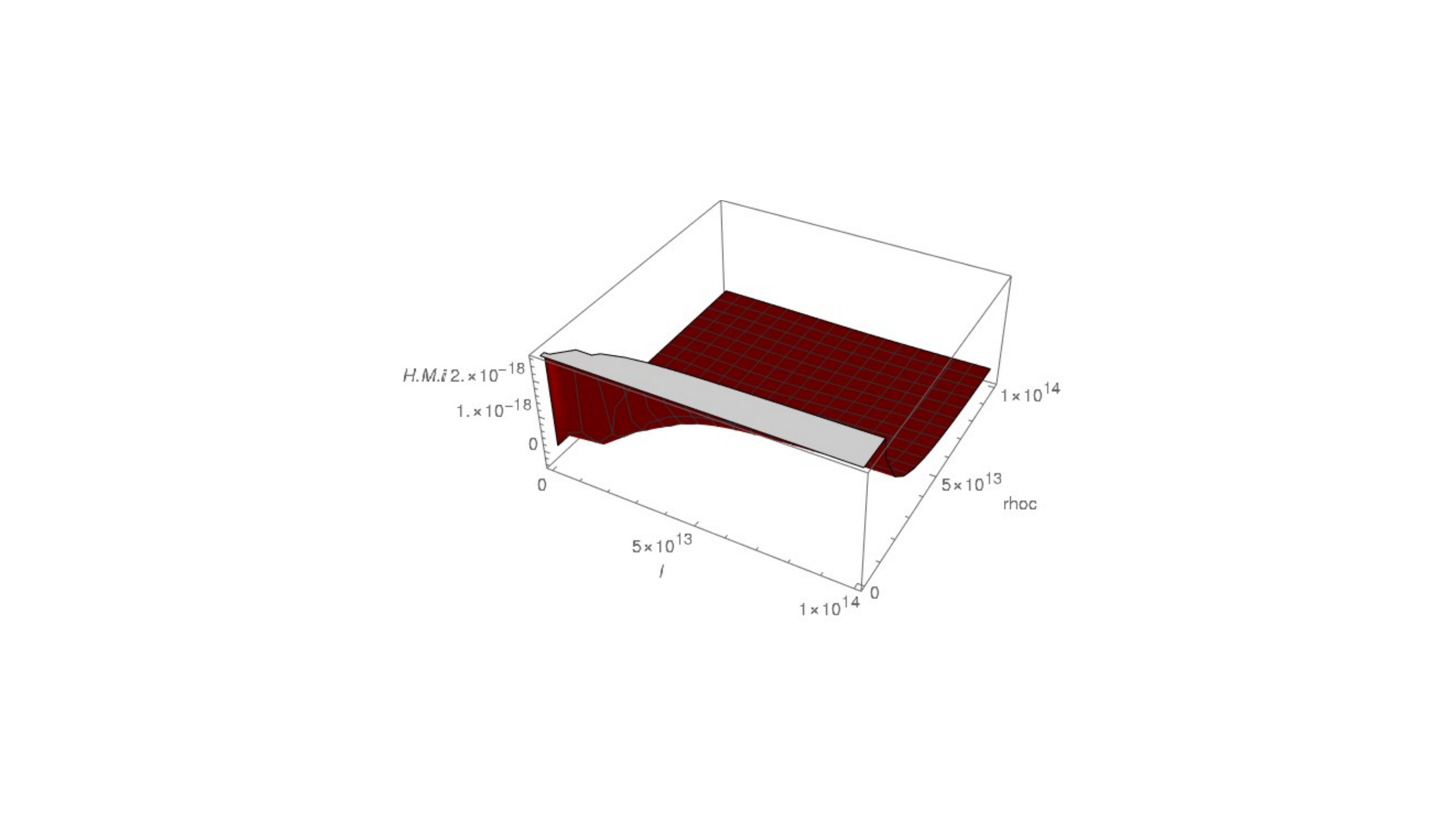}
\includegraphics[width=.55\textwidth]{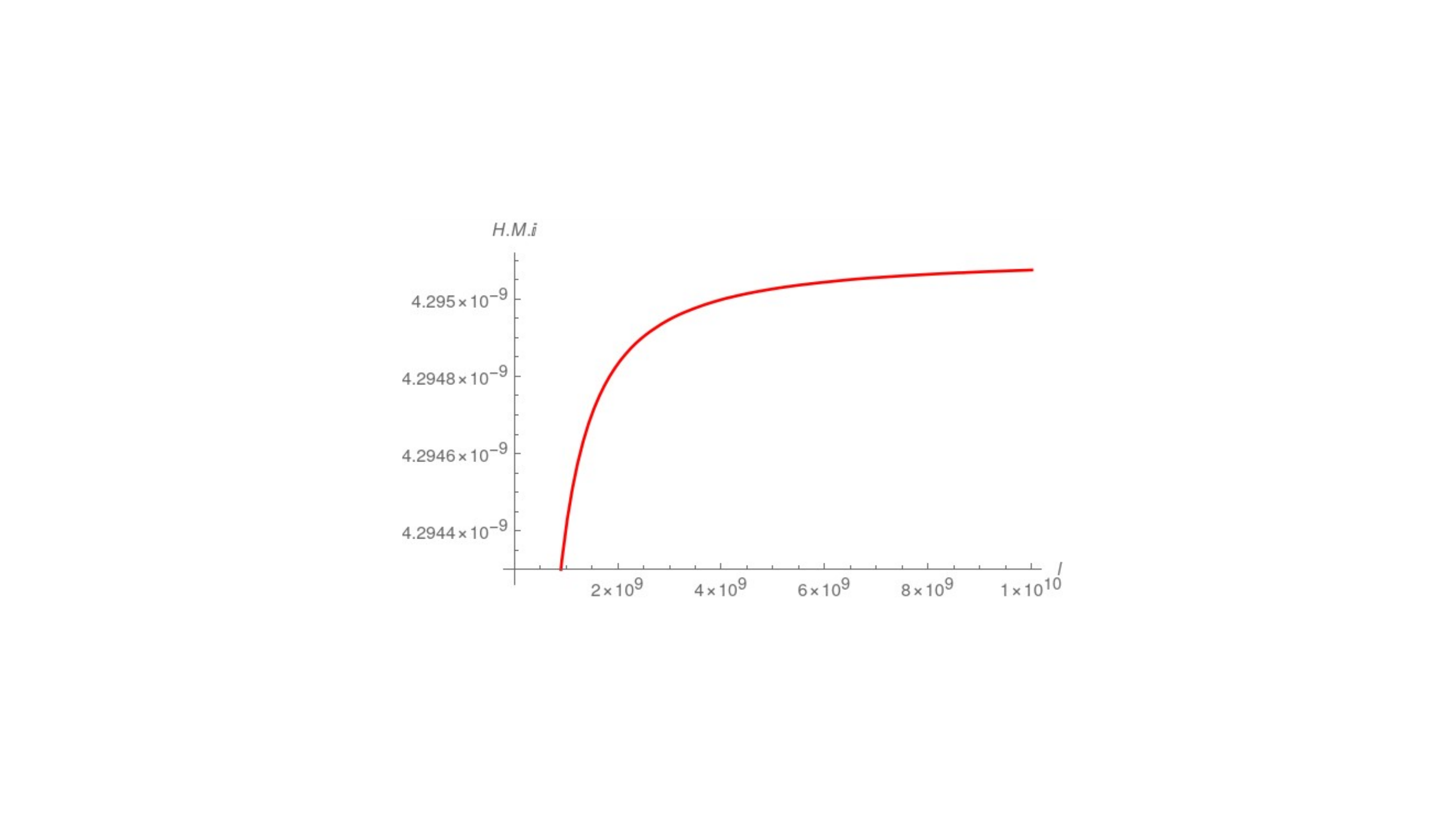}	
\includegraphics[width=.55\textwidth]{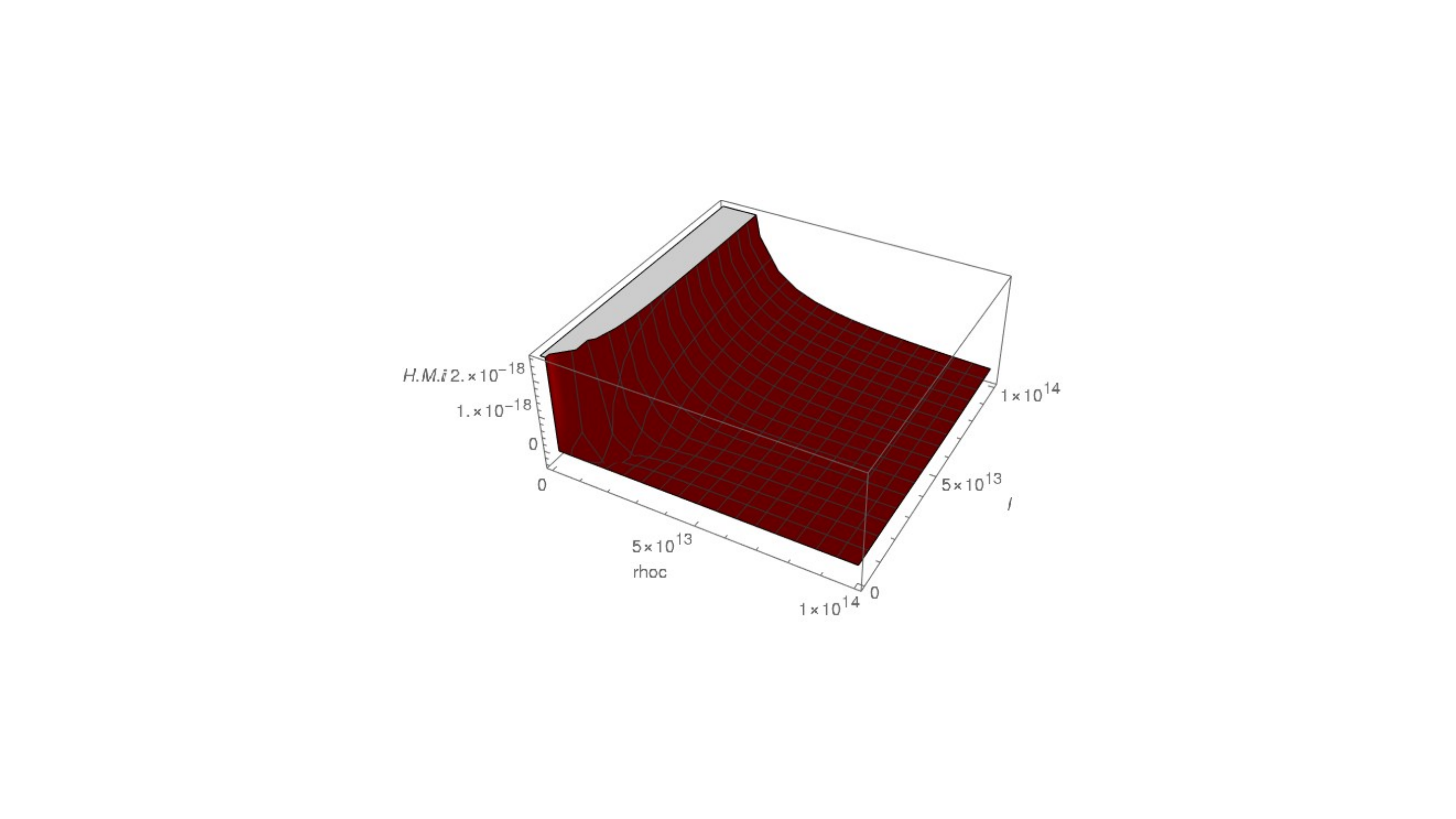}
\includegraphics[width=.55\textwidth]{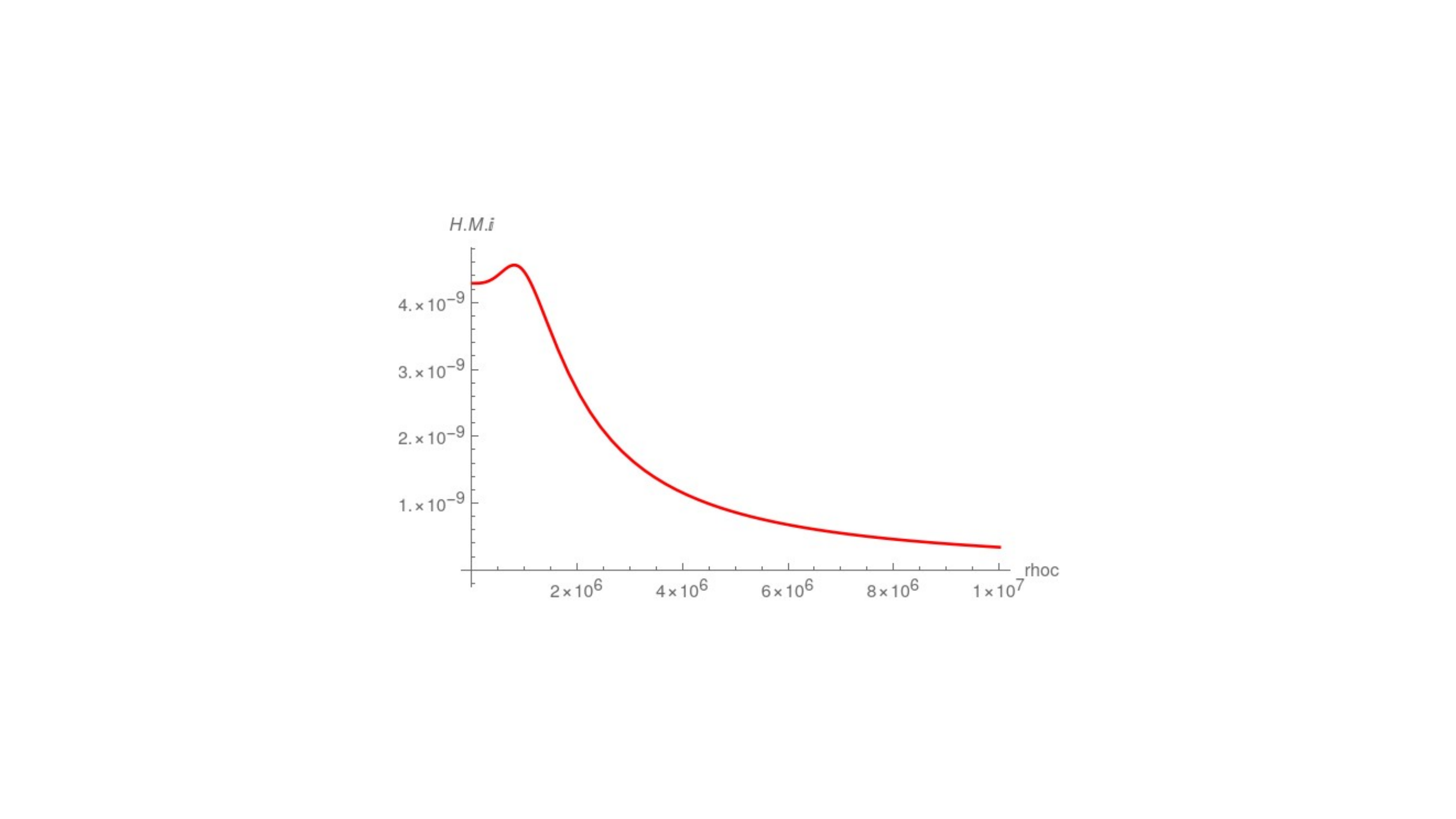}
\includegraphics[width=.55\textwidth]{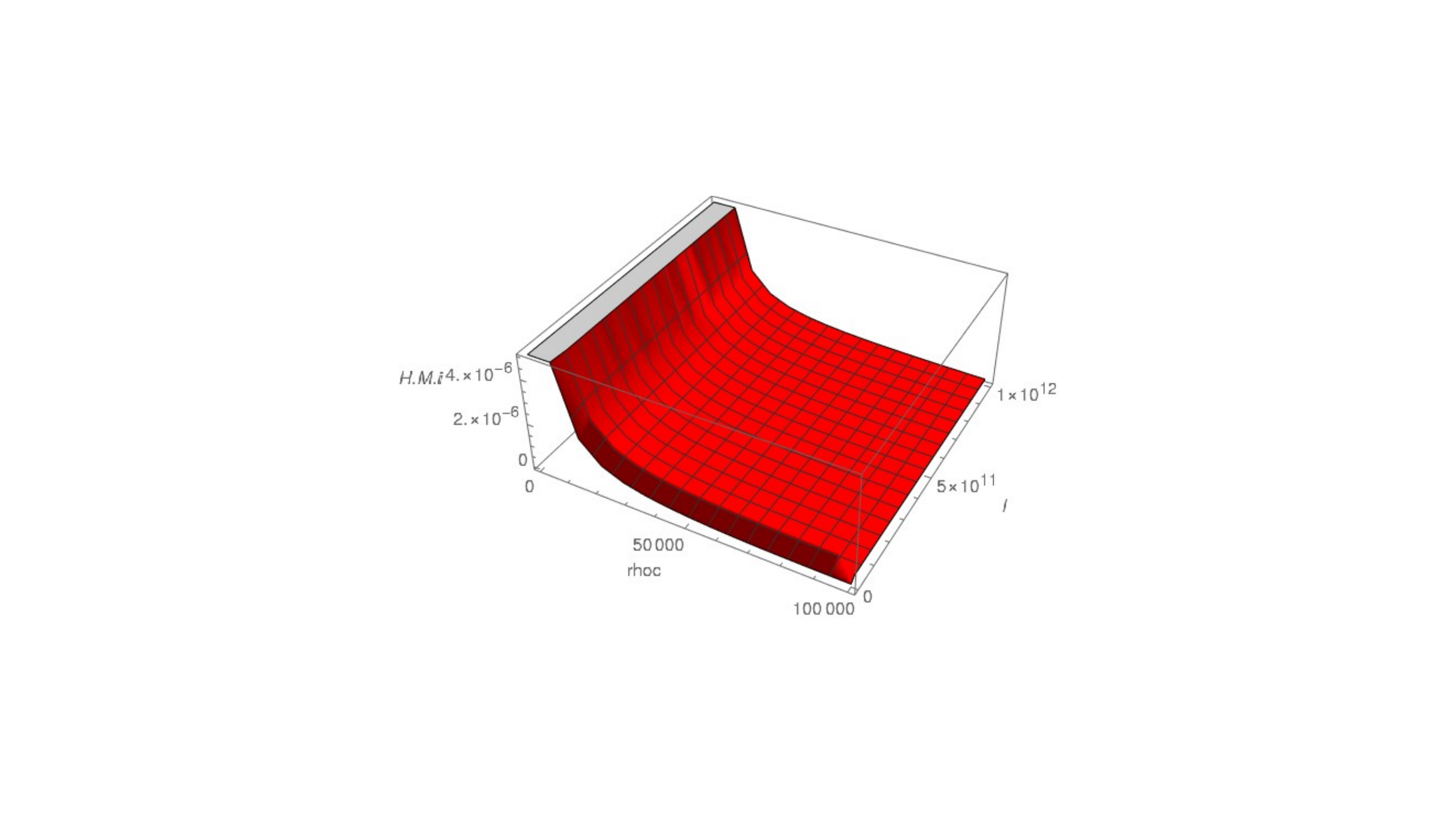}
\includegraphics[width=.55\textwidth]{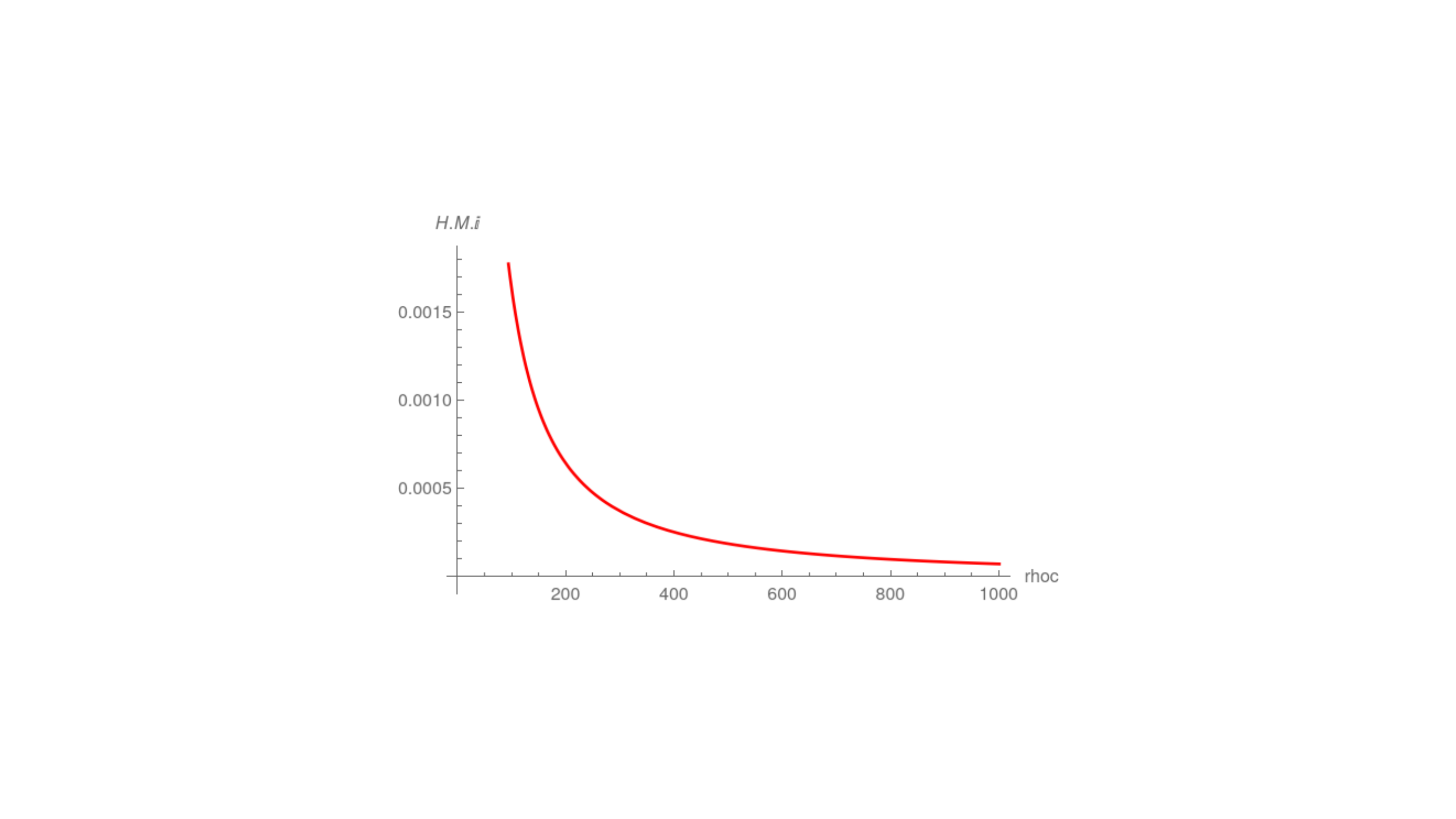}
\caption{(First row) \,\,: (left)\,:\, H.M.I as a function of $(l,\rho_c)$ for  $h = (10)^6$ \,,\, (right)\, : \, H.M.I  as a function of l with $\rho_c = 1$,  $h = (10)^6$  \,:\,  Both the plots are showing when we increase l gradually from zero, for fixed h,   at certain point given in terms of $h_{\rm crit}$, H.M.I is undergoing a first order phase transition and also for a given $\rho_c$,   H.M.I increases with the increase of l
  \quad;\quad (Second row) \,:\,  ( left)\, : \, H.M.I  as a function of $(\rho_c , l)$ for  $h = (10)^6$\,, (right) \,:\, H.M.I  as a function of $\rho_c$  for $l = {(10)}^{10}$,$h = {(10)}^6$,  this value of l is chosen to probe $l>> \rho_c$ regime where for $\rho_c >> l$ H.M.I is zero \, :\,  Both the 3D and 2D plots are showing, for a given l, H.M.I falls with the increase of cut off $\rho_c$ and goes to zero for $\rho_c >> l $ regime.  Also for nonzero h,    H.M.I is finite at $\rho_c = 0$
 \quad;\quad (Last row) \,:\,  ( left)\, : \, H.M.I  as a function of $(\rho_c , l)$ for  $h = 0 $\,, (right) \,:\, H.M.I  as a function of $\rho_c$  for $l = {10}^{10}$,$h = 0$,   \, :\,  Both the 3D and 2D plots are showing for $h = 0$  H.M.I diverge at $\rho_c = 0$ as expected
}
\label{hmibasic7by3}
\end{figure}

`

\begin{figure}[H]
\begin{center}
\textbf{ For $ d - \theta > 1$, H.M.I as a function of h,l  }
\end{center}
\vskip2mm
\includegraphics[width=.65\textwidth]{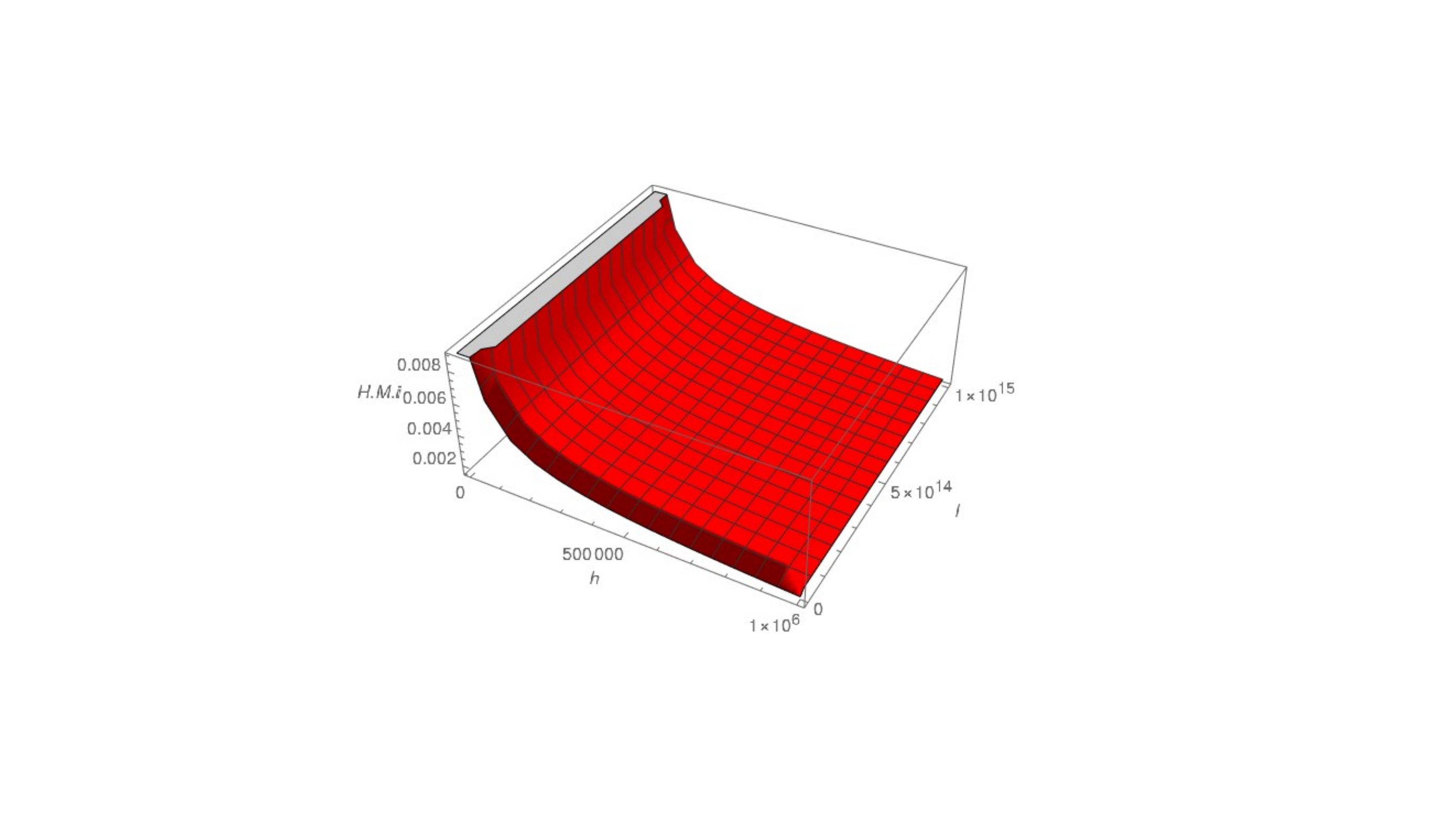}
\includegraphics[width=.65\textwidth]{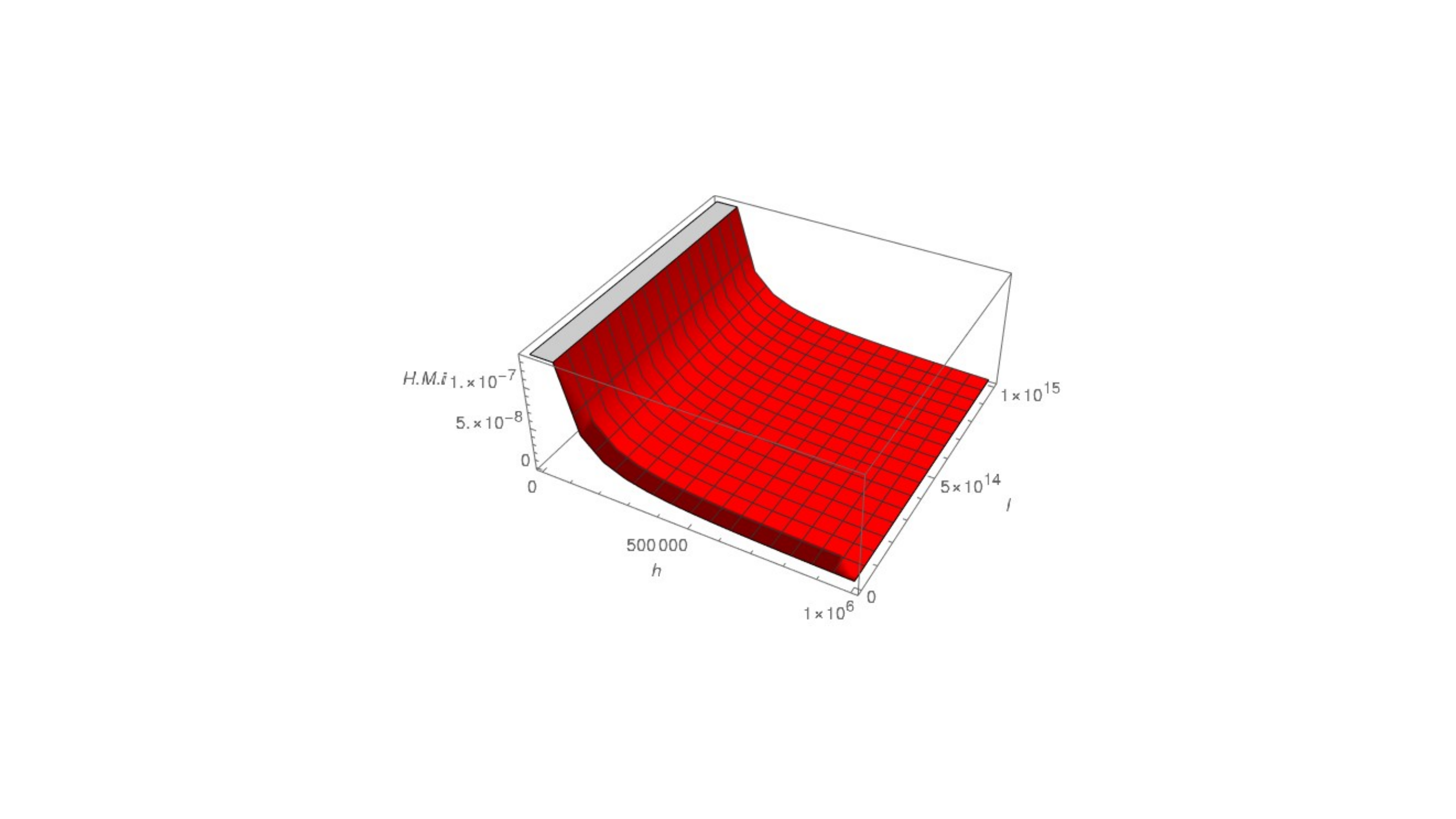}	

\caption{H.M.I is plotted as a function of (h,l)  for  $\rho_c = 10 $\,,\, (left) \,,\, for $d - \theta = {\frac{3}{2}}$ \,,\, (right) \,,\, for \,$d - \theta = {\frac{7}{3}}$\,,\. both the plots are showing for a given l,\, H.M.I decreases with the increase of h and for a given h, H.M.I increases with the increase of l.  
}
\label{hmibasich1122}
\end{figure}

\begin{figure}[H]
\begin{center}
\textbf{H.M.I as a function of ($\rho_c$ , h ) for fixed l  }
\end{center}
\vskip2mm
\includegraphics[width=.65\textwidth]{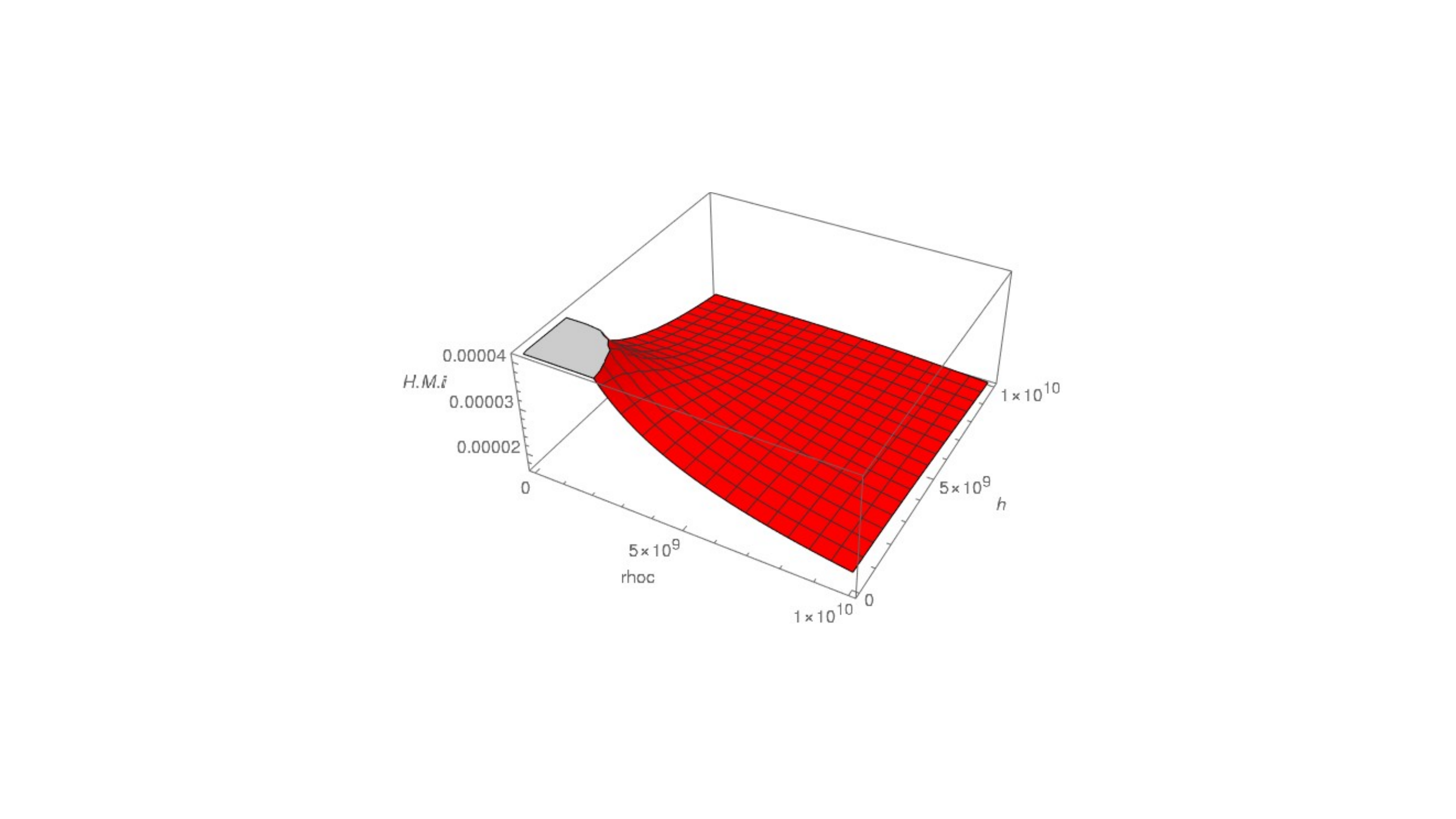}
\includegraphics[width=.65\textwidth]{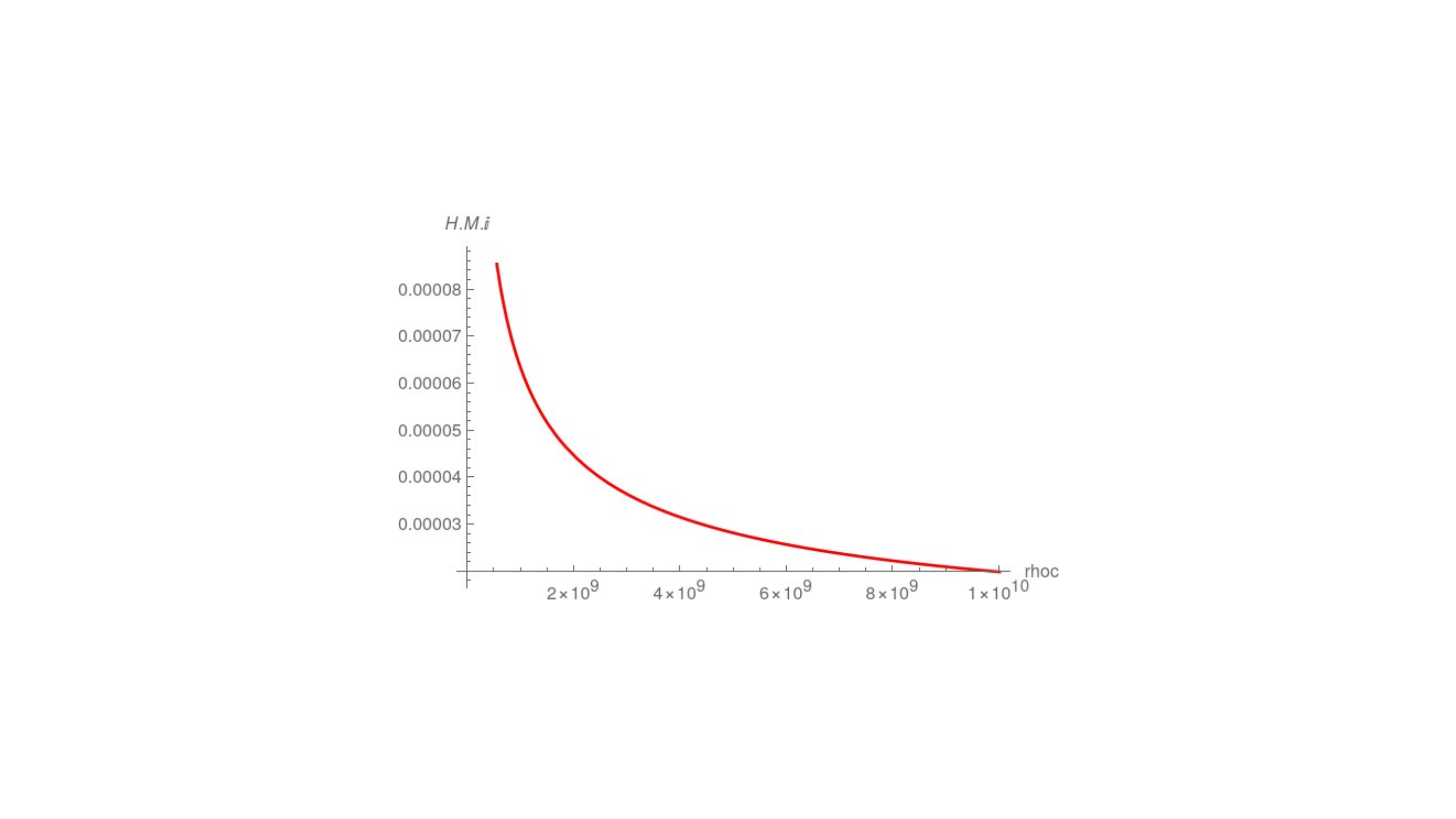}	
\includegraphics[width=.65\textwidth]{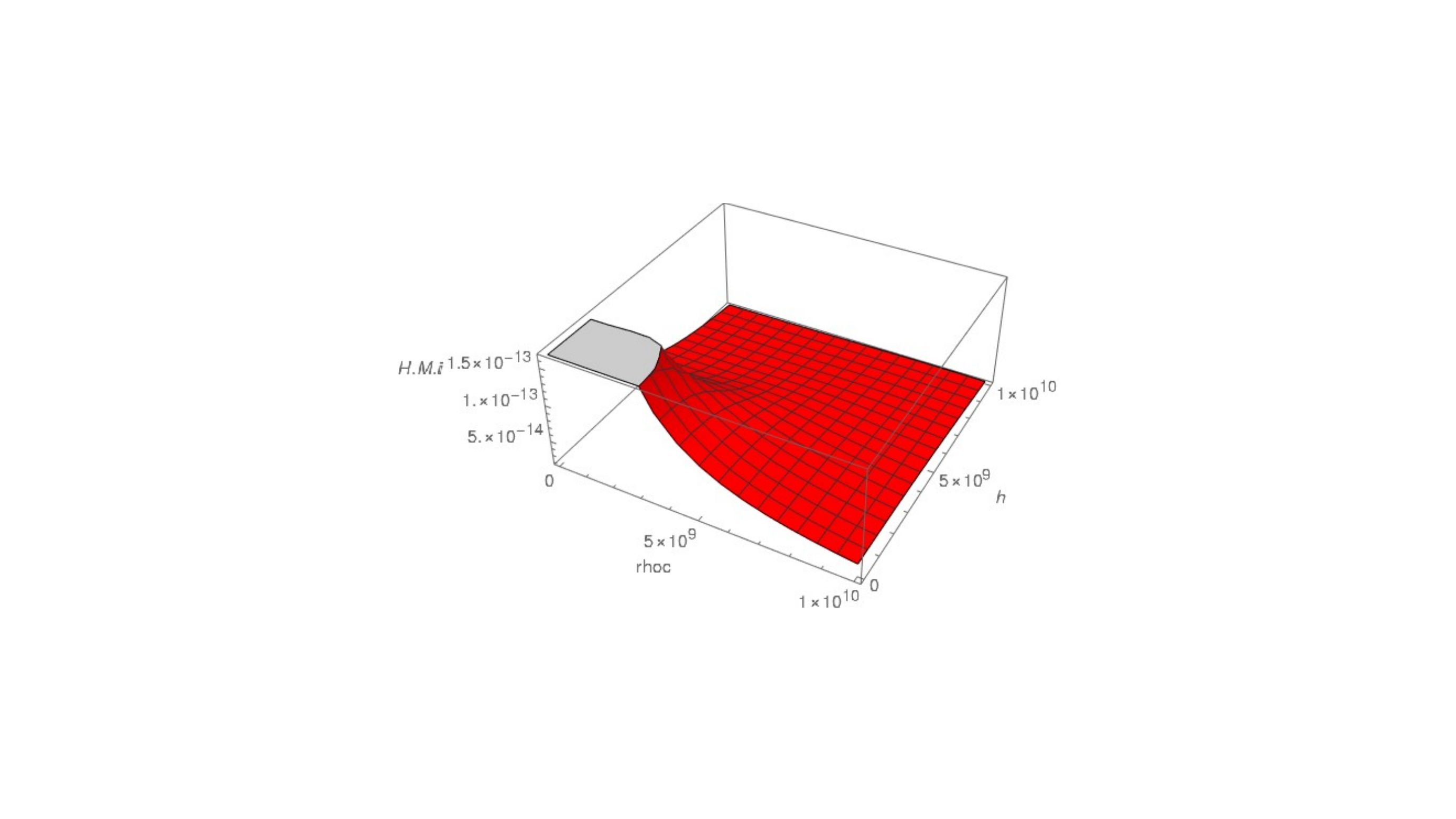}
\includegraphics[width=.65\textwidth]{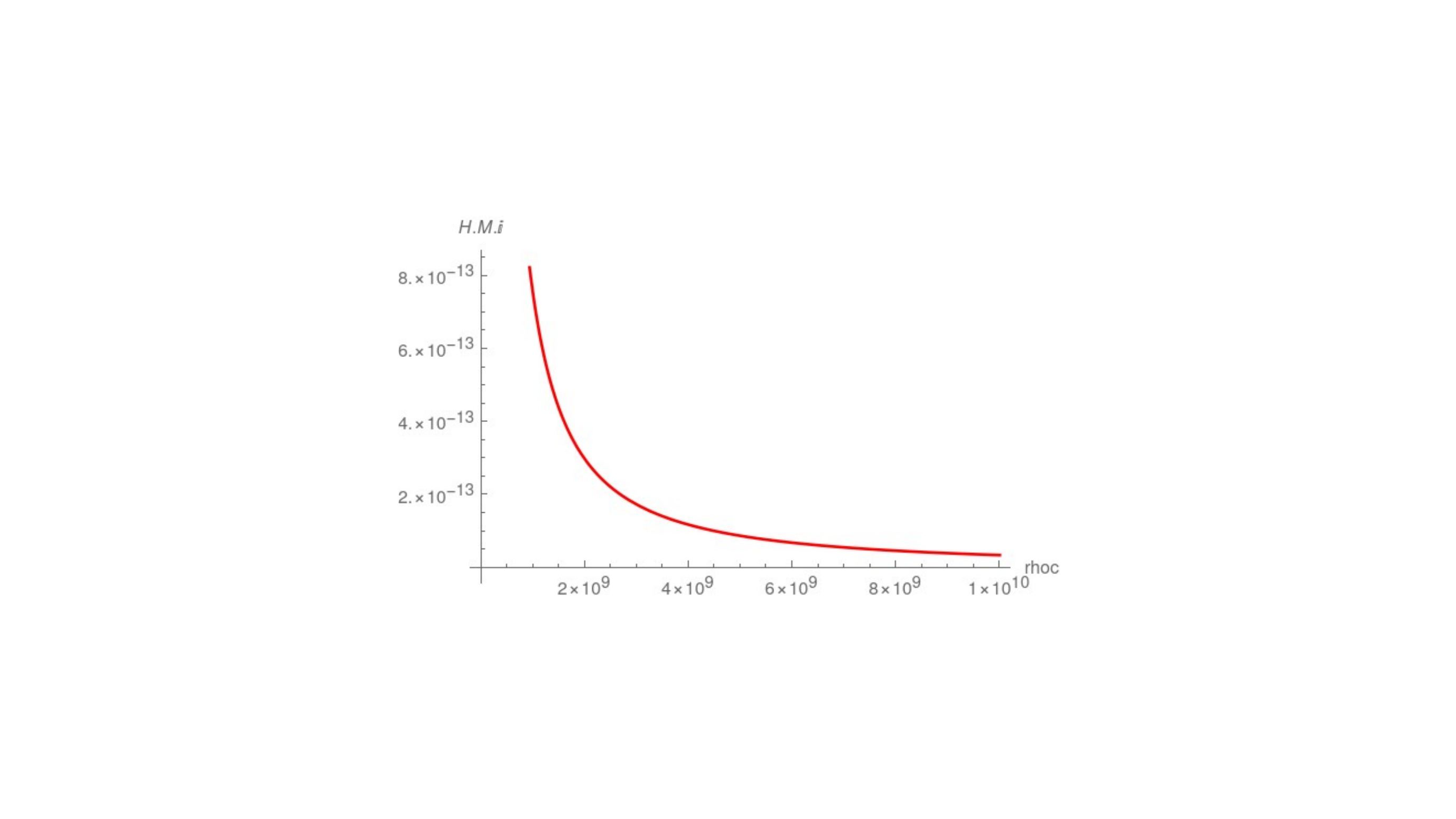}
\caption{(First row) \,\,:\,\,(left) \,:\, H.M.I as a function of $\rho_c$,h for $d - \theta = {\frac{3}{2}}$,  $l = {10}^{17}	$\,\,,\,\, (right)\,:\, H.M.I as a function of $\rho_c$ for $l= {10}^{10}, $h=0$	$,  \quad;\quad (Last row) \,:\,  ( left)\, : \, H.M.I plotted as a function of $(\rho_c , l)$ \, ,\,  $d - \theta= {\frac{7}{3}}$  \,\,,(right)\, :\, H.M.I plotted as a function of $\rho_c$  for $l = {10}^{10}$, $ h=0$ this value of l is chosen to probe $l>> \rho_c$ regime where for $\rho_c >> l$ H.M.I is zero \, :\,  Both the 3D and 2D plots  for both the values of $d - \theta$ are showing, for a given l, H.M.I falls with the increase of cut off $\rho_c$ and goes to zero for $\rho_c >> l $ regime. Also for $h = 0$ for zero cut off, H.M.I diverge , as expected but the very moment we make cut-off finite, H.M.I also become finite
}
\label{hmirhoch3by2}
\end{figure}

\begin{figure}[H]
\begin{center}
\textbf{ For $ d - \theta > 1$ \,: \,  The evolution of H.M.I with $ d - \theta$  }
\end{center}
\vskip2mm
\includegraphics[width=.65\textwidth]{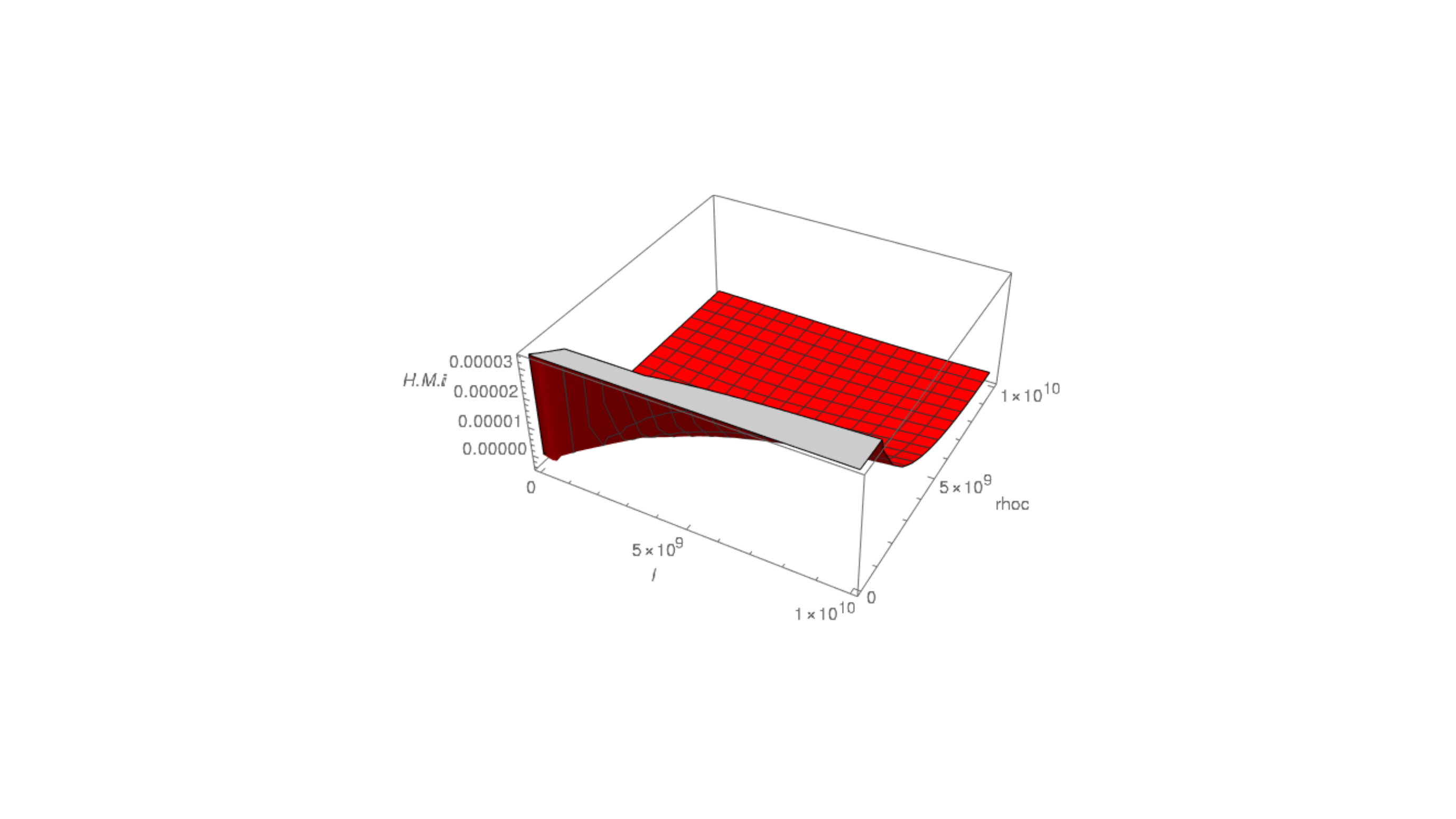}
\includegraphics[width=.65\textwidth]{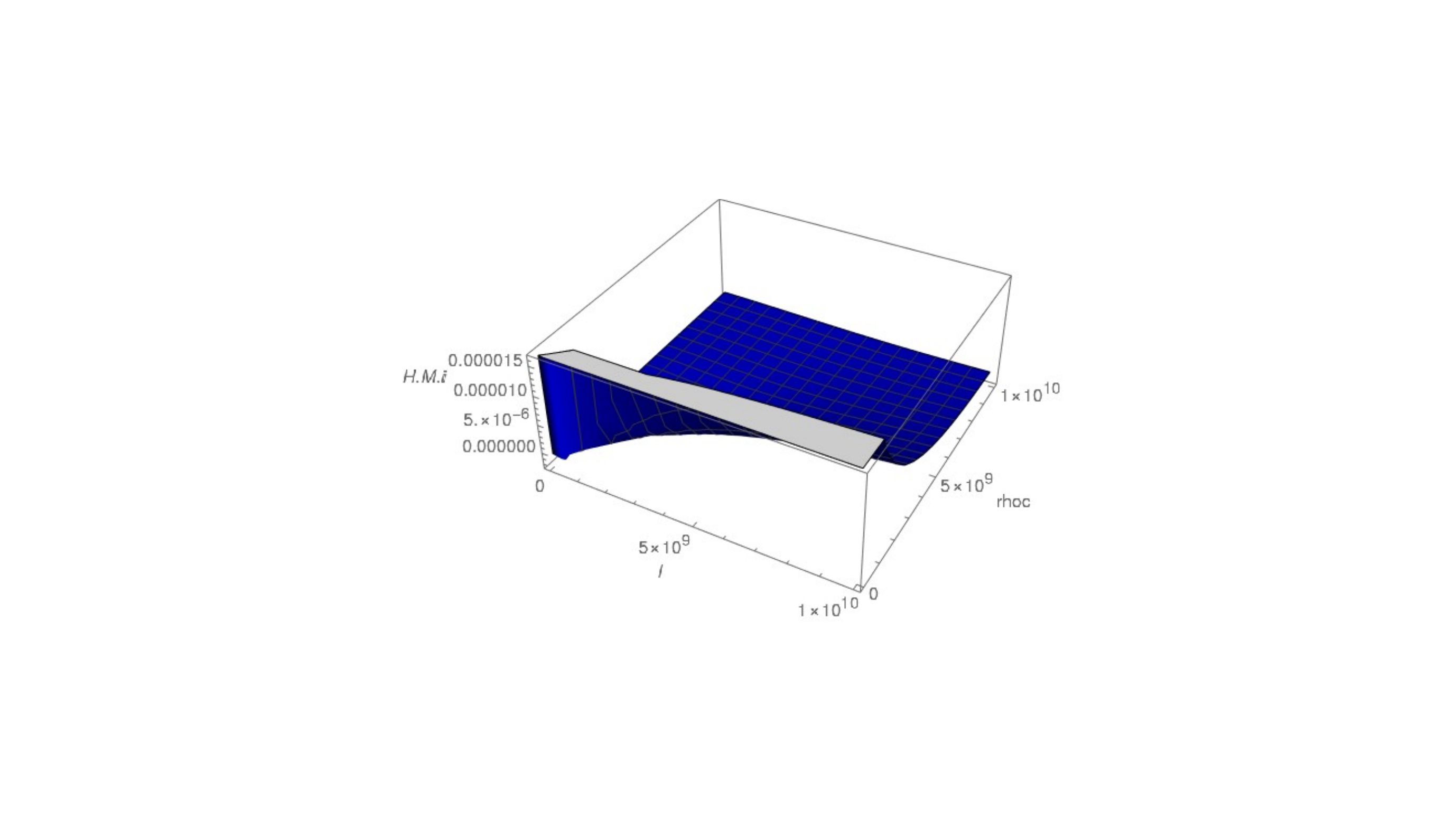}	
\includegraphics[width=.65\textwidth]{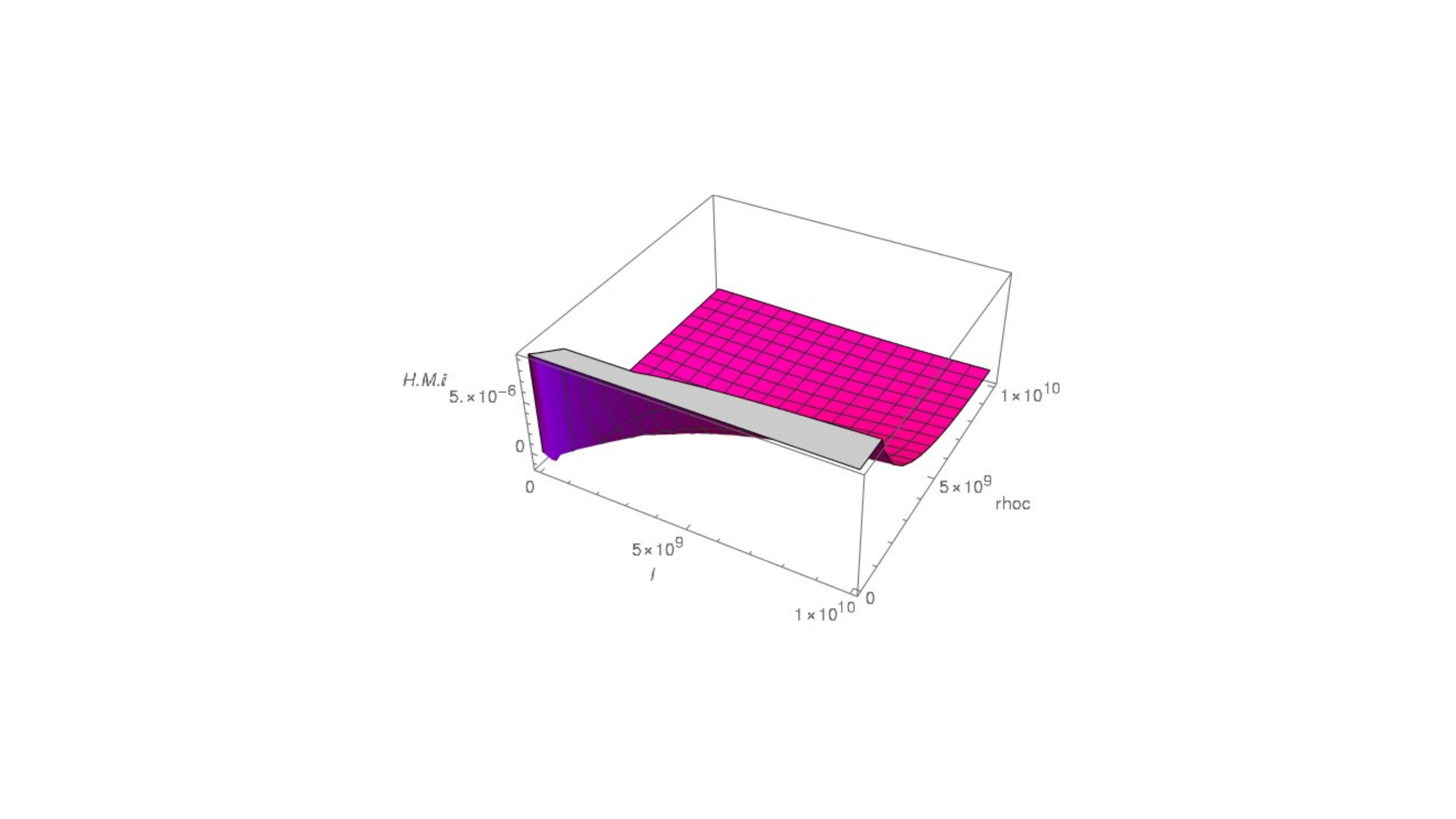}
\includegraphics[width=.65\textwidth]{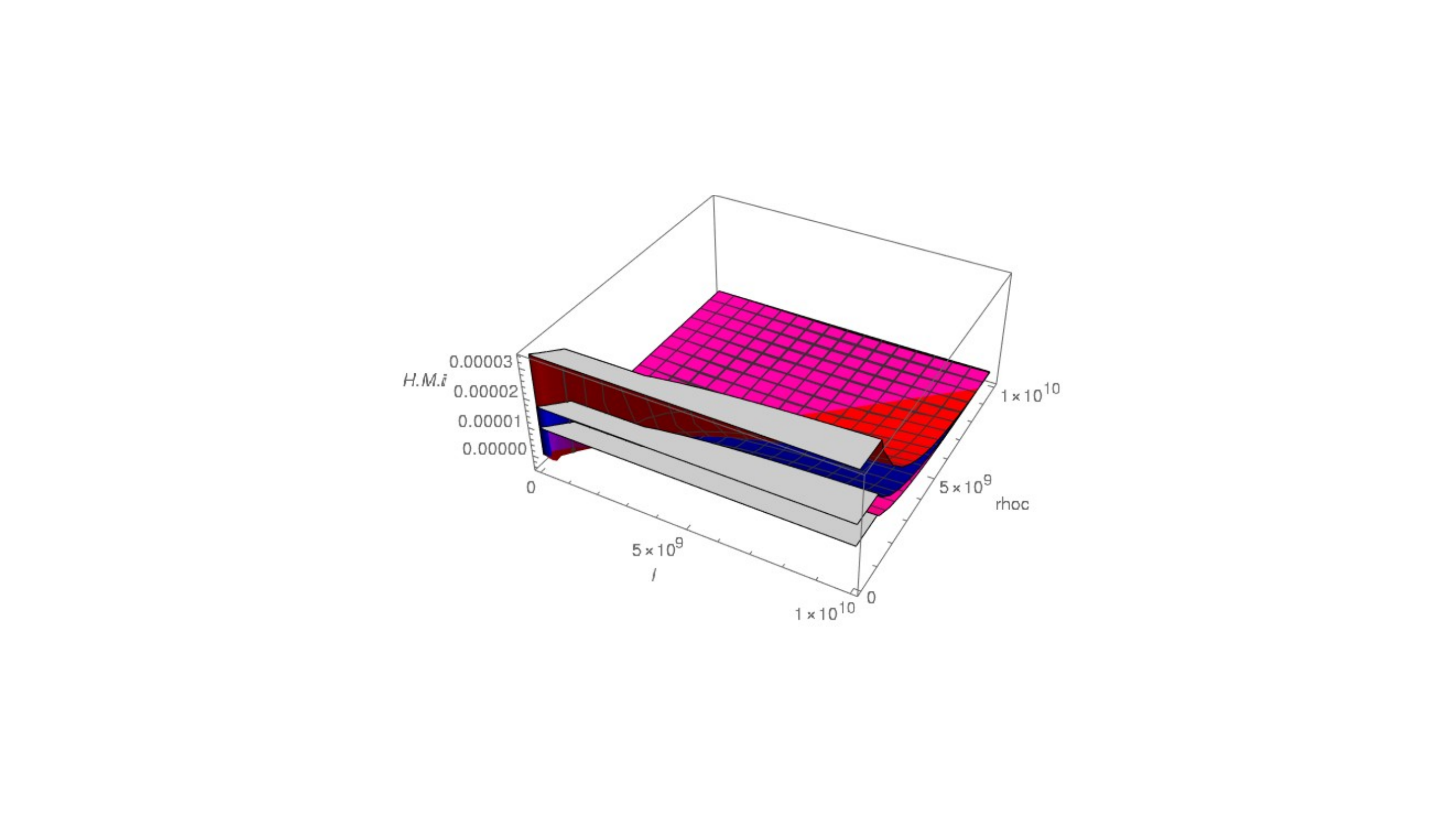}
\includegraphics[width=.65\textwidth]{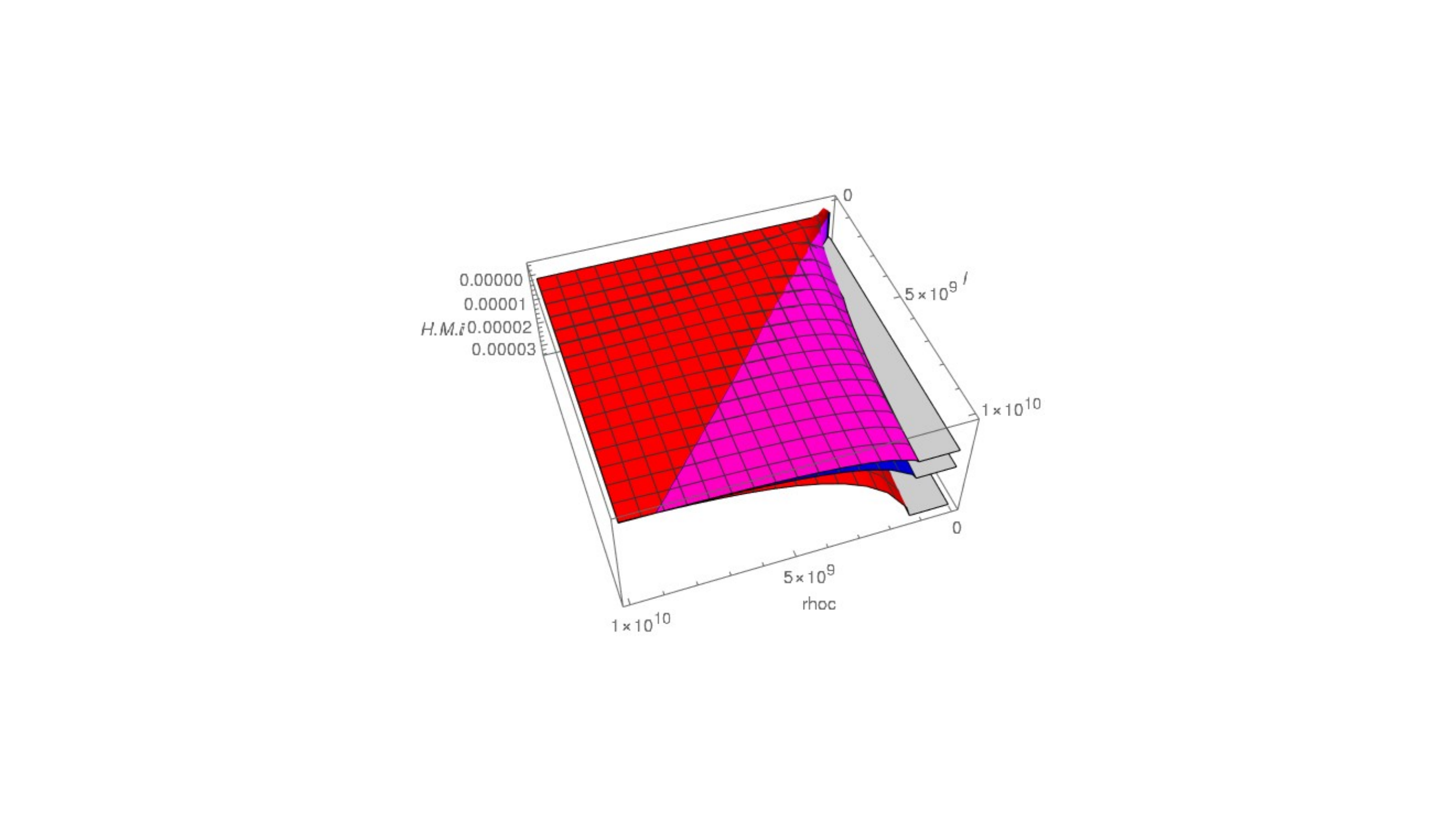}	
\includegraphics[width=.65\textwidth]{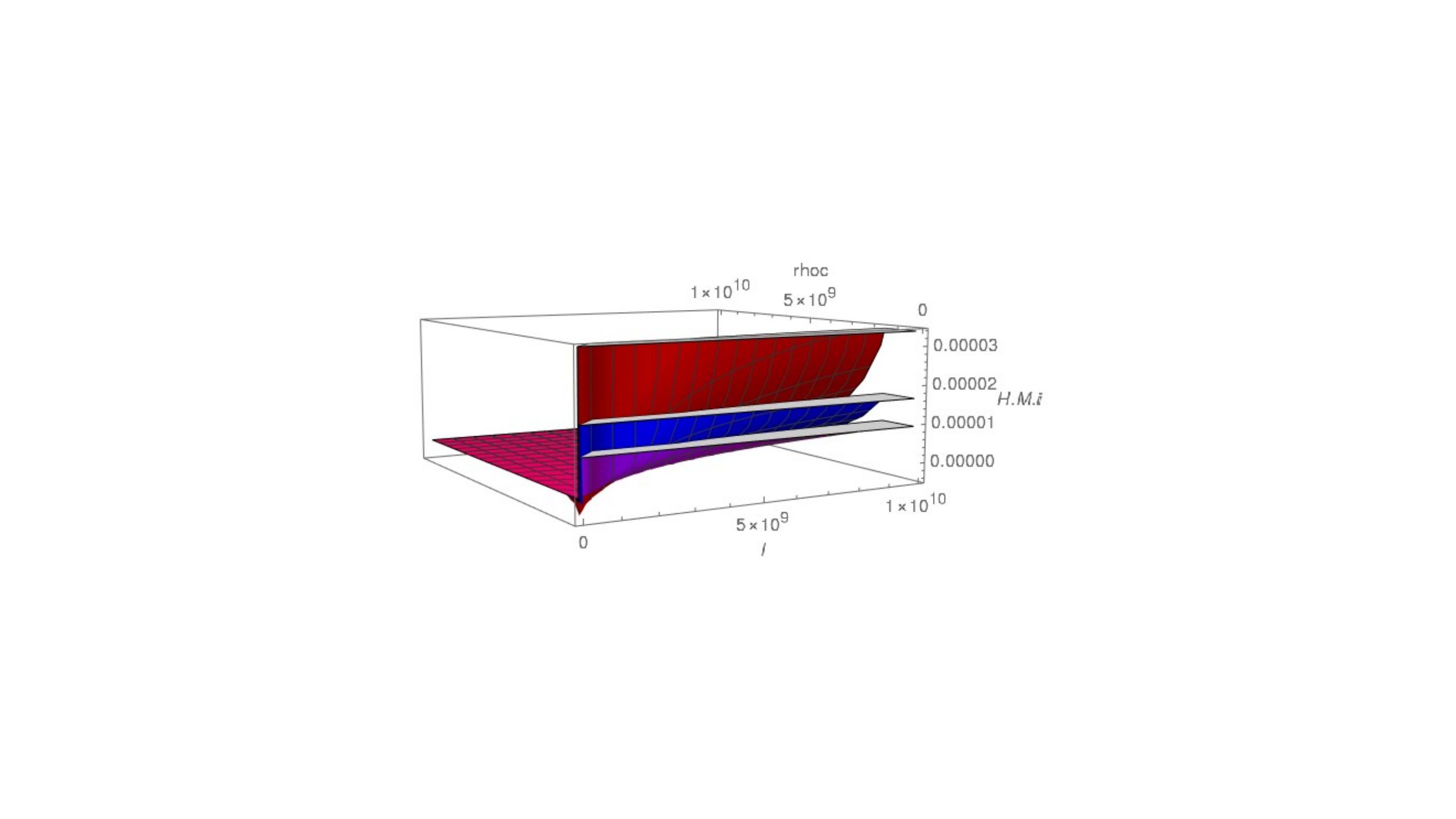}	
\caption{(First row) \,\,:\,\,( From  left to right) \,,\, H.M.I plotted as a function of $(l,\rho_c)$ for $ h = {10}^{6}$ for $d - \theta = 1.500\,,\, d - \theta = 1.533 $,
  \quad;\quad (Second row) \,:\,  ( left)\, : \, HEE plotted as a function of $(l,\rho_c)$ \, ,\,  $d - \theta= 1.566$ \,\,,\,\, (right) \,:\, The frontview of the  overlap of the three, showing that for $l >> \rho_c$ H.M.I decreases with the increase of $d - \theta$
 \, ,\,   \quad;\quad (Last row ) ,\, ( left )\, : \,  \,:\, The backside view of the overlap of the three, showing for $\rho_c >> l$ the H.M.I for different $d - \theta$   actually merges, supporting the fact that for a fixed l, H.M.I decreases with the increase of $\rho_c$ and ultimately falls towards zero.\,\,;\,\. (right)\,:\,  The sideview of the overlap of the three showing that $h_{\rm crit}$  decreases with the increase of $d - \theta$ }
\label{hmievbgreater1}
\end{figure}

\subsection{H.M.I for $d - \theta < 1$}

We recall the expression of H.M.I for $d - \theta < 1$ given by (\ref{entropylrhoc2}).     Substituting the same in the expression of HMI  for two strips of length l,  in their connected phase (\ref{H}),  we obtain the expression of HMI for $d - \theta > 1$

Consequently we write H.M.I,for $d - \theta < 1$ as
\ber
 & & H.M.I\n
 &=&   2{\frac{    R^{d-1}    L^{d-1} {\left(  {\left( {\left({\frac{A_{10} l}{ 2 }}\right)}^{  \left( d - \theta + 1 \right)} + {\left( \rho_c \right)}^{  \left( d - \theta + 1 \right)} \right)}^{\frac{1}{(d - \theta + 1)}}   \right)}^{\theta - d +1} \,\, {{}_2 F_1}   \left\lbrack {\frac{1}{2}}, {\frac{1}{2}}( - 1 +{ \frac{1}{d - \theta }}), {\frac{1}{2}}(1 +{ \frac{1}{d - \theta }}) , 1 \right \rbrack }{  4 G_N (\theta + 1- d )    }}\n
                      &- & 2 R^{d-1}     \left({\rho_c}\right)^{ \theta +1- d} {\frac{   L^{d-1}     \,\, {{}_2 F_1}   \left\lbrack {\frac{1}{2}}, {\frac{1}{2}}( - 1 +{ \frac{1}{d - c}}), {\frac{1}{2}}(1 +{ \frac{1}{d - \theta}}) , \left({\frac{\rho_c}{\left\lbrack  {\left( {\left({\frac{A_{10} l}{ 2 }}\right\rbrack}^{  \left( d - \theta + 1 \right\rbrack} + {\left( \rho_c \right)}^{  \left( d - \theta + 1 \right)} \right)}^{\frac{1}{(d - \theta + 1)}}  \right)}}\right)^{2(d - \theta)}  \right \rbrack }{  4 G_N( \theta - d  + 1)   }}\n
  &- &   {\frac{     R^{d-1}   L^{d-1} {\left(  {\left( {\left({\frac{A_{10} (2l + h)}{ 2 }}\right)}^{  \left( d - \theta + 1 \right)} + {\left( \rho_c \right)}^{  \left( d - \theta + 1 \right)} \right)}^{\frac{1}{(d - \theta + 1)}}   \right)}^{\theta - d +1} \,\, {{}_2 F_1}   \left\lbrack {\frac{1}{2}}, {\frac{1}{2}}( - 1 +{ \frac{1}{d - \theta }}), {\frac{1}{2}}(1 +{ \frac{1}{d - \theta }}) , 1 \right \rbrack }{  4 G_N (\theta + 1- d )    }}\n
                      &+ & R^{d-1}    \left({\rho_c}\right)^{ \theta +1- d} {\frac{  L^{d-1}       \,\, {{}_2 F_1}   \left\lbrack {\frac{1}{2}}, {\frac{1}{2}}( - 1 +{ \frac{1}{d - c}}), {\frac{1}{2}}(1 +{ \frac{1}{d - \theta}}) , \left({\frac{\rho_c}{\left\lbrack  {\left( {\left({\frac{A_{10} (2l + h)}{ 2 }}\right\rbrack}^{  \left( d - \theta + 1 \right\rbrack} + {\left( \rho_c \right)}^{  \left( d - \theta + 1 \right)} \right)}^{\frac{1}{(d - \theta + 1)}}  \right)}}\right)^{2(d - \theta)}  \right \rbrack }{  4 G_N( \theta - d  + 1)   }}\n
&- &   {\frac{    R^{d-1}    L^{d-1} {\left(  {\left( {\left({\frac{A_{10} ( h)}{ 2 }}\right)}^{  \left( d - \theta + 1 \right)} + {\left( \rho_c \right)}^{  \left( d - \theta + 1 \right)} \right)}^{\frac{1}{(d - \theta + 1)}}   \right)}^{\theta - d +1} \,\, {{}_2 F_1}   \left\lbrack {\frac{1}{2}}, {\frac{1}{2}}( - 1 +{ \frac{1}{d - \theta }}), {\frac{1}{2}}(1 +{ \frac{1}{d - \theta }}) , 1 \right \rbrack }{  4 G_N (\theta + 1- d )    }}\n
                      &+ & R^{d-1}   \left({\rho_c}\right)^{ \theta +1- d} {\frac{   L^{d-1}      \,\, {{}_2 F_1}   \left\lbrack {\frac{1}{2}}, {\frac{1}{2}}( - 1 +{ \frac{1}{d - c}}), {\frac{1}{2}}(1 +{ \frac{1}{d - \theta}}) , \left({\frac{\rho_c}{\left\lbrack  {\left( {\left({\frac{A_{10} ( h)}{ 2 }}\right\rbrack}^{  \left( d - \theta + 1 \right\rbrack} + {\left( \rho_c \right)}^{  \left( d - \theta + 1 \right)} \right)}^{\frac{1}{(d - \theta + 1)}}  \right)}}\right)^{2(d - \theta)}  \right \rbrack }{  4 G_N( \theta - d  + 1)   }}
\la{hmiblessthan1}
\eer

As before,  the global expression for the turning point, is exact in the regime $l>> \rho_c$ and $\rho_c>> l$ only,  so here we will use (\ref{hmibgreaterthan1}),  obtain the 3D and 2D plots of H.M.I vs $(l,\rho_c)$ for some fixed h and different $ d- \theta$ , study its feature in the regime  $l>> \rho_c$ and $\rho_c>> l$, while we understand in the rest of the regime in $(l,\rho_c)$ plane it is an interpolating functiom between these two regime.  We will also obtain the plot  of H.M.I as a function of h.  Here all the relevant plots are shown in (\ref{hmibasic4by9}, \ref{hmibasic1by3}, \ref{hmibasichlless}, \ref{hmirhoch4by5} ,  \ref{hmievbless1}).

\begin{figure}[H]
\begin{center}
\textbf{ For $ d - \theta < 1$,  with  $ d - \theta  = {\frac{4}{9}}$\,: \,H.M.I vs $(l,\rho_c) $, H.M.I vs l and H.M.I vs $\rho_c$ plots for fixed h  }
\end{center}
\vskip2mm
\includegraphics[width=.55\textwidth]{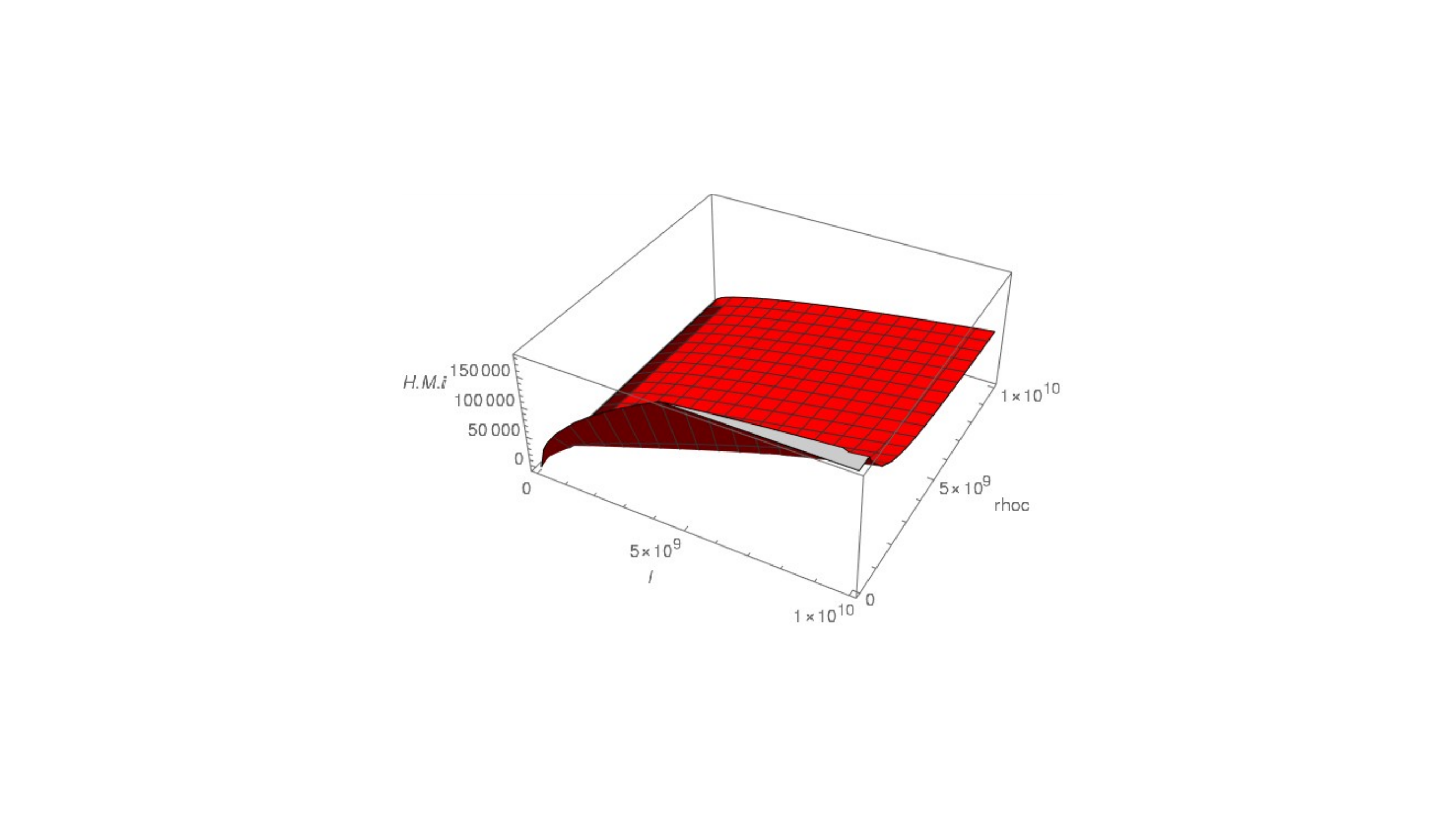}
\includegraphics[width=.55\textwidth]{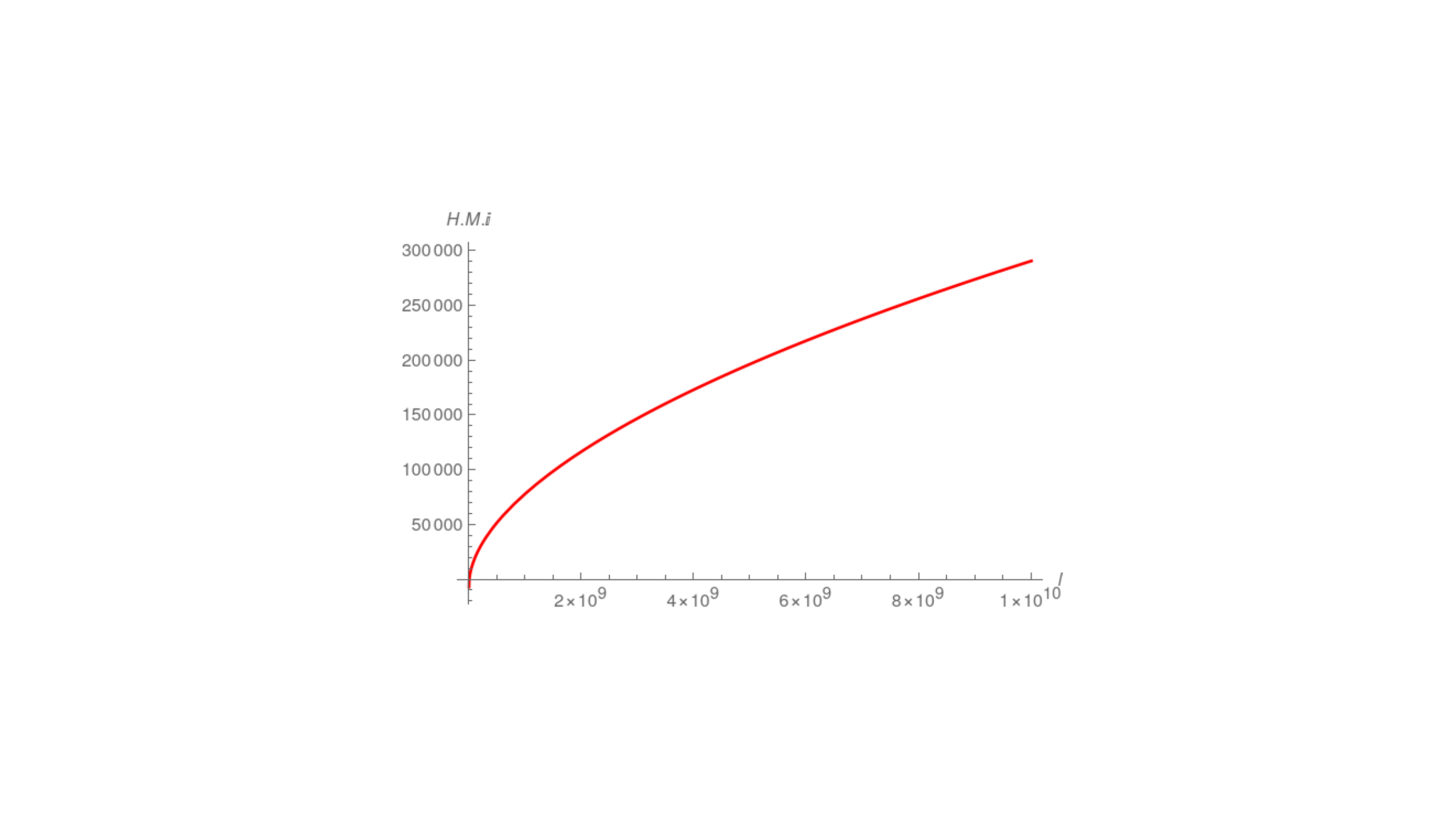}	
\includegraphics[width=.55\textwidth]{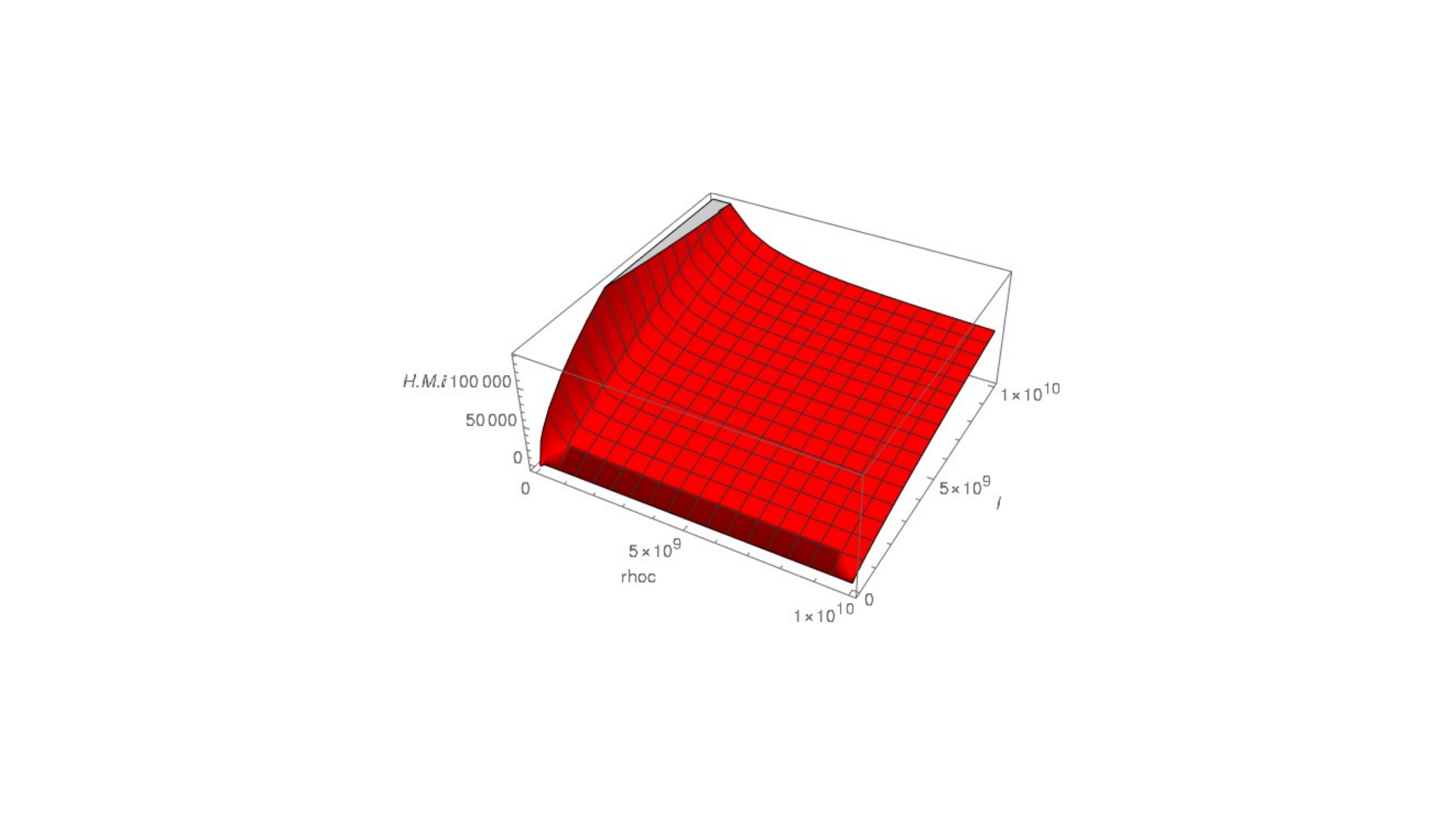}
\includegraphics[width=.55\textwidth]{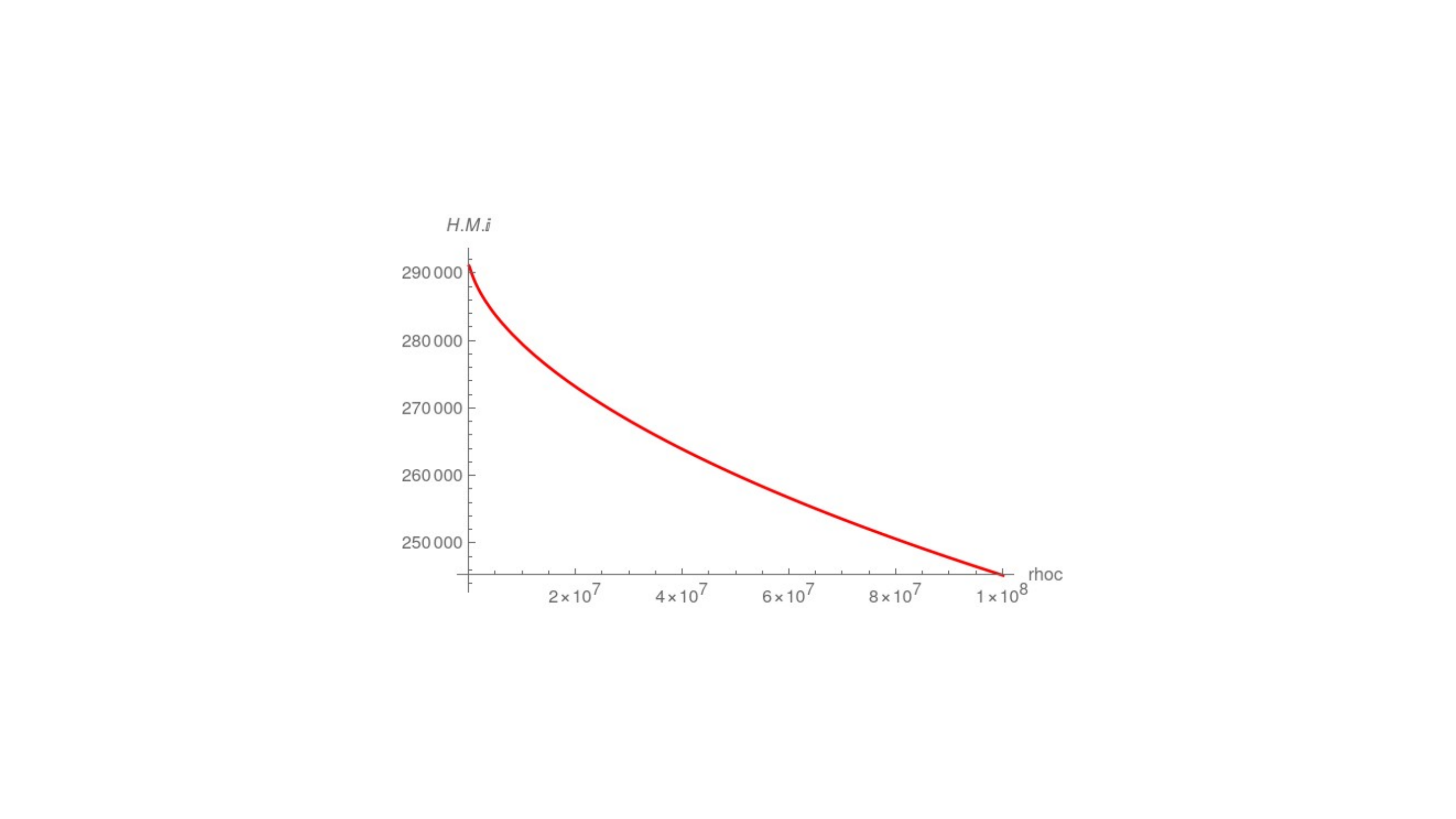}
\includegraphics[width=.55\textwidth]{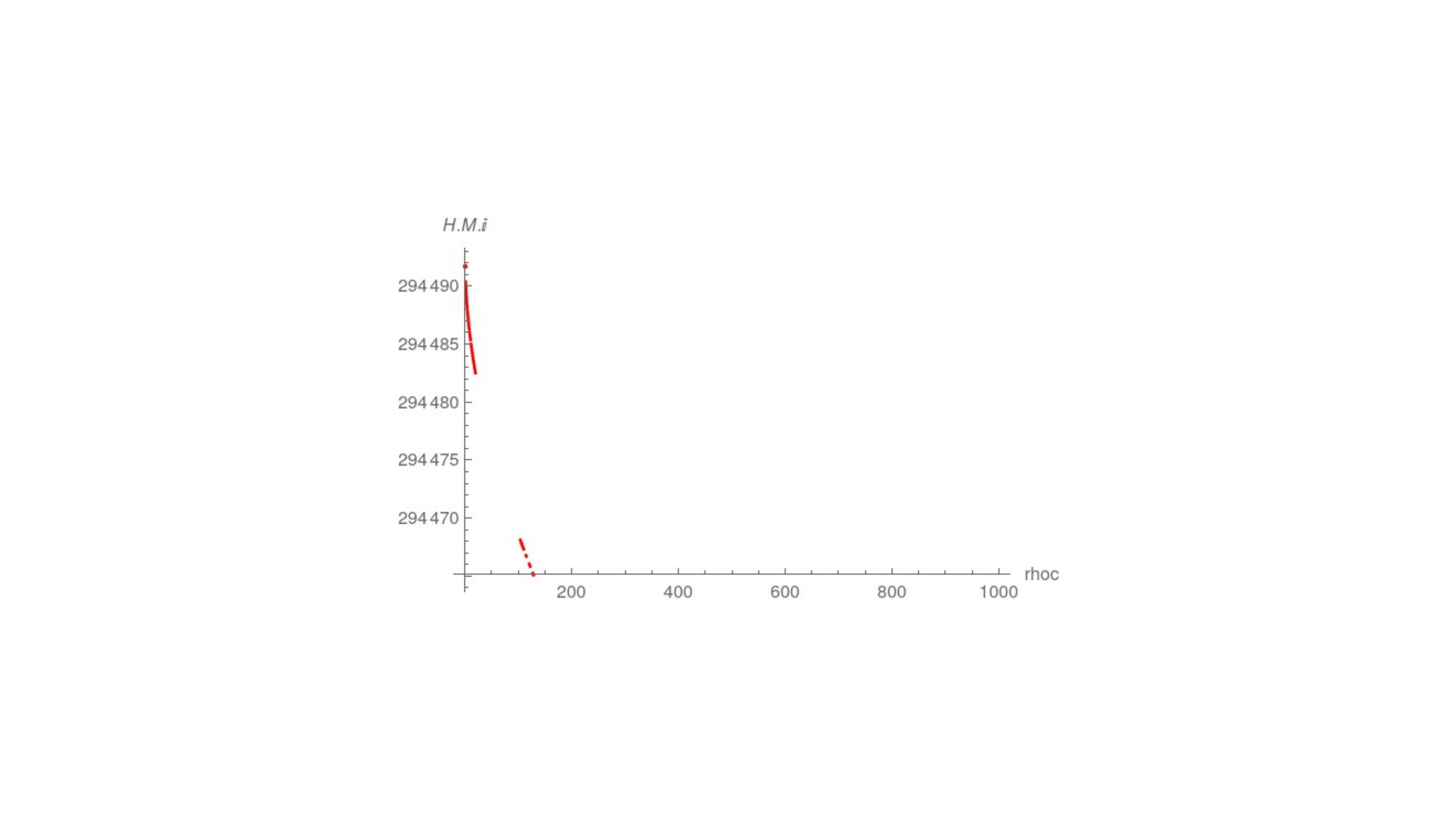}
\caption{(First row) \,\,: (left)\,:\, H.M.I as a function of $(l,\rho_c)$ for  $h = (10)^6$ \,,\, (right)\, : \, H.M.I  as a function of l with $\rho_c = 1$,  $h = (10)^6$  \,:\,  Both the plots are showing when we increase l gradually from zero, for fixed h,   at certain point given in terms of $h_{\rm crit}$, H.M.I is undergoing a first order phase transition and also for a given $\rho_c$,   H.M.I increases with the increase of l
  \quad;\quad (Second row) \,:\,  ( left)\, : \, H.M.I  as a function of $(\rho_c , l)$ for  $h = (10)^6$\,, (right) \,:\, H.M.I  as a function of $\rho_c$  for $l = {(10)}^{10}$,$h = {(10)}^6$,  this value of l is chosen to probe $l>> \rho_c$ regime where for $\rho_c >> l$,   H.M.I is zero  \, :\,  Both the 3D and 2D plots are showing, for a given l, H.M.I falls with the increase of cut off $\rho_c$ and goes to zero for $\rho_c >> l $ regime.  Also for nonzero h,    H.M.I is finite at $\rho_c = 0$
 \quad;\quad (Last row) \,:\,  H.M.I  as a function of $\rho_c$  for $l = {10}^{10}$,$h = 0$,   \, :\,   the  2D plot is showing for $h = 0$  H.M.I diverge at $\rho_c = 0$ as expected
}
\label{hmibasic4by9}
\end{figure}

\begin{figure}[H]
\begin{center}
\textbf{ For $ d - \theta < 1$, with  $ d - \theta  = {\frac{1}{3}} $,   H.M.I vs $(l,\rho_c) $, H.M.I vs l and H.M.I vs $\rho_c$ plots for fixed h  }
\end{center}
\vskip2mm
\includegraphics[width=.55\textwidth]{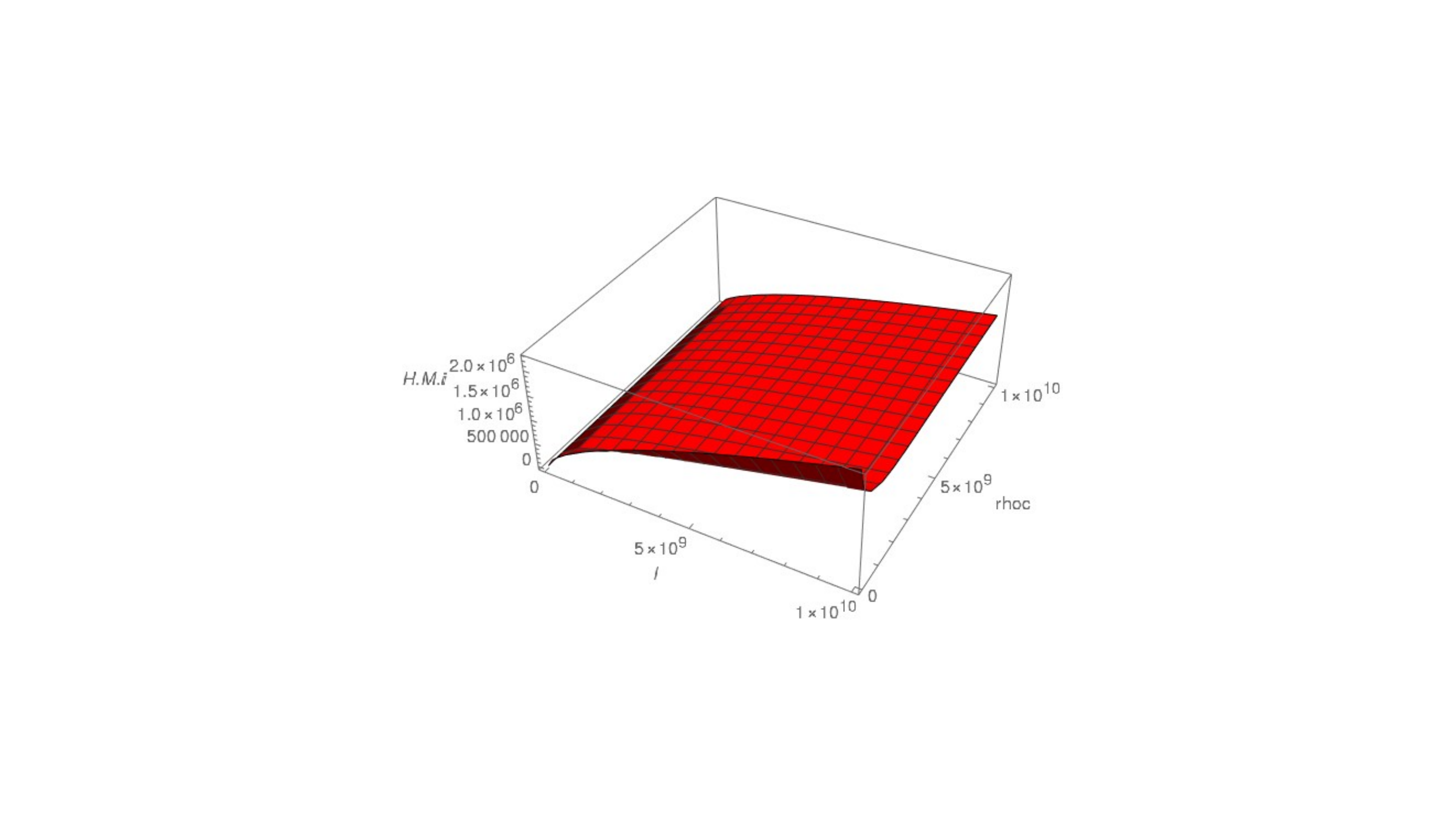}
\includegraphics[width=.55\textwidth]{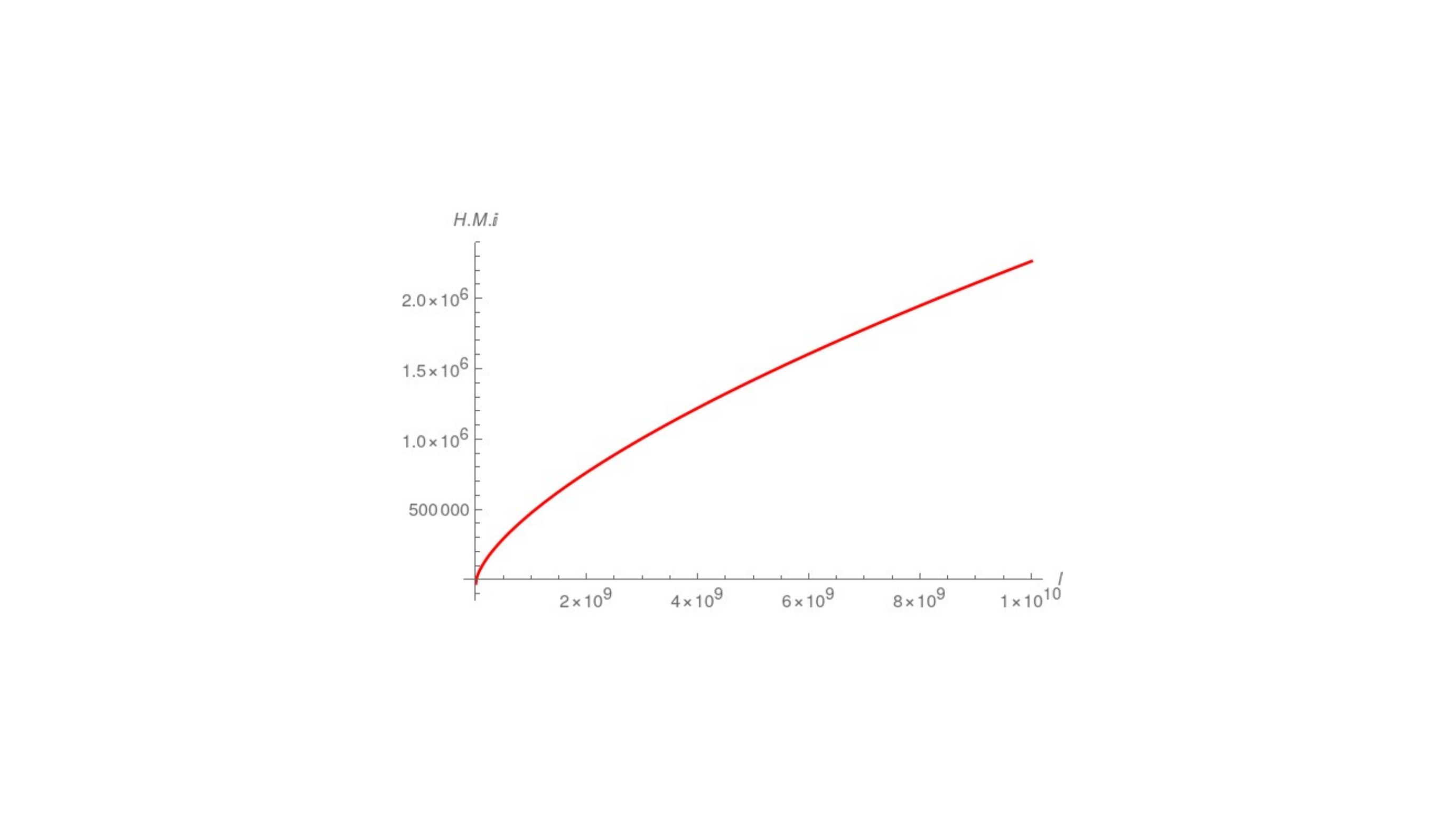}	
\includegraphics[width=.55\textwidth]{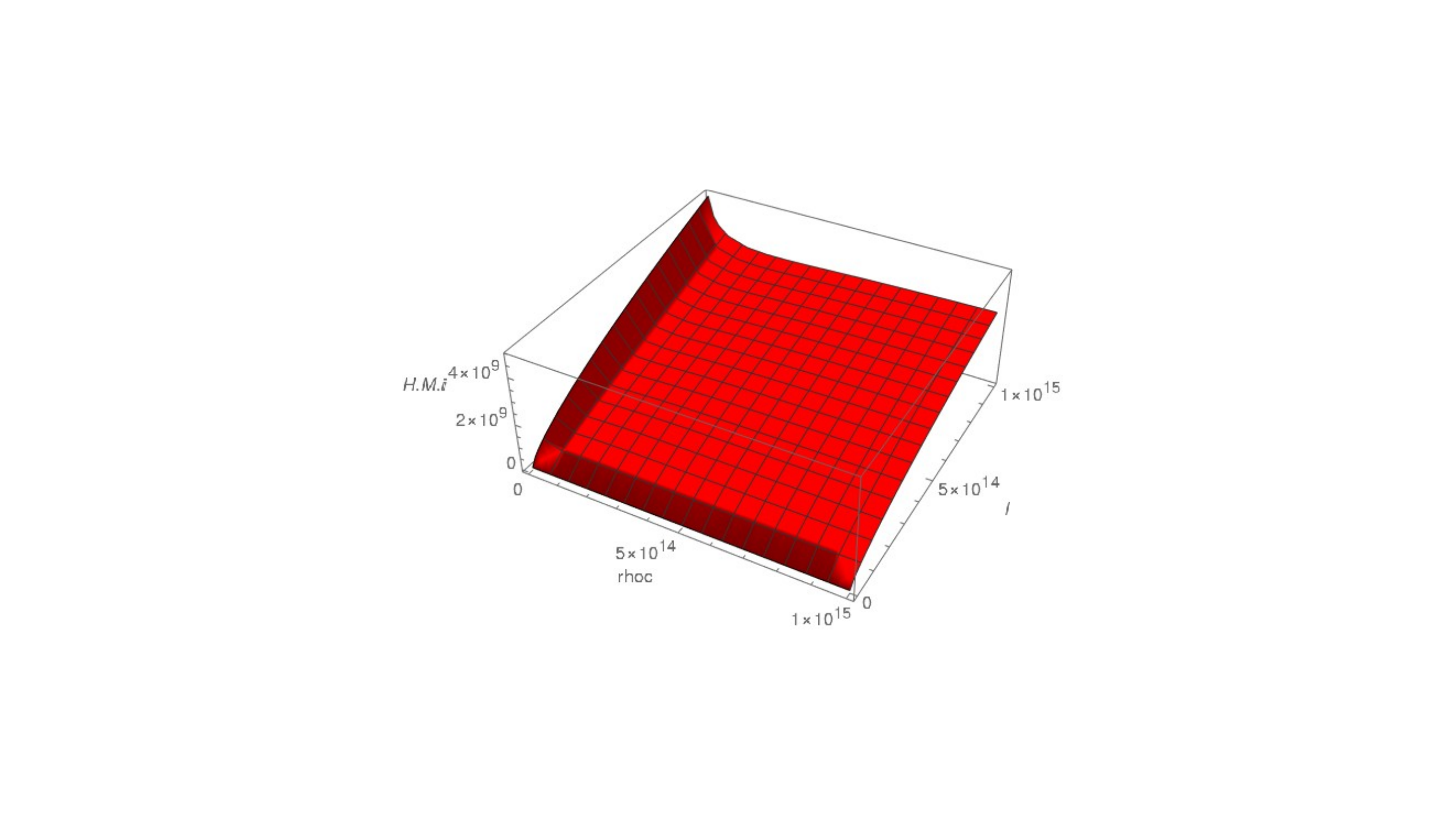}
\includegraphics[width=.55\textwidth]{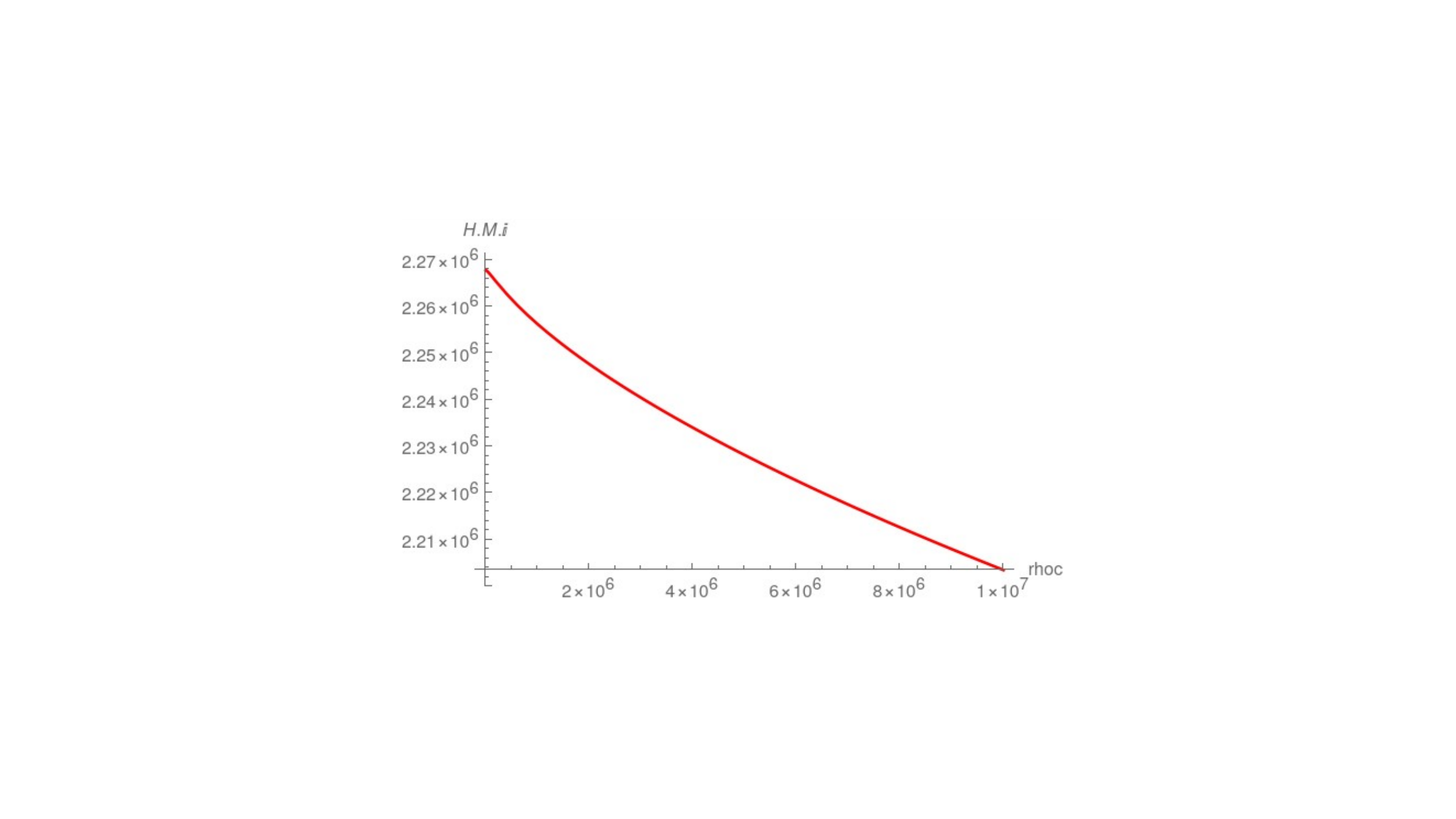}
\includegraphics[width=.55\textwidth]{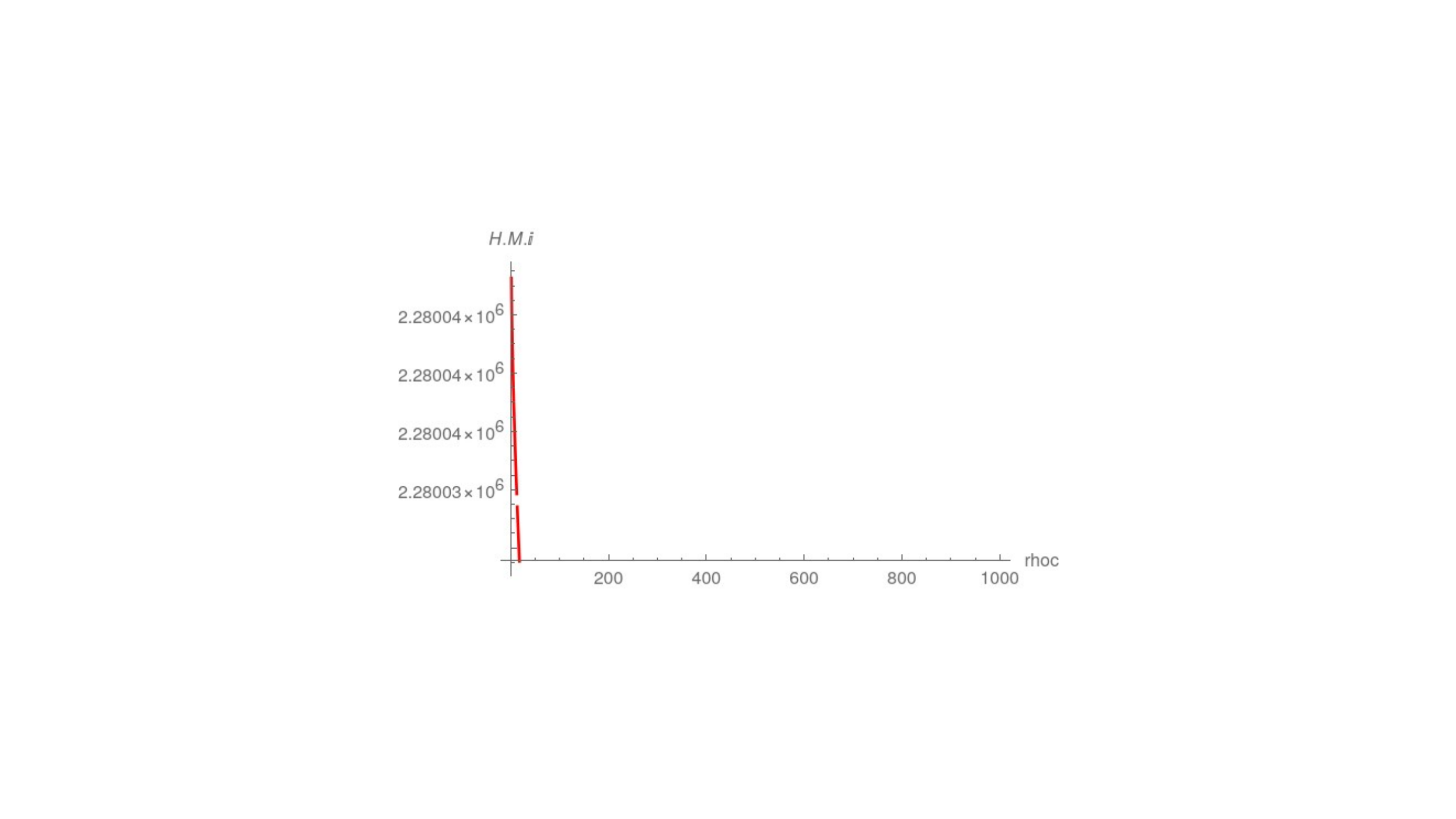}
\caption{(First row) \,\,: (left)\,:\, H.M.I as a function of $(l,\rho_c)$ for  $h = (10)^6$ \,,\, (right)\, : \, H.M.I  as a function of l with $\rho_c = 1$,  $h = (10)^6$  \,:\,  Both the plots are showing when we increase l gradually from zero, for fixed h,   at certain point given in terms of $h_{\rm crit}$, H.M.I is undergoing a first order phase transition and also for a given $\rho_c$,   H.M.I increases with the increase of l
  \quad;\quad (Second row) \,:\,  ( left)\, : \, H.M.I  as a function of $(\rho_c , l)$ for  $h = (10)^6$\,, (right) \,:\, H.M.I  as a function of $\rho_c$  for $l = {(10)}^{10}$,$h = {(10)}^6$,  this value of l is chosen to probe $l>> \rho_c$ regime where for $\rho_c >> l$,   H.M.I is zero  \, :\,  Both the 3D and 2D plots are showing, for a given l, H.M.I falls with the increase of cut off $\rho_c$ and goes to zero for $\rho_c >> l $ regime.  Also for nonzero h,    H.M.I is finite at $\rho_c = 0$
 \quad;\quad (Last row)  \,:\, H.M.I  as a function of $\rho_c$  for $l = {10}^{10}$,$h = 0$,   \, :\, the  2D plot is  showing for $h = 0$  H.M.I diverge at $\rho_c = 0$ as expected}
\label{hmibasic1by3}
\end{figure}

\begin{figure}[H]
\begin{center}
\textbf{ For $ d - \theta < 1$, H.M.I as a function of h,l   for fixed $\rho_c$}
\end{center}
\vskip2mm
\includegraphics[width=.65\textwidth]{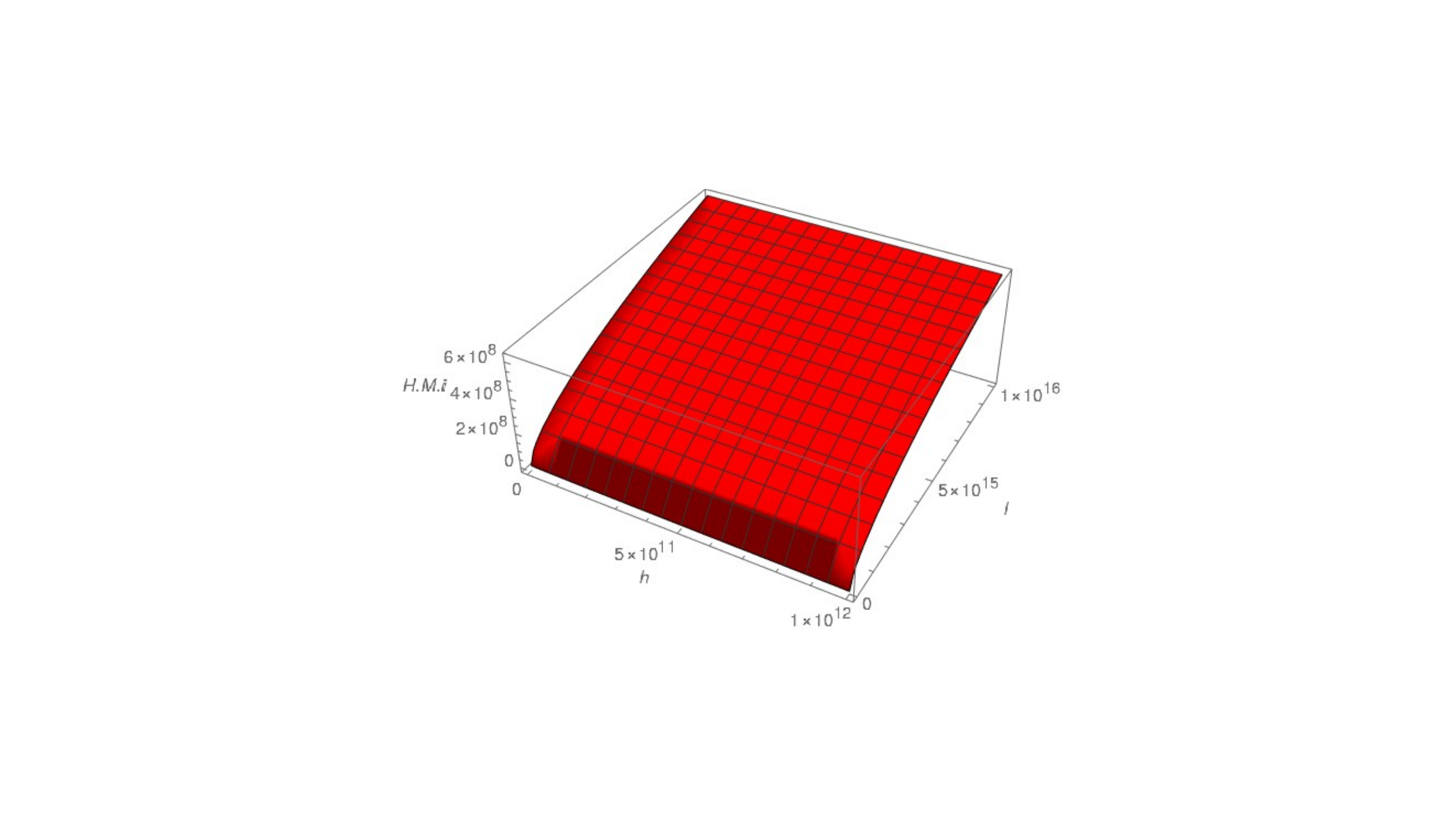}
\includegraphics[width=.65\textwidth]{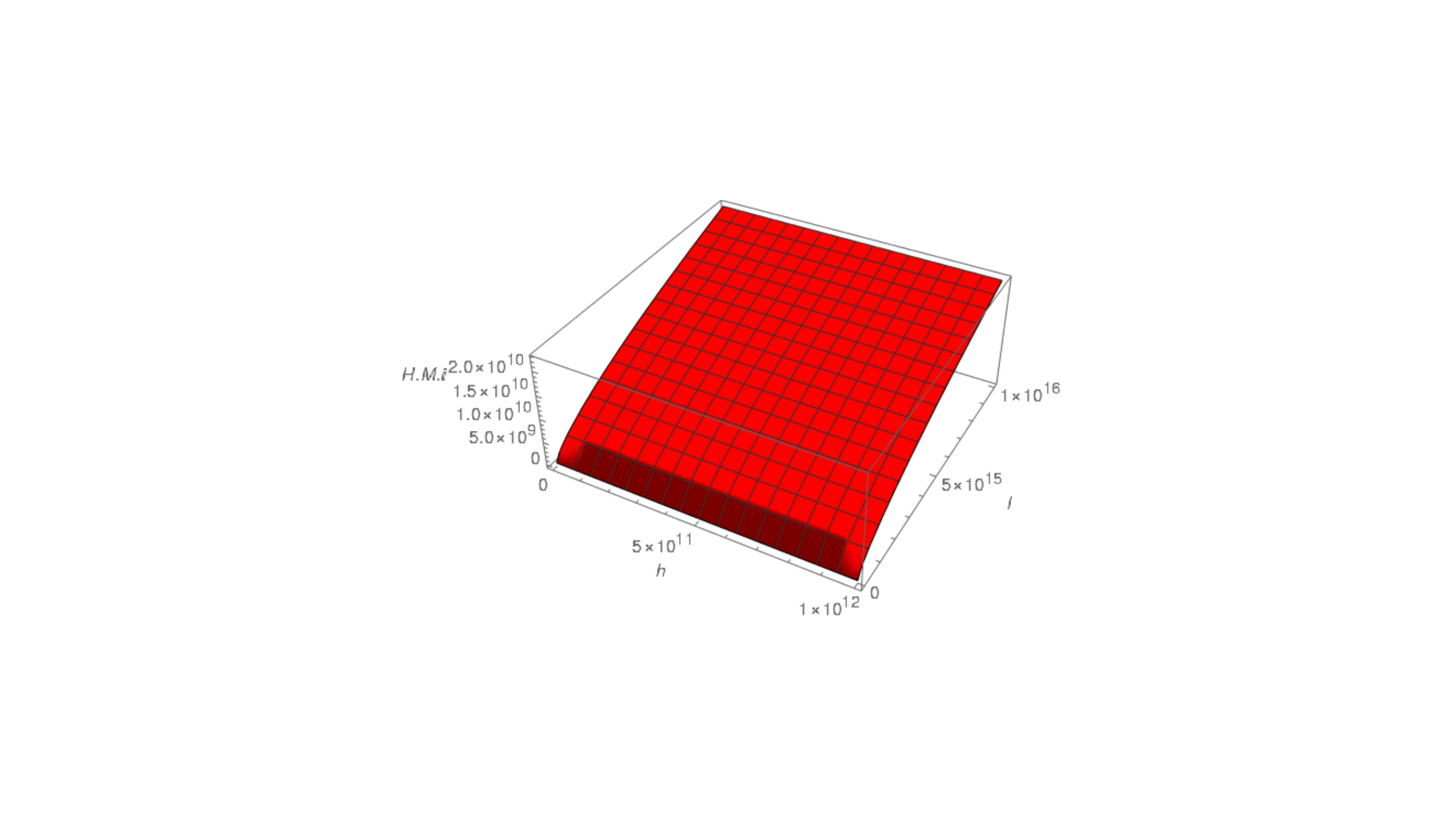}	

\caption{H.M.I is plotted as a function of (h,l)\,,\, (left) \,,\, for $d - \theta = {\frac{4}{9}}$ \,,\, (right) \,,\, for \,$d - \theta = {\frac{1}{3}}$\,,\., $\rho_c = 100$,  both the plots are showing for for a given h, H.M.I increases with the increase of l.  This is also showing the interesting feature that at $h = 0$,  H,M.I	 is finite,  when the cut off is nonzero}
\label{hmibasichlless}
\end{figure}

\begin{figure}[H]
\begin{center}
\textbf{H.M.I as a function of ($\rho_c$ , h )  for fixed l  }
\end{center}
\vskip2mm
\includegraphics[width=.65\textwidth]{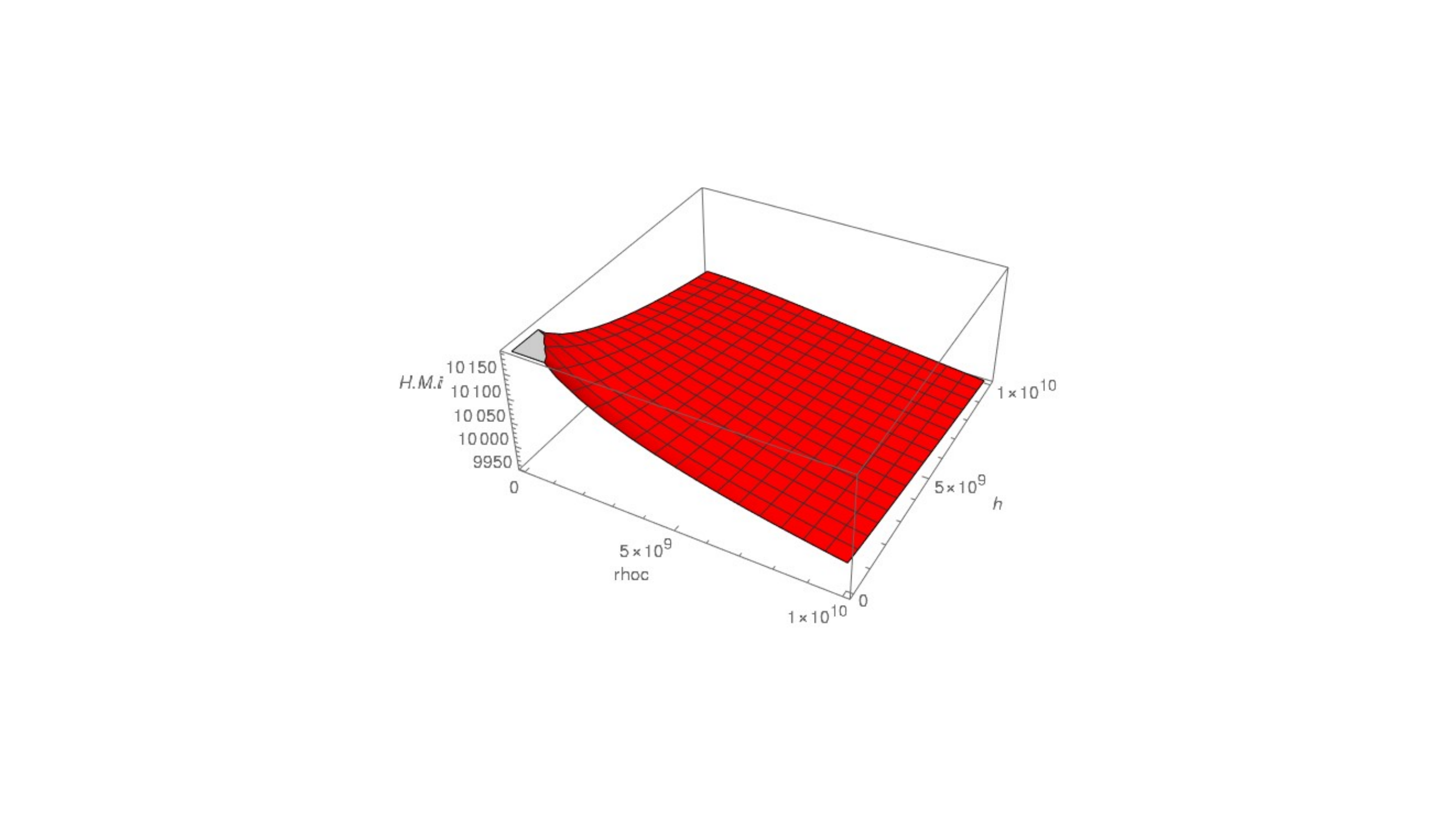}
\includegraphics[width=.65\textwidth]{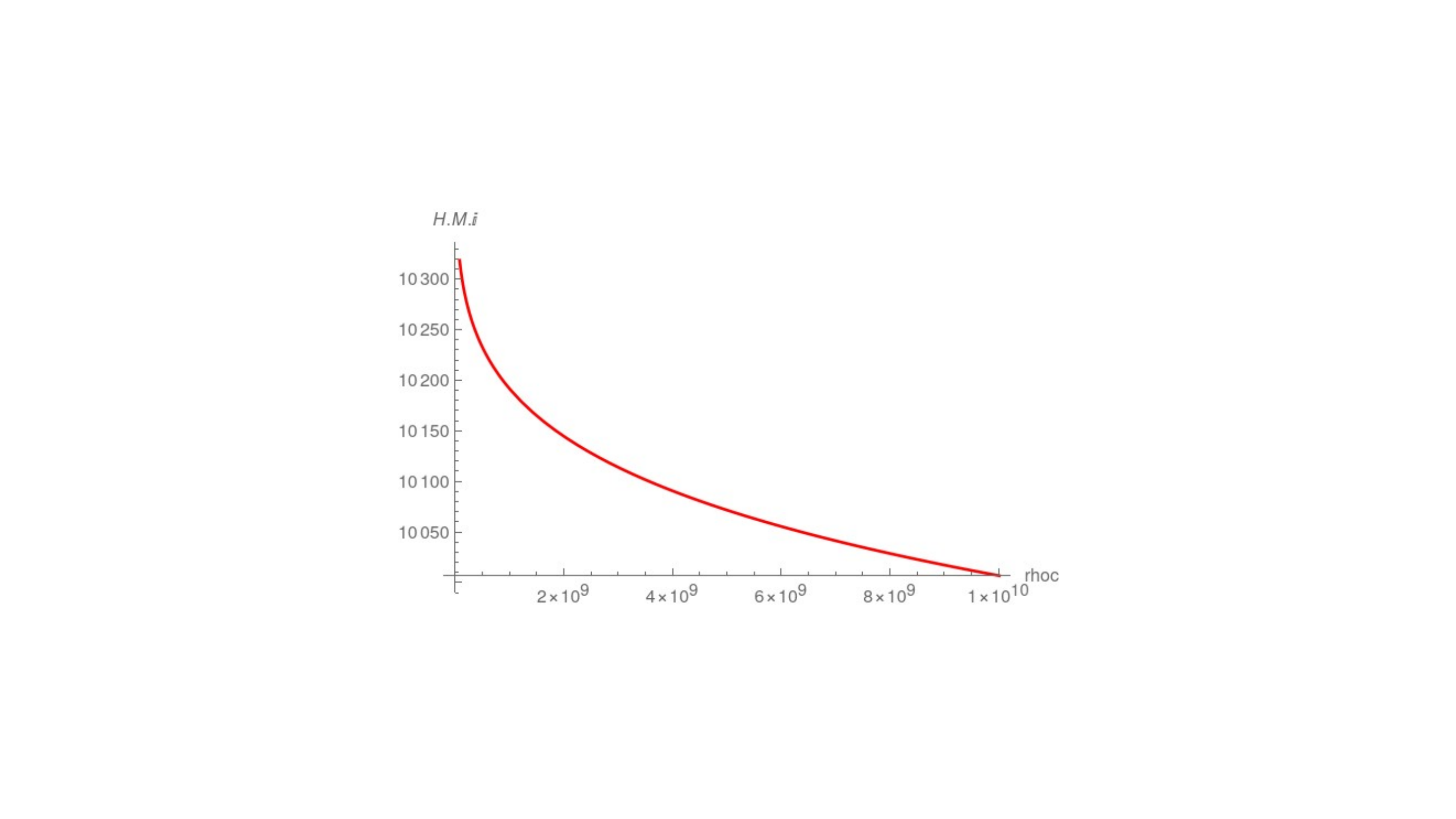}	
\includegraphics[width=.65\textwidth]{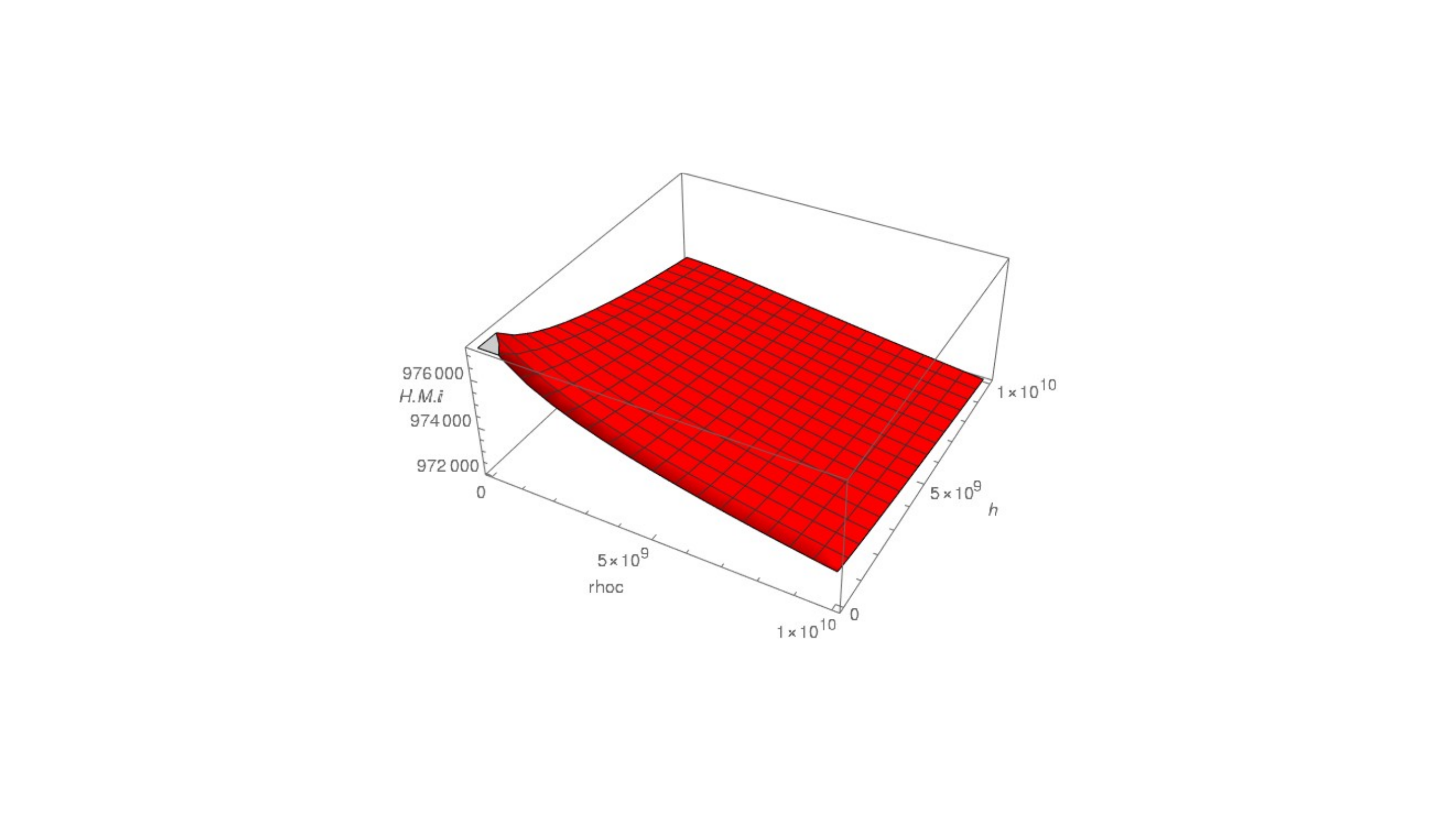}
\includegraphics[width=.65\textwidth]{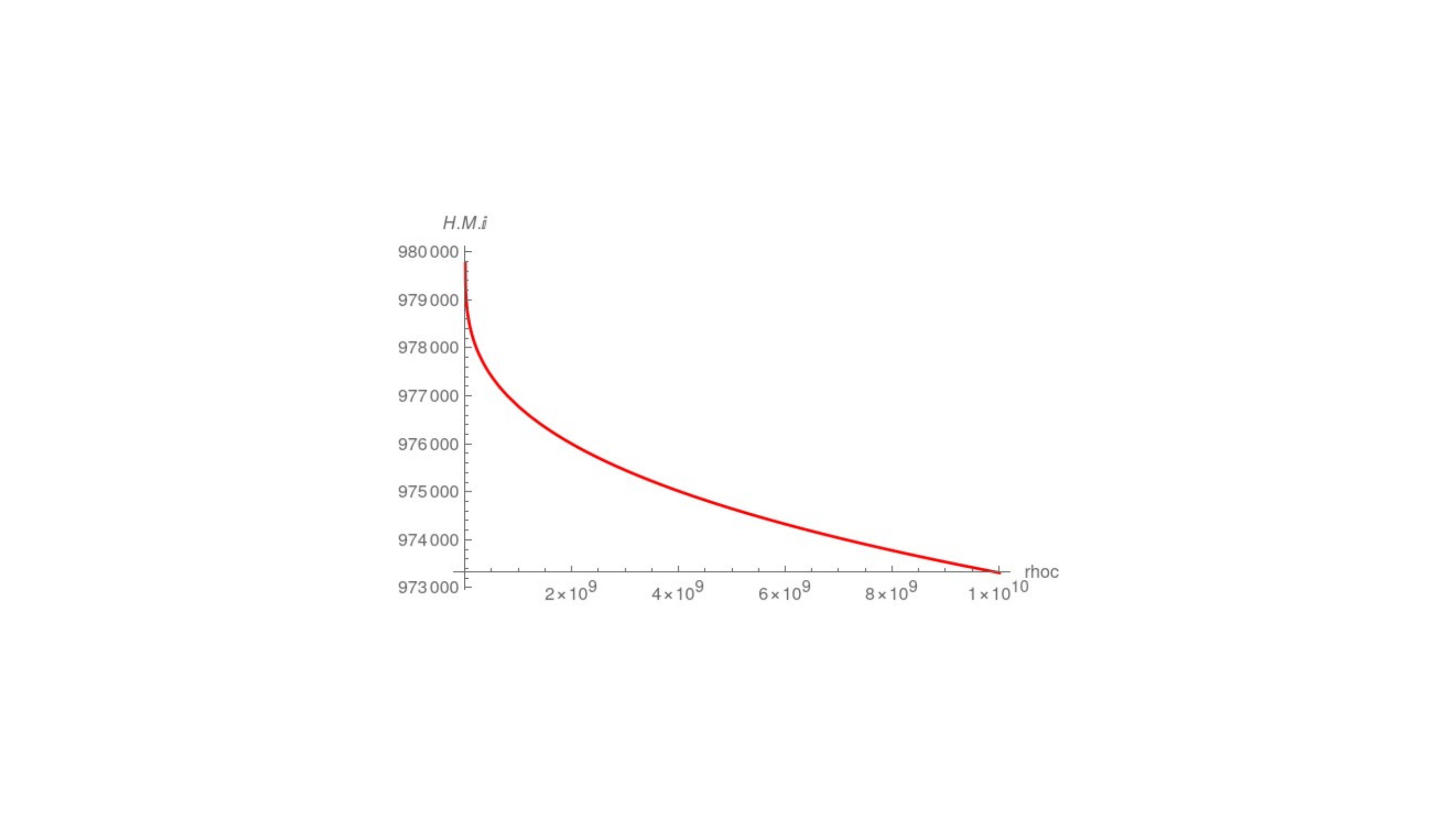}
\caption{(First row) \,\,:\,\,(left) \,:\, H.M.I as a function of $\rho_c$,h for $d - \theta = {\frac{4}{5}}$,  $l = {10}^{17}	$\,\,,\,\, (right)\,:\, H.M.I as a function of $\rho_c$ for $l= {10}^{17},   $h=0$	$,  \quad;\quad (Last row) \,:\,  ( left)\, : \, H.M.I plotted as a function of $(\rho_c , l)$ \, ,\,  $d - \theta= {\frac{2}{3}}$  \,\,,(right)\, :\, H.M.I plotted as a function of $\rho_c$  for $l = {10}^{17}$, $ h=0$ this value of l is chosen to probe $l>> \rho_c$ regime where for $\rho_c >> l$ H.M.I is zero \, :\,  Both the 3D and 2D plots  for both the values of $d - \theta$ are showing, for a given l, H.M.I falls with the increase of cut off $\rho_c$ and goes to zero for $\rho_c >> l $ regime. Also for $h = 0$,   for zero cut off, H.M.I diverge , as expected but the very moment we make cut-off finite, H.M.I also become finite
}
\label{hmirhoch4by5}
\end{figure}

\begin{figure}[H]
\begin{center}
\textbf{ For $ d - \theta < 1$ \,: \,  The evolution of H.M.I with $ d - \theta$  for fixed   $ h = {10}^{6}$ }
\end{center}
\vskip2mm
\includegraphics[width=.65\textwidth]{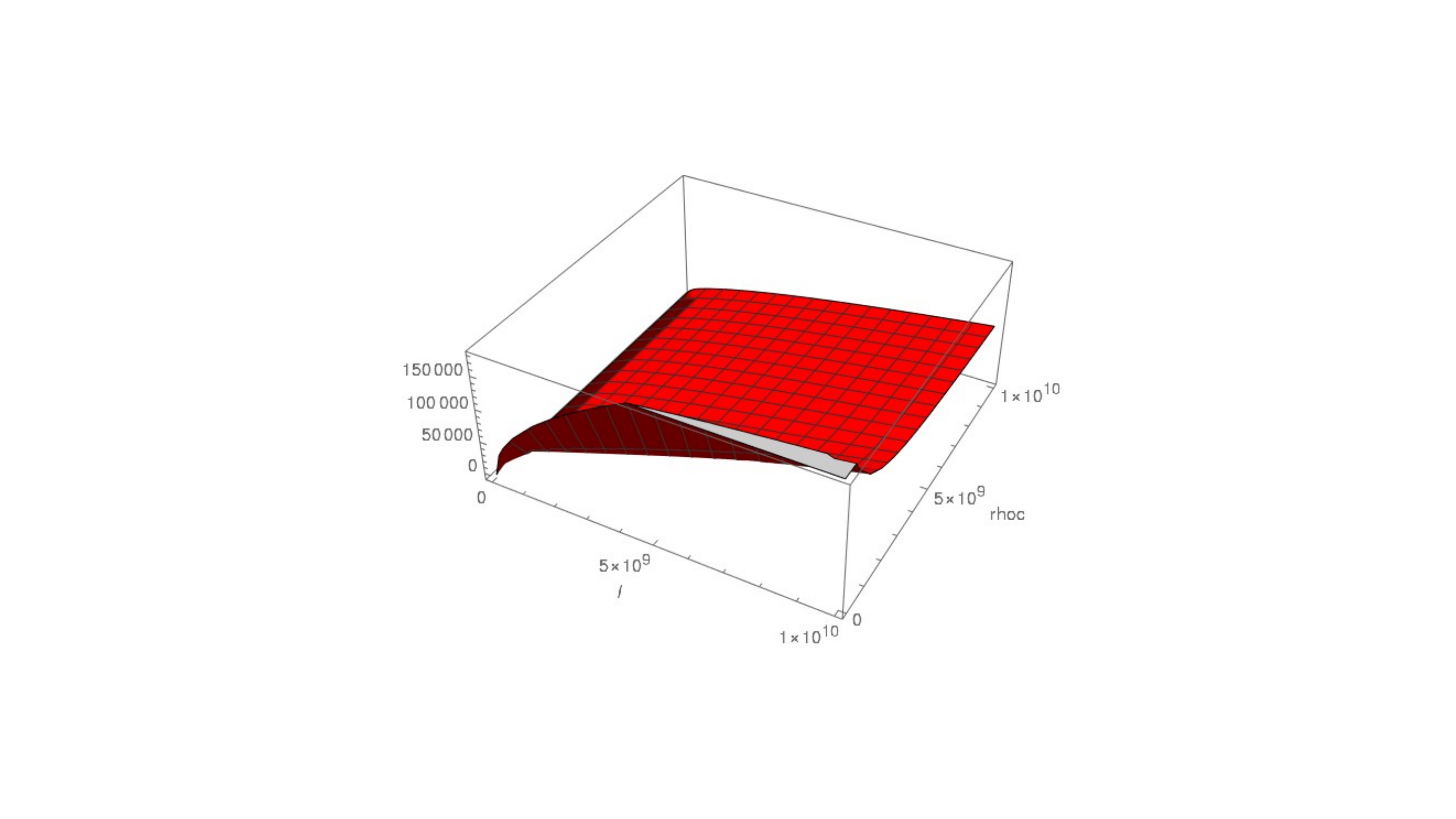}
\includegraphics[width=.65\textwidth]{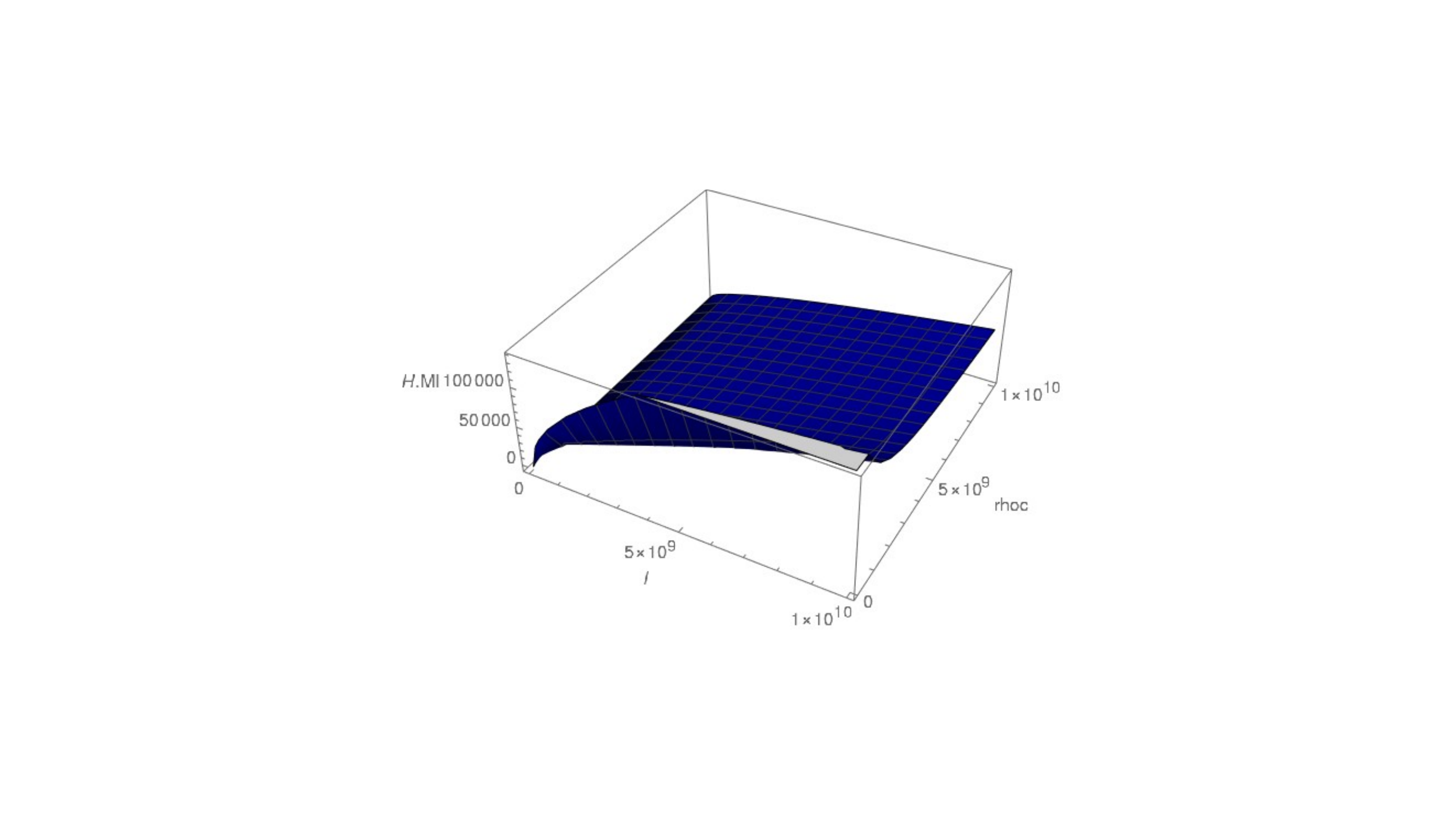}	
\includegraphics[width=.65\textwidth]{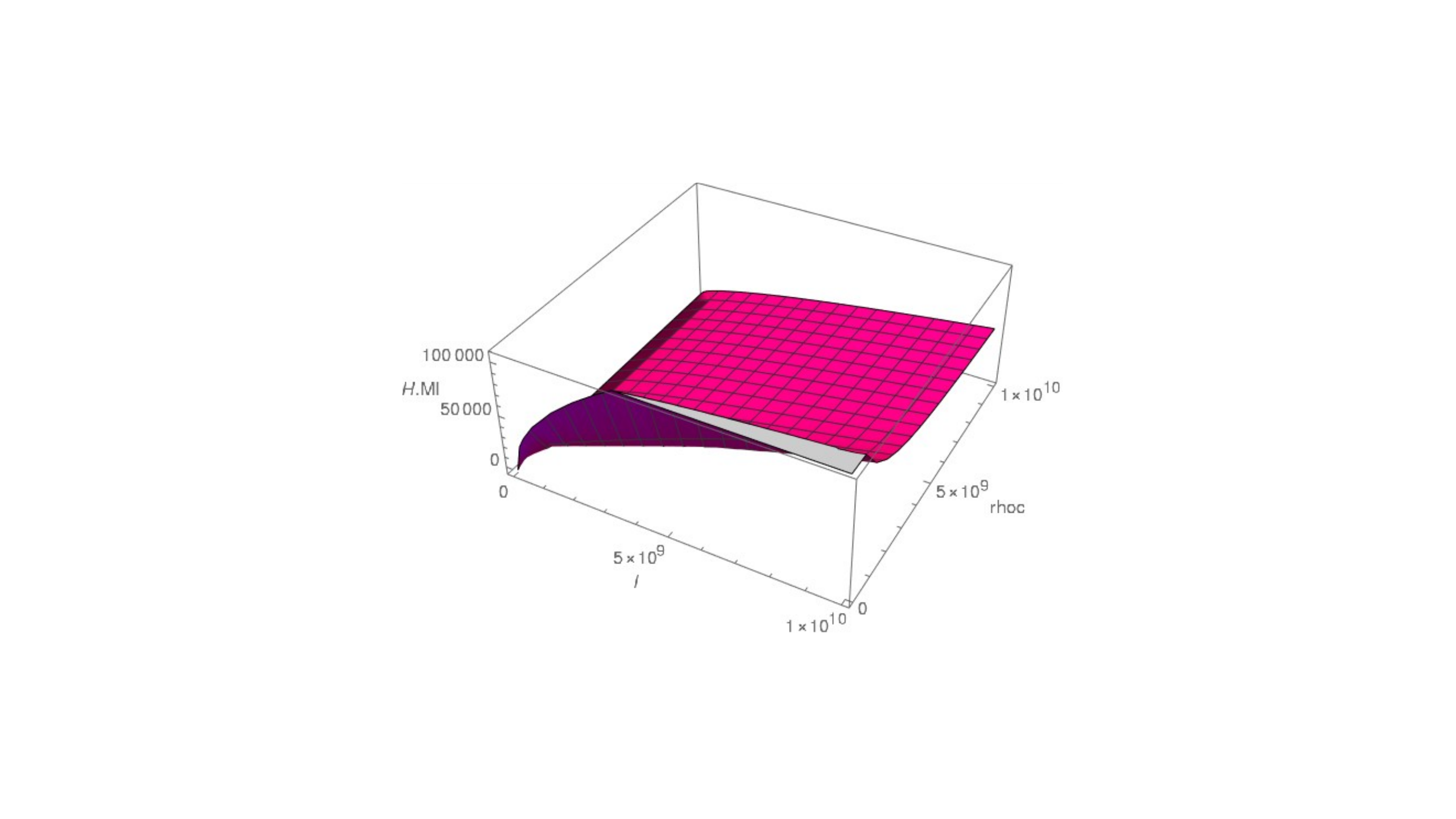}
\includegraphics[width=.65\textwidth]{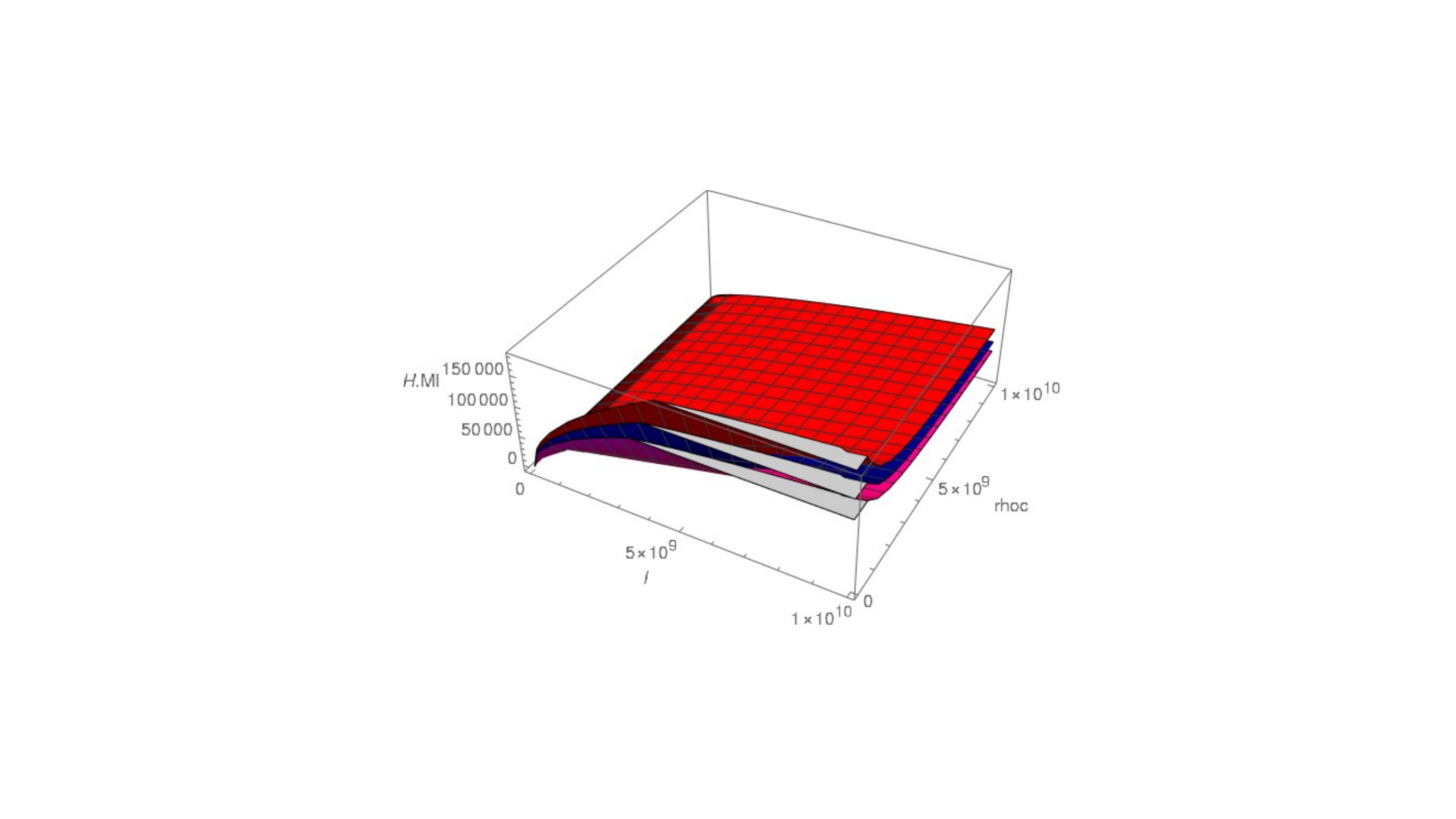}
\includegraphics[width=.65\textwidth]{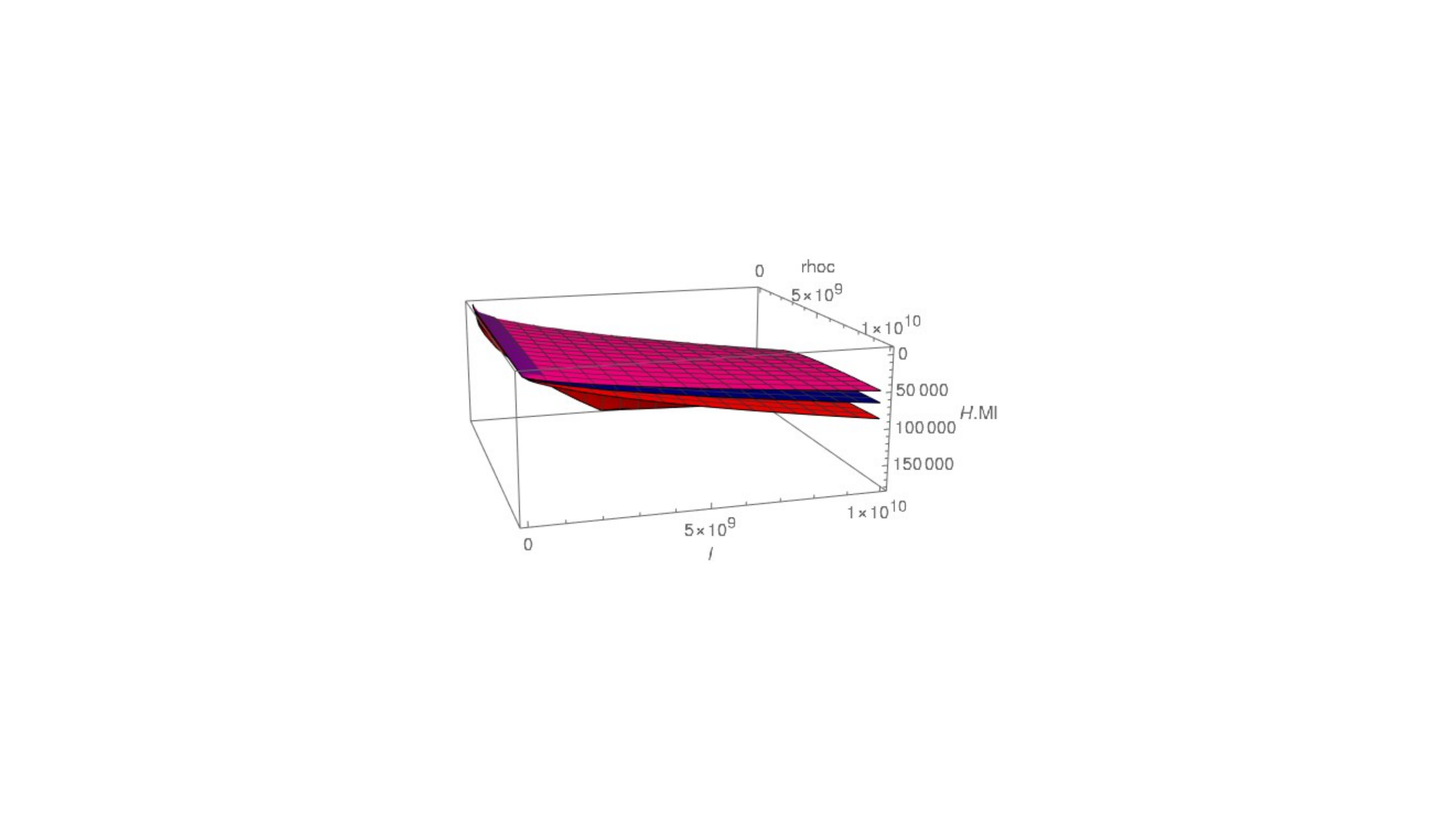}	
\includegraphics[width=.65\textwidth]{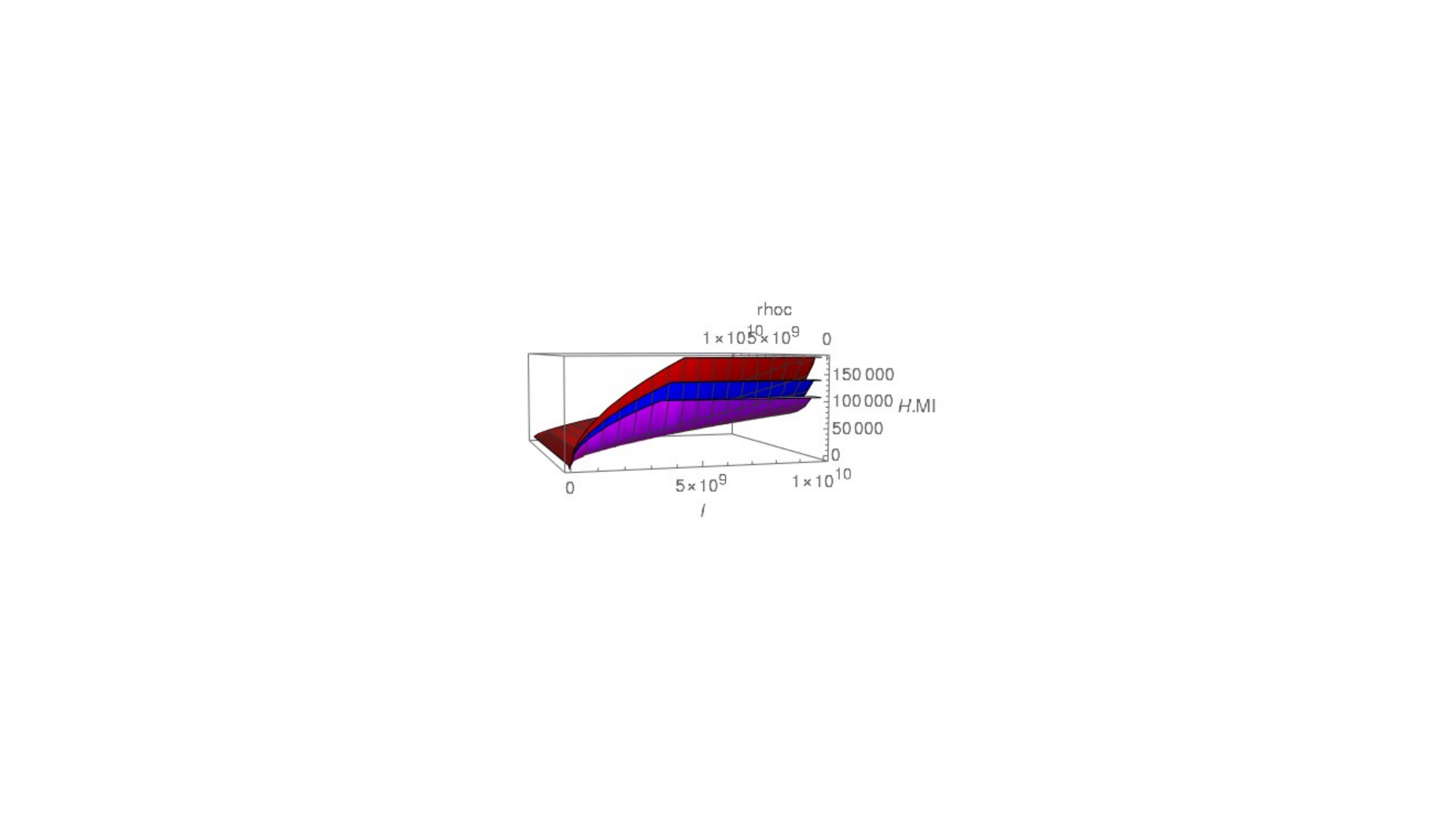}	
\caption{(First row) \,\,:\,\,( From  left to right) \,,\, H.M.I plotted as a function of $(l,\rho_c)$ for $d - \theta = 0.444\,,\, d - \theta = 0.455$,
  \quad;\quad (Second row) \,:\,  ( left)\, : \, H.M.I  plotted as a function of $(l,\rho_c)$ \, ,\,  $d - \theta= 0.466$ \,\,,\,\, (right) \,:\, The frontview of the  overlap of the three, showing that for $l >> \rho_c$ H.M.I decreases with the increase of $d - \theta$
 \, ,\,   \quad;\quad (Last row ) ,\, ( left )\, : \,  \,:\, The backside view of the overlap of the three, showing for $\rho_c >> l$ the H.M.I for different $d - \theta$   actually merges, supporting the fact that for a fixed l, H.M.I decreases with the increase of $\rho_c$ and ultimately falls towards zero.\,\,;\,\. (right)\,:\,  The sideview of the overlap of the three showing that $h_{\rm crit}$  decreases with the increase of $d - \theta$ }
\label{hmievbless1}
\end{figure}

Here the plots (\ref{hmib11}, \ref{hmib1hl}). are showing, H.M.I at $ b= 1$ bearimg the same properties as $b \ne 1$

\section{Entanglement wedge cross section at finite radial cut off}

In this section we are going to evaluate the entanglemen wedge cross section of two strips,  each of the length l, placed along the boundary of a Hyperscaling violating geometry with finite radial cut off, as given in (\ref{finalmetric}), along any of the x-axis, placed both side of the origin, symmetrically.  To reach to an expression of $E_W$, we go back to the original definition of  $E_W$ as defined in \cite{taka}.   To start with let us consider  (\ref{figm}), which is symbollically describing the geometrical configuration
\vskip1mm
 Bulk M: d+1 dimension
\vskip0.5mm
Boundary :  $\partial M$ d dimension
\vskip0.5mm
Boundary is a sum of disjoint manifold 
\vskip0.5mm
$$ \partial M  = N_1 \cup N_2 \cup ...\cup N_n$$
\vskip0.5mm
Subsystem A
$$A = A_1 \cup A_2 \cup ....A_n$$.

\begin{figure}[H]
 \includegraphics[width= 0.5 \textwidth, trim={2cm 12cm 4cm 2cm,clip}]{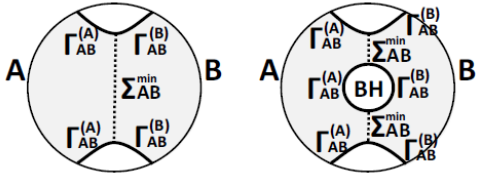}
 \caption{entanglement of purification }
 \label{figm}
 \end{figure}

In (\ref{figm}) the combination A,B, describes a biparite system, describing a physical system with mixed state, while geometrically they are defining the subregion of $ \partial M $.  Now for the subregion A, we can think of codimension 2 surface surface $\Gamma_A$, which satisfy  $\partial A = \partial \Gamma_A$  and are 
homologous to A and minimal area of A will give HEE.  Same about the subregion B.  Now when we bring A and B somewhat close, the entangling surface topology will change and both A and B will share a common entangling surface as indicated in Fig.(\ref{figm}).   Now $\Gamma_A$ is homologous to A implies that there exist a bulk 
region  $R_A$ such that $\partial R_A = \Gamma_A \cup A $.  Now as established in \cite{gravdualdensity3},   $R_A$ is a Cauchy slice and  the domain of dependence of $ R_A $
gives the entanglement wedge.  Accordingly in Fig.(\ref{figm}), we have the entanglement wedge is given by $ M_{AB}$, with

\be
\partial M_{AB} = A \cup B \cup {\Gamma_{AB}^{\rm min}}
\la{entanglementwedge}
\ee.

We divide $\Gamma_{AB}$ into two parts
\be
 \Gamma_{AB} = \Gamma_{AB}^A + \Gamma_{AB}^{B}
\label{definition}
\ee

Define 
\be
{\tilde{\Gamma}}_A = A \cup \Gamma_A^{(AB)} \quad ; \quad    {\tilde{\Gamma}}^B = B \cup \Gamma_B^{(AB)} 
\label{difficult}
\ee

Clearly, from the symmetry of the system, the HEE of these two regimes, $ S( \rho_{\left({\tilde{\Gamma}}_A \right)}) $ and 
$ S( \rho_{\left({\tilde{\Gamma}}_B \right)}) $, must be equal.  Clearly from (\ref{entanglementwedge}) it is evident that

\be
\partial M_{AB} = {\tilde{\Gamma}}^A \cup  {\tilde{\Gamma}}^B
\label{jj}
\ee

Now one can find a surface  $\Sigma_{AB}$ such that

\be
\partial \Sigma_{\rm min}^{AB} = \partial{ \tilde{\Gamma}}^A =  \partial{ \tilde{\Gamma}}^B
\label{partial}
\ee

and also $ \Sigma_{\rm min}^{AB} $ is homologous to ${\tilde{\Gamma}}_A$ and ${\tilde{\Gamma}}_B $.  The entanglement wedge cross section is defined as 
So finally we define  $E_W$ as 
\be
E_W (\rho_{AB}) =  \min_{\Gamma^{(A)}_{AB} \subset \Gamma^{\rm min}_{AB} } = \left\lbrack {\frac{A(\Sigma^{\rm min}_{AB})}{4G_N}}\right\rbrack
\label{cross}
\ee

Now for our case of two strips, placed symmetrically on any x-axis (\ref{finalmetric}), around origin (\ref{figm}) is being realized as (\ref{figmnew})

\begin{figure}[H]
 \includegraphics[width=.65\textwidth]{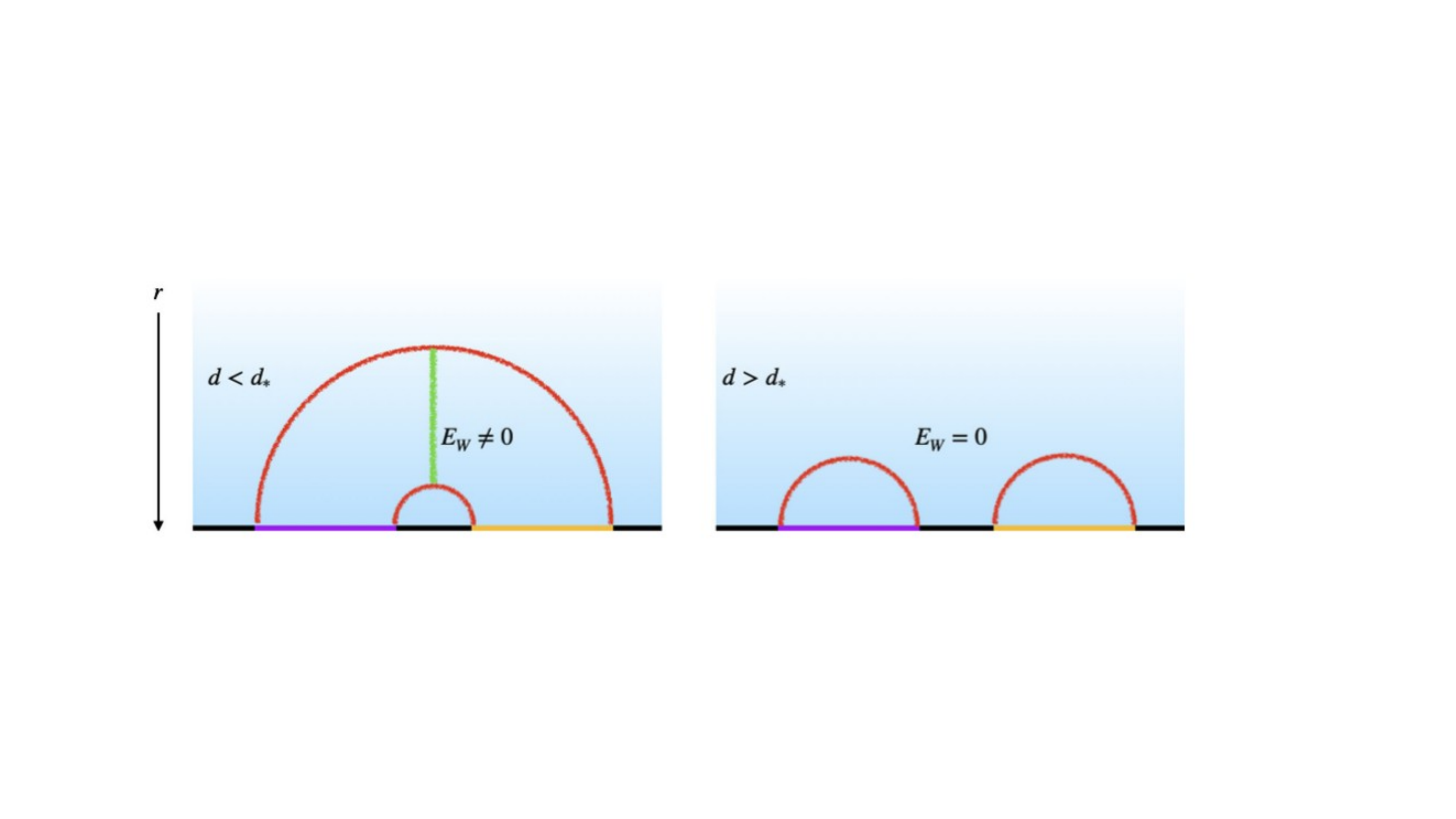}
 \caption{entanglement of purification }
 \label{figmnew}
 \end{figure} 

We need to evalute $\Sigma_{AB}^{\rm min}$ for our system, i.e two stips, each of length l, spaced both side of the origin along $x_i$ axis, as indicated in the figure Fig.(\ref{figmnew}).      Since we have started from the symmetric configuration. i.e two strips are being placed at equal distance from origin, along $x_i$ axis,   and also the entangling surface is given by $ x_i = x_i (r) $,  So clearly $\Sigma_{AB}^{\rm min}$ will run along the radial direction from the tip of the entangling surface $\Gamma_h$ to the tip of the other entangling surface $\Gamma(2l + h)$ where the tip must be given by the repectictive turning point.  So, by using the fact that, 
$\Sigma_{AB}^{\rm min}$
 must be fixed in the time direction, we must have from the metric (\ref{finalmetric}) that $E_W$ is given by

\ber  
E_W &=& {\frac{R^d L^{d - 1}}{4G_N}} \int_{\rho_0 ( h )}^{ \rho_0 (2l + h)}  {\frac{d\rho}{\rho^{d - \theta}}}\n
     &=& {\frac{1}{1 - d + \theta}}   {\frac{R^d L^{d - 1}}{4G_N}}  \left\lbrack  {\left\lbrace\rho_0 (2l + h) \right\rbrace}^{1 - d + \theta}   -  {\left\lbrace\rho_0 ( h) \right\rbrace}^{1 - d + \theta}\right\rbrack \quad;\quad {\rm for}\,\, d - \theta \ne 1
\la{Ew}
\eer
For $ d - \theta = 1 $,  we have

\ber  
E_W &=& {\frac{R^d L^{d - 1}}{4G_N}} \int_{\rho_0 ( h )}^{ \rho_0 (2l + h)}  {\frac{d\rho}{\rho^{d - \theta}}}\n
     &=&   {\frac{R }{4G_N}} \int_{\rho_0 ( h )}^{ \rho_0 (2l + h)}  {\frac{d\rho}{\rho}}\n     
      &=& {\frac{R }{4G_N}} \log\left\lbrack {\frac{\rho_0 (2l + h)} {\rho_0 ( h )}}  \right\rbrack
\la{Ewzero}
\eer

Now we will come to the expected properties of EWCS in the presence of a cut off..

Now coming to the the properties of EWCS first we consider the expression of EoP given by (\ref{eopdefinition}).  Since this is basically EE coming from the entanglement of the original mixed state and the purifying state and when the original state is the pure one it reduces to the Von Neumann entropy of the same.    
Clearly then,  from the field theory point of view, as we discussed in the section 4,  it will increases with l and decreases with the increase of the cut off
 $\rho_c$.  Again, from the field theory point of view,  EWCS will show a discontinuous phase transition, when the system is taking transition from a connected to a disconnected phase,   i.e  exactly where the mutual information is   undergoing a  first order phase transition,  as indicated in  Fig.(\ref{figmnew}).    Consequently
for a given $(l,\rho_c)$,  EWCS will decrease when the separation between the two strips,  h  will increase.  Finally,  exactly at the point of transition,  EWCS  is expected to show a sudden fall from its finite value  to zero(for disconnected phase), which indicates a discontinuous phase transition!  
\vskip0.5mm  
In the gravity side, for a fixed h,l,  if we increase $\rho_c$,  since the tip of the entangling surface associated with h.  given by $\Gamma_h$ will grow faster than tip of the $\Gamma_{2l + h}$ ,  as evident from     (\ref{ultimaterho0nonvanishing}) and also intutively, from Fig.(\ref{entanglingsurface})   that due to increase of cut off 
the effective reduction in   $  \Gamma_{2l + h}$  is much more than $\Gamma_h$,  so EWCS will decrease with the increase of cut off!  Also for a fixed $(l,\rho_c)$,  if we increase the separation h.  again from (\ref{ultimaterho0nonvanishing}) is evident that since the tip of $ \Gamma_h $ will grow faster than the tip of 
 $ {\Gamma_{2l + h} }$,  so EWCS will fall with the increase of h and become minimum at the point of transition.   However  unlike HMI which is zero at the point of transition,   it is evident that EWCS will remain finite at the point of transition.  It will be evident from the discussion of our case (\ref{ewcsb1})    that EWCS will become zero only at $l = 0$ and so remain finite at the point of transition.  So for EWCS,  it would be a discontinuous phase transition and exactly at the point of transition,  there would be a sudden drop in EWCS from its finite value to zero  ( as schematically shown in (\ref{figmnew}) !   Also, without cut-off,   EWCS is an UV finite quantity as long as the objects A and B are separated!  When they are being combined, i.e share a common boundary, EWCS will diverge!  The scenario changes in the presence of the cut off as either evident from (\ref{Ew}, \ref{Ewzero}).    Now in the presence of cut-off,  since it will never allow the tip of the entangling surface given by $\Gamma_h$ to go to zero or in other words even when the separation h vanishes,   it will effectively act as regulator to make  EWCS non-diverging!  However at zero cut off , we must find the divergence!  

 Moreover EWCS shows certain inequalities \cite{ewcsinequalities},  one of which is   
\be
EWCS \ge {\frac{1}{2}}\cdot HMI
\la{inequal1}
\ee
 This inequality saturates,  only when the concerned physical system goes to the pure state!  Now for very large cut-off, since EE, MI as well as EWCS will go to zero
for  $\rho_c >> l$,  so at that regime   the system can thought as it is in  its pure state and consequently at $\rho_c >> l$ the inequality is expected to saturate where in the other regime $l>> \rho_c$,  where the mixed state persists, it would be a normal inequality!

Also,  since in the connected phase we have l >> h,  so from (\ref{ultimaterho0nonvanishing}) it is evident that for $l >> \rho_c$ regime we will have the most dominent term in  the expression of EWCS  (\ref{Ew}) is given by $ \sim {\left\lbrace\rho_0 (2l + h) \right\rbrace}^{1 - d + \theta} \sim l^{1 - d + \theta}$
  so that EWCS, in this regime is expected to decrease with the increase of $(d - \theta)$ in spite of the decrease of the denominator, given by $ 1 - ( d- \theta)$ 
which will be less effective in large l regime!  On the otherhand,  once we move to the regime $\rho_c >> l$ regime. then with $l >> h$ for connected phase, we will have the dominant contribution of EWCS in this regime will be given by zero (as follows on substitution of (\ref{ultimaterho0nonvanishing}) in (\ref{Ew})) and remain invariant on variation of $d - \theta$!    Combining the two aspect,  one can see that when we define the global expression over $(l, \rho_c)$ plane,  EWCS will decrease with the increase of $(d - \theta)$,  when the difference in their expression for different  $(d - \theta)$ will be maximum in $l >> \rho_c $ regime and minimum in $\rho_c >> l$ regime, where they all merge together to zero!  Indeed this is the behaviour of EWCS we are going to establish in the next few sections! 

\vskip0.5mm

Finally,  coming to the question of the impact of the global symmetry (\ref{2dsymmetry}) on EWCS, we can just extend it for the scaling of h as well as  in the case of HMI and can write,  following  (\ref{Ew}, \ref{Ewzero})
\be
(l, h, \rho_c) \rightarrow (kl, kh, k\rho_c) \Rightarrow {E_W}(kl, kh, k\rho_c) = {k}^{1 - (d - \theta)}  {E_W}(l, h, \rho_c) 
\la{symmetryewcs}
\ee

Here finally we must comment on the fact that when the $\rho_c = 0$, the symmetry,  as expressed in (\ref{symmetryewcs}),  do hold 
as shown in \cite{lifshitz, chargedbrane }.   However there it was a normal mathematical  functional dependence and not really corresponds to any spacetime symmetry with the geometrical origin.   This fact actually supports the view of our discussion in section 3 that  the emergent bulk symmetry can really be viewed as generalization of boundary scaling symmetry or vice versa.

Here in the next three subsections we will obtain the plots for EWCS for $d - \theta = 1$, $d - \theta > 1$, $d - \theta < 1$ and show that these properties are indeed holds for the complete parameter regime of $d, \theta$ with $ \theta \le d $   and show all the above properties hold!.   However one drawback of our plots is since we are unable to determine $h_{\rm crit}$ because of complicated structure of H.M.I and consequently cannot tell about the exact phase transition point,  so here we will present the plots for the connected phase only  with extrapolation to $l = 0$!

\subsection{EWCS for $d-\theta =  1$}

Here, for $d - \theta = 1$, we can obtain the expression, following (\ref{Ewzero}),  where we substitute the expression for $\rho_0$ from (\ref{dthetra1} ).

\ber  
E_W &=& {\frac{R^d L^{d - 1}}{4G_N}} \int_{\rho_0 ( h )}^{ \rho_0 (2l + h)}  {\frac{d\rho}{\rho^{d - \theta}}}\n
     &=&   {\frac{R^d L^{d - 1}}{4G_N}} \int_{\rho_0 ( h )}^{ \rho_0 (2l + h)}  {\frac{d\rho}{\rho}}\n     
      &=& {\frac{ R^d L^{d - 1} }{4G_N}} \log\left\lbrack {\frac{\rho_0 (2l + h)} {\rho_0 ( h )}}  \right\rbrack
\la{Ewzeronew12123}
\eer

Here we first present the plot for $E_W $  vs $(l,\rho_c)$ plot  to understand its basic nature Fig.(\ref{ewcsb1}), whether the inequality $E_W  \ge {\frac{1}{2}} \left( H.M.I\right)$ is satisfied and it indeed saturates in $\rho_c >> l $ regime,   we have explored this fact in in Fig,(\ref{ewcshmib1ineauality}  ),   the variation of      $E_W$   with (h, l)  for a given cut off is given in Fig.(\ref{ewcsb1hl}) and finally we obtain  plot for $E_W$ vs $(\rho_c. h)$
 to check whether we have divergence at $h = 0$ for zero cut off   in  (\ref{ewcsb1rhoch}).  Since we failed to find $h_{\rm crit}$ because of the complication of the expression of HMI so here we will show the plots of EWCS for its connected phase only.  EWCS, unlike HMI, posses a finite value at the point of transition and then suddenly drops to zero so that it shows a discontious phase transition.  In the connected phase EWCS becomes zero exactly $l = 0$.  Here we extrapolate our plots for connected phase upto that point.

\begin{figure}[H]
\begin{center}
\textbf{ For $d - \theta =1$, EWCS vs $(l,\rho_c)$ plot  for fixed h for the connected phase, extrapolated to $l = 0$ }
\end{center}
\vskip2mm
\includegraphics[width=.65\textwidth]{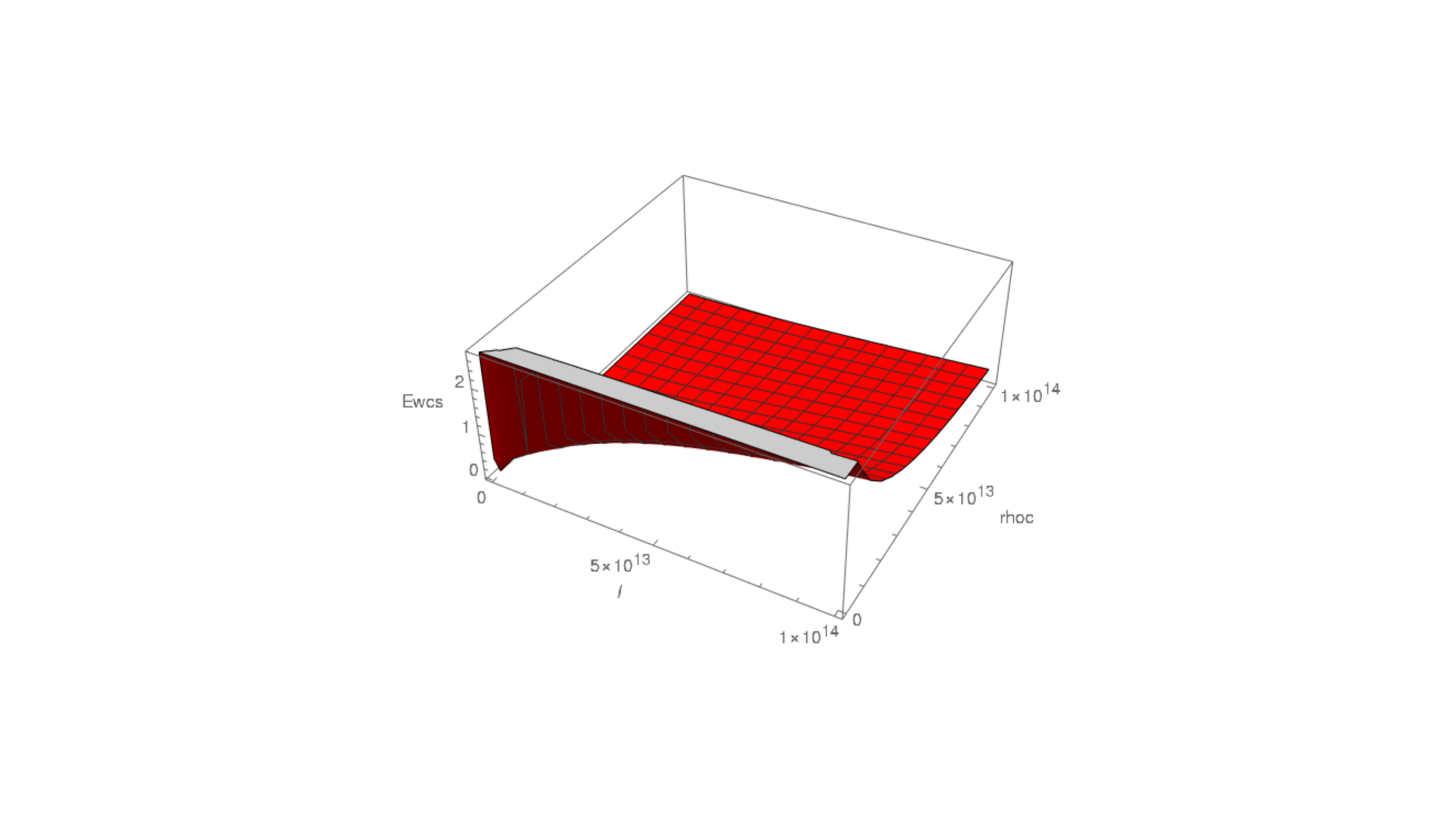}
\includegraphics[width=.65\textwidth]{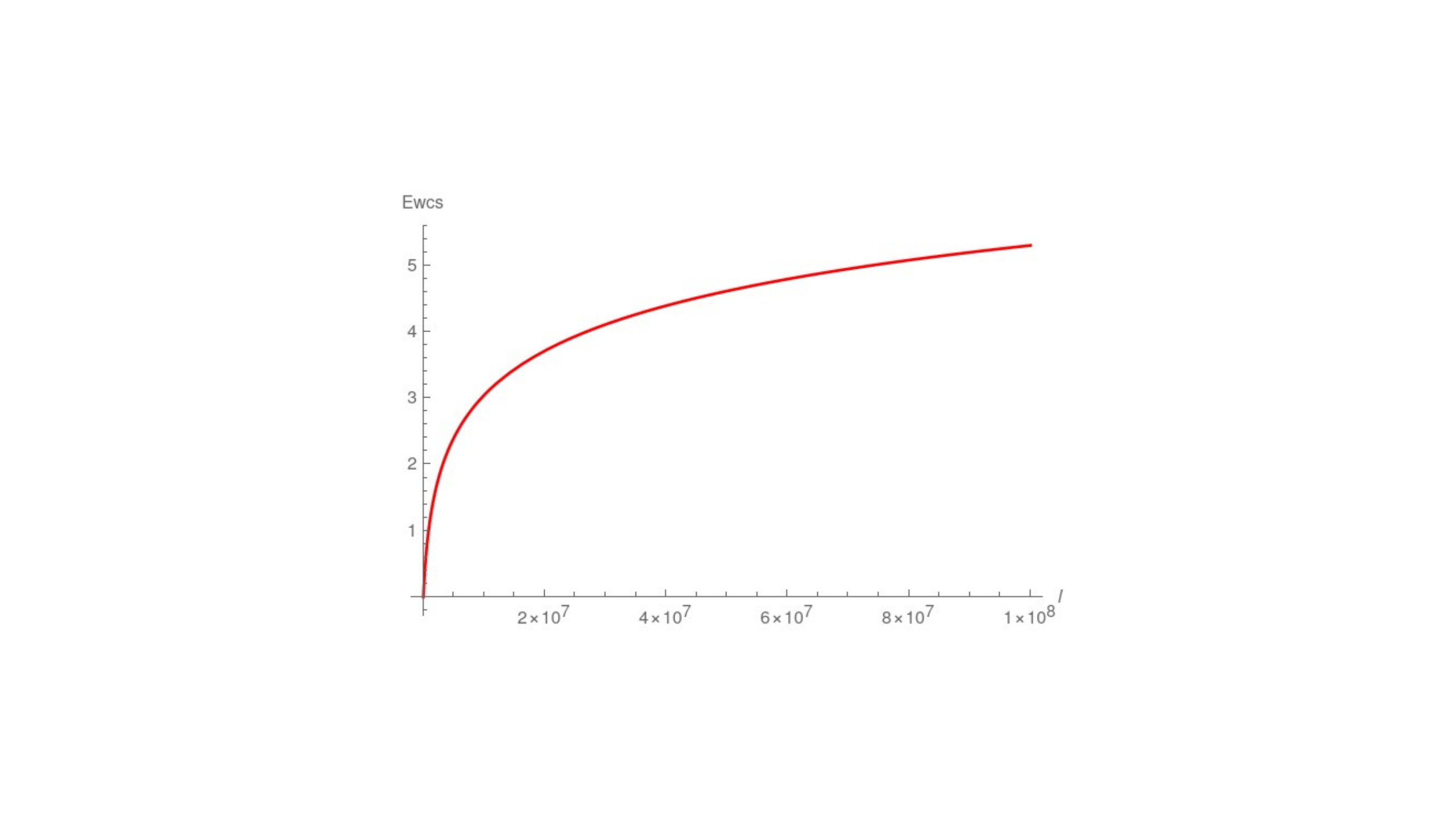}
\includegraphics[width=.65\textwidth]{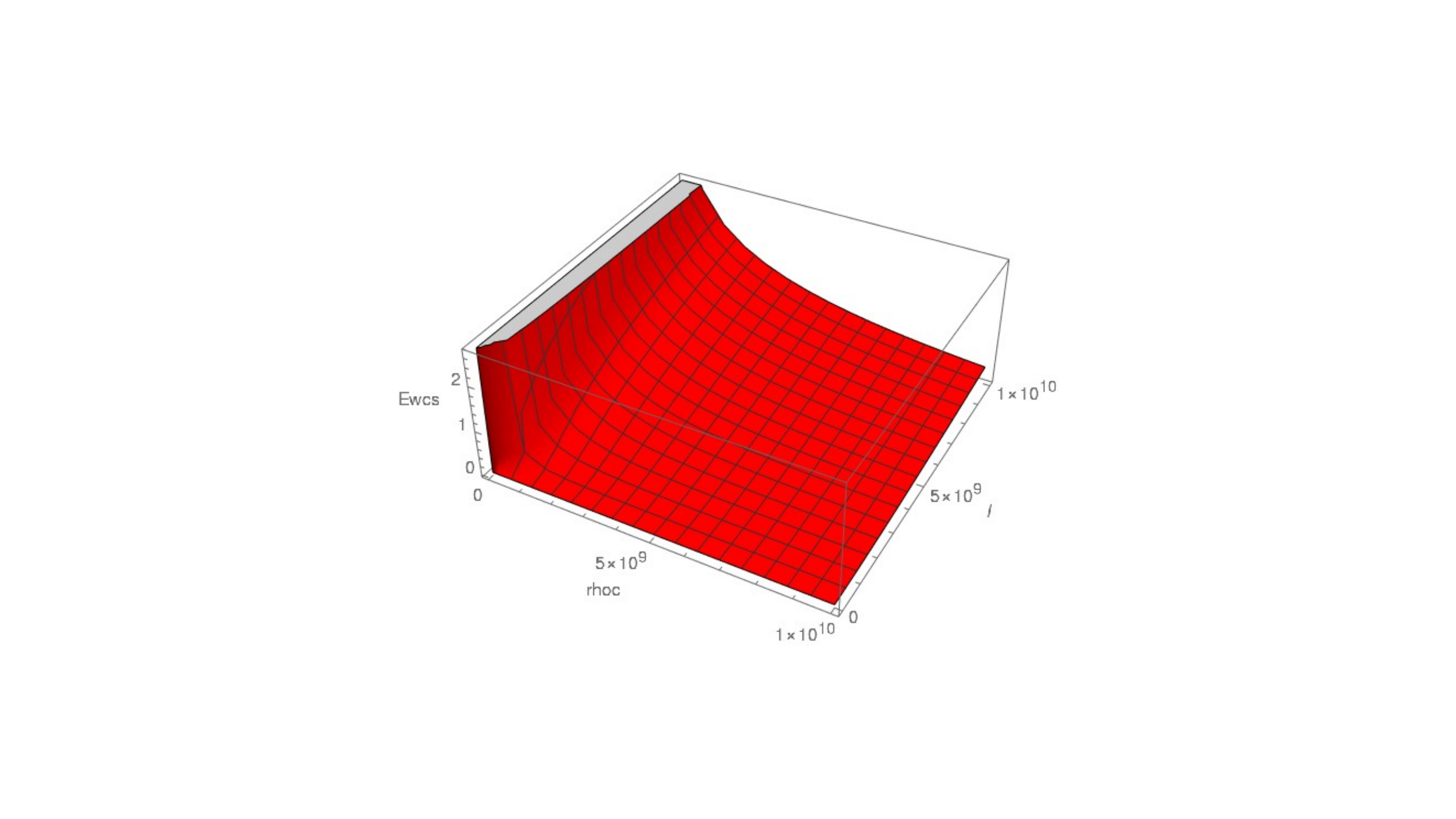}
\includegraphics[width=.65\textwidth]{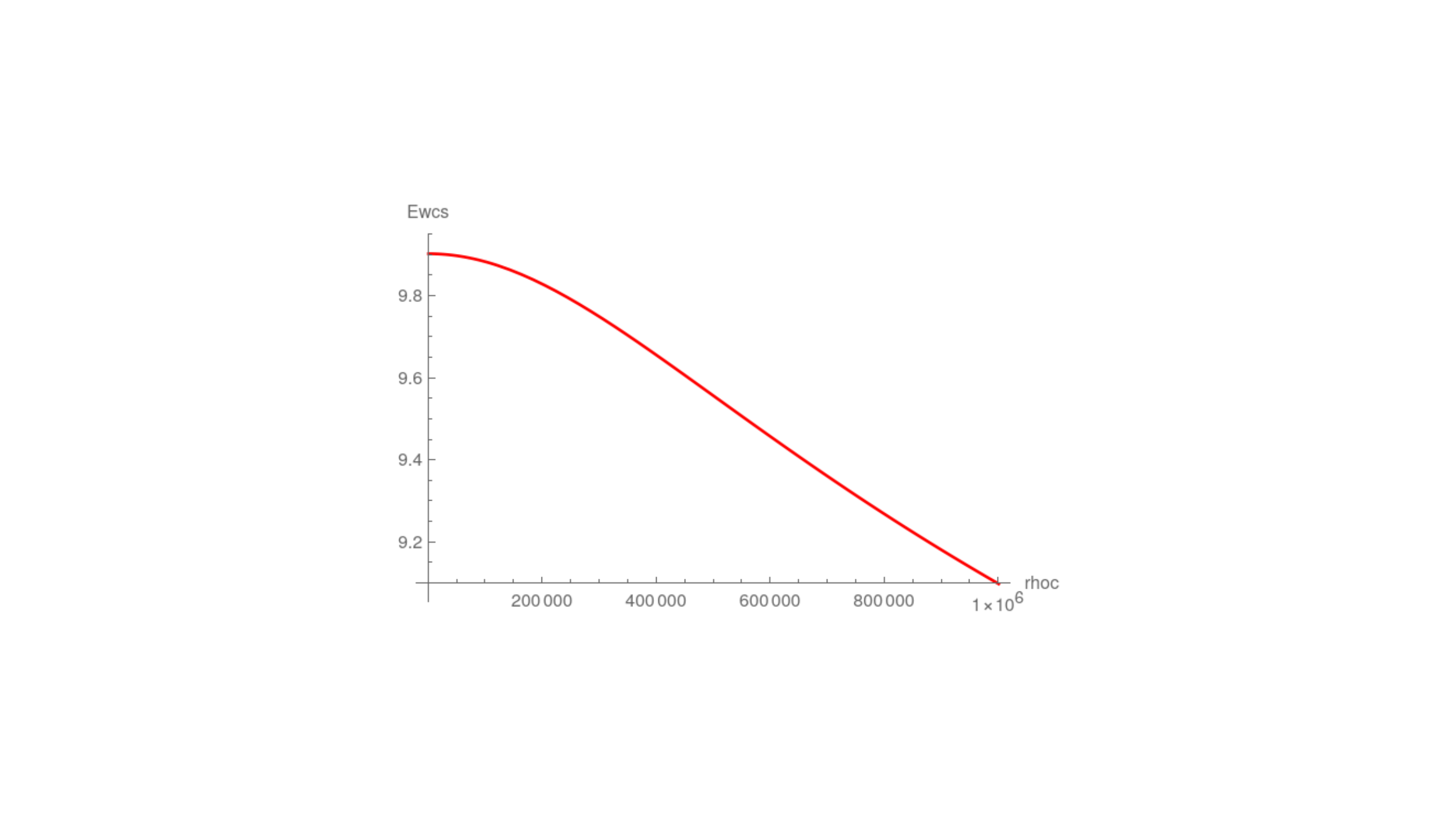}
\includegraphics[width=.65\textwidth]{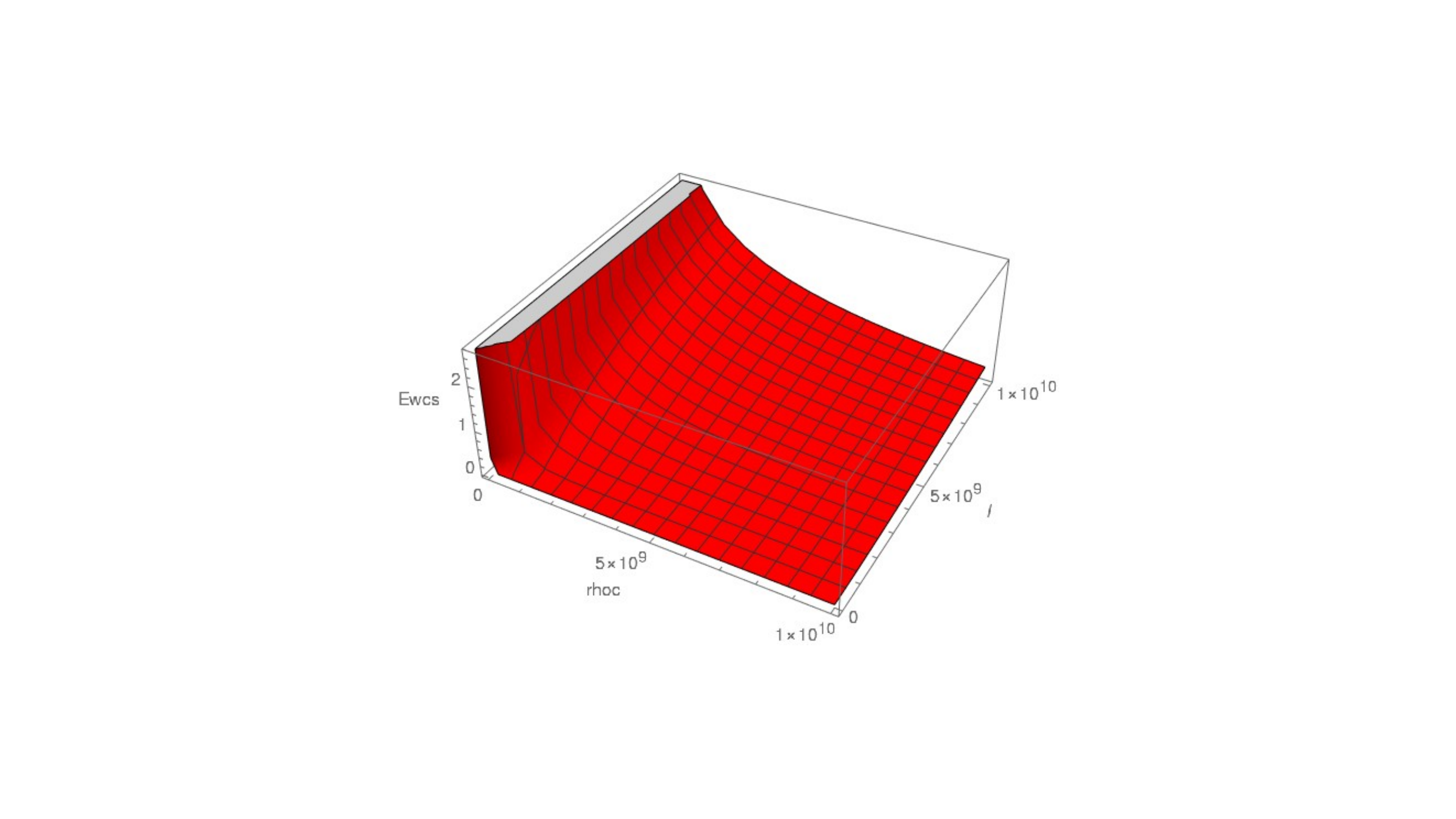}
\includegraphics[width=.65\textwidth]{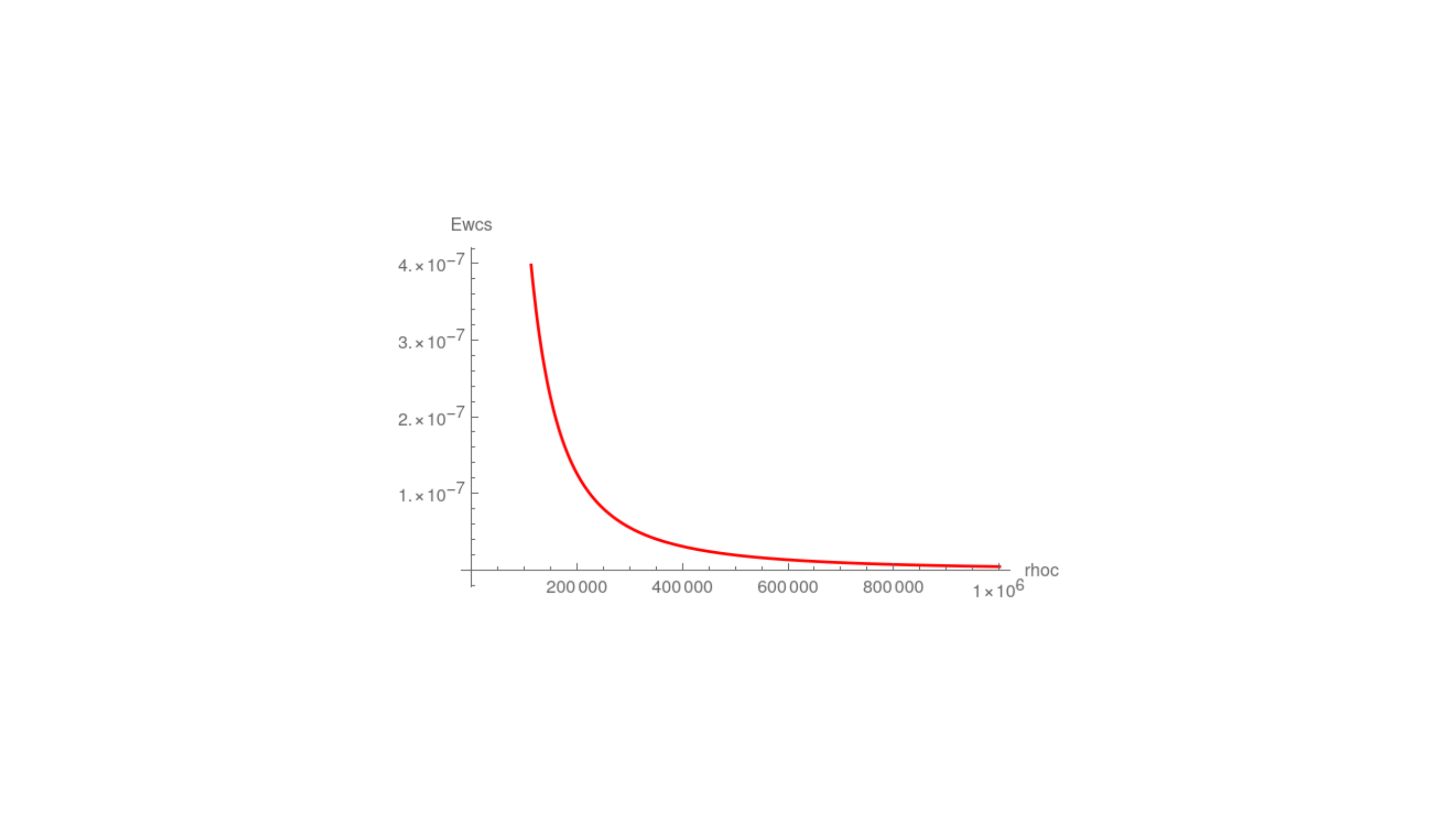}
\caption{ (First row) \,\,: (left)\,:\, EWCS as a function of $(l,\rho_c)$ for  $h = (10)^6$ \,,\, (right)\, : \, EWCS  as a function of l with $\rho_c = 10000$,  $h = (10)^6$  \,:\,  Both the plots are showing, in the connected phase EWCS increases with l, decreases with $\rho_c$ 
\quad;\quad (Second row) \,:\,  ( left)\, : \, EWCS  as a function of $(\rho_c , l)$ for  $h = (10)^6$\,, (right) \,:\, EWCS  as a function of $\rho_c$ 
 for $l = {(10)}^{10}$,   $h = {(10)}^6$,  this value of l is chosen to probe $l>> \rho_c$ regime where for $\rho_c >> l$ EWCS is zero \, :\,  Both the 3D and 2D plots are showing, for a given l,  EWCS   falls with the increase of cut off $\rho_c$ and goes to zero for $\rho_c >> l $ regime.  Also for nonzero h,    
EWCS is finite at $\rho_c = 0$
 \quad;\quad (Last row) \,:\,  ( left)\, : \, EWCS  as a function of $(\rho_c , l)$ for  $h = 0 $\,, (right) \,:\, EWCS  as a function of $\rho_c$  for $l = 100 $,  $h = 0$,   \, :\,  Both the 3D and 2D plots are showing for $h = 0$,    EWCS  diverge at $\rho_c = 0$ as expected     }
\la{ewcsb1}
\end{figure}

Next in order to establish the inequality  $E_W  \ge {\frac{1}{2}} \left( H.M.I\right)$, we consider  the expression of H.M.I from Fig.(\ref{hmib11}) which was defined over $h = {10}^6$ and consider the overlap of the  $EWCS-l-\rho_c$ and ${\frac{1}{2}}$ $(H.M.I)-l-\rho_c$ plot,  to see, whether the inequalituy holds:
 Fig.(\ref{ewcshmib1ineauality})

\begin{figure}[H]
\begin{center}
\textbf{ For $d - \theta =1$, to show the EWCS-H.M.I inequality holds,   $h = {10}^6$  }
\end{center}
\vskip2mm
\includegraphics[width=.65\textwidth]{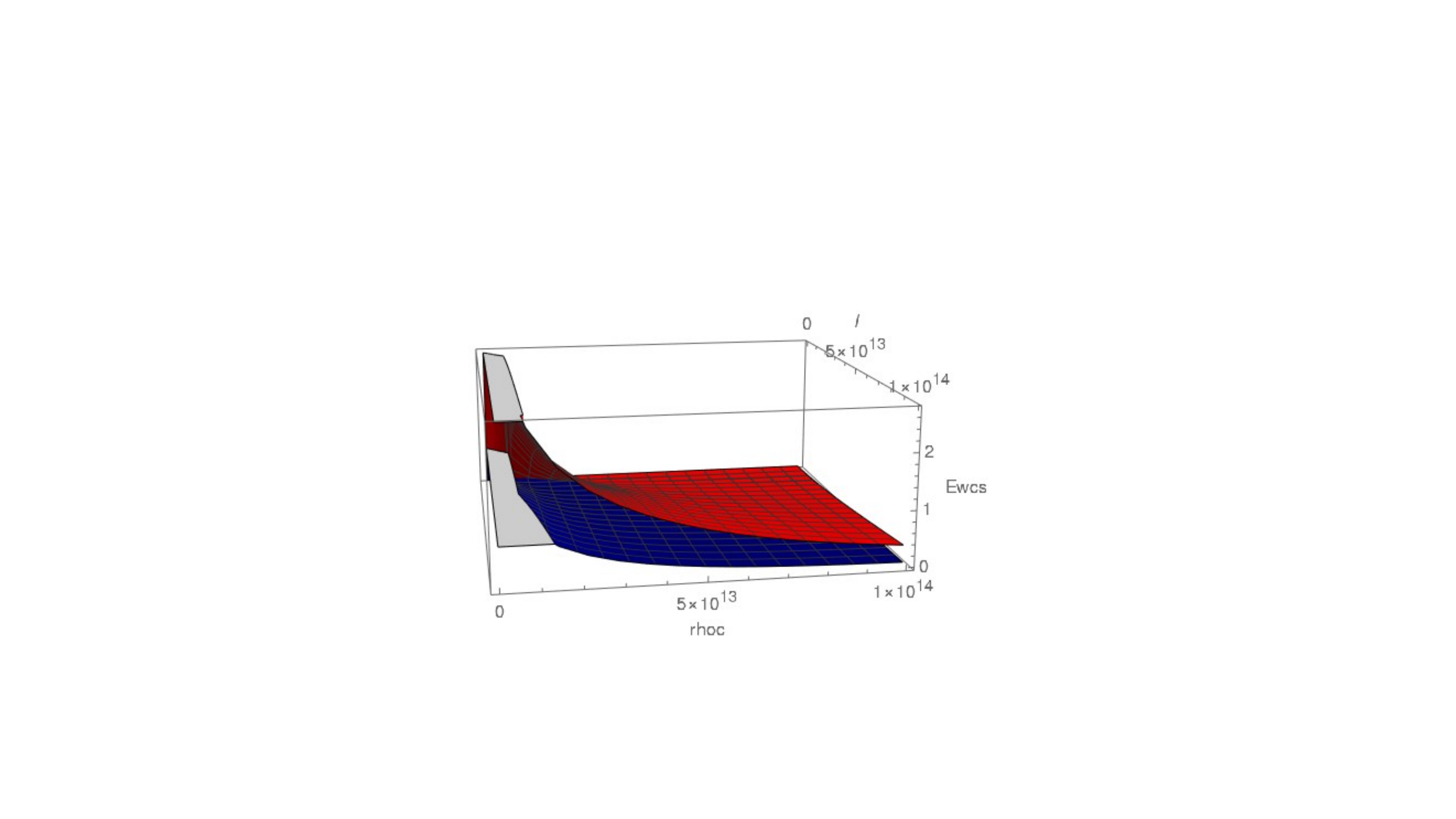}
\includegraphics[width=.65\textwidth]{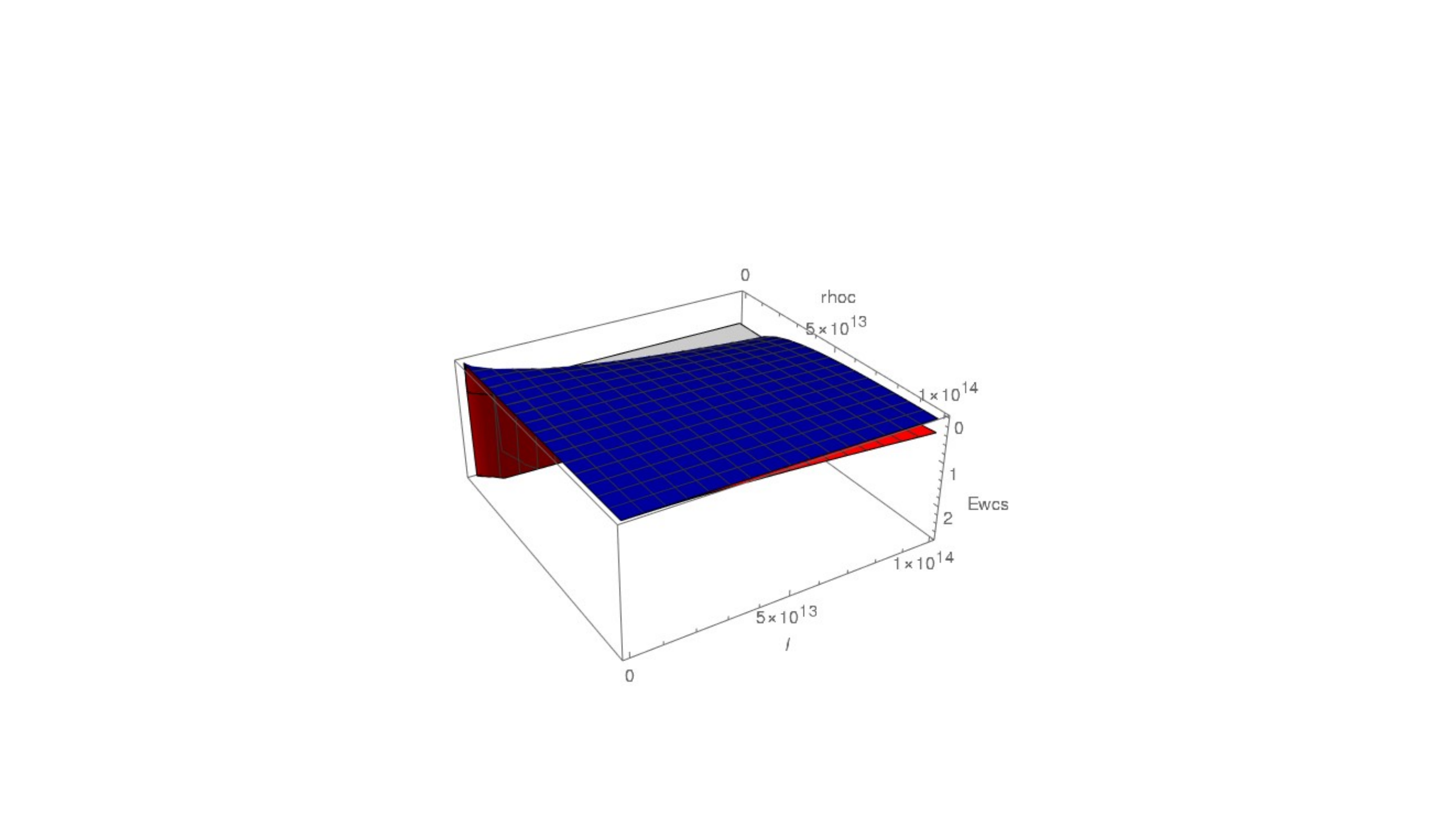}
\caption{The overlap of $EWCS-l-\rho_c$ (in Red) and ${\frac{1}{2}} (H.M.I)-l-\rho_c$ (in Blue) plots are being considered,showing \,,\, (left)\,:\, The  frontview of the overlap plot showing for $l>> \rho_c$ the inequality holds,  \, , \, (right) \,:\, The backside view of the overlap plots  showing for the regime $\rho_c >> l $ the two merges, i.e the inequality saturates.   }
\la{ewcshmib1ineauality}
\end{figure}

Finally to study the behaviour of EWCS  as a function of h, we consider Fig,(\ref{ewcsb1hl})

\begin{figure}[H]
\begin{center}
\textbf{ For $ d - \theta = 1$, EWCS as a function of h,l,  for the connected phase, extrapolated to $l = 0$    }
\end{center}
\vskip2mm
\includegraphics[width=.65\textwidth]{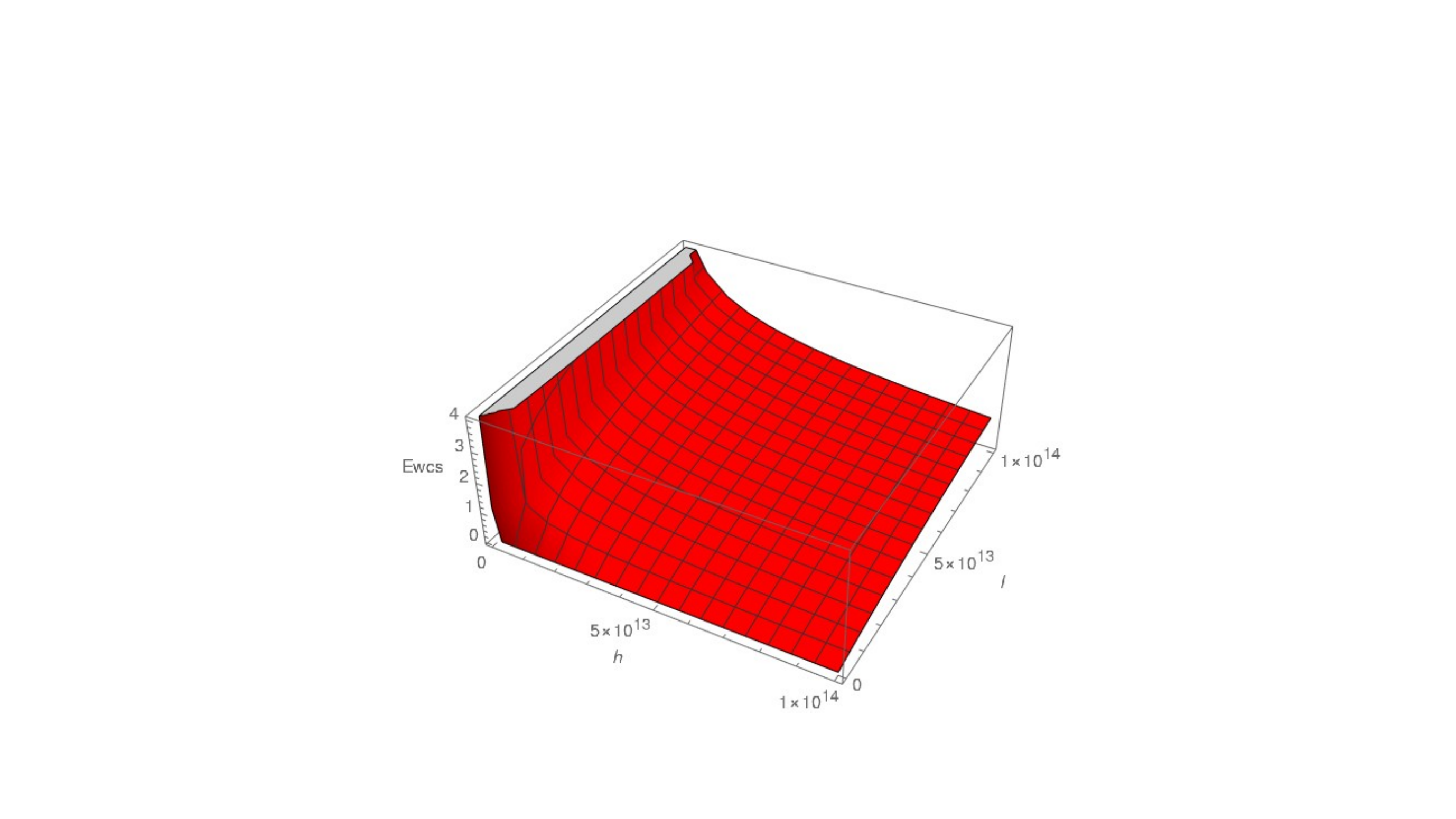}
	
\caption{EWCS for $d - \theta = 1$ is plotted as a function of (h,l)\,,\, \,,\, $\rho_c = 100$,   the plot is showing for for a given h, EWCS  increases with the increase of l and for a given l, for fixed $\rho_c$ they fall with the increase of h
}

\label{ewcsb1hl}
\end{figure}

\begin{figure}[H]
\begin{center}
\textbf{ For $d - \theta = 1$, EWCS as a function of $(\rho_c , h)$  for the connected phase, extrapolated to $l = 0$  }
\end{center}
\vskip2mm
\includegraphics[width=.65\textwidth]{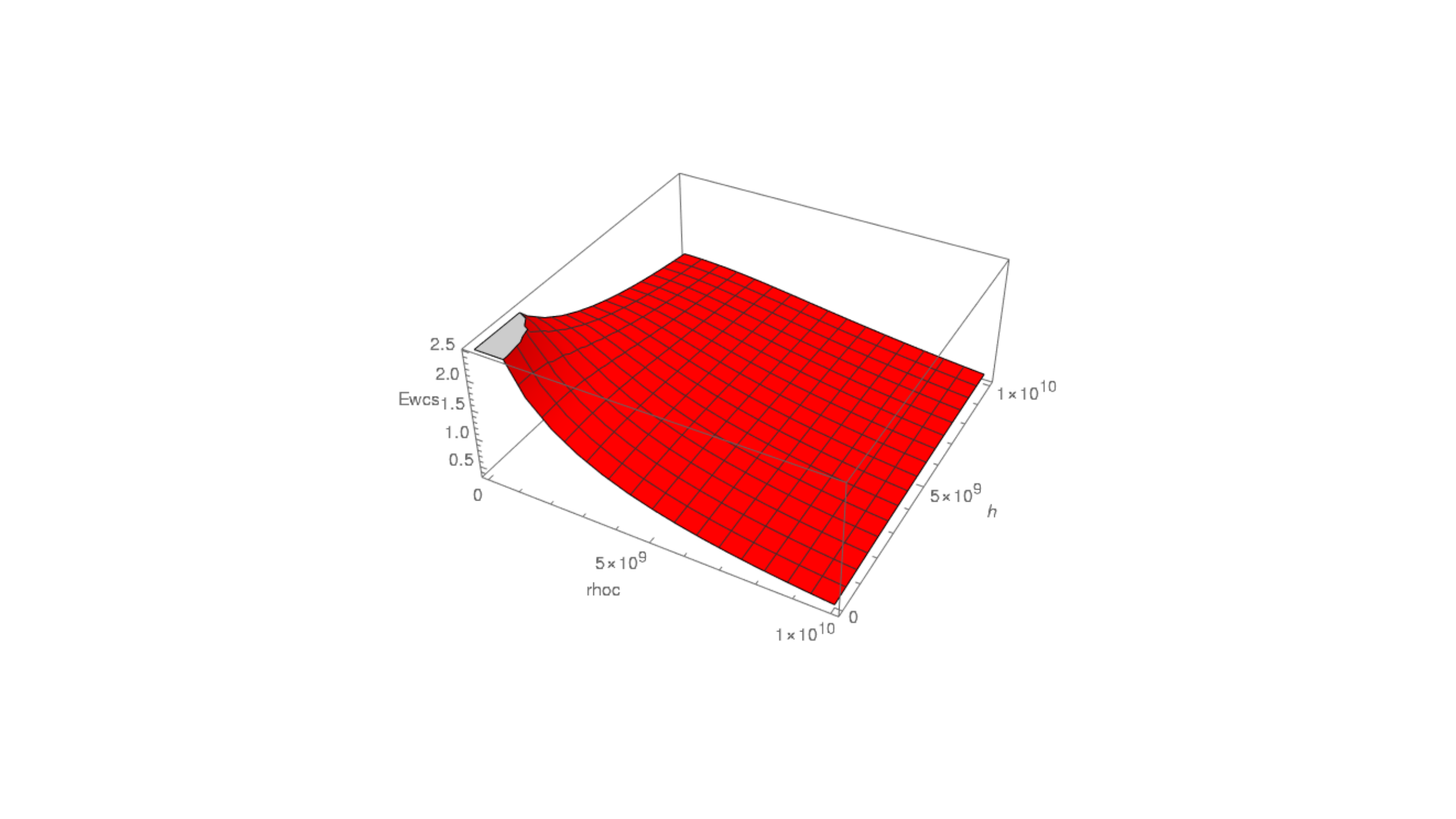}

\caption{  The plot is  showing EWCS falls with increase of both  $(\rho_c, h)$,   diverge at $h=0$ for zero cut-off but become finite at $h = 0$ for nonzero cut off }
\la{ewcsb1rhoch}
\end{figure}

  \subsection{EWCS for $d - \theta > 1$ }
Here we will consider  the expression of EWCS from (\ref{Ew}), substituting the expression for $\rho_0$  from (\ref{ultimaterho0nonvanishing}),  study its different properties as mentioned in the beginning of this section.  
  We will present it through Fig.(\ref{ewcsbasic3by2}, \ref{ewcsbasic7by3}, \ref{ewcsbgreaterthan1hl} , \ref{ewcsbgreaterthan1rhoch}, 
\ref{ewcsevbgreater1} ).  Since we failed to find $h_{\rm crit}$ because of the complication of the expression of HMI so here we will show the plots of EWCS for its connected phase only.  EWCS, unlike HMI, posses a finite value at the point of transition and then suddenly drops to zero so that it shows a discontious phase transition.  In the connected phase EWCS becomes zero exactly $l = 0$.  Here we extrapolate our plots for connected phase upto that point.

\begin{figure}[H]
\begin{center}
\textbf{ For   $ d - \theta  = {\frac{3}{2}}$\,: \, Ewcs vs $(l,\rho_c)$, Ewcs vs l, Ewcs vs $\rho_c$ graph for fixed h for the connected phase, extrapolated to $l = 0$ }
\end{center}

\includegraphics[width=.45\textwidth]{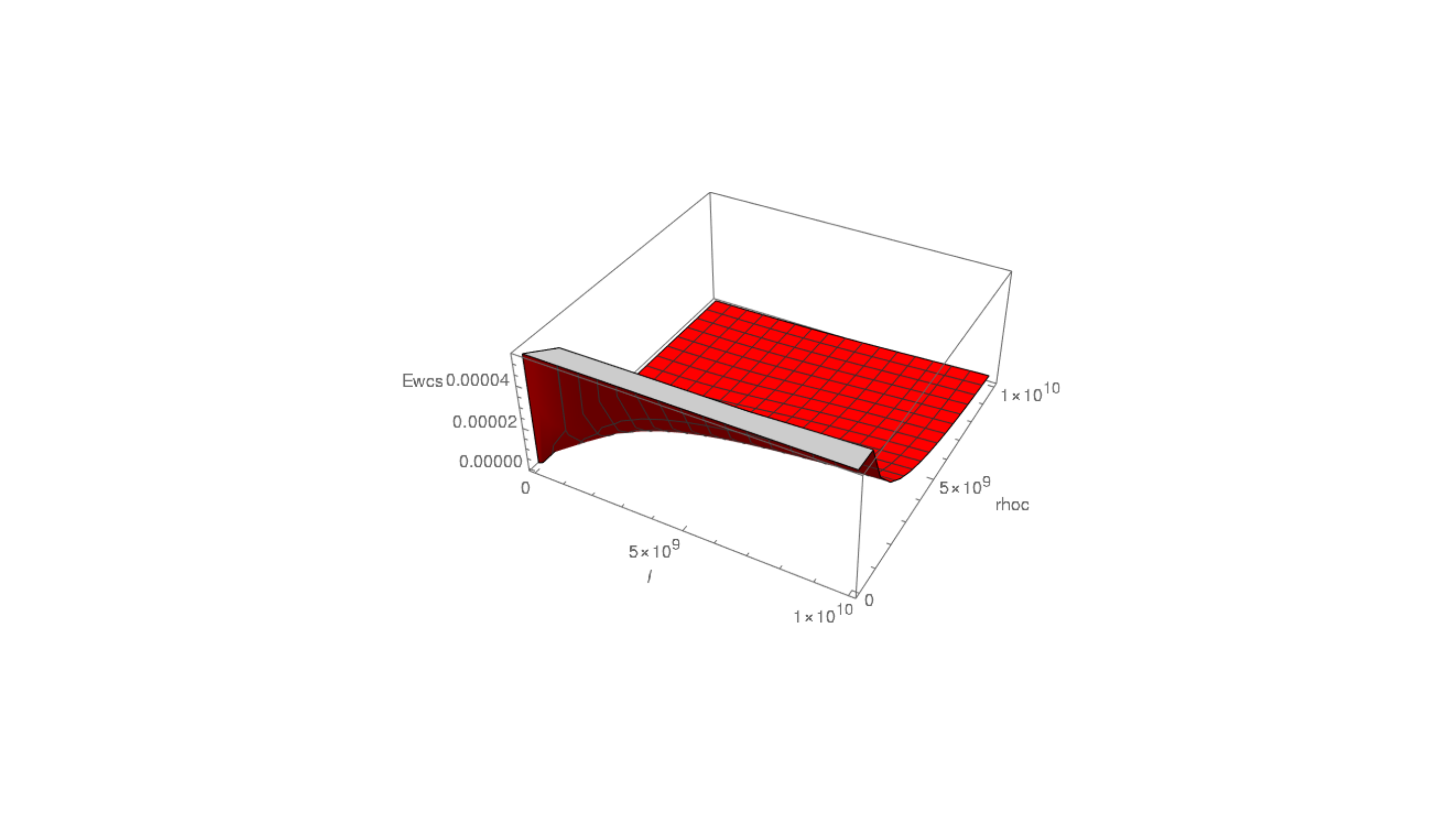}
\includegraphics[width=.45\textwidth]{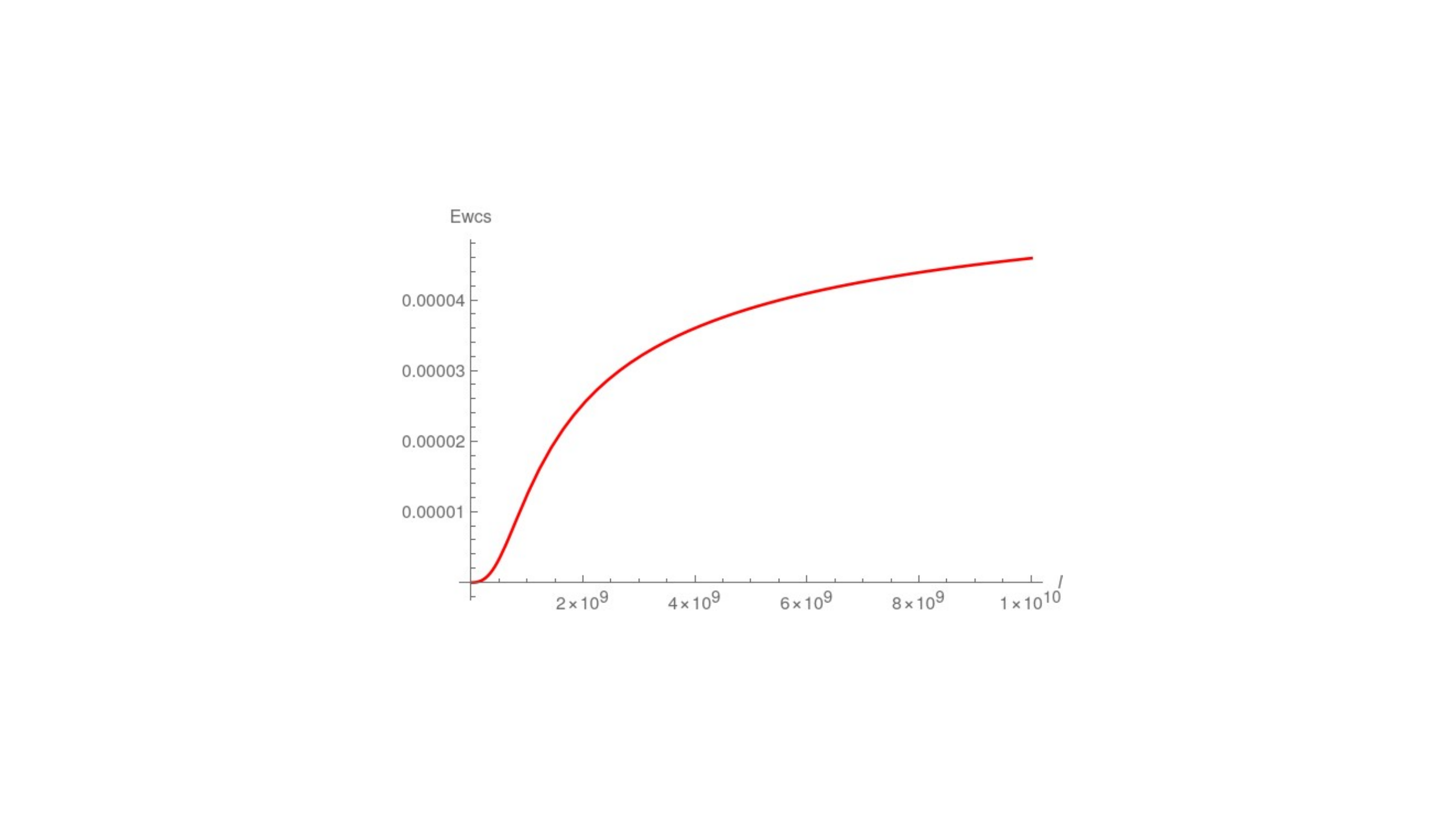}	
\includegraphics[width=.45\textwidth]{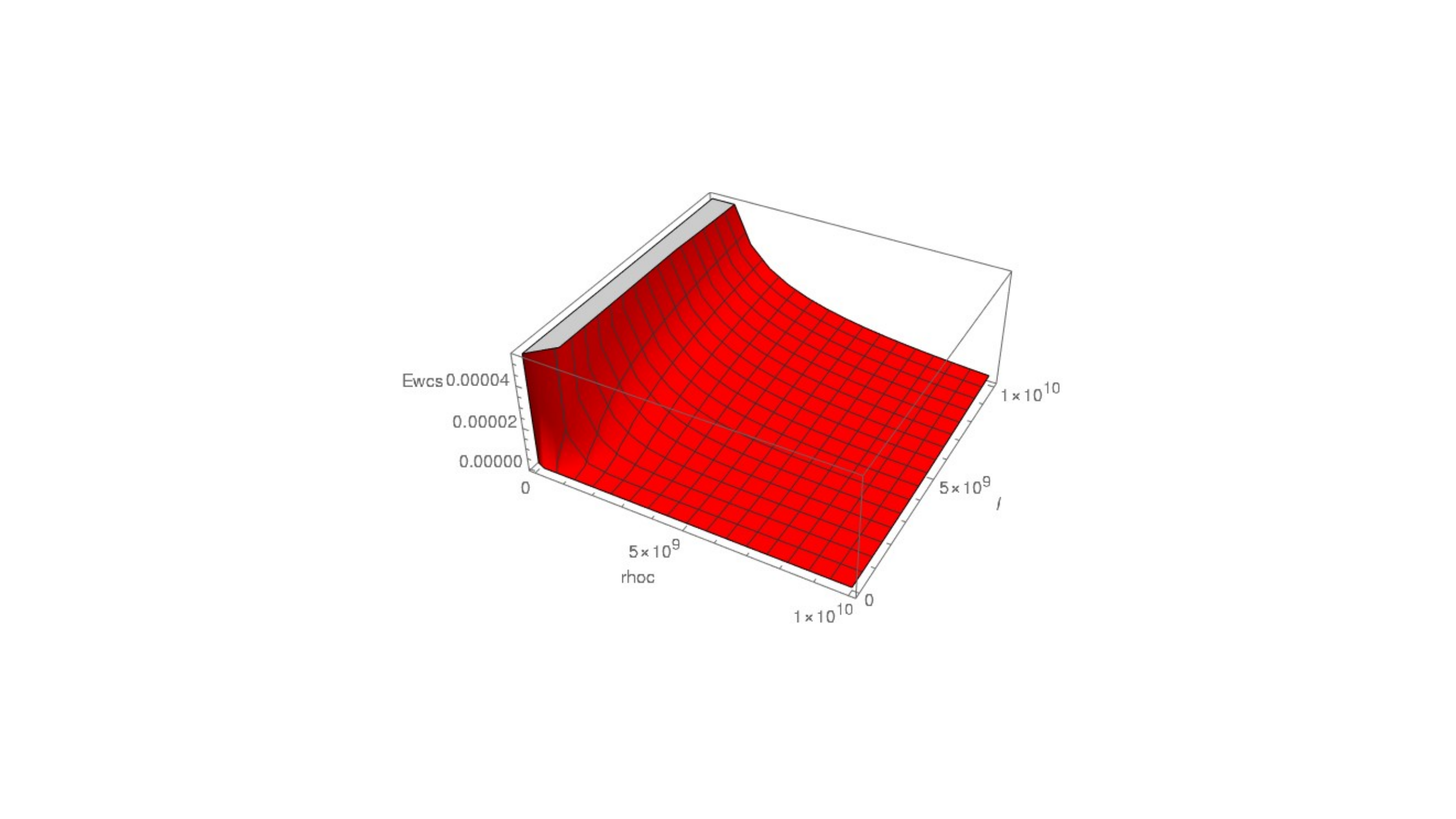}
\includegraphics[width=.45\textwidth]{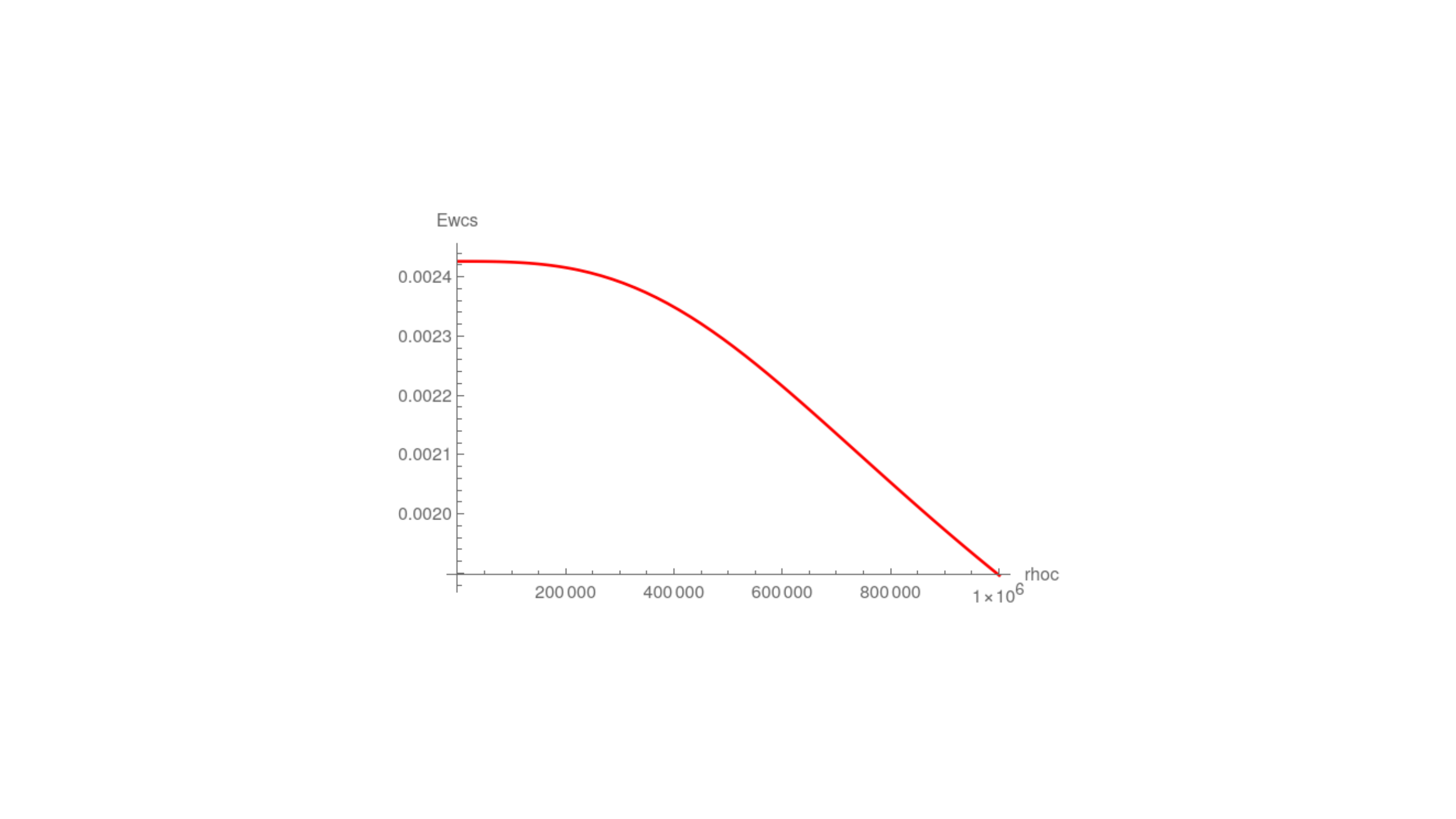}
\includegraphics[width=.45\textwidth]{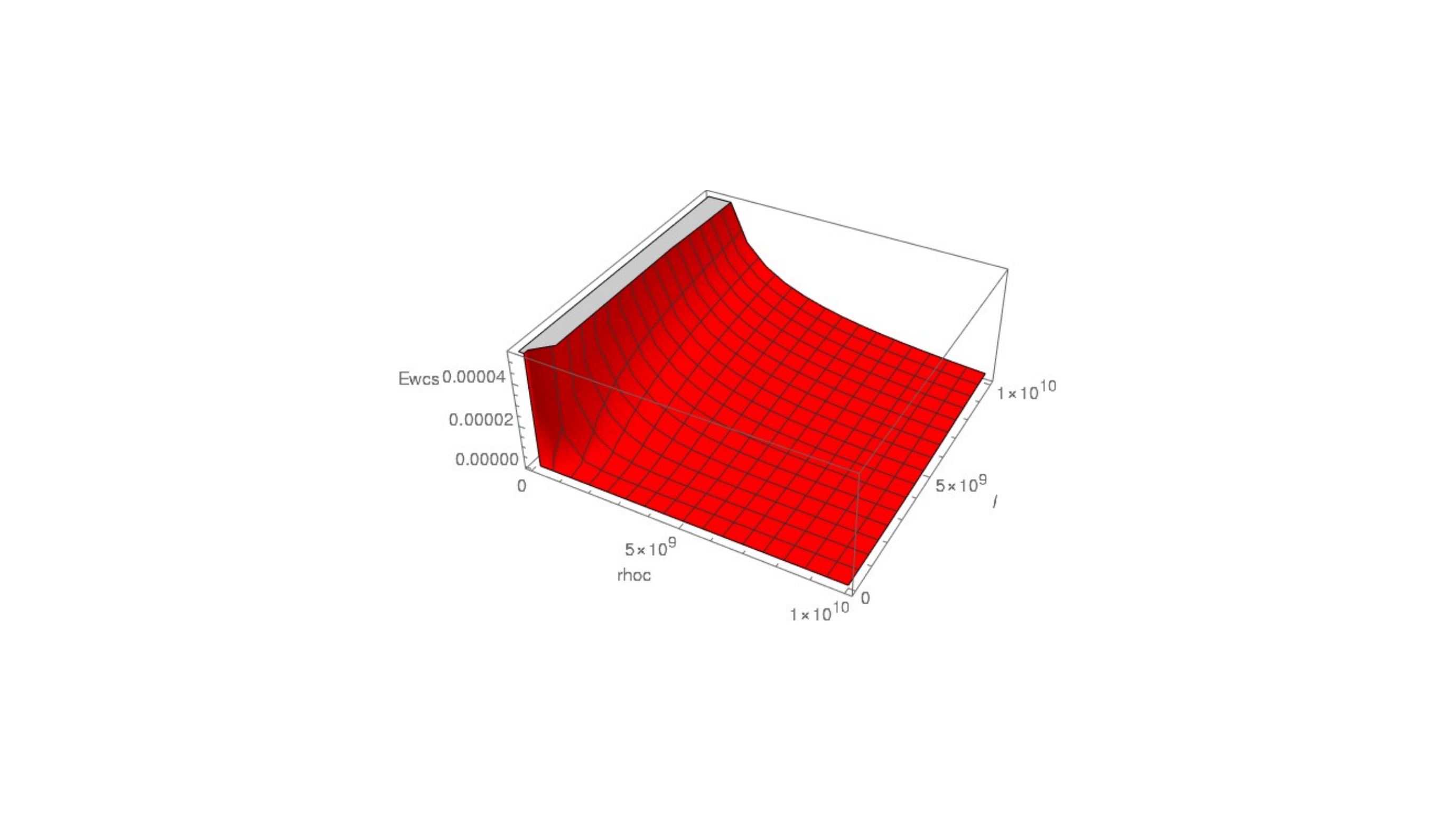}
\includegraphics[width=.45\textwidth]{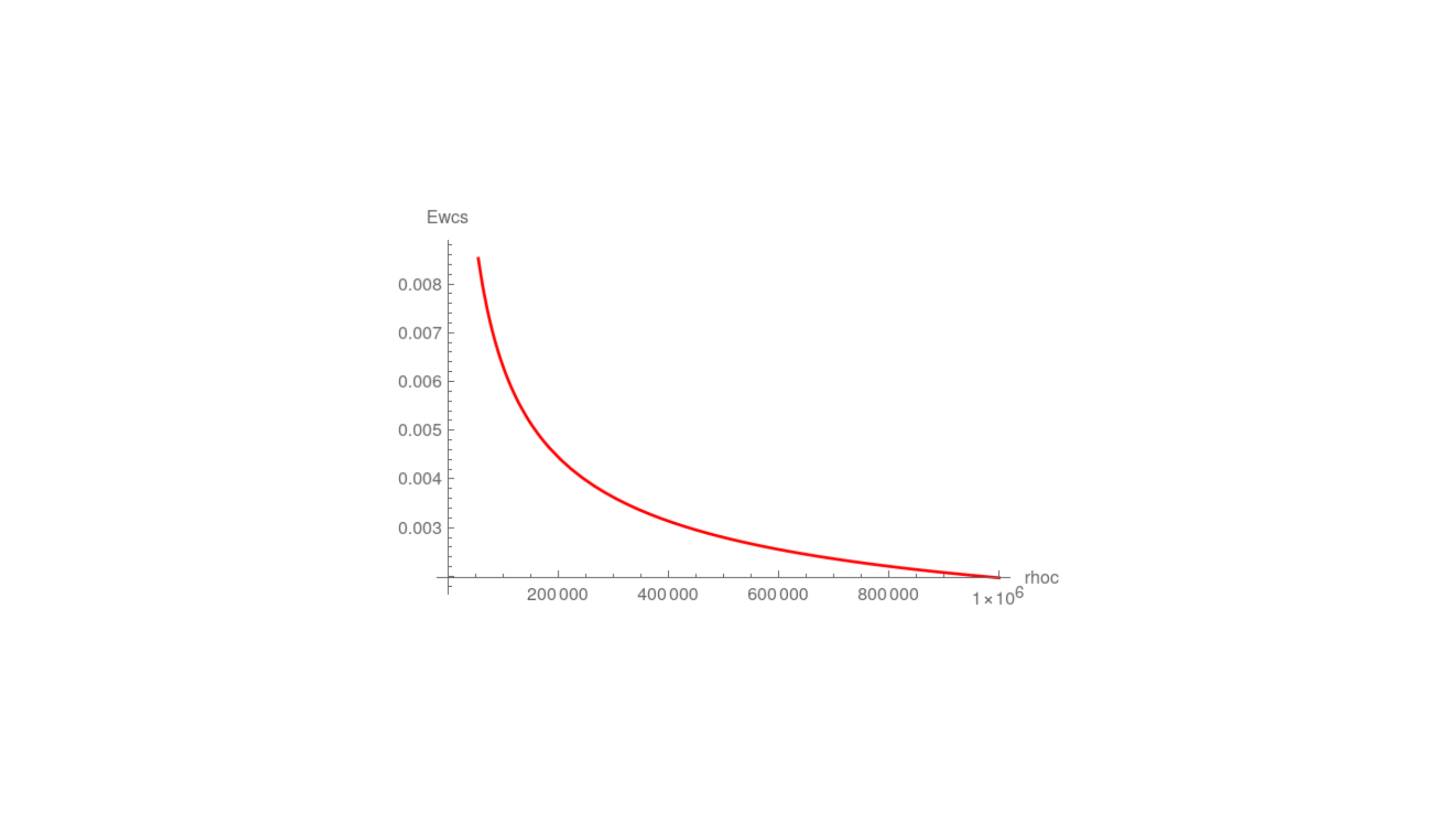}
\includegraphics[width=.45\textwidth]{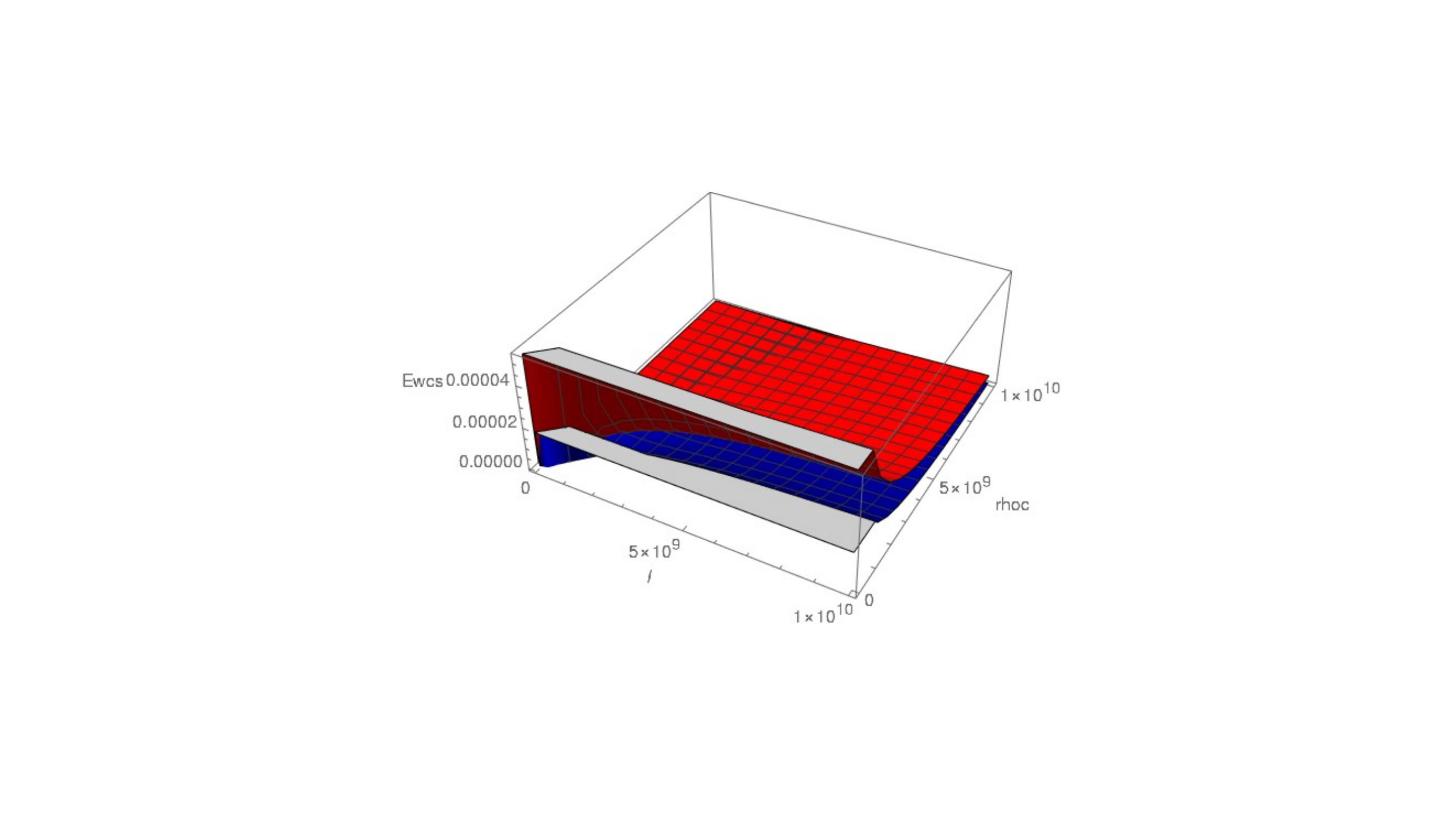}
\includegraphics[width=.45\textwidth]{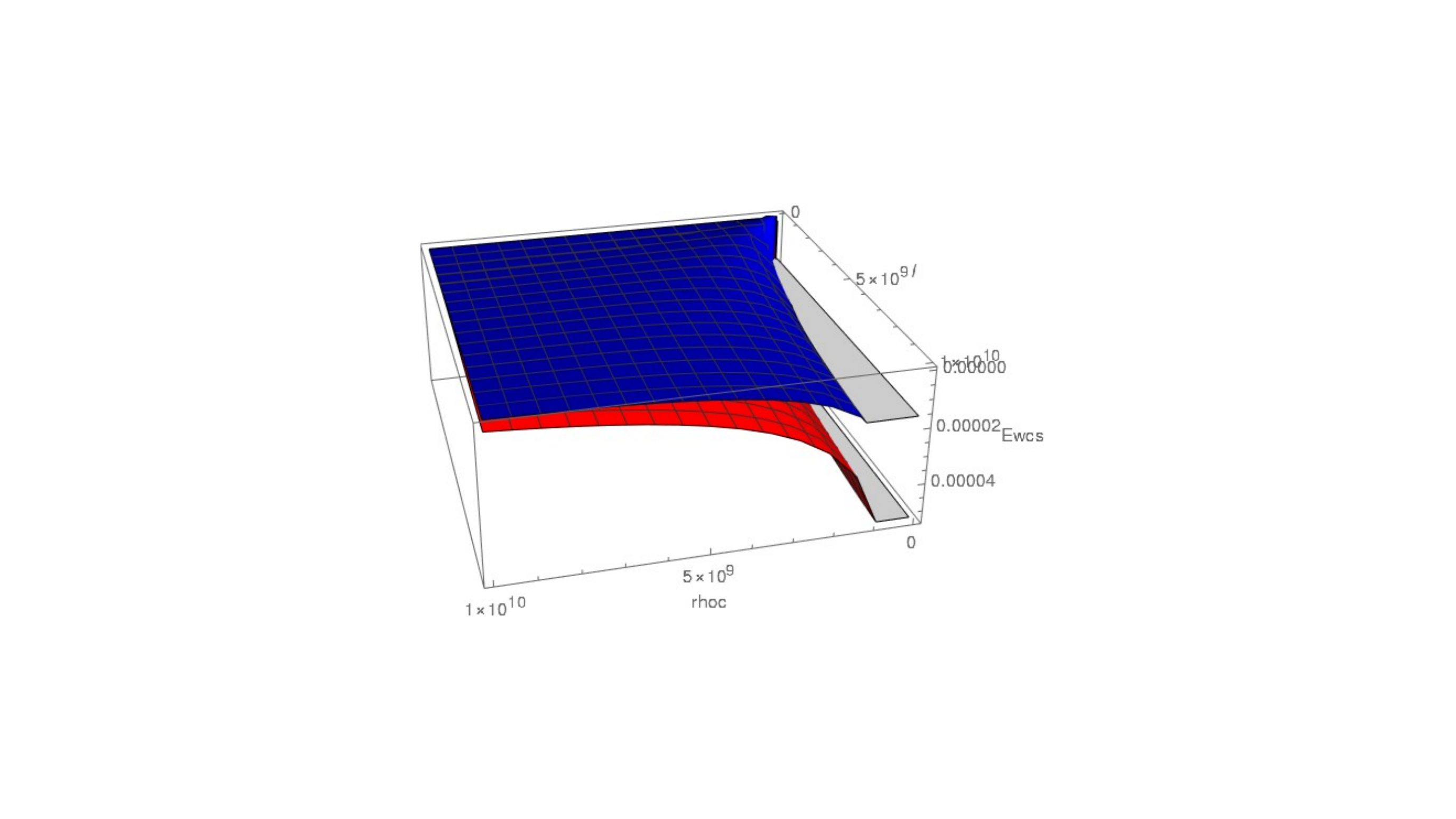}
\caption{(First row) \,\,:\,\,(   left ) \,,\, EWCS    as a function of $(l,\rho_c)$  in its connected phase(extrapolated to $l = 0$)  for $h = (10)^6$ \, (right),\,:\, , Ewcs  as a function of l for  with $\rho_c = {10}^9$, $h = (10)^6$ \,:\,  Both the plots  showing, EWCS increases with the increase of l.  .
  \,;\, (Second row) \,:\,  ( left)\, : \, EWCS  as a function of $(\rho_c,l)$ for $h = (10)^6$ \, ,\,    \,\,,\,\,  \,\, , (right) \,:\, EWCS plotted as a function of $\rho_c$  for $l = {10}^{10}$, $h = (10)^6$ \, :\,  We see, EWCS falls with the increase of cut off $\rho_c$ and goes to zero for $\rho_c >> l $ regime and finite at zero cut off for non zero h \,,\,
(Third row) \,:\,  ( left)\, : \, EWCS  as a function of $(\rho_c,l)$  for $h = 0$, \,, (right) \,:\, EWCS  as a function of $\rho_c$  for $l = {10}^{10}$, for $h=0$,  \, :\, Here unlike the case of nonzero cut off we see that Ewcs diverges for $h = 0$ for zero cut off.
\,,],(Last row)\,\,:\,\,   The overlap of EWCS-l-$\rho_C$ plot and 
${\frac{1}{2}} \left( H.M.I\right)$-l-$\rho_c$ plot considered to see whether the inequality  $EWCS \ge {\frac{1}{2}} \left( H.M.I\right)$ holds, with EWCS in red and ${\frac{1}{2}} \left( H.M.I\right)$ in blue \,: \, (left)\,:\, The frontview of  overlap plot, showing for $l >>\rho_c$ inequality satisfies \, (right) \, The backsideview of  overlap plot, showing for $\rho_c >> l$, inequality saturates 
}
\label{ewcsbasic3by2}
\end{figure}

\begin{figure}[H]
\begin{center}
\textbf{ For   $ d - \theta  = {\frac{7}{3}}$\,: \, Ewcs vs $(l,\rho_c)$, Ewcs vs l, Ewcs vs $\rho_c$ graph for fixed h for the connected phase, extrapolated to $l = 0$ }
\end{center}
\vskip2mm
\includegraphics[width=.45\textwidth]{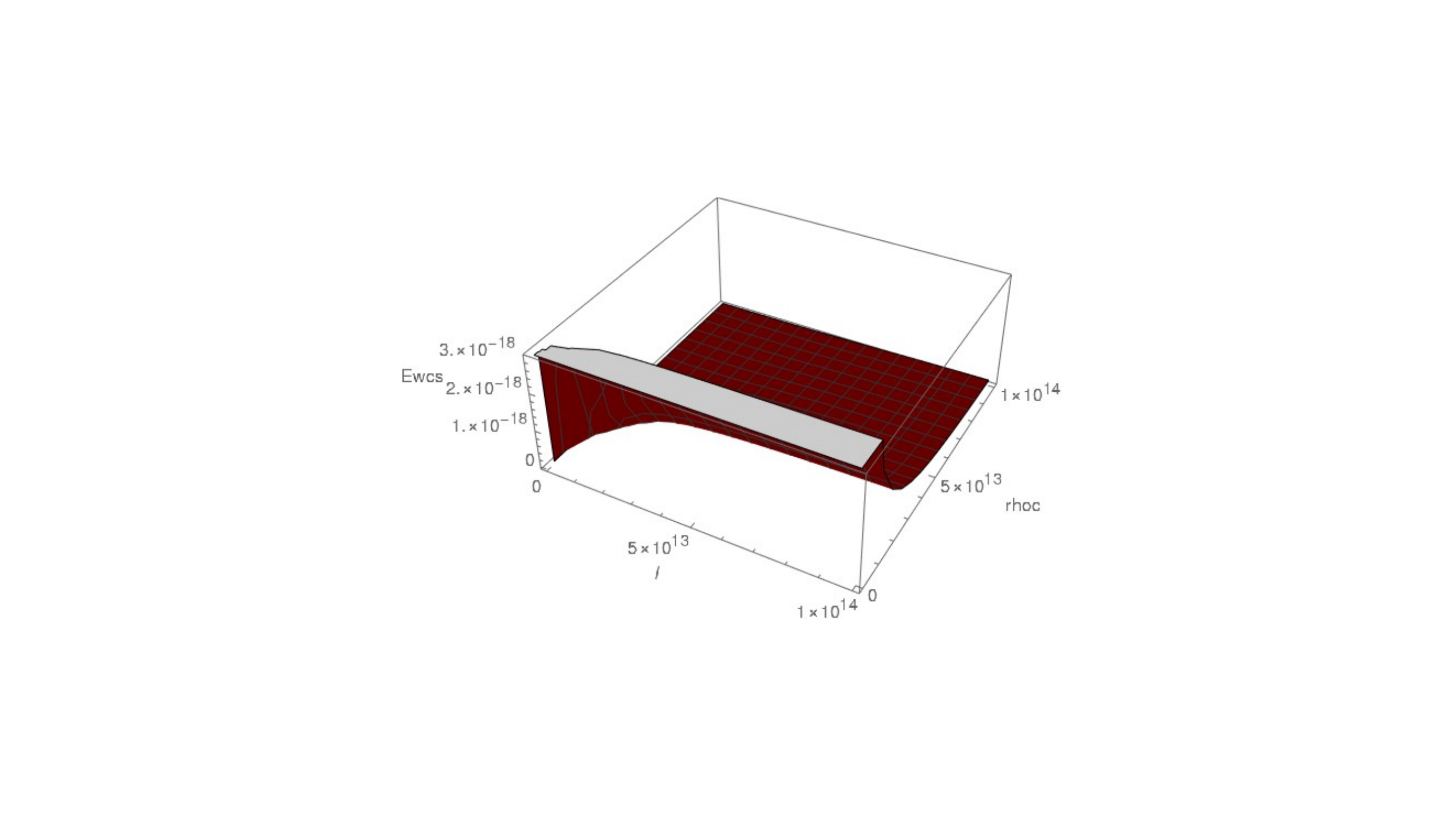}
\includegraphics[width=.45\textwidth]{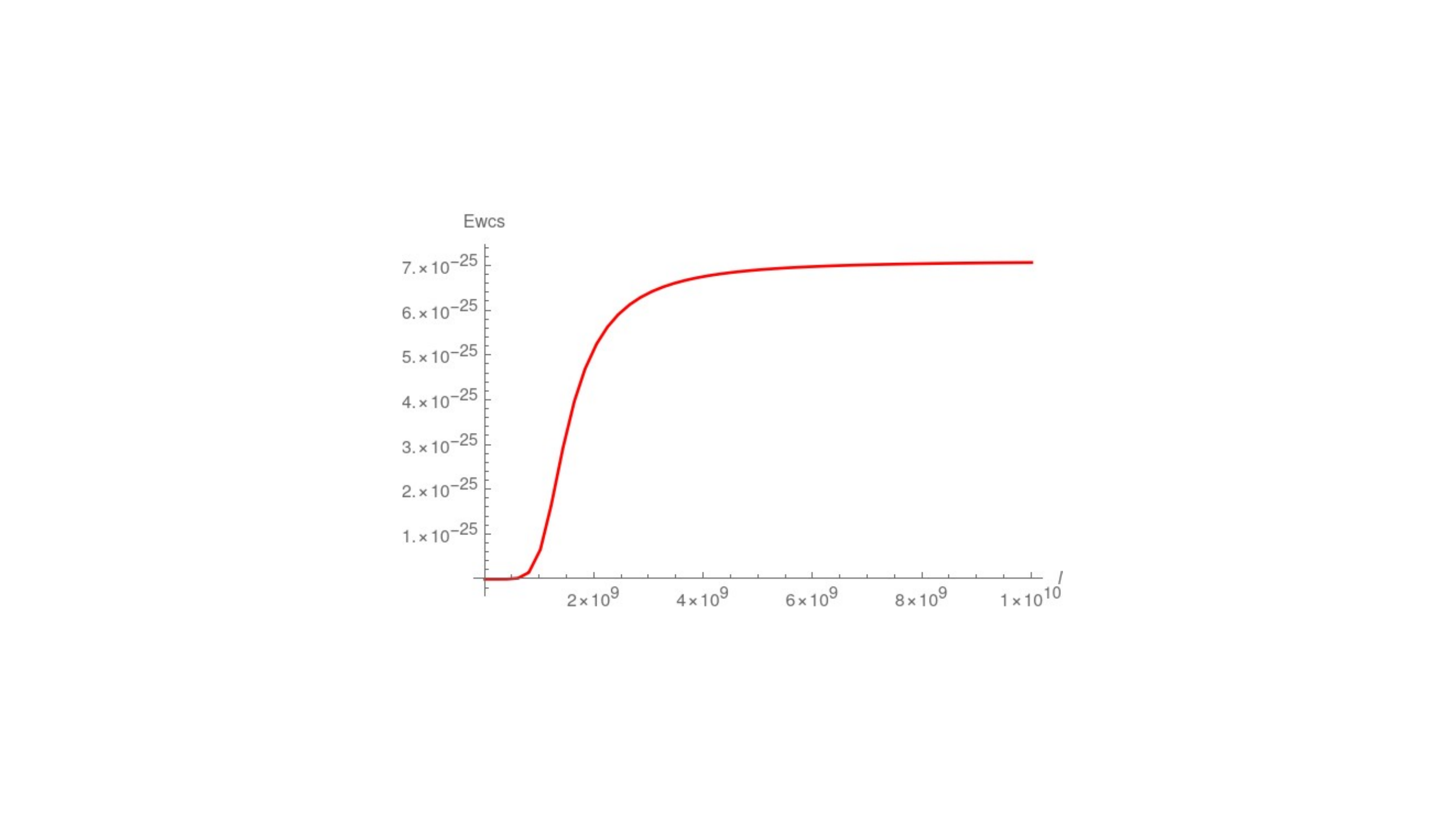}	
\includegraphics[width=.45\textwidth]{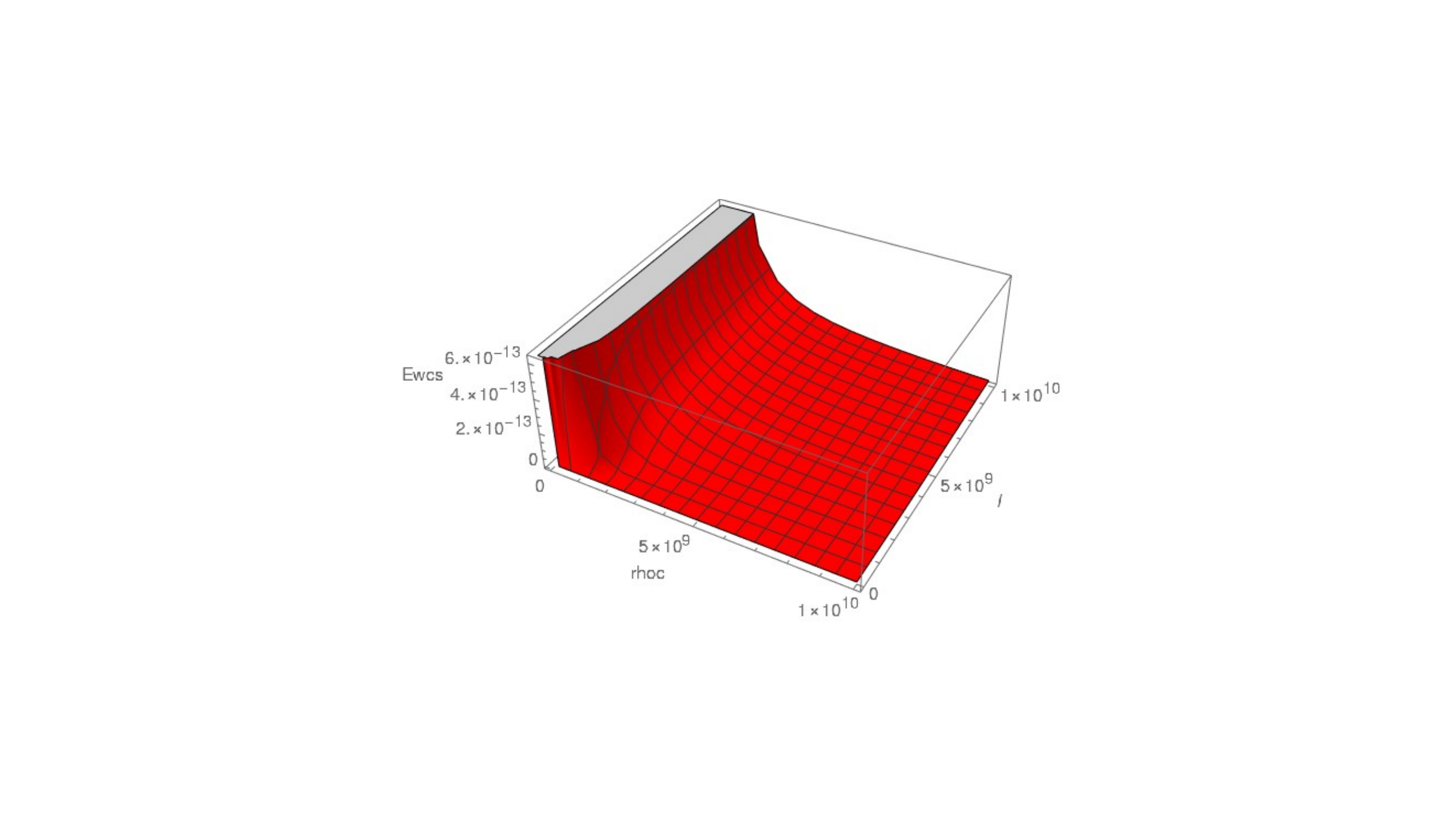}
\includegraphics[width=.45\textwidth]{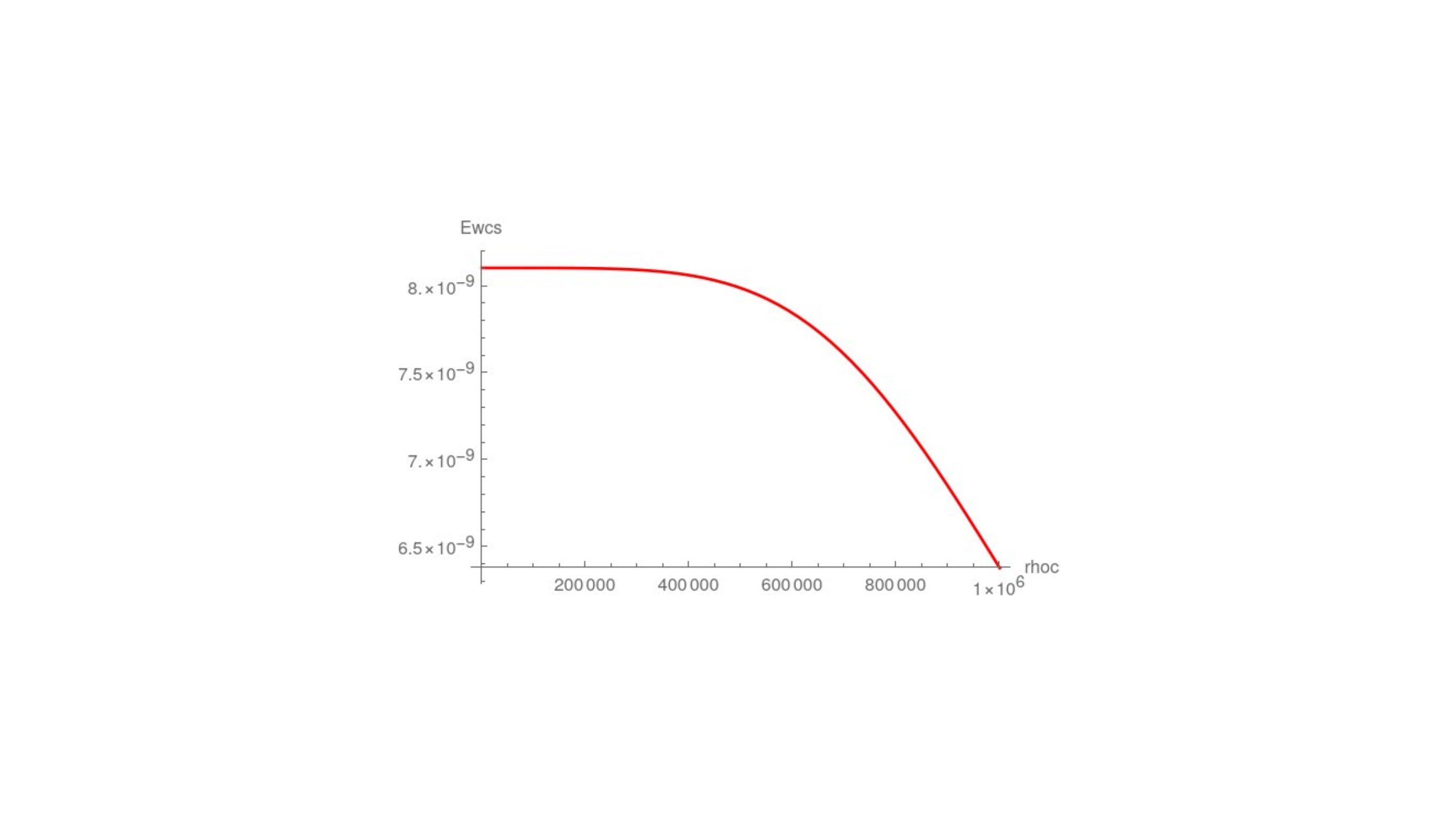}
\includegraphics[width=.45\textwidth]{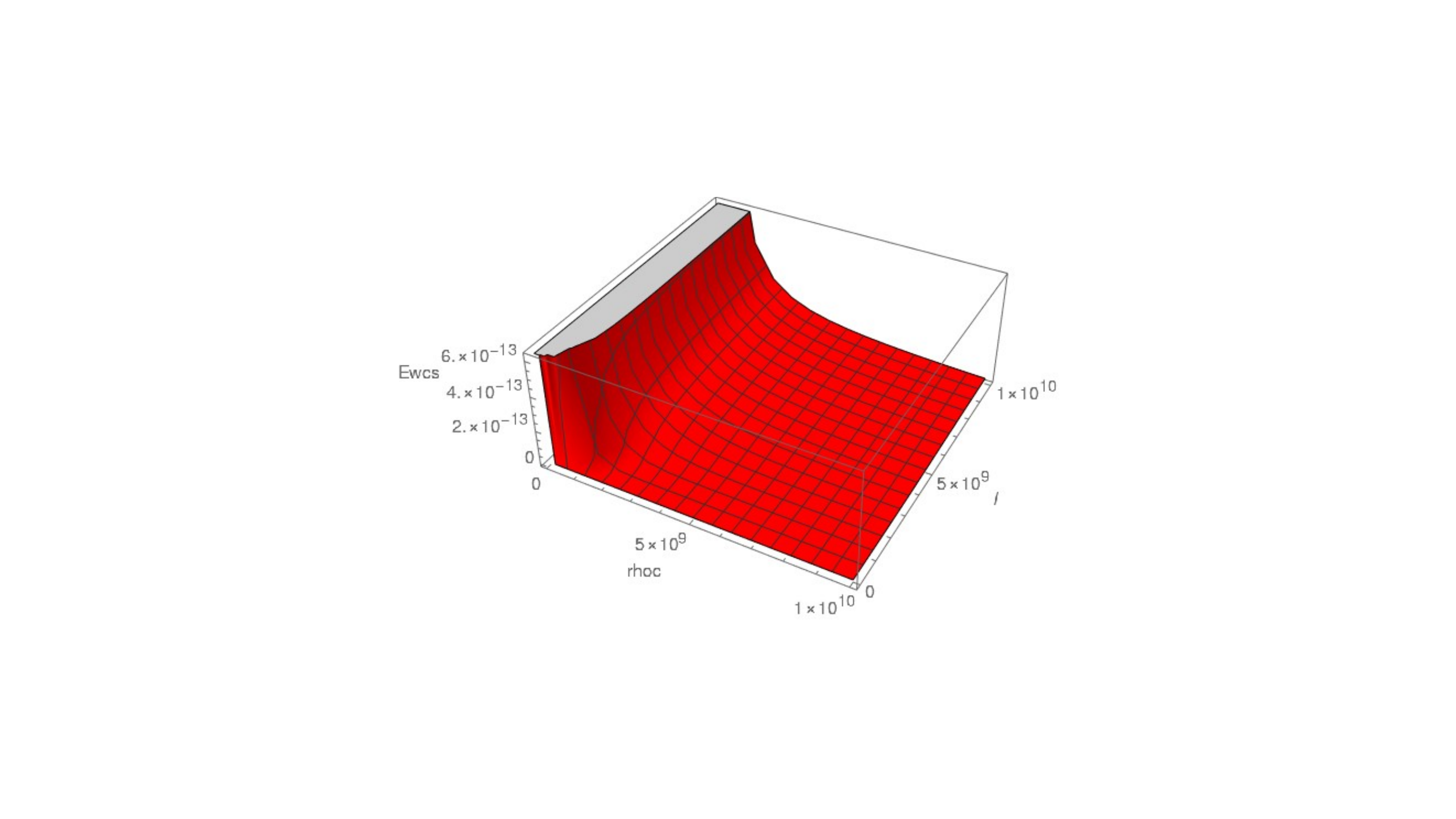}
\includegraphics[width=.45\textwidth]{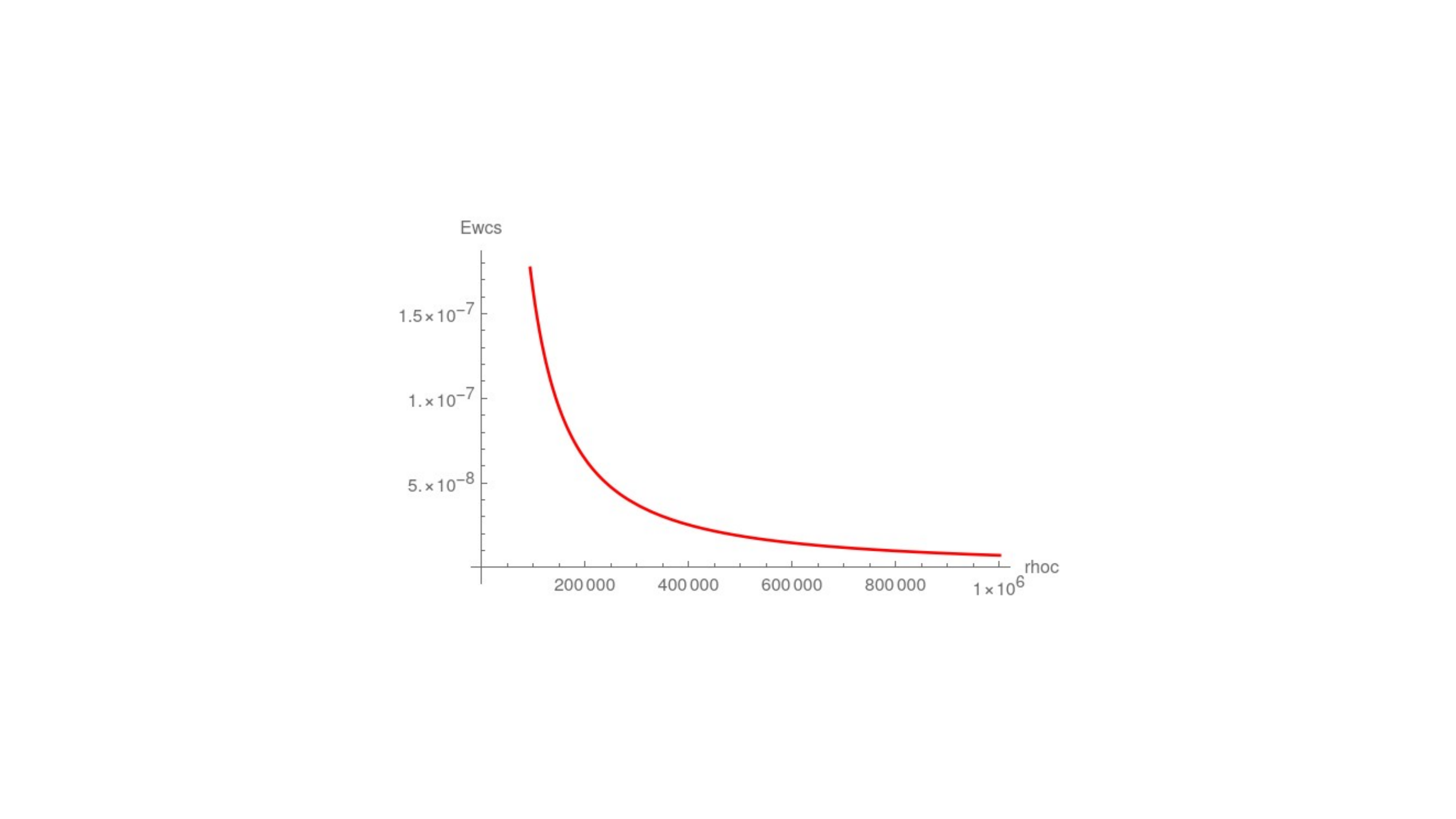}
\includegraphics[width=.45\textwidth]{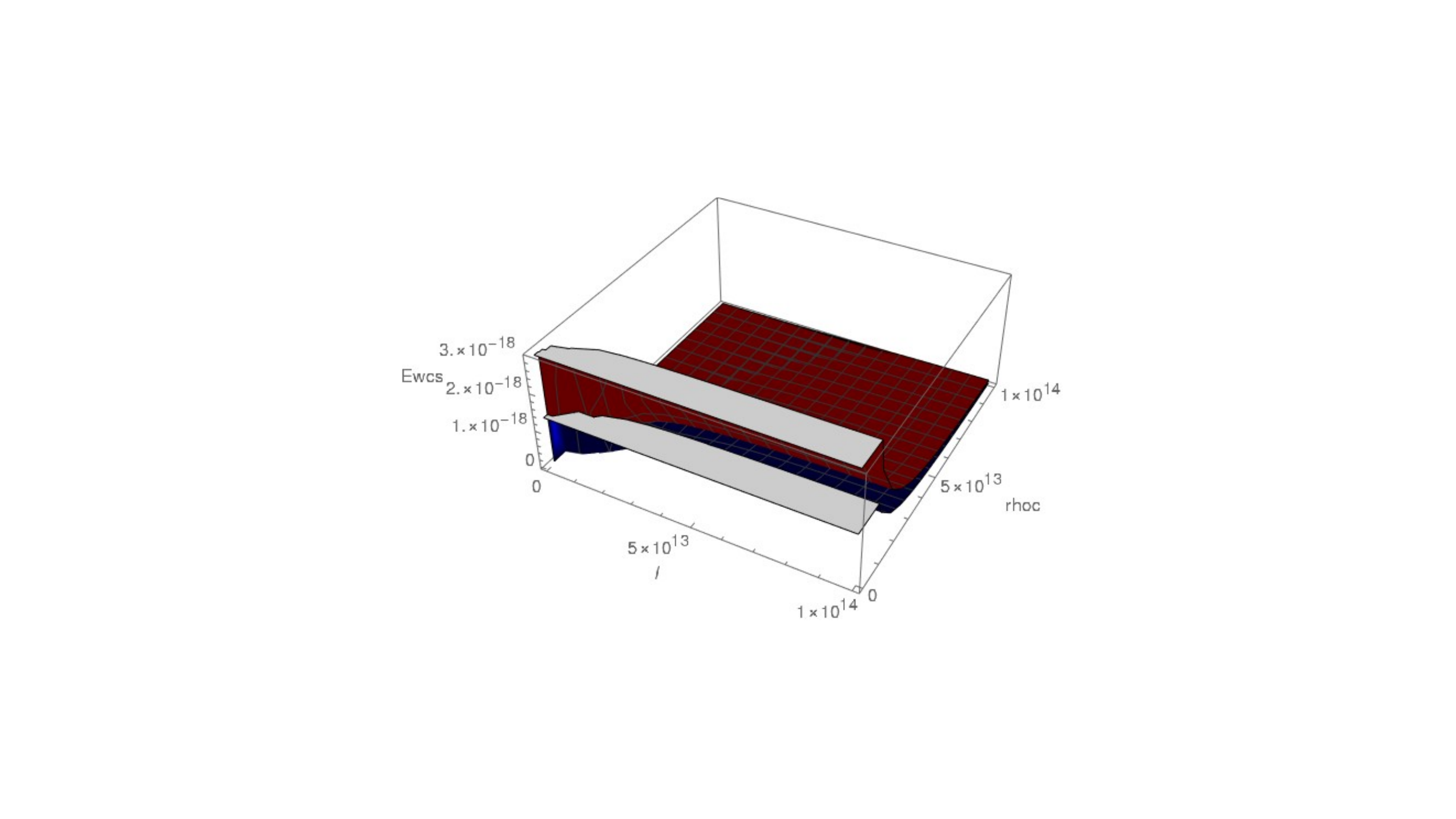}
\includegraphics[width=.45\textwidth]{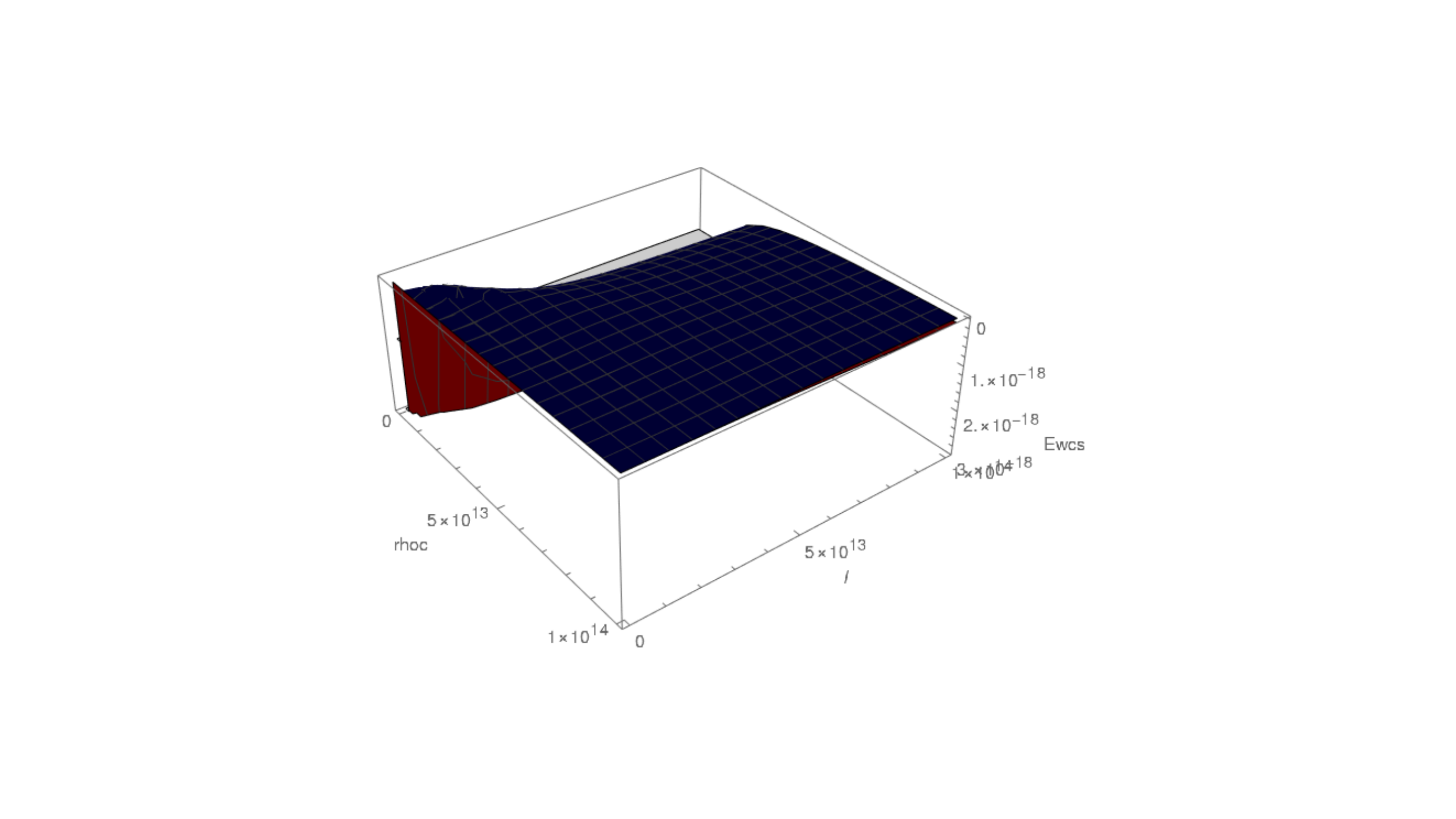}
\caption{(First row) \,\,:\,\,(   left ) \,,\, EWCS vs  $(l,\rho_c)$ plot in its connected phase(extrapolated to $l = 0$) \, (right),\,:\, , Ewcs plotted as a function of l with $\rho_c = (3.16)\times {10}^9 $ \,:\,  Both the plots  showing that  EWCS increases with the increase of l 
 \,;\, (Second row) \,:\,  ( left)\, : \, EWCS plotted as a function of $(\rho_c,l)$ for $h = (10)^6$, \, (right) \,:\, EWCS plotted as a function of $\rho_c$  for $l = {10}^{10}$, $h = (10)^6$ \, :\,  We see, for a given l, EWCS falls with the increase of cut off $\rho_c$ and goes to zero for $\rho_c >> l $ regime  and  finite at zero cut off for non zero h \,,\, (Third row) \,:\,  ( left)\, : \, EWCS  as a function of $(\rho_c,l)$  for $h = 0$     \,,\,   , (right) \,:\, EWCS  as a function of $\rho_c$  for $l = {10}^{10}$, for $h=0$,  \, :\, Here unlike the case of nonzero cut off we see that Ewcs diverges for $h = 0$ for zero cut off. \,,\,(Last row)\,\,:\,\,   The overlap of EWCS-l-$\rho_C$ plot and 
${\frac{1}{2}} \left( H.M.I\right)$-l-$\rho_c$ plot considered to see whether the inequality  $EWCS \ge {\frac{1}{2}} \left( H.M.I\right)$ holds, with EWCS in red and ${\frac{1}{2}} \left( H.M.I\right)$ in blue \,: \, (left)\,:\, The frontview of  overlap plot, showing for $l >>\rho_c$ inequality satisfies \, (right) \, The backsideview of  overlap plot, showing for $\rho_c >> l$, inequality saturates 
}
\label{ewcsbasic7by3}
\end{figure}

\begin{figure}[H]
\begin{center}
\textbf{ For $d - \theta > 1$, EWCS vs $(h, l)$ plot  for its connected phase,  extrapolated to $l = 0$ }
\end{center}
\vskip2mm
\includegraphics[width=.65\textwidth]{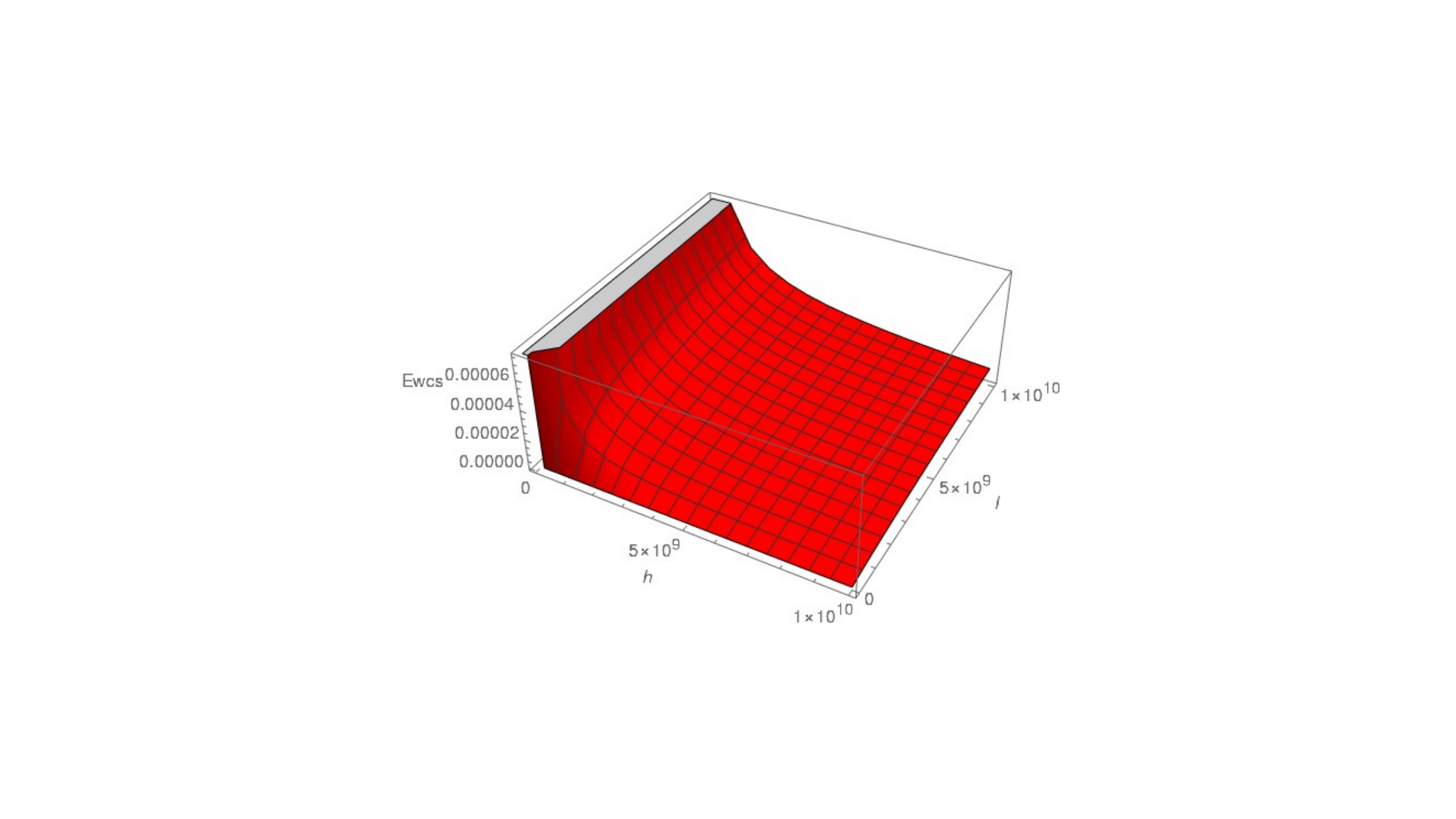}
\includegraphics[width=.65\textwidth]{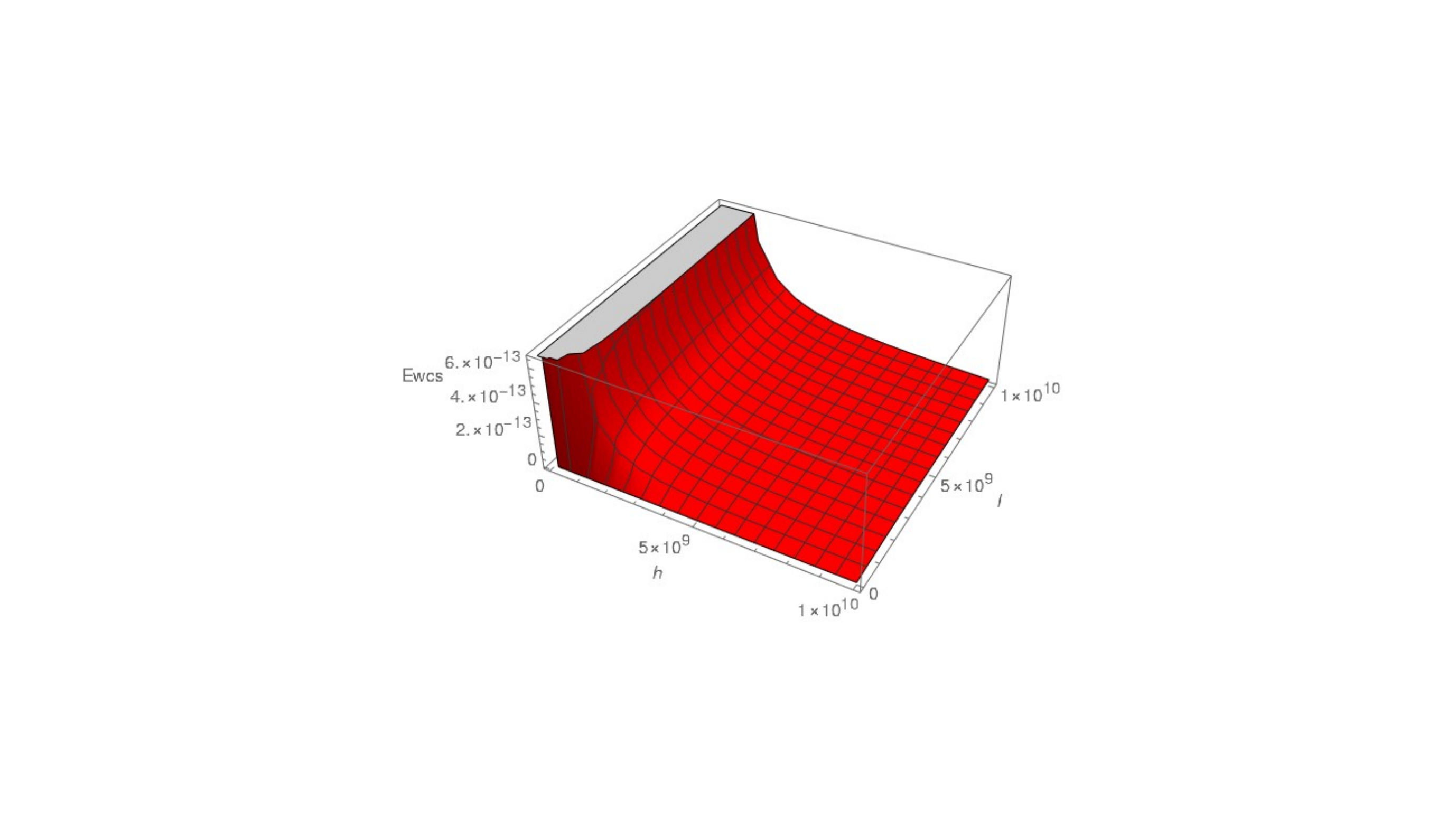}
\caption{ (left) \,:\,EWCS  vs $(h,l )$  plot for $\rho_c = 100$\, (left) \, $d - \theta = {\frac{3}{2}}$, \, , \, (right) $d - \theta = {\frac{7}{3}}$ \,: \, showing EWCS falls with increase of  h, for a given l,with $\rho_c$ fixed.  It is also showing for a given h,EWCS  increses with l  }
\la{ewcsbgreaterthan1hl}
\end{figure}

\begin{figure}[H]
\begin{center}
\textbf{ For $d - \theta > 1$, EWCS as a function of $(\rho_c , h)$  for its connected phase,  extrapolated to $l = 0$}
\end{center}
\vskip2mm
\includegraphics[width=.65\textwidth]{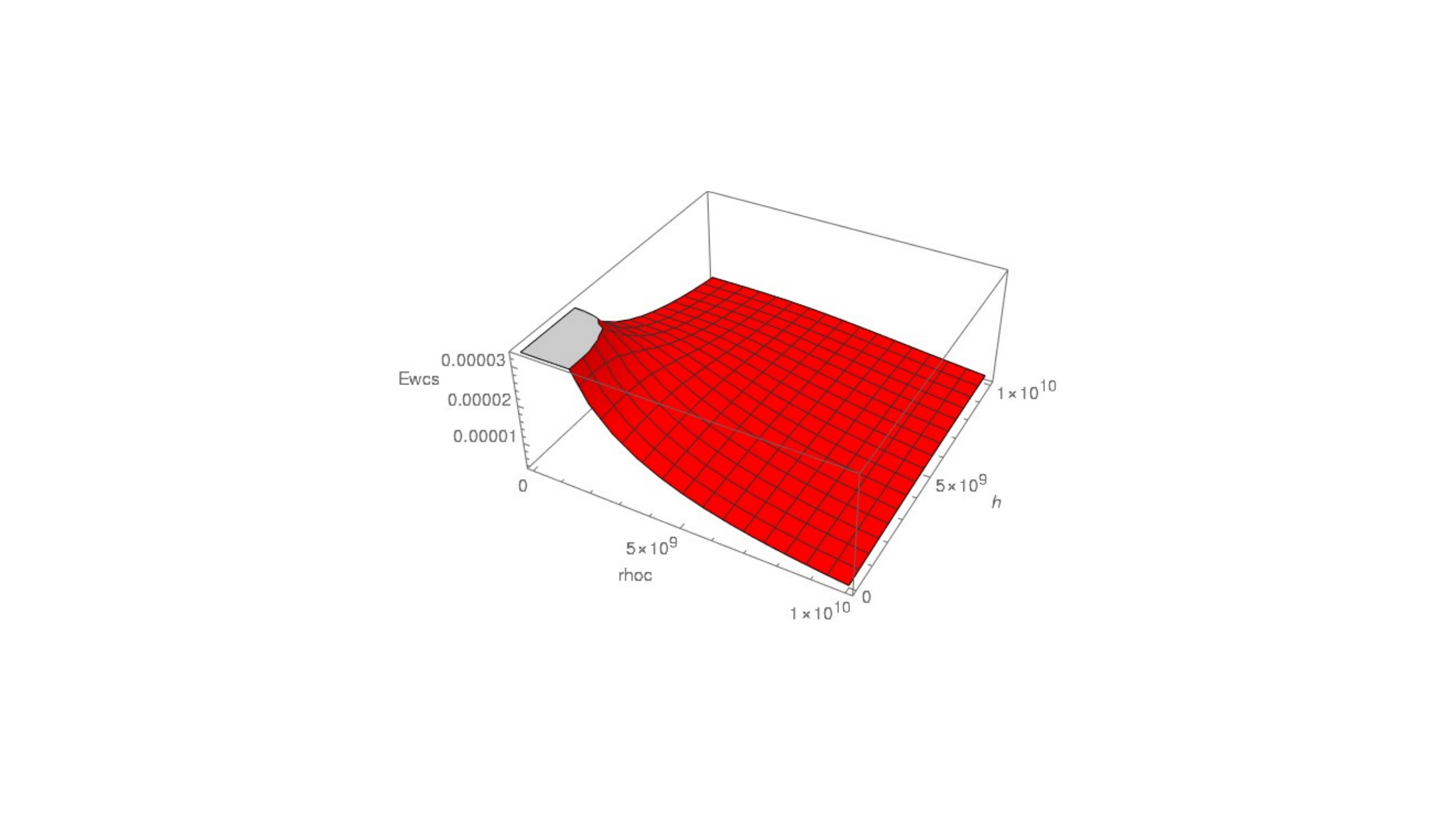}
\includegraphics[width=.65\textwidth]{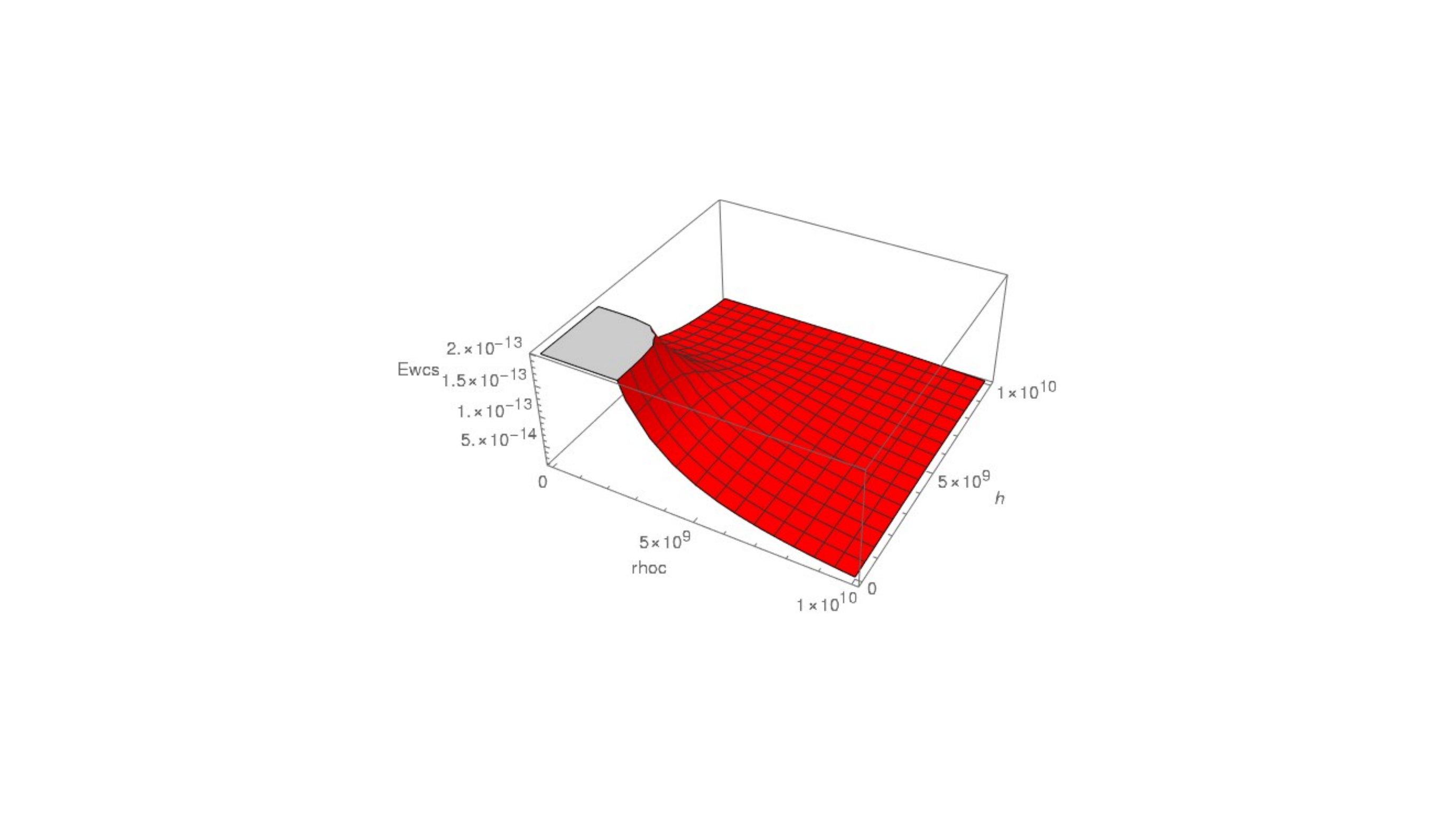}
\caption{ (left) \,:\,EWCS  vs $(\rho_c,h)$  plot for $l = {10}^{10}$\, (left) \, $d - \theta = {\frac{3}{2}}$, \, , \, (right) $d - \theta = {\frac{7}{3}}$ \,: \, showing EWCS falls with increase of both  $(\rho_c, h)$,   diverge at $h=0$ for zero cut-off but become finite at $h = 0$ for nonzero cut off }
\la{ewcsbgreaterthan1rhoch}
\end{figure}

\begin{figure}[H]
\begin{center}
\textbf{ For $ d - \theta > 1$ \,: \,  The evolution of EWCS with $ d - \theta$  for its connected phase,  extrapolated to $l = 0$ }
\end{center}
\vskip2mm
\includegraphics[width=.65\textwidth]{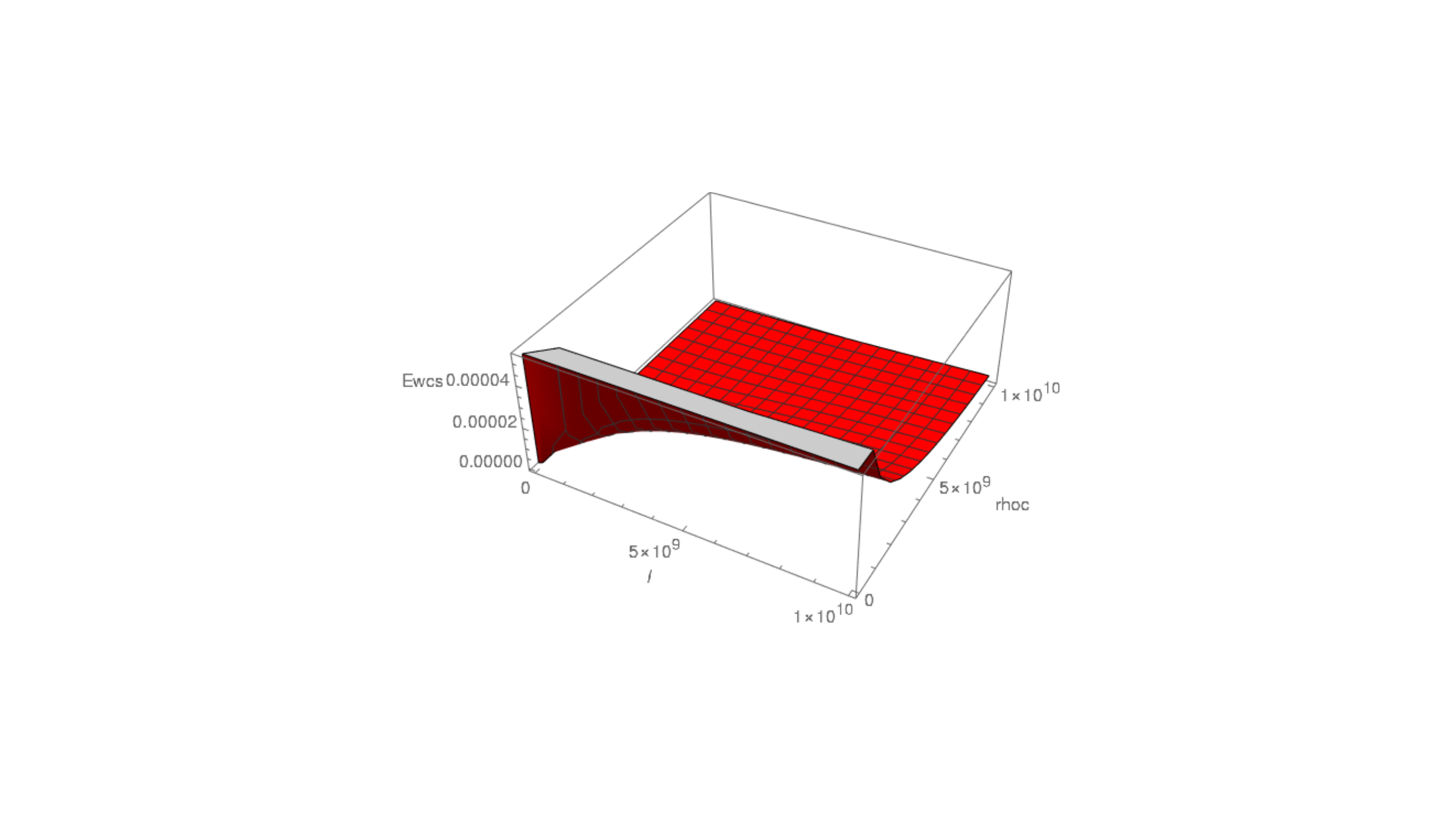}
\includegraphics[width=.65\textwidth]{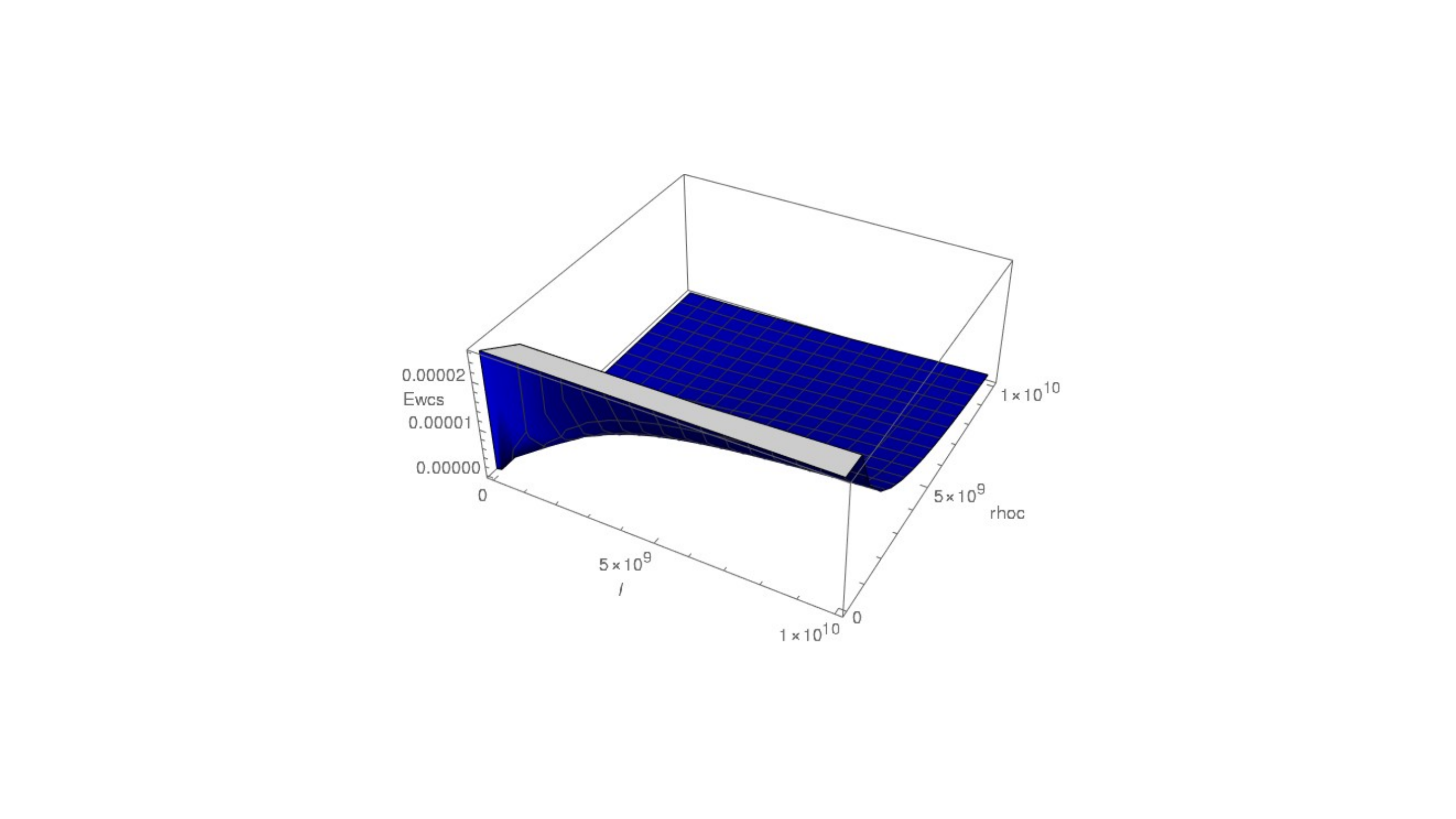}	
\includegraphics[width=.65\textwidth]{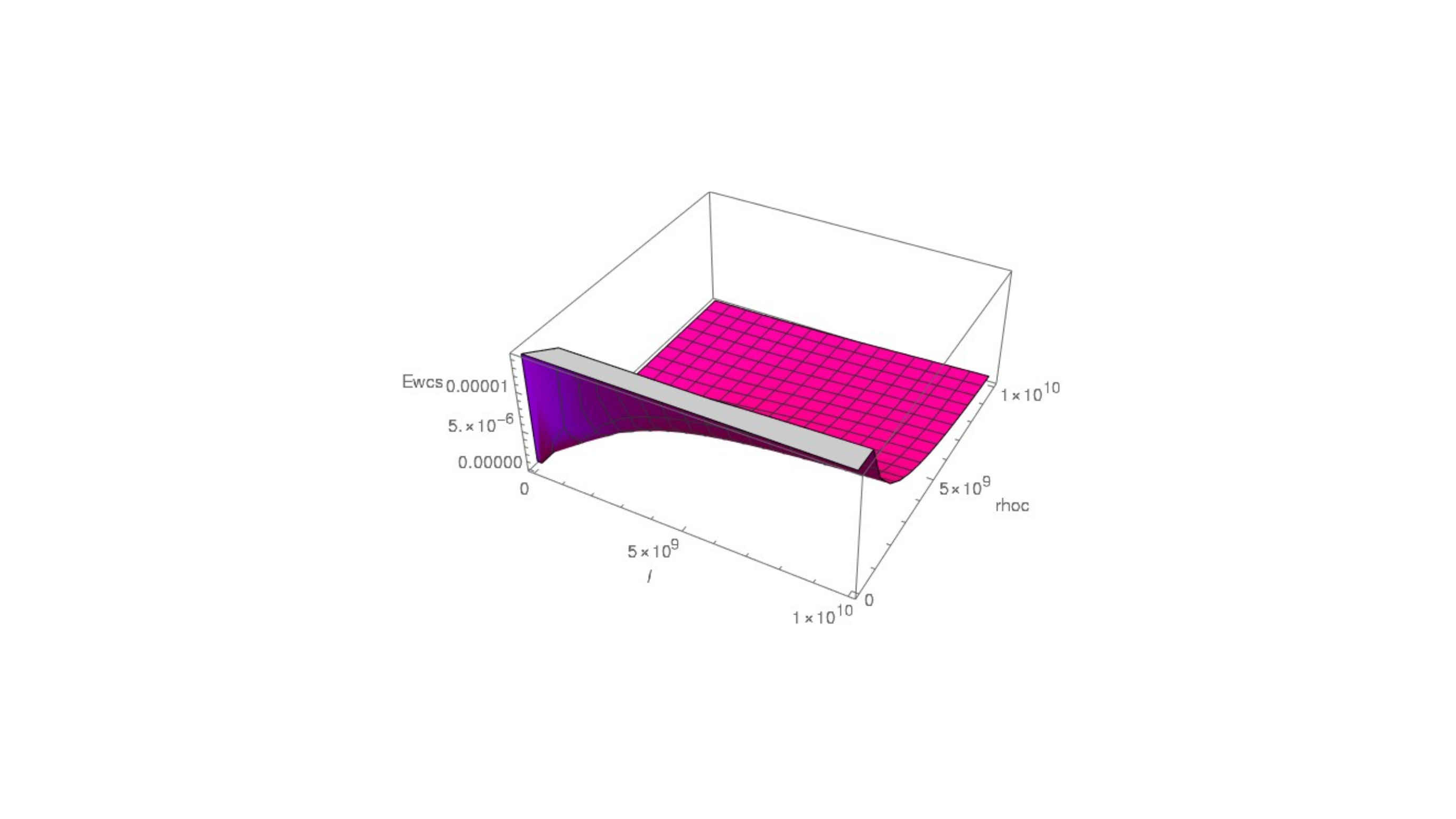}
\includegraphics[width=.65\textwidth]{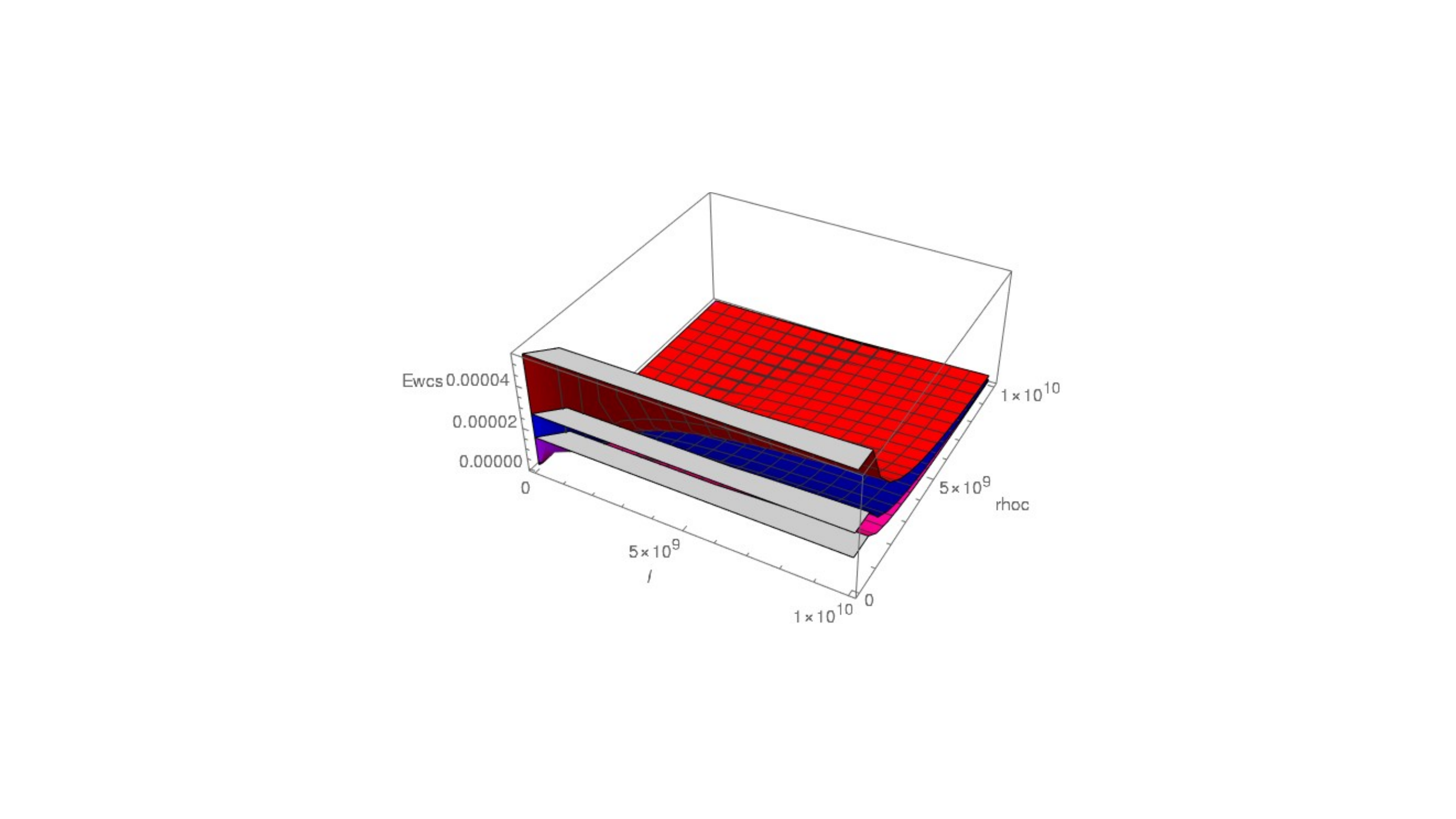}
\includegraphics[width=.65\textwidth]{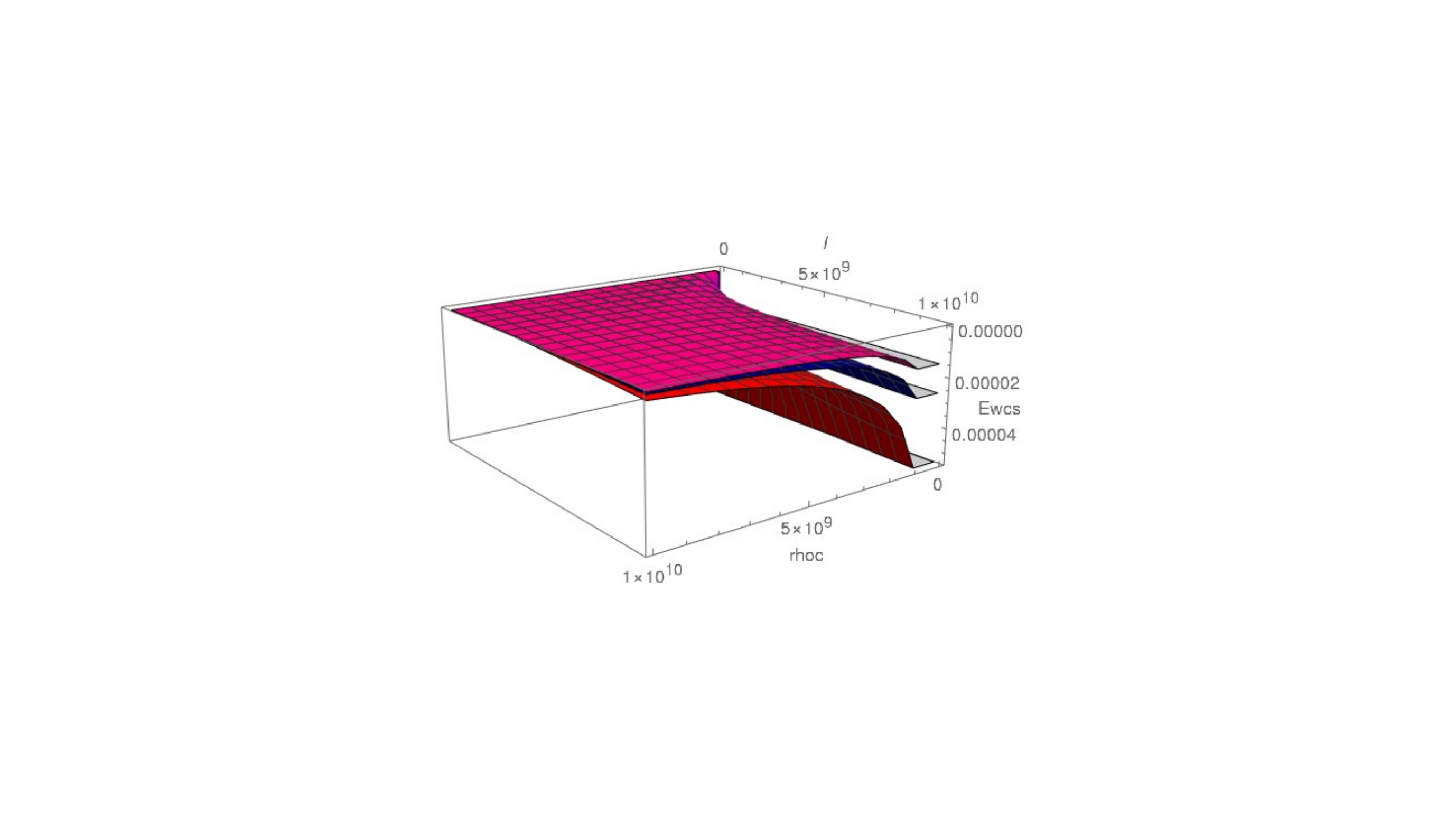}	
\includegraphics[width=.65\textwidth]{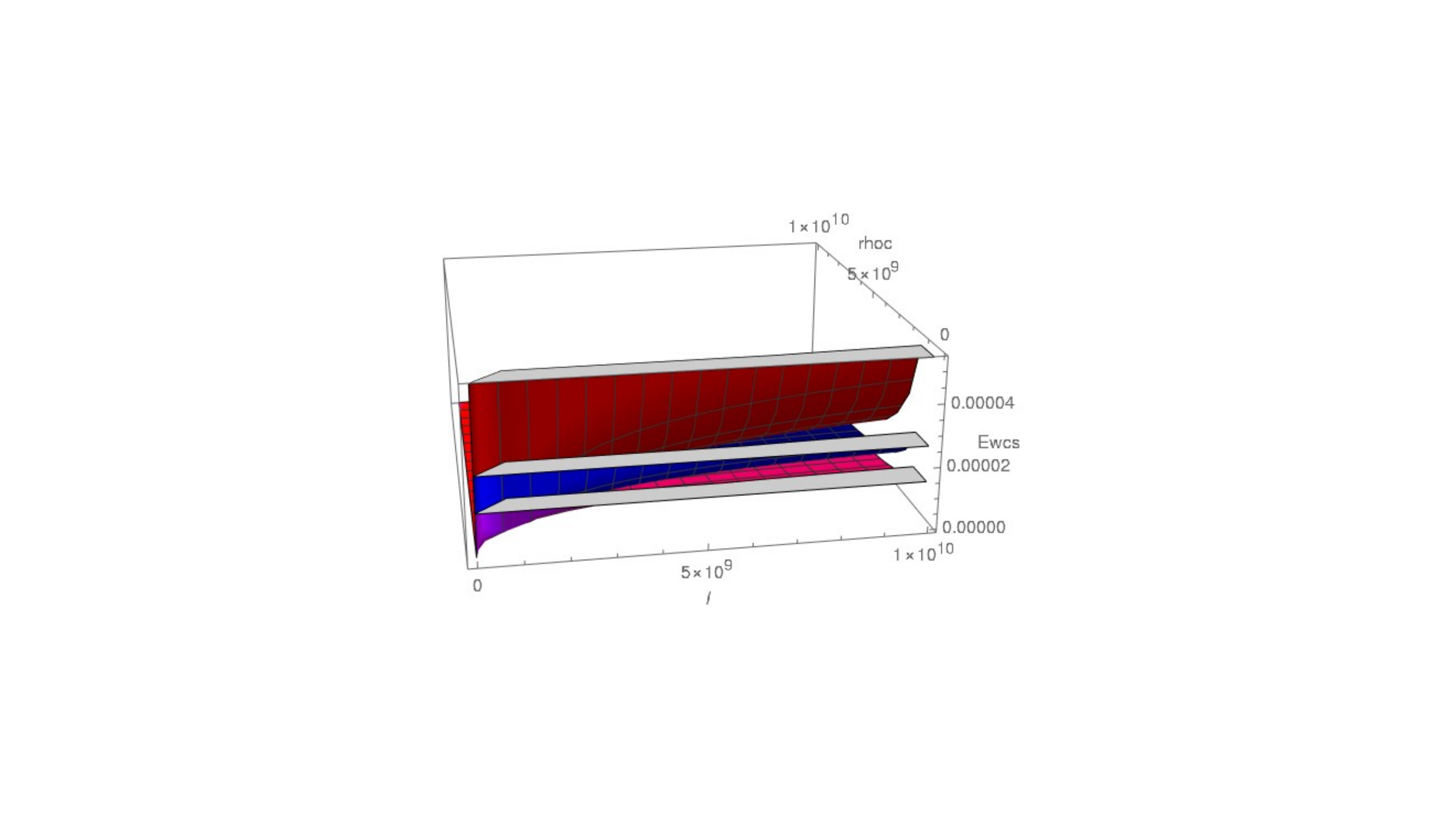}	
\caption{(First row) \,\,:\,\,( From  left to right) \,,\, EWCS plotted as a function of $(l,\rho_c)$ for $d - \theta = 1.500\,,\, d - \theta = 1.533 $,
  \quad;\quad (Second row) \,:\,  ( left)\, : \, EWCS plotted as a function of $(l,\rho_c)$ \, ,\,  $d - \theta= 1.566$ \,\,,\,\, (right) \,:\, The frontview of the  overlap of the three, showing that for $l >> \rho_c$ EWCS  decreases with the increase of $d - \theta$
 \, ,\,   \quad;\quad (Last row ) ,\, ( left )\, : \,  \,:\, The backside view of the overlap of the three, showing for $\rho_c >> l$ the EWCS for different $d - \theta$   actually merges, supporting the fact that for a fixed l, EWCS decreases with the increase of $\rho_c$ and ultimately falls towards zero.\,\,;\,\. (right)\,:\,  The sideview of the overlap of the three showing that EWCS in its connected phase smoothly reach to zero  at $l = 0$(extrapolated to $l = 0$) for all $d - \theta$}
\label{ewcsevbgreater1}
\end{figure}

\subsection{EWCS for $d - \theta < 1$}
Here we substitute the expression of $\rho_0$ for $ d - \theta < 1$ from (\ref{ultimaterho0nonvanishing}) in (\ref{Ew}) to  obtain the expression  of EWCS and study its different properties. 
Here in this section we will show that, $E_W$, as we have constructed here, satisfies the basic properties.  We will present it through Fig.(\ref{ewcsbasic4by9}, \ref{ewcsbasic1by3},  \ref{ewcsblessthan1hl}, \ref{ewcslessthan1rhoch},  \ref{ewcsevblessthan1} ).  Since we failed to find $h_{\rm crit}$ because of the complication of the expression of HMI so here we will show the plots of EWCS for its connected phase only.  EWCS, unlike HMI, posses a finite value at the point of transition and then suddenly drops to zero so that it shows a discontious phase transition.  In the connected phase EWCS becomes zero exactly $l = 0$.  Here we extrapolate our plots for connected phase upto that point.

\begin{figure}[H]
\begin{center}
\textbf{For   $ d - \theta  = {\frac{4}{9}}$\,: \, Ewcs vs $(l,\rho_c)$, Ewcs vs l, Ewcs vs $\rho_c$ graph for fixed h  for the connected phase, extrapolated to $l = 0$   }
\end{center}
\vskip2mm
\includegraphics[width=.40\textwidth]{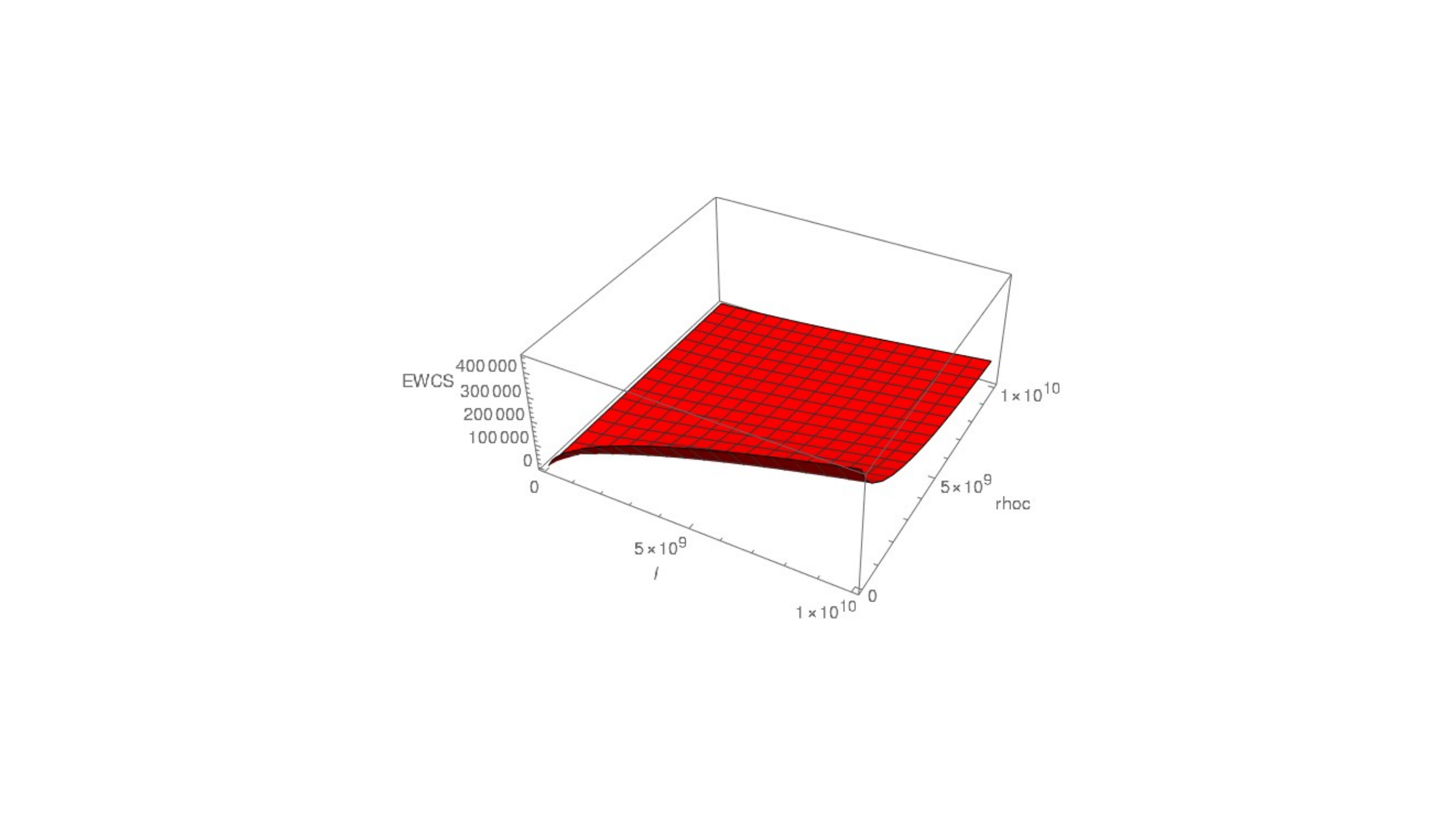}
\includegraphics[width=.40\textwidth]{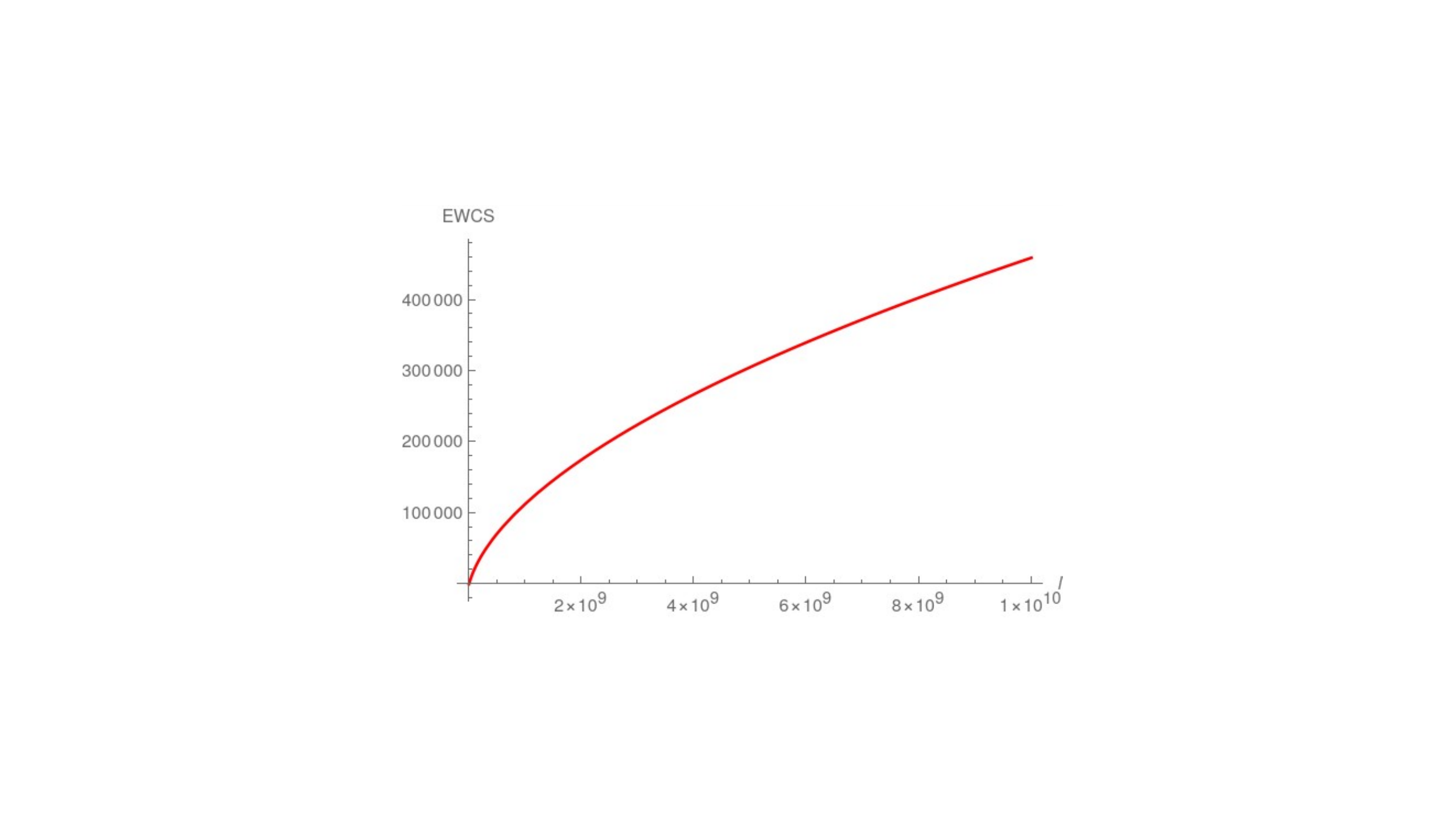}	
\includegraphics[width=.40\textwidth]{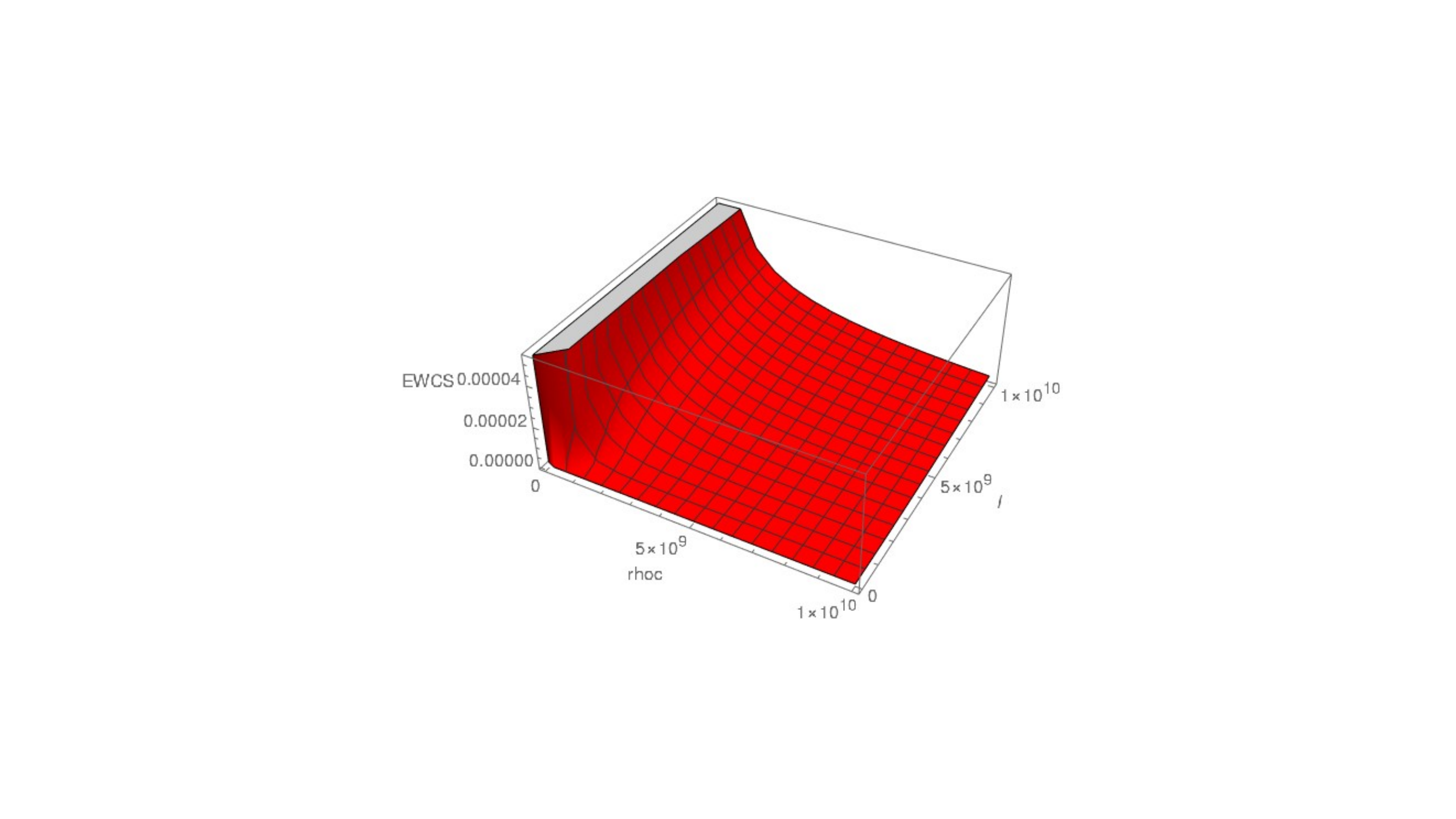}
\includegraphics[width=.40\textwidth]{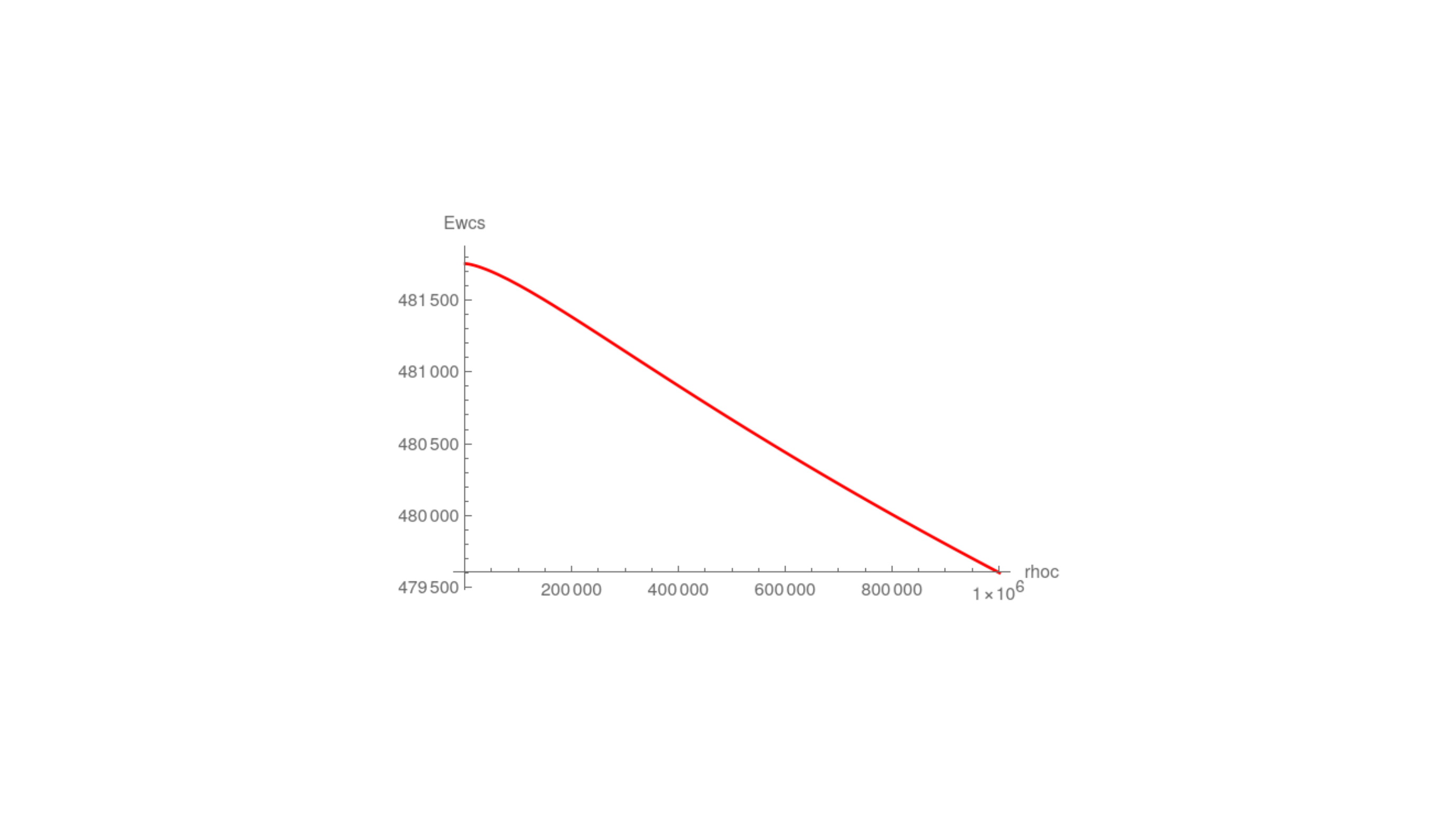}
\includegraphics[width=.40\textwidth]{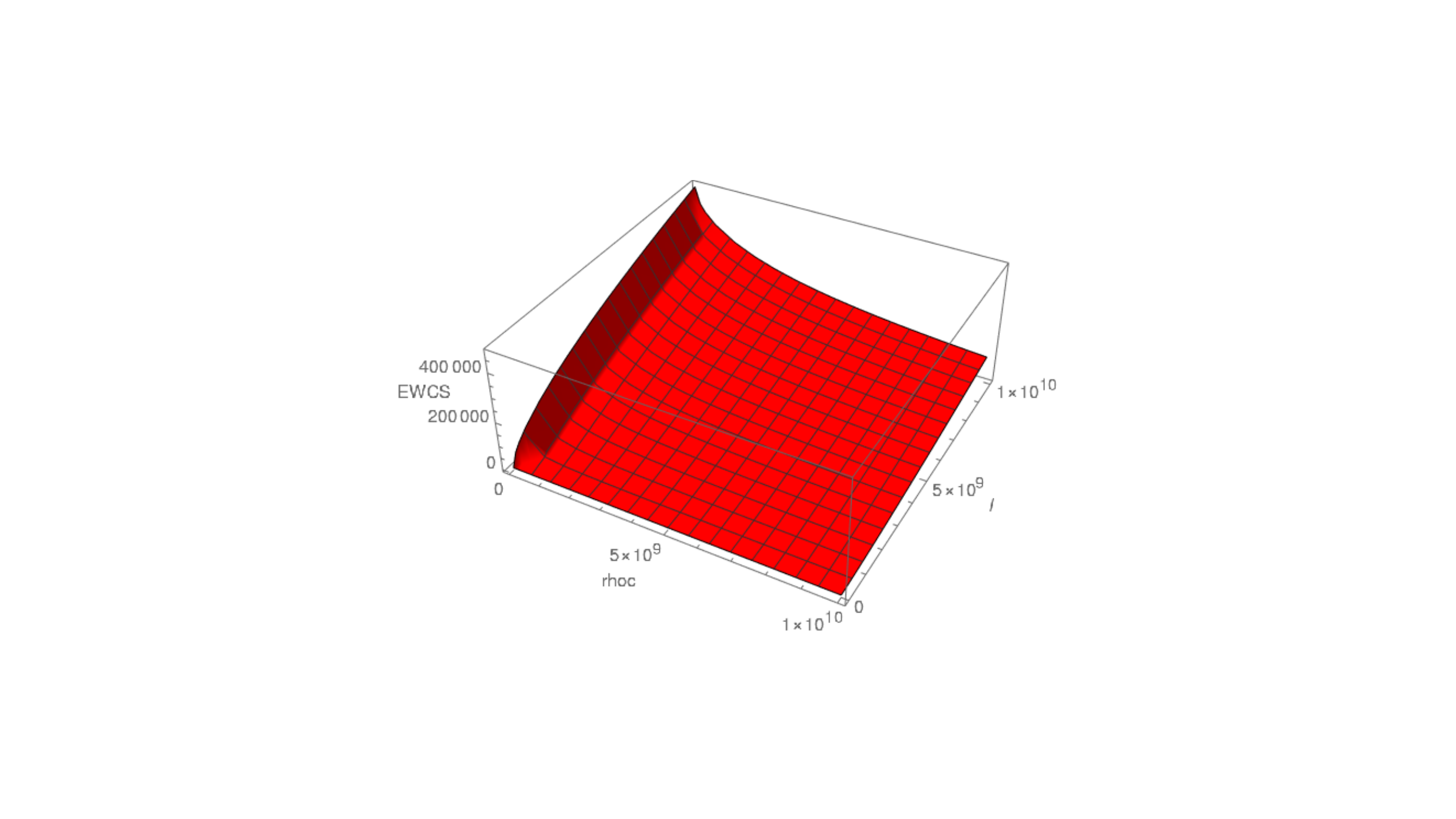}
\includegraphics[width=.40\textwidth]{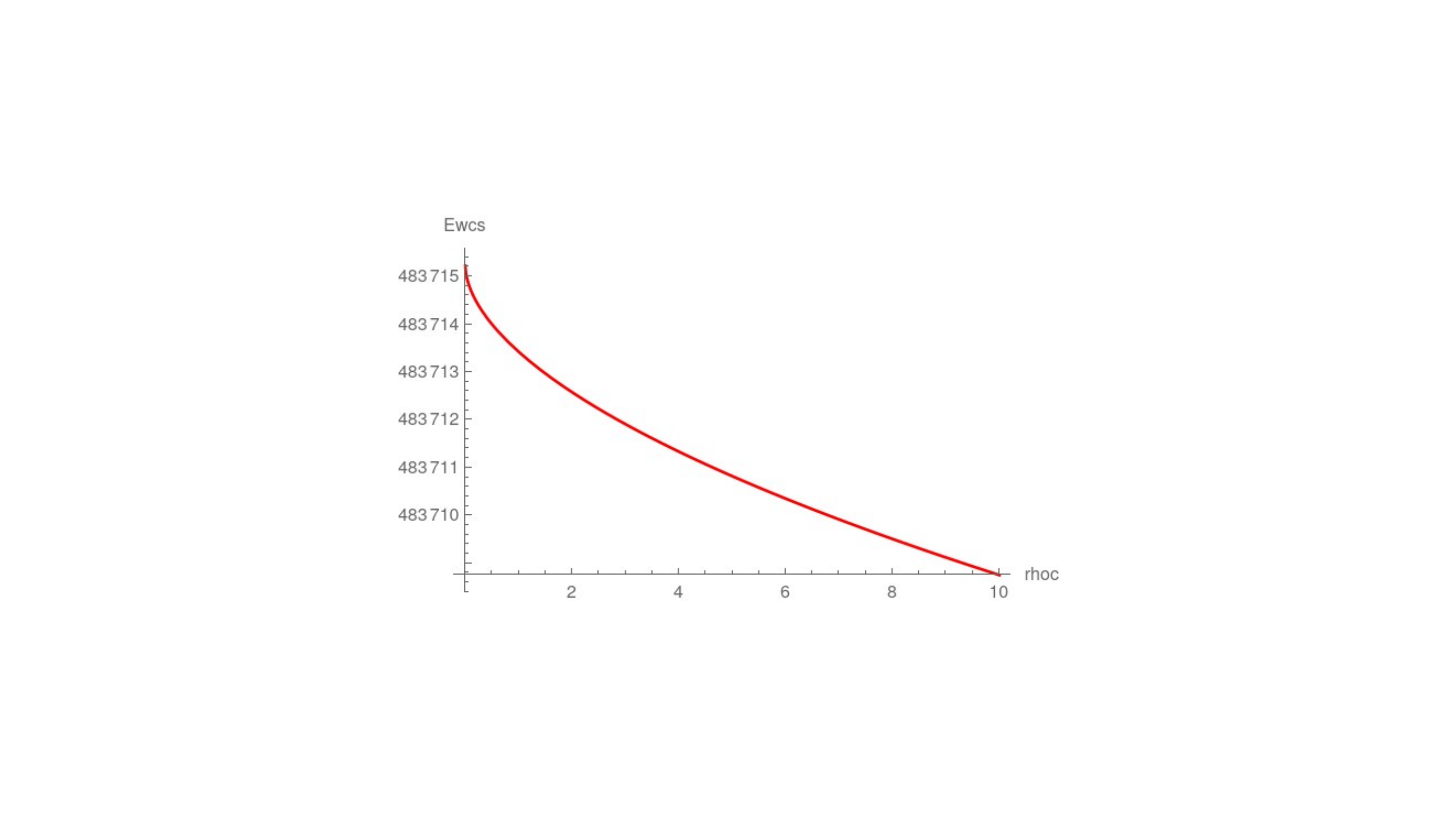}
\includegraphics[width=.40\textwidth]{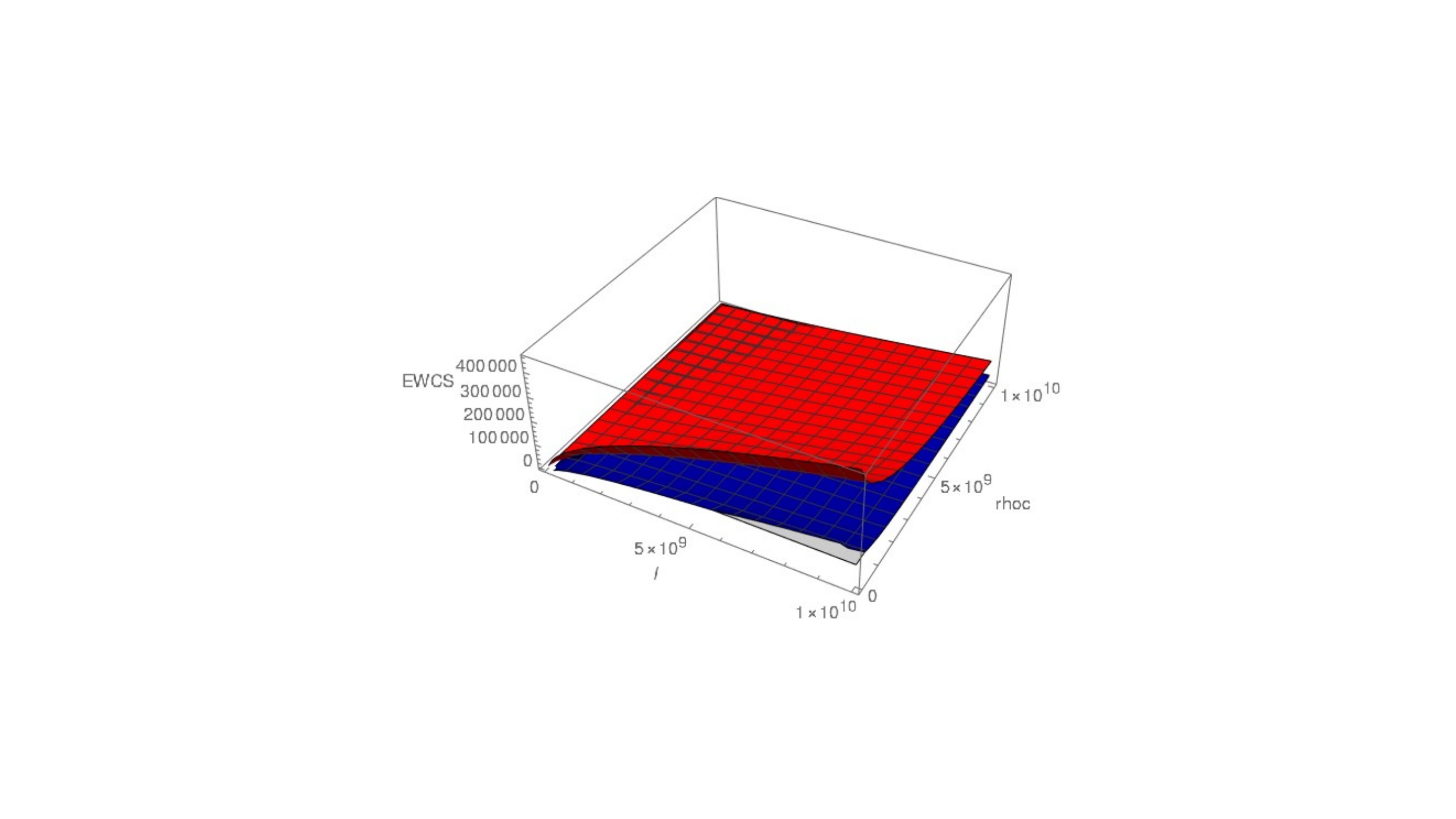}
\includegraphics[width=.40\textwidth]{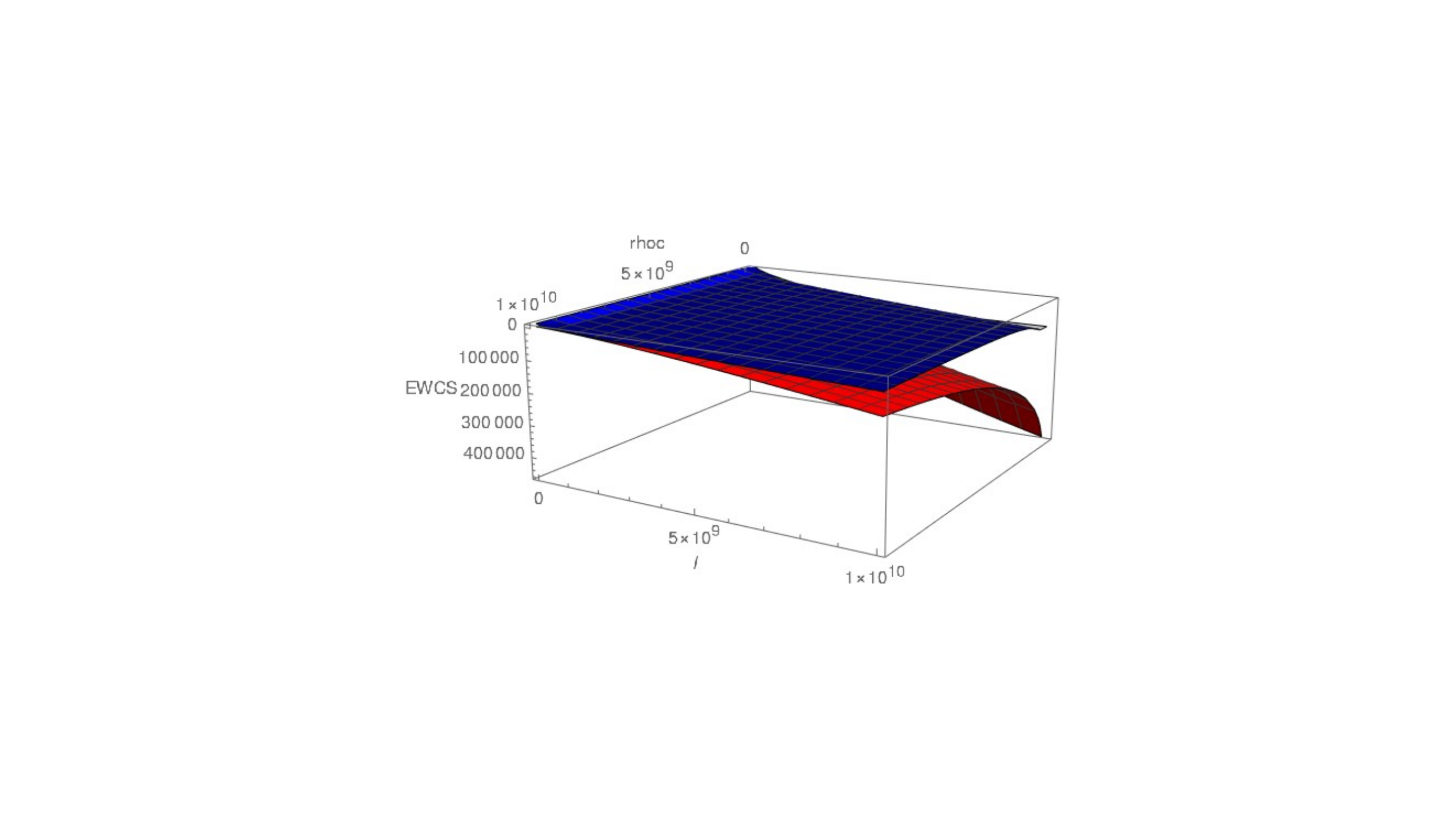}
\caption{(First row) \,\,:\,\,(   left ) \,,\, EWCS is  plotted as a function of $(l,\rho_c)$ \, (right),\,:\, , Ewcs plotted as a function of l with $\rho_c = {10}$ \,:\,  Both the plots  showing, EWCS increases with the increase of l 
  \,;\,   (Second row) \,:\,  ( left)\, : \, EWCS plotted as a function of $(\rho_c,l)$ for $h = (10)^6$, \,  (right) \,:\, EWCS plotted as a function of $\rho_c$  for $l = {10}^{10}$, $h = (10)^6$ \, :\,  We see, for a given l, EWCS falls with the increase of cut off $\rho_c$ and goes to zero for $\rho_c >> l $ regime  and  finite at zero cut off for non zero h \,,\, (Third row)\,:\,( left)\, : \, EWCS  as a function of $(\rho_c,l)$  for $h = 0$     \,,\,   , (right) \,:\, EWCS  as a function of $\rho_c$  for $l = {10}^{10}$, for $h=0$,  \, :\, Here unlike the case of nonzero cut off we see that Ewcs, for $l>>\rho_c$    diverges for $h = 0$ for zero cut off and the 2D plot is taken over very short range of $\rho_c$ to see this divergence explicitly\,,\,(Last row)\,\,:\,\,   The overlap of EWCS-l-$\rho_C$ plot and 
${\frac{1}{2}} \left( H.M.I\right)$-l-$\rho_c$ plot considered to see whether the inequality  $EWCS \ge {\frac{1}{2}} \left( H.M.I\right)$ holds, with EWCS in red and ${\frac{1}{2}} \left( H.M.I\right)$ in blue \,: \, (left)\,:\, The frontview of  overlap plot, showing for $l >>\rho_c$ inequality satisfies \, (right) \, The backsideview of  overlap plot, showing for $\rho_c >> l$, inequality saturates }
\label{ewcsbasic4by9}
\end{figure}

\begin{figure}[H]
\begin{center}
\textbf{ For   $ d - \theta  = {\frac{1}{3}}$\,: \, Ewcs vs $(l,\rho_c)$, Ewcs vs l, Ewcs vs $\rho_c$ graph for fixed h for the connected phase, extrapolated to $l = 0$  }
\end{center}
\vskip2mm
\includegraphics[width=.40\textwidth]{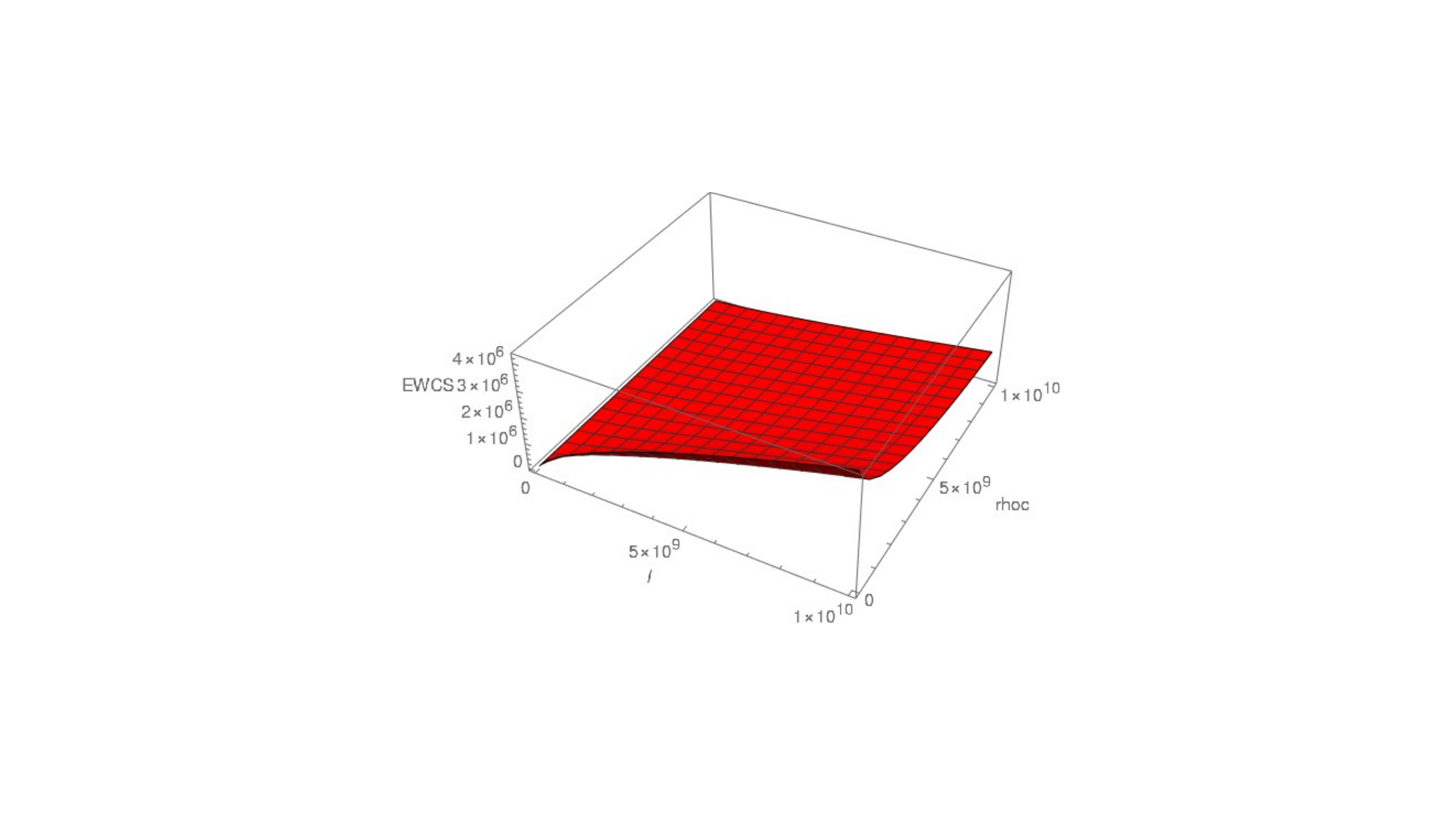}
\includegraphics[width=.40\textwidth]{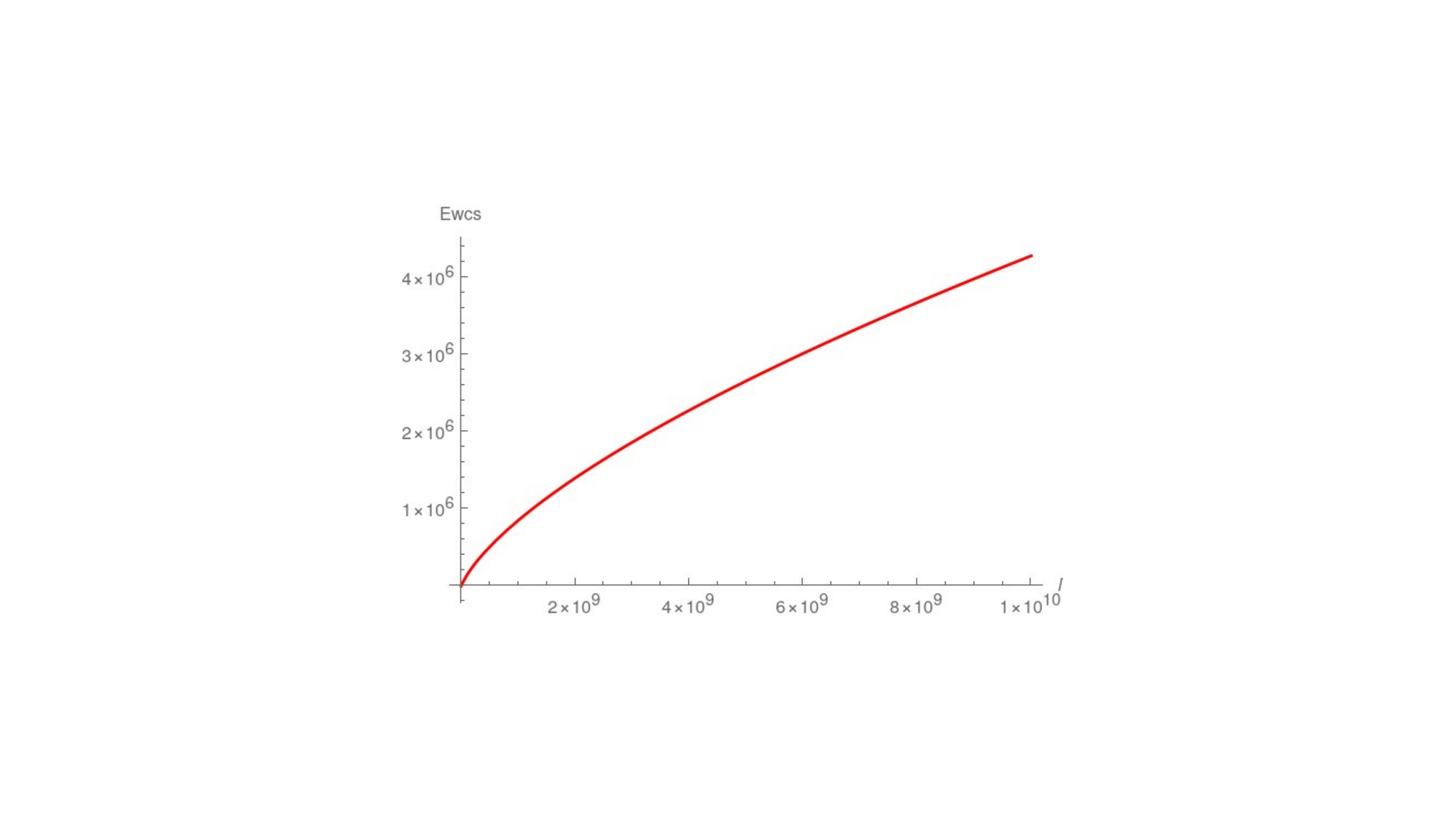}	
\includegraphics[width=.40\textwidth]{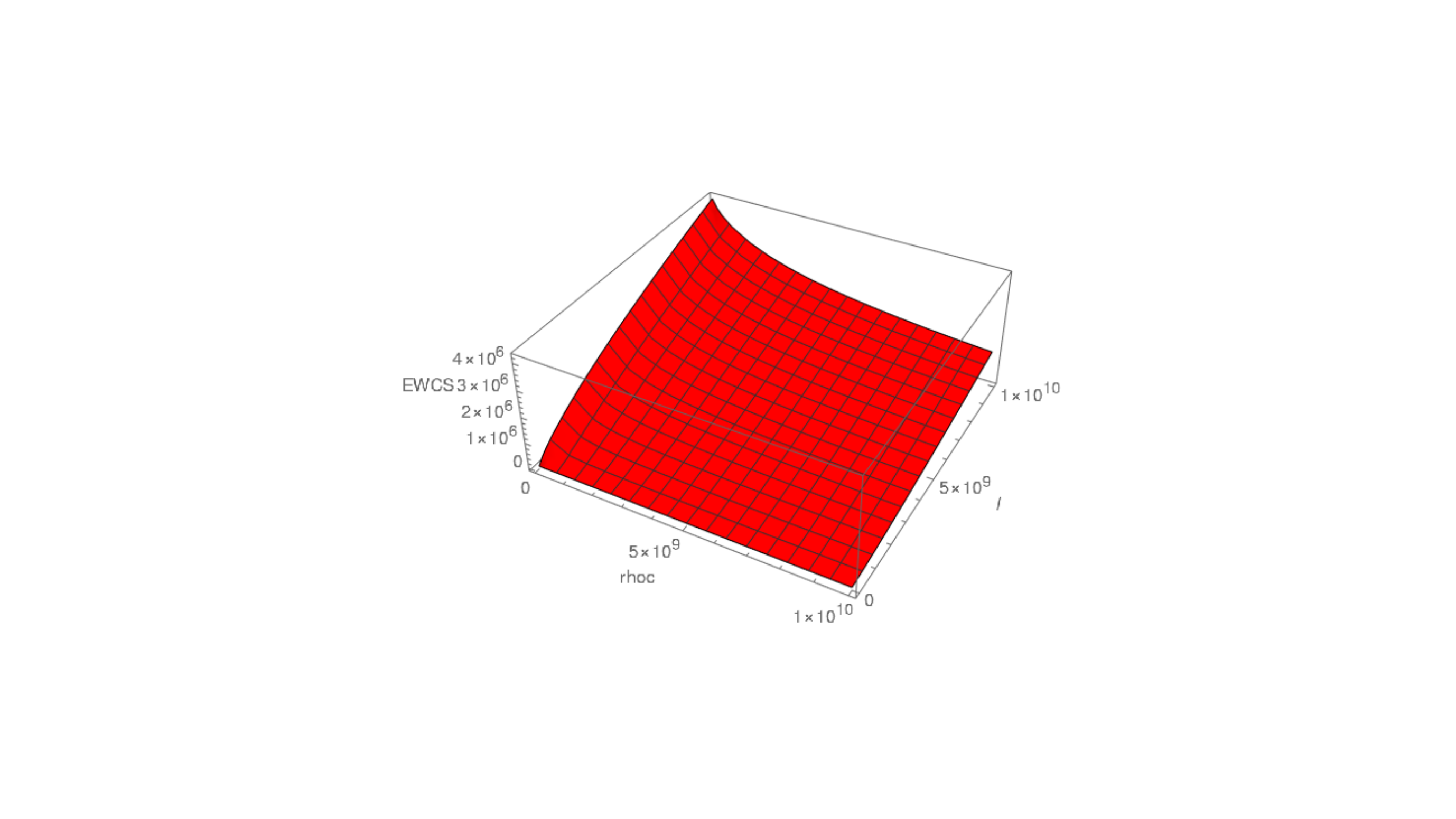}
\includegraphics[width=.40\textwidth]{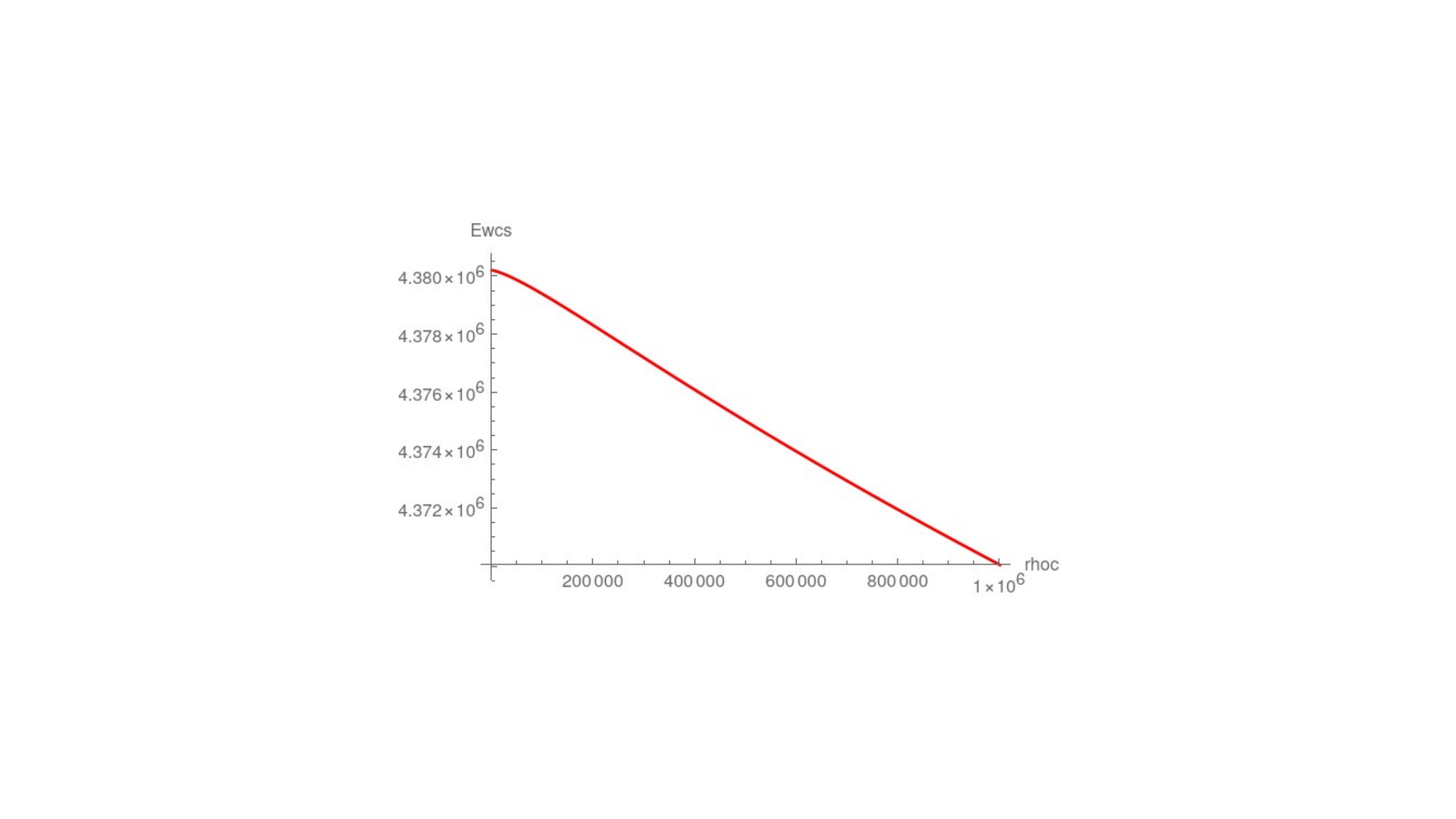}
\includegraphics[width=.40\textwidth]{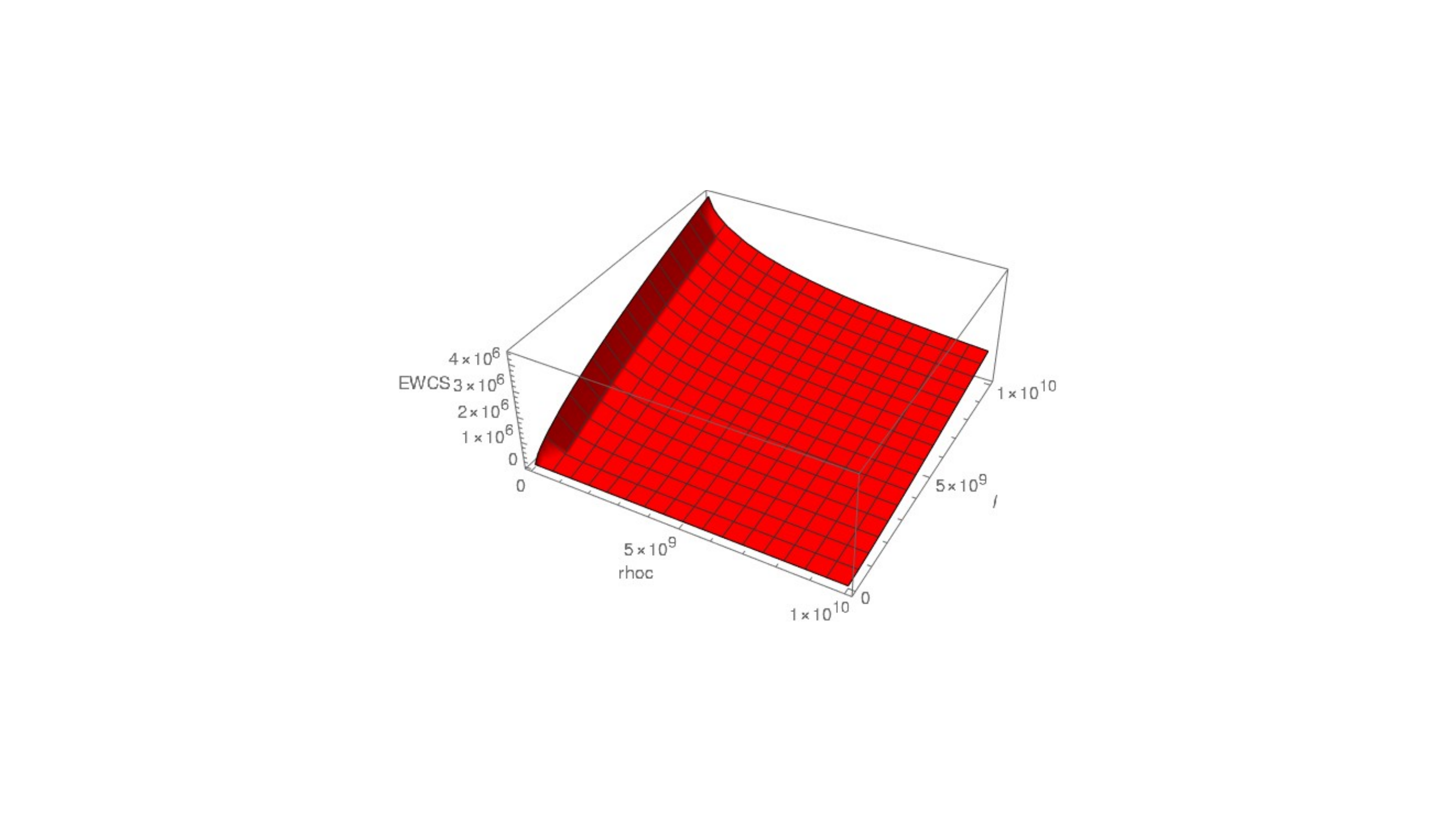}
\includegraphics[width=.40\textwidth]{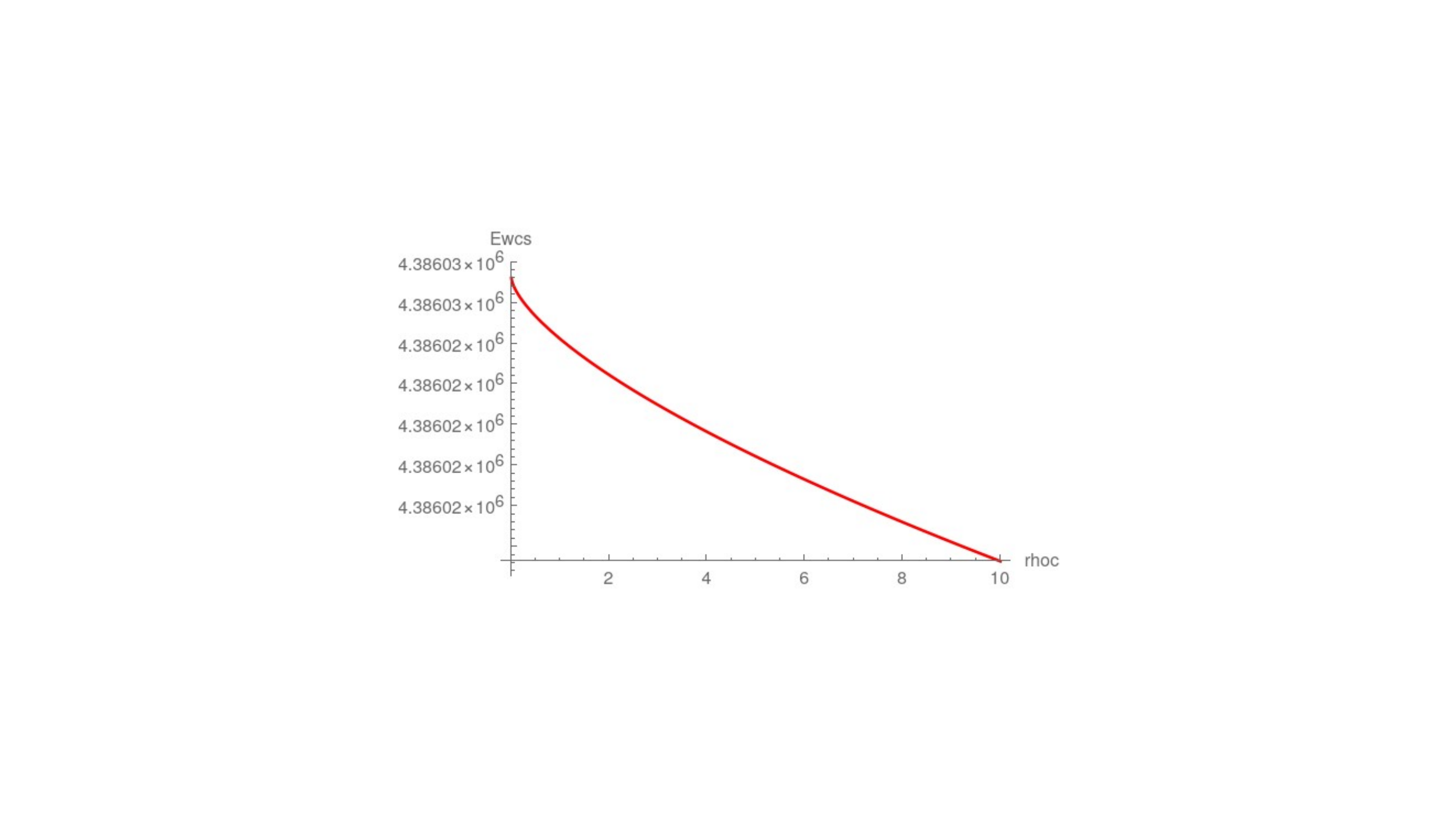}
\includegraphics[width=.40\textwidth]{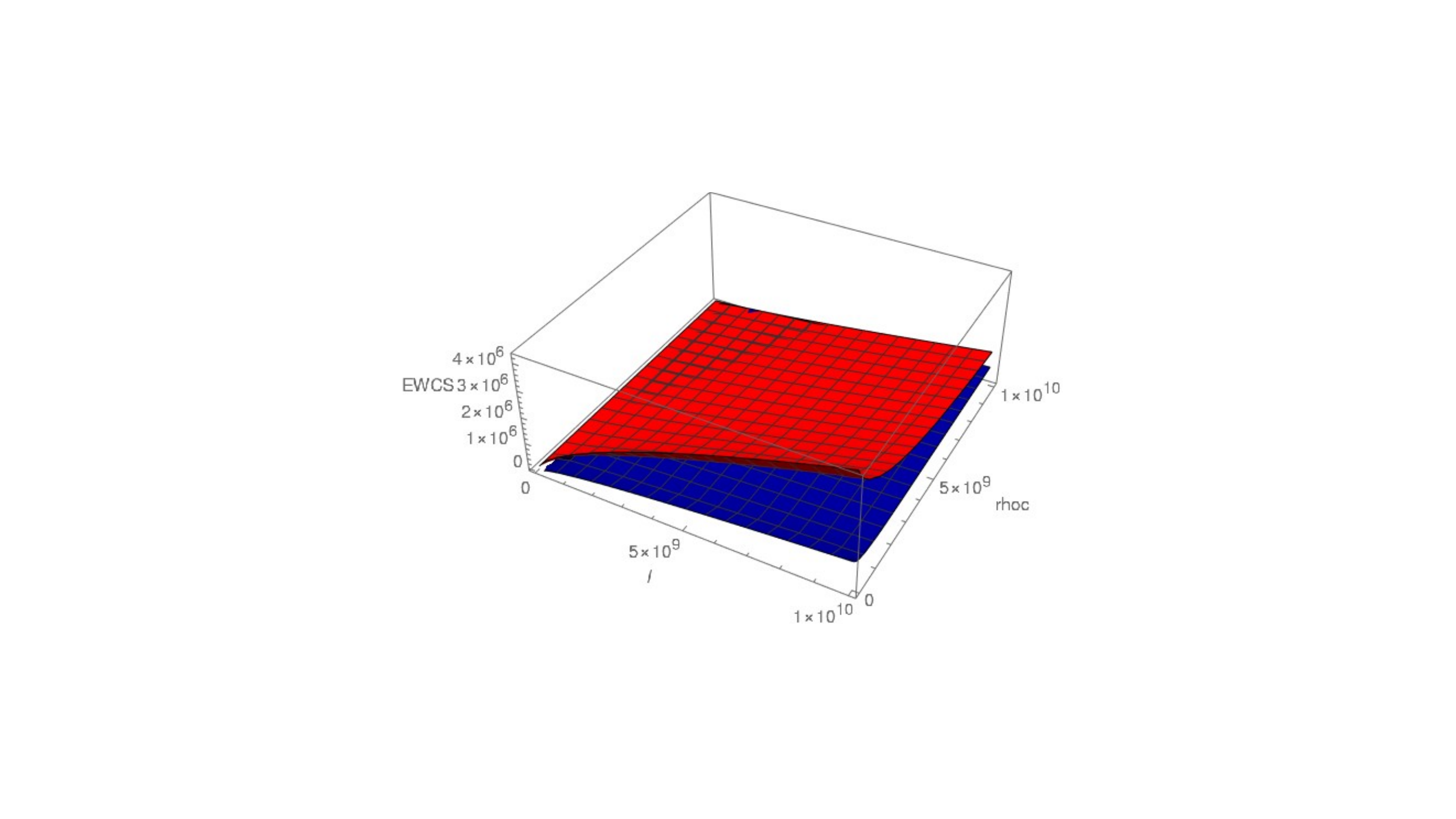}
\includegraphics[width=.40\textwidth]{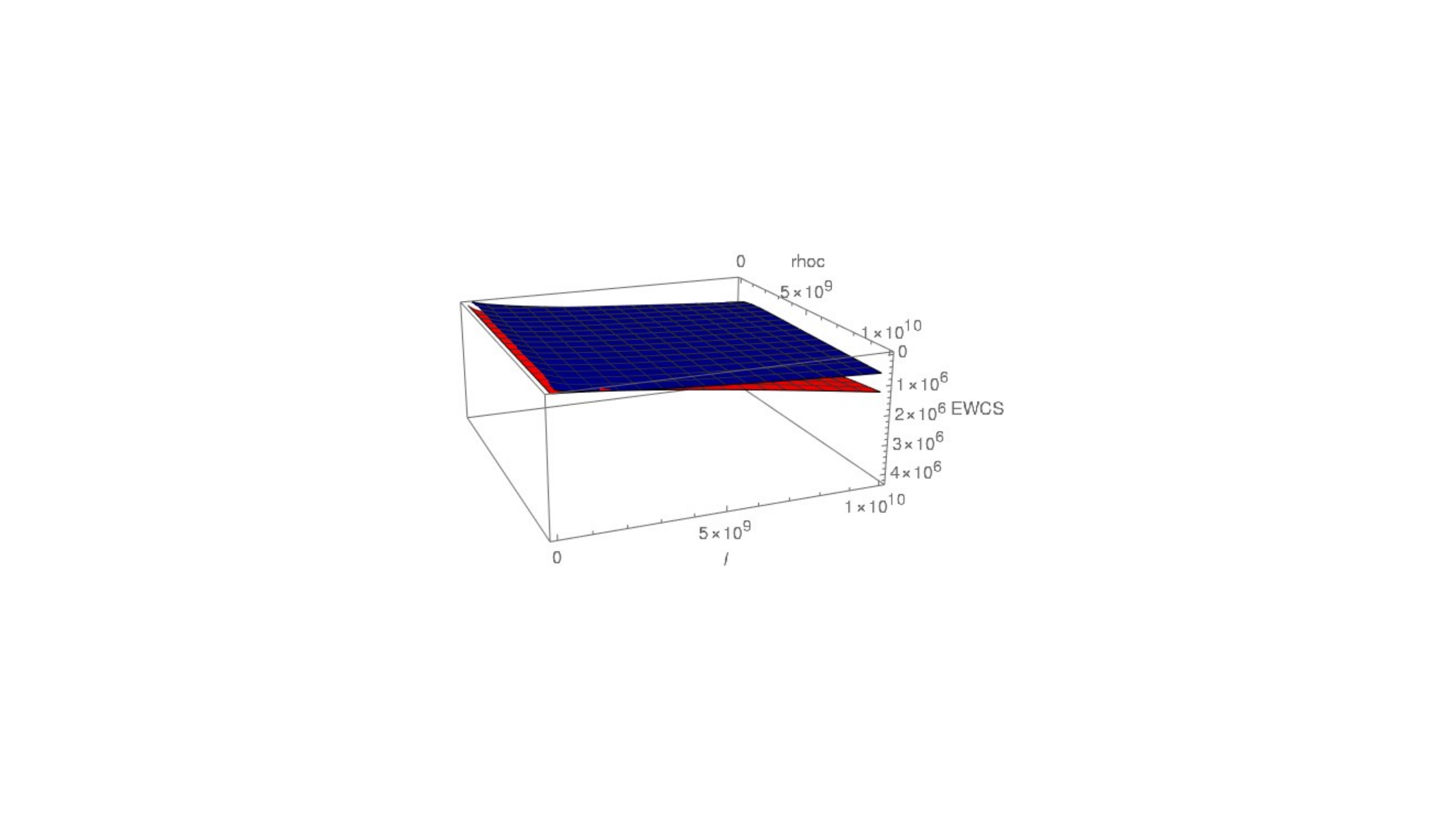}
\caption{(First row) \,\,:\,\,(   left ) \,,\, EWCS is  plotted as a function of $(l,\rho_c)$ \, (right),\,:\, , Ewcs plotted as a function of l with $\rho_c =  10$ \,:\,  Both the plots  showing that  EWCS increases with the increase of l 
  \,;\, (Second row) \,:\,  ( left)\, : \, EWCS plotted as a function of $(\rho_c,l)$ for $h = (10)^6$, \,  (right) \,:\, EWCS plotted as a function of $\rho_c$  for $l = {10}^{10}$, $h = (10)^6$ \, :\,  We see, for a given l, EWCS falls with the increase of cut off $\rho_c$ and goes to zero for $\rho_c >> l $ regime  and  finite at zero cut off for non zero h \,,\, (Third row) \,:\,  ( left)\, : \, EWCS  as a function of $(\rho_c,l)$  for $h = 0$     \,,\,   , (right) \,:\, EWCS  as a function of $\rho_c$  for $l = {10}^{10}$, for $h=0$,  \, :\, Here unlike the case of nonzero cut off we see that Ewcs, for $l>>\rho_c$    diverges for $h = 0$ for zero cut off and the 2D plot is taken over very short range of $\rho_c$ to see this divergence explicitly
. \,,\,(Last row)\,\,:\,\,   The overlap of EWCS-l-$\rho_C$ plot and 
${\frac{1}{2}} \left( H.M.I\right)$-l-$\rho_c$ plot considered to see whether the inequality  $EWCS \ge {\frac{1}{2}} \left( H.M.I\right)$ holds, with EWCS in red and ${\frac{1}{2}} \left( H.M.I\right)$ in blue \,: \, (left)\,:\, The frontview of  overlap plot, showing for $l >>\rho_c$ inequality satisfies \, (right) \, The backsideview of  overlap plot, showing for $\rho_c >> l$, inequality saturates 
}
\label{ewcsbasic1by3}
\end{figure}

\begin{figure}[H]
\begin{center}
\textbf{ For $d - \theta < 1$, EWCS vs $(h, l)$ plot  for very long range, with $\rho_c = 100$  }for the connected phase, extrapolated to $l = 0$ 
\end{center}
\vskip2mm
\includegraphics[width=.65\textwidth]{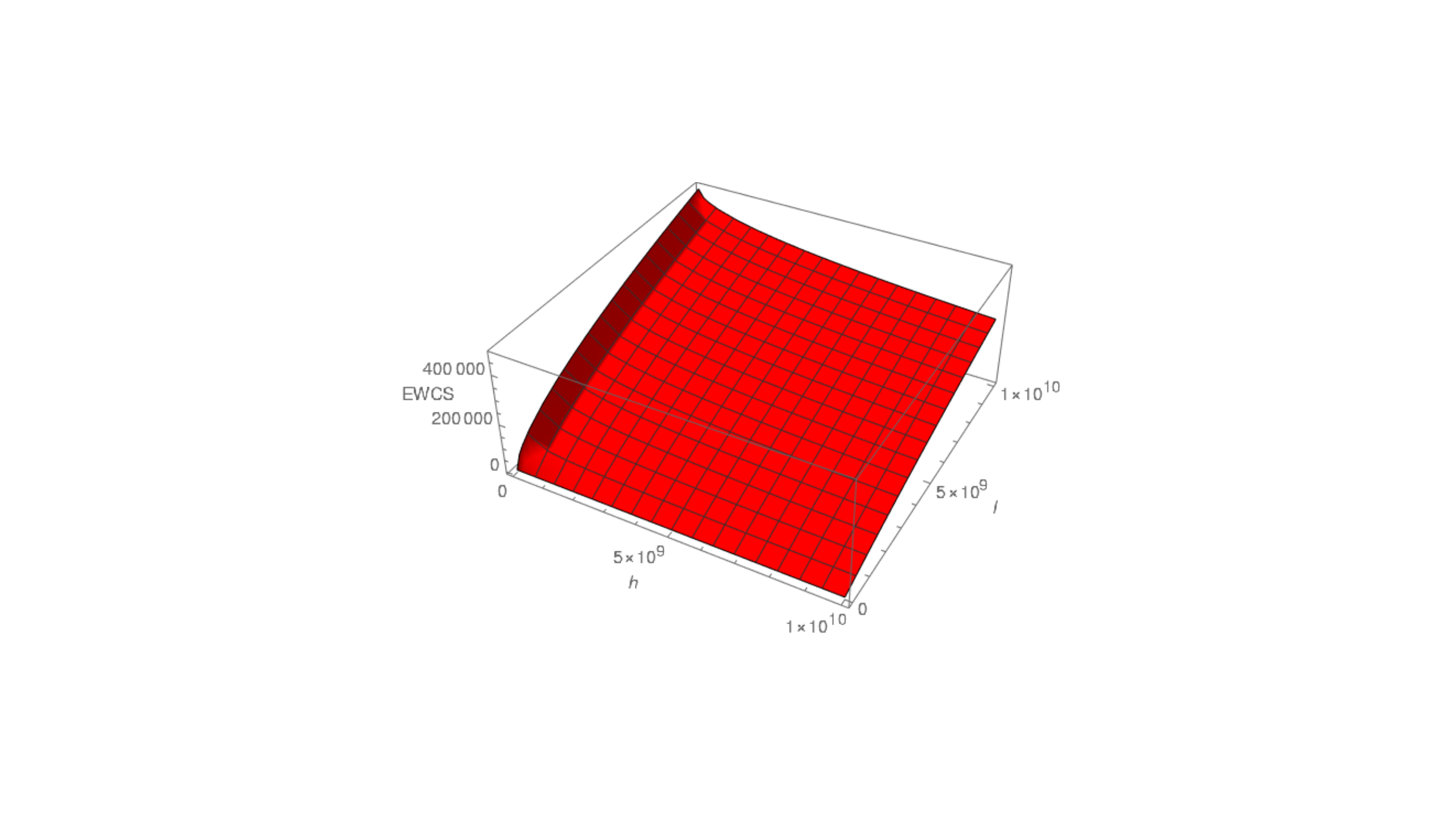}
\includegraphics[width=.65\textwidth]{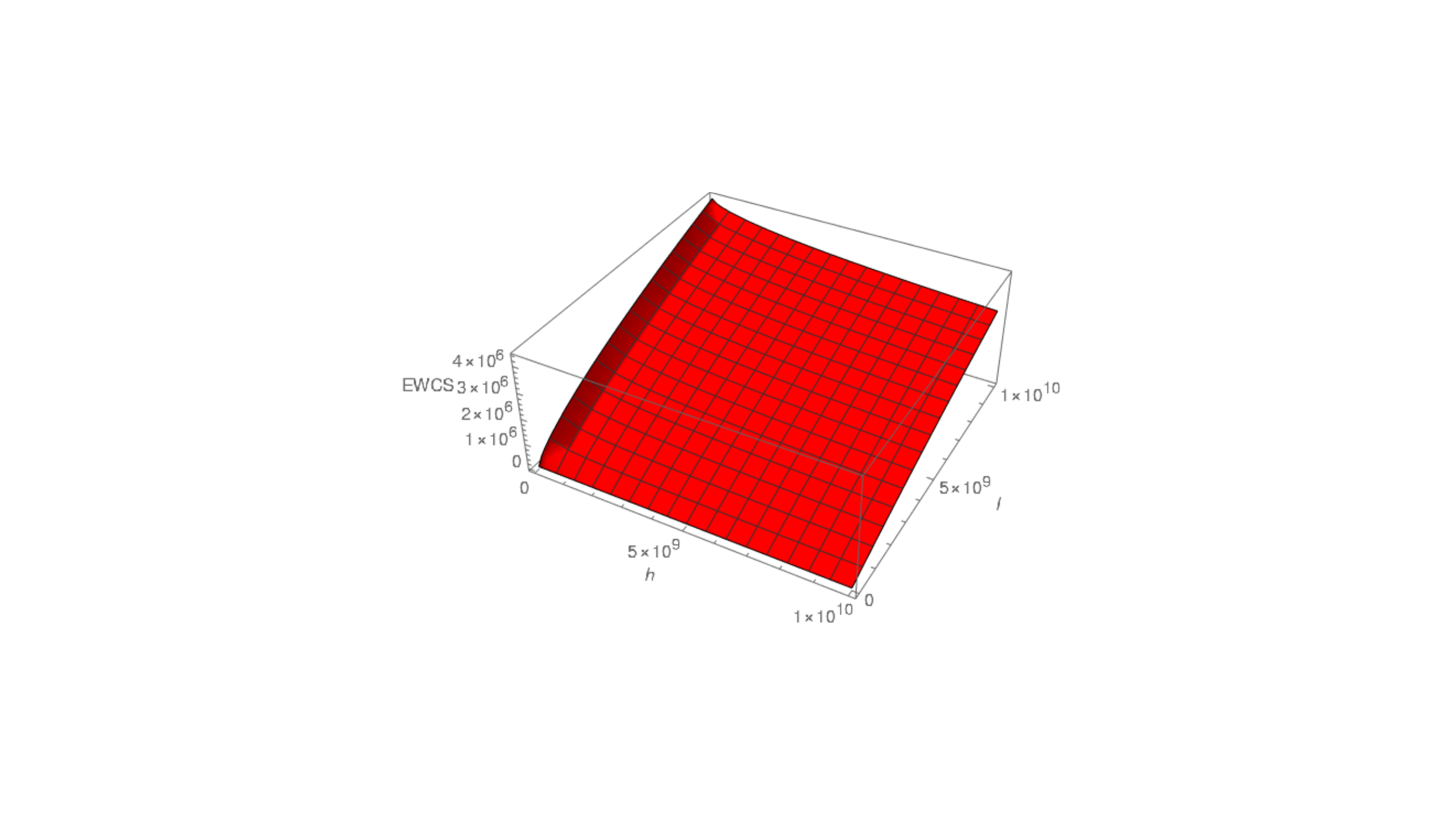}
\caption{ (left) \,:\,EWCS  vs $(h,l )$  plot for $d - \theta = {\frac{4}{9}}$, \, , \, (right) $d - \theta = {\frac{1}{3}}$ \,: \, showing EWCS falls with increase of  h, for a given l,with $\rho_c$ fixed.  It is also showing for a given h,EWCS  increses with l  }
\la{ewcsblessthan1hl}
\end{figure}

\begin{figure}[H]
\begin{center}
\textbf{ For $d - \theta < 1$, EWCS as a function of $(\rho_c , h)$ for the connected phase, extrapolated to $l = 0$  }
\end{center}
\vskip2mm
\includegraphics[width=.65\textwidth]{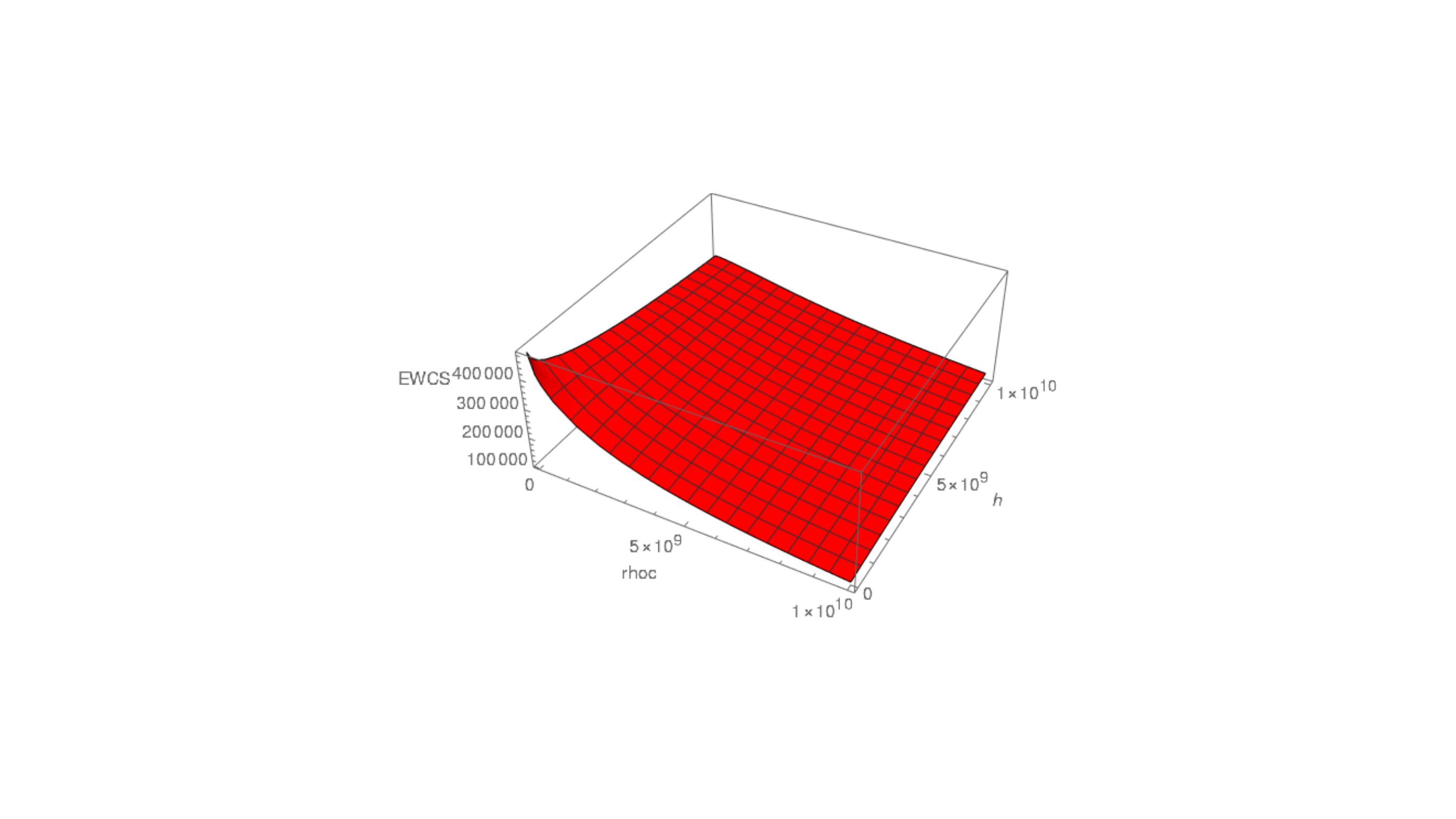}
\includegraphics[width=.65\textwidth]{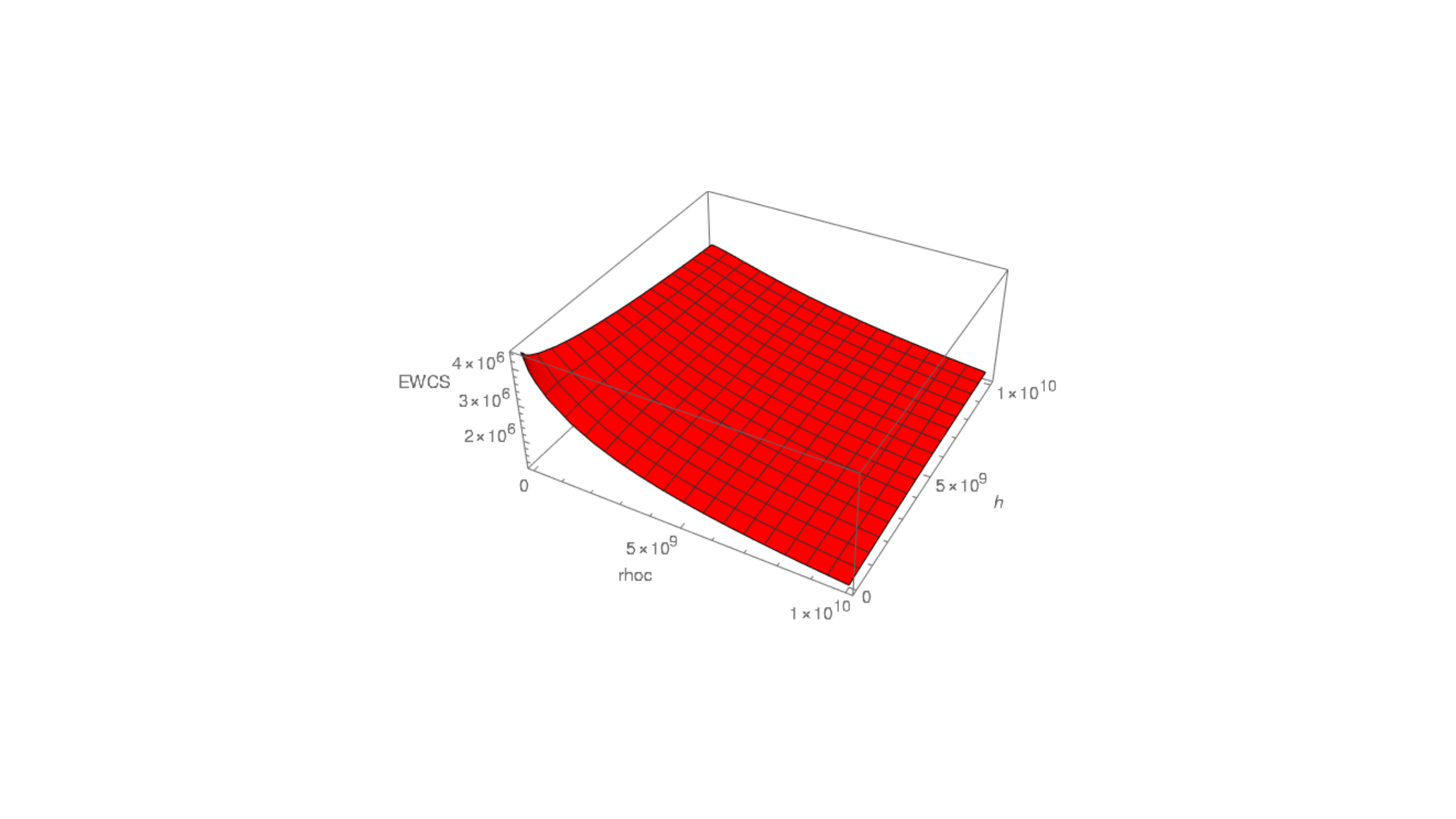}
\caption{ (left) \,:\,EWCS  vs $(\rho_c,h)$  plot for $l = {10}^{10}$\, (left) \, $d - \theta = {\frac{4}{9}}$, \, , \, (right) $d - \theta = {\frac{1}{3}}$ \,: \, showing EWCS falls with increase of both  $(\rho_c, h)$,   diverge at $h=0$ for zero cut-off but become finite at $h = 0$ for nonzero cut off }
\la{ewcslessthan1rhoch}
\end{figure}

\begin{figure}[H]
\begin{center}
\textbf{ For $ d - \theta < 1$ \,: \,  The evolution of EWCS with $ d - \theta$ with  $ h = {10}^{6}$  }
\end{center}
\vskip2mm
\includegraphics[width=.65\textwidth]{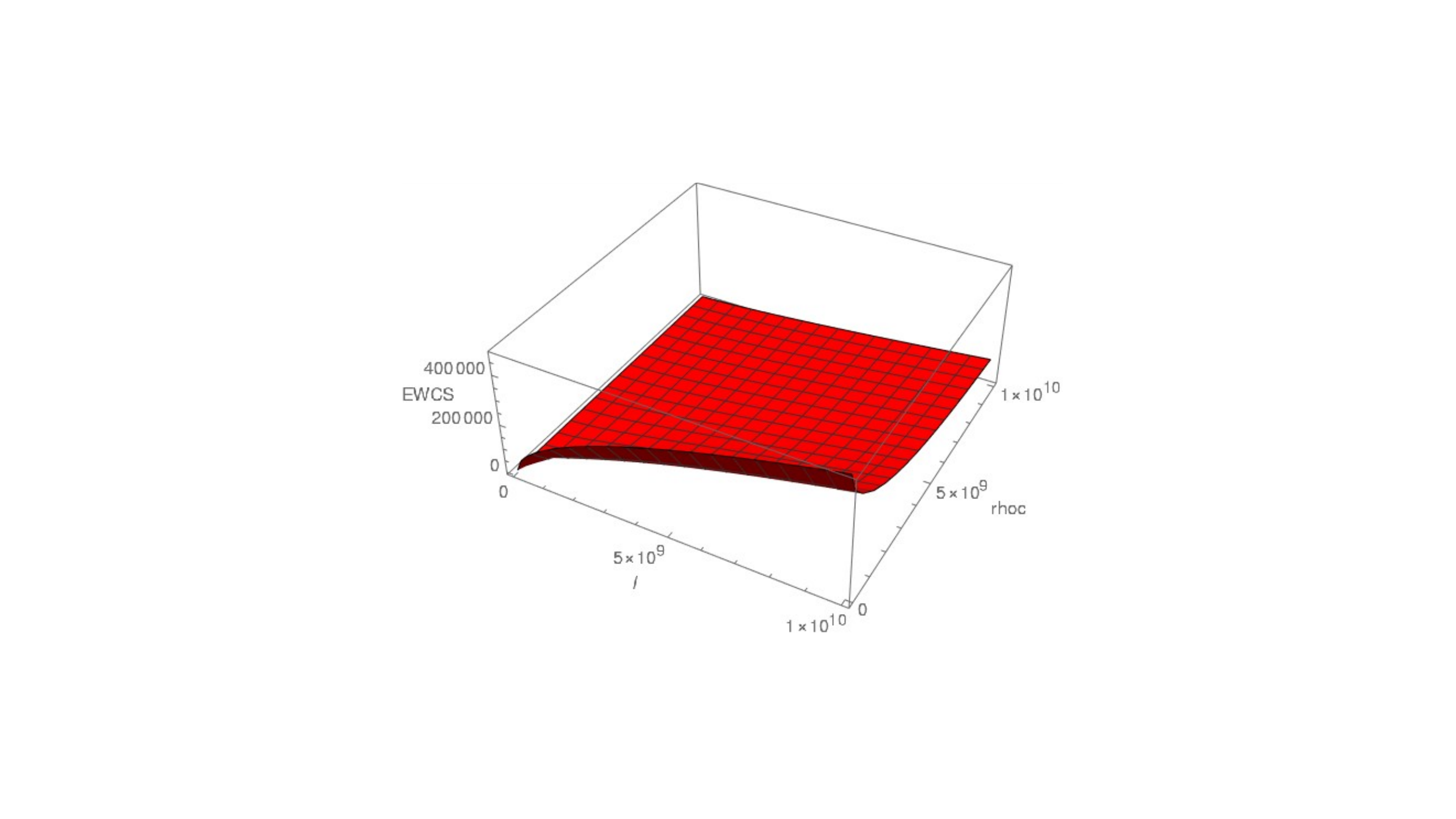}
\includegraphics[width=.65\textwidth]{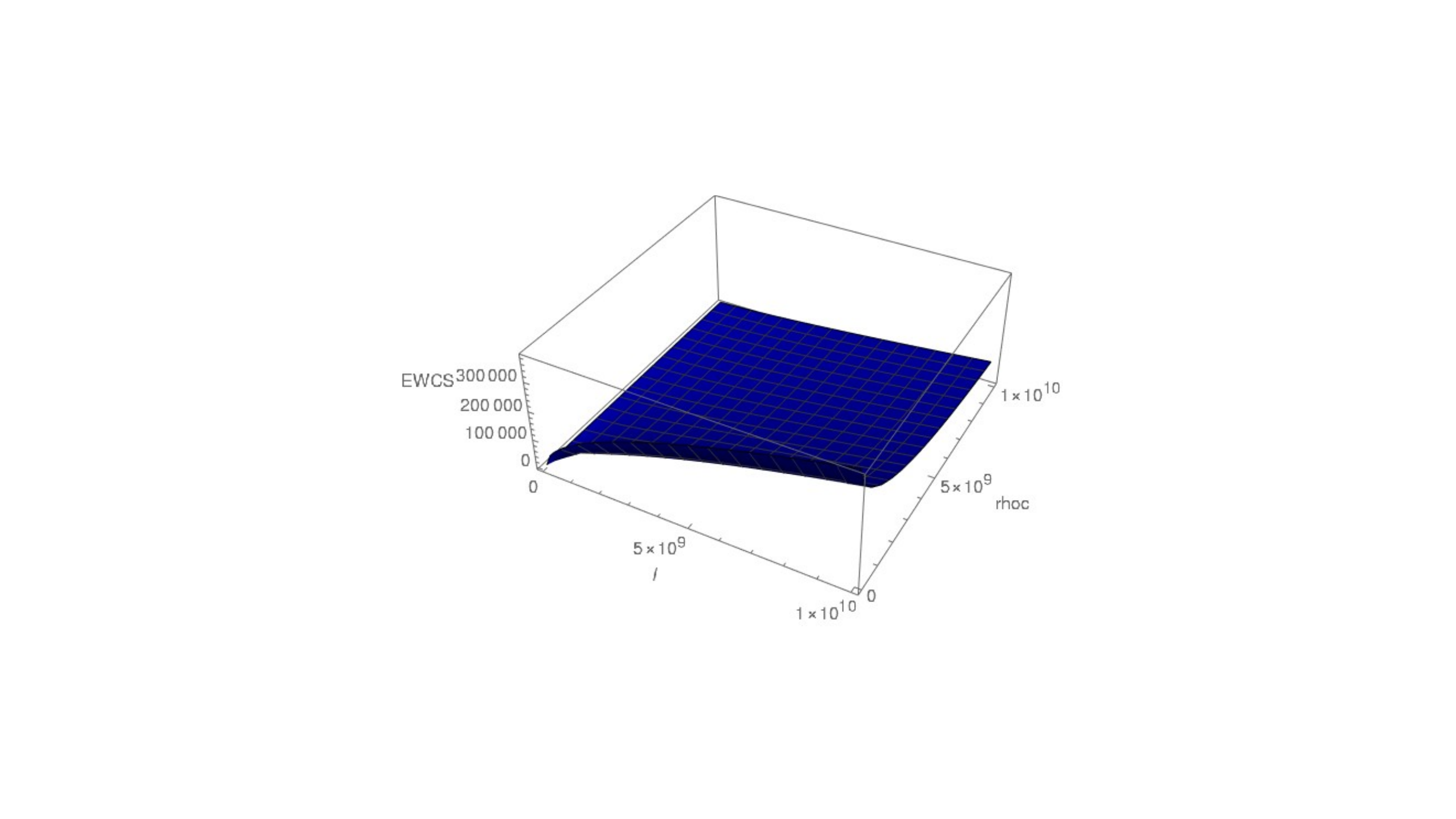}	
\includegraphics[width=.65\textwidth]{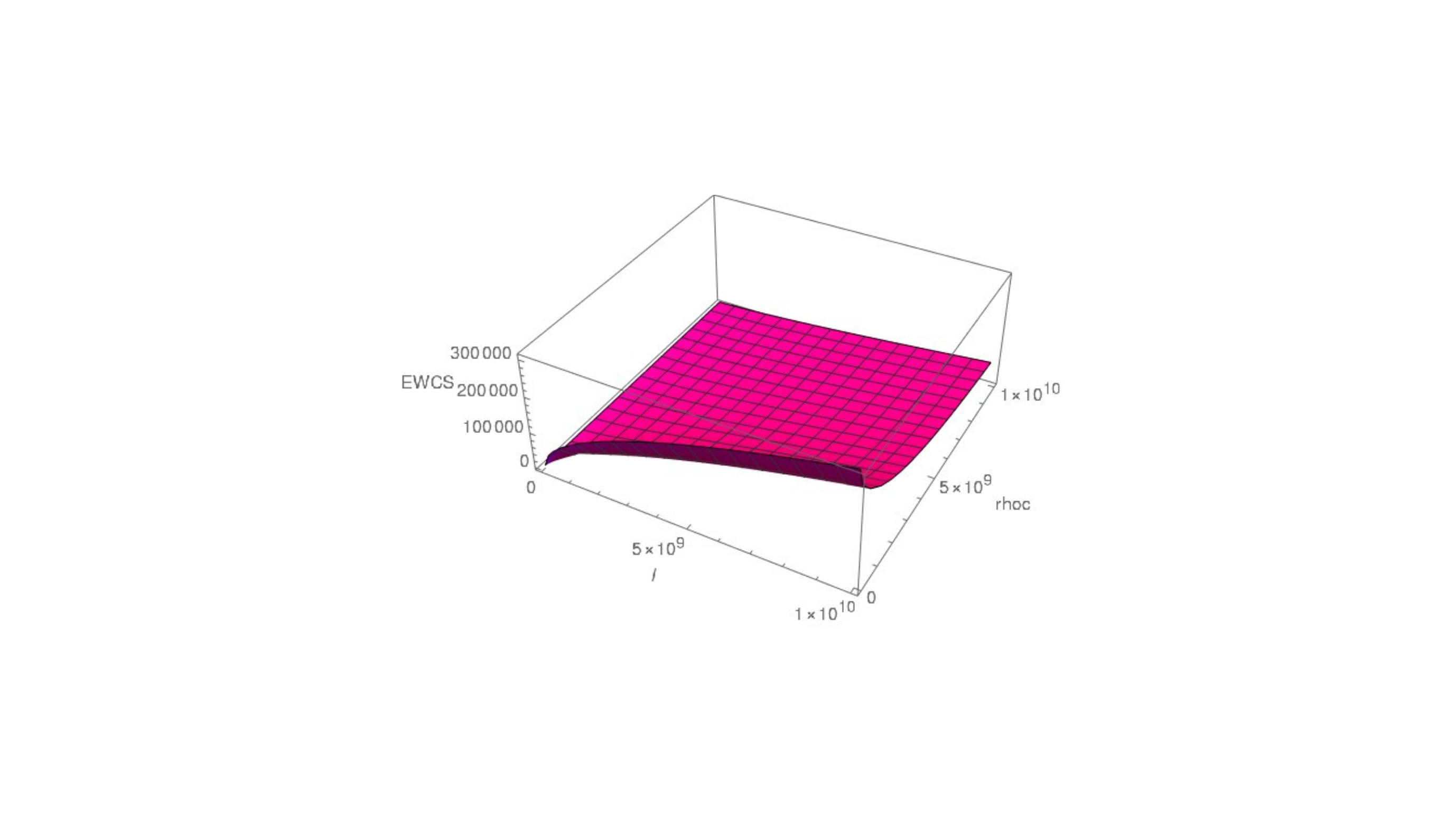}
\includegraphics[width=.65\textwidth]{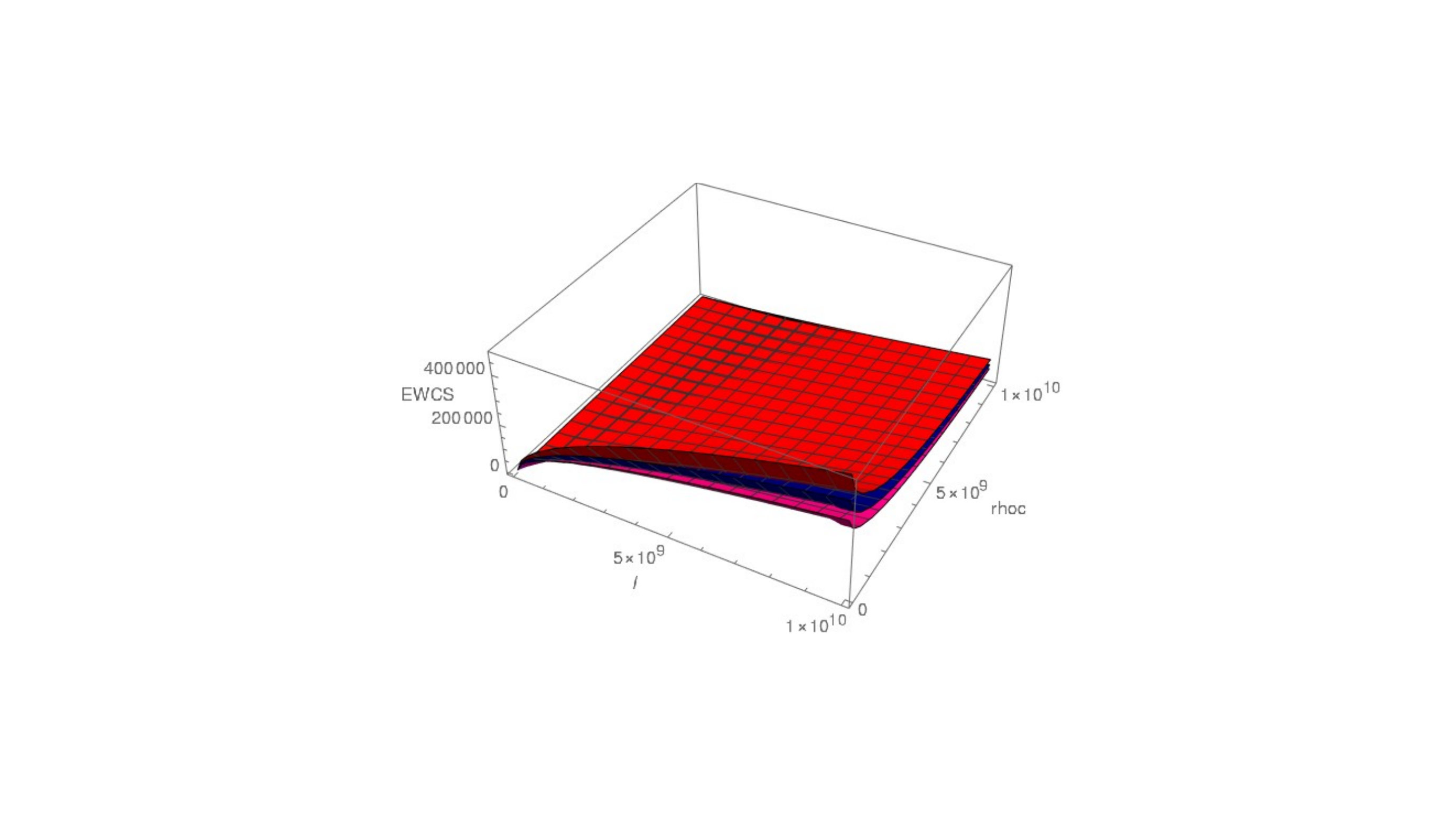}
\includegraphics[width=.65\textwidth]{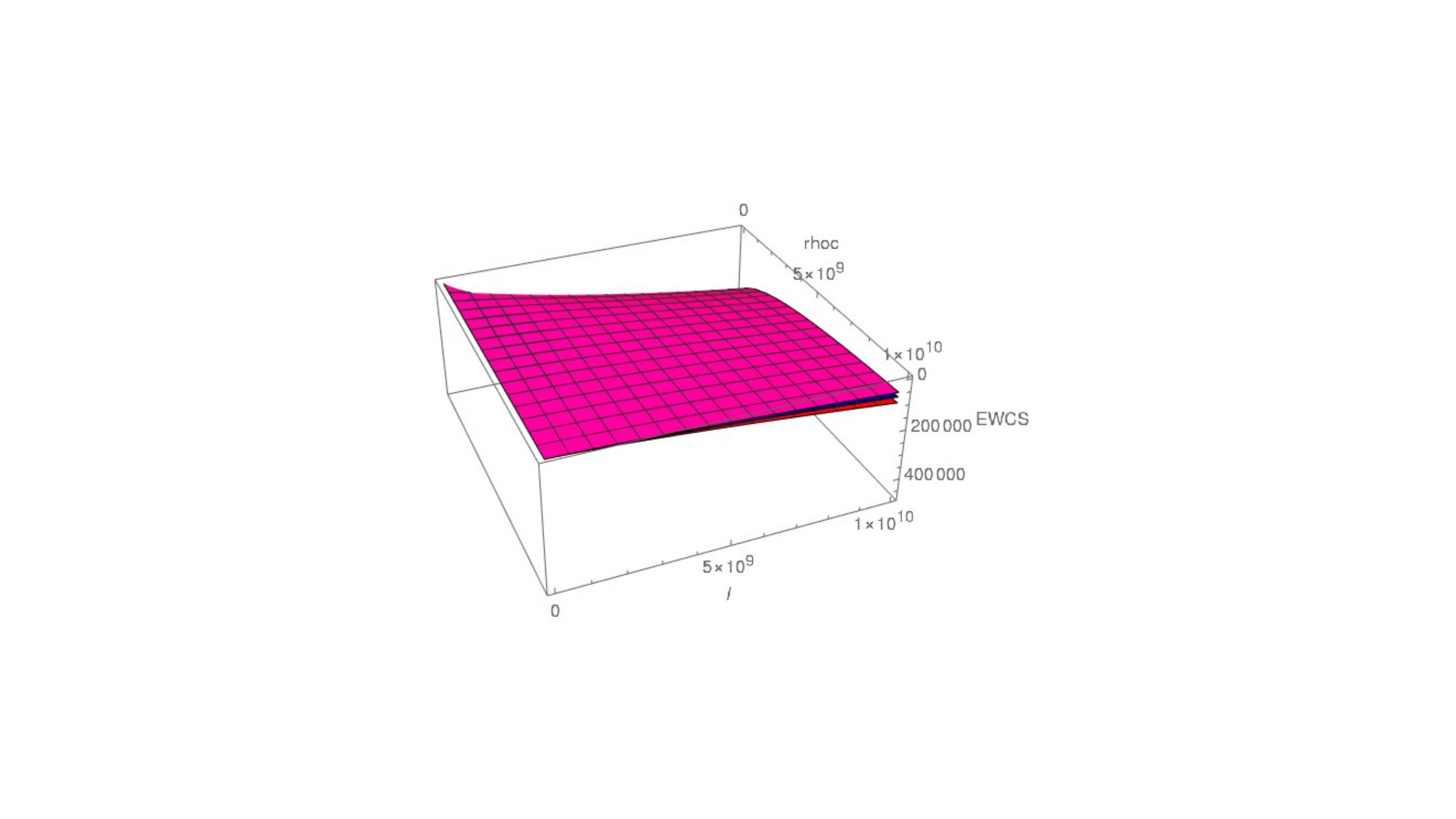}	
\includegraphics[width=.65\textwidth]{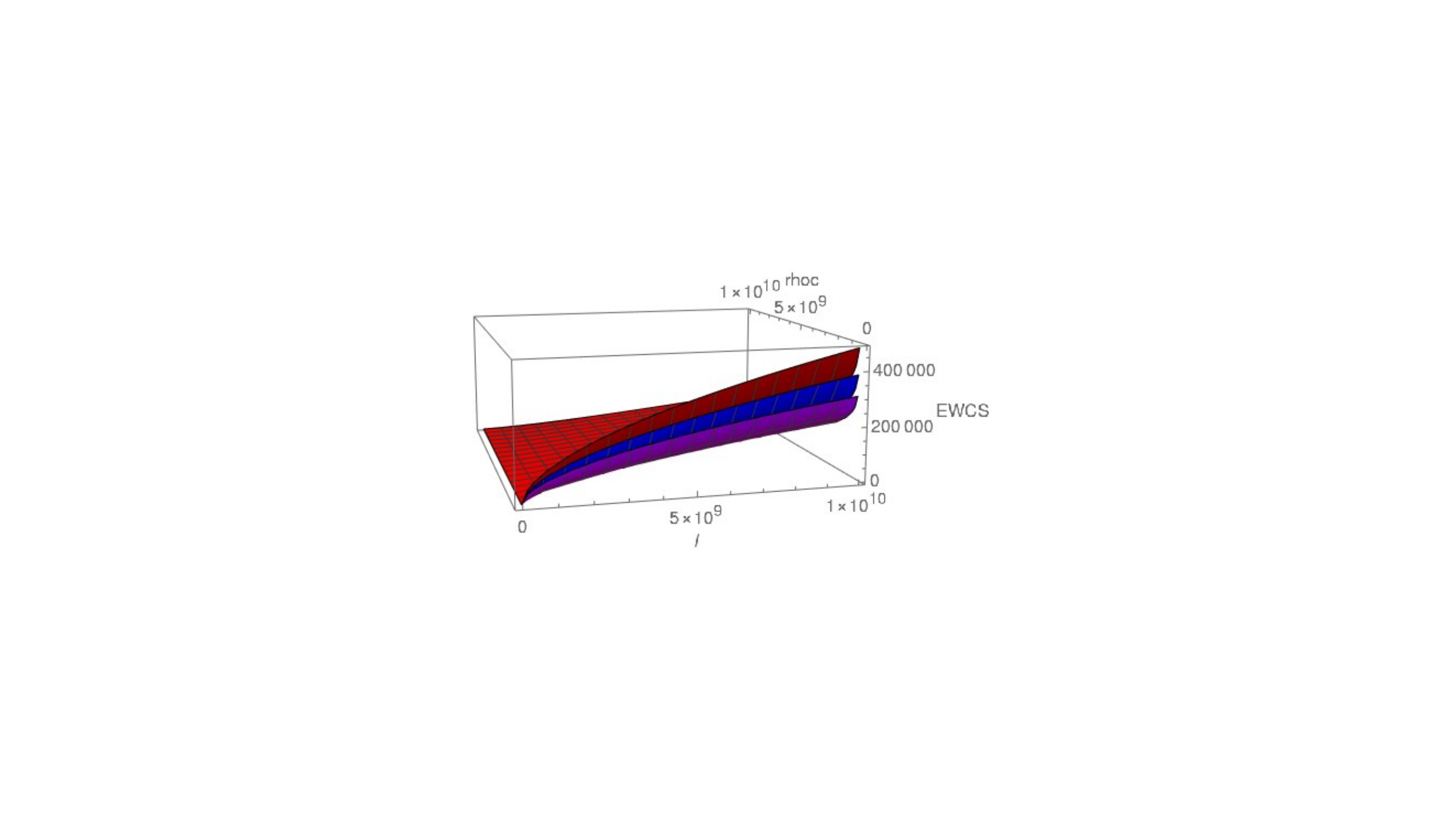}	
\caption{(First row) \,\,:\,\,( From  left to right) \,,\, EWCS plotted as a function of $(l,\rho_c)$ for $d - \theta = 0.444\,,\, d - \theta =  $,0.455
  \quad;\quad (Second row) \,:\,  ( left)\, : \, EWCS plotted as a function of $(l,\rho_c)$ \, ,\,  $d - \theta= 0.466$ \,\,,\,\, (right) \,:\, The frontview of the  overlap of the three, showing that for $l >> \rho_c$ EWCS  decreases with the increase of $d - \theta$
 \, ,\,   \quad;\quad (Last row ) ,\, ( left )\, : \,  \,:\, The backside view of the overlap of the three, showing for $\rho_c >> l$ the EWCS for different $d - \theta$   actually merges, supporting the fact that for a fixed l, EWCS decreases with the increase of $\rho_c$ and ultimately falls towards zero.\,\,;\,\. (right)\,:\,  The sideview of the overlap of the there showing that EWCS in its connected phase smoothly reach to zero  at $l = 0$(extrapolated to $l = 0$)  for all $d - \theta$ }
\label{ewcsevblessthan1}
\end{figure}

\section{Conclusion and the outlook}

In this work,  we have constructed  and studied some measures of quantum entanglement of some physical system  defined on the boundary of some  spacetime regime  where the field theory on the boundary is described by $T{\overline{T}}$ deformed CFT.  We have the dual  bulk geometry is given by the Hyperscaling violating geometry with most general dimension d  and Hyper scaling violating factor. $\theta$ with  $\theta \le d$,  which according to the statement of AdS/CFT  duality  should have a finite radial cut-off $\rho_c$,   with $\rho_c$ is a function of  $T{\overline{T}}$   deformation coefficient $\mu$.     We consider  the  scenario where the boundary system can be in pure or mixed state,  and    consider the boundary subregion in the form of the single and disjoint interval with the form of strip and studied Entanglement of entropy, Mutual information
and the Entanglement of purification. Given the fact that $T{\overline{T}}$ deformation is a solvable irrevalent deformation we can expect that the boundary physical system should have a smooth description over the complete parameter-regime $(\mu, l)$ with $\mu$ is the $T{\overline{T}}$  deformation coefficient and l is the length of the strip,  which implies we are supposed to have the expression of the measure of all these quantum entanglement,   globally defined over complete 2D parameter regime of $(\mu,l)$.  Since the gravity dual of  $T{\overline{T}}$  deformation in the boundary theory have an interpretation over the bulk geometry with the finite radial cut off $\rho_c$, so in the gravity dual of such boundary system we are expected to have these measures of these quantum entanglements, globally defined over 2D $(l, \rho_c)$ plane!  Here we considered time independent boundary system   and consequently tried to obtain these quantum measures from the formal RT prescription.  Here we consider zero temperature case only.  However we found the greatest barrier to obtain such global description is that the governing equation for the turning point can neither solvable exactly or at perturbative level for a global solution over 2D $(l,\rho_c)$ plane.  In order to obtain the same first we consider speculative approach,  found it is possible to have a globally defined solution for the turning point $\rho_0(l,\rho_c)$,  for the complete parameter regime of $(d,\theta)$ which can be exactly determined in the regime $l >> \rho_c$ and $\rho_c >> l$ only.  Next, interestingly we found the underlying theory posses a global symmetry structure,  which is although not the symmetry of the original theory but it emerges on application of RT formalism to such geometry 
for the complete parameter regime of $d, \theta$  with the symmetry transition is irrespective of $d, \theta$.  We found  this global synnetry along with  the two 
global boundary conditions and the other consistency conditions,  can completely fix the complete global structure of the turning point $\rho_0 (l,\rho_c )$,  which we found to consist of three parts.  We have shown, out of these three category of terms,  one of them can be uniquely determined and if we consider the particular regime 
$ l>> \rho_c $ and $ \rho_c >> l $, then over this regime,  this is much much more dominating over two others and so can be considered as effectively giving the global solution of the turning point $\rho_0 (l, \rho_c)$ over this regime.  In the rest of the regime of $(l, \rho_c)$ plane,  the expression of $\rho_0 (l, \rho_c)$,  can be thought as an interpolating function between these two regime,  where all our consistency-chechecking plots are showing, that even these interpolating function is quite close to the exact one,  particularly for $\sim d - \theta > {\frac{1}{5}} $.  We have constructed three measures of the quantum entanglement as we mentioned from gravity,  namely Holographic Entanglement of Entropy,   Holographic Mutual Information,  and the holographic dual of the Entanglement of Purification which is given by Entanglement Wedge Cross Section.   Next both from the field theory point of view and from the dual gravity theory we have somewhat intutively,  without any 
explicit calculation, have argued about the expected properties of these three measures of the quantum entanglement where given the fact that our expression of the turning point $\rho_0(l, \rho_c)$ is exact in the regime $l >> \rho_c$ and $\rho_c >> l$ and even in the rest of the regime iof 2D $(l,\rho_c)$ 
 is close to the exact one, one can expect that these measures of the quantum entanglement which are globally defined over 2D $(l, \rho_c)$ plane(basically the first quadrant of it),  should show all these expected properties over the complete regime!  Indeed we have found,  each and every expected properties,  including the expected fall of these quantum-measures along with the increase of the cut-off $\rho_c$ ( or equivalently the increase in  $T{\overline{T}}$  deformation coefficient $\mu$), for two disjoint boundary intervals in the form of the strips, the vanishing of  UV divergence of HMI and EWCS at zero separation in the presence of nonzero cut off and the vanishing of UV divergence of EE in the presence of nonzero cut off,  the expected fall of HMI and EWCS with the increase of the separation between these two disjoint intervals even in the presence of nonzero cut off and their eventual phase transition from the connected to disconnected  phase,  all holds.
Also  EWCS satisfies an inequality,  t.e, it is always  $ EWCS  \ge {\frac{1}{2}} HMI $, and this inequality saturates only when the system is in pure state,   Given that, we can expect that in $ l >> \rho_c$ this inequality will hold prominently and more we move towards $\rho_c >> l$ regime these two will be closer and eventually at $\rho_c >> l$,  they two will merge i.e this inequality will saturates because of the fact that at  $\rho_c >> l$ all these quantum measures are falling towards zero,   indicating that the system is in pure state!   Most interestingly,  we have found from our 3D plots  of these quantum measures vs $(l,\rho_c)$,  that this is indeed the case!  Finally, since the expression of HEE is too complicated,  cannot really give much insight about the quantum entanglement of the physical system!,  so here we have provided the most simplified expression of HEE  in $ l >> \rho_c$ and  $\rho_c >> l$ and have shown with 3D plots that these two can exactly merge with the exact expression of HEE over the respective regime!  We have further shown that all these quantum-measures increases with the decrease of $d - \theta$!  These all proves the validity of our global expression of the turning point and consequently the global expression of all these measures of quantum-entanglement over 2D$(l,\rho_c)$ plane in gravity side and consequently over 2D parameter regime $(\mu,l)$ in the boundary theory!   We have also discussed the impact of the global symmetry structure on all these quantum measures and also discussed about the possible spacetime origin of such  symmetry, although subjected to further study!
\vskip0.5mm
As the future course of work, we must need to evaluate all the field theory dual of all these quantum measures and compare them with the one as we have obtained from gravity! It is also interesting to study the origin of the global symmetry structure in more depth., as we have  found here!  Next,  since we have mentioned that all these quantum-measures  are the expression of the respective measure in the leading order in  $G_N$  corresponds to which in the dual boundary theory,  if it is a large N gauge theory,  corresponds to $O(N^2)$ term it is interesting to study the first order quantum correction and to see whether the global symmetry as we have found here, still holds!   It is also interesting to study these quantum measures in time-independent cases and that way to study the covariant generalization of our theory! 
Finally it is also interesting to study the finite temperature theory and the duality aspect which is actually our work in progress and we hope to report on these
 issues in the near future!

\section*{Acknowledgment}

I  would like to thank Mohsen Alishahiha very much for suggesting the original research problem where the present project is the part of the same and also for his most  valuable suggestions  and comments during the course of this work!  I am really very much indebted to him.  My sincere thanks to  Salomeh-Khoeini-Moghaddam and  Mohd. Reza Tanhayi for collaboration at the early stage of the project.   I am very much greatly indebted to Kalyana Rama for very useful and extensive discussion during the course of the work.  I would like to express my very much deep sense of gratitude to Partha Mukhopadhyay for all his kind help for which I could pursue this research.   I would also like to express my deep sense of gratitude to my authorities, HEP group, IMSC,  for all their kind  help,   which enabled me to do this research.  This work is supported by DAE research fellowship,  India.

\appendix{\noindent {}}
\section{Appendix}
\setcounter{equation}{0}
\subsection{ Determination of the global solution of the turning point from the speculative approach}

Here we continue our analysis based on speculation as we started in section 2.2, proceed with the expressions of turning point (\ref{firstexpectation}, \ref{secondexpectation}).   Next we consider (\ref{tanmoy}).  In order to write it in terms of incomplete $\beta$ function , we consider the following

 \ber
& &{{}_2 F_1}   \left\lbrack {\frac{1}{2}}, {\frac{1}{2}}(1 +{ \frac{1}{d - \theta}}), {\frac{1}{2}}(3 +{ \frac{1}{d - \theta}}) , \left({\frac{\rho_c}{\rho_0}}\right)^{2(d - \theta)}\right\rbrack\n
&=& \displaystyle\sum_{k = 0}^\infty {\frac{\left( {\frac{1}{2}}\right)_k  \left( {\frac{1}{2}}(1 +{ \frac{1}{d - \theta}})\right)_k }{\left({\frac{1}{2}}(3 +{ \frac{1}{d - \theta}})\right)_k  k!}}   \left({\frac{\rho_c}{\rho_0}}\right)^{2 k (d - \theta)}
 \la{hypergeometricfirst}
 \eer    
with
\be
 (a)_n = 1, \quad  \quad {\rm for} \quad   n=0 \quad  \quad; \quad  \quad (a)_n = a(a+1)(a+2).......(a +n -1) ,\quad  \quad {\rm for} \quad n > 0\,  ,
 \la{hypergeometriccondition}
 \ee

 Now following (\ref{hypergeometriccondition})  we have
 \ber
{\frac{\left( {\frac{1}{2}}(1 +{ \frac{1}{d - \theta}})\right)_k}{\left({\frac{1}{2}}(3 +{ \frac{1}{d - \theta}})\right)_k }}
 &=& {\frac{{\frac{1}{2}}(1 + \left({ \frac{1}{d - \theta}} \right)}{ {\frac{1}{2}}(1 + (\left({ \frac{1}{d - \theta}}\right) + k }}                                                                                                                                                                                                                                                                                                                                                               
 \la{dingdong}
 \eer
 
 Substituting (\ref{dingdong})  in (\ref{tanmoy}), we get
 \newpage

\ber
{\frac{l}{2}} &=& -  {\frac{ \rho_0  \,\, {{}_2 F_1}   \left\lbrack {\frac{1}{2}}, {\frac{1}{2}}(1 +{ \frac{1}{d - \theta }}), {\frac{1}{2}}(3 +{ \frac{1}{d - \theta }}) , 1 \right \rbrack }{  \theta - d  - 1   }}\n
                      &+& \left({\frac{\rho_c}{\rho_0}}\right)^{d - \theta +1} {\frac{ \rho_0  \,\, {{}_2 F_1}   \left\lbrack {\frac{1}{2}}, {\frac{1}{2}}(1 +{ \frac{1}{d - \theta}}), {\frac{1}{2}}(3 +{ \frac{1}{d - \theta}}) , \left({\frac{\rho_c}{\rho_0}}\right)^{2(d - \theta)}  \right \rbrack }{  \theta - d  - 1   }} \n
&=& -  \left(  1 \right)^{d - \theta +1} {\frac{ \rho_0 }{\theta - d  - 1 }} \displaystyle\sum_{k = 0}^\infty{\frac{ \left( {\frac{1}{2}}\right)_k  \left( {\frac{1}{2}}(1 +{ \frac{1}{d - \theta}})\right)_k }{\left({\frac{1}{2}}(3 +{ \frac{1}{d - \theta}})\right)_k  k!}}   \left(1 \right)^{2 k (d - \theta)}
   \n
 &+& \left({\frac{\rho_c}{\rho_0}}\right)^{d - \theta +1} {\frac{ \rho_0}{ \theta - d  - 1  }}   \displaystyle\sum_{k = 0}^\infty {\frac{\left( {\frac{1}{2}}\right)_k  \left( {\frac{1}{2}}(1 +{ \frac{1}{d - \theta}})\right)_k }{\left({\frac{1}{2}}(3 +{ \frac{1}{d - \theta}})\right)_k  k!}}   \left({\frac{\rho_c}{\rho_0}}\right)^{2 k (d - \theta)}\n 
 &=& -  \left(  1 \right)^{d - \theta +1} {\frac{ \rho_0 }{\theta - d  - 1 }} \displaystyle\sum_{k = 0}^\infty{\frac{ \left( {\frac{1}{2}}\right)_k  \left( {\frac{1}{2}}(1 +{ \frac{1}{d - \theta}})\right)_k }{\left({\frac{1}{2}}(3 +{ \frac{1}{d - \theta}})\right)_k  k!}}   \left(1 \right)^{2 k (d - \theta)}
   \n
 &+& \left({\frac{\rho_c}{\rho_0}}\right)^{d - \theta +1} {\frac{ \rho_0}{ \theta - d  - 1  }}   \displaystyle\sum_{k = 0}^\infty {\frac{\left( {\frac{1}{2}}\right)_k  \left( {\frac{1}{2}}(1 +{ \frac{1}{d - \theta}})\right)_k }{\left({\frac{1}{2}}(3 +{ \frac{1}{d - \theta}})\right)_k  k!}}   \left({\frac{\rho_c}{\rho_0}}\right)^{2 k (d - \theta)}\n 
 &=& -  \left(  1 \right)^{d - \theta +1} {\frac{ \rho_0 }{\theta - d  - 1 }} \displaystyle\sum_{k = 0}^\infty{\frac{ \left( {\frac{1}{2}}\right)_k   {\frac{1}{2}}\left(1 +{ \frac{1}{d - \theta}}\right)   }{\left({\frac{1}{2}}(1 + { \frac{1}{d - \theta}}) + k\right) k!}}   \left(1 \right)^{2 k (d - \theta)}
   \n
 &+& \left({\frac{\rho_c}{\rho_0}}\right)^{d - \theta +1} {\frac{ \rho_0}{ \theta - d  - 1  }}   \displaystyle\sum_{k = 0}^\infty {\frac{\left( {\frac{1}{2}}\right)_k  {\frac{1}{2}}\left(1 +{ \frac{1}{d - \theta}}\right) }{  \left({\frac{1}{2}}(1 + { \frac{1}{d - \theta}}) + k\right)  k!}}   \left({\frac{\rho_c}{\rho_0}}\right)^{2 k (d - \theta)}\n 
&=&  \left(  1 \right)^{2(d - \theta). {\frac{1}{2}}\left(1 +{ \frac{1}{d - \theta}}\right)  } {\frac{ \rho_0 }{2(d - \theta)  }} \displaystyle\sum_{k = 0}^\infty{\frac{ \left( {\frac{1}{2}}\right)_k    }{\left({\frac{1}{2}}(1 + { \frac{1}{d - \theta}}) + k\right) k!}}  \left\lbrace \left(1 \right)^{2  (d - \theta)}\right\rbrace^k  \n
 &-& \left({\frac{\rho_c}{\rho_0}}\right)^{2(d - \theta).{\frac{1}{2}}\left(1 +{ \frac{1}{d - \theta}}\right) } {\frac{ \rho_0}{2(d - \theta)   }}   \displaystyle\sum_{k = 0}^\infty {\frac{\left( {\frac{1}{2}}\right)_k  }{  \left({\frac{1}{2}}(1 + { \frac{1}{d - \theta}}) + k\right)  k!}}   \left\lbrace\left({\frac{\rho_c}{\rho_0}}\right)^{2  (d - \theta)}\right\rbrace^k\n 
 &=&   {\frac{ \rho_0 }{2(d - \theta)  }} \left\lbrack B\left( \left(  1 \right)^{2(d - \theta)}; \,{\frac{1}{2}}\left(1 +{ \frac{1}{d - \theta}}\right)\, , {\frac{1}{2}}  \right) \, - \, B\left( \left( {\frac{\rho_c}{\rho_0}}  \right)^{2(d - \theta)}; \,{\frac{1}{2}}\left(1 +{ \frac{1}{d - \theta}} \right)\, ,  {\frac{1}{2}}  \right) \right\rbrack\, , \n         
\la{sandesh}
\eer

We substitute (\ref{sandesh}) in (\ref{tanmoy}) so that the governing equation for $\rho_0 (l,\rho_c)$ can be reexpressed as

\ber
{\frac{A_{10} l}{2}}  + {\frac{A_{10}  \rho_0 }{2(d - \theta)  }} B\left( \left( {\frac{\rho_c}{\rho_0}}  \right)^{2(d - \theta)}; \,{\frac{1}{2}}\left(1 +{ \frac{1}{d - \theta}} \right) , {\frac{1}{2}}  \right)
  &=&   {\frac{A_{10} \rho_0 }{2(d - \theta)  }} B\left( \left(  1 \right)^{2(d - \theta)}; \,{\frac{1}{2}}\left(1 +{ \frac{1}{d - \theta}}\right)\, , {\frac{1}{2}}  \right) \,    \n         
\la{betarelation}
\eer

Now we first consider the case $d - \theta > 1$ where we are going to show  (\ref{firstexpectation}) will hold for $l >> \rho_c$ and for $\rho_c >> l$ region 

To proceed, first we consider 2nd term of the l.h.s of the above equation
 \ber
& & {\frac{A_{10} \rho_0 }{2(d - \theta)  }} B\left( \left( {\frac{\rho_c}{\rho_0}}  \right)^{2(d - \theta)}; \,{\frac{1}{2}}\left(1 +{ \frac{1}{d - \theta}} \right) , {\frac{1}{2}}  \right)\n
 &=& {\frac{A_{10} \rho_0 }{2(d - \theta)  }}  \int_0^{\left({\frac{\rho_c}{\rho_0}}\right)^{2(d-\theta)}} dt \,\,  t^{\left\lbrace{\frac{1}{2}}\left(1 +{ \frac{1}{(d - \theta)}} \right) -1\right\rbrace} \left(1 -t\right)^{-\frac{1}{2}};\quad \quad {\rm with}\quad{\rm substitution}\quad (t)^{\frac{1}{2}} = u \n
 &=&{\frac{ 2A_{10}\rho_0 }{2(d - \theta)  }}\int_0^{\left({\frac{\rho_c}{\rho_0}}\right)^{(d-\theta)}} \,\, du\, u\,u^{\left\lbrace\left(1 +{ \frac{1}{(d - \theta)}} \right) -2\right\rbrace} \left(1 - u^2\right)^{-\frac{1}{2}};\quad \quad {\rm with}\quad{\rm substitution}\quad  u = \sin\theta_1 \n
 &=&  {\frac{ 2A_{10} \rho_0 }{2(d - \theta)  }}  \int_0^{\left({\frac{\rho_c}{\rho_0}}\right)^{(d-\theta)}} d  \left({\sin} \theta_1 \right)\,\,  \left({\sin} \theta_1 \right)^{\left\lbrace\left(1 +{ \frac{1}{d - \theta}} \right) -2\right\rbrace}.\sin\theta_1 {\frac{1}{{\cos}\theta_1}}\n
 &=& {\frac{A_{10} \rho_0 }{(d - \theta)  }}  \int_0^{\left({\frac{\rho_c}{\rho_0}}\right)^{(d-\theta)}} d  \left({\sin} \theta_1 \right)\,\,  \left({\sin} \theta_1 \right)^{\left\lbrace\left({ \frac{1}{d - \theta}} \right) \right\rbrace}.  {\frac{1}{{\cos}\theta_1}}\n
  \la{beta2nd}
 \eer

 Next consider strip width l as another term in l.h.s in (\ref {betarelation}) , when we can write
\ber
A_{10}{\frac{l}{2}} &=& \int_0^{\frac{A_{10}l}{2}} dl_1  ;\quad \quad {\rm with}\quad{\rm substitution}\quad l_1 = \rho_0 \left\lbrace\cos\theta_2\right\rbrace^{\frac{1}{d - \theta}}\n
{\frac{A_{10} l}{2}} &=& {\frac{\rho_0}{ d -\theta}}  \int_{ 0}^{{\left\lbrace {\frac{A_{10} l}{2\rho_0}}\right\rbrace}^{d-\theta}} \left( \cos\theta_2\right)^{\left\lbrace{\frac{1}{d - \theta}}\right\rbrace - 1} d(\cos\theta_2)\n
{\frac{A_{10} l}{2}} &=& {\frac{\rho_0}{ d -\theta}}  \int_{ 0}^{{\left\lbrace {\frac{A_{1`0} l}{2\rho_0}}\right\rbrace}^{d-\theta}} \left( \cos\theta_2\right)^{\left\lbrace{\frac{1}{d - \theta}}\right\rbrace }  {\frac{1}{{\cos}\theta_2}}d(\cos\theta_2)\n
\la{lrelation}
\eer

Now while one can always substitute (\ref{beta2nd}) and (\ref{lrelation}) in l.h.s of (\ref{betarelation}), while the r.h.s of the (\ref{betarelation})
is given by, incomplete $\beta$ function with argument 1, i.e ordinary $\beta$ function which can be expressed in the same way as in (\ref{beta2nd}) with upper limit replaced by one
\ber
& &  {\frac{A_{10} \rho_0 }{2(d - \theta)  }} B\left( \left(  1 \right)^{2(d - \theta)}; \,{\frac{1}{2}}\left(1 +{ \frac{1}{d - \theta}}\right)\, , {\frac{1}{2}}  \right) \n
  &=& {\frac{A_{10} \rho_0 }{2(d - \theta)  }}  \int_0^1 dt \,\,  t^{\left\lbrace{\frac{1}{2}}\left(1 +{ \frac{1}{(d - \theta)}} \right) -1\right\rbrace} \left(1 -t\right)^{-\frac{1}{2}};\quad \quad {\rm with}\quad{\rm substitution}\quad (t)^{\frac{1}{2}} = u \n
 &=& {\frac{A_{10} \rho_0 }{(d - \theta)  }}  \int_0^1 \,\, du\, u\,u^{\left\lbrace\left(1 +{ \frac{1}{(d - \theta_1)}} \right) -2\right\rbrace} \left(1 - u^2\right)^{-\frac{1}{2}};\quad \quad {\rm with}\quad{\rm substitution}\quad  u = \sin\theta_1 \n
&=& {\frac{ A_{10}\rho_0 }{(d - \theta)  }}  \int_0^1 d  \left({\sin} \theta_1 \right)\,\,  \left({\sin} \theta_1 \right)^{\left\lbrace\left(1 +{ \frac{1}{d - \theta}} \right) -2\right\rbrace}.\sin\theta_1 {\frac{1}{{\cos}\theta_1}}\n
 &=& {\frac{ A_{10}\rho_0 }{(d - \theta)  }}  \int_0^1 d  \left({\sin} \theta_1 \right)\,\,  \left({\sin} \theta_1 \right)^{\left\lbrace\left({ \frac{1}{d - \theta}} \right) \right\rbrace}.  {\frac{1}{{\cos}\theta_1}}\n
 &=&  {\frac{A_{10} \rho_0 }{(d - \theta)  }}  \int_0^{\left({\frac{\rho_c}{\rho_0}}\right)^{(d-\theta)}}  d  \left({\sin} \theta_1 \right)\,\,  \left({\sin} \theta_1 \right)^{\left\lbrace\left({ \frac{1}{d - \theta}} \right) \right\rbrace}.  {\frac{1}{{\cos}\theta_1}}\n
&+& {\frac{A_{10} \rho_0 }{(d - \theta)  }}  \int_{\left({\frac{\rho_c}{\rho_0}}\right)^{(d-\theta)}}^1  d  \left({\sin} \theta_1 \right)\,\,  \left({\sin} \theta_1 \right)^{\left\lbrace\left({ \frac{1}{d - \theta}} \right) \right\rbrace}.  {\frac{1}{{\cos}\theta_1}}
\la{r.h.s}
\eer

Now substituting (\ref{beta2nd}) and  (\ref{lrelation}) in l.h.s of  (\ref{betarelation}) and (\ref{r.h.s}) in r.h.s of  (\ref{betarelation})  we get

\ber
 & & {\frac{A_{10} \rho_0 }{(d - \theta)  }}  \int_0^{\left({\frac{\rho_c}{\rho_0}}\right)^{(d-\theta)}} d  \left({\sin} \theta_1 \right)\,\,  \left({\sin} \theta_1 \right)^{\left\lbrace\left({ \frac{1}{d - \theta}} \right) \right\rbrace}.  {\frac{1}{{\cos}\theta_1}}\n
 &+& {\frac{\rho_0}{ d -\theta}}  \int_{ 0}^{{\left\lbrace {\frac{A_{10} l}{2\rho_0}}\right\rbrace}^{d-\theta}} \left( \cos\theta_2\right)^{\left\lbrace{\frac{1}{d - \theta}}\right\rbrace }  {\frac{1}{{\cos}\theta_2}}d(\cos\theta_2)\n
 &=& {\frac{A_{10} \rho_0 }{(d - \theta)  }}  \int_0^{\left({\frac{\rho_c}{\rho_0}}\right)^{(d-\theta)}}  d  \left({\sin} \theta_1 \right)\,\,  \left({\sin} \theta_1 \right)^{\left\lbrace\left({ \frac{1}{d - \theta}} \right) \right\rbrace}.  {\frac{1}{{\cos}\theta_1}}\n
&+& {\frac{A_{10} \rho_0 }{(d - \theta)  }}  \int_{\left({\frac{\rho_c}{\rho_0}}\right)^{(d-\theta)}}^1  d  \left({\sin} \theta_1 \right)\,\,  \left({\sin} \theta_1 \right)^{\left\lbrace\left({ \frac{1}{d - \theta}} \right) \right\rbrace}.  {\frac{1}{{\cos}\theta_1}}
\la{todefine}
\eer

Now in the above  (\ref {todefine}) while the first integral is trivially equal both side, we need to equate the 2nd integral in both l.h.s and r.h.s, which is

\ber
& & {\frac{\rho_0}{ d -\theta}}  \int_{ 0}^{{\left\lbrace {\frac{A_{10} l}{2\rho_0}}\right\rbrace}^{d-\theta}} \left( \cos\theta_2\right)^{\left\lbrace{\frac{1}{d - \theta}}\right\rbrace }  {\frac{1}{{\cos}\theta_2}}d(\cos\theta_2)\n
 &=& {\frac{A_{10} \rho_0 }{(d - \theta)  }}  \int_{\left({\frac{\rho_c}{\rho_0}}\right)^{(d-\theta)}}^1  d  \left({\sin} \theta_1 \right)\,\,  \left({\sin} \theta_1 \right)^{\left\lbrace\left({ \frac{1}{d - \theta}} \right) \right\rbrace}.  {\frac{1}{{\cos}\theta_1}}
\la{2ndintegral}
\eer

Now to compare the limit of the integrals both side, we must need to transfer the variable in the integral in r.h.s from $\sin\theta_1$ to $\cos\theta_1$ which redefines the above relation as

\ber
& &  \int_{ 0}^{{\left\lbrace {\frac{A_{10} l}{2\rho_0}}\right\rbrace}^{d-\theta}}\left\lbrack \left( \cos\theta_2\right)^{\left\lbrace{\frac{1}{d - \theta}} - 1\right\rbrace }\right\rbrack  d(\cos\theta_2) \n
&=& A_{10} \int^{\sqrt{1 - \left\lbrace{\left({\frac{\rho_c}{\rho_0}}\right)^{(d-\theta)}}\right\rbrace^2}}_{ 0} d  \left({\cos\theta_1}\right) \left\lbrack \left({\sin} \theta_1 \right)^{\left\lbrace\left({ \frac{1}{d - \theta}} \right)  - 1\right\rbrace}.\right\rbrack
\la{new}
\eer

So the above required equality in (\ref{new}) is clearly showing that there is no simple relation exist between the upper limits of the integral that is between ${\left({\frac{\rho_c}{\rho_0}}\right)}^{(d-\theta)}$ and ${\left({\frac{A_{10} l}{2\rho_0}}\right)}^{d-\theta}$, for finite $\rho_c$ finite l except the trivial case with $d-\theta =1$. So, to proceed with, we consider the case $ l >> \rho_c$ and $\rho_c >> l$.

\subsection{Considering the limit $l >> \rho_c$ and $\rho_c >> l$}

First consider the exact point, $\rho_c = 0\, ,   {\frac{A_{10} l}{2\rho_0}} = 1$
That is at this point we exactly have  the upper limit of both the integral  in l.h.s  and r.h.s  in (\ref{new})  given by an equality

\be
{\sqrt{1 - \left\lbrace{\left({\frac{\rho_c}{\rho_0}}\right)^{(d-\theta)}}\right\rbrace^2}} = \left\lbrace {\frac{A_{10} l}{2\rho_0}}\right\rbrace^{(d - \theta)} = 1
\la{first point}
\ee

We verify, both the l.h.s and r.h.s of (\ref{new}) are equal at this limit

\ber
{\rm l.h.s} &=& \int_{ 0}^{1}\left\lbrack \left( \cos\theta_2\right)^{\left\lbrace{\frac{1}{d - \theta}} - 1\right\rbrace }\right\rbrack  d(\cos\theta_2) = d -\theta\n
{\rm r.h.s} &=& A_{10} \int_{0}^1  d  \left({\cos\theta_1}\right) \left\lbrack \left({\sin} \theta_1 \right)^{\left\lbrace\left({ \frac{1}{d - \theta}} \right)  - 1\right\rbrace}.\right\rbrack = A_{10} . {\frac{{\sqrt{\pi}}  \Gamma\left\lbrack{\frac{1}{2}}\left( 1 + {\frac{1}{d-\theta}} \right) \right\rbrack}{2\Gamma\left\lbrack 1 + {\frac{1}{2(d-\theta)}} \right\rbrack}}\n
\Rightarrow {\rm r.h.s} &=&   \left\lbrack {\frac{\Gamma\left\lbrack{\frac{1}{2(d - \theta)}}  \right\rbrack}{{\sqrt{\pi}}\Gamma\left\lbrack {\frac{1}{2}}\left(1 + {\frac{1}{d- \theta}}\right) \right\rbrack}}\right\rbrack .   {\frac{{\sqrt{\pi}}  \Gamma\left\lbrack{\frac{1}{2}}\left( 1 + {\frac{1}{d-\theta}} \right) \right\rbrack}{2.{\frac{1}{2(d-\theta)}} .\Gamma\left\lbrack  {\frac{1}{2(d-\theta)}} \right\rbrack}}  = d - \theta  \n
\Rightarrow  {\rm l.h.s} &=& {\rm r.h.s}\,,
 \la{bothl.hg.s.r.h.s}
\eer

where we have used the expression of $A_{10}$ from (\ref{proportionalityconstantgen}) and used the $\Gamma$ function property $\Gamma(1 + x) = x\Gamma(x)$.
Next we ask the question, can we have a finite region around the extreme point ${\frac{\rho_c}{\rho_0}} = 0 \,\, , \,\, {\frac{A_{10} l}{2\rho_0}} = 1$ where the upper limit of both the integral in l.h.s and r.h.s  in (\ref{new})  are actually equal to each other,  and we also have  have l.h.s = r.h.s in (\ref{new}).   In other words, can we have a finite region around the extreme points, where  we can expect,  a solution of (\ref{new})  exist?
Let us reframe the question in the following way:

Let us define the variable
\be
{\left\lbrace {\frac{A_{10} l}{2\rho_0}}\right\rbrace}^{d-\theta} = z_1 \quad ; \quad  {\left({\frac{\rho_c}{\rho_0}}\right)^{(d-\theta)}}  = z_2
\la{definevariable}
\ee

In this variable, equating  the upper limit of both r.h.s and l.h.s in (\ref{new}) imples a circle
\be
z_1^{2(d - \theta)} + z_2^{2(d - \theta)} = 1\quad; \quad {\frac{dz_1}{dz_2}}  = - {\frac{z_2}{z_1}} \Rightarrow   {\frac{dz_1}{dz_2}} |_{\rho_c = 0 \,;\,, {\frac{A_{10} l}{2\rho_0}} = 1} = 0
\la{circle}
\ee

Let us write the l.h.s and r.h.s of (\ref{new}) in the form

\ber
& & f_1\left (z_1 \right) =   \int_{ 0}^{{\left\lbrace {\frac{A_{10} l}{2\rho_0}}\right\rbrace}^{d-\theta}}\left\lbrack \left( \cos\theta_2\right)^{\left\lbrace{\frac{1}{d - \theta}} - 1\right\rbrace }\right\rbrack  d(\cos\theta_2) \n
& & f_2\left(z_2 \right) =   {A_{10}} \int^{\sqrt{1 - \left\lbrace{\left({\frac{\rho_c}{\rho_0}}\right)^{(d-\theta)}}\right\rbrace^2}}_{ 0}      \n
d\left({\cos\theta_1}\right) \left\lbrack \left({\sin} \theta_1 \right)^{\left\lbrace\left({ \frac{1}{d - \theta}} \right)  - 1\right\rbrace}.\right\rbrack
& & {\rm with} \quad  z_1 = z_1(z_2)
\la{charam}
\eer

Next we define 
\be
g(z)= f_1(\sqrt{1 - z}) - f_2(\sqrt{1 - z})
\la{gz}	
\ee

So our original question is being reframed as, can we have a finte region around $z = 0$ where we have $g(z) = 0$, where exactly at $z = 1$, ,  we have found it to be zero? Here we present two plots in support of that in Fig,(\ref{gzz}). Clearly $ z = 0$ with $ z = {\frac{\rho_c}{\rho+_0}}$, denotes the regime $l>> \rho_c$.  	Clearly the other regime $\rho_c >> l$ will be denoted by $z = 1$.  The plots  in Fig,(\ref{gzz})  are showing that in both of these regimes  our speculation-based global solution for turning point, given by (\ref{firstexpectation}) holds, for $ d - \theta > 1$!

In the section 3 we have also cross checked these solutions for $\rho_0(l,\rho_c)$ from (\ref{firstexpectation})  by substituting it in (\ref{tanmoy}) which holds for the above nentioned regime for $d - \theta > 1$ but does not hold for $d - \theta < 1$.

\begin{figure}[H]
\begin{center}
\textbf{ For $ d - \theta > 1$, g(z) vs z plot showing the validity of our speculation-based global solution for the turning point $\rho_0(l,\rho_c)$ in the regime $l>> \rho_c$ and $\rho_c  >> l$ }
\end{center}
\vskip2mm
\includegraphics[width=.65\textwidth]{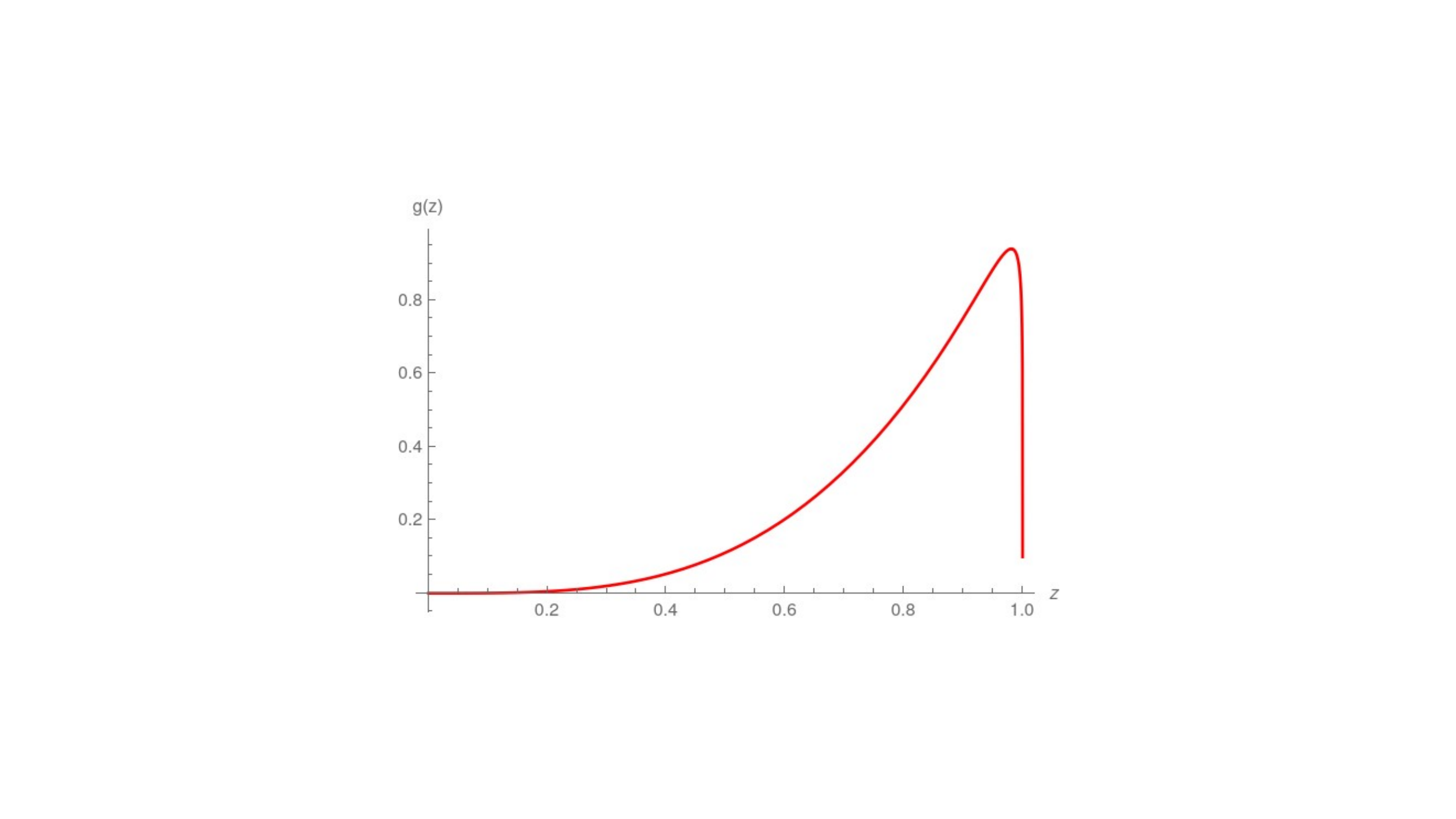}
\includegraphics[width=.65\textwidth]{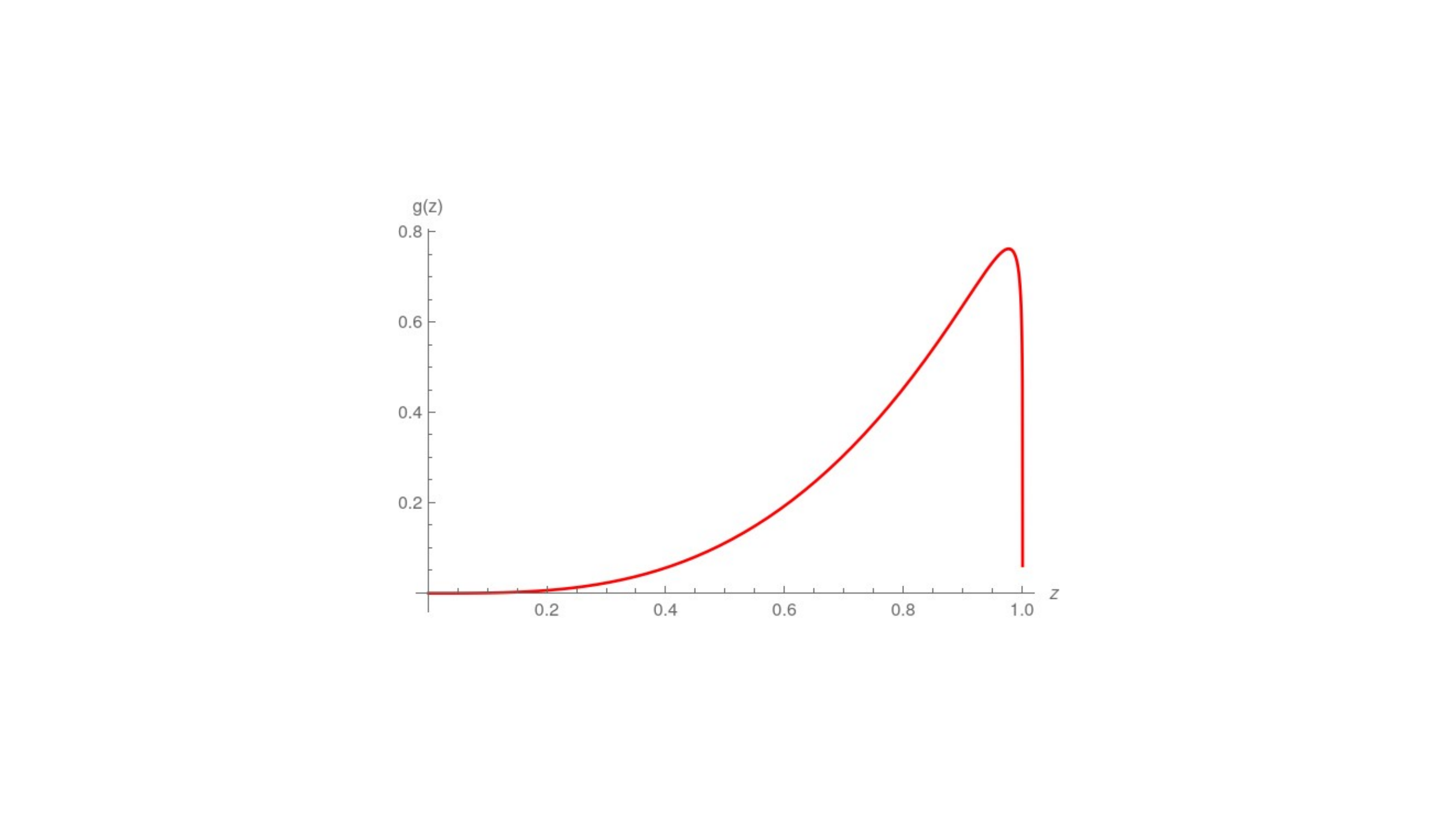}	

\caption{g(z) vs  z plot \, : \, (left) \.:\. For $ d - \theta = {\frac{5}{2}}$\, ,\, (right)\, : \, $ d - \theta = {\frac{9}{4}}$ showing there exist a finite region around $z = 0$ where our solution for $\rho_0(l,\rho_c)$ based on speculation do hold.  These plots are also showing that there exist a narrow region around z=1 where our solution for $\rho_0(l,\rho_c)$ based on speculation do hold
}
\label{gzz}
\end{figure}

We have found the other speculation-based global solution for the turning point as given in (\ref{secondexpectation}) holds for the regime $d - \theta < 1$ as we have shown  in Fig,(\ref{gzzless}).

\begin{figure}[H]
\begin{center}
\textbf{ For $ d - \theta < 1$, g(z) vs z plot showing the validity of our speculation-based global solution for the turning point $\rho_0(l,\rho_c)$ in the regime $l>> \rho_c$ and $\rho_c  >> l$ }
\end{center}
\vskip2mm
\includegraphics[width=.65\textwidth]{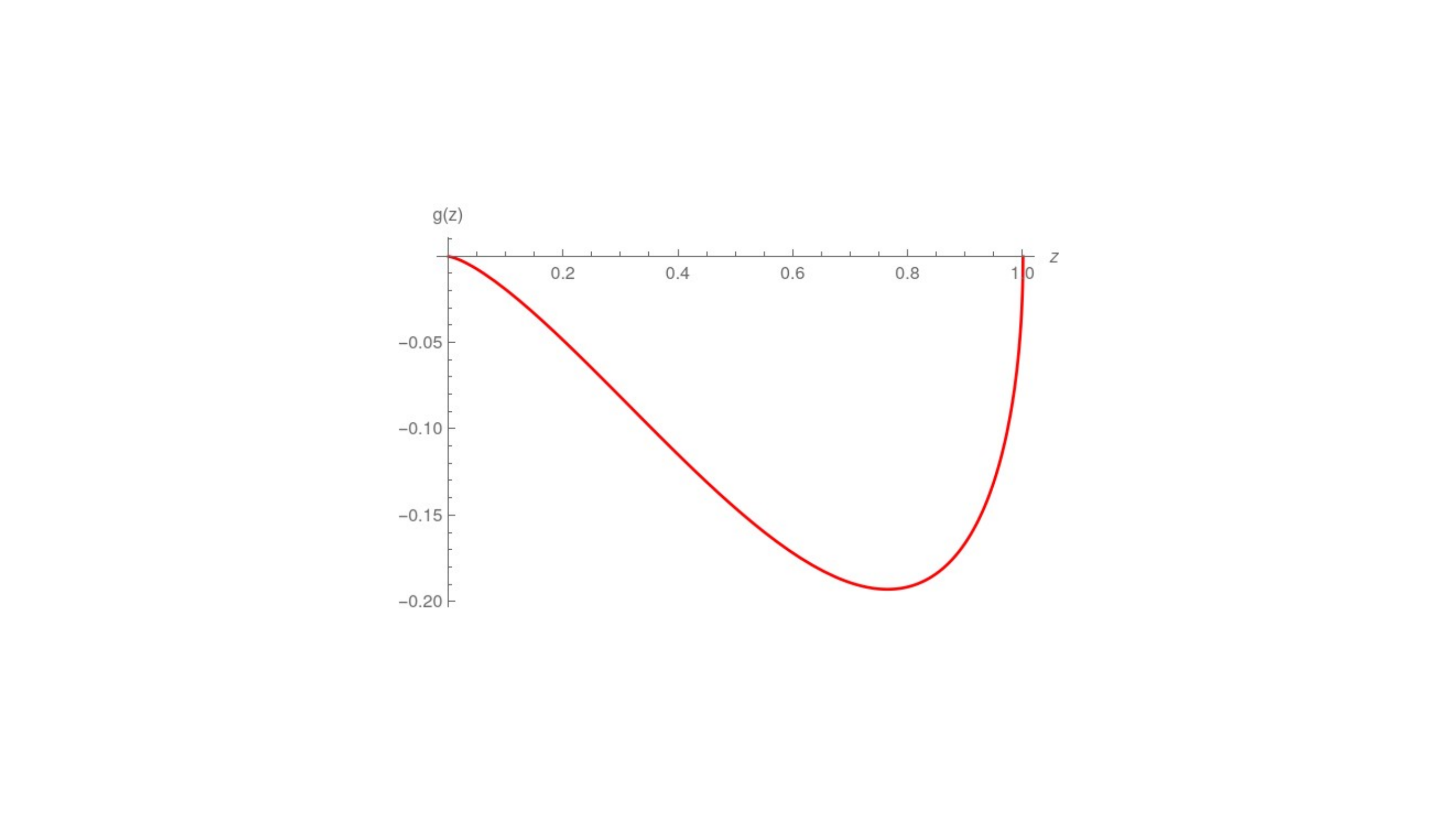}
\includegraphics[width=.65\textwidth]{gzz2by5.pdf}	

\caption{g(z) vs  z plot \, : \, (left) \.:\. For $ d - \theta = {\frac{2}{5}}$\, ,\, (right)\, : \, $ d - \theta = {\frac{4}{9}}$ showing there exist a narrow region around $z = 0$  and $z=1$ which denotes the regime $l >> \rho_c$  and $\rho_c >> l$ regime respectively where our speculation-based global solution for the turning point holds
}
\label{gzzless}
\end{figure}

\end{document}